\documentclass[a4paper,12pt]{book}

\usepackage[utf8x]{inputenc}
\usepackage[left=2cm,right=2cm,top=2cm,bottom=3cm]{geometry}

\usepackage{etoolbox}

\usepackage[linktoc=all,hidelinks]{hyperref}
\usepackage{amsmath}
\usepackage{amsfonts}
\usepackage{amssymb}
\usepackage{appendix}
\usepackage{graphicx}
\usepackage{afterpage}
\usepackage[nottoc,numbib]{tocbibind}

\usepackage{empheq,slashed}
\usepackage{verbatim,slashed}
\numberwithin{equation}{section}
\usepackage{hyperref,setspace}
\usepackage{tikz-cd}
\usepackage{mathrsfs}
\usepackage[numbers,sort&compress]{natbib}
\usepackage[nottoc]{tocbibind}

\usepackage{empheq,wrapfig}
\usepackage{dsfont}

\newcommand{\pl}{\partial}

\newcommand{\be}{\begin{align}}
\newcommand{\ee}{\end{align}}

\usepackage{enumitem}

\newcommand{\mm}{{\ensuremath{{\mu}}}}

\newcommand{\aAt}{{\ensuremath{\mathtt{A}}}}
\newcommand{\aBt}{{\ensuremath{\mathtt{B}}}}

\newcommand{\aA}{{\ensuremath{\mathcal{A}}}}
\newcommand{\aB}{{\ensuremath{\mathcal{B}}}}

\newcommand{\fud}[2]{{}^{#1}{}_{#2}\,}
\newcommand{\fdu}[2]{{}_{#1}{}^{#2}\,}
\newcommand{\fudu}[3]{{}^{#1}{}_{#2}{}^{#3}\,}

\newcommand{\fdud}[3]{{}_{#1}{}^{#2}{}_{#3}\,}

\newcommand{\bry}{{{\bar{y}}}}
\newcommand{\brz}{{{\bar{z}}}}

\newcommand{\hs}{{\mathfrak{hs}}}

\newcommand{\hhbar}{{\lambda}}
\newcommand{\tr}{{\mathrm{tr}}}
\newcommand{\pluk}{{\boldsymbol{p}}}

\DeclareMathOperator{\sign}{sign}

\newcommand{\action}[2]{{\left\langle\vphantom{#2}#1\,\right|\left.\vphantom{#1}#2\right\rangle}}
\newcommand{\scalar}[2]{{\left\langle\vphantom{#2}#1\right.;\left.\vphantom{#1}#2\right\rangle}}

\newcommand{\besubeqs}{\begin{subequations}}
\newcommand{\esubeqs}{\end{subequations}}

\usepackage{tensor}
\usepackage{todonotes}

\renewcommand{\bar}[1]{\overline{#1}}

\newtheorem{theorem}{\fbox{\color{violet}{Theorem}}}[section]

\newcommand{\TikzRect}[2]{\filldraw[color=black,fill=red]  (#1-\R,#2-\R) rectangle (#1+\R,#2+\R);}

\newcommand{\TikzRectG}[2]{\filldraw[color=black,fill=green]  (#1-\R,#2-\R) rectangle (#1+\R,#2+\R);}

\newcommand{\FIELDS}{
\begin{tikzpicture}[scale=0.7]
\tikzset{%
  >=latex, 
  inner sep=0pt,%
  outer sep=2pt,%
  mark coordinate/.style={inner sep=0pt,outer sep=0pt,minimum size=3pt,
    fill=black,circle}%
}
\def\R{0.15}
\def\mar{0.15}

\draw[->,black,thick] (0,0) -- (0,9) node[left]{$\# A$ };
\draw[->,black,thick] (0,0) -- (11,0) node[right]{$\# A'$} ;

\filldraw[color=black,fill=green]  (12,6) circle (\R);
\node[right] at (12.3,6) {one-forms, $\omega$};

\TikzRectG{12}{7}  \TikzRect{12.5}{7}  
\node[right] at (12.7,7) {zero-forms, $C$};

\filldraw[color=black,fill=green]  (4,0) circle (\R);
\filldraw[color=black,fill=green]  (3,1) circle (\R);
\filldraw[color=black,fill=green]  (2,2) circle (\R);
\filldraw[color=black,fill=green]  (1,3) circle (\R);
\filldraw[color=black,fill=green]  (0,4) circle (\R);
\draw[green,thick] (4,0) -- (0,4);

    \draw[-latex,thick] (4.5,5.5) node[right,scale=1.0]{$\omega^{A(2s-2)}$}
        to[out=-180,in=50] (+\mar, 4);
        
    \draw[-latex,thick] (4.5,8) node[right,scale=1.0]{$\Psi^{A(2s)}$, $h=-s$ Weyl tensor}
        to[out=-120,in=10] (+\mar,5);

    \draw[-latex,thick] (8.5,1.5) node[right,scale=1.0]{$C^{A'(2s)}$, $h=+s$ Weyl tensor}
        to[out=180,in=-45] (5, -\mar);
        
    \draw[rounded corners] ( -0.5 , 3.5 ) rectangle (0.5, 5.5) {};        

    \draw[-latex,thick] (2,-1) node[left,scale=1.0]{$\omega^{A'(2s-2)}$}
        to[out=0,in=-90] (4, -\mar);

    \TikzRect{0}{5}
    \TikzRect{1}{6}
    \TikzRect{2}{7}
    \TikzRect{3}{8}
    \draw[red,thick] (0,5) -- (4,9);

    \TikzRectG{5}{0}
    \TikzRectG{6}{1}
    \TikzRectG{7}{2}
    \TikzRectG{8}{3}
    \draw[green,thick] (5,0) -- (9,4);

    \filldraw[color=black,fill=black]  (4.5-1.5*\R,-\R) rectangle (4.5+1.5*\R,+\R);

    \draw[-latex,thick] (4.5,3.5) node[above,scale=1.0]{\large $\mathcal{V}(e,e,C)$-cocycle}
        to[out=-90,in=90] (4.5, 0.5-\mar);

\end{tikzpicture}}

\newcommand{\firstCircle}{%
  \tikz[
  baseline=-0.5ex,
    text height=1.5ex,
    text depth=.25ex
    ]{
  \begin{scope}[rotate=-60]
  \draw[thick] (0,0) circle (0.6cm);
  \coordinate [label=above right: { $\,\,$}] (B) at (90:0.55);
  \end{scope}}%
}

\newcommand{\secondCircle}{%
  \raisebox{0.2ex}{\tikz[
  baseline=-0.5ex,
    text height=1.5ex,
    text depth=.25ex
    ]{\begin{scope}[rotate=-60]
  \draw[thick] (0,0) circle (0.6cm);
    \coordinate [label=above left: { $a$}] (B) at (213:0.57);
    \coordinate [label=above right: { $c$}] (B) at (90:0.45);
    \coordinate [label=below: { $b$}] (B) at (-30:0.6);
    \draw[ultra thick, blue ] (210:0.6) -- (0,0);
    \draw[ultra thick, blue ] (-30:0.6) -- (0,0);
    \draw[ultra thick, blue ] (90:0.6) -- (0,0);
    \end{scope}}}%
}

\newcommand{\thirdCircle}{%
  \tikz[baseline=-0.6ex]{\begin{scope}[rotate=-60]
  \draw[thick] (0,0) circle (0.6cm);
    \draw[-][red, thin](60:0.6) .. controls (70: 0.49)..  (90:0.45); 
    \draw[-][red, thin](120:0.6) .. controls (115: 0.4)..  (90:0.3); 
    \draw[-][red, thin](30:0.6) .. controls (30: 0.45)..  (90:0.15); 
    \draw[-][red, thin](240:0.6) .. controls (230: 0.49)..  (210:0.45); 
    \draw[-][red, thin](210:0.3) .. controls (150: 0.5)..  (150:0.6); 
    \draw[-][red, thin](210:0.15) .. controls (260: 0.45).. (260:0.6);
    \draw[-][red, thin](-30:0.3) .. controls (-67: 0.37)..  (-80:0.6);  
    \draw[-][red, thin](-30:0.45) .. controls (-10: 0.5)..  (0:0.6); 
    \coordinate [label=above left: { $a$}] (B) at (213:0.57);
    \coordinate [label=above right: { $c$}] (B) at (90:0.45);
    \coordinate [label=below: { $b$}] (B) at (-30:0.6);
    \draw[ultra thick, blue ] (210:0.6) -- (0,0);
    \draw[ultra thick, blue ] (-30:0.6) -- (0,0);
    \draw[ultra thick, blue ] (90:0.6) -- (0,0);
    \end{scope}}%
}

\newcommand{\fourthCircle}{%
  \tikz[baseline=-0.6ex]{\begin{scope}[rotate=-60]
  \draw[thick] (0,0) circle (0.6cm);
  \draw[-latex] (150:0.6) -- (150:0.85);
    \draw[-][red, thin](60:0.6) .. controls (70: 0.49)..  (90:0.45); 
    \draw[-][red, thin](120:0.6) .. controls (115: 0.4)..  (90:0.3); 
    \draw[-][red, thin](30:0.6) .. controls (30: 0.45)..  (90:0.15); 
    \draw[-][red, thin](240:0.6) .. controls (230: 0.49)..  (210:0.45); 
    \draw[-][red, thin](210:0.3) .. controls (150: 0.5)..  (150:0.6); 
    \draw[-][red, thin](210:0.15) .. controls (260: 0.45).. (260:0.6);
    \draw[-][red, thin](-30:0.3) .. controls (-67: 0.37)..  (-80:0.6);  
    \draw[-][red, thin](-30:0.45) .. controls (-10: 0.5)..  (0:0.6); 
    \coordinate [label=above left: { $a$}] (B) at (213:0.57);
    \coordinate [label=above right: { $c$}] (B) at (90:0.45);
    \coordinate [label=below: { $b$}] (B) at (-30:0.6);
    \draw[ultra thick, blue ] (210:0.6) -- (0,0);
    \draw[ultra thick, blue ] (-30:0.6) -- (0,0);
    \draw[ultra thick, blue ] (90:0.6) -- (0,0);
    \end{scope}}%
}

\newcommand{\fifthCircle}{%
  \tikz[baseline=-0.6ex]{\begin{scope}[rotate=-60]
  \draw[thick] (0,0) circle (0.6cm);
  \draw[-latex] (150:0.6) -- (150:0.85);
    \draw[-][red, thin](60:0.6) .. controls (70: 0.49)..  (90:0.45); 
    \draw[-][red, thin](120:0.6) .. controls (115: 0.4)..  (90:0.3); 
    \draw[-][red, thin](30:0.6) .. controls (30: 0.45)..  (90:0.15); 
    \draw[-][red, thin](240:0.6) .. controls (230: 0.49)..  (210:0.45); 
    \draw[-][red, thin](210:0.3) .. controls (150: 0.5)..  (150:0.6); 
    \draw[-][red, thin](210:0.15) .. controls (260: 0.45).. (260:0.6);
    \draw[-][red, thin](-30:0.3) .. controls (-67: 0.37)..  (-80:0.6);  
    \draw[-][red, thin](-30:0.45) .. controls (-10: 0.5)..  (0:0.6); 
    \coordinate [label=above left: { $a$}] (B) at (213:0.57);
    \coordinate [label=above right: { $c$}] (B) at (90:0.45);
    \coordinate [label=below: { $b$}] (B) at (-30:0.6);
    \draw[ultra thick, blue ] (210:0.6) -- (0,0);
    \draw[ultra thick, blue ] (-30:0.6) -- (0,0);
    \draw[ultra thick, blue ] (90:0.6) -- (0,0);
    \coordinate [label=left: { $\delta_1$}] (B) at (279:0.5);
    \coordinate [label=below right: { $\gamma_3$}] (B) at (30:0.45);
    \coordinate [label=right: { $\gamma_1$}] (B) at (60:0.45);
    \coordinate [label=above right: { $\gamma_2$}] (B) at (140:0.44);
    \coordinate [label=below right: { $\beta_1$}] (B) at (-15:0.45);
    \coordinate [label=left: { $\alpha_1$}] (B) at (240:0.52);
    \coordinate [label=below left: { $\beta_2$}] (B) at (-60:0.45);
    \coordinate [label=above left: { $\alpha_2$}] (B) at (148:0.55);
    \end{scope}}%
}

\newtheorem{lemma}[theorem]{Lemma}

\newtheorem{definition}[theorem]{Definition}

\usepackage[all,cmtip]{xy}

\numberwithin{equation}{section}

\begin{document}

\thispagestyle{empty}

\includegraphics[width=30mm]{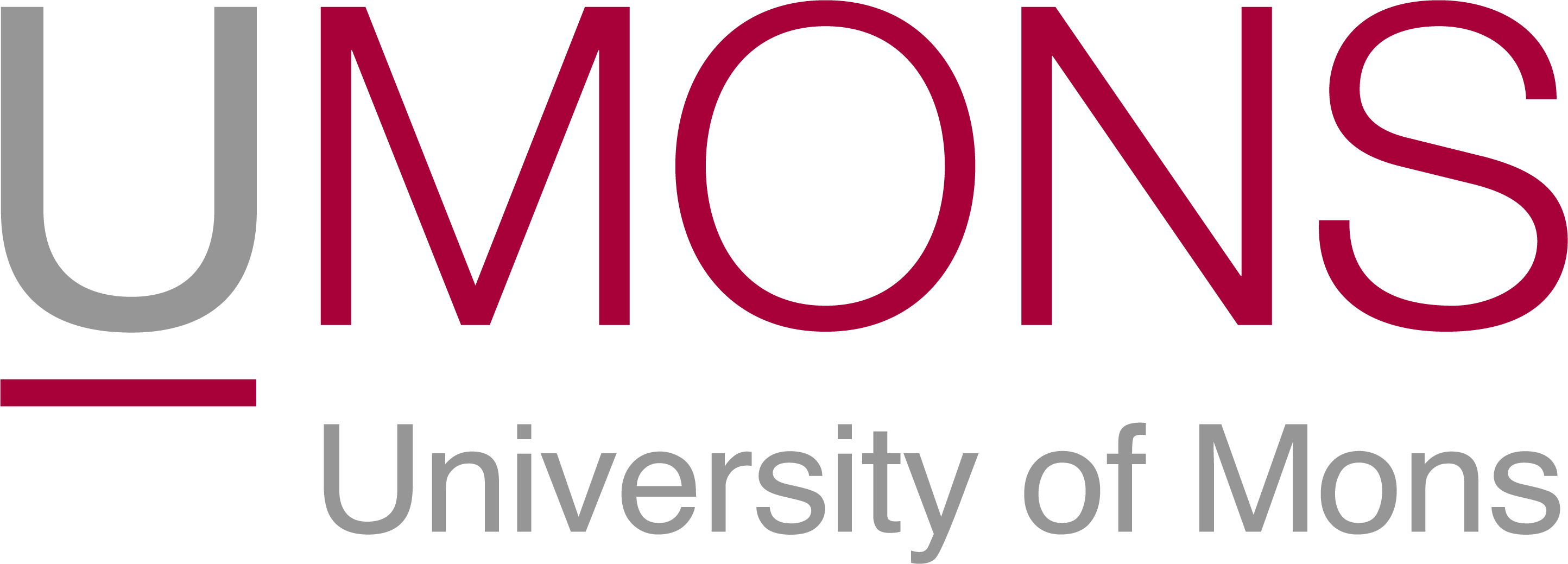}
\hfill
\includegraphics[width=30mm]{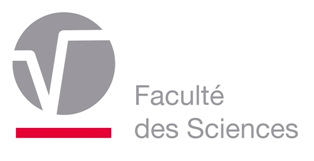}

\begin{center}

\vspace*{20mm}
\hrule
\vspace{10mm}
{\Large \sc Chiral Higher Spin Gravity From Strong Homotopy Algebra}
\vspace{10mm}
\hrule

\vspace{15mm}
{\large {Richard van Dongen}}

\vspace{5mm}
{Service de Physique de l’Univers, Champs et Gravitation\\
Facult\'e des Sciences\\
Université de Mons}

\vspace{5mm}
{11 July 2025}

\vspace{20mm}
{\sc Th\`ese pr\'esent\'ee en vue de l’obtention\\
du grade de Doctorat en sciences}

\vspace{20mm}
{\sc Promoteur de th\`ese}
\begin{align}
    \text{Evgeny Skvortsov} &\quad \text{Universit\'e de Mons, Belgique} \nonumber
\end{align}

{\sc Co-promoteur de th\'ese}
\begin{align}
    \text{Thomas Basile} &\quad \text{Universit\'e de Mons, Belgique} \nonumber
\end{align}

\vspace{5mm}
\centering{\sc Membres du jury}
\begin{align}
\text{Evgeny Skvortsov} &\quad \text{Universit\'e de Mons, Belgique} \nonumber\\
\text{Nicolas Boulanger} &\quad \text{Universit\'e de Mons, Belgique} \nonumber\\
\text{Andrea Campoleoni} &\quad \text{Universit\'e de Mons, Belgique} \nonumber\\
\text{Xavier Bekaert} &\quad \text{Universit\'{e} de Tours, France} \nonumber\\
\text{Carlo Iazeolla} &\quad \text{Guglielmo Marconi University, Italie} \nonumber
\end{align}

\end{center}

\chapter*{Acknowledgements}
\addcontentsline{toc}{chapter}{Acknowledgements}
\pagenumbering{roman}

This thesis presents a collection of results obtained during my four-year PhD in Mons, conducted under the supervision of Evgeny Skvortsov. His unwavering support, impeccable guidance, and excellent ability to suggest fruitful research directions were of immense value to both this thesis and the beginning of my academic career. His way of pushing me to get results has raised my personal standards. I remain truly humbled and deeply grateful for his initial interest in collaborating with me in 2021. Joining him in Mons and pursuing a PhD under his supervision gave me the safe space to develop, learn and grow -- both in my research and as a researcher.

The second-most influential person behind this thesis is my former housemate, Dazza van Dongen–Schouten, who -- over the course of this PhD -- also became my wife (though, for the record, that development was not officially part of the doctoral programme). I am aware of the enormous sacrifice she made in moving to a new country for me, only to end up living with someone who often spent evenings and weekends buried in calculations, muttering "I almost got the result" or "I need to add one more thing to complete the paper". Her constant support and timely distractions made sure I kept taking good care of myself. Me being allowed to explain my work to her helped me in verbalizing my writings and brought me to new ideas.  

I am also grateful to my collaborators Evgeny Skvortsov, Alexey Sharapov and Arsenii Sukhanov. Their dedication, insightful suggestions and many inspiring discussions were instrumental in shaping this work. Working with them taught me how to co-create, collaborate and communicate my research effectively. I want to thank you for an immensely valuable and deeply enjoyable period. 

In addition to those already mentioned, I would also like to thank my fellow PhD students -- Shailesh Dhasmana, Mattia Serrani, Josh O'Connor, and Arsenii Sukhanov -- with whom I shared many enjoyable moments, engaging discussions, and unforgettable memories at summer schools and conferences. All your critical ways of thinking, giving feedback and challenging my ideas taught me to always be open for a discussion. I am glad and honored that there have been instances where the favor could be returned. 

I would also like to thank my co-supervisor, Thomas Basile, for his valuable comments on the draft version of this thesis. His feedback significantly improved the clarity and quality of the final text.

Lastly, I would like to thank the complete staff of the unit. Without you there would have been no support, facilities and opportunities. You have created a truly enjoyable, safe and supportive space with open minded people without prejudices.

\vfill

\paragraph{Funding.} This research project was supported by the European Research Council (ERC) under the European Union’s Horizon 2020 research and innovation programme, grant number 101002551.

\chapter*{Abstract}
\addcontentsline{toc}{chapter}{Abstract}

In this thesis, we derive the equations of motion of Chiral Higher Spin Gravity (HiSGRA) in terms of its underlying $L_\infty$-algebra. Chiral HiSGRA contains self-dual Yang-Mills and self-dual gravity as closed subsectors, which themselves form closed subsectors of Yang-Mills and general relativity. We begin by constructing a covariant formulation for self-dual Yang-Mills and self-dual gravity, and subsequently extend this construction to the full Chiral Higher Spin Gravity.

Remarkably, the $L_\infty$-algebra is constructed from an $A_\infty$-algebra of pre-Calabi-Yau type, suggesting a deep connection to non-commutative deformation quantization. The structure maps of the resulting $L_\infty$-algebra are expressed as integrals of a simple exponential over convex polygons in $\mathbb{R}^2$. The existence of this covariant and coordinate independent formulation of chiral HiSGRA demonstrates, via the AdS/CFT correspondence, that $O(N)$ vector models possess a closed chiral subsector.

Finally, we prove that the $A_\infty$-algebra follows from Stokes' theorem -- a crucial feature of the known formality theorems. To this end, we construct integration spaces that generalize convex polygons to $\mathbb{R}^3$, and are intimately connected to positive Grassmanians. This Stokes-based derivation points towards a novel generalization of Kontsevich' formality theorem to the non-commutative setting.

\setcounter{tocdepth}{2}
\tableofcontents
\addcontentsline{toc}{chapter}{Contents}

\chapter{Introduction}
\label{chap:intro}
\pagenumbering{arabic}

\section{Motivation}

\begin{wrapfigure}{r}{0.4\textwidth}
    \centering
    \includegraphics[width=0.4\textwidth]{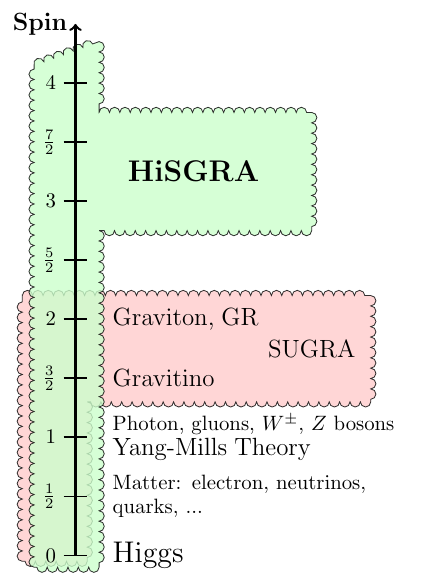}
    \caption{The standard model and general relativity are described by fields up to spin $2$. In addition to this lower-spin sector, HiSGRA also incorporates higher spins. Figure adopted from \cite{Bekaert:2022poo}}
    \label{fig:spinChart}
\end{wrapfigure}
Today, our best understanding of the universe is formulated in the remarkably successful standard model and theory of general relativity. The latter describes the dynamics of spacetime by means of differential geometry, giving rise to gravity and various other effects. Meanwhile, the standard model explains the behaviour of the fundamental particles that everything in the universe consists of in the framework of quantum field theory. It consists of a collection of interacting particles, or rather, quantum fields, of spin up to $1$. In the spin $0$ sector, it describes the Higgs field, which is responsible for giving mass to other fundamental particles. Matter, such as the quarks and electrons that constitute atoms, live in the spin $1/2$ sector, while the electromagnetic, the weak and the strong nuclear force are mediated by spin $1$ fields, i.e. photons, gluons, $W^\pm$ and $Z$ bosons. This is neatly summarized in Figure \ref{fig:spinChart}. 
Mathematically, spin labels the irreducible representations of the Lorentz algebra (or more precisely, its double cover). From a group-theoretic perspective, it can take arbitrary integer and half-integer values. However, all elementary particles observed in nature so far are bounded by spin-$1$. Still, it has been theorized that if general relativity could be incorporated into the framework of quantum field theory, it would be mediated by the graviton, a massless spin-$2$ particle \cite{weinberg1965infrared}. Although gravitons have not been observed so far, the vast majority of the physics community believes that their existence -- and with it, a quantum field theory description of general relativity -- could be key to understanding profound questions regarding black hole singularities, dark matter, dark energy, and more.

From the theoretical side, the existence of irreducible representations alone is not sufficient to be physically relevant; one still needs to construct a consistent, interacting theory containing them in the spectrum. This has proven to be particularly difficult for spin higher than $2$. In fact, various no-go theorems, including the famous Weinberg's low-energy theorem \cite{weinberg1965infrared} and Coleman-Mandula theorem \cite{Coleman:1967ad}, have ruled out the existence of such theories in specific settings. We will discuss these theorems in more detail later. However, these no-go theorems often require certain assumptions, such as a vanishing cosmological constant, analyticity of the S-matrix, a finite spectrum, unitarity, etc. HiSGRA (higher spin gravity) is the study of theories that contain the graviton and at least one interacting massless higher spin ($s \geq 2$) field in its spectrum and it probes the theories that are viable when some assumptions in these no-go theorems are relaxed.

As was already touched upon briefly, a longstanding problem in physics is the construction of a theory of quantum gravity. This entails the inclusion of the graviton, a massless spin-2 particle, in the framework of quantum field theory. Naive attempts to quantize the gravitational field inevitably lead to a perturbatively non-renormalizable theory. Over the last decades, various resolutions have been proposed, including supersymmetry and string theory. Interestingly, both possess the remarkable feature of canceling UV divergences, which greatly improves the UV behaviour of gravity \cite{VanNieuwenhuizen:1981ae,nicolai2014quantum}. These cancellations occur in supersymmetric theories due to the extended symmetry, while in string theory, among other reasons, this is due to the presence of an infinite tower of massive higher spin fields. 

HiSGRA in four spacetime dimensions employs both strategies: its spectrum contains an infinite tower of massless higher spin fields\footnote{It was long believed that an infinite tower of massless states is required in $4d$ HiSGRa. However, it was recently found that chiral HiSGRA contains many consistent truncations of chiral HiSGRA with a finite number of fields \cite{serrani2025classification}.}, which, as a consequence of being massless, constitute a gauge algebra of an underlying higher spin symmetry. The philosophy is simple: the infinite-dimensional higher spin symmetry constrains the theory to such an extent that only a few (or even no) counterterms are allowed. It is therefore believed that HiSGRA provides a consistent, and potentially UV-finite, toy model of quantum gravity. Recent studies have shown promising cancellations of UV divergences in various HiSGRA theories \cite{Ponomarev:2019ltz,Skvortsov:2018jea,Joung:2015eny,Skvortsov:2020wtf,Skvortsov:2020gpn}.

The reach of HiSGRA is not limited to quantum gravity alone; it also provides novel insights into conformal field theories, including the celebrated Ising model, which remains not yet solved exactly in $3D$. This is achieved by means of holography. It was conjectured in \cite{Klebanov:2002ja} that HiSGRA is holographically dual to Chern-Simons vector models. In the large-$N$ limit, the conformal symmetry is extended to an exact, infinite-dimensional \textit{higher spin symmetry} and these models correspond to free CFTs \cite{Eastwood:2002su,maldacena2013constraining,boulanger2013uniqueness}. In this case, the higher spin symmetry is strong enough to fix all correlation functions \cite{alba2016constraining,alba2013constraining,Colombo:2012jx,didenko2013exact,didenko2013exact,bonezzi2017noncommutative}. For finite $N$, interactions are turned on and the higher spin symmetry is broken, leading to the non-conservation of higher spin currents. However, this non-conservation occurs in a controlled manner, due to the sparse spectrum of Chern-Simons vector models. Specifically, this ensures that the non-conservation is expressed in terms of composite operators built from higher spin currents themselves. This is often referred to as \textit{slightly broken higher spin symmetry} \cite{Maldacena:2012sf}. Early results suggest that slightly broken higher spin symmetry is still strong enough to fix three-point and four-point functions \cite{Maldacena:2012sf,giombi2017higher,Turiaci:2018nua,Skvortsov:2018uru,Li:2019twz,Kalloor:2019xjb,silva2021four,Jain:2021gwa}.

The $3D$ Chern Simons vector models exhibit a couple of notable dualities, among which the conjectured $3D$ bosonization duality \cite{giombi2012chern,Li:2019twz,aharony2012correlation,Aharony:2015mjs,Karch:2016sxi,Seiberg:2016gmd}. This duality states that the correlation functions for models describing interacting fermions coincide with those of interacting bosons, i.e. the observables in these theories are insensitive to their microscopic description. Slightly broken higher spin symmetry is expected to be constraining enough to fix all correlation functions and to express them in terms of higher spin invariants \cite{giombi2017higher,Li:2019twz}.

The development of HiSGRA also has implications for pure mathematics. Exact higher spin symmetry has been proposed to result from the \textit{deformation quantization} of Poisson manifolds. Kontsevich' celebrated formality theorem guarantees the existence of an associative star-product on every finite-dimensional Poisson manifold, with its deformation controlled by the Poisson structure \cite{Kontsevich:1997vb}. Many higher spin algebras\footnote{Higher spin algebras were first described in \cite{Fradkin:1987ks} and \cite{Konstein:1989ij}.} can be realized as particular examples of such star-product algebras \cite{Sharapov:2019vyd}, establishing a deep link between deformation quantization and higher spin theory.

Interacting HiSGRA, however, must be constructed from a more general structure. Proposed candidates include the deformation quantization of Poisson orbifolds and the non-commutative deformation quantization of Poisson manifolds. The latter is a non-commutative deformation of the associative star-product algebras and naturally leads to the construction of $A_\infty$-algebras, from which $L_\infty$-algebras can be obtained that describe a field theory. These are examples of \textit{strong homotopy algebras}. The explicit construction of HiSGRA as its underlying $L_\infty$-algebra may offer guidance to the generalization of Kontsevich' formality theorem.

Another field that is deeply related to HiSGRA is string (field) theory. In this framework, strings excitations form an infinite tower of massive higher spin fields. In the tension-less limit, these fields become massless and are conjectured to be described by a HiSGRA \cite{gross1987high,gross1988string,Bonelli:2003kh,Sagnotti:2003qa,Sagnotti:2011qp}. Alternatively, string (field) theory can be thought of as a spontaneously broken phase of a HiSGRA. Not only may enforcing the connection between string (field) theory and higher spin theory result in insight into why string theory is UV-finite, but it would also provide a limit of string (field) theory that allows for significantly more manageable computations.

\subsection{History of HiSGRA}

Although HiSGRA holds far-reaching promises, the construction of these theories has proven to be remarkably difficult because of their extremely constraining nature. This is not surprising, as a multitude of no-go theorems even go as far as disproving the existence of HiSGRA, albeit under specific assumptions. Yet, over the past few decades, the development of HiSGRA has seen substantial progress, with the construction of several concrete theories. To better understand the technical challenges and the key successes of HiSGRA, let us give a brief historical account of its development.

\begin{itemize}
\item In 1939, Wigner classified the irreducible unitary representations of the Poincar\'{e} group, laying the foundations for massless and massive higher spin fields \cite{Wigner:1939cj}. In the same year, a consistent set of free equations of motion for massive fields of arbitrary spin propagating in Minkowski space were obtained by Fierz and Pauli \cite{FierzSUGRA}. Due to the need to include a variety of auxiliary fields to obtain a consistent Lagrangian formulation for these free equations, this was only found in 1974 by Singh and Hagen \cite{Singh:1974qz}.

\item In 1978, a gauge-invariant set of free equations of motion describing massless tensor fields of any spin was found by Fronsdal, generalizing lower spin gauge theories \cite{Fronsdal:1978rb}. While this was first obtained in flat space, Fronsdal lifted this result to (A)dS shortly after, opening up the possibility to evade the no-go theorems.

\item Despite these significant achievements, several no-go theorems\footnote{See \cite{Bekaert:2010hw} for a modern discussion on no-go theorems.} emerged between the 1960s and 1980s that impose severe restrictions on interacting HiSGRA in flat space. The most well-known of these are the Coleman-Mandula theorem \cite{Coleman:1967ad} and Weinberg's low-energy theorem \cite{weinberg1965infrared}.

Weinberg's theorem proves that long-range interactions involving massless fields of spin $s>2$ are inconsistent, effectively implying that the interaction strength must vanish. The Coleman-Mandula theorem states that the symmetry group of a non-trivial, analytic, unitary $S$-matrix with a finite number of particle types below any mass threshold is the direct product of the Poincar\'{e} group and the internal symmetry group. In other words, the symmetry group cannot mix spacetime and internal symmetries, as is required for HiSGRA.

Other examples of no-go theorems -- these particular ones formulated in the 1970s and 1980s -- include the Weinberg-Witten theorem and the Aragone Deser argument. The former should be combined with Weinberg's low-energy theorem. This states that massless higher spin fields couple minimally to gravitons at low energies, which in turn is prohibited by the Weinberg-Witten theorem \cite{Weinberg:1980kq}. In the same vein, Aragone and Deser studied a spin-$5/2$ particle coupled minimally to gravity and showed that this leads to inconsistencies due to the breakdown of gauge-invariance -- a result later generalized to arbitrary spin \cite{Aragone:1979hx,aragone1971constraints}.

These theorems, however, are based on various assumptions, such as: the spectrum of the theory contains a finite amount of particle types below any mass, higher spin particles interact via minimal coupling, the $S$-matrix is analytic, the theory is unitary and local. When any of these assumptions is lifted -- for example, breaking unitarity or introducing an infinite tower of fields -- the no-go theorems no longer directly apply and may be circumvented, opening a door for the construction of a consistent formulation of interacting HiSGRA. However, avoiding a number of assumptions does not have to immediately lead to any consistent theory by itself.

Let us note that it has long been thought that moving to (A)dS is a way to avoid all the no-go theorems. However, Weinberg/Coleman-Mandula theorems have a direct AdS analog \cite{Maldacena:2011jn,Boulanger:2013zza}. In fact, thanks to AdS/CFT correspondence the proof of some no-go results in flat space, the most notable being \cite{Bekaert:2010hp}, can be streamlined \cite{Bekaert:2010hp,Bekaert:2015tva,Maldacena:2015iua,Sleight:2017pcz,Ponomarev:2017nrr,Ponomarev:2017qab,Neiman:2023orj}. At present, there do not seem to be any invariant statements that would prefer (A)dS space to the flat one, see e.g. discussion in \cite{Krasnov:2021nsq}. For example, the two-derivative gravitational interactions of massless higher-spin fields exist both in the $4d$ flat \cite{boulanger2013uniqueness,Bengtsson:1983pd,Bengtsson:1986kh} and $(A)dS_4$ \cite{Metsaev:2018xip} spacetimes. However, they cannot be written in terms of the standard Fronsdal field: only the $(2s-2)$-derivative can be constructed in the flat space and a fixed linear combination of the $2$-derivative and $(2s-2)$-derivative ones can be constructed in $AdS_4$ \cite{Fradkin:1986ka,Fradkin:1987ks}. The latter just shows that certain statements made within the Fronsdal approach to higher spin fields do not have an invariant meaning.

\item It was not until the 1980s that (some) interactions of higher-spin fields were studied, see e.g. \cite{Bengtsson:1983pg,Berends:1984rq,Fradkin:1986ka,Fradkin:1987ks}. Around the same time, the Vasiliev equation were introduced \cite{vasiliev1990consistent}, with generalizations to $3d$ \cite{vasiliev2003nonlinear} and arbitrary dimensions \cite{prokushkin1999higher}. These equations provide a formal framework for studying interactions in higher spin theories, though they are not sufficiently restrictive to uniquely determine them. Essentially, the equations allow one to parameterize in a neat way all possible gauge-invariant and, in general, nonlocal interaction vertices together with, in general, nonlocal field redefinitions. Therefore, being the most general ansatz for interactions, the equations hide infinitely many free parameters.\footnote{For example, it is well-known that the Noether procedure is empty or, essentially, anything is a solution provided the locality is abandoned \cite{Barnich:1993vg}. Equivalently, in the light-cone gauge ``any'' function is a Hamiltonian provided the boost generators are allowed to be nonlocal. } Most notably, the Vasiliev equations are built using the unfolding formalism \cite{Vasiliev:1988sa}, rather than on the conventional construction from local Lagrangians. This formalism will be discussed in greater detail later in this thesis.

The fields described in the Vasiliev equations live in an extended non-commutative space where a star-product algebra defines the higher spin algebra. The construction of the Vasiliev equations has led to a wide array of technical tools and profound conceptual insights, many of which will play an important role in this thesis. However, one of the main challenges of the Vasiliev formalism is that the interactions are manifestly non-local\footnote{Instead of trying to directly impose locality, which does not seem to be even possible for the complete theory (with Chiral HiSGRA being a notable exception that can be associated to the self-dual truncation of the complete HiSGRA) one can try to extract some physical observables directly at the level of the given $L_\infty$-algebra, e.g. \cite{Sharapov:2020quq}, or  to resort to even more general ideas, e.g. \cite{Sezgin:2011hq, Bonezzi:2016ttk,DeFilippi:2019jqq,DeFilippi:2021xon,diaz2024fractional,diaz2024fractional}.}, see for instance \cite{Boulanger:2015ova} for a discussion. As a result, it is difficult to extract conventional quantum field theory observables from the theory, such as the (holographic) $S$-matrix.
\end{itemize}

Other interacting HiSGRAs are conformal HiSGRA, which generalizes conformal gravity \cite{Segal:2002gd}; topological ($3D$) HiSGRA, which generalizes the conventional formulation of gravity as Chern-Simons theory with gauge algebra $\mathfrak{sl}(2,\mathbb{R})\oplus \mathfrak{sl}(2,\mathbb{R})$ to $\mathfrak{sl}(N,\mathbb{R})\oplus \mathfrak{sl}(N,\mathbb{R})$, with $N$ the highest spin present in the theory \cite{Bergshoeff:1989ns,Blencowe:1988gj,Henneaux:2010xg,Campoleoni:2010zq,Pope:1989vj,Fradkin:1989xt,Grigoriev:2019xmp}; and \textit{chiral HiSGRA} \cite{Skvortsov:2020wtf,Skvortsov:2018jea,Ponomarev:2016lrm}, which generalizes and unifies self-dual Yang-Mills and self-dual gravity with their higher spin extensions. It is the latter theory that we investigate in this thesis and we will elaborate on it in more detail later in the thesis.

\section{Goals and structure of the thesis}

The thesis follows two lines of research:
\begin{itemize}
    \item The first is the construction of a covariant formulation of chiral HiSGRA in terms of its underlying $L_\infty$-algebra, using the methods provided by the unfolding approach. In a few words, we exploit the integrability of chiral HiSGRA to obtain the cubic structure map of this $L_\infty$-algebra. From there, we employ homological perturbation theory (HPT) to extend this result to higher order structure maps. We start by applying this procedure to the self-dual theories, i.e. self-dual Yang-Mills and self-dual gravity, for which the $L_\infty$-algebra is just a differential graded Lie algebra. These cases simplify greatly, because for them the higher structure maps vanish.
    
    This line is based on \cite{SDFDA,SDFDA2,Sharapov:2022faa,Sharapov:2022wpz,sharapov2023more}.

    \item The second concerns a connection to a putative extension of Kontsevich' formality theorem. We uncover this by expressing the theory's $L_\infty$-relations as Stokes' theorem. Historically, the results were obtained using homological perturbation theory, but they will be presented diagrammatically.

    This line is based on \cite{sharapov2024strong}.
\end{itemize}

The thesis is structured as follows. In Chapter \ref{chap:intro}, we start by reviewing some mathematical methods that were used in the thesis, after which we offer a self-contained summary of the papers this thesis is based on, omitting only technical details and computations. In Chapter \ref{chap:SD}, we construct the $L_\infty$-algebra for the self-dual theories SDYM and SDGR. This result is generalized in Chapter \ref{chap:cubic}, where we obtain the cubic interaction for chiral HiSGRA. In Chapter \ref{chap:all}, we employ homological perturbation theory to derive all higher-point vertices for chiral HiSGRA. The second line of research is captured in Chapter \ref{chap:Stokes}, where we prove the $A_\infty$-relations underlying chiral HiSGRA using Stokes' theorem. Chapter \ref{chap:conclusion} concludes the main part of the thesis with an overview of the results obtained in the thesis and a brief discussion. Finally, in Appendix \ref{app:notation} we introduce some relevant notation and conventions and Appendix \ref{app:selfdual}, \ref{app:cubic} and \ref{app:all} contain appendices relevant for Chapter \ref{chap:SD}, \ref{chap:cubic} and \ref{chap:all}, respectively.

\section{Mathematical framework}
\label{sec:mathFram}

Throughout this thesis, we make extensive use of concepts and techniques from strong homotopy algebras, homological perturbation theory, and deformation quantization. In this section, we aim to provide a brief and sufficient introduction to these topics, that will allow the reader to read the remainder of the paper. More details on how homological perturbation theory is applied in the thesis can be found in Appendix \ref{app:hpt} and Section \ref{sec:allVertices}. 

For more information on strong homotopy algebras, we refer to \cite{kraft2024introduction,jurvco2019algebras}. A discussion on homological perturbation theory applied to HiSGRA can be found in \cite{Li:2018rnc} and \cite{bordemann2008deformation,waldmann2016recent} contain comprehensive reviews on deformation quantization.

\subsection{Differential graded structures}

The definition of strong homotopy algebras requires a prior understanding of differential graded algebras and related structures, which we now discuss.

\begin{definition}
    Consider the vector spaces $V_n$ over the field $\mathbb{C}$ with degree $n\in \mathbb{Z}$. Then, the direct sum $V=\bigoplus_{n \in \mathbb{Z}}V_n$ is called a $\mathbb{Z}$-graded vector space. An element $v_n\in V_n$ is called homogeneous and is said to be of degree $n$, denoted $|v_n|=n$.
\end{definition}

\begin{definition}
    A degree-shifted vector space $V[l]$, $l\in\mathbb{Z}$,  is obtained by shifting the degrees of the linear subspaces of $V$ according to
    \begin{align}
        V[l]&=\bigoplus_{k\in\mathbb{Z}}(V[l])_k \,, & (V[l])_k &= V_{k+l} \,.
    \end{align}
\end{definition}
Graded vector spaces are the starting point of the structures that appear in strong homotopy algebras.
\begin{definition}
    A differential graded vector space is a $\mathbb{Z}$-graded vector space $V$ together with a differential $d: V\rightarrow V$ for which $d(V_n) \subset V_{n+1}$ and satisfies $d^2=0$. The collection $(V,d)$ is called a cochain complex.
\end{definition}
Equipped with an associative product, we obtain the structure of a differential graded algebra.
\begin{definition}
    A differential graded algebra (DGA) is a differential graded vector space $V$ endowed with a product $m: V \otimes V \rightarrow V$ of degree $0$ that satisfies, given homogeneous elements $v_i\in V$,
    \begin{itemize}
        \item associativity: $m(v_1,m(v_2,v_3))=m(m(v_1,v_2),v_3)$;
        \item graded Leibniz rule: $dm(v_1,v_2)=m(dv_1,v_2)+(-1)^{|v_1|}m(v_1,dv_2)$.
    \end{itemize}
\end{definition}
Replacing the associative product with a graded Lie bracket gives rise to a differential graded Lie algebra.
\begin{definition}
    A differential graded Lie algebra (DGLA) is a differential graded vector space $V$ endowed with a bracket $[-,-]:V_n\otimes V_m \rightarrow V_{n+m}$ of degree $0$ that satisfies
    \begin{itemize}
        \item graded skew-symmetry $[v_1,v_2]=-(-1)^{|v_1||v_2|}[v_2,v_1]$;
        \item graded Jacobi identity $[v_1,[v_2,v_3]]=[[v_1,v_2],v_3]+(-1)^{|v_1||v_2|}[v_2,[v_1,v_3]]$;
        \item graded Leibniz rule $d[v_1,v_2]=[dv_1,v_2]+(-1)^{|v_1|}[v_1,dv_2]$.
    \end{itemize}
\end{definition}
Adding commutativity, we obtain a differential graded commutative algebra.
\begin{definition}
    A differential graded commutative algebra (DGCA) is a DGA whose product is graded commutative, meaning it satisfies 
    \begin{align}
        m(v_1,v_2)=(-1)^{|v_1||v_2|}m(v_2,v_1)
    \end{align}
    for all homogeneous elements $v_1$, $v_2$.
\end{definition}

The sign obtained from swapping multiple homogeneous elements is called the Koszul sign.
\begin{definition}
    Let $v_1,\dots,v_n$ be homogeneous element of a graded vector space, each with degree $v_i$. Let $S_n$ denote the permutation group, i.e. the group of all permutations of the set $\{1,\dots,n\}$. Given a permutation $\sigma\in S$, the Koszul sign $\epsilon(\sigma;v_1,\dots,v_n)$ is defined by:
    \begin{align}
        v_{\sigma(1)}\otimes\dots\otimes v_{\sigma(n)} = \epsilon(\sigma;v_1,\dots,v_n)v_1\otimes\dots\otimes v_n \,,
    \end{align}
    where the sign $\epsilon(\sigma;v_1,\dots,v_n)$ arises from swapping homogeneous elements according to
    \begin{align}
        m(v_1,v_2)=(-1)^{|v_1||v_2|}m(v_2,v_1) \,.
    \end{align}
\end{definition}

\subsection{Strong homotopy algebras}

Strong homotopy algebras, being generalizations of differential graded algebras, include the what is called $A_\infty$- and $L_\infty$- algebras, which we focus on in this thesis. More specifically, the former is a generalization of the notion of a differential graded associative algebra (DGA) and the latter generalizes differential graded Lie algebras (DGLA). Both admit three equivalent definitions; one algebraic, one coalgebraic and one geometric. While the first of these is easier to understand, the last two provide deeper insights and are more useful to relate to field theory. 

We will present the definitions in the order mentioned above. However, we choose to present only the algebraic definition for both the $A_\infty$- and $L_\infty$-algebra and to discuss the other two definitions only for $L_\infty$-algebras, in order to avoid too much repetition. Subsequently, we will discuss the connection between $L_\infty$-algebras and field theories.

Roughly speaking, an $A_\infty$-algebra generalizes DGAs by relaxing the requirement that the bilinear product is associative. This failure of associativity is controlled by a trilinear map. Higher associativity constraints are imposed, with, again, a non-associative piece that is controlled by multilinear maps. Formally, an $A_\infty$-algebra contains infinitely many higher order products. A DGA is an example of an $A_\infty$-algebra for which all higher (trilinear and beyond) maps vanish. Similarly, an $L_\infty$-algebra allows for the failure of the graded Jacobi identity in a controlled manner by introducing higher brackets. DGAs and DGLAs are examples of $A_\infty$-algebras and $L_\infty$-algebras, respectively, for which all higher (trilinear and beyond) maps vanish.

\begin{definition}[algebraic definition]
    An $A_\infty$-algebra is a collection $(V,m_1,m_2, \dots)$ of a $\mathbb{Z}$-graded vector space $V=\bigoplus_{n}V_n$ over the field\footnote{In general, it can be defined over any field $\mathbb{K}$, but we restrict ourselves to $\mathbb{C}$ for all practical purposes. The same is true for $L_\infty$-algebras.} $\mathbb{C}$ and linear products $m_k: V^{\otimes k} \rightarrow V$ of degree $2-k$ $(k\geq 1)$, satisfying the $A_\infty$-relations
    \begin{align}
        \sum_{r=0}^{n-1}\sum_{k=1}^{n-r}(-1)^{rk+n-k-r}m_{n-k+1}\circ(\mathds{1}^{\otimes r}\otimes m_k \otimes \mathds{1}^{\otimes(n-k-r)}) &= 0 \,, & n \geq 1\,. 
    \end{align}
\end{definition}
To better understand this structure, it is instructive to evaluate the $A_\infty$-relations for $n \leq 3$. For this, we define homogeneous elements $v_i \in V_i \subset V$, with degree $|v_i|$.
\begin{itemize}
    \item $n=1$: This gives the relation 
    \begin{align*}
        m_1\circ m_1(v_1) =0 \,,
    \end{align*}
    which implies that $m_1$ is a differential on $V$.
    \item $n=2$: We find 
    \begin{align*}
        m_1\circ m_2(v_1,v_2) = m_2(m_1(v_1),v_2) + (-1)^{|v_1|}m_2(v_1,m_1(v_2)) \,,
    \end{align*}
    i.e. the graded Leibniz rule, meaning that $m_1$ is a derivation on $V$ with respect to the product $m_2$.
    \item $n=3$: The relations give 
    \begin{align*}
        m_2(&m_2(v_1,v_2),v_3) - m_2(v_1,m_2(v_2,v_3)) = m_1\circ m_3(v_1,v_2,v_3) +\\
        &+m_3(m_1(v_1),m_2,m_3) +(-1)^{|v_1|} m_3(v_1,m_1(v_2),v_3)+ (-1)^{|v_1|+|v_2|}m_3(v_1,v_2,m_1(v_3)) \,.
    \end{align*}
    The associativity fails to hold due to the presence of the map $m_3$. 
\end{itemize}
It is now easy to see that a DGA $(V,d,\,\cdot\,)$, i.e. $m_1=d$, $m_2(v_1,v_2)= v_1\cdot v_2$ and $m_{\geq 3}=0$, is an example of an $A_\infty$-algebra.

The name strong homotopy algebra reflects that -- in the case of an $A_\infty$-algebra -- the failure of associativity is captured by a homotopy. This can be made precise using the following notion:

\begin{definition}
    Let $f,g:V\rightarrow W$ be two morphisms of degree $0$ between the cochain complexes $(V,d_V)$ and $(W,d_W)$, meaning $d_W f=f d_V$ and $d_W g = g d_V$. A homotopy is a morphism $h: V \rightarrow W$ of degree $-1$ between $V$ and $W$ such that $f-g=h\circ d_V+d_W\circ h$.
\end{definition}
Now consider the associator $Ass(m_2)=m_2(m_2(\bullet,\bullet),\bullet)-m_2(\bullet,m_2(\bullet,\bullet))$. The $A_\infty$-relation for $n=3$ becomes
\begin{align}
    Ass(m_2)-0=d_V\circ m_3 + m_3\circ d_{V^{\otimes 3}}\,,
\end{align}
with $d_V=m_1$ and $d_{V^{\otimes 3}}$ is the extension of $m_1$ to $V^{\otimes 3}$ via the Leibniz rule. This equations show that $Ass(m_2)$ is homotopic to $0$, with $m_3$ providing the homotopy. The higher ($n > 3$) $A_\infty$-relations lead to higher homotopies in a similar fashion. This observation justifies the name strong homotopy algebra.

\begin{definition}[algebraic definition]
    An $L_\infty$-algebra is a collection $(\mathfrak{g},l_1,l_2,\dots)$ of a $\mathbb{Z}$-graded vector space $\mathfrak{g}$ over the field $\mathbb{C}$ and linear products $l_k: \mathfrak{g}^{\wedge k} \rightarrow \mathfrak{g}$ of degree $2-k$ $(k\geq 1)$ that are graded skew-symmetric $l_n(v_{\sigma(1)},\dots,v_{\sigma(n)})=(-1)^\sigma\epsilon(\sigma,v)l_n(v_1,\dots,v_n)$ and satisfy the $L_\infty$-relations
    \begin{align}
        \sum_{k=1}^{n}(-1)^k\sum_{\sigma\in \text{Unsh(k,n-k)}}(-1)^\sigma \epsilon(\sigma,v)l_{n-k+1}(l_k(v_{\sigma(1)},\dots, v_{\sigma(k)}),v_{\sigma(k+1)},\dots,v_{\sigma(n)}) = 0\,.
    \end{align}
    Here, $\epsilon(\sigma, v)$ is the Koszul sign of the permutation $\sigma$ and the $\text{Unsh}(k,n-k)$ subset of the permutation group $(k,n-k)$-unshuffle. An unshuffle $\tau\in \text{Unsh}(k,n-k)$ is a permutation such that its inverse $\tau^{-1}\in \text{Sh}(k,n-k)$ is a shuffle. The latter group consists of the permutations that satisfy 
    \begin{align*}
        \sigma(1) &< \sigma(2) < \dots < \sigma(k), & \sigma(k+1) &< \dots < \sigma(n)\,.
    \end{align*}
\end{definition}
Again, we analyze the $L_\infty$-relations for $n\leq 3$. Let $x_i \in V_i \subset V$ be homogeneous elements.
\begin{itemize}
    \item $n=1$: We find the relation
    \begin{align}
        l_1 \circ l_1(x_1) = 0 \,,
    \end{align}
    which means that $l_1$ is a differential on $\mathfrak{g}$.
    \item $n=2$: This gives
    \begin{align}
        l_1\circ l_2(x_1,x_2) = l_2(l_1(x_1),x_2) + (-1)^{|x_1|}l_2(x_1,l_1(x_2)) \,,
    \end{align}
    which implies that $l_1$ is a derivation on $\mathfrak{g}$ with respect to $l_2$.
    \item $n=3$: this yields
    \begin{align}
        \begin{aligned}
            &l_2(l_2(x_1,x_2),x_3)+(-1)^{(|x_1|+|x_2|)|x_3|}l_2(l_2(x_3,x_1),x_2)+(-1)^{(|x_2|+|x_3|)|x_1|}l_2(l_2(x_2,x_3),x_1)=\\
            &l_1\circ l_3(x_1,x_2,x_3)+l_3(l_1(x_1),x_2,x_3)+(-1)^{|x_1|}l_3(x_1,l_1(x_2),x_3)+\\
            &+(-1)^{|x_1|+|x_2|}l_3(x_1,x_2,l_1(x_3)) \,,
        \end{aligned}
    \end{align}
    which shows that the graded Jacobi identity is satisfied up to a term controlled by $l_3$.
\end{itemize}
Similarly to the $A_\infty$-algebra, we note that the DGLA $(\mathfrak{g},d,[-,-])$ is an example of an $L_\infty$-algebra with $l_1=d$, $l_2(x_1,x_2)=[x_1,x_2]$ and $l_{\geq 3}=0$. Moreover, one can always obtain an $L_\infty$-algebra through the graded anti-symmetrization of an $A_\infty$-algebra, comparable to how a Lie algebra is obtained from an associative one. Analogously, not all $L_\infty$-algebras can be obtained from $A_\infty$-algebras.

With this we understand that $A_\infty/L_\infty$-algebras generalize differential graded structures. Let us now turn to the other two definitions of $L_\infty$-algebras.

The second definition of an $L_\infty$-algebra uses the notion of graded coalgebras.
\begin{definition}
    A graded coalgebra is a graded vector space $C=\bigoplus_{n\in\mathbb{Z}}C_n$, equipped with a comultiplication map of degree $0$,
    \begin{align}
        \Delta: C\rightarrow C \otimes C \,,
    \end{align}
    that is coassociative, meaning it satisfies
    \begin{align}
        (\Delta \otimes \mathrm{id}) \circ \Delta = (\mathrm{id}\otimes \Delta)\circ \Delta \,.
    \end{align}
\end{definition}
Analogously to differential graded algebras, one can enrich a graded coalgebra with a differential.
\begin{definition}
    A differential graded coalgebra is a graded coalgebra, together with a map $b : C \rightarrow C$ of degree $1$, that satisfies the co-Leibniz rule
    \begin{align}
        \Delta \circ b = (b \otimes \mathrm{id} + \mathrm{id} \otimes b) \circ \Delta
    \end{align}
    and satisfies $b^2=0$.
\end{definition}

An example of a differential graded coalgebra is a tensor coalgebra of a differential graded vector space $V$,
\begin{align}
    T^c(V)=\bigoplus_{n \geq 0}V^{\otimes n}
\end{align}
with comultiplication
\begin{align}
    \Delta(v_1\otimes \dots \otimes v_n) = \sum_{i=0}^n(v_1\otimes \dots \otimes v_i)\otimes (v_{i+1}\otimes \dots \otimes v_n)
\end{align}
for homogeneous elements $v_1,\dots, v_n \in V$. In the following, a  major role is played by the reduced tensor algebra, which we denote by $\bar{T}^c(V)=\bigoplus_{n=1}^{\infty}V^{\otimes n}$.

\begin{definition}[coalgebraic definition]
    An $L_\infty$-structure on $\mathfrak{g}$ is encoded by a coderivation $b$ of degree $1$ on the reduced symmetric coalgebra of $\mathfrak{g}$,  $\bar{S}^c(\mathfrak{g}[1])=\bigoplus_{n \geq 1}\mathfrak{g}^{\odot n}$, satisfying $b^2=0$. The coderivation is given by a set of maps $b_k: \mathfrak{g}^{\otimes k} \rightarrow \mathfrak{g}$ for $k\geq 1.$
\end{definition}

Given the condition $b^2=0$, it can be proven that the components $b_k$ are isomorphic to the maps $m_k$ satisfing the $L_\infty$-relations. See \cite{Li:2018rnc,kraft2024introduction} for a more detailed discussion.

\begin{definition}[geometric definition] \label{def:geometric}
    An $L_\infty$-algebra on $\mathfrak{g}$ is encoded by a homological vector field $Q$ of degree $1$ on the graded manifold $\mathfrak{g}[1]$, called a $Q$-manifold, satisfying $Q^2=0$.
\end{definition}

A physical theory does not have a single unique $L_\infty$-algebra associated to it. It is therefore useful to define the following, 

\begin{definition}
    Let $\phi: \bar{S}^c(\mathfrak{g}[1]) \rightarrow \bar{S}^c(\mathfrak{g}'[1])$ be a morphism of graded coalgebras composed of a collection of maps $\phi_n: S^{n}(\mathfrak{g}[1]) \rightarrow \mathfrak{g}'[1]$, such that $\phi \circ b = b' \circ \phi$. Then, $\phi: (\mathfrak{g},b) \rightarrow (\mathfrak{g}',b')$ is an $L_\infty$-morphism between these two $L_\infty$-algebras.
\end{definition}

In general, $L_\infty$-morphisms are too general to preserve the physical data of a theory. Instead, one should consider the slightly more specific notion of $L_\infty$-quasi-isomorphisms. Physically, they allow one to relate two formulations of equivalent field theories with different field content, e.g. field theories that are equivalent after integrating out fields.
\begin{definition}
    Consider two cochain complexes $(V,d_V)$ and $(W,d_W)$ and a morphism of complexes $f: V\rightarrow W$, i.e. degree-$0$ morphism satisfying $d_Wf=fd_V$. Then $f$ is called a quasi-isomorphism of complexes if the induced morhpisms on the cohomology
    \begin{align}
        H^{n}(f): H^n(V,d_V) \rightarrow H^n(W,d_W)
    \end{align}
    are isomorphisms of all degrees $n$.
\end{definition}

\begin{definition}
    An $L_\infty$-quasi-isomorphism is an $L_\infty$-morphism $\phi$ for which the first structure map $\phi_1$ is a quasi-isomorphism of complexes.
\end{definition}

In the BV-BRST treatment of gauge theories, one can construct an $L_\infty$-algebra $L=\bigoplus L_n$ as an extension of the BV-BRST complex. This $L_\infty$-algebra contains all physical data captured by the theory, such as gauge symmetries $\in L_{\leq 0}$, classical fields $\in L_1$, equations of motion $\in L_2$, and Noether identities $\in L_{\geq 3}$.

One particularly useful $L_\infty$-algebra is the minimal model. In a sense, it can be viewed as the ``smallest'' $L_\infty$-algebra that contains all physical information of a field theory. It is related to the $L_\infty$-algebra obtained from the BV-BRST complex by an $L_\infty$-quasi-isomorphism.
\begin{definition}
    An $L_\infty$-algebra is minimal (also known as minimal model) if the first structure map vanishes, $l_1=0$.
\end{definition}

The $L_\infty$-algebra we are after in this thesis is constructed from definition \ref{def:geometric}. Consider a (non-negatively) graded supermanifold $\mathcal{N}$ with coordinates $X^A$ equipped with a homological vector field $Q$ of degree $1$, i.e.
\begin{align} \label{nilpotency}
    [Q,Q]&=0 \qquad\Leftrightarrow \qquad Q^{\mathcal{B}}\frac{\partial}{\partial X^{\mathcal{B}}}Q^\mathcal{A}=0\,.
\end{align}
The pair $(\mathcal{N},Q)$ is then considered to be a differential graded manifold. The equations of motion can then be written as a sigma model
\begin{align} \label{sigmamodel}
    d\Phi=Q(\Phi) \,,
\end{align}
where the form-fields $\Phi=\Phi(x,dx)$ are maps $\Phi: T[1]\mathcal{M} \rightarrow \mathcal{N}$ from the degree $1$ shifted tangent bundle of a spacetime manifold $\mathcal{M}$ to the supermanifold $\mathcal{N}$. The maps $\Phi^A$ are pullbacks of the coordinates $X^A$. Equations of this form are then said to form a free differential algebra (FDA) \cite{Sullivan77}.

In practice, the form-fields $\Phi(x,dx)$ are the fields present in the spectrum of a given field theory. For traditional gauge theories\footnote{Exceptions include $p$-form gauge fields.}, the spectrum splits in two sectors composed of zero-forms $\omega$ and one-forms $C$. The former typically contain the gauge fields, while the latter contain the (components of) the field strengths. For these fields, the equation of motion \ref{sigmamodel} yields the equations
\begin{align} \label{toSolve}
    \begin{aligned}
        d\omega &= \mathcal{V}(\omega,\omega) + \mathcal{V}(\omega,\omega,C) + \mathcal{V}(\omega,\omega,C,C)+\dots\,,\\
        dC &= \mathcal{U}(\omega,C) + \mathcal{U}(\omega,C,C) + \dots \,,
    \end{aligned}
\end{align}
which lead to $L_\infty$-relations when applying the de Rham differential $d$ and imposing the integrability condition $d^2=0$. This is equivalent to the nilpotency condition \eqref{nilpotency} and constitutes $L_\infty$-relations that describe the chiral HiSGRA.

\subsection{Higher spin algebra and its deformation}
It is well known that the higher spin algebra typically admits a star-product realization, see \cite{Didenko:2014dwa} for a review. In order to motivate this, let us recall the standard definition of the higher spin algebra $\mathfrak{hs}$ on $AdS_4$. By the definition of HiSGRA, the spectrum of the theory contains a graviton, which is described by an $\mathfrak{so}(3,2)$-connection. Hence, we start with the universal enveloping algebra $U(\mathfrak{so}(3,2))$, which can be viewed as a module over itself. However, this space is too large -- it contains many elements that do no correspond to higher spin symmetry generators. Therefore the higher spin algebra $\mathfrak{hs}$ is defined as the quotient of the universal enveloping algebra by some ideal $\mathcal{I}$,
\begin{align}
    \mathfrak{hs}=\frac{U(\mathfrak{so}(3,2))}{\mathcal{I}} \,.
\end{align}
Typically, the ideal $\mathcal{I}$ is the annihilator of the singleton representation inside the universal enveloping algebra $U(\mathfrak{so}(3,2))$. In other words, $\mathcal{I}$ is the two-sided ideal generated by all elements of $U(\mathfrak{so}(3,2))$ that act trivially on the singleton module.

Alternatively, $\mathfrak{hs}$ can be realized without specifying the ideal $\mathcal{I}$ explicitly, by constructing it as a deformation of $U(\mathfrak{so}(3,2))$ that preserves the Jacobi identity. This happens naturally if $\mathfrak{hs}$ acts on the space of functions $\mathbb{C}[y,\bar{y}]$, where $y^A$ and $\bar{y}^{A'}$ are $\mathfrak{sl}(2,\mathbb{C})$ spinors, and the Lie bracket is the commutator of some associative product, i.e.
\begin{align}
    [f,g]_\star(y,\bar{y})=f(y,\bar{y})\star g(y,\bar{y})-g(y,\bar{y})\star f(y,\bar{y})
\end{align}
and
\begin{align}\label{assIntro}
    f(y,\bar{y})\star (g(y,\bar{y})\star h(y,\bar{y}))= (f(y,\bar{y})\star g(y,\bar{y}))\star h(y,\bar{y})\,.
\end{align}

The star-product can be written as
\begin{align} \label{starIntro}
    a\star b = ab + \hbar \phi_1(a,b) + \hbar^2\phi_2(a,b)+\dots\,,
\end{align}
where the point-wise product is denoted by juxtaposition and $\phi_i$ are bidifferential operators acting on $y^A$ and $\bar{y}^{A'}$. The fact that $\star$ must satisfy \eqref{assIntro}, implies that $\{-,-\}$ is a Poisson bracket with $\{a,b\}=\frac{a\star b-b\star a}{\hbar}\Big|_{\hbar=0}$. \textit{Deformation quantization} is the field of study that considers exactly such star products. In fact, Kontsevich proved in his celebrated formality theorem \cite{Konts} that every Poisson manifold admits a deformation of its algebra of functions that is determined by its Poisson structure in terms of a star-product. The simplest non-trivial example is the Moyal-Weyl star-product that is obtained from a constant non-degenerate Poisson structure, i.e. proportional to $\epsilon^{AB}$.\footnote{See Appendix \ref{app:notation} for the definition.}

However, it was proven in \cite{maldacena2013constraining} that exact higher spin symmetry leads to a free CFT on the boundary of $AdS_4$. Indeed, we will shortly see that the current star-product gives only the first structure map in the first line of \eqref{toSolve}, which encodes only information of the free HiSGRA.

It is well-known that higher spin algebras are rigid -- they cannot be deformed smoothly. As was shown in \cite{Sharapov:2018kjz}, the $\mathbb{Z}_2$-extension of the higher spin algebra $\mathfrak{hs}\rtimes \mathbb{Z}_2$ does admit a one-parameter family of deformations and is described by an $L_\infty$-algebra that is constructed from an $A_\infty$-algebra. This turns out to be related to \textit{non-commutative deformation quantization}. In short, consider the deformation of an associative product that is controlled by a formal parameter $\hbar$. This is given by \eqref{starIntro}. We then promote the deformation parameter $\hbar$ to an element of $\mathfrak{hs}$.\footnote{A subtle, but crucial, difference is that we have $\hbar\in\mathfrak{hs}^\star$, which is inspired by the free higher spin action. This will be explained in \ref{subsec:cubic}.} The power series in \eqref{starIntro} now provides genuine multilinear maps, and is able to produce the right-hand side of \eqref{toSolve}.

Thus, the $\mathbb{Z}_2$-extension of $\mathfrak{hs}$ is given by multilinear maps $m_k(a,b,\hbar,\dots,\hbar)$ that do not satisfy the (higher order) Jacobi identity, and its failure is controlled by other structure maps. This is very reminiscent of slightly broken higher spin symmetry, where the non-conservation of higher spin currents is controlled by other higher spin currents. This resemblance is precisely the reason why $A_\infty$-algebras are expected to capture the essence of slightly broken higher spin symmetry.

\begin{wrapfigure}{r}{0.4\textwidth}
    \centering
    \includegraphics[width=0.4\textwidth]{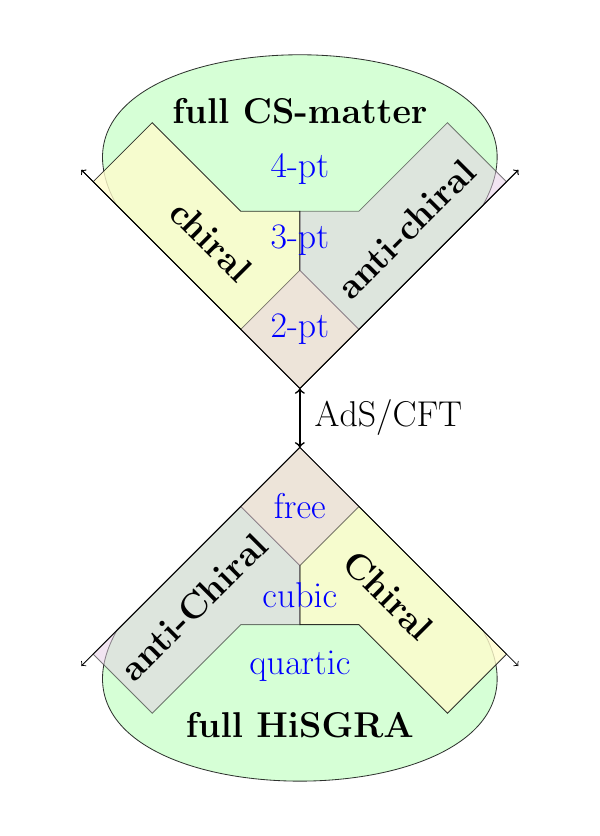}
    \caption{(Anti-)Chiral HiSGRA is a closed subsector of a full, parity-invariant HiSGRA. It should be dual to a closed subsector of Chern-Simons vector models.}
    \label{fig:sector}
\end{wrapfigure}

A powerful method for explicitly constructing such an $A_\infty$-algebra is \textit{homological perturbation theory} (HPT). Here, we will briefly discuss the logical flow of how this is achieved. For more details on how HPT is applied in this thesis, we refer to Appendices \ref{app:homo} and \ref{app:hpt} and to Section \ref{sec:allVertices}. A detailed review of HPT applied to HiSGRA is found in \cite{Li:2018rnc}.

In a nutshell, HPT is a means of transferring algebraic data from one cochain complex to another. It also allows one to transfer algebraic data between cochain complexes built from tensor coalgebras and a codifferentials, which are equivalent to an $A_\infty$-algebra. The homological perturbation lemma then provides an explicit method for deriving a deformed codifferential in one cochain complex by deforming the codifferential in the other. In particular, one cochain complex is composed of the tensor coalgebra of the cohomology of $\mathfrak{hs}$ with zero differential and the other is the tensor coalgebra of a larger graded vector space of which the cohomology of $\mathfrak{hs}$ is a subspace, together with some nonzero differential. One then perturbs this differential and perturbatively obtains an $A_\infty$-algebra with $m_1=0$ on the other cochain complex through the homological perturbation lemma. This is the minimal model we are interested in.

\section{Chiral HiSGRA and self-dual theories}
\label{sec:summary}

As was briefly mentioned, chiral HiSGRA contains self-dual Yang-Mills, self-dual gravity and their higher spin extensions as consistent truncations. Similarly to these theories, chiral HiSGRA breaks parity invariance. At the same time, chiral HiSGRA itself is thought to be a consistent truncation of some putative parity-invariant HiSGRA, see Figure \ref{fig:sector}. This diagram shows that through the AdS/CFT correspondence, chiral HiSGRA is thought to be dual to a chiral subsector of Chern-Simons vector models.

Not only did the construction of chiral HiSGRA offer a new example of a HiSGRA, it also showed that different formulations of HiSGRA can lead to more (or less) restrictive conditions on the theory. More specifically, it was found that chiral HiSGRA in the light-cone gauge allows more types of interaction than the covariant, metric-like formulation of HiSGRA. Here, the name metric-like refers to the fact that higher spin fields are represented as tensors, generalizing the notion of the metric. Another formulation that we have already encountered is the unfolding formalism. In this approach, one expresses the equations of motion as a first-order differential equation of wedge-products of the fields, which are forms valued in finite-dimensional irreducible representations of the Lorentz algebra. This approach generalizes the frame-like formulation of general relativity.

The foundation for chiral HiSGRA was laid by Metsaev in the early 1990s, when he classified all cubic interactions of massless higher spin fields in flat space and found many -- though not all -- such interactions in (A)dS, by working in the light cone-gauge \cite{metsaev1991s}. Much later, in 2016, Ponomarev and Skvortsov observed that this classification contains two consistent sets of interactions \cite{Ponomarev:2016lrm}: interactions of fields with helicity $\lambda_1$, $\lambda_2$ and $\lambda_3$, with the sum $\lambda_1+\lambda_2+\lambda_3 \geq 0$ leads to the what is called chiral HiSGRA, while $\lambda_1+\lambda_2+\lambda_3 \leq 0$ gives rise to the anti-chiral HiSGRA. The vertex for the chiral theory reads
\begin{align}
    V_{chiral} &= \sum_{\lambda_1,\lambda_2,\lambda_3} C_{\lambda_1,\lambda_2,\lambda_3}V_{\lambda_1,\lambda_2,\lambda_3} \,, & C_{\lambda_1,\lambda_2,\lambda_3} &= \frac{\kappa (l_p)^{\lambda_1+\lambda_2+\lambda_3-1}}{\Gamma(\lambda_1+\lambda_2+\lambda_3)} \,,
\end{align}
which has all its structure constants fixed in terms of the dimensionless constant $\kappa$ and a constant of dimension length $l_p$, which may be chosen to be the Planck length. Here, $V_{\lambda_1,\lambda_2,\lambda_3}$ captures the kinematics of the scattering and it gives the amplitude
\begin{align}
    V_{\lambda_1,\lambda_2,\lambda_3}\Big|_{on-shell}\sim [12]^{\lambda_1+\lambda_2-\lambda_3}[13]^{\lambda_1-\lambda_2+\lambda_3}[23]^{-\lambda_1+\lambda_2+\lambda_3}
\end{align}
for complex momenta. The antichiral vertex can be obtained simply by complex conjugation.

The theory is named (anti-)chiral, because it treats positive and negative helicities asymmetrically. Interestingly, the theory truncates at cubic order; no higher order interactions are present. Even more strikingly, the results hold both in flat space and in (A)dS. In flat space, chiral HiSGRA cleverly evades traditional no-go theorems in several ways: some no-go theorems rely on gauge invariant methods, whereas chiral HiSGRA was constructed in the light-cone gauge; the theory contains an infinite tower of massless fields; it is non-unitary; and despite the existence of non-trivial interactions, the $S$-matrix is trivial. Recently, a covariant action has been constructed for free higher spin fields \cite{Krasnov:2021nsq}, which will be a building block of this thesis, and it has been generalized to the partially massless setting \cite{basile2023chiral}.

\subsection{SDYM and SDGR}

Self-dual theories, including self-dual Yang-Mills (SDYM) and self-dual gravity (SDGR) and their higher spin extensions HS-SDYM and HS-SDGRA, are closed subsectors of the full theories. This means that solutions to the self-dual theories are solutions in the full ones and that scattering amplitudes coincide. In many aspects, the self-dual theories are simpler theories than their full theory counterparts. For example, self-dual theories possess various powerful properties: they are integrable, UV-finite and one-loop exact. Moreover, self-dual theories allow one to perform a perturbative expansions of the full theory around a self-dual background, rather than a flat one. As a consequence, it is often fruitful to study the self-dual version of a theory, as it can expose underlying structures and provide simpler computational methods for certain problems.

\subsubsection{SDYM}

Let us show the perturbative expansion of Yang-Mills around its self-dual sector as an example. In this derivation and in the remainder of the thesis, we will be working with two-component spinor language, which is well-suited for $4d$-theories and offers many advantages for self-dual theories in particular. Our notation is explained in Appendix \ref{app:notation} and more details can be found in \cite{penroserindler}. Let us still mention the most important definitions. 

Let the gauge field $A=A^a_{AA'}t_ae^{AA'}$, with $e_{AA'}$ the vielbein, be valued in some Lie algebra with a non-degenerate invariant bilinear form with Lie algebra generator $t_a$. The generator can be thought of as $t_a\in \text{Mat}_N$. For simplicity, we will not write the generators from now on. We can decompose the field strength into its self-dual and anti-self-dual components $F_{AA}$ and $F_{A'A'}$, respectively. We adopt the convention that spin tensors with repeated indices are symmetrized, e.g. $F_{AA}$ is equivalent to $\frac{1}{2}(F_{AB}+F_{BA})$, up to relabeling of indices. We choose the field $F^{AA}$ to carry negative helicity degrees of freedom and $F^{A'A'}$ to carry positive helicity ones. The (anti-)self-duality condition states
\begin{align}
    (\star F)_{AA} &= F_{AA} \,,& (\star F)_{A'A'} &= -F_{A'A'}\,.
\end{align}
The decomposition of the field strength reads
\begin{align}
    F=dA-A\wedge A=H^{AA}F_{AA}+H^{A'A'}F_{A'A'} \,.
\end{align}
Here, $H^{AA}$ and $H^{A'A'}$ are the self-dual and anti-self-dual $2$-form basis, respectively. They are composed of two vielbeins: $H^{AA}=e\fud{A}{A'}\wedge e^{AA'}$ and $H^{A'A'}=e\fdu{A}{A'}\wedge e^{AA'}$. The wedge-products between the two-form basis elements read
\begin{align}
    \begin{aligned}
        H_{AA}\wedge H_{BB}&=\epsilon_{AB}\epsilon_{AB}\text{dVol}_4\,, & H_{AA}\wedge H_{A'A'} &=0 \,,\\
        H_{A'A'}\wedge H_{B'B'}&=-\epsilon_{A'B'}\epsilon_{A'B'}\text{dVol}_4  \,,
    \end{aligned}
\end{align}
where the volume for $\text{dVol}_4$ is chosen with a convenient normalization.

The Yang-Mills action reads
\begin{align}
    S[F]= -\frac{1}{2g^2} \tr \int F\wedge \star F = -\frac{1}{4g^2}\text{tr}\int (F^{AA}F_{AA}+F^{A'A'}F_{A'A'})\text{dVol}_4 \,.
\end{align}
A topological term, sometimes called the $\theta$-term, can be added to the action. Due to its topological nature, it does not affect any perturbative properties of the theory. This term reads
\begin{align}
    S_{\text{top}} = \text{tr}\int F\wedge F = \text{tr}\int (F^{AA}F_{AA}-F^{A'A'}F_{A'A'})\text{dVol}_4 \,.
\end{align}
Adding the right amount of the topological term allows one to express the action only in the self-dual field strength,
\begin{align} \label{almostSDYM}
    S[F]-\frac{1}{4g^2}S_{\text{top}} = -\frac{1}{2g^2}\text{tr}\int F^{AA}F_{AA} \text{dVol}_4\,.
\end{align}
which is perturbatively equivalent to the Yang-Mills action.

In order to derive the self-dual action, consider the action
\begin{align} \label{actionSub}
    S[F,\Psi]= \tr\int \Psi^{AA}F_{AA}\text{dVol}_4+\frac{g^2}{2} \tr\int \Psi^{AA}\Psi_{AA}\text{dVol}_4 \,,
\end{align}
where the symmetric field $\Psi^{AA}$ is a Lagrange multiplier. Varying with respect to the fields, yields
\begin{align} \label{eomAlmostSDYM1}
    D\fdu{A}{A'}\Psi^{AB}&=0 \,, & F_{AA} &=-g^2\Psi_{AA}\,,
\end{align}
with $D=e^{AA'}D_{AA'}=\nabla-[A,\bullet]$ the gauge and Lorentz covariant derivative. The Lorentz covariant derivative $\nabla$ acts on a generic spin-tensor $T^{A(n),A(m)}$ as
\begin{align}
    \nabla T^{A(n),A'(m)}=dT^{A(n),A'(m)}- n\omega\fud{A}{B}T^{A(n-1)B,A'(m)}+m\omega\fud{A'}{B'}T^{A(n),A'(m-1)B'}
\end{align}
and gives
\begin{align*}
    \nabla^2T^{A(n),A'(m)}&=-n  H\fud{A}{B}T^{A(n-1)B,A'(m)}-m  H\fud{A'}{B'}T^{A(n),A'(m-1)B'} \,.
\end{align*}
on a background with constant curvature with cosmological constant $\Lambda$.

The second equation in \eqref{eomAlmostSDYM1} is purely algebraic, meaning that $\Psi_{AA}$ is an auxiliary field. Substituting it back into the action \eqref{actionSub} returns \eqref{almostSDYM}, so this is still perturbatively equivalent to the Yang-Mills action. The self-dual limit is obtained in the limit $g\rightarrow 0$, where we have $F_{AB}=0$. While it is the self-dual component of the field strength that vanishes -- and so the field strength is now anti-self-dual --, we still refer to this as the self-dual theory out of convenience. The last equation in \eqref{eomAlmostSDYM1} indicates that the degrees of freedom described by $F_{AA}$ do not vanish in the self-dual limit; they are carried over to the field $\Psi^{AA}$. To better illustrate this, we can write the decomposition of the field strength two-form as
\begin{align}
    F=-g^2H^{AA}\Psi_{AA}+H^{A'A'}F_{A'A'}\,.
\end{align}
Now, the field $F_{A'A'}$ carries positive helicity degrees of freedom and $\Psi^{AA}$ negative helicity ones.
Thus, in the self-dual limit, the negative helicity degrees of freedom survive, but they simply decouple from the field strength and the gauge fields. 

Although through different methods, self-dual gravity can be related to general relativity using the Plebanksi action \cite{Plebanski:1977zz}. For chiral HiSGRA it is still an open question whether it can be related to a parity-invariant completion of the theory, because such a theory has not yet been constructed. However, as discussed before, such a theory is conjectured to exist through the AdS/CFT correspondence, so it is likely that the above-mentioned relation generalizes to higher spins.

\paragraph{FDA for SDYM in flat space.} Here we set $\Lambda=0$. The starting point for the FDA is given by
\begin{align} \label{eomAlmostSDYM}
    D\fdu{A}{A'}\Psi^{AB}&=0 \,, & F_{AA} &=0\,,
\end{align}
i.e. the self-dual limit of \eqref{eomAlmostSDYM1}, together with the Bianchi identity
\begin{align} \label{bianchi}
    DF=H^{B'B'}\wedge DF_{B'B'}=0 \,.
\end{align}
The strategy is to find solutions to \eqref{eomAlmostSDYM} and \eqref{bianchi} that are parametrized by new fields that encode the derivatives of the original fields. For example, the first equation in \eqref{eomAlmostSDYM} does not require the derivative of $\Psi^{AA}$ to vanish completely. We can decompose $D^{BB'}\Psi^{AA}$ into its symmetric and antisymmetric components. This yields
\begin{align}
    D^{AA'}\Psi^{BC}=D^{(A|A'}\Psi^{B|C)}+\frac{1}{2}\epsilon^{AC}D\fdu{D}{A'}\Psi^{BD}\,,
\end{align}
where $X^{(A}Y^{B)}=\frac{1}{2}(X^AY^B+X^BY^A)$ indicates symmetrization and $X^AY^B-X^BY^A=\epsilon^{AB}X_CY^C$ is used to extract the antisymmetric component. Clearly, \eqref{eomAlmostSDYM} does not impose any condition on the symmetric component, while it forces the anti-symmetric one to vanish. We then introduce a new field $\Psi^{AAA,A'}$ to parametrize the symmetric component and we get
\begin{align}
    D\Psi^{AA}=e_{BB'}\Psi^{AAB,B'}\,.
\end{align}
To both this equation and \eqref{bianchi}, one can apply $D$ and employ
\begin{align}
    D^2 \chi=-[F,\chi]=-\bar{H}^{B'B'}[\bar{F}_{B'B'},\chi]\,.
\end{align}
The full FDA can be constructed by repeatedly imposing the Bianchi identity to constrain the new fields. One should keep in mind that these fields are not independent; they encode the derivatives of the physical field $\Psi^{AA}$.  

The field content of the free (Abelian) and interacting theory is the same, because both versions of the theory contain the same amount of local degrees of freedom. The field theory of the free theory is given by
\begin{align}
    dF^{A(k),A'(k+2)}=e_{BB'}F^{A(k)B,A'(k+2)B'}\,,
\end{align}
\begin{align}
    d\Psi^{A(k+2),A'(k)}=e_{BB'}\Psi^{A(k+2)B,A'(k)B'}\,.
\end{align}

In order to have a genuine FDA, \eqref{sigmamodel} implies that the vielbein and (anti-)self-dual components of the spin-connection are part of the field content. The field content is described by maps $\Phi: T[1]\mathcal{M}\rightarrow \mathcal{N}$, see \eqref{sigmamodel}. It is summarized by
\begin{align*}
    \mathcal{N}&: && 
    \begin{aligned}
        1&: e^{AA'}\,, \omega^{AB}\,, \omega^{A'B'} \,, A \,,\\
        0&: F^{A(k),A'(k+2)}\,, \Psi^{A(k+2),A'(k)}\,, k=0,1,2,...
    \end{aligned}
\end{align*}
and the FDA takes the form
\besubeqs
\begin{align}
    \begin{aligned}
        d e^{AA'}&= \omega\fud{A}{B}\wedge
        e^{BA'}+\omega\fud{A'}{B'}\wedge e^{A B'} \,,\\  
        d\omega^{AB}&= \omega\fud{A}{C}\wedge \omega^{BC} \,,\\
        d\omega^{A'B'}&= \omega\fud{A'}{C'}\wedge \omega^{B'C'} \,,\\
        dA&= AA + H_{B'B'}F^{B'B'} \,,\\
        dF&= l_2(\omega,F)+l_2(A,F)+l_2(e,F)+l_3(e,F,F) \,,\\
        d\Psi&= l_2(\omega,\Psi)+l_2(A,\Psi)+l_2(e,\Psi) +l_3(e,F,\Psi) \,.
    \end{aligned}
\end{align}
\esubeqs
The operations 
\begin{align}
    \begin{aligned}
    l_2(\omega,F)&= k\omega\fud{A}{B}F^{A(k-1)B,A'(k+2)}+(k+2)\omega\fud{A'}{B'}F^{A(k),A'(k+1)B'} \,,\\
    l_2(A,F)&=[A,F]
    \end{aligned}
\end{align}
define the bilinear maps $l_2$ corresponding to parts of the usual Lorentz and gauge covariant derivatives. To simplify the FDA, they can be absorbed into $D=\nabla-[A,\bullet]$. The term $H_{B'B'}F^{B'B'}$ is a specific tri-linear map $l_3(e,e,F)$. The FDA becomes
\besubeqs
\begin{align}
    \begin{aligned}
        \nabla e^{AA'}&= 0 \,, & \nabla^2&=0 \,,\\
        dA&= AA + H_{B'B'}F^{B'B'} \,, \\
        DF&= l_2(e,F)+l_3(e,F,F) \,, \\
        D\Psi&= l_2(e,\Psi)+l_3(e,F,\Psi) \,.
    \end{aligned}
\end{align}
\esubeqs
The first line implies that we are in flat space and the spin-connection vanishes. The associated $L_\infty$-relations are
\begin{align}
    \begin{aligned}
        D^2 F +l_2(e, DF) +l_3(e,DF,F)+l_3(e,F,DF)&\equiv0 \,,
        \\
        D^2\Psi+l_2(e,D\Psi)+l_3(e,DF,\Psi)+l_3(e,F,D\Psi)&\equiv0
    \end{aligned}
\end{align}
and decompose into
\begin{align} 
    \begin{aligned}
        l_2(e, l_2(e,F))&\equiv0  \,, \\
        -[H_{BB}F^{BB}, F] +l_2(e, l_3(e,F,F)) +l_3(e,l_2(e,F),F)+l_3(e,F,l_2(e,F))&\equiv0  \,, \\
        l_3(e,l_3(e,F,F),F)+l_3(e,F,l_3(e,F,F))&\equiv0 \,, \\
        l_2(e,l_2(e,\Psi))&\equiv0 \,, \\
        -[H_{BB}F^{BB},\Psi]+l_2(e,l_3(e,F,\Psi))+l_3(e,l_2(e,F),\Psi)+l_3(e,F,l_2(e,\Psi))&\equiv0  \,,\\
        l_3(e,l_3(e,F,F),\Psi)+l_3(e,F,l_3(e,F,\Psi))&\equiv0 \,.
    \end{aligned}
\end{align}

The full solution to the interacting theory is given by
\begin{align} \boxed{
    \begin{aligned}
    D F_{A(k),A'(k+2)}&=e^{BB'}F_{A(k)B,A'(k+2)B'}\\
    &-e\fdu{A}{B'}\sum_{n=0}^{k-1}{{\tfrac{2}{(n+1)!}\tfrac{(k+2)!}{(k-n-1)!(k-n+1)(k+1)}}}[F_{A(n),A'(n+1)B'},F_{A(k-n-1),A'(k-n+1)}]\,.
    \end{aligned}}
\end{align}
and
\begin{align}\boxed{
\begin{aligned}
    D\Psi_{A(k+2),A'(k)}&=e^{CC'}\Psi_{A(k+2)C,A'(k)C'}\\
        &-e\fdu{A}{C'}\sum_{n=0}^{k-1}\tfrac{2}{(n+1)!}\tfrac{k-n+2}{k+3}\tfrac{k!}{(k-n-1)!}[F_{A(n),A'(n+1)C'},\Psi_{A(k-n+1),A'(k-n-1)}]\\
        &+e\fdu{A}{C'}\sum_{n=0}^{k-2}\tfrac{2}{(n+2)!}\tfrac{n+1}{k+3}\tfrac{k!}{(k-n-2)!}[F_{A(n),A'(n+2)},\Psi_{A(k-n+1),A'(k-n-2)C'}]\,.
\end{aligned}}
\end{align}
It was also verified that all higher order structure maps vanish.

\paragraph{FDA for SDYM on background with constant curvature.}  The free Maxwell equations on a background with constant curvature rewritten as an FDA read \cite{Vasiliev:1986td}
\besubeqs 
\begin{align}
    dA&=H^{B'B'}F_{B'B'} +\epsilon H^{BB} \Psi_{BB} \,, \\
\label{eqMaxwBAIntro}
    \nabla F^{A(k),A'(k+2)}&=e_{BB'}F^{A(k)B,A'(k+2)B'} +k(k+2)\Lambda e^{AA'} F^{A(k-1), A'(k+1)} \,, \\
\label{eqMaxwCAIntro}
    \nabla \Psi^{A(k+2),A'(k)}&=e_{CC'}\Psi^{A(k+2)C,A'(k)C'}+k(k+2)\Lambda e^{AA'} \Psi^{A(k+1),A'(k-1)} \,,
\end{align}
\esubeqs
where $\Lambda$ is the cosmological constant, which we will set to $\Lambda=1$, and $\epsilon$ allows one to take the self-dual limit. On general grounds, the FDA for SDYM, together with the covariant constancy condition on the vielbein, reads
\begin{align*}
    \nabla e^{AA'}&= 0 \,, \\ 
    dA&= AA + H_{B'B'}F^{B'B'} \,, \\
    DF&= l_2(e,F)+\tilde{l}_2(e,F)+l_3(e,F,F) \,, \\
    D\Psi&= l_2(e,\Psi)+\tilde{l}_2(e,\Psi)+l_3(e,F,\Psi)\,,
\end{align*}
where $\tilde{l}_2$ encodes the corrections to the free equations \eqref{eqMaxwBAIntro} and \eqref{eqMaxwCAIntro}. The gravitational curvature gives
\begin{align*}
    \nabla^2T^{A(n),A'(m)}&=-n  H\fud{A}{B}T^{A(n-1)B,A'(m)}-m  H\fud{A'}{B'}T^{A(n),A'(m-1)B'} \,.
\end{align*}
Since the new FDA is a deformation of the one in flat space, the only new $L_\infty$-relations are
\besubeqs
\begin{align} 
    \tilde{l}_2(e,l_3(e,F,F))+l_3(e,\tilde{l}_2(e,F),F)+l_3(e,F,\tilde{l}_2(e,F))&=0 \,,  \\
    \tilde{l}_2(e,l_3(e,F,\Psi))+l_3(e,\tilde{l}_2(e,F),\Psi)+l_3(e,F,\tilde{l}_2(e,\Psi))&=0  \,.
\end{align}
\esubeqs
This system is solved by
\besubeqs
    \begin{align} \boxed{
        \begin{aligned}
            D& F_{A(k),A'(k+2)}=e^{BB'}F_{A(k)B,A'(k+2)B'}+k(k+2)e_{AA'}F_{A(k-1),A'(k+1)}\\
            &\qquad-e\fdu{A}{B'}\sum_{n=0}^{k-1}\tfrac{2}{(n+1)!}\tfrac{(k+2)!}{(k-n-1)!(k-n+1)(k+1)}[F_{A(n),A'(n+1)B'},F_{A(k-n-1),A'(k-n+1)}]\,,
        \end{aligned}}
    \end{align}
    \begin{align} \boxed{
        \begin{aligned}
        D\Psi_{A(k+2),A'(k)}&=e^{CC'}\Psi_{A(k+2)C,A'(k)C'}+k(k+2)e_{AA'}\Psi_{A(k+1),A'(k-1)}\\
        &-e\fdu{A}{C'}\sum_{n=0}^{k-1}\tfrac{2}{(n+1)!}\tfrac{k-n+2}{k+3}\tfrac{k!}{(k-n-1)!}[F_{A(n),A'(n+1)C'},\Psi_{A(k-n+1),A'(k-n-1)}]\\
        &+e\fdu{A}{C'}\sum_{n=0}^{k-2}\tfrac{2}{(n+2)!}\tfrac{n+1}{k+3}\tfrac{k!}{(k-n-2)!}[F_{A(n),A'(n+2)},\Psi_{A(k-n+1),A'(k-n-2)C'}]\,.
        \end{aligned}}
    \end{align}
\esubeqs
Similarly to flat space, it was verified that all higher order maps vanish.

\subsubsection{SDGR}

\paragraph{FDA for SDGR.} Self-dual gravity with vanishing cosmological constant can be formulated with the help of two fields \cite{Krasnov:2021cva}: a one-form $\omega^{AA}$ and a zero-form $\Psi^{A(4)}$. They can be obtained from the Plebanksi action \cite{Plebanski:1977zz}, which reads
\begin{align}
    \int \Psi^{A(4)}\wedge F_{AA} \wedge F_{AA} \,,
\end{align}
where we identify $F_{AA}=d\omega_{AA}$. The equations of motion are
\begin{align} \label{FF&PsiF}
    F_{AA} \wedge F_{AA}&=0 \,, &  d\Psi^{AABB}\wedge F_{BB}&=0 \,.
\end{align}
The solution to the former equation in flat space is $d\omega_{AA}=H_{AA}$, where $e^{AA'}=dx^{AA'}$. Repeated application of $d$ to this equation and to the second equation in \eqref{FF&PsiF}, yields the linearized FDA
\begin{align} 
    &\begin{aligned}
        d \omega^{AA}&=e\fud{A}{B'}\wedge e^{AB'}\,,\\
        d e^{AA'}&=\omega\fud{A}{B}\wedge e^{BA'} \,,\\
        d \omega^{A'A'}&=\omega\fud{A'}{B'}\wedge\omega^{A'B'}+H_{M'M'}C\fud{M'M'A'A'} \,, \\
        d\Psi^{A(4)}&= e_{BB'}\Psi^{A(4)B,B'}\,, 
    \end{aligned}
\end{align}
where $C^{A'(4)}$ is the anti-self-dual component of the Weyl tensor. We observe that gravitational corrections are only described by the anti-self-dual field $\omega^{A'A'}$. The free equations for helicity $\pm 2$ fields are \cite{Penrose:1965am}
\begin{align}
    \nabla\fdu{B}{A'}\Psi^{A(3)B}&=0 \,, & \nabla\fud{A}{B'}C^{A'(3)B'}&=0\,,
\end{align}
and can be rewritten in the FDA form as \cite{Vasiliev:1986td}
\begin{align}
    \nabla C^{A(k),A'(k+4)}&=e_{CC'}C^{A(k)C,A'(k+4)C'} \,, & \nabla \Psi^{A(k+4),A'(k)}&=e_{CC'}\Psi^{A(k+4)C,A'(k)C'} \,.
\end{align} 
It follows that the field content of the FDA, which again constitute the coordinates on the supermanifold $\mathcal{N}$, is
\begin{align}
    \mathcal{N}&: && 
    \begin{aligned}
        1&: \omega^{A'B'}\,,e^{AA'}\,, \omega^{AB} \,,\\\
        0&: C^{A(k),A'(k+4)}\,, \Psi^{A(k+4),A'(k)}\,, k=0,1,2,...
    \end{aligned}
\end{align}
We define the covariant derivative $\nabla=d-\omega$, which lacks the $\omega^{A'B'}$-part. For an arbitrary spin-tensor $T^{A(n),A'(m)}$ we get
\begin{align} 
    \nabla^2T^{A(n),A'(m)}=-mH_{M'M'}C\fud{M'M'A'}{B'}T^{A(n),B'A'(m-1)} \,.
\end{align}
These $L_\infty$-relations then read
\begin{align}
    \begin{aligned}
        -(k+4) H_{MM}C\fud{MMA}{B}C^{A(k+3)B,A'(k)}+l_2(e,\nabla C)+l_3(e,\nabla C,C)+l_3(e,C,\nabla C)&=0 \,, \\
        -k H_{MM}C\fud{MMA}{B}\Psi^{A(k-1)B,A'(k+4)}+l_2(e,\nabla\Psi)+l_3(e,\nabla C,\Psi)+l_3(e,C,\nabla\Psi)&=0\,,
    \end{aligned}
\end{align}
and decompose into
\begin{align} 
        &l_2(e,l_2(e,C))=0\,, \\
           &l_3(e,l_3(e,C,C),C)+l_3(e,C,l_3(e,C,C))=0 \,, \\
       & l_2(e,l_2(e,\Psi))=0  \,, \\     
       & l_3(e,l_3(e,C,C),\Psi)+l_3(e,C,l_3(e,C,\Psi))=0  \,,\\
        &\begin{aligned}
            -(k+4) H_{MM}C\fud{MMA}{B}C^{A(k+3)B,A'(k)}&+l_2(e,l_3(e,C,C))       \\
            &+l_3(e,l_2(e,C),C)+l_3(e,C,l_2(e,C))=0  \,,
        \end{aligned}\\
        &\begin{aligned}
           -kH_{MM}C\fud{MMA}{B}\Psi^{A(k-1)B,A'(k+4)}&+l_2(e,l_3(e,C,\Psi))\\&+l_3(e,l_2(e,C),\Psi)+l_3(e,C,l_2(e,\Psi))=0\,. 
        \end{aligned}
\end{align}
The solution to this system is found to be given by
\begin{align} \boxed{
    \begin{aligned}
        \nabla C_{A(k),A'(k+4)}&=e^{CC'}C_{A(k)C,A'(k+4)C'}\\
        &+\sum_{n=0}^{k-1}\tfrac{2}{(n+2)!}\tfrac{(k+4)!(k-n)}{(k-n+2)!(k+1)}e\fdu{A}{C'}C\fdu{A(n),A'(n+2)C'}{D'}C_{A(k-n-1),A'(k-n+2)D'}
    \end{aligned}}
\end{align}
and
\begin{align}  \boxed{
    \begin{aligned}
        \nabla\Psi_{A(k+4),A'(k)}&=e^{CC'}\Psi_{A(k+4)C,A'(k)C'}\\
        &+\sum_{n=0}^{k}\tfrac{2}{(n+2)!}\tfrac{k!}{(k-n-2)!}\tfrac{k-n+4}{k+5}e\fdu{A}{C'}C\fdu{A(n),A'(n+2)C'}{D'}\Psi_{A(k-n+3),A'(k-n-2)D'}\\
        &-\sum_{n=0}^{k}\tfrac{2}{(n+2)!}\tfrac{k!}{(k-n-3)!}\tfrac{n+1}{(k+5)(n+3)}e\fdu{A}{C'}C\fdu{A(n),A'(n+3)}{D'}\Psi_{A(k-n+3),A'(k-n-3)C'D'}\,.
    \end{aligned}}
\end{align}
Again, it was proved that the FDA is consistent without higher order structure maps.

\subsection{Cubic interactions for Chiral HiSGRA}
\label{subsec:cubic}
\begin{figure}[h!]
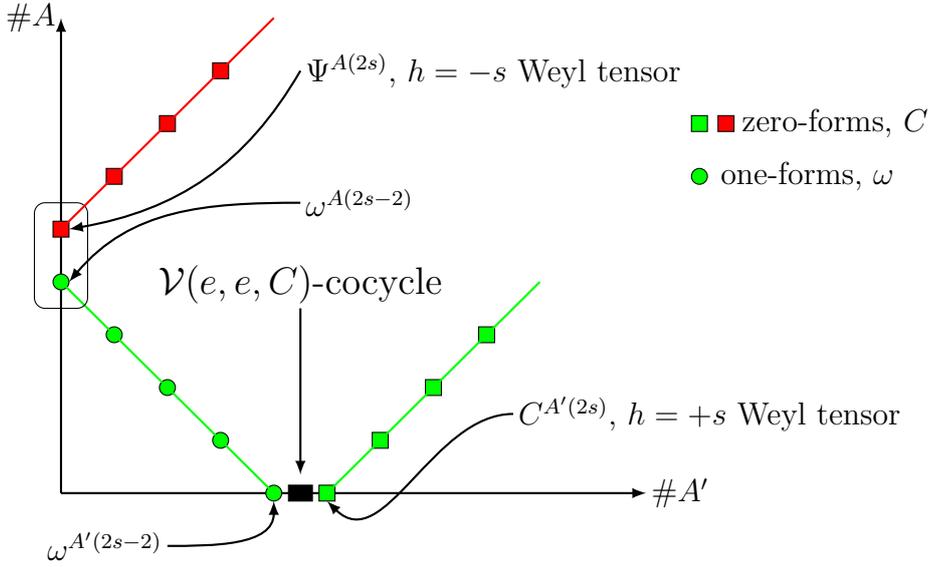

\FIELDS
  \caption{Field content of HiSGRA. Along the axes we have the number of unprimed/primed indices on a spin-tensor. The black square shows a cocycle that links the one-form sector to zero-forms (at the free level it relates two fields for each spin's subsystem). The two fields in the rounded rectangle enter the free action. The rest of the fields encode derivatives thereof in a coordinate invariant and background independent way. The solid lines link pairwise the fields that `talk' to each other in the free equations. }\label{fig:figure1Into}
\end{figure}

We start by constructing the cubic interaction vertex on a background with arbitrary cosmological constant. We relate to this the parameter $\lambda=\sqrt{|\Lambda|}$. The free action for chiral HiSGRA is
\begin{align}\label{actionIntro}
    S= \int \Psi^{A(2s)}\wedge H_{AA}\wedge \nabla \omega_{A(2s-2)}\,,
\end{align}
with $\omega_{A(2s-2)}=\Phi_{A(2s-2)B,B'}e^{BB'}$ encoding the higher spin gauge potential. The fields take values in a Lie algebra with a non-degenerate invariant bilinear form, so they can be assumed to take values in $\text{Mat}_N$. The action is easily seen to generalize the SDYM and SDGR actions. The equations of motion read
\begin{align}
    \nabla \Psi^{A(2s)}\wedge H_{AA}&=0\,, && H^{AA}\wedge \nabla \omega^{A(2s-2)}=0\,.
\end{align}
Following the same procedure as for SDYM and SDGR, the FDA of the free chiral theory is ($n=2s-2$)
\begin{align}
    \begin{aligned}
        \nabla \omega^{A(n)}&= e\fud{A}{B'} \omega^{A(n-1),B'}\,,\\
        \nabla \omega^{A(n-i),A'(i)}&= e\fud{A}{B'} \omega^{A(n-i-1),A'(i)B'}+\lambda e\fdu{B}{A'}\omega^{A(n-i)B,A'(i-1)}\,, && i=1,...,n-1\,,\\
        \nabla \omega^{A'(n)}&= H_{B'B'} C^{A'(n)B'B'}+\lambda e\fdu{B}{A'}\omega^{B,A'(n-1)}\,,\\
        \nabla C^{A'(n+k+2)}&= e_{BB'} C^{B,A'(n+k+2)B'}\,,\\
        \nabla C^{A(k),A'(n+k+2)}&= e_{BB'} C^{A(k)B,A'(n+k+2)B'}+\lambda e^{AA'}C^{A(k-1),A'(n+k-1)}\,, && k=1,2,...\,,\\
        \nabla \Psi^{A(n+2)}&= e_{BB'} \Psi^{A(n+2)B,B'}\,,\\
        \nabla \Psi^{A(n+k+2),A'(k)}&= e_{BB'} \Psi^{A(k+n+2)B,A'(k)B'}+\lambda e^{AA'}\Psi^{A(n+k+1),A'(k-1)}\,, && k=1,2,...
    \end{aligned}
\end{align}
and after renaming $\Psi^{A(n),A'(m)} \rightarrow C^{A(n),A'(m)}$, where $n-m=2s,$ the field content reads
\begin{align}
    \begin{aligned}
        h&=+s: && \omega^{A(2s-2-k),A'(k)}\,, C^{A(i),A'(2s+i)} \,, \qquad k=0,...,2s-2\,,\quad i=0,1,2,... \,,\\
        h&=-s: && C^{A(2s+i),A'(i)}\,,\qquad i=0,1,2,... \,,\\
        h&=0: && C^{A(i),A'(i)}\,,\qquad i=0,1,2,...\,.
    \end{aligned}
\end{align}
The only physical fields are the ones present in the action, while all other fields are derivatives thereof. The field content is visually represented in Figure \ref{fig:figure1Into}.

It will prove useful to collect all fields in generating functions,
\begin{align}
    \begin{aligned}
    \omega(y,\bry)&= \sum_{n,m}\tfrac{1}{n!m!} \omega_{A(n),A'(m)}\, y^A...y^A\, \bry^{A'}...\bry^{A'} \,,\\
    C(y,\bar{y})&=\sum_{n,m}\tfrac{1}{n!m!} C_{A(n),A'(m)}\, y^A...y^A\, \bry^{A'}...\bry^{A'} \,.
    \end{aligned}
\end{align}
The free equations become
\begin{align} \label{nablaOmega&nablaC}
    \begin{aligned}
    \nabla\omega &= e^{BB'}(\lambda \bar{y}_{B'}\partial_B +y_{B} \pl_{B'}) \omega +H^{B'B'} \pl_{B'}\pl_{B'}C(y=0,\bry)\,,\\
    \nabla C&= e^{BB'}(\lambda y_B\bar{y}_{B'}-\pl_B \pl_{B'}) C\,.
    \end{aligned}
\end{align}
Decomposing the covariant derivative $\nabla$, we obtain
\begin{align}
    d\omega &= \mathcal{V}(\omega_0, \omega) +\mathcal{V}(\omega_0,\omega_0,C)\,,& 
    d C&= \mathcal{U}(\omega_0, C)\,,
\end{align}
which is the linearized version of \eqref{toSolve}, where the spin-$2$ sector of $\omega$ is set to its background value $\omega_0$.

This allows us to relate the equations in \eqref{nablaOmega&nablaC} to boundary conditions for the structure maps of the $L_\infty$-algebra. These boundary conditions read
\begin{align}\label{bcIntro}
    \begin{aligned}
        \mathcal{V}(e,\omega)+\mathcal{V}(\omega,e)&=e^{CC'}(\lambda \bar{y}_{C'}\partial_C + y_C \pl_{C'}) \omega\,,\\
        \mathcal{U}(e,C)+\mathcal{U}(C,e)&=e^{CC'} (\lambda y_C\bar{y}_{C'}-\pl_C \pl_{C'}) C\,,\\
        \mathcal{V}(e,e,C)&= e\fdu{B}{C'}e^{BC'} \pl_{C'} \pl_{C'} C(y=0,\bry)\,,
    \end{aligned}
\end{align}
with $e=e^{AA'}y_A\bar{y}_{A'}$.

The structure maps take on the form of poly-diffential operators
\begin{align}
    \mathcal{V}(f_1,...,f_n)&= \mathcal{V}(y, \pl_1,...,\pl_2)\, f_1(y_1)...f_n(y_n) \Big|_{y_i=0}\,,
\end{align}
where $f_i$'s are $\omega$'s or $C$'s. We have explicitly indicated dependence of $y$'s; the $\bry$'s can be treated similarly. Our notation can be summarized as follows: (i) we abbreviate $y^{A}\equiv p_0^{A}$, $\pl^{y_i}_{A}\equiv p_{A}^i$, $\bry^{A'}\equiv q_0^{A'}$, $\pl^{\bry_i}_{A'}\equiv q_{A'}^i$; (ii) contractions $p_{ij}\equiv p_i \cdot p_j\equiv -\epsilon_{AB}p^A_{i}p_{j}^B=p^A_{i}p_{jA}$ are done in such a way that $\exp[p_0\cdot p_i]f(y_i)=f(y_i+y)$; (iii) Lorentz invariance forbids to mix primed and unprimed indices; (iv) all indices must be contracted with $\epsilon_{AB}$ or $\epsilon_{A'B'}$; (v) explicit arguments $y_i$ in $f$'s and the symbol  $|_{y_i=0}$ are omitted. Importantly, all operators are assumed to be local, i.e. they send polynomials to polynomials.

The variables $y$ and $\bar{y}$ are treated unequally and the structure map can be written as
\begin{align}
    \mathcal{V}(f_1,...,f_n)&= v(f_1'(y),..., f_n'(y)) \otimes f_1''(\bry)\star... \star f_n''(\bry)  \,,
\end{align}
where $f_i(y,\bry)=f_i'(y) \otimes f_i''(\bry)$. All $\bry$-dependent factors are multiplied via the star-product:
\begin{align}
    f_1''(\bry)\star ...\star f_n''(\bry)&= \exp{\left[\sum_{0=i<j=n} r_i \cdot r_j\right]} f_1''(\bry_1)...f_n''(\bry_n)\Big|_{\bry_i=0}\,.
\end{align}
Thus, we have the Weyl algebra $A_1$ acting on the $\bar{y}$'s.

The goal is to solve the $L_\infty$-relations \eqref{toSolve} with the boundary conditions \eqref{bcIntro}. To illustrate the procedure, let us go through a few examples. At the lowest order in $C$ we find
\begin{align}
    \mathcal{V}(\mathcal{V}(\omega,\omega),\omega)-\mathcal{V}(\omega,\mathcal{V}(\omega,\omega))=0 \,.
\end{align}
This simply means that $\omega \in \mathfrak{hs}$ and the first structure map $\mathcal{V}(\omega,\omega)$ is an associative product in the higher spin algebra $\mathfrak{hs}$. The structure map is found to be
\begin{empheq}[box=\fbox]{align} 
     \mathcal{V}(\omega,\omega)&=\exp{[p_{01}+p_{02}]} \,.
\end{empheq}
To improve readability, we omitted the component of the structure map acting on the $\bar{y}$ and we also omitted the field this poly-differential operator acts on in the right-hand side. This result shows that the structure map $\mathcal{V}(\bullet,\bullet)$ is the commutative product on the commutative algebra on functions $\mathbb{C}[y]$. The higher spin algebra thus reads $\mathfrak{hs}=\mathbb{C}[y]\otimes A_1\otimes \text{Mat}_N$.

The structure map $\mathcal{U}(\omega,C)$ obeys
\begin{align}
    \mathcal{U}(\mathcal{V}(\omega,\omega),C)-\mathcal{U}(\omega,\mathcal{U}(\omega,C))=0 \,,
\end{align}
which implies that $\mathcal{U}(\bullet,\bullet)$ defines a representation of the algebra, so the fields $C$ reside in the module of $\mathfrak{hs}$. In fact, the action \eqref{actionIntro} suggests that this module is $\mathfrak{hs}^\star$, as the action is a functional $S:\mathfrak{hs}\otimes\mathfrak{hs}^\star \rightarrow\mathbb{R}$.

Because the fields $\omega$ and $C$ take values in $\text{Mat}_N$, we must distinguish between different orderings of the fields. We write
\begin{align}
    \mathcal{U}(\omega,C)=\mathcal{U}_1(\omega,C)+\mathcal{U}_2(C,\omega)\,,
\end{align}
where $\mathcal{U}_1(\omega,\bullet)$ and $\mathcal{U}_2(\bullet,\omega)$ encode the left and right action of the higher spin algebra $\mathfrak{hs}$ on the bimodule, respectively. These maps are now structure maps of an underlying $A_\infty$-algebra, from which our $L_\infty$-algebra is obtained by graded anti-symmetrizing its structure maps. The integrability condition gives
\begin{align}
    \begin{aligned}
    \mathcal{U}_1(\mathcal{V}(\omega,\omega),C)-\mathcal{U}(\omega,\mathcal{U}_1(\omega,C)) &=0 \,,\\
    \mathcal{U}_2(\mathcal{U}_1(\omega,C),\omega)-\mathcal{U}_1(\omega,\mathcal{U}_2(C,\omega)) &=0 \,,\\
    \mathcal{U}_2(C,\mathcal{V}(\omega,\omega))+\mathcal{U}_2(\mathcal{U}_2(C,\omega),\omega) &=0 \,,
    \end{aligned}
\end{align}
which are exactly the defining relations of the bimodule over $\mathfrak{hs}$. The solution is given by
\begin{equation}\boxed{
    \begin{aligned}
        &\mathcal{U}_1(\omega,C)=\exp{[p_{02}+p_{12}]} \,,\\
        &\mathcal{U}_2(C,\omega)=-\exp{[p_{01}-p_{12}]} \,,
    \end{aligned}}
\end{equation}
which realizes the left and right coadjoint action of $\mathfrak{hs}$ on its dual space $\mathfrak{hs}^\star$.

The cubic map $\mathcal{V}(\omega,\omega,C)$ is decomposed into
\begin{align}
    \mathcal{V}(\omega,\omega,C)=\mathcal{V}_1(\omega,\omega,C)+\mathcal{V}_2(\omega,C,\omega)+\mathcal{V}_3(C,\omega,\omega)    \,.
\end{align}
The consistency conditions are obtained in the same manner as before and the solution is given by
\begin{empheq}[box=\fbox]{align}\label{cubicResult}
    \mathcal{V}_1(\omega,\omega,C)&: && +p_{12} \int_{\Delta_2}\exp[\left(1-t_1\right) p_{01}+\left(1-t_2\right) p_{02}+t_1 p_{13}+t_2 p_{23}]\,, \\
    \mathcal{V}_2(\omega,C,\omega)&: &&\begin{aligned}
       -p_{13}& \int_{\Delta_2}\exp[\left(1-t_2\right) p_{01}+\left(1-t_1\right) p_{03}+t_2 p_{12}-t_1 p_{23}]+\\
       -&p_{13} \int_{\Delta_2}\exp[\left(1-t_1\right) p_{01}+\left(1-t_2\right) p_{03}+t_1 p_{12}-t_2 p_{23}]\,,
    \end{aligned}
    \\
    \mathcal{V}_3(C,\omega,\omega)&: && +p_{23} \int_{\Delta_2}\exp[\left(1-t_2\right) p_{02}+\left(1-t_1\right) p_{03}-t_2 p_{12}-t_1 p_{13}]\,.
\end{empheq}
Here, $\Delta_2$ is the $2$-simplex described by $0 \leq t_1 \leq t_2 \leq 1$.

Similarly, we split the structure map $\mathcal{U}(\omega,C,C)$ into
\begin{align}
    \mathcal{U}(\omega,C,C)=\mathcal{U}_1(\omega,C,C)+\mathcal{U}_2(C,\omega,C)+\mathcal{U}_3(C,C,\omega)   \,. 
\end{align}
The consistency conditions require them to take the form
\begin{empheq}[box=\fbox]{align}
    \mathcal{U}_1(\omega,C,C)&: && +p_{01} \int_{\Delta_2}\exp[\left(1-t_2\right) p_{02}+t_2 p_{03}+\left(1-t_1\right) p_{12}+t_1 p_{13}]\,, \\
    \mathcal{U}_2(C,\omega,C)&: &&\begin{aligned}
       -p_{02}& \int_{\Delta_2}\exp[t_2 p_{01}+\left(1-t_2\right) p_{03}-t_1 p_{12}+\left(1-t_1\right) p_{23}]+\\
       -&p_{02} \int_{\Delta_2}\exp[t_1 p_{01}+\left(1-t_1\right) p_{03}-t_2 p_{12}+\left(1-t_2\right) p_{23}]\,,
    \end{aligned}
    \\
    \mathcal{U}_3(C,C,\omega)&: && +p_{03}\, \int_{\Delta_2}\exp[\left(1-t_1\right) p_{01}+t_1 p_{02}+\left(t_2-1\right) p_{13}-t_2 p_{23}]\,.
\end{empheq}

The structure maps that were obtained can be verified by using them to compute the cubic tree level scattering amplitude, as is done in Appendix \ref{app:amplitude}. These calculations confirm that the structure maps are correct. They also survive another, independent test: the FDA obtained for the FDA for SDYM and SDGR are found from Taylor expansion of the structure maps. Moreover, they are directly related to the FFS cocycle obtained in \cite{FFS}.

While chiral HiSGRA in light-cone gauge only admits cubic interactions, there is no guarantee that the covariant formulation in terms of the underlying $L_\infty$-algebra truncates at cubic level. In fact, it is easy to write down the next order of $L_\infty$-relations and check if they are consistent with vanishing quartic terms and find that this is not the case. This implies that the higher order maps are required to ensure Lorentz covariance; this covariance is manifestly broken in light-cone gauge. Another improvement of the obtained structure maps would be the generalization to define them on any background with constant curvature. Both enhancements will be made in the next section.

\subsection{Higher order interactions}
While it is easy to verify that the $L_\infty$-relations fail for vanishing quartic structure maps, solving them by brute force turns out to be an inefficient strategy. Instead, we will rely on HPT to obtain all higher order structure maps. The cubic map provides a perturbation away from the free theory, from which a HPT recipe can be cooked up. The more general details of HPT are given in Appendix \ref{app:hpt} and the recipe is elaborated on more in Appendix \ref{app:homo}. Here we will only demonstrate how to build structure maps on a background with arbitrary cosmological constant $\Lambda$. Instead of the cosmological constant itself, we will use the parameter $\lambda=\sqrt{|\Lambda|}$ and refer to this as the cosmological constant for convenience.

\paragraph{Homological perturbation recipe.}
The main idea behind the HPT approach is that we enlarge the algebra of polynomial functions $\mathbb{C}[y^A]$ to $\mathbb{C}[y^A,z^A,dz^A]$, with exterior differential $d_z$ in $z$. Instead of deforming $\mathfrak{hs}$ directly, we deform the DGA $(\mathbb{C}[y^A,z^A,dz^A],d_z)$ and map this deformation back to $\mathfrak{hs}$.  The product in $\mathbb{C}[y^A,z^A,dz^A]$ is the composition of the wedge product and the star-product
\begin{align}
    (f\star g) (Y)=\exp(Y^a\partial^1_a+Y^a\partial^2_a+\Omega^{ab}\partial^1_a\partial^2_b)f(Y_1)g(Y_2)|_{Y_{1,2}=0} \,,
\end{align}
where $Y^a\equiv (y^A,z^A)$ and the matrix $\Omega^{ab}$ is given by
\begin{align}
    \Omega^{ab}=-\begin{pmatrix}
        \lambda \epsilon & \epsilon \\
        -\epsilon & 0
    \end{pmatrix} \,.
\end{align}
Alternatively, it is more convenient for our purposes to write the star-product in the integral form 
\begin{align}
    (f\star g)(y,z)&= \int du\,dv\,dp\,dq\,f(y+u,z+v) g(y+q,z+p) \exp(v\cdot q-u\cdot p +\lambda\, q\cdot u)\,.
\end{align}
Using the Poincar\'e lemma, we also define the homotopy operator $h$. Given a cochain complex $(V,d)$, a contracting homotopy operator is a morphism of degree $-1$ $h:V\rightarrow V$, such that 
\begin{align} \label{ch}
    dh+hd=\text{id}_V \,.
\end{align}
It acts on one-forms $f^{(1)}=dz^Af^{(1)}_A(z)$ and two-forms $f^{(2)}=\frac{1}{2}f^{(2)}(z)\epsilon_{AB}dz^A\wedge dz^B$ as
\begin{align}
    f^{(1)}&=h[f^{(2)}]=dz^A z_{A}\int_0^1 t dt f^{(2)}(tz) \,, & f^{(0)}&=h[f^{(1)}]=z^A\int_{0}^{1} dt f_{A}^{(1)}(tz)
\end{align}
and annihilates zero-forms $f^{(0)}$, i.e. $h[f^{(0)}]=0$, see also \cite{Didenko:2014dwa}.

We will construct the structure maps using connected graphs. The star-product, here denoted by $\mu$, represents a trivalent vertex with two incoming legs and one outgoing. Internal lines correspond to the homotopy operator. Each diagram may be decorated by one or two $\omega$'s and arbitrarily many $C$'s. Graphs with two $\omega$'s depict $\mathcal{V}$ structure maps, while graphs with one $\omega$ represent $\mathcal{U}$ structure maps. All $C$'s enter via $\Lambda[C]=h[C\diamond \frac{1}{2}\varkappa dz^2]$, with $\varkappa=\exp[z^Ay_A]$ and
\begin{align}
    C\diamond g(z,y) = g(z,y+p_i)C(y_i)\Big|_{y_i=0} \,.
\end{align}
In order to construct bigger trees, it is convenient to label the fields. We write $\alpha_i\in\mathbb{A}_0$ for the $C$ fields and $a,b\in\mathbb{A}_1$ for the $\omega$ fields, where $\mathbb{A}_0$ and $\mathbb{A}_1$ are the zero-form and one-form sector of the $A_\infty$-algebra to be constructed via HPT, respectively. Concretely, the $A_\infty$-algebra is $\hat{\mathbb{A}}=\big(\mathbb{A}_1\oplus \mathbb{A}_0\big)\otimes A_1\otimes \text{Mat}_M$ with $\mathbb{A}_1=\mathbb{C}[y][-1]$ and $\mathbb{A}_0=\mathbb{C}[y]^\star\cong \mathbb{C}[[\partial_y]]$. We expect this structure to deform in (A)dS. Structure maps are obtained by computing all admissible graphs with the corresponding order of fields.
As an example, we consider the vertex $\mathcal{V}(\omega,\omega,C)$. We find
$$
   \mathcal{V}_1(\omega,\omega,C)=\omega(y) \star h[ \omega(y) \star \Lambda[C] ]|_{z=0}= \begin{tikzcd}[column sep=small,row sep=small]
   & {}& \\
    & \mu\arrow[u]  & \\
    \omega\arrow[ur]  & & \mu\arrow[ul, "h" ']   & \\
    & \omega \arrow[ur]& &\Lambda[C]\arrow[ul] \,,
\end{tikzcd}
$$
its mirror image
$$
   \mathcal{V}_3(C,\omega,\omega)= h[\Lambda[C] \star  \omega(y)  ] \star \omega(y)|_{z=0} = \begin{tikzcd}[column sep=small,row sep=small]
   && {} \\
    && \mu\arrow[u]   \\
    &\mu\arrow[ur, "h"]  & & \omega \arrow[ul]    \\
    \Lambda[C] \arrow[ur]& &\omega\arrow[ul] &&
\end{tikzcd}
$$
and the middle vertex receives contributions from two graphs
\begin{align*}
    \mathcal{V}_2(\omega,C,\omega)&=\omega(y) \star h[  \Lambda[C] \star\omega(y) ]|_{z=0}+ h[  \omega(y)\star  \Lambda[C]] \star \omega(y)|_{z=0}=
\end{align*}
$$
=\begin{tikzcd}[column sep=small,row sep=small]
   & {}& \\
    & \mu\arrow[u]  & \\
    \omega\arrow[ur]  & & \mu\arrow[ul, "h" ']   & \\
    & \Lambda[C] \arrow[ur]& &\omega \arrow[ul]
\end{tikzcd} \oplus\begin{tikzcd}[column sep=small,row sep=small]
   && {} \\
    && \mu\arrow[u]   \\
    &\mu\arrow[ur, "h"]  & & \omega \arrow[ul]    \\
    \omega \arrow[ur]& &\Lambda[C]\arrow[ul] &&
\end{tikzcd}   
$$

Let us get familiar with the procedure by considering the construction of $\mathcal{V}(\omega,\omega,C)$ as an example. We start with
\begin{align}
    \Lambda[C] = dz^A z_A \int_0^1 dt \,t\varkappa(tz,y+p_3)\alpha(y_3)\Big|_{y_i=0} \,.
\end{align}
Next, we have
\begin{align}
    b(y)\star\Lambda[C] = dz^A(z_A+p_A^2)e^{yp_2}\int_0^1 dt \,t\varkappa(tz+tp_2,y+p_3+\lambda p_2)b(y_2)\alpha(y_3)\Big|_{y_i=0}
\end{align}
and then we apply the homotopy operator,
\begin{align}
    h[b(y)\star\Lambda[C]] = (z\cdot p_2)e^{yp_2}\int_0^1 dt'\int_0^1 dt \,t\varkappa(tt'z+tp_2,y+p_3+\lambda p_2)b(y_2)\alpha(y_3)\Big|_{y_i=0}\,.
\end{align}
Lastly, we take another star-product and set $z=0$. We get
\begin{align}
    a\star h[b\star \Lambda[C]]\Big|_{z=0} = p_{12}e^{p_1+y_2}\int_0^1dt' \int_0^1dt\,t\varkappa(tt'p_1+tp_2,y+p_3+\lambda p_1+\lambda p_2)a(y_1)b(y_2)\alpha(y_3)\Big|_{y_i=0}\,.
\end{align}
We then perform a change of integration variables $u=t\,t'$ and $v=t$ and we obtain the extension of $\mathcal{V}(\omega,\omega,C)$ in \eqref{cubicResult} to arbitrary $\lambda$.

Explicitly, for the bilinear maps we find
\begin{align} 
    \begin{aligned}
    \mathcal{V}(\omega,\omega) &= \exp[p_{01} + p_{02} + \lambda p_{12}] \,, \\
    \mathcal{U}(\omega,C) &= \exp[p_{02} + p_{12} + \lambda p_{02} ] \,, \\
    \mathcal{U}(C,\omega) &= - \exp[-p_{12} + p_{02} - \lambda p_{01}] \,.
    \end{aligned}
\end{align}
This defines the sectors $\mathbb{A}_1=A_\lambda[-1]$ and $\mathbb{A}_0=A_\lambda^\star$ of the $A_\infty$-algebra underlying chiral HiSGRA in (A)dS. Here, the algebra $A_\lambda$ interpolates between the commutative algebra $A_0=\mathbb{C}[y]$ and the Weyl algebra $A_1$.

\paragraph{Quartic vertex.} Before we give the general solution, let us present the quartic vertex. It was checked that they satisfy the next order in the $L_\infty$-relations. Here, there are $6$ different ways to order the arguments $\omega^2 C^2$. We find
{\allowdisplaybreaks
\begin{align*}
    \mathcal{V}_1(\omega,\omega,C,C)&=(p_{12})^2\int_{\mathcal{D}_1}\exp((1-u_1-u_2)p_{01}+(1-v_1-v_2)p_{02}+u_1 p_{13}+u_2 p_{14}+v_1 p_{23}+v_2 p_{24}+\\
    &+\lambda p_{12}(1+u_1+u_2-v_1-v_2+u_1 v_2-u_2 v_1))\,,\\
    \mathcal{V}_2(\omega,C,\omega,C)&=-(p_{13})^2\int_{\mathcal{D}_1}\exp(p_{01}(1-u_1-u_2)+(1-v_1-v_2)p_{03}+u_2 p_{12}+u_1 p_{14}-v_2 p_{23}+v_1 p_{34}+\\
    &+\lambda p_{13}(1+u_1-u_2-v_1-v_2-u_1 v_2+u_2 v_1))\\
    &-(p_{13})^2\int_{\mathcal{D}_1}\exp(p_{01}(1-u_1-u_2)+(1-v_1-v_2)p_{03}+u_1 p_{12}+u_2 p_{14}-v_1 p_{23}+v_2 p_{34}+\\
    &+\lambda p_{13}(1-u_1+u_2-v_1-v_2+u_1 v_2-u_2 v_1)) + \\
    &-(p_{13})^2\int_{\mathcal{D}_2}\exp((1-u^R-v^L)p_{01}+(1-u^L-v^R)p_{03}+v^L p_{12}+u^R p_{14}-u^L p_{23}+v^R p_{34}\\
    &+\lambda p_{13}(1-u^L+u^R-v^L-v^R-u^L u^R+v^L v^R)) \,,\\
    \mathcal{V}_3(\omega,C,C,\omega)&=(p_{14})^2\int_{\mathcal{D}_1}\exp((1-u_1-u_2)p_{01}+(1-v_1-v_2)p_{04}+u_2 p_{12}+u_1 p_{13}-v_2 p_{24}-v_1 p_{34}+\\
    &+\lambda p_{14}(1-u_1-u_2-v_1-v_2-u_1 v_2+u_2 v_1))-\\
    &+(p_{14})^2\int_{\mathcal{D}_1}\exp((1-v_1-v_2)p_{01}+(1-u_1-u_2)p_{04}+v_1 p_{12}+v_2 p_{13}-u_1 p_{24}-u_2 p_{34}+\\
    &+\lambda p_{14}(1-u_1-u_2-v_1-v_2-u_1 v_2+u_2 v_1))-\\
    &+(p_{14})^2\int_{\mathcal{D}_2}\exp((1-u^R-v^L)p_{01}+(1-u^L-v^R)p_{04}+v^L p_{12}+u^R p_{13}-u^L p_{24}-v^R p_{34}\\
    &+\lambda p_{14}(1-u^L-u^R-v^L-v^R-u^L u^R+v^L v^R)) \,,\\
    \mathcal{V}_4(C,\omega,\omega,C)&=(p_{23})^2\int_{\mathcal{D}_2}\exp((1-u^R-v^L)p_{02}+(1-u^L-v^R)p_{03}-v^L p_{12}-u^L p_{13}+u^R p_{24}+v^R p_{34}\\
    &+\lambda p_{23}(1+u^L+u^R-v^L-v^R-u^L u^R+v^L v^R))\,,\\
    \mathcal{V}_5(C,\omega,C,\omega)&=-(p_{24})^2\int_{\mathcal{D}_1}\exp((1-v_1-v_2)p_{02}+(1-u_1-u_2)p_{04}-v_2 p_{12}-u_2 p_{14}+v_1 p_{23}-u_1 p_{34}+\\
    &+\lambda p_{24}(1-u_1+u_2-v_1-v_2+u_1 v_2-u_2 v_1))+\\
    &-(p_{24})^2\int_{\mathcal{D}_1}\exp((1-v_1-v_2)p_{02}+(1-u_1-u_2)p_{04}-v_1 p_{12}-u_1 p_{14}+v_2 p_{23}-u_2 p_{34}+\\
    &+\lambda p_{24}(1+u_1-u_2-v_1-v_2-u_1 v_2+u_2 v_1))+\\
    &-(p_{24})^2\int_{\mathcal{D}_2}\exp((1-u^R-v^L)p_{02}+(1-u^L-v^R)p_{04}-v^L p_{12}-u^L p_{14}+u^R p_{23}-v^R p_{34}\\
    &+\lambda p_{24}(1+u^L-u^R-v^L-v^R-u^L u^R+v^L v^R))\,,\\
    \mathcal{V}_6(C,C,\omega,\omega)&=(p_{34})^2\int_{\mathcal{D}_1}\exp((1-v_1-v_2)p_{03}+(1-u_1-u_2)p_{04}-v_2 p_{13}-u_2 p_{14}-v_1 p_{23}-u_1 p_{24}+\\
    &+\lambda p_{34}(1+u_1+u_2-v_1-v_2+u_1 v_2-u_2 v_1))\,,\\
\end{align*}}\noindent
with integration variables
\begin{align*}
    u_1&\equiv\frac{t_1t_2(1-t_3)t_4}{1-t_1t_2t_3}\,, & v_1&\equiv\frac{t_1(1-t_2t_3)}{1-t_1t_2t_3} \,,\\
    u_2&\equiv\frac{(1-t_1t_2)t_3t_4}{1-t_1t_2t_3}\,, & v_2&\equiv\frac{(1-t_1)t_3}{1-t_1t_2t_3} \,,
\end{align*}
which belong to some integration domain $\mathcal{D}_1$ and
\begin{align*}
    u^L&\equiv\frac{t_1t_2(1-t_3)}{1-t_1t_2t_3t_4}\,, & v^L&\equiv\frac{t_1(1-t_2t_3t_4)}{1-t_1t_2t_3t_4}\,,\\
    u^R&\equiv\frac{(1-t_1)t_3t_4}{1-t_1t_2t_3t_4}\,, & v^R&\equiv\frac{t_3(1-t_1t_2t_4)}{1-t_1t_2t_3t_4} 
\end{align*}
parametrizing the domain $\mathcal{D}_2$. In order to identify these domains, we invert the above relations and use $0 \leq t_i \leq 1$. For $\mathcal{D}_1$ we find
\begin{align*}
    t_1&=\frac{u_2v_1(1-v_1-v_2)+u_1v_2(v_1+v_2)}{u_1v_2+u_2(1-v_1-v_2)} \,, & t_3&=\frac{v_2}{1-v_1}  \,, \\
    t_2&=\frac{u_1v_2}{u_2v_1(1-v_1-v_2)+u_1v_2(v_1+v_2)} \,, & t_4&=u_1+u_2\frac{1-v_1}{v_2} \,.
\end{align*}
It was found that the coordinates $u_i$ and $v_i$ belong to the interval $[0,1]$ and that they satisfy
\begin{align*}
    0\leq v_2 \leq 1\,,\qquad 0\leq u_1\leq v_1\leq 1-v_2\,,\qquad \frac{u_1}{v_1}\leq \frac{u_2}{v_2}\leq \frac{1-u_1}{1-v_1} \,.
\end{align*}
As a result, the integrals should be evaluated as
\begin{align*}
    \int_{\mathcal{D}_1}&\equiv\int_0^1 dv_2 \int_0^{1-v_2} dv_1 \int_0^{v_1} du_1 \int_{\frac{u_1 v_2}{v_1}}^{v_2\frac{1-u_1}{1-v_1}} du_2 \,.
\end{align*}
For $\mathcal{D}_2$ we find the inverted relations
\begin{align*}
    t_1&=\frac{u^Lu^R-v^Lv^R+v^L}{1-v^R} \,, & t_3&=\frac{u^Lu^R-v^Lv^R+v^R}{1-v^L} \,, \\
    t_2&=\frac{u^L}{u^Lu^R-v^Lv^R+v^L} \,, & t_4&=\frac{u^R}{u^Lu^R-v^Lv^R+v^R} \,.
\end{align*}
Again, the coordinates $u^{L/R}_i$ and $v^{L/r}_i$ belong to $[0,1]$. The other restrictions are
\begin{align*}
    &0 \leq u^L \leq 1 \,, & &0 \leq u^L \leq v^L \leq 1-u^R \,,\\
    &\frac{u^L}{v^L} \leq \frac{1-v^R}{1-u^R} \,, & &\frac{u^R}{v^R} \leq \frac{1-v^L}{1-u^L}
\end{align*}
and the integration is evaluated as
\begin{align*}
    \int _{\mathcal{D}_2}&\equiv \int_0^1 du^L \int_0^{1-u^L}du^R \int_{u^L}^{1-u^R}dv^L \int_{u^R\frac{1-u^L}{1-v^L}}^{1-\frac{u^L(1-u^R)}{v^L}} dv^R \,.
\end{align*}

\paragraph{All orders.} Here we outline the procedure used to construct all vertices of chiral HiSGRA for a fixed ordering of the arguments, as obtained via the HPT method described above. One should keep in mind that all trees that have the same ordering should be taken into account.

A generic tree $T$, with $n$ leaves on the left branch and $m$ leaves on the right branch, that represents a $\mathcal{V}$ vertex takes the form of the left panel in Figure \ref{fig:treesIntro}. The right panel shows the tree we refer to as the base tree $T_0$. We construct the $A_\infty$ structure map $\mathcal{V}(\dots , C, \dots, \omega,\dots,\omega,\dots, C,\dots)$ from the graph as follows. We assign two-dimensional vectors $\vec q_i=(u_i,v_i)$, $\vec r_i  =  (p_{1,i},p_{2,i})$ to $\alpha_i\in\mathbb{A}_0$ and $\vec r_{m+n+1} = (p_{01},p_{02})$, $\vec q_{m+n+1} = (1-\sum_{i=1}^{m+n}u_i, 1-\sum_{i=1}^{m+n} v_i)$ to $\alpha_0\in\mathbb{A}_0$. We also introduce the matrices 
\begin{align}
    P_{T_0}&=(\vec 0,\vec 0,\vec r_1,\dots,\vec r_{m+n}, \vec r_{m+n+1})\,, &
    Q_{T_0}&= (-\vec e_1,-\vec e_2,\vec q_1,\dots \vec q_{m+n},\vec q_{m+n+1})\,,
\end{align}
belonging to the base tree $T_0$, where $\vec e_1 = \begin{pmatrix}
    1\\
    0
\end{pmatrix}$, $\vec e_2 = \begin{pmatrix}
    0\\
    1
\end{pmatrix}$.
The base tree $T_0$ yields the expression
\begin{align}
\begin{aligned}
       (p_{12})^{n+m}\int_{\mathbb{V}_{n+m}} \exp\Big[ P_{T_0}Q_{T_0}^t+\lambda p_{12}|Q_{T_0}| \Big]\,, 
\end{aligned}
\end{align}
where $|Q_T|$ is the sum of minors of $Q_T$. This tree is understood to yield the vertex $\mathcal{V}(\omega,\omega,C,\dots,C)$ when acting on
\begin{align} \label{abc1Intro}
    a(y_1)b(y_2)c_1(y_3)\dots c_{m+n}(y_{m+n+2})|_{y_i}=0
\end{align}
for $a,b\in \mathbb{A}_1$ and $\alpha_i\in \mathbb{A}_0$. The configuration space $\mathbb{V}_{n+m}$ is given by the chain of inequalities
\begin{align}\label{inequalitiesIntro}
    \frac{u_1}{v_1} &\leq \frac{u_2}{v_2} \leq \dots \leq \frac{u_{n+m+1}}{v_{n+m+1}} \,, & u_{n+m+1}\equiv 1-\sum_{i=1}^{n+m}u_i \,, && v_{n+m+1} &= 1-\sum_{i=1}^{n+m}v_i \,,
\end{align}
where $0 \leq u_i \leq 1$ and $0 \leq v_i \leq 1$ for $i=1,2,\dots,n+m+1$.
\begin{figure}[!ht]
\centering

\caption{ A generic tree $T$ in the left panel with elements of $\mathbb{A}_0$ attached left and right arbitrarily and the `base' tree $T_0$ in the right panel with only elements of $\mathbb{A}_0$ attached to the right on the right branch. $T$ can be obtained form $T_0$ through flipping $\alpha_i$'s to the left of the right branch and/or shifting them to the left branch.}\label{fig:treesIntro}
\end{figure}
A generic tree $T$ can be obtained from $T_0$ through two types of operations: (i) flipping $\alpha_i$ to the left of the right branch and (ii) a counterclockwise shift of all $\alpha_i$'s along the cord connecting $a$ and $b$. Importantly, in the latter case $\alpha_0$ also moves along the cord, while another $\alpha_i$ takes its place. To express the symbol corresponding to $T$ we define $P_T = (\vec 0,\vec 0, \vec r_1,\dots,\vec{r}_m,-\vec r_{m+1},\vec{r}_{m+2},\dots,\vec r_{n+m}, -\vec r_{m+n+1})$. We also define a matrix $Q_T$ by filling up its columns, starting with $\vec e_1$, corresponding to $a$ in Fig.\ref{fig:treesIntro} and from there on with $\vec a_i$ following through the tree counterclockwise. As an example, for the tree in the left panel of Fig.\ref{fig:treesIntro} this looks like 
\begin{align}
    Q_T&=(-\vec q_a,\vec{q}_{n+m+1},\vec q_{n+m-1},\dots,\vec q_4,\vec q_2,-\vec q_b,\vec q_1,\vec q_3,\dots,\vec q_{m+1},\dots,\vec q_{m+n})\,.
\end{align}
The cosmological term for a generic tree is then given by $\lambda p_{12} |Q_T|$, where $|Q_T|$ is the sum of minors of $Q_T$. 

For vertices, the labels on $p_i$ and the corresponding arguments $y_i$ of $a$, $b$ and the $\alpha_j$'s are read off from the tree from left to right. Since we have labeled them from bottom right to top left, we require a permutation $\sigma_T$ that relabels the $p_i$'s and $y_i$'s accordingly. Moreover, $\sigma_T$ also shuffles the elements in \eqref{abc1Intro} corresponding to their respective position in the tree $T$. A generic tree with cosmological constant contributes to a vertex by
\begin{align}
    s_T\sigma_T (p_{12})^{m+n}\int_{\mathbb{V}_{m+n}} \exp(\text{tr}[P_T Q^t]+\lambda p_{12}|Q_T|) a(y_1)b(y_2)\alpha_1(y_3)\dots \alpha_{y_{m+n}}(y_{m+n+2})|_{y_i=0} \,.
\end{align}
Here, $s_T=(-1)^k$ and $k$ is the number of zero-forms $C$ in between the two $\omega$'s. The sign $\sigma_T$ is the combination of the sign factor we get by evaluating the product of two branches with a sign coming from homological perturbation theory. Finally, a vertex with generic ordering of its arguments is constructed by adding all trees with that ordering. 

The $\mathcal{U}$ structure maps can also be constructed using HPT. Alternatively, one can construct a non-degenerate pairing
\begin{align}
    \langle f|g \rangle = \exp[p_{12}]f(y_1)g(y_2)\Big|_{y_1=y_2=0}\,,
\end{align}
with $f\in \mathbb{A}_1$ and $g\in \mathbb{A}_0$. This yields the duality maps $\langle fg|h\rangle = \langle f|\mathcal{U}_1(g,h) \rangle$ and $\langle fg|h \rangle = - \langle g| \mathcal{U}_2(h,f)$, where $\langle f|g\rangle = \langle g|f(-y) \rangle$. This can directly be used to related $\mathcal{V}$ and $\mathcal{U}$ structure maps by
\begin{align} 
    \langle \mathcal{V}_1(a,b,\alpha_1,\dots,\alpha_n)|\alpha_{n+1}\rangle=\langle a|\mathcal{U}_1(b,\alpha_1,\dots,\alpha_n,\alpha_{n+1})\rangle \,.
\end{align}
The explicit expression for the $\mathcal{U}$ structure map can be found in \ref{sec:uvertices}, where it is observed that the two methods agree.

Not only does the duality map provide a means of deriving the $\mathcal{U}$ structure maps, it also identifies and relates classes of both $\mathcal{V}$ and $\mathcal{U}$ structure maps. For example, all $\mathcal{V}$ structure maps with the same total number of elements of $\mathbb{A}_0$ and the same number of elements of $\alpha_i\in\mathbb{A}_0$ between $a,b \in \mathbb{A}_1$ are related to each other by the commuting diagram
\begin{align}
    &\langle \mathcal{V}(\alpha_{j+1},\dots,\alpha_n,a,\alpha_1,\dots,\alpha_i,b,\alpha_{i+1}\dots,\alpha_{j-1}|\alpha_j\rangle = \\
    &\qquad\qquad\langle \mathcal{V}(\alpha_{j-k+1},\dots,\alpha_n,a,\alpha_1,\dots,\alpha_i,b,\alpha_{i+1},\dots,\alpha_{j-k-1}|\alpha_{j-k}\rangle  \,.
\end{align}
Lastly, we have the duality map among $\mathcal{U}$ structure maps,
\begin{align}
    \langle a|\mathcal{U}_{n+2}(\alpha_1,\dots,\alpha_{n+1},b)\rangle &= - \langle b|\mathcal{U}_1(a,\alpha_1,\dots,\alpha_{n+1})\rangle \,.
\end{align}
The duality maps are visually depicted by
\begin{center}
\begin{tikzpicture}
  \matrix (m)
    [
      matrix of math nodes,
      row sep    = 3em,
      column sep = 4em
    ]
    {
      \mathcal{V}(a,b,\alpha_1,\dots,\alpha_n)              & \mathcal{U}(b,\alpha_1,\dots,\alpha_{n+1}) \\
      \mathcal{V}(\alpha_1,\dots,\alpha_n,a,b) & \mathcal{U}(\alpha_1,\dots,\alpha_{n+1},b)            \\
    };
  \path
    (m-1-1) edge [<->] node [left] {$\mathcal{V}$-$\mathcal{V}$} (m-2-1)
    (m-1-1.east |- m-1-2)
      edge [<->] node [above] {$\mathcal{V}$-$\mathcal{U}$} (m-1-2)
    (m-2-1.east) edge [<->] node [below] {$\mathcal{V}$-$\mathcal{U}$} (m-2-2)
    (m-1-2) edge [<->] node [right] {$\mathcal{U}$-$\mathcal{U}$} (m-2-2);
\end{tikzpicture}
\end{center}
As a result, the duality maps hugely reduce the number of structure maps that are to be computed.

\paragraph{Locality.} A crucial observation is that the structure maps $\mathcal{V}$ and $\mathcal{U}$ are manifestly local up to all orders. This is a result of the absence of the operators $p_{ij}$ in the exponential. To see this, we must remember that all structure maps contain a hidden star-product on the primed indices, which contains contractions between the $C$'s. Taylor expanding this gives an infinite series in powers of $q_{ij}$. Since descendants encode derivatives of the physical fields, we have $p_{ij}q_{ij} \leftrightarrow \nabla^{AA'}\nabla_{AA'}$. So a structure map is called local if it does not have any contractions between the $C(y)$'s.

\paragraph{Minimality.} The $L_\infty$ structure maps admit some form of minimality in the sense that they are almost zero. For $\lambda=0$ and Abelianizing the theory, i.e. the fields are valued in $\text{Mat}_1$, we find
\begin{align*}
    \begin{aligned}
        \sum_{k+m+n=N} \mathcal{V}(\alpha_{1}\ldots,\alpha_{k}, b,\alpha_{k+1},\ldots,\alpha_{k+m},a, \alpha_{k+m+1},\ldots,\alpha_{k+m+n}) &= 0 \,, \quad \text{for} \quad N \geq 1\,,\\
        \sum_{m+n=N} \mathcal{U}(\alpha_{1}\ldots,\alpha_{m}, b,\alpha_{m+1},\ldots,\alpha_{m+n}) &= 0 \,, \quad \text{for} \quad N\geq 2\,.
    \end{aligned}
\end{align*}
This means that if we assume the $A_\infty$-algebra to be $\mathbb{C}[y] \otimes A_1 \otimes \text{Mat}_1$ (i.e. fields on flat space without matrix values), the structure maps of the associated $L_\infty$-algebra vanish, apart from $\mathcal{V}(\omega,\omega)$ and $\mathcal{U}(\omega,C)$, which describe the free theory. This is to be expected, since we are now taking the graded anti-symmetrization of a commutative structure.
\subsubsection{Configuration space}
The expressions for the structure maps as obtained from HPT are far from the ones we have presented here. Originally, the integration domains are simply hypercubes $[0,1]^{2n}$ with $n$ the number of $C$'s, as HPT only produces integrals over the line $[0,1]$. On the other hand, the integrands quickly become unmanageable even for low order maps. We found an appropriate change of variables that trades off the simplicity of the integration domain for a manageable integrand. This leaves us with the integration domains $\mathbb{V}_n$, also called configuration space, described in \eqref{inequalitiesIntro}. 

\begin{figure}
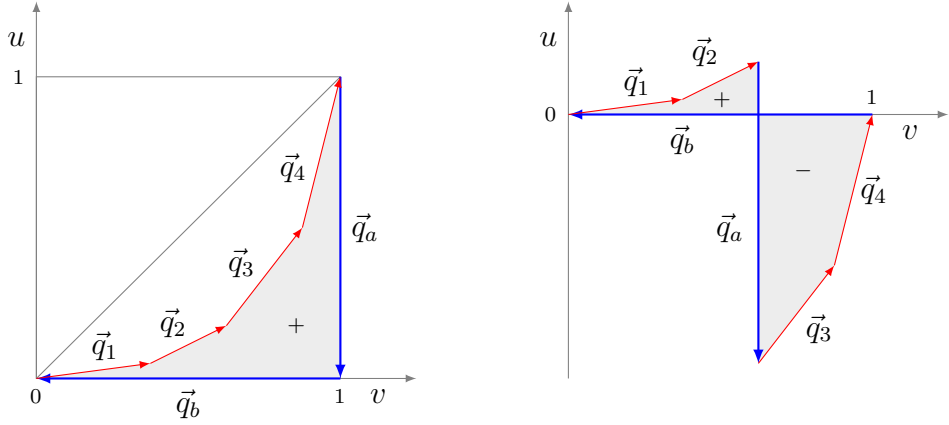

    \centering

\caption{In the left panel a swallowtail constructed from the vectors $(\vec{q}_1,\vec{q}_2,\vec{q}_3,\vec{q}_4,\vec{q}_a,\vec{q}_b)$ and in the right panel a self-intersecting polygon constructed from $(\vec{q}_1,\vec{q}_2,\vec{q}_a,\vec{q}_3,\vec{q}_4,\vec{q}_b)$. They are associated to the quintic structure maps, respectively.}\label{fig:quarticIntro}
\end{figure}

While $\mathbb{V}_n$ has a more complicated structure than the hypercube, it still admits a neat geometric interpretation. An example, the configuration space $\mathbb{V}_3$ that is associated to the quintic structure maps, is represented in the left panel of Figure \ref{fig:quarticIntro}. The configuration space is obtained by connecting the vectors $\vec{q}_a,\vec{q}_b,\vec{q}_i\in\mathbb{R}^2$ in the order $Q=(\vec{q}_a,\vec{q}_b,\vec{q}_1,\vec{q}_2,\vec{q}_3)$. This leads to a closed polygon with six vertices, of which three are fixed at $(0,0)$, $(1,0)$, and $(0,1)$. The configuration space consists of the remaining three vertices in the 'bulk' of the unit square. As a defining feature of the configuration space, \eqref{inequalitiesIntro} only allows these vertices to be positioned such that the upper boundary of the shaded region is concave. In other words, the complement of the shaded region under the diagonal is convex. We refer to the concave regions as \textit{swallowtails}. In Appendix \ref{app:allorders} we prove that the swallowtails are compact and $\mathbb{V}_n \subset [0,1]^{2n} \subset \mathbb{R}^{2n}$.

The above is easily generalizable. The configuration space $\mathbb{V}_n$ shows up in the expression for the structure maps $\mathcal{V}(\omega,\omega,C,\dots,C)$ with $n$ $C$'s. One draws a closed polygon in the order $Q=(\vec{q}_a,\vec{q}_b,\vec{q}_1,\dots,\vec{q}_{n+1})$. This yields a polygon with $n+3$ vertices, of which $3$ are fixed. The configuration space consists of the remaining $n$ vertices that are restricted to form a swallowtail.

Another interesting finding is that the volume of the shaded region in the left panel of Figure \ref{fig:quarticIntro} is equal to half the term proportional to the cosmological constant $\lambda$ in the structure maps corresponding to the base tree with any number of $C$'s. 

This is not the case for trees with other topologies, however. As an alternative, the cosmological term is related to a self-intersecting polygon, such as the one in the right panel of \ref{fig:quarticIntro}. In general, the (possibly self-intersecting) polygon associated to a tree $T$ is obtained from connecting the vectors $\vec{q}_a$, $\vec{q}_b$ and $\vec{q}_i$ as they appear in $Q_T$. If a polygon is self-intersecting, the shaded region splits up in two separate regions, each of which are themselves swallowtails. The shaded region above (below) the horizontal axis provides a positive (negative) contribution to the oriented area of the polygon. The cosmological term in any tree $T$ is then given by half the oriented area of the polygon constructed from $Q_T$. This statement has been made explicit, as the cosmological term is proportional to the sum of minors of $Q_T$.

When representing the structure maps of the $A_\infty$-algebra that underpins chiral HiSGRA in the most efficient coordinates, we encounter a deeper geometric layer in terms of swallowtails. This hints that there might be an alternative way to construct other HiSGRAs by choosing different configuration spaces, akin to Kontsevich' formality theorem \cite{Kontsevich:1997vb}. Another observation is made in Section \ref{sec:swallowtail}. Here, the swallowtails are reformulated as positive Grassmannians, which have been recently showing up in statistical physics, integrable models and scattering amplitudes. A recent account of this topic can be found in \cite{PGr, williams2021positive}.

\subsubsection{Pre-Calabi-Yau algebra}

From the above construction, the $A_\infty$-algebra underlying chiral HiSGRA is now revealed to be $\hat{\mathbb{A}}=\mathbb{A}\otimes B$, where $B$ is the unital associative algebra $B=A_1\otimes \text{Mat}_N$ and the what is called minimal $A_\infty$-algebra $\mathbb{A}$ splits up as $\mathbb{A}=\mathbb{A}_0\oplus\mathbb{A}_1=A_\lambda^\star\oplus A_\lambda[-1]$. As discussed in Appendix \ref{CY}, this is a pre-Calabi-Yau algebra of degree $2$. Pre-Calabi-Yau algebras are a particular type of $A_\infty$-algebra that offer precisely the type of duality maps that we used to relate different structure maps to each other. Interestingly, pre-Calabi-Yau algebras are related to deformation quantization of non-commutative manifolds and often show up in string field theory \cite{kontsevich2021pre}.

\section{Stokes' theorem}
The formulation of chiral HiSGRA based on its $A_\infty$-algebra raises the questions (i) what structures in deformation quantization and non-commutative geometry gives rise to the structures in chiral HiSGRA and (ii) is there a more general formality theorem of which the (Shoikhet-Tsygan-)Kontsevich formality is a particular example and that gives chiral HiSGRA? In this section, we summarize the results obtained in Chapter \ref{chap:Stokes}, where we proved the $A_\infty$-relations using Stokes' theorem, as is also done in the known formality theorems. Together with the similarities between the structure maps obtained in the previous sections and Kontsevich' recipe to find star-products, this hints at the extension of the known formality theorems as mentioned above.

\begin{figure}[h!]
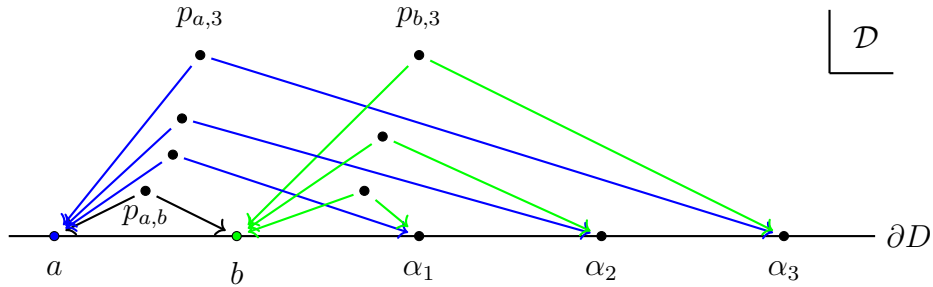

\centering

\caption{A typical Kontsevich-like graph contributing to a quintic structure map. $D$ is the upper half-plane and $\pl D$ is its boundary.}\label{formalityIntro}
\end{figure}

Namely, like the bi-differential structure maps that constitute the Kontsevich star-product, the structure maps can be represented after Taylor expansion by sums over graphs $\Gamma$ as 
\begin{align} \label{taylorInto}
    m_n(f_1,\dots,f_n) &= \sum_{\Gamma}w_\Gamma\mathcal{W}_\Gamma(f_1\otimes \dots \otimes f_n) \,, & f_i &\in\mathbb{A}\,.
\end{align}
Here, $w_\Gamma$ are weights and $\mathcal{W}_\Gamma$ are poly-differential operators. The main difference from Kontsevich is that the weights are expressed in terms of integrals over some configuration space $C_\Gamma$ of concave polygons, i.e. swallowtails. As for the Moyal-Weyl case, the graphs are built from simple wedges that represent $p_{\bullet,\bullet}$, excluding $p_{ij}$, see Figure \ref{formalityIntro}. Completing the analogy with the Moyal-Weyl star-product, we then have the simplest non-trivial Poisson structure $\epsilon^{AB}$.

The $A_\infty$-relations have the schematic form
\begin{align} \label{AinftyIntro}
    \sum_{i,j}\pm m_i(\bullet,\dots,m_j(\bullet,\dots,\bullet),\dots,\bullet) = 0\,,
\end{align}
which is a 'product' of structure maps. In order to prove the $A_\infty$-relations using Stokes' theorem, we construct a configuration space $\mathcal{C}$ and a closed form $\Omega$ on $\mathcal{C}$, such that the boundary
\begin{align}
    \partial \mathcal{C} = \sum_{\Gamma,\Gamma'}\mathcal{C}_\Gamma\times \mathcal{C}_{\Gamma'}
\end{align}
reproduces the product space of the configuration spaces of \eqref{taylorInto}. These will be chosen such that $\Omega$ evaluated on $\partial C$ exactly reproduces \eqref{AinftyIntro}. With $\Omega$ restricted to $\partial C$ providing the $A_\infty$-terms, while also being closed, this yields the (schematic) equivalence
\begin{align} \label{schematicIntro}
    0&=\int_\mathcal{C} d\Omega=\int_{\partial \mathcal{C}}\Omega \,, & \Longleftrightarrow & & A_\infty\text{-relations} \,.
\end{align}
This section follows Chapter \ref{chap:Stokes} closely, only omitting some details. However, we will provide instructions for drawing diagrams and extracting expressions from them. We have no choice but to write these instructions in roughly the same way.

\subsection{The recipe}
The initial calculations we performed were heavily based on HPT. However, this led to enormously involved expressions, so we built a diagrammatic representation for the proof. The diagrammatic representation is fully self-consistent, so it renders the initial calculations obsolete.

\subsubsection{Structure maps}

\begin{figure}[!ht]
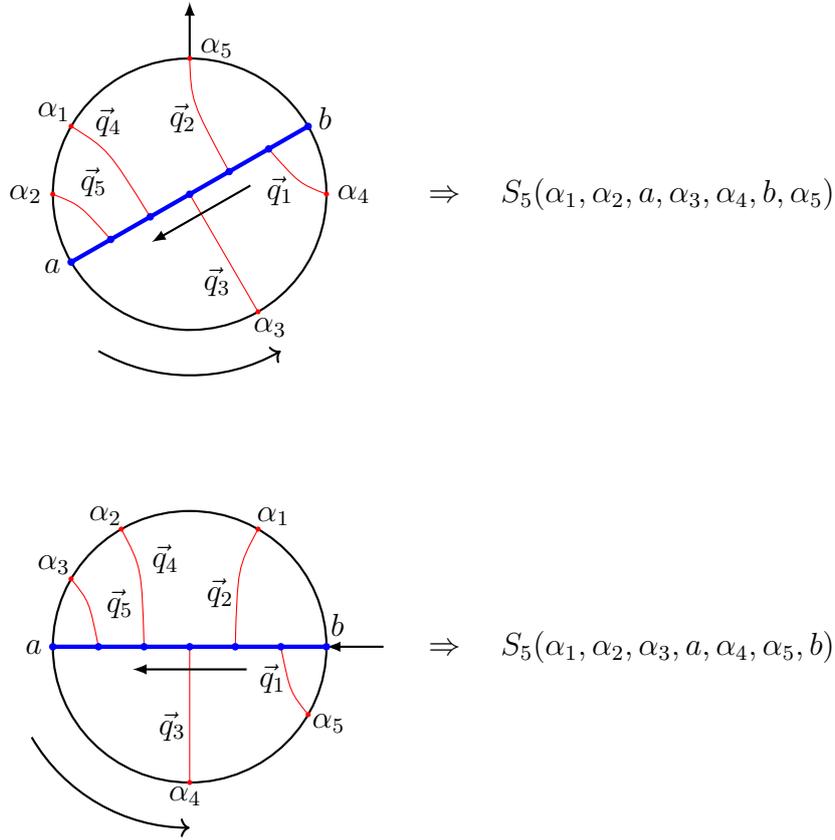

\centering

\caption{Two  disk diagrams that contribute to $S_{5}$.}\label{fig:treeCircle}
\end{figure}

Before we lay out the recipe for building diagrams and extracting the closed form $\Omega$ and the configuration space $C$ out of them, we propose an alternative diagrammatic representation of the trees we encounter in the previous section. This will also turn the diagrams required for the Stokes' theorem proof a natural generalization. Since the expressions built out of the trees were constructed by reading off the ordering of the fields counterclockwise, it is natural to wrap the trees inside a circle, see Figure \ref{fig:treeCircle}.

Each disk diagram consists of a circle with a diameter, trivalent vertices, lines from the diameter to the boundary and an arrow on one of the fields on the boundary. Lines cannot intersect each other or themselves. The arrow points away from the disk if it is attached to a field belonging to $\mathbb{A}_0$ and otherwise it is pointing inwards. The disk will always be rotated such that the arrow is found on the northern semicircle or on the points where the boundary and the diameter intersect.

It turns out to be useful to assign to each disk diagram a structure map
\begin{align*}
    S_N(\alpha_1,\dots,\alpha_k,a,\alpha_{k+1},\dots,\alpha_{k+m},b,\alpha_{k+m+n}) \,,
\end{align*}
where $\alpha_i\in\mathbb{A}_0$ and $a,b\in\mathbb{A}_1$. The total number of elements of $\mathbb{A}_0$ is\footnote{We also use the symbol $N$ in $\text{Mat}_N$. These uses are unrelated and should be distinguished from context. We avoid using matrix-valued $N$ in any explicit calculations, so no confusion should arise} $N=k+m+n$. Only later will we extract the $A_\infty$ structure maps $m_{N+1}$, with $N+1$ arguments, from $S_N$. The ordering of the arguments of $S_N$ is determined by the order in which they appear on the boundary of the disk diagram in the counterclockwise direction, starting from the arrow. This is called the \textit{boundary ordering}. The labels on the $\alpha_i$ are also assigned according to the boundary ordering. One assigns the vectors $\vec{q}_i=(u_i,v_i)$ to the red lines with $i$ increasing in the direction from $b$ to $a$ along the diameter. This is called the \textit{bulk ordering}. We also assign $\vec{q}_a=(-1,0)$ and $\vec{q}_b=(0,-1)$ to the points $a$ and $b$, respectively. The vectors forms a closed polygon, i.e.
\begin{align}
    \vec{q}_1+\dots+\vec{q}_N+\vec{q}_a+\vec{q}_b=0
\end{align}
We also require $u_i,v_i \geq 0$. Next, we assign a 'time' $t_i=u_i/v_i$ at which we pass through the $i$-th vertex following the bulk ordering. The chronological ordering $t_i \leq t_j$ for $i \leq j$ then gives
\begin{align}
    0 \leq \frac{u_1}{v_2} \leq \frac{u_2}{v_2} \leq \dots \leq \frac{u_N}{v_N} \leq \infty \,.
\end{align}
Together, this describes the integration domain $\mathbb{V}_{N-1}$.

With regard to extracting the integrand out of a disk diagram $D$, one defines a $2\times (N+2)$ array $Q_D$, which is composed of vectors $\vec{q}_i$ according to the boundary ordering, i.e.
\begin{align}
    Q_D &= \begin{pmatrix}
        \vec{q}_{i_1},\dots,\vec{q}_{i_{N+2}} 
    \end{pmatrix}\,, \quad \text{with} \quad  i_k \in\{a,b,1,2,\dots,N\}\,.
\end{align}
The labels on the vectors are also named according to the boundary ordering. To simplify expressions, we will often refer to the \textit{canonical ordering} associated with $S_N(a,b,\alpha_1,\dots,\alpha_N)$, which gives
\begin{align}
    Q_D = \begin{pmatrix}
        \vec{q}_a,\vec{q}_b,\vec{q}_{1},\dots,\vec{q}_{N+2}
    \end{pmatrix}\,.
\end{align}
This corresponds to the base tree $T_0$ in the previous section. We also define a $2\times(N+2)$ array
\begin{equation}
    P_D=( \vec{r}_{1},\ldots , \vec{r}_{k},\vec{r}_a,\vec{r}_{k+1},\dots,\vec{r}_{k +m},\vec{r}_b,\vec{r}_{k+m+1},\dots,\vec{r}_{k+m+n})\,,
\end{equation}
where $\vec{r}_i=\begin{pmatrix}
    p_{1,i},p_{2,i}
\end{pmatrix}$ for $i=1,\dots,N$ and $\vec{r}_a=\vec{r}_b=\begin{pmatrix}
    0,0
\end{pmatrix}$.

Lastly, we define $s_D$ to be the number of $\alpha_i$'s in the southern semicircle in diagram $D$. We then define the integrand
\begin{align}
    I_D=s_D(p_{a,b})^{N-1}\exp[\tr(P_DQ_D^t)+\lambda|Q_D|p_{a,b}]
\end{align}
and the structure map $S_N$ then reads
\begin{align}
    S_N=\sum_{D}\int_{\mathbb{V}_{N-1}}I_D\,,
\end{align}
where the sum is over all admissable disk diagrams with the same ordering of arguments.

The structure map $m_{N+1}$ can be computed from $S_N$ using the natural pairing
\begin{align}
    \langle \bullet,\bullet \rangle : \mathbb{A}_1\otimes \mathbb{A}_0 \rightarrow\mathbb{C}\,.
\end{align}
For example, given a map $S_N(\dots)=\langle \mathcal{V}(\dots),\alpha\rangle$, one may remove the last $\alpha$ and obtain the $\mathcal{V}$ structure map. Alternatively, for $f(y_1)\in\mathbb{A}_1$ and $g(y_2)\in\mathbb{A}_0$, one may use
\begin{align} 
    \langle f(y_1) , g(y_2) \rangle = - \langle g(y_2) , f(y_1) \rangle = \exp[p_{1,2}]f(y_1)g(y_2)|_{y_1=y_2=0}
\end{align}
to extract the $\mathcal{V}$ and $\mathcal{U}$ structure maps from $S_N$. This procedure yields the same structure maps as the original procedure in the previous section.

\subsubsection{Recipe for Stokes' theorem proof}
\begin{figure}[!ht]
\centering
\begin{tikzpicture}[scale=0.3]
\draw[thick](0,0) circle (4);
\draw[thick](10,0) circle (4);
\draw [thick, -Latex] (6,0) -- (4,0);
 \draw (-6.5,0) node[gray, scale=5]{$\Sigma$\;} ;
 \coordinate [label=below: { $\mathcal{U}$}] (B) at (0,1);
  \coordinate [label=below: { $\mathcal{V}$}] (B) at (10,1);
   \coordinate [label=below: { ${}$}] (B) at (-7.5,-2.5);
   \draw[thick, ->] (10,5) arc (90:10:5);  
   \draw[thick, ->] (0,5) arc (90:10:5);  
    \filldraw [red] (90:4)+(10,0) circle (3pt);     
  \filldraw [red] (30:4)+(10,0) circle (3pt);  
   \filldraw [blue] (60:4)+(10,0) circle (5pt);      
   \filldraw [blue] (-120:4)+(10,0) circle (5pt);     
  \filldraw [red] (0:4)+(10,0) circle (3pt);    
   \filldraw [red] (-30:4)+(10,0) circle (3pt);    
    \filldraw [red] (-75:4)+(10,0) circle (3pt);     
\filldraw [red] (-150:4)+(10,0) circle (3pt);       
    \filldraw [red] (135:4)+(10,0) circle (3pt); 
    \filldraw [red] (180:4)+(10,0) circle (3pt); 
      \filldraw [red] (90:4) circle (3pt);     
  \filldraw [red] (30:4) circle (3pt);  
   \filldraw [red] (60:4) circle (3pt);      
   \filldraw [red] (-120:4) circle (3pt);     
  \filldraw [blue] (0:4)circle (5pt);    
   \filldraw [red] (-30:4) circle (3pt);    
    \filldraw [red] (-75:4) circle (3pt);     
\filldraw [red] (-150:4) circle (3pt);       
    \filldraw [red] (135:4) circle (3pt); 
    \filldraw [blue] (180:4) circle (5pt);   
     \draw (17,0) node[black, scale=1]{$=0$} ;
    \end{tikzpicture}
    \caption{Graphical representation for the $A_\infty$-relations. }\label{SD4Intro}    
    \end{figure}
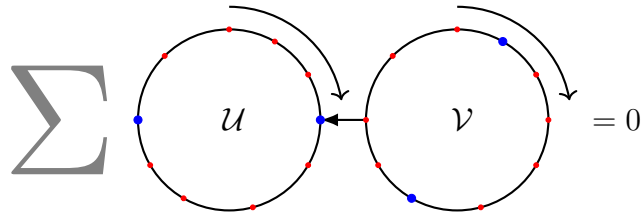

The schematic form of the Stokes' theorem proof as given in \eqref{schematicIntro} can be made more precise by
\begin{align} \label{moreDetailIntro}
    \begin{aligned}
        0 = \sum \int_{\mathbb{W}_{k,l,m,n}} d\Omega_{k,l,m,n}^a(y) + d&\Omega_{k,l,m,n}^c(y) = \sum \int_{\partial\mathbb{W}_{k,l,m,n}} \Omega_{k,l,m,n}^a(y) + \Omega_{k,l,m,n}^c(y)\\
        &\Updownarrow\\
        A_{\infty}&\text{-relations}
    \end{aligned}
\end{align}
where $k+m+l+n=N$. Here, $\mathbb{W}_{k,l,m,n}$ are integration spaces and $\Omega_{k,l,m,n}^a(y)$ and $\Omega_{k,l,m,n}^c(y)$ are closed differential forms, called \textit{potentials}, with values in multi-linear maps
\begin{align*}
    \Omega_{k,l,m,n}^{a,c}(y) : T^k\mathbb{A}_0 \otimes \mathbb{A}_1 \otimes T^l\mathbb{A}_0 \otimes \mathbb{A}_1 \otimes T^m\mathbb{A}_0 \otimes \mathbb{A}_1 \otimes T^n\mathbb{A}_0 \rightarrow \mathbb{A}_1\,.
\end{align*}
The potentials take $a,b,c\in\mathbb{A}_1$ and $c_i\in\mathbb{A}_0$ as arguments and the labels $k,l,m,n$ indicate how the $\alpha_i$'s are distributed between the elements of $\mathbb{A}_1$. As will become clear soon, there exist two potentials $\Omega^a$ and $\Omega^c$ whose ordering can never mix. Expressed in terms of the disk diagrams for structure maps $\mathcal{V}$ and $\mathcal{U}$, the $A_\infty$-relations take on the diagrammatical form that is shown in Figure \ref{SD4Intro}, which represents the insertion of one structure map into an other.

We must also consider configuration spaces $\mathbb{W}_{k,l,m,n}\subset \mathbb{R}^{2N+1}$ with $\text{dim}(\mathbb{W}_{k,l,m,n})=2N+1$. The dimension is determined by the fact that the boundary $\partial\mathbb{W}_{k,l,m,n}$ should be the product space of two swallowtails. Again, we build disk diagrams from which the potentials and configuration spaces can be derived.

The following is nearly identical to what is discussed in Chapter \ref{chap:Stokes}. Still, we believe it is too important not to mention here. The recipe to construct disk diagrams for the potentials and configuration space is:
\begin{itemize}[labelsep=1em,leftmargin=2em]
    \item[\firstCircle] Consider a circle. The interior will be referred to as the \textit{bulk} and the circle as the \textit{boundary}.
    \item[\secondCircle] Choose three distinct points on the boundary and label them $a,b,c$ counterclockwise. Consider the point at the center of the bulk, now called \textit{junction}, and connect this to each of the points $a,b,c$ by blue lines. These points correspond to elements of $\mathbb{A}_1$. The lines are called $a$-leg, $b$-leg and $c$-leg, correspondingly.
    \item[\thirdCircle] Draw any number of red lines connecting these legs to the boundary at either side of the legs. Lines are not allowed to intersect. Their endpoints at the boundary correspond to elements of $\mathbb{V}_0$.
    \item[\fourthCircle] Connect an arrow to one of the vertices on the boundary of the disk between $a$ and $c$, pointing away from the disk, i.e. there has to be one marked point on the boundary. If the arrow is connected to a vertex connected to the $a$-leg, the potential that can be extracted using \eqref{potPairing} is $\Omega_{k,l,m,n}^a$, while $\Omega_{k,l,m,n}^c$ can be found when it is connected to the $c$-leg.
    \item[\fifthCircle] Label the points at the boundary that are connected to red lines $\alpha_i, \beta_i, \gamma_i, \delta_i$ if the lines emanate from the $a$-leg, $b$-leg, $c$-leg or are in between the red line connected to the arrow and the junction, respectively, and $i$ increases from the boundary to the junction. This way if the arrow is attached to an argument belonging to the $a$-leg, the arguments after the arrow are labelled $\delta_i$ and those in between $a$ and including the arrow are named $\alpha_i$. The subscripts $k,l,m,n$ on the potentials count the number of points with labels $\alpha_i, \beta_i, \gamma_i, \delta_i$, respectively, disregarding the label associated with the arrow.
    \item The diagram must contain at least one element of $\mathbb{A}_0$, connected to the piece of the boundary between $a$ and $c$, which can be attached to either the $a$-leg or the $c$-leg. If the diagram contains more than one element of $\mathbb{A}_0$, they have to be attached to at least two different legs.
\end{itemize}
From the diagram we can read off the expressions for the scalars
\begin{align}
    \begin{aligned}
        &\langle \Omega^a_{k,l,m,n},\alpha_{k+1}\rangle & \text{and} && \langle \Omega^c_{k,l,m,n},\gamma_{m+1}\rangle \,,
    \end{aligned}
\end{align}
depending on the name of the field that the arrow connects to. If the arrow is connected through a line to the $a$-leg ($c$-leg), we call this an $a$-diagram ($c$-diagram). From here the potentials and configurations space can be obtained in the same way as the structure maps $\mathcal{V}$ and $\mathcal{U}$ were found from $S_N$. It is easy to see that a $c$-diagram can always be obtained from reversing the boundary ordering on an $a$-diagram. This gives
\begin{align}
    \Omega_{k,l,m,n}^a \quad\leftrightarrow \quad \Omega_{m,l,k,n}^c \,, \quad \text{with} \quad a\leftrightarrow c\,.
\end{align}
Thus, the potential $\Omega_{k,l,m,n}^a$ and $\Omega_{m,l,k,n}^c$ are accompanied by the same integration domain $\mathbb{W}_{k,l,m,n}$.

To the $a$-leg, $b$-leg and $c$-leg we assign the vectors $\vec{q}_{a}=(-1,0,0)$, $\vec{q}_{b}=(0,-1,0)$ and $\vec{q}_{c}=(0,0,-1)$, respectively. Otherwise they are $\vec{q}_{a,i}=(u_i^a,v_i^a,w_i^a)$, $\vec{q}_{b,i}=(u_i^b,v_i^b,w_i^b)$ and $\vec{q}_{c,i}=(u_i^c,v_i^c,w_i^c)$ for the elements of $\mathbb{A}_0$ in the order that they are attached to the $a$-leg, $b$-leg or $c$-leg from the boundary inwards. The labels increase from boundary to junction. We now assign a time $t_i^{uv}=u_i^\bullet/v_i^\bullet$, $t_i^{uw}=u_i^\bullet/w_i^\bullet$ and $t_i^{vw}=v_i^\bullet/w_i^\bullet$ and impose a chronological ordering along three paths in the bulk of the disk diagram.
\begin{itemize}
    \item Path 1: One moves from $b$ to $c$, after which one continues to $a$. This imposes chronological ordering in time $t_i^{uv}$.
    \item Path 2: One moves from $c$ to $b$, after which one continues to $a$. This imposes chronological ordering in time $t_i^{uw}$.
    \item Path 3: One moves from $c$ to $a$, after which one continues to $b$. This imposes chronological ordering in time $t_i^{vw}$.
\end{itemize}

    \begin{figure}[!ht]
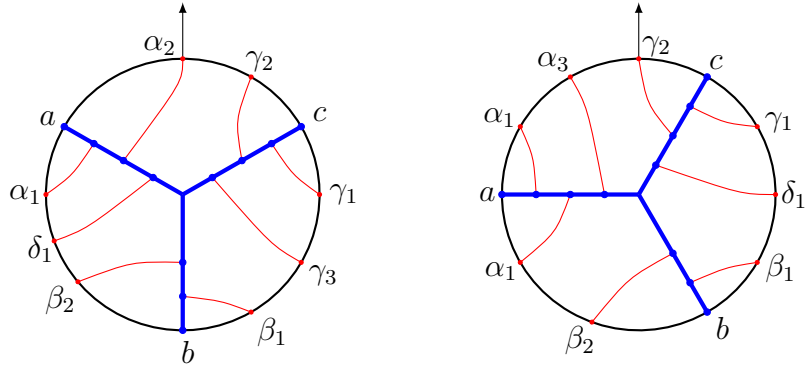

    \centering
 \, 
    \caption{The disk diagrams for $\langle\Omega_{1,2,3,1}^a(a,\alpha_1,\delta_1,\beta_2,b,\beta_1,\gamma_3,\gamma_1,c,\gamma_2),\alpha_2 \rangle$ and $\langle \Omega_{3,2,1,1}^c(\alpha_3,\alpha_1,a,\alpha_2,\beta_2,b,\beta_1,\delta_1,\gamma_1,c),\gamma_2\rangle$ on the left and right, respectively.}\label{SD5Intro}
    \end{figure}
    The paths are shown in Figure \ref{pathsIntro}. Together with the condition that all $u$, $v$ and $w$ variables take values between $0$ and $1$, the chronological order shows the configuration space $\mathbb{W}_{k,l,m,n}$ to be
\begin{align} \label{ineqsIntro}
    \begin{aligned}
        0 \leq& u_i^\bullet,v_i^\bullet,w_i^\bullet \leq 1 \,, \quad \sum (u_i^\bullet,v_i^\bullet,w_i^\bullet)=(1,1,1) \,, \quad  \frac{v_i^a}{w_i^a}\frac{w_i^b}{u_i^b}\frac{u_i^c}{v_i^c}=1\,,\\
        0 \leq& \frac{u_1^b}{v_1^b} \leq \dots \leq \frac{u_{l}^b}{v_{l}^b} \leq \frac{u_m^c}{v_m^c} = \dots = \frac{u_1^c}{v_1^c} \leq \frac{u_{k+n+1}^a}{v_{k+n+1}^a} \leq \dots \leq \frac{u_1^a}{v_1^a}\leq \infty\,,\\
       0 \leq& \frac{u_1^c}{w_1^c} \leq \dots \leq \frac{u_m^c}{w_m^c} \leq \frac{u_{l}^b}{w_{l}^b} = \dots = \frac{u_1^b}{w_1^b} \leq \frac{u_{k+n+1}^a}{w_{k+n+1}^a} \leq \dots \leq \frac{u_1^a}{w_1^a}\leq \infty\,,\\
        0 \leq& \frac{v_1^c}{w_1^c} \leq \dots \leq \frac{v_m^c}{w_m^c} \leq \frac{v_{k+n+1}^a}{w_{k+n+1}^a} = \dots = \frac{v_1^a}{w_1^a}\leq \frac{v_{l}^b}{w_{l}^b} \leq \dots \leq \frac{v_1^b}{w_1^b} \leq \infty \,.
    \end{aligned}
\end{align}
Here, the $\bullet$ stands for $a$, $b$ or $c$. A naive counting of the dimension of the configuration space gives the wrong dimension. After all, the configuration space is parametrized by the $3N+3$ coordinates $u$, $v$ and $w$. However, one can see in Figure \ref{pathsIntro} that the paths run over some legs twice, which leads to the equalities
\begin{align}
    \frac{u_m^c}{v_m^c} \leq \dots \leq \frac{u_1^c}{v_1^c} \leq \frac{u_1^c}{v_1^c} \leq \dots \leq \frac{u_m^c}{v_m^c} \quad \Rightarrow \quad  \frac{u_m^c}{v_m^c} = \dots = \frac{u_1^c}{v_1^c} \,.
\end{align}
The second relation in \eqref{ineqsIntro} is called the \textit{closure constraint} and it restricts $3$ more coordinates. The third relation fixes one more. 
Together, this fixes exactly the right amount of coordinates, such that $\text{dim}(\mathbb{W}_{k,l,m,n})=2N+1$.

The chains of equalities can be parametrized by $\alpha$, $\frac{1}{\beta}$ and $\gamma$ according to
\begin{align}
    \begin{aligned}
        \alpha&=\frac{u_i^c}{v_i^c} \,, & \frac{1}{\beta} &= \frac{v_i^a}{w_i^a} \,, & \gamma &= \frac{u_i^b}{w_i^b} \,,
    \end{aligned}
\end{align}
which in turn characterizes the planes in which the vectors $\vec{q}_i\in\mathbb{R}^3$ lie. The third relation in the first line of \eqref{ineqsIntro} relates the planes by
\begin{align}
    \gamma=\frac{\alpha}{\beta}\,.
\end{align}

\begin{figure}[h!]
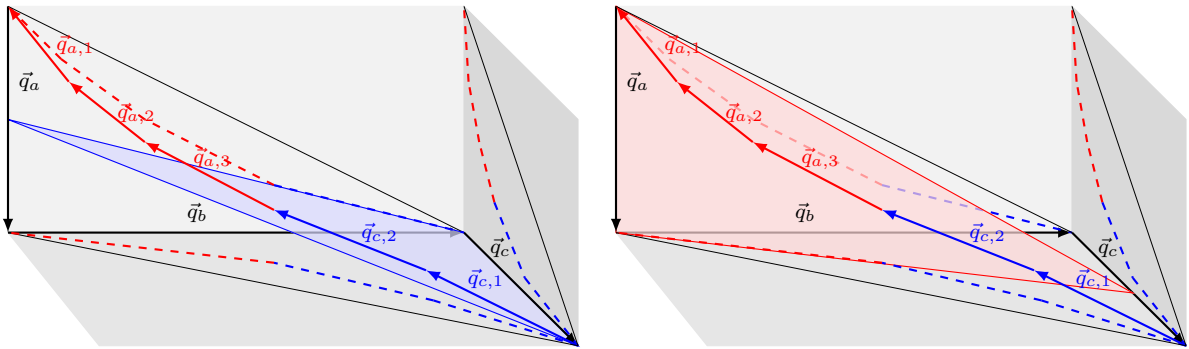


\begin{center}




\end{center}

\caption{On the left (right) the blue (red) shaded region depicts the plane in which the vectors $\vec{q}_{c,i}$ ($\vec{q}_{a,i}$) lie in $\mathbb{W}_{0,0,2,2}$. The scale in the $v$-direction was doubled to accentuate the details in the pictures.}\label{coplanIntro}
\end{figure}

A special configuration space $\mathbb{W}_{0,0,m,n}$ is associated to the what is called \textit{left-ordered} disk diagrams. They give rise to potentials of the form $\Omega^{\bullet}_{0,0,m,n}(a,b,c,\gamma_1,\dots,\gamma_m,\delta_1,\dots,\delta_n)$. The configuration space $\mathbb{W}_{0,0,m,n}$ admits a nice visualization $\mathbb{R}^3$ in a way that it extends the swallowtails in $\mathbb{R}^2$, see Figure \ref{coplanIntro}. We refer to the closed polygons $(\vec{q}_{a},\vec{q}_b,\vec{q}_c,\vec{q}_{c,1},\dots,\vec{q}_{c,m},\vec{q}_{a,n+1},\dots,\vec{q}_{a,1})$ as maximally concave polygons, because their projections onto the $uv$-, $uw$- and $vw$-plane are swallowtails. The equalities in the $uv$-chain \eqref{ineqsIntro} enforce all vectors $\vec{q}_{c,i}$ to be coplanar and the equalities in the $vw$-plane require the vectors $\vec{q}_{a,i}$ to be coplanar. This is also depicted in Figure \ref{coplanIntro}.

\begin{figure}
    \centering
    %
    \caption{Paths 1 - 3 from left to right, leading to the time ordering of times $t^{uv}_i$, $t^{uw}_i$ and $t^{vw}_i$, respectively.}
    \label{pathsIntro}
\end{figure}

The potentials turns out to be found easiest by considering a master space $\mathbb{U}_N\subset\mathbb{R}^{3N}$ described by
\begin{align} 
        0 &\leq u_i,v_i,w_i \leq 1 \,, & &\sum_{i=1}^{N+1}(u_i,v_i,w_i)=(1,1,1)\,.
\end{align}
On this space we define the master potential
\begin{align*}
    \Omega_N^{a} =\mu I_D \,,
\end{align*}
with $\mu$ the measure
\begin{align} 
    \begin{aligned}
        \mu =& \mu_1\wedge\dots\wedge\mu_N \,,\\
        \mu_{i} =&  p_{a,b} du_i \wedge dv_i + p_{a,c} du_i \wedge dw_i + p_{b,c} dv_i \wedge dw_i \,,\\
    \end{aligned}
\end{align}
and $I_D$ the integrand
\begin{align} 
    I_D = s_D\exp[\text{Tr}[P_D Q_D^T] + \lambda (p_{a,b}|Q_D^{12}|+p_{a,c}|Q_D^{13}|+p_{b,c}|Q_D^{23}|)] \,.
\end{align}
$Q_D$ is an array filled with $q$-vectors according to the boundary ordering and $s_D$ is a sign that will be explained soon. $P_D$ is an array filled with $r$-vectors, also matching the boundary ordering. The $r$-vectors are $\vec{r}_i=\begin{pmatrix}
    p_{a,i},p_{b,i},p_{c,i}
\end{pmatrix}$ with $i=1,\dots, N$ for elements of $\mathbb{A}_0$ and $\vec{r}_a=\begin{pmatrix}
    -1,0,0
\end{pmatrix}$, $\vec{r}_b=\begin{pmatrix}
    0,-1,0
\end{pmatrix}$ and $\vec{r}_c=\begin{pmatrix}
    0,0,-1
\end{pmatrix}$ for elements of $\mathbb{A}_1$.
The potential $\Omega^a_{k,l,m,n}$ is obtained by restricting the master potential to $\mathbb{W}_{k,l,m,n}$, i.e.
\begin{align}
    \Omega_{k,l,m,n}^{a} = \Omega_N^{a}\Big|_{\mathbb{W}_{k,l,m,n}} \,.
\end{align}
From there, $\Omega^c_{k,l,m,n}$ can be found.

\begin{figure}
    \centering
    \begin{tikzpicture}[scale=0.3]
        \draw[thick](0,0) circle (6);
        \draw[ultra thick, blue](0,0) -- (150:6);
        \draw[ultra thick, blue](0,0) -- (30:6);
        \draw[ultra thick, blue](0,0) -- (270:6);
        \draw[red,thin] (150:3) .. controls (90:5) .. (90:6);
        \draw[dashed, red](0,0) -- (0,6);
        \draw[dashed, red](0,0) -- (-30:6);
        \draw[dashed, red](0,0) -- (210:6);
        \draw[-Latex] (0,6) -- (0,8.5);
        \filldraw [red!45!white, fill opacity=0.7] (0:0cm) -- (30:6cm) arc(30:-30:6cm) -- cycle;
        \filldraw [red!45!white, fill opacity=0.7] (0:0cm) -- (-90:6cm) arc(-90:-150:6cm) -- cycle;
        \filldraw [red!45!white, fill opacity=0.7] (0:0cm) -- (150:6cm) arc(150:210:6cm) -- cycle;
        \coordinate[label=above left : $a$] (B) at (150:6);
        \coordinate[label=above right : $c$] (B) at (30:6);
        \coordinate[label=below : $b$] (B) at (270:6);

        \begin{scope}[xshift=20cm]
        \draw[thick](0,0) circle (6);
        \draw[ultra thick, blue](0,0) -- (150:6);
        \draw[ultra thick, blue](0,0) -- (30:6);
        \draw[ultra thick, blue](0,0) -- (270:6);
        \draw[red,thin] (30:3) .. controls (90:5) .. (90:6);
        \draw[dashed, red](0,0) -- (0,6);
        \draw[dashed, red](0,0) -- (-30:6);
        \draw[dashed, red](0,0) -- (210:6);
        \draw[-Latex] (0,6) -- (0,8.5);
        \filldraw [red!45!white, fill opacity=0.7] (0:0cm) -- (-30:6cm) arc(-30:30:6cm) -- cycle;
        \filldraw [red!45!white, fill opacity=0.7] (0:0cm) -- (-30:6cm) arc(-30:-90:6cm) -- cycle;
        \filldraw [red!45!white, fill opacity=0.7] (0:0cm) -- (150:6cm) arc(150:210:6cm) -- cycle;
        \coordinate[label=above left : $a$] (B) at (150:6);
        \coordinate[label=above right : $c$] (B) at (30:6);
        \coordinate[label=below : $b$] (B) at (270:6);
        \end{scope}
    \end{tikzpicture}
    \caption{The sign $s_D$ of $\Omega_{k,l,m,n}^a$ and $\Omega_{k,l,m,n}^c$ is determined by the number of red lines in the shaded regions in the left and right diagram, respectively.}
    \label{signIntro}
\end{figure}
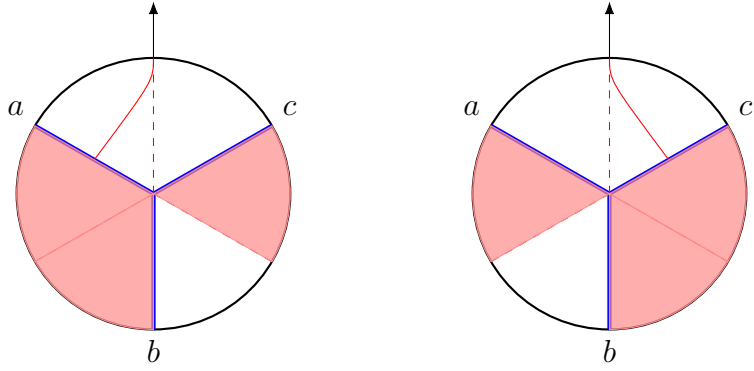

The sign $s_D$ of the potential $\Omega^a_{k,l,m,n}$ and $s'_D$ for $\Omega^c_{k,l,m,n}$ are given by
\begin{align}
    \begin{aligned}
        s_D &= (-1)^M  & \text{and} & & s'_D &= (-1)^{M+1} \,,
    \end{aligned}
\end{align}
where $M$ is the number of red lines in the shaded regions in the left (right) panel of Figure \ref{signIntro} for the diagram associated with $\Omega^a_{k,l,m,n}$ ($\Omega^c_{k,l,m,n}$).

\subsubsection{Boundaries}
\begin{figure}
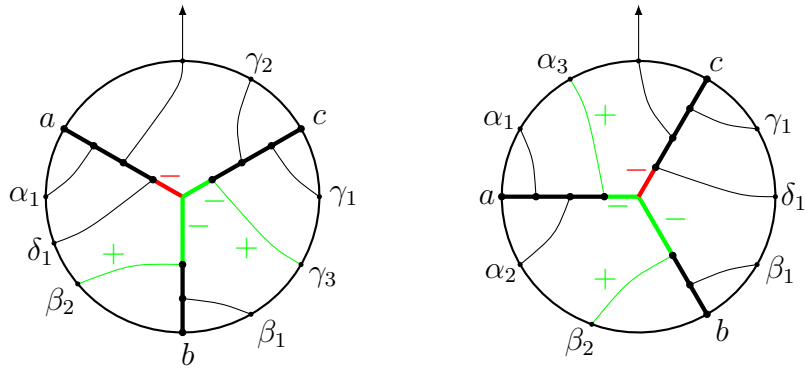

\centering

\caption{Disk diagrams that show which line segments correspond to boundaries that yield non-zero expressions for potentials of the type $\Omega_{k,l,m,n}^a$ and $\Omega_{k,l,m,n}^c$ on the left and right, respectively. The diagrams also show the signs that the boundaries are accompanied with.} \label{signsIntro}
\end{figure}

The boundary of $\mathbb{W}_{k,l,m,n}$ is a union of many boundary components $P_i$, i.e. $\partial\mathbb{W}_{k,l,m,n}=\cup_i P_i$. Each $P_i$ is obtained by saturating one inequality in \eqref{ineqsIntro}. We differentiate between various types of boundary components.
\begin{itemize}
    \item $A_\infty$-components: on these boundary components, one retrieves a term from the $A_\infty$-relations.
    \item Gluing term: the potential are non-zero on these boundary components, but do not produce $A_\infty$-terms. Instead, there is another disk diagram with a different topology that evaluates to the same value (up to a minus sign) on one of its boundary components. The contributions then cancel pairwise.
    \item Zero measure: on these boundary components the measure vanishes and so do the potentials. This gives no contribution.
    \item Higher codimension coboundary: sometimes saturating inequalities in \eqref{ineqsIntro} has implications on other inequalities and fixes more than one coordinate, thereby leading to higher codimension boundaries. This gives no contribution. To apply Stokes' theorem, we are only interested in codimension $1$ boundaries.
\end{itemize}
As mentioned above, the limits are reached by saturating an inequality in \eqref{ineqsIntro}. In the HPT language this corresponds to evaluating an integral on its upper or lower integration bound. This allows us to formulate a diagrammatic visualization of disk diagrams on the corresponding boundaries. All internal lines, except for the outmost segment of the $a$-leg, $b$-leg and $c$-leg contain an integration, due to the presence of the homotopy operator, and give rise to two boundary terms. Most of these fall into the classes of vanishing boundary contributions that were discussed above. The internal lines that yield non-vanishing boundary terms are represented by green and red lines in Figure \ref{signsIntro}. The green/red lines correspond to evaluating the potential on the upper/lower bound and they are accompanied by the sign with which these terms appear.

For the non-vanishing boundary terms, we should devise a visual representation for transforming a disk diagram into a nested disk diagram, see Figure \ref{SD4Intro}, in which the $A_\infty$-relations are represented. To this end, we distinguish between the boundaries related to (i) the line segments on the $a$-leg, $b$-leg, and $c$-leg and (ii) the line segments connected to elements $\alpha_i$, $\beta_i$, and $\gamma_i$.

In the former case, the chord segment shrinks to a point and the closed curved line is moved to the junction and then it moves past the junction to another leg, such that it is still connected to the same arc of the circle and lines do not intersect. Then, the leg at which we evaluate the boundary is split off and forms a disk diagram itself, which is connected to the a curved line in the original disk diagram by an arrow pointing toward the diameter of the new disk diagram. If moving the curved line over the junction results in two legs having no curved lines attached to them, both legs are separated from the disk diagram, forming a new one. Then, a curved line in the new disk diagram is connected by an arrow with the diameter of the original one. Lastly, if moving the curved line over the junction does not lead to an admissible nested diagram, the curved line remains at the junction and this becomes a gluing term. On the other hand, if the boundary is evaluated on a curved line, the leg onto which this line is attached splits off and forms a new disk diagram with a curved line in the original disk diagram connected to the new diagram by an arrow pointing to its diameter.

\begin{figure}[h!]
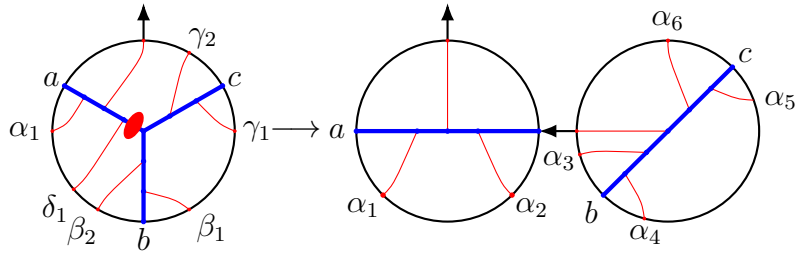

        \centering

        \caption{An example of a boundary term contributing to $\mathcal{V}(a,\alpha_1,\alpha_2,\mathcal{V}(\alpha_3,b,\alpha_4,\alpha_5,c,\alpha_6))$. After splitting the disk diagram in two, one relabels the elements of $\mathbb{A}_0$, as to make them correspond to the arguments of the structure maps $\mathcal{V}$.}
        \label{fig:1}
    \end{figure}

An example of this procedure is given in Figure \ref{fig:1}, which gives rise to nested structure maps of type $\mathcal{V}(\dots,\mathcal{V}(\dots),\dots)$. This boundary is obtained by setting $u_i^\bullet=0$, where $\bullet$ stands for $a$, $b$ or $c$.

\begin{figure}[h!]
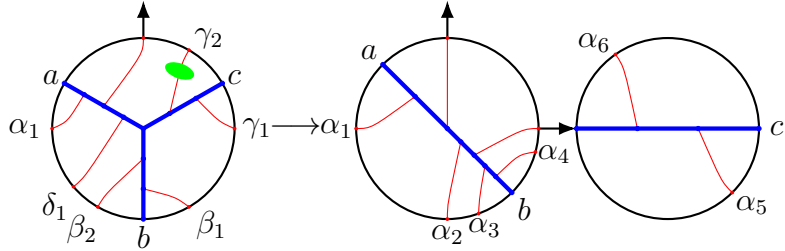

        \centering

        \caption{An example of a boundary term contributing to $\mathcal{V}(a,\alpha_1,\alpha_2,\alpha_3,b,\alpha_4,\mathcal{U}(\alpha_5,c,\alpha_6))$.}
        \label{fig:2}
    \end{figure}

Another example is displayed in Figure \ref{fig:2}, which gives rise to nested structure maps of type $\mathcal{V}(\dots,\mathcal{U}(\dots),\dots)$. This boundary is reached by setting $w_i^\bullet=0$.

\begin{figure}[h!]
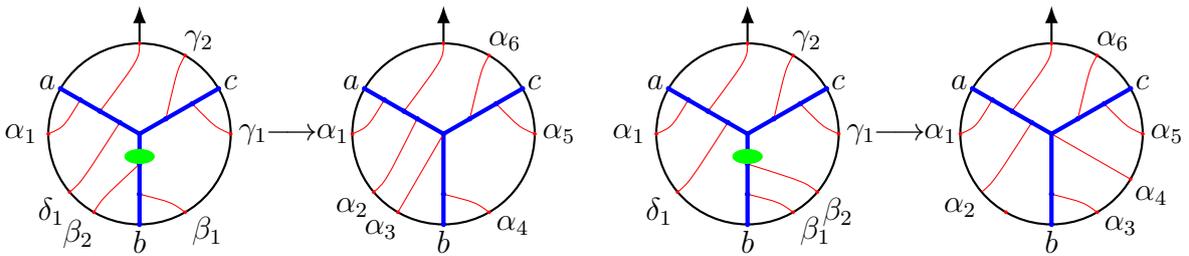

       \centering
       %
       \caption{Two examples of a boundary term contributing to a gluing term, with both orientations of $\beta_2$.}
       \label{fig:3}
   \end{figure}

Lastly, a gluing term is depicted in Figure \ref{fig:3}. This boundary term is obtained by setting $\frac{u_m^c}{v_m^c}=\frac{u_l^b}{v_l^b}$.

Now that the recipe has been explained, all that is left to do is to see if the terms in the $A_\infty$-relations are indeed recovered correctly by evaluating the potentials on the boundary components of $\mathbb{W}_{k,l,m,n}$ without producing additional terms. We leave the details to chapter \ref{chap:Stokes} and only state the result: the $A_\infty$-relations underlying chiral HiSGRA can be proven by Stokes' theorem for any value of the cosmological constant and this positive result strongly hints at some extension of the (Shoikhet-Tsygan-) Kontsevich formality theorem.

\chapter{Minimal model for self-dual theories}
\label{chap:SD}

In this chapter, we derive the Free Differential Algebra (FDA) for self-dual Yang-Mills and self-dual gravity. The content is entirely based on \cite{SDFDA}, co-authored with Evgeny Skvortsov, and published in the \textit{Journal of High Energy Physics}.

Note that, due to changes introduced in later works, the primed and unprimed spinor indices are swapped in this chapter and the next compared to the introduction and the other chapters.

\section{Introduction}
Self-dual theories have a number of remarkable properties that make them very useful toy models in general and first order approximations to more complicated theories: 
(a) self-dual theories are closed subsectors of the corresponding complete theories; (b) as a result, all solutions of self-dual theories are solutions of the full ones; (c) all amplitudes of self-dual theories are also amplitudes of the full ones; (d) self-dual theories are integrable; (e) self-dual theories are finite and one-loop exact; (f) existence of a self-dual truncation allows one to rearrange the perturbation theory in a nontrivial way, e.g. to represent Yang-Mills theory as expansion over self-dual rather than flat backgrounds; (g) tools from twistor theory are very-well adapted to self-dual theories, see e.g. \cite{Penrose:1976js,Ward:1977ta,Atiyah:1978wi,Chalmers:1996rq,Mason:1991rf,Witten:2003nn,Berkovits:2004jj,Atiyah:2017erd}. We are interested in constructing $L_\infty$-algebras of the simplest self-dual theories: SDYM and SDGR, to uncover their algebraic structure.  

There is a hierarchy of $L_\infty$-algebras that originate from (quantum) field theories and string field theory, see e.g. \cite{Zwiebach:1992ie,Gaberdiel:1997ia,Kajiura:2003ax,Lada:1992wc,Alexandrov:1995kv,Barnich:2004cr,Hohm:2017pnh,jurvco2019algebras}. The simplest $L_\infty$-algebras emerge from a re-interpretation of the BV-BRST formalism: upon expanding the master action in ghosts and anti-fields one finds multilinear maps that obey $L_\infty$-relations. Another $L_\infty$-algebra emerges from the jet space version of the BV-BRST formulation of a given gauge theory \cite{Barnich:1994db,Barnich:1994mt,Barnich:2010sw,Grigoriev:2012xg,Grigoriev:2019ojp}. Such $L_\infty$ is especially useful when investigating various properties of this gauge theory systematically, e.g. classification of deformations of the action, or the question of possible anomalies \cite{Barnich:1994db,Barnich:1994mt}. Given an $L_\infty$-algebra one can consider various equivalent reductions. The smallest possible quasi-isomorphic algebra is the minimal model, which still captures all the relevant properties of the field theory. Another closely related algebraic structure is Free Differential Algebra \cite{Sullivan77}, which emerges as the sigma-model based on the minimal model. 

In this letter we construct the minimal models for self-dual Yang-Mills and self-dual gravity theories. As a starting point we take the Chalmers-Siegel action \cite{Chalmers:1996rq} for SDYM and the recently constructed action for SDGR with vanishing cosmological constant \cite{Krasnov:2021cva}, which is equivalent to other actions in the literature \cite{Siegel:1992wd,AbouZeid:2005dg}.  

Our general motivation stems from several possible applications, where we hope to understand from the algebraic, $L_\infty$, point of view: (i) integrability of self-dual theories; (ii) the double-copy relations, see \cite{Bern:2008qj,Bern:2010ue} and \cite{Campiglia:2021srh,Borsten:2021hua} for the recent results in this direction. Also, the results serve as a starting point for covariantization \cite{SDFDA2} of Chiral Higher Spin Gravity \cite{Metsaev:1991mt,Metsaev:1991nb,Ponomarev:2016lrm,Ponomarev:2017nrr,Skvortsov:2018jea,Skvortsov:2020wtf,Skvortsov:2020gpn}.

We begin with a short review of relation between $L_\infty$ and field theory and then proceed to SDYM and SDGR, respectively, with some technicalities left to appendices.

\section{Minimal models}
\label{sec:FDA1}
As it was already sketched in the introduction, given any (gauge) field theory in the BV-BRST language it is natural to consider its jet space extension \cite{Brandt:1997iu,Brandt:1996mh,Barnich:2010sw,Grigoriev:2012xg,Grigoriev:2019ojp}, which is what is done when investigating the local BRST-cohomology \cite{Barnich:1994db,Barnich:1994mt}. The jet space extension leads to a rather big $L_\infty$-algebra, better say to a $Q$-manifold provided global issues are taken into account. Various $Q$-cohomology groups correspond to all physically relevant quantities, e.g. deformations/interactions, anomalies, charges, etc., see e.g. \cite{Barnich:1994db,Barnich:1994mt}. For every $L_\infty$-algebra there always exists a (usually much smaller) $L_\infty$-algebra, known as the minimal model, see e.g. \cite{Huebschmann,Grigoriev:2019ojp,Grigoriev:2020lzu}, that contains the same information --- it is said to be quasi-isomorphic.\footnote{There is also another, 'quantum', minimal model \cite{Arvanitakis:2019ald} --- the $L_\infty$-algebra given by 1PI correlation functions. }  Some care is needed to prove the same statement for field theories \cite{Barnich:2009jy,Grigoriev:2019ojp}, where relevant $L_\infty$-algebras are necessarily infinite-dimensional. Minimal models were first introduced by Sullivan \cite{Sullivan77} in the context of differential graded algebras to study rational homotopy theory. We construct such minimal models for SDYM and SDGR.

Given any non-negatively graded supermanifold $\mathcal{N}$ equipped with a homological vector field $Q$, $QQ=0$, e.g. given by the minimal model, one can write down a sigma-model \cite{Barnich:2010sw}:
\begin{align*}
    d \Phi&= Q(\Phi)  \,,
\end{align*}
where $\Phi\equiv \Phi^\aA$ are maps $\Pi T\mathcal{M} \rightarrow \mathcal{N}$ from the exterior algebra of differential forms on a spacetime manifold $\mathcal{M}$ to $\mathcal{N}$. Together with natural gauge symmetries the sigma-model is equivalent to the classical equations of motion of the initial field theory \cite{Barnich:2010sw,Grigoriev:2012xg,Grigoriev:2019ojp}, thereby having the form of a Free Differential Algebra, see \cite{Sullivan77} for exact definitions.\footnote{FDA was introduced by Sullivan and applied to problems in topology. Later, FDA's sneaked into physics in the context of supersymmetry and supergravity \cite{vanNieuwenhuizen:1982zf,DAuria:1980cmy} and, even later, applied to construct formally consistent deformations of the FDA for free higher spin fields \cite{Vasiliev:1988sa}. } In this thesis we adopt a more pragmatic point of view on minimal models: we seek for the classical equations of motion as an FDA \cite{Vasiliev:1988sa}. If $\Phi=\{ \Phi^\aA\}$ are coordinates on $\mathcal{N}$, then $Q=Q^\aA \pl/\pl \Phi^\aA$ and 
\begin{align*}
    Q^2&=0 &&\Longleftrightarrow && Q^\aB \tfrac{\pl}{\pl \Phi^\aB} Q^\aA=0\,.
\end{align*}
This condition is equivalent to the Frobenius integrability of the field equations, i.e. the equations are formally consistent. The $L_\infty$-relations emerge by Taylor expanding $QQ=0$ at a stationary point of $Q$ \cite{Alexandrov:1995kv}. By abuse of notation we always denote coordinates on $\mathcal{N}$ and the corresponding fields by the same symbols. For a large class of field theories $\mathcal{N}$ has coordinates of degree-one and degree-zero to be associated with gauge connection(s) $A$ and with some matter-like zero-forms $L$. The simplest FDA with this data reads  
\begin{align*}
    dA&=\tfrac12[A,A]\,,& dL&=\rho(A)L\,,
\end{align*}
and is equivalent to $A$ taking values in some Lie algebra and to $L$ taking values in its module $\rho$. We consider these equations free. In particular, it is easy to solve them locally in the pure gauge form, e.g. $A=g^{-1} dg$. The most general deformation of the free equations here-above that is consistent with the form-degree counting reads
\begin{align*}
     dA&=l_2(A,A)+l_3(A,A,L)+l_4(A,A,L,L)+\ldots=F_A(A;L)\,,\\
     dL&=l_2(A,L)+l_3(A,L,L)+\ldots=F_L(A;L)\,.     
\end{align*}
Our strategy for each of the cases, SDYM and SDGR, is to start off with an action, rewrite the variational equations of motion in the 'almost' FDA form, where 'almost' means that at each step the equations/$Q$-structure will require new fields/coordinates on $\mathcal{N}$ be introduced. At the end of the day we find the complete $Q$. Interacting field theories are defined modulo admissible field redefinitions (those that do not change the $S$-matrix). We found a field frame where no structure maps higher than $l_3(\bullet,\bullet,\bullet)$ are needed for SDYM and SDGR, which also fixes all field redefinitions.\footnote{This is a key difference with respect to \cite{Vasiliev:1988sa}, where locality and field redefinitions are not taken into account \cite{Boulanger:2015ova,Skvortsov:2015lja}, which results in a general ansatz for interactions rather than a concrete theory.  }

\section{SDYM}
\label{sec:SDYM}

\subsection{Action, initial data}

The theory can be formulated  \cite{Chalmers:1996rq}  with two fields:\footnote{We use almost exclusively the two-component spinor language, which is well-suited for $4d$-theories. A short compendium can be found in Appendix \ref{app:notation}. A classical source is \cite{penroserindler}. The most important fact about our notation is that symmetric or to be symmetrized indices can be denoted by the same letter. Also, $A(k)\equiv A_1...A_k$.  } the usual one-form gauge potential $A\equiv A_\mu \, dx^\mu \equiv A_\mu^a \, dx^\mu\, T_a$ and a zero-form $\Psi^{AB}\equiv\Psi^{BA}$, $\Psi^{AB}\equiv \Psi^{AB; a}\, T_a$. Here $T_a$ are generators of some Lie algebra with a non-degenerate invariant bilinear form. We usually suppress form indices and $dx$'s, as well as the Lie algebra indices. In practice, it is convenient to think of generators $T_a$ as of taking values in some matrix algebra and assume $A$ and $\Psi^{AB}$ to take values in $\mathrm{Mat}_N$, with matrix indices again suppressed. The action reads\footnote{Here, see also appendix \ref{app:notation}, $H^{AB}\equiv H^{BA}$, $H^{A'B'}\equiv H^{B'A'}$ is the basis of self-dual two-forms, $H^{AB}\equiv e\fud{A}{C'}\wedge e^{BC'}$, \textit{idem.} for $H^{A'B'}$. Vierbein one-form is $e^{AA'}$. }
\begin{align} \label{actionSDYMchap}
    S_{SDYM}&=\mathrm{tr} \int \Psi^{A'B'} \wedge H_{A'B'} \wedge F \,,
\end{align}
where $F=dA-A\wedge A$ and we prefer to omit $\wedge$-symbol. The equations of motion imply
\begin{align} \label{SDYMeq}
    F_{A'B'}&=0 \,, & D\fud{A}{B'}\Psi^{A'B'}&=0 \,,
\end{align}
where $D\equiv dx^{AA'}\,D_{AA'}\equiv \nabla- [A,\bullet]$ is the gauge and Lorentz covariant derivative. We also used the decomposition of $F$ into (anti)self-dual parts
\begin{equation*}
    F=H^{BB}F_{BB}+H^{B'B'}F_{B'B'} \,.
\end{equation*}
We can rewrite the variational equations as
\begin{align}\label{F&Psi}
    dA-AA&=H^{BB}F_{BB}\,, & D \Psi^{A'B'}&= e_{CC'}\Psi^{C,A'B'C'} \,,
\end{align}
which is the starting point for constructing the corresponding $L_\infty$-algebra. The first equation simply states that $F_{A'B'}=0$ and, hence, connection $A$ is self-dual. Therefore, only the self-dual part may not be trivial and it is parameterized by $F^{AB}$. A simple consequence is the Bianchi identity for $F^{AB}$. In the second equation we introduced a field $\Psi^{A,A'B'C'}$ that parameterizes the first derivative of $\Psi$ that is consistent with \eqref{SDYMeq}, i.e. it corresponds to a coordinate on the on-shell jet of $\Psi^{A'B'}$.\footnote{Equations of motion for free fields of arbitrary spin can be recast into the FDA form \cite{Vasiliev:1986td}. The on-shell jet is very easy to describe in spinorial language \cite{penroserindler}. } 

The problem is, therefore, to find a completion of \eqref{F&Psi}, which requires an infinite set of coordinates on $\mathcal{N}$ and $Q$ defined on them in such a way that $QQ=0$. The first few terms of $Q$ and $\mathcal{N}$ are already clear from \eqref{F&Psi}. The on-shell jet space is also well-known \cite{penroserindler}. It is the same as for the free theory where we turned off non-Abelian Yang-Mills groups that result in non-linearities. That the coordinates on $\mathcal{N}$ are the same for the free and interacting theories is due to the requirement for them to have the same number of local degrees of freedom. 

\paragraph{Coordinates, on-shell jet.} The coordinates on $\mathcal{N}$ are: degree-one $A$; degree-zero $F^{A(k+2),A'(k)}$ and $\Psi^{A(k),A'(k+2)}$, $k=0,1,2,...$. The free equations, i.e. (self-dual) Maxwell equations on Minkowski space, can be written as  \cite{Vasiliev:1986td}
\begin{align}\label{eqMaxwA}
    dA&=H^{BB}F_{BB} +\epsilon H^{B'B'} \Psi_{B'B'} \,,
\end{align}
which just defines $F^{AB}$ and $\Psi^{A'B'}$ as (anti)-self-dual components of $dA$. The Bianchi identities imply 
\begin{align}\label{eqMaxwB}
    dF^{A(k+2),A'(k)}=e_{BB'}F^{A(k+2)B,A'(k)B'}\,,
\end{align}
and a similar chain of equations for the field $\Psi$
\begin{align}\label{eqMaxwC}
    d\Psi^{A(k),A'(k+2)}=e_{CC'}\Psi^{A(k)C,A'(k+2)C'}\,.
\end{align}
The system \eqref{eqMaxwA}, \eqref{eqMaxwB}, \eqref{eqMaxwC} is equivalent to Maxwell equations, i.e. no self-dual truncation has yet been taken. The free SDYM equations are obtained by setting $\epsilon=0$ in \eqref{eqMaxwA}, while no other modifications are needed. What erasing $\Psi^{A'B'}$ from \eqref{eqMaxwA} does is that it makes the anti-selfdual part of $dA$ vanish. The $\Psi$-subsystem \eqref{eqMaxwC} decouples and describes the second degree of freedom (say, helicity $-1$). The first equations in \eqref{eqMaxwB} and \eqref{eqMaxwC} are equivalent to the well-known \cite{Penrose:1965am}
\begin{align*}
    D\fdu{A}{B'}F^{AB}&=0 \,, & D\fud{A}{B'}\Psi^{A'B'}&=0\,,
\end{align*}
and describe helicity $+1$ and $-1$ degrees of freedom. Subsystems \eqref{eqMaxwB} and \eqref{eqMaxwC} are closed and identical to each other (upon swapping primed and unprimed indices). What makes them different is that only the physical degree of freedom carried by $F$ gets embedded into $A$ once we set $\epsilon=0$. There is no change in the number of physical degrees of freedom in the $\epsilon=0$ limit.  

\paragraph{General form.} In order to have a genuine FDA we should incorporate the background gravitational fields: vierbein $e^{AA'}$ and the (anti)-self-dual components $\omega^{AB}$, $\omega^{A'B'}$ of spin-connection. Finally, we have
\begin{align*}
    \mathcal{N}&: && 
    \begin{aligned}
        1&: e^{AA'}\,, \omega^{AB}\,, \omega^{A'B'} \,, A \,,\\
        0&: F^{A(k+2),A'(k)}\,, \Psi^{A(k),A'(k+2)}\,, k=0,1,2,...
    \end{aligned}
\end{align*}
We will prove below that the complete $L_\infty$-algebra of SDYM can be cast into the following simple form:
\besubeqs
\begin{align*}
    d e^{AA'}&= \omega\fud{A}{B}\wedge
    e^{BA'}+\omega\fud{A'}{B'}\wedge e^{A B'} \,,\\  
    d\omega^{AB}&= \omega\fud{A}{C}\wedge \omega^{BC} \,,\\
    d\omega^{A'B'}&= \omega\fud{A'}{C'}\wedge \omega^{B'C'} \,,\\
    dA&= AA + H_{BB}F^{BB} \,,\\
    dF&= l_2(\omega,F)+l_2(A,F)+l_2(e,F)+l_3(e,F,F) \,,\\
    d\Psi&= l_2(\omega,\Psi)+l_2(A,\Psi)+l_2(e,\Psi) +l_3(e,F,\Psi) \,.
\end{align*}
\esubeqs
Some of the maps above are self-evident, e.g. $l_2(\omega,\bullet)$ and $l_2(A,\bullet)$ are parts of the usual Lorentz and gauge covariant derivatives. $H_{BB}F^{BB}$ is a specific tri-linear map $l_3(e,e,F)$. 
Introducing the standard Lorentz covariant derivative $\nabla$ and appending it with the gauge part $[A,\bullet]$ we define $D=\nabla -[A,\bullet]$. The equations reduce to
\besubeqs
\begin{align*}
    \nabla e^{AA'}&= 0 \,, & \nabla^2&=0 \,,\\
    dA&= AA + H_{BB}F^{BB} \,, \\
    DF&= l_2(e,F)+l_3(e,F,F) \,, \\
    D\Psi&= l_2(e,\Psi)+l_3(e,F,\Psi) \,.
\end{align*}
\esubeqs
The first line is equivalent to living in Minkowski space. Covariant derivative $D$ allows us to absorb $l_2(A,F)=[A,F]$ and $l_2(A,\Psi)=[A,\Psi]$. The $L_\infty$-structure relations are equivalent to (i) $e$, $\omega$ being a flat connection of Poincare algebra; (ii) a bit more nontrivial $L_\infty$-relations that follow from
\begin{align*}
    D^2 F +l_2(e, DF) +l_3(e,DF,F)+l_3(e,F,DF)&\equiv0 \,,
    \\
    D^2\Psi+l_2(e,D\Psi)+l_3(e,DF,\Psi)+l_3(e,F,D\Psi)&\equiv0
\end{align*}
and decompose into
{\allowdisplaybreaks
\besubeqs \label{SDYMstasheff}
\begin{align} 
    l_2(e, l_2(e,F))&\equiv0 \label{SDYMstasheffF1} \,, \\
    -[H_{BB}F^{BB}, F] +l_2(e, l_3(e,F,F)) +l_3(e,l_2(e,F),F)+l_3(e,F,l_2(e,F))&\equiv0 \label{SDYMstasheffF2} \,, \\
    l_3(e,l_3(e,F,F),F)+l_3(e,F,l_3(e,F,F))&\equiv0 \label{SDYMstasheffF3} \,, \\
    l_2(e,l_2(e,\Psi))&\equiv0 \label{SDYMstafhessPsi1} \,, \\
    -[H_{BB}F^{BB},\Psi]+l_2(e,l_3(e,F,\Psi))+l_3(e,l_2(e,F),\Psi)+l_3(e,F,l_2(e,\Psi))&\equiv0 \label{SDYMstafhessPsi2} \,,\\
    l_3(e,l_3(e,F,F),\Psi)+l_3(e,F,l_3(e,F,\Psi))&\equiv0 \label{SDYMstafhessPsi3}\,.
\end{align}
\esubeqs}
The first and the fourth relations are guaranteed by the free equations of motion. 

\subsection{FDA, flat space}
\label{sec:flatSDYM}
\paragraph{Appetizer.} Firstly, let us explain why a non-linear completion of \eqref{eqMaxwA}, \eqref{eqMaxwB}, \eqref{eqMaxwC} is necessary. The root of the nonlinear completion is in the fact that $D^2\neq0$ and, for a 
field $\chi$ in representation $\rho$ of the Yang-Mills algebra we find $D^2\chi=-\rho(F)\chi$. In the adjoint representation, one gets (matrix/Lie algebra indices are implicit)
\begin{align*}
	D^2\chi=-[F,\chi]=-H_{BB}[F^{BB},\chi] \,.
\end{align*}
Therefore, the Bianchi identity for the first equation in the $F$-subsystem
\begin{align*}
	D F^{AA}=e_{BB'}F^{AAB,B'}
\end{align*}
leads to 
\begin{align} \label{nablasquaredF}
	D^2 F^{AA}=-H_{BB}[F^{BB},F^{AA}]=-e_{BB'}\wedge D F^{AAB,B'}\,.
\end{align}
The aim is to use the above equation to obtain $D F^{AAA,A'}$. Matching the indices and imposing that $D F^{AAA,A'}$ is a one-form, one may take the ansatz
\begin{align*}
    \begin{aligned}
	   D F^{AAA,A'}&=e_{CC'}F^{AAAC,A'C'}+\alpha e\fdu{C}{A'}[F^{AC},F^{AA}]+\beta e\fdu{C}{A'}[F^{AA},F^{AC}]\\
	   &+\gamma e^{AA'}[F\fud{A}{C},F^{AC}] \,.
	 \end{aligned}
\end{align*}
The Fierz identity \eqref{Fierz} and the anti-symmetry of the commutator reduce this to
\begin{equation*}
    D F^{AAA,A'}=e_{CC'}F^{AAAC,A'C'}+\alpha_{00} e\fdu{C}{A'}[F^{AC},F^{AA}]\,,
\end{equation*}
where the label on $\alpha_{00}$ was added for future convenience. Upon contraction with $e_{BB'}$ this yields\footnote{The first term can be rewritten as $e_{BB'}\wedge e_{CC'}F^{AABC,B'C'}=\tfrac{1}{2}(H_{BC}\epsilon_{B'C'}+\epsilon_{BC}H_{B'C'})F^{AABC,B'C'}$  and vanishes as the contracted indices are symmetrized in $F^{AABC,B'C'}$ and anti-symmetrized in the $\epsilon$'s. The last term must be zero, because $e_{BB'}\wedge e\fdu{C}{B'}=\tfrac{1}{2}H_{BC}$ is symmetric in $B,C$, whereas the commutator is anti-symmetric. }
\begin{align} \label{e-nablaF3}
    \begin{aligned}
	    e_{BB'}\wedge D F^{AAB,B'}&=0+\tfrac{1}{3}\alpha_{00} e_{BB'}\wedge e\fdu{C}{B'}[F^{BC},F^{AA}]+\tfrac{2}{3}\alpha_{00}e_{BB'}\wedge e\fdu{C}{B'}[F^{AC},F^{AB}]\\
	    &=\tfrac{1}{3}\alpha_{00} H_{BB}[F^{BB},F^{AA}] \,,
    \end{aligned}
\end{align}
where \eqref{e-wedge-e} was used. Comparing this to \eqref{nablasquaredF}, one obtains the solution 
\begin{align} \label{DF3}
    D F^{AAA,A'}=e_{CC'}F^{AAAC,A'C'}+3e\fdu{C}{A'}[F^{AC},F^{AA}]\,,
\end{align}
where the first term on the r.h.s. is there due to the free equations. Similarly, taking the covariant derivative of the above result yields another consistency equation. Following the same steps as before, one finds
\begin{align*}
    \begin{aligned}
	D^2 F^{AAA,A'}&=-H_{BB}[F^{BB},F^{AAA,A'}]=-e_{BB'}\wedge D F^{AAAB,A'B'}\\
	&-3e\fdu{C}{A'}\wedge[D F^{AC},F^{AA}]-3e\fdu{C}{A'}[F^{AC},D F^{AA}]\,,
	\end{aligned}
\end{align*}
which results in
\begin{align} \label{e-DF4}
    \begin{aligned}
    e_{BB'}\wedge D F^{AAAB,A'B'}&=H_{BB}[F^{BB},F^{AAA,A'}]-\tfrac{3}{2}H_{BB}[F^{AA},F^{ABB,A'}]
    \\
    &+\tfrac{3}{2}H_{BB}[F^{AB},F^{AAB,A'}]+\tfrac{3}{2}H\fdu{B'}{A'}[F^{AB},F\fudu{AA}{B}{,B'}]\,.
    \end{aligned}
\end{align}
The minimal ansatz for $DF^{AAAA,A'A'}$ reads
\begin{align*}
    \begin{aligned}
    D F^{AAAA,A'A'}&=e_{CC'}F^{AAAAC,A'A'C'}+\alpha_{02} e\fdu{C}{A'}[F^{AC},F^{AAA,A'}]\\
    &+\alpha_{12} e\fdu{C}{A'}[F^{AAC,A'},F^{AA}]\,.
    \end{aligned}
\end{align*}
We contract this with $e_{BB'}$ to find
\begin{align}
    \begin{aligned}
        e_{BB'}\wedge DF^{AAAB,A'B'} &= \tfrac{3\alpha_{02}}{16}H_{BB}[F^{BB},F^{AAA,A'}]+(\tfrac{9\alpha_{02}}{16}-\tfrac{3\alpha_{12}}{8})H_{BB}[F^{AB},F^{AAB,A'}] \\
        &-\tfrac{3\alpha_{12}}{8}H_{BB}[F^{AA},F^{ABB,A'}] + (\tfrac{3\alpha_{02}}{16}+\tfrac{\alpha_{12}}{8})H\fdu{B'}{A'}[F^{AB},F\fudu{AA}{B}{,B'}] \,.
    \end{aligned}
\end{align}
We compare this to \eqref{e-DF4} to obtain the result
\begin{align*}
    \begin{aligned}
        D F^{AAAA,A'A'}&=e_{CC'}F^{AAAAC,A'A'C'}+\tfrac{16}{3}e\fdu{C}{A'}[F^{AC},F^{AAA,A'}]\\
        &+4e\fdu{C}{A'}[F^{AAC,A'},F^{AA}]\,.
    \end{aligned}
\end{align*}
The procedure presented above is nothing more than the practical realisation of solving the $L_\infty$-relation \eqref{SDYMstasheffF2}. This procedure will be generalized next. 

\paragraph{Main course, $\boldsymbol{F}$-sector.} By looking at the first few equations in the system it is easy to come up with an ansatz:
\begin{align}\label{spin1ansatz}
\begin{aligned}
        D F_{A(k+2),A'(k)}&=e^{BB'}F_{A(k+2)B,A'(k)B'}\\
        &+\sum_{n=0}^{k-1}\alpha_{nk}e\fud{B}{A'}[F_{A(n+1)B,A'(n)},F_{A(k-n+1),A'(k-n-1)}] \,,   
\end{aligned}
\end{align}
for any $k\geq 0$. This ansatz makes use of the fact that $D F_{A(k+2),A'(k)}$ should be a one-form, which requires the presence of $e^{BB'}$ and it matches the number of (un)-primed indices. In any non-linear theory there is always a freedom to perform field redefinitions. We have also fixed the redefinitions by requiring that there are no index contractions between $F$ in $[F,F]$. Terms with contracted indices can easily be introduced by field-redefinitions. Our ansatz contains only the terms that are necessary to ensure consistency and, thereby, is the minimal one.  

Taking the covariant derivative of the ansatz yields
\begin{align} \label{DsquaredF}
    \begin{aligned}
        D^2F_{A(k+2),A'(k)}=&-H^{BB}[F_{BB},F_{A(k+2),A'(k)}]=-e^{BB'}\wedge DF_{A(k+2)B,A'(k)B'}\\
        &-e\fud{B}{A'}\wedge\sum_{n=0}^{k-1}\alpha_{nk}[DF_{A(n+1)B,A'(n)},F_{A(k-n+1),A'(k-n-1)}] \\
        &-e\fud{B}{A'}\wedge\sum_{n=0}^{k-1}\alpha_{nk}[F_{A(n+1)B,A'(n)},DF_{A(k-n+1),A'(k-n-1)}] \,.
    \end{aligned}
\end{align}
and considering only terms quadratic in $F$ gives\footnote{In the third term we have made the anti-symmetry of the commutator explicit by writing $[X,Y]=\tfrac{1}{2}([X,Y]-[Y,X])$ and renaming the dummy indices accordingly. This automatically gets rid of terms that vanish because of symmetry reasons, like the last term in the middle expression of \eqref{e-nablaF3}. As the summation now runs up to $n=k$, the coefficient $\alpha_{kk}$ shows up, so we set $\alpha_{kk}=0$ by hand since it was not present in the ansatz.}
\begin{align} \label{e-nabla-F1}
    \begin{aligned}
    e^{BB'}\wedge D F_{A(k+2)B,A'(k)B'}&=H^{BB}[F_{BB},F_{A(k+2),A'(k)}]\\
    &-\tfrac{1}{2}H^{BB}\sum_{n=0}^{k-1}\alpha_{nk}[F_{A(n+1)BB,A'(n+1)},F_{A(k-n+1),A'(k-n-1)}]\\
    &-\tfrac{1}{4}H^{BB}\sum_{n=0}^{k}(\alpha_{nk}-\alpha_{(k-n)k})[F_{A(n+1)B,A'(n)},F_{A(k-n+1)B,A'(k-n)}]\\
    &+\tfrac{1}{2}H\fud{B'}{A'}\sum_{n=0}^{k-1}\alpha_{nk}[F\fdud{A(n+1)}{B}{,A'(n)},F_{A(k-n+1)B,A'(k-n-1)B'}] \,,
    \end{aligned}
\end{align}
where terms cubic in $F$ are ignored for now. Alternatively, we contract $e^{BB'}$ with $D F_{A(k+3),A'(k+1)}$ to obtain
\allowdisplaybreaks{
\begin{align*}
    &e^{BB'}\wedge D F_{A(k+2)B,A'(k)B'}=-\tfrac{1}{2}H^{BB}\alpha_{0(k+1)}\tfrac{k+2}{(k+3)(k+1)}[F_{BB},F_{A(k+2),A'(k)}]\\
    &-\tfrac{1}{2}H^{BB}\sum_{n=0}^{k-1}\alpha_{(n+1)(k+1)}\tfrac{(n+2)(k+2)}{(k+3)(k+1)}[F_{A(n+1)BB,A'(n+1)},F_{A(k-n+1),A'(k-n-1)}]\\
    &-\tfrac{1}{4}H^{BB}\sum_{n=0}^{k}(\alpha_{n(k+1)}\tfrac{(k-n+2)(k+2)}{(k+3)(k+1)}-\alpha_{(k-n)(k+1)})\tfrac{(n+2)(k+2)}{(k+3)(k+1)})[F_{A(n+1)B,A'(n)},F_{A(k-n+1)B,A'(k-n)}]\\
    &+\tfrac{1}{2}H\fud{B'}{A'}\sum_{n=0}^{k}(\alpha_{(k-n)(k+1)}\tfrac{(n+2)(k-n)}{(k+3)(k+1)}+\alpha_{n(k+1)}\tfrac{(k-n+2)(k-n)}{(k+3)(k+1)})[F\fdud{A(n+1)}{B}{,A'(n)},F_{A(k-n+1)B,A'(k-n-1)B'}] \,.
\end{align*}}
Comparing this with \eqref{e-nabla-F1} results in the following system of recurrence relations:
\besubeqs
    \begin{align*}
        0&=\alpha_{0k}+\tfrac{2k(k+2)}{k+1}\,,\\
        0&=\alpha_{(n+1)(k+1)}\tfrac{(n+2)(k+2)}{(k+3)(k+1)}-\alpha_{nk}\,,\\
        0&=\alpha_{n(k+1)}\tfrac{(k-n+2)(k+2)}{(k+3)(k+1)}-\alpha_{(k-n)(k+1)}\tfrac{(n+2)(k+2)}{(k+3)(k+1)}-\alpha_{nk}+\alpha_{(k-n)k}\,,\\
        0&=\alpha_{(k-n)(k+1)}\tfrac{(n+2)(k-n)}{(k+3)(k+1)}+\alpha_{n(k+1)}\tfrac{(k-n+2)(k-n)}{(k+3)(k+1)}-\alpha_{nk}\,.
    \end{align*}
\esubeqs
This system is over-determined, but, nevertheless,  is solved by
\begin{align*}
    \alpha_{nk}=-\tfrac{2}{(n+1)!}\tfrac{(k+2)!}{(k-n-1)!(k-n+1)(k+1)}\,.
\end{align*}
The full solution reads
\begin{align} \label{spin1sol}\boxed{
    \begin{aligned}
    D F_{A(k+2),A'(k)}&=e^{BB'}F_{A(k+2)B,A'(k)B'}\\
    &-e\fud{B}{A'}\sum_{n=0}^{k-1}{{\tfrac{2}{(n+1)!}\tfrac{(k+2)!}{(k-n-1)!(k-n+1)(k+1)}}}[F_{A(n+1)B,A'(n)},F_{A(k-n+1),A'(k-n-1)}]\,.
    \end{aligned}}
\end{align}
It was assumed that the ansatz only contains linear and quadratic terms in $F$. The fact that terms cubic in $F$ vanish in \eqref{e-nabla-F1} is proved in Appendix \ref{app:sdymTruncation}. This confirms the $L_\infty$-relation in \eqref{SDYMstasheffF3} and it implies that $D F_{A(k+2),A'(k)}$ indeed truncates at quadratic order.

\paragraph{Main course, $\boldsymbol{\Psi}$-sector.} 
As was clear from the $L_\infty$-relations in \eqref{SDYMstasheff}, the non-linear extension of the $\Psi$-sector is different from the $F$-sector. The minimal ansatz for $D\Psi_{A(k),A'(k+2)}$ is slightly more involved as it reads
\begin{align}\label{ansatzpsi}
\begin{aligned}
    D\Psi_{A(k),A'(k+2)}&=e^{CC'}\Psi_{A(k)C,A'(k+2)C'}+\sum_{n=0}^{k-1}\beta_{nk}e\fud{C}{A'}[F_{A(n+1)C,A'(n)},\Psi_{A(k-n-1),A'(k-n+1)}]\\
    &+\sum_{n=0}^{k-2}\gamma_{nk}e\fud{C}{A'}[F_{A(n+2),A'(n)},\Psi_{A(k-n-2)C,A'(k-n+1)}]\,.
\end{aligned}
\end{align}
We follow the same steps as for the $F$-sector: we write the Bianchi identity for the ansatz above and as a parallel calculation we contract $e^{BB'}$ with $\Psi_{A(k+1),A'(k+3)}$ to obtain two expressions for $e^{BB'}\wedge D \Psi_{A(k)B,A'(k+2)B'}$ and compare them. This provides us with a system of recurrence relations for $\beta_{nk}$ and $\gamma_{nk}$. The details of the calculation are left for Appendix \ref{app:SDYMPsi}. The system is solved by
\begin{align*}
    \beta_{nk}&=-\tfrac{2}{(n+1)!}\tfrac{k-n+2}{k+3}\tfrac{k!}{(k-n-1)!}\,, & \gamma_{nk}&=\tfrac{2}{(n+2)!}\tfrac{n+1}{k+3}\tfrac{k!}{(k-n-2)!}\,.
\end{align*}
The full solution reads
\begin{align}\label{solpsi}\boxed{
\begin{aligned}
    D\Psi_{A(k),A'(k+2)}&=e^{CC'}\Psi_{A(k)C,A'(k+2)C'}\\
        &-e\fud{C}{A'}\sum_{n=0}^{k-1}\tfrac{2}{(n+1)!}\tfrac{k-n+2}{k+3}\tfrac{k!}{(k-n-1)!}[F_{A(n+1)C,A'(n)},\Psi_{A(k-n-1),A'(k-n+1)}]\\
        &+e\fud{C}{A'}\sum_{n=0}^{k-2}\tfrac{2}{(n+2)!}\tfrac{n+1}{k+3}\tfrac{k!}{(k-n-2)!}[F_{A(n+2),A'(n)},\Psi_{A(k-n-2)C,A'(k-n+1)}]\,.
\end{aligned}}
\end{align}
In Appendix \ref{app:sdymTruncation} we show that this solution ensures consistency of the $L_\infty$-relation in \eqref{SDYMstafhessPsi3}, i.e. the above solution does not require higher order corrections.

\paragraph{Summary.}
SDYM can be cast in the form of an $L_\infty$-algebra. This gives rise to three $L_\infty$-relations for the $F$-sector and the $\Psi$-sector of SDYM, see \eqref{SDYMstasheff}. The first of each gives rise to the free equation for $DF_{A(k+2),A'(k)}$ and $D\Psi_{A(k),A'(k+2)}$. The second $L_\infty$-relation can be solved to obtain the quadratic piece of the non-linear extension in both sectors, which are proportional to $[F,F]$ and $[F,\Psi]$, respectively. In particular, the coefficients can be found by writing down the minimal ans{\"a}tze \eqref{spin1ansatz} and \eqref{ansatzpsi} and checking their Bianchi identities. This yields two expressions for $e^{BB'}F_{A(k+2)B,A'(k)B'}$ and $e^{BB'}\Psi_{A(k)B,A'(k+2)B'}$. Comparing them gives rise to a system of recurrence relations, whose solution gives the final results \eqref{spin1sol} and \eqref{solpsi}, i.e. the boxed equations above. Furthermore, the third $L_\infty$-relation ensures that the system is closed, i.e. there are no higher order corrections. It is proved that these relation are indeed satisfied for the obtained solutions and hence the expressions we have found are the complete non-linear extensions for the two sectors.

An interesting follow up would be to consider the higher spin extensions of SDYM \cite{Ponomarev:2017nrr,Krasnov:2021nsq} and the supersymmetric higher spin extensions constructed in \cite{Devchand:1996gv}.

\subsection{FDA, constant curvature space}
\label{sec:SDYMcurved}
As a simple modification of SDYM on Minkowski background we can consider a constant curvature background, i.e. de Sitter or anti-de Sitter spaces. The action is the same. Let us first recall that the free Maxwell equations on a constant curvature background rewritten as an FDA read \cite{Vasiliev:1986td}
\besubeqs \label{Maxw}
\begin{align}\label{eqMaxwAA}
    dA&=H^{BB}F_{BB} +\epsilon H^{B'B'} \Psi_{B'B'} \,, \\
\label{eqMaxwBA}
    \nabla F^{A(k+2),A'(k)}&=e_{BB'}F^{A(k+2)B,A'(k)B'} +k(k+2)\Lambda e^{AA'} F^{A(k+1), A'(k-1)} \,, \\
\label{eqMaxwCA}
    \nabla \Psi^{A(k),A'(k+2)}&=e_{CC'}\Psi^{A(k)C,A'(k+2)C'}+k(k+2)\Lambda e^{AA'} \Psi^{A(k-1),A'(k+1)} \,.
\end{align}
\esubeqs
The only difference is the presence of new $e^{AA'}$-terms that are consistent on their own and do not require any other modifications. It is also convenient to set $\Lambda=1$ in what follows. The $L_\infty$-algebra for SDYM on a constant background is given by
\begin{align*}
    d e^{AA'}&= \omega\fud{A}{B}\wedge e^{BA'}+\omega\fud{A'}{B'}\wedge e^{A B'} \,, \\
    d\omega^{AB}&= \omega\fud{A}{C}\wedge \omega^{BC}+  H^{AB} \,, \\
    d\omega^{A'B'}&= \omega\fud{A'}{C'}\wedge \omega^{B'C'}+  H^{A'B'} \,, \\
    dA&=  AA + H_{BB}F^{BB} \,, \\
    dF&= l_2(\omega,F) + l_2(A,F)+l_2(e,F)+\tilde{l}_2(e,F)+l_3(e,F,F) \,, \\
    d\Psi&= l_2(\omega,\Psi) + l_2(A,\Psi)+l_2(e,\Psi)+\tilde{l}_2(e,\Psi) +l_3(e,F,\Psi)\,,
\end{align*}
where $\tilde{l}_2$ encodes the gravitational correction to the free equations \eqref{eqMaxwBA} and \eqref{eqMaxwCA}. The contributions $$l_2(\omega,F)= (k+2)\omega\fud{A}{B}F^{A(k+1)B,A'(k)}+k\omega\fud{A'}{B'}F^{A(k+2),B'A'(k-1)}$$ and $l_2(A,F)=[A,F]$ (and similarly for $\Psi$) can be absorbed into the covariant derivative $D=\nabla-[A,\bullet]$. As a result, the relations can be rewritten as
\begin{align*}
    \nabla e^{AA'}&= 0 \,, \\ 
    dA&= AA + H_{BB}F^{BB} \,, \\
    DF&= l_2(e,F)+\tilde{l}_2(e,F)+l_3(e,F,F) \,, \\
    D\Psi&= l_2(e,\Psi)+\tilde{l}_2(e,\Psi)+l_3(e,F,\Psi)\,.
\end{align*}
As different from $\nabla^2=0$ in flat space, in a constant curvature background we have for any spin-tensor $T^{A(n),A'(m)}$
\begin{align*}
    \nabla^2T^{A(n),A'(m)}&=-n  H\fud{A}{B}T^{A(n-1)B,A'(m)}-m  H\fud{A'}{B'}T^{A(n),A'(m-1)B'} \,.
\end{align*}
The $L_\infty$-relations of the sought for $L_\infty$-algebra read
\besubeqs
    \begin{align}
    -[H_{BB}F^{BB},F]+l_2(e,DF)+\tilde{l}_2(e,DF)+l_3(e,DF,F)+l_3(e,F,DF)&\equiv0 \,, \\
    -[H_{BB}F^{BB},\Psi]+l_2(e,D\Psi)+\tilde{l}_2(e,D\Psi)+l_3(e,DF,\Psi)+l_3(e,F,D\Psi)&\equiv0 \,.
    \end{align}
\esubeqs
Since $\tilde{l}$ can be viewed as a deformation of the previously found FDA, all terms without $\tilde{l}$ vanish already. The remaining nontrivial relations read
\besubeqs\label{SDYMstasheffLambda}
\begin{align} 
    \tilde{l}_2(e,l_3(e,F,F))+l_3(e,\tilde{l}_2(e,F),F)+l_3(e,F,\tilde{l}_2(e,F))&=0 \,, \label{stasheffLambdaF} \\
    \tilde{l}_2(e,l_3(e,F,\Psi))+l_3(e,\tilde{l}_2(e,F),\Psi)+l_3(e,F,\tilde{l}_2(e,\Psi))&=0 \label{stasheffLambdaPsi} \,,
\end{align}
\esubeqs
where we ignore terms quadratic in the cosmological constant. These relations are satisfied automatically. 
A proof of this given in Appendix \ref{app:sdymLambda}. Consequently, on a constant curvature gravitational background we obtain
\besubeqs
    \begin{align} \label{curvedFsol}\boxed{
        \begin{aligned}
            D& F_{A(k+2),A'(k)}=e^{BB'}F_{A(k+2)B,A'(k)B'}+k(k+2)e_{AA'}F_{A(k+1),A'(k-1)}\\
            &\qquad-e\fud{B}{A'}\sum_{n=0}^{k-1}\tfrac{2}{(n+1)!}\tfrac{(k+2)!}{(k-n-1)!(k-n+1)(k+1)}[F_{A(n+1)B,A'(n)},F_{A(k-n+1),A'(k-n-1)}]\,,
        \end{aligned}}
    \end{align}
    \begin{align} \label{curvedPsisol}\boxed{
        \begin{aligned}
        D\Psi_{A(k),A'(k+2)}&=e^{CC'}\Psi_{A(k)C,A'(k+2)C'}+k(k+2)e_{AA'}\Psi_{A(k-1),A'(k+1)}\\
        &-e\fud{C}{A'}\sum_{n=0}^{k-1}\tfrac{2}{(n+1)!}\tfrac{k-n+2}{k+3}\tfrac{k!}{(k-n-1)!}[F_{A(n+1)C,A'(n)},\Psi_{A(k-n-1),A'(k-n+1)}]\\
        &+e\fud{C}{A'}\sum_{n=0}^{k-2}\tfrac{2}{(n+2)!}\tfrac{n+1}{k+3}\tfrac{k!}{(k-n-2)!}[F_{A(n+2),A'(n)},\Psi_{A(k-n-2)C,A'(k-n+1)}]\,.
        \end{aligned}}
    \end{align}
\esubeqs

\paragraph{Summary.} We constructed the $L_\infty$-algebra of SDYM on a constant curvature background and derived the corresponding $L_\infty$-relations. The free Maxwell equations on a constant curvature background in terms of an FDA, \eqref{Maxw}, are well-known in the literature and solve the first $L_\infty$-relation of both the $F$-sector and $\Psi$-sector. In section \ref{sec:flatSDYM} we computed the non-linear extension of $DF_{A(k+2),A'(k)}$ and $D\Psi_{A(k),A'(k+2)}$ on a flat background. In the second $L_\infty$-relation of each sector we see an interplay between the gravitational contribution of the free equations and the non-linear extension on flat space. We demonstrated that the second $L_\infty$-relation for both sectors decomposes into the flat space $L_\infty$-relation and a new relation containing the gravitational contributions in such a way that the latter does not contribute to the quadratic order in $DF_{A(k+2),A'(k)}$ and $D\Psi_{A(k),A'(k+2)}$. The third $L_\infty$-relation then contains no gravitational contribution and remains satisfied. The complete non-linear extension of both sectors are only modified in the linear terms according to the free equations and are shown in the boxed equation \eqref{curvedFsol} and \eqref{curvedPsisol} above.

\section{SDGR}
\label{sec:SDGR}

\subsection{Action, initial data}

Self-dual gravity with vanishing cosmological constant can be formulated with the help of two fields \cite{Krasnov:2021cva}: one-form $\omega^{A'B'}$ and zero-form $\Psi^{A'B'C'D'}$. The action reads
\begin{align}\label{flsd}
    \int \Psi^{A'B'C'D'}\wedge d\omega_{A'B'} \wedge d\omega_{C'D'} \,.
\end{align}
The equations of motion are ($F^{A'B'}=d\omega^{A'B'}$)
\begin{align}\label{fleq}
    F_{(A'B'} \wedge F_{C'D')}&=0 \,, &  d\Psi^{A'B'C'D'}\wedge F_{A'B'}&=0 \,.
\end{align}
One-form $\omega^{A'B'}$ looks like the anti-self-dual part of the Lorentz spin-connection, but it is not. The curvature $F_{A'B'}$ for $\omega^{A'B'}$ lacks the "$\omega\omega$"-part. Nevertheless, this interpretation is not very far from the reality since action \eqref{flsd} can be understood as a limit of that for self-dual gravity with cosmological constant \cite{Krasnov:2016emc}. In the latter $F^{A'B'}=d\omega^{A'B'}-\omega\fud{A'}{C'}\wedge \omega^{C'B'}$ is the canonical one and the limit is to drop the $\omega\omega$-part. 

Minkowski space is a special solution of \eqref{fleq}: $\omega_0^{A'A'}=x\fdu{C}{A'} dx^{CA'}$ such that $d\omega_0^{A'B'}=H^{A'B'}$, where $H^{A'B'}$ is built from the Minkowski's space vierbein $e^{AA'}=dx^{AA'}$, $H^{A'B'}\equiv e\fdu{C}{A'}\wedge e^{CB'}$ and its conjugate is $H^{AB}\equiv e\fud{A}{C'}\wedge e^{BC'}$. One can easily write down the first few equations of the FDA that corresponds to variational equations \eqref{fleq}:\footnote{As a side remark, let us write the curvature for $so(3,2)\sim sp(4)$, which is relevant for anti-de Sitter space (they correspond to Lorentz generators $L_{A'A'}$, $L_{AA}$ and to translations $P_{AA'}$):
\besubeqs
\begin{align*}
    d \omega^{AA}-\omega\fud{A}{C}\wedge\omega^{CB}- e\fud{A}{B'}\wedge e^{AB'}&=R^{AA}\,,\\
    d e^{AA'}-\omega\fud{A'}{B'}\wedge e^{AB'}-\omega\fud{A}{B}\wedge e^{BA'} &=T^{AA'}\,,\\
    d \omega^{A'A'}-\omega\fud{A'}{C'}\wedge\omega^{C'B'}- e\fdu{B}{A'}\wedge e^{BA'}&=R^{A'A'}\,,
\end{align*}
\esubeqs
The gauge algebra for the SDGR with zero scalar curvature can be understood as a limit of $so(3,2)$-algebra where $L_{A'A'}$ become abelian \cite{Krasnov:2021cva}. 
}
\besubeqs\label{gaugesect}
\begin{align} \label{RandPsi}
    &\qquad\begin{aligned}
     d \omega^{A'A'}&=e\fdu{B}{A'}\wedge e^{BA'}\,,\\
    d e^{AA'}&=\omega\fud{A}{B}\wedge e^{BA'} \,,\\
    d \omega^{AA}&=\omega\fud{A}{C}\wedge\omega^{CA}+H_{MM}C\fud{MMAA} \,, 
    \end{aligned}\\
    &\,\, d\Psi^{A'A'A'A'}= e_{BB'}\Psi^{B,A'A'A'A'B'}\,. \label{firstpsi}
\end{align}
\esubeqs
The main idea is to identify the right gauge algebra \cite{Krasnov:2021cva}. This is the starting for constructing the $L_\infty$-algebra. The first equation of \eqref{RandPsi} implies that the gravitational degrees of freedom fully reside in the anti-self-dual part. The last equation of \eqref{RandPsi} identifies the only nonvanishing part of the curvature with the self-dual Weyl tensor $C^{ABCD}$, $R\fud{A}{B}=H_{MM}C\fud{MMA}{B} $. As a result one obtains a Bianchi identity for $R\fud{A}{B}$. Eq. \eqref{firstpsi} introduces a new field $\Psi^{A,A'B'C'D'E'}$, which parameterizes the first derivative of $\Psi$ and is contained in the on-shell jet of $\Psi^{A'B'C'D'}$. Similarly to SDYM we aim to find a completion of \eqref{RandPsi} and we need to define an infinite set of coordinates on $\mathcal{N}$ and $Q$ such that $QQ=0$. 

\paragraph{Coordinates, on-shell jet.} Coordinates on supermanifold $\mathcal{N}$ coincide with those of the free massless spin-two field, i.e. with \cite{Vasiliev:1986td} and \cite{penroserindler}. Indeed, the set of one-forms turned out to be the same, while the zero-forms begin with (anti)-self-dual components of Weyl tensor and are just the on-shell nontrivial derivatives of those. Therefore, the coordinates on $\mathcal{N}$ are: degree-one $\omega^{AB}$, $e^{AA'}$ and $\omega^{A'B'}$; degree-zero $C^{A(k+4),A'(k)}$ and $\Psi^{A(k+4),A'(k)}$, $k=0,1,2,...$. A similar discussion follows as for SDYM. In particular, the free equations for helicity $\pm 2$ fields are \cite{Penrose:1965am}
\begin{align*}
    \nabla\fud{A}{B'}\Psi^{A'B'C'D'}&=0 \,, & \nabla\fdu{A}{B'}C^{ABCD}&=0\,,
\end{align*}
and can be rewritten in the FDA form as \cite{Vasiliev:1986td}
\begin{align}\label{freespintwo}
    \nabla C^{A(k+4),A'(k)}&=e_{CC'}C^{A(k+4)C,A'(k)C'} \,, & \nabla \Psi^{A(k),A'(k+4)}&=e_{CC'}\Psi^{A(k)C,A'(k+4)C'} \,.
\end{align}
One needs to supplement these equations with the free limit of \eqref{gaugesect}. Our problem is to find a nonlinear completion of \eqref{freespintwo} that is consistent with \eqref{gaugesect}.

\paragraph{General form.} 
The supermanifold $\mathcal{N}$ has coordinates
\begin{align*}
    \mathcal{N}&: && 
    \begin{aligned}
        1&: \omega^{A'B'}\,,e^{AA'}\,, \omega^{AB} \,,\\\
        0&: C^{A(k+4),A'(k)}\,, \Psi^{A(k),A'(k+4)}\,, k=0,1,2,...
    \end{aligned}
\end{align*}
Now, we try to reformulate the theory in the $L_\infty$-form. Given the data above and our desire to truncate the FDA at $l_3(\bullet,\bullet,\bullet)$, we write 
\begin{align*}
     d \omega^{A'A'}&=e\fdu{B}{A'}\wedge e^{BA'}\,,\\
    d e^{AA'}&=\omega\fud{A}{B}\wedge e^{BA'} \,,\\
    d \omega^{AA}&=\omega\fud{A}{C}\wedge\omega^{CA}+H_{MM}C\fud{MMAA} \,, \\
    dC&= l_2(\omega,C)+l_2(e,C)+l_3(e,C,C) \,, \\
    d\Psi&= l_2(\omega,\Psi)+l_2(e,\Psi) +l_3(e,C,\Psi)\,.
\end{align*}
We define the covariant derivative $\nabla=d-\omega$, which lacks the $\omega^{A'B'}$-part. For an arbitrary spin-tensor $T^{A(n),A'(m)}$ we get
\begin{align} \label{nablasquared}
    \nabla^2T^{A(n),A'(m)}=-nH_{MM}C\fud{MMA}{B}T^{BA(n-1),A'(m)} \,.
\end{align}
The covariant derivative allows one to absorb the terms $l_2(\omega,C)$ and $l_2(\omega,\Psi)$ and we can write
    \begin{align*}
        \nabla C&=l_2(e,C)+l_3(e,C,C) \,, &
        \nabla \Psi&=l_2(e,\Psi)+l_3(e,C,\Psi)\,.
    \end{align*}
This gives rise to the $L_\infty$-relations for SDGR, which read
\begin{align*}
    \begin{aligned}
        -(k+4) H_{MM}C\fud{MMA}{B}C^{A(k+3)B,A'(k)}+l_2(e,\nabla C)+l_3(e,\nabla C,C)+l_3(e,C,\nabla C)&=0 \,, \\
        -k H_{MM}C\fud{MMA}{B}\Psi^{A(k-1)B,A'(k+4)}+l_2(e,\nabla\Psi)+l_3(e,\nabla C,\Psi)+l_3(e,C,\nabla\Psi)&=0\,,
    \end{aligned}
\end{align*}
and decompose into
\besubeqs \label{SDGRstasheff}
    \begin{align} 
        &l_2(e,l_2(e,C))=0 \label{stasheffC1} \,, \\
           &l_3(e,l_3(e,C,C),C)+l_3(e,C,l_3(e,C,C))=0 \label{stasheffC3} \,, \\
       & l_2(e,l_2(e,\Psi))=0 \label{stasheffSDGRPsi1} \,, \\     
       & l_3(e,l_3(e,C,C),\Psi)+l_3(e,C,l_3(e,C,\Psi))=0 \label{stasheffSDGRPsi3} \,,\\
        &\begin{aligned}
            -(k+4) H_{MM}C\fud{MMA}{B}C^{A(k+3)B,A'(k)}&+l_2(e,l_3(e,C,C))       \\
            &+l_3(e,l_2(e,C),C)+l_3(e,C,l_2(e,C))=0  \,, \label{stasheffC2}
        \end{aligned}\\
        &\begin{aligned}
           -kH_{MM}C\fud{MMA}{B}\Psi^{A(k-1)B,A'(k+4)}&+l_2(e,l_3(e,C,\Psi))\\&+l_3(e,l_2(e,C),\Psi)+l_3(e,C,l_2(e,\Psi))=0\,. \label{stasheffSDGRPsi2} 
        \end{aligned}
    \end{align}
\esubeqs

\subsection{FDA}
\label{sec:SDGRflat}
\paragraph{Appetizer.}
Let us first illustrate our approach by presenting the source of the non-linear extension with an explicit example. We follow roughly the same steps as for SDYM, though some subtle differences arise. The most important ones come from the commutativity of the $C$'s and the additional contraction of unprimed indices that we will see shortly.

The Bianchi identity for the curvature, $\nabla R_{AA}=0$ implies
\begin{align*}
    \nabla C_{AAAA}=e^{BB'}C_{AAAB,B'} \,.
\end{align*}
Its own Bianchi identity via \eqref{nablasquared} imposes
\begin{align*}
    \nabla^2C^{AAAA}=-e^{BB'}\wedge\nabla C_{AAAAB,B'}=4H^{BB}C\fdu{ABB}{D}C_{AAAD} \,.
\end{align*}
We need to construct an ansatz for $\nabla C_{AAAAA,A'}$. Commutativity of the $C$'s and the Fierz identity allow us to construct the minimal ansatz as
\begin{align} \label{eq:nablaC}
    \nabla C_{AAAAA,A'}=e^{CC'}C_{AAAAAB,A'C'}+a_{01}e\fud{C}{A'}C\fdu{AAC}{D}C_{AAAD}\,.
\end{align}
Contracting the ansatz with $e^{BB'}$ yields
\begin{align*}
    \begin{aligned}
        e^{BB'}\wedge\nabla C_{AAAAB,B'}&=-\tfrac{2a_{01}}{5}H^{BB}C\fdu{ABB}{D}C_{AAAD}-\tfrac{3a_{01}}{5}H^{BB}C\fdu{AAB}{D}C_{AABD}\\
        &=-\tfrac{2a_{01}}{5}H^{BB}C\fdu{ABB}{D}C_{AAAD}\,.
    \end{aligned}
\end{align*}
One term is dropped, as commuting the two $C$'s and raising/lowering the contracted indices tells us that this term vanishes. Comparing the result with \eqref{eq:nablaC} yields the solution
\begin{align*}
    \nabla C_{AAAAA,A'}=e^{CC'}C_{AAAAAC,A'C'}+10e\fud{C}{A'}C\fdu{AAC}{D}C_{AAAD}\,.
\end{align*}
The procedure that we have followed is a practical realisation of solving the $L_\infty$-relation \eqref{stasheffC2}. This procedure will be generalized next.

\paragraph{Main course, $\boldsymbol{C}$-sector.} Using the same criteria as before we propose the minimal ansatz
\begin{align*}
    \nabla C_{A(k+4),A'(k)}=e^{CC'}C_{A(k+4)C,A'(k)C'}+\sum_{n=0}^{k-1}a_{nk}e\fud{C}{A'}C\fdud{A(n+2)C}{D}{,A'(n)}C_{A(k-n+2)D,A'(k-n-1)}\,.
\end{align*}
Taking another derivative leads to
\begin{align} \label{nablasquaredC}
    \begin{aligned}
        \nabla^2C_{A(k+4),A'(k)}&=(k+4)H^{BB'}C\fdu{ABB}{D}C_{A(k+3)D,A'(k)}=-e^{CC'}\wedge\nabla C_{A(k+4)C,A'(k)C'}\\
        &-\sum_{n=1}^{k-1}a_{nk}e\fud{C}{A'}\wedge\nabla C\fdud{A(n+2)C}{D}{,A'(n)}C_{A(k-n+2)D,A'(k-n-1)}\\
        &-\sum_{n=0}^{k-2}a_{nk}e\fud{C}{A'}\wedge C\fdud{A(n+2)C}{D}{,A'(n)}\nabla C_{A(k-n+2)D,A'(k-n-1)}\,.
    \end{aligned}
\end{align}
Considering only terms quadratic in $C$ yields
\begin{align} \label{eC}
    \begin{aligned}
        e^{CC'}\wedge\nabla C_{A(k+4)C,A'(k)C'}&=-(k+4)H^{BB}C\fdu{ABB}{D}C_{A(k+3)D,A'(k)}\\
        &-\tfrac{1}{2}H^{BB}\sum_{n=0}^{k-1}a_{nk}C\fdud{A(n+2)BB}{D}{,A'(n+1)}C_{A(k-n+2)D,A'(k-n-1)}\\
        &-\tfrac{1}{2}H^{BB}\sum_{n=0}^{k}(\tfrac{a_{nk}}{2}-\tfrac{a_{(k-n)k}}{2})C\fdud{A(n+2)B}{D}{,A'(n)}C_{A(k-n+2)BD,A'(k-n)}\\
        &-\tfrac{1}{2}H\fdu{A'}{B'}\sum_{n=0}^{k-1}a_{nk}C\fdud{A(n+2)B}{D}{,A'(n)}C\fdud{A(k-n+2)}{B}{D,A'(k-n-1)B'}\,,
    \end{aligned}
\end{align}
where in the third line we made the anti-commuting property of the $C$'s explicit, together with the anti-symmetry of the spinorial inner product. At the same time we contract $e^{BB'}$ with $\nabla C_{A(k+5),A'(k+1)}$ to obtain
\begin{align*}
    \begin{aligned}
    e^{BB'}&\wedge\nabla C_{A(k+4)B,A'(k)B'}=-H^{BB}\tfrac{k+2}{(k+5)(k+1)}a_{0(k+1)}C\fdu{ABB}{D}C_{A(k+3)D,A'(k)}\\
    &-\tfrac{1}{2}H^{BB}\sum_{n=0}^{k-1}\tfrac{(k+2)(n+3)}{(k+5)(k+1)}a_{(n+1)(k+1)}C\fdud{A(n+2)BB}{D}{,A'(n+1)}C_{A(k-n+2)D,A'(k-n-1)}\\
    &-\tfrac{1}{4}H^{BB}\sum_{n}^{k}(\tfrac{(k+2)(k-n+3)}{(k+5)(k+1)}a_{n(k+1)}-\tfrac{(k+2)(n+3)}{(k+5)(k+1)}a_{(k-n)(k+1)})\\
    &\times C\fdud{A(n+2)B}{D}{,A'(n)}C_{A(k-n+2)BD,A'(k-n)}\\
    &-\tfrac{1}{2}H\fdu{A'}{B'}\sum_{n=0}^{k-1}(\tfrac{(k-n)(n+3)}{(k+5)(k+1)}a_{(k-n)(k+1)}+\tfrac{(k-n)(k-n+3)}{(k+5)(k+1)}a_{n(k+1)})\\
    &\times C\fdud{A(n+2)B}{D}{,A'(n)}C\fdud{A(k-n+2)}{B}{D,A'(k-n-1)B'} \,.
    \end{aligned}
\end{align*}
Comparing this expression with \eqref{eC} brings about the following system of recurrence relations:
    \begin{align*}
        0&=a_{0k}-\tfrac{(k+4)(k+3)k}{k+1} \,, \\
        0&=a_{(n+1)(k+1)}-\tfrac{(k+5)(k+1)}{(k+2)(n+3)}a_{nk} \,, \\
        0&=\tfrac{(k+2)(k-n+3)}{(k+5)(k+1)}a_{n(k+1)}-\tfrac{(k+2)(n+3)}{(k+5)(k+1)}a_{(k-n)(k+1)}-a_{nk}+a_{(k-n)k} \,, \\
        0&=a_{nk}-\tfrac{(k-n)(n+3)}{(k+5)(k+1)}a_{(k-n)(k+1)}-\tfrac{(k-n)(k-n+3)}{(k+5)(k+1)}a_{n(k+1)} \,.
    \end{align*}
This over-determined system is solved by
\begin{align*}
    a_{nk}=\tfrac{2}{(n+2)!}\tfrac{(k+4)!(k-n)}{(k-n+2)!(k+1)}
\end{align*}
and the full solution reads\footnote{A closely related problem was addressed in \cite{Vasiliev:1989xz}, which is to find an FDA form of the full gravity to the next to the leading order (the problem to find the complete minimal model for gravity does not seem to admit a solution in a closed form, even though it does always exist as a matter of principle). It would be interesting to understand what \cite{Vasiliev:1989xz} describes since it does not coincide with the FDA of SDGR with (non)-vanishing cosmological constant. The physical degrees of freedom are the same though.}
\begin{align} \label{Csol} \boxed{
    \begin{aligned}
        \nabla C_{A(k+4),A'(k)}&=e^{CC'}C_{A(k+4)C,A'(k)C'}\\
        &+\sum_{n=0}^{k-1}\tfrac{2}{(n+2)!}\tfrac{(k+4)!(k-n)}{(k-n+2)!(k+1)}e\fud{C}{A'}C\fdud{A(n+2)C}{D}{,A'(n)}C_{A(k-n+2)D,A'(k-n-1)}\,.
    \end{aligned}}
\end{align}
In appendix \ref{app:sdgrTruncation} we prove that this solution is complete, i.e. no higher order terms arise.

\paragraph{Main course, $\boldsymbol{\Psi}$-sector.} For the $\Psi$-sector we follow a similar approach. The minimal ansatz reads
\begin{align} \label{eq:SDGRPsiansatz}
    \begin{aligned}
        \nabla\Psi_{A(k),A'(k+4)}&=e^{CC'}\Psi_{A(k)C,A'(k+4)C'}+\sum_{n=0}^{k}b_{nk}e\fud{C}{A'}C\fdud{A(n+2)C}{D}{,A'(n)}\Psi_{A(k-n-2)D,A'(k-n+3)}\\
        &+\sum_{n=0}^{k}c_{nk}e\fud{C}{A'}C\fdud{A(n+3)}{D}{,A'(n)}\Psi_{A(k-n-3)CD,A'(k-n+3)}\,.
    \end{aligned}
\end{align}
The details of the calculations are left to Appendix \ref{app:SDGRPsi}, but the approach is as follows: we take the covariant derivative of the ansatz above. We also contract $e^{BB'}$ with $\nabla\Psi_{A(k+1),A'(k+5)}$. Both will give us an expression for $e^{BB'}\wedge\nabla\Psi_{A(k)B,A'(k+4)B'}$ and we compare them. This results in a system of recurrence relations, which is solved by
    \begin{align*}
        b_{nk}&=\tfrac{2}{(n+2)!}\tfrac{k!}{(k-n-2)!}\tfrac{k-n+4}{k+5} \,, &
        c_{nk}&=-\tfrac{2}{(n+2)!}\tfrac{k!}{(k-n-3)!}\tfrac{n+1}{(k+5)(n+3)}\,,
    \end{align*}
and the solution in the $\Psi$-sector reads
\begin{align} \label{psisolSDGR} \boxed{
    \begin{aligned}
        \nabla\Psi_{A(k),A'(k+4)}&=e^{CC'}\Psi_{A(k)C,A'(k+4)C'}\\
        &+\sum_{n=0}^{k}\tfrac{2}{(n+2)!}\tfrac{k!}{(k-n-2)!}\tfrac{k-n+4}{k+5}e\fud{C}{A'}C\fdud{A(n+2)C}{D}{,A'(n)}\Psi_{A(k-n-2)D,A'(k-n+3)}\\
        &-\sum_{n=0}^{k}\tfrac{2}{(n+2)!}\tfrac{k!}{(k-n-3)!}\tfrac{n+1}{(k+5)(n+3)}e\fud{C}{A'}C\fdud{A(n+3)}{D}{,A'(n)}\Psi_{A(k-n-3)CD,A'(k-n+3)}\,.
    \end{aligned}}
\end{align}
We prove in Appendix \ref{app:sdgrTruncation} that $\nabla\Psi_{A(k),A'(k+4)}$ is consistent as it is and does not require higher order terms. 

\paragraph{Summary.}
We rewrote SDGR as an $L_\infty$-algebra. This gives rise to three $L_\infty$-relations for the $C$-sector and the $\Psi$-sector, see \eqref{SDGRstasheff}. Solving the first relation of each sector yields the free equations for $\nabla C^{A(k+4),A'(k)}$ and $\nabla \Psi^{A(k),A'(k+4)}$. We constructed a minimal ansatz for a non-linear extension of the free equations and used the second $L_\infty$-relation to determine its structure. The results are shown in the boxes expressions above, \eqref{Csol} and \eqref{psisolSDGR}. The third $L_\infty$-relation is found to be satisfied for the obtained solutions, which implies that the minimal ansatz is sufficient to solve the whole system.

An interesting followup of this project is construct FDA for SDGR in the constant-curvature background. The action of this theory \cite{Krasnov:2016emc} is even more natural 
\begin{align*}
    \int \Psi^{A'B'C'D'}\wedge F_{A'B'} \wedge F_{C'D'}\,,
\end{align*}
where $F^{A'B'}=d\omega^{A'B'}-\omega\fud{A'}{C'}\wedge \omega^{C'B'}$. However, it is more nonlinear, featuring quartic terms (the quintic one vanishes). A simpler problem is to consider the higher spin extensions of SDGR \cite{Ponomarev:2017nrr,Krasnov:2021nsq} with vanishing cosmological constant.

\section{Conclusions and Discussion}
Since every (gauge) field theory defines and is defined by its minimal model, a certain $L_\infty$-algebra, our general motivation is to first understand how various properties of field theories, e.g. integrability, asymptotic symmetries, conserved charges, actions, anomalies etc., can be understood in the known cases and derived from this $L_\infty$-algebra in the cases where this information is yet unavailable. For example, it would be interesting to understand the Ward construction of Yang-Mills instantons \cite{Ward:1977ta} from the $L_\infty$ point of view.

As we have reviewed in section \ref{sec:FDA1}, the minimal model can naturally be associated to any gauge theory and it is the smallest $L_\infty$-algebra that captures all local BRST cohomology of this field theory. However, the minimal model is usually difficult to construct explicitly. Apart from this paper, the only available examples where minimal models were explicitly constructed are (a) Chern-Simons theory, which is just $dA=AA$ and, for that reason, is hard to consider this as a genuine example of a minimal model (nevertheless, this toy model was quite useful to prove that all matter-free higher spin gravities in $3d$ are of Chern-Simons form \cite{Grigoriev:2020lzu}); (b) another example is discussed in \cite{Vasiliev:1989xz} and is closely related to the SDGR FDA of the present paper. It is tempting to argue that minimal models can explicitly be constructed only for theories that feature some kind of (hidden) simplicity, e.g. they are integrable like Chern-Simons theory, SDYM and SDGR.

All physically relevant local information about a given theory is encoded in its minimal model via the $Q$-cohomology. For example, conserved  charges correspond to elements of $H(Q)$ of spacetime form degree $3$ and field space degree $0$ with values in the trivial module, while the action functional corresponds to elements of $H(Q)$ of spacetime form degree $4$ and field space degree $0$ with values in the trivial module. A more complicated example is the presymplectic structure $\Omega_{\aA \aB}\, \delta\Phi^\aA \wedge \delta \Phi^\aB$ which is a two-form on the field space and is a degree $(d-1)$ form from the spacetime point of view. It corresponds to $H(Q,\Lambda^2( \mathcal{N}))$, where the action of $Q$ is understood as Lie derivative $L_Q$ along $Q$ that is defined canonically on $(p,q)$-tensors on $\mathcal{N}$, see \cite{Alkalaev:2013hta,Grigoriev:2016wmk,Sharapov:2020quq,Sharapov:2021drr} for more detail and examples. As the last example, $Q$-cohomology with values in vector fields, $H(Q,T^{1,0}(\mathcal{N}))$, corresponds to spactime $0$-forms and field space $1$-forms. It is responsible for deformations of $Q$ itself, i.e. it classifies possible interactions. It is worth noting that $Q$-cohomology can often be computed without having to know the minimal model explicitly. The latter is an additional bonus that should be a signal of integrability.

\chapter{Cubic interactions in chiral HiSGRA}
\label{chap:cubic}

In this chapter, we obtain the cubic interaction vertex for chiral HiSGRA. The content is entirely based on \cite{SDFDA2}, co-authored with Evgeny Skvortsov, and published in the \textit{Physical Review D}.

Note that, due to changes introduced in later works, the primed and unprimed spinor indices, and equivalently $y$ and $\bar{y}$, are swapped in this chapter and the previous compared to the introduction and the other chapters.

\section{Introduction}

Higher Spin Gravities (HiSGRA) are defined to be the smallest possible extensions of gravity with massless fields of arbitrary spin. While there are good reasons to expect higher spin states to play an important role in general, e.g. string theory, the masslessness should imitate the high energy behavior and, for that reason, HiSGRA can be interesting probes of the quantum gravity problems since some of the issues can become visible already at the classical level. Indeed, it is quite challenging to construct HiSGRA due to massless higher spin fields facing numerous issues. As a result, all concrete HiSGRA's available at the moment are quite peculiar: topological models in $3d$ with (partially)-massless and conformal fields \cite{Blencowe:1988gj,Bergshoeff:1989ns,Campoleoni:2010zq,Henneaux:2010xg,Pope:1989vj,Fradkin:1989xt,Grigoriev:2019xmp}; $4d$ conformal HiSGRA  \cite{Segal:2002gd,Tseytlin:2002gz,Bekaert:2010ky} that is a higher spin extension of Weyl gravity; Chiral HiSGRA \cite{Metsaev:1991mt,Metsaev:1991nb,Ponomarev:2016lrm,Skvortsov:2018jea,Skvortsov:2020wtf}  and its truncations \cite{Ponomarev:2017nrr,Krasnov:2021nsq}.\footnote{There are also other interesting recent ideas, e.g. \cite{deMelloKoch:2018ivk,Aharony:2020omh} and \cite{Sperling:2017dts,Tran:2021ukl,Steinacker:2022jjv}.} In this paper we covariantize the interactions of Chiral HiSGRA. 

Chiral HiSGRA is easy to describe due to its simplicity --- interactions stop at the cubic level in the action. It is built from the standard cubic interactions, even though the formulation available before this thesis is in the light-cone gauge. It is advantageous that the light-cone gauge and the spinor-helicity formalism are closely related \cite{Chalmers:1998jb,Chakrabarti:2005ny,Chakrabarti:2006mb,Bengtsson:2016jfk,Ponomarev:2016cwi}. As is well-known \cite{Bengtsson:1986kh,Benincasa:2011pg}, the Lorentz invariance fixes cubic amplitudes  $V_{\lambda_1,\lambda_2,\lambda_3}$ and for any triplet of helicities $\lambda_1+\lambda_2+\lambda_3>0$ there is the unique vertex and the corresponding amplitude:
\begin{align}\label{genericV1}
   V_{\lambda_1,\lambda_2,\lambda_3}\Big|_{\text{on-shell}} \sim 
    [12]^{\lambda_1+\lambda_2-\lambda_3}[23]^{\lambda_2+\lambda_3-\lambda_1}[13]^{\lambda_1+\lambda_3-\lambda_2}\,.
\end{align}
Chiral Theory can be defined as a unique combination of vertices \cite{Metsaev:1991mt,Metsaev:1991nb,Ponomarev:2016lrm} that (a) contains at least one nontrivial self-interaction of a higher spin state with itself; (b) leads to a Lorentz-invariant theory; (c) does not require higher order contact vertices. These assumptions imply that the spectrum of the theory has to contain massless fields of all spins $s=0,1,2,...$, i.e. helicities $\lambda \in(-\infty, +\infty)$ and all coupling constants are uniquely fixed to be
\begin{align}\label{eq:magicalcoupling1}
    V_{\text{Chiral}}&= \sum_{\lambda_1,\lambda_2,\lambda_3}  C_{\lambda_1,\lambda_2,\lambda_3}V_{\lambda_1,\lambda_2,\lambda_3}\,, && C_{\lambda_1,\lambda_2,\lambda_3}=\frac{\kappa\,(l_p)^{\lambda_1+\lambda_2+\lambda_3-1}}{\Gamma(\lambda_1+\lambda_2+\lambda_3)}\,.
\end{align}
Here, $l_p$ is a constant of dimension length, e.g. Planck length, and $\kappa$ is an arbitrary dimensionless constant. In principle, there exists the $\phi^3$-vertex, i.e. $\lambda_i=0$, but it is not present in Chiral Theory. We also see that the $\Gamma$-function restricts the range of summation to  $\lambda_1+\lambda_2+\lambda_3>0$. All such vertices are present. For example, one has the half of the usual $+2,+2,-2$ Einstein-Hilbert vertex and, provided the Yang-Mills groups are turned on, the Yang-Mills interaction $+1,+1,-1$. Importantly, the higher derivative corrections are also needed, e.g. the half of the Goroff-Sagnotti counterterm \cite{Goroff:1985th}, which is $+2,+2,+2$. Such higher derivative terms originate from string theory as well, e.g. \cite{Metsaev:1986yb}. 

It was shown that the tree-level amplitudes vanish on-shell \cite{Skvortsov:2018jea,Skvortsov:2020wtf}. At one-loop there are no UV divergences and the one-loop amplitudes are proportional to the all helicity plus amplitudes of QCD or self-dual Yang-Mills (SDYM) at one loop \cite{Skvortsov:2020gpn}. They also have a higher spin kinematical factor and a factor of the total number of degree of freedom $\sum_\lambda 1$. The latter is infinite and, as in any QFT with infinitely many fields, see e.g. \cite{Fradkin:1982kf}, has to be given a prescription for. A great deal of vacuum one-loop results \cite{Gopakumar:2011qs,Tseytlin:2013jya,Giombi:2013fka,Giombi:2014yra,Beccaria:2014jxa,Beccaria:2014xda,Beccaria:2015vaa,Gunaydin:2016amv,Bae:2016rgm,Skvortsov:2017ldz} suggest that this has to be regularized to zero.   

The power of the light-cone gauge is in that it excludes unphysical degrees of freedom and evades ambiguities of covariant (gauge) descriptions. However, many interesting questions, e.g. nontrivial backgrounds, exact solutions, higher order quantum corrections, are easier to tackle within a covariant description. Until recently a subtlety has been that Chiral Theory requires all vertices \eqref{genericV1}, some of which cannot be written locally within the most common covariant approach to higher spin fields \cite{Conde:2016izb}, where a massless spin-$s$ field is represented by a symmetric rank-$s$ tensor $\Phi_{\mu_1...\mu_s}$. This puzzle has been resolved in \cite{Krasnov:2021nsq}, where it was shown that the most basic problematic interactions of higher spin fields --- Yang-Mills and gravitational --- can easily be constructed by employing the covariant field variables discovered first in Twistor Theory \cite{Atiyah:1979iu,Hitchin:1980hp, Eastwood:1981jy,Woodhouse:1985id}. This should not be surprising since Chiral Theory was shown to admit a formulation similar to self-dual Yang-Mills and self-dual gravity \cite{Ponomarev:2017nrr} and Twistor techniques are most natural for self-dual theories.  

In the present paper we extend these results to Chiral Theory and construct its minimal model or,  equivalently, its classical equations of motion as a Free Differential Algebra \cite{Sullivan77} to next-to-leading-order (NLO). In other words, Chiral Theory can be written as a sigma-model $ d \Phi= Q(\Phi) $, where $\Phi$ are maps from $\Pi T\mathcal{M}$ (the algebra of differential forms on a manifold $\mathcal{M}$) to another supermanifold $\mathcal{N}$ equipped with a homological vector field $Q$, $QQ=0$. All essential information about a given theory, e.g. action, anomalies, etc., is encoded in its minimal model as the $Q$-cohomology \cite{Barnich:2009jy,Grigoriev:2019ojp}. Therefore, the results of the paper can be used to investigate the quantum properties of Chiral Theory, as well as to construct an action and look for classical solutions.  

The paper is organized as follows. After a brief introduction in section \ref{sec:mm} into Free Differential Algebras (FDA) and minimal models, we give in section \ref{sec:initial1} a concise overview of \cite{Krasnov:2021nsq} where covariant actions for the higher spin extensions of self-dual Yang-Mills and self-dual Gravity were constructed. These results give important hints on how to extend them to Chiral Theory, which these two classes are contractions of \cite{Ponomarev:2017nrr}. To find the right gauge algebra (higher spin algebra) is the first step and it was done in \cite{Krasnov:2021cva}. We then proceed in section \ref{sec:FDA1} to the main part and construct $L_\infty$-structure maps/interaction vertices. We also check that some three-point amplitudes \eqref{eq:magicalcoupling1} are correctly reproduced. The latter means that the FDA incorporates all the physically relevant information at next-to-leading order (NLO). There are still some higher structure maps to be found that are required for the complete covariantization of Chiral Theory.

\section{Minimal Models}
\label{sec:mm}
There is a very useful $L_\infty$-algebra, better say a $Q$-manifold, that can naturally be associated to any (gauge) theory and encodes all relevant information about it, which is called the minimal model. As is explained in \cite{Brandt:1997iu,Brandt:1996mh,Barnich:1994db,Barnich:1994mt,Barnich:2010sw,Grigoriev:2012xg,Grigoriev:2019ojp,Grigoriev:2020lzu}, one begins with the jet space BV-BRST formulation of a given (gauge) theory. This way one gets a huge $L_\infty$-algebra which has been quite useful in the analysis of numerous problems in (quantum) field theories, see e.g. \cite{Barnich:1994db,Barnich:1994mt}. One can then consider various equivalent reductions of this algebra that are quasi-isomorphic to it. An important step is to take a usually much smaller equivalent $L_\infty$-algebra, called its minimal model.  The minimal model is, in some sense, the smallest possible $L_\infty$-algebra associated to a given field theory. Nevertheless, modulo the usual topological issues, it contains the full information about invariants, conserved currents, actions, counterterms, anomalies, etc. of the initial field theory \cite{Barnich:2009jy,Grigoriev:2019ojp}. 

Given a BRST complex that is non-negatively graded, e.g. the minimal model, one can consider an associated sigma model whose fields are coordinates on the above $Q$-manifold \cite{Barnich:2010sw}:  
\begin{align}\label{mevenmore1}
    d \Phi&= Q(\Phi) \,.
\end{align}
Here, $\Phi\equiv \Phi(x,dx)$ are maps $\Pi T\mathcal{M} \rightarrow \mathcal{N}$ from the exterior algebra of differential forms on a spacetime manifold $\mathcal{M}$ to a supermanifold $\mathcal{N}$ that is equipped with a homological vector field $Q$, $QQ=0$. Equations \eqref{mevenmore1} and their natural gauge symmetries are equivalent to the initial field theory,\footnote{In general, the equations describe the parameterized version of the initial gauge field theory \cite{Barnich:2010sw}.} thereby providing its reformulation as a Free Differential Algebra.\footnote{Sullivan introduced Free Differential Algebras in \cite{Sullivan77} together with minimal models in the case of differential graded Lie algebras. FDA were re-introduced into physics \cite{vanNieuwenhuizen:1982zf,DAuria:1980cmy} in the supergravity context and a bit later in the higher spin gravity context in \cite{Vasiliev:1988sa}. }

If $\Phi^\aA$ are coordinates on $\mathcal{N}$, $QQ=0$ is equivalent to \eqref{mevenmore1} being formally consistent (that is $dd=0$ does not lead to any algebraic constraints on the fields), which can be rewritten as
\begin{align}
    Q^2&=0 &&\Longleftrightarrow && Q^\aB \frac{\pl}{\pl \Phi^\aB} Q^\aA=0\,.
\end{align}
The latter condition, when Taylor expanded in $\Phi$, is equivalent to the $L_\infty$-relations \cite{Stasheff,Alexandrov:1995kv} that define an $L_\infty$-algebra. This shows that FDA, $L_\infty$ and $Q$-manifolds are all closely related.  
In many practical applications of minimal models, e.g. gauge field theories including gravity,\footnote{For some of the supergravities forms of higher degree need to be introduced.}  coordinates on the formal graded manifold $\mathcal{N}$ consist of two subsets: degree-one and degree-zero. We denote the coordinates and, then, the corresponding fields $\omega$ and $C$, respectively. From the spacetime point of view $\omega$ becomes a one-form connection of some Lie algebra and zero-form $C$ becomes a matter field taking values in some representation $\rho$. The simplest system one can write
\begin{align}
    d\omega &=\tfrac12[\omega,\omega]\,,& dC&=\rho(\omega)C \,,
\end{align}
consists of the flatness condition for $\omega$ and of the covariant constancy equation on $C$. These two equations will describe a background and the physical degrees of freedom propagating on it. The most general non-linear deformation reads\footnote{It was first proposed in \cite{Vasiliev:1988sa} to look for Higher Spin Gravities in the form of an FDA. However, it is important to constrain the vertices by further conditions: (a) to restrict to a basis of independent interaction vertices (otherwise one and the same interaction can be present in infinitely many equivalent but differently looking forms); (b) to impose some form of locality (otherwise any deformation can be completed with higher orders \cite{Barnich:1993vg}, or, in the light-cone gauge, any function can serve as a Hamiltonian unless we care about locality of the boost generators). All these issues are present \cite{Boulanger:2015ova,Skvortsov:2015lja} in \cite{Vasiliev:1988sa}. Therefore, unless (a) and (b) are taken into account $Q$ just gives the most general ansatz for interactions consistent with symmetries rather than any concrete theory. These issues are under control in the present paper.  }
\begin{equation}
    \begin{array}{rcl}
         d\omega&=&l_2(\omega,\omega)+l_3(\omega,\omega,C)+l_4(\omega,\omega,C,C)+\ldots\,,\\
    dC&=&l_2(\omega,C)+l_3(\omega,C,C)+\ldots\,.
    \end{array}
\end{equation}
This algebraic structure can also be identified as a Lie algebroid. Here the initial data ---  Lie algebra and its module --- are encoded in the bilinear maps $l_2(\omega,\omega)$ and $l_2(\omega,C)$, respectively. The higher spin algebra for Chiral Theory was guessed in \cite{Krasnov:2021cva} based on its truncation to the self-dual gravity sector. The module structure is easy to identify, see below. The problem is to find the higher order vertices. In the paper we determine $l_3(\bullet,\bullet,\bullet)$.

\section{HS-SDYM and HS-SDGR}
\label{sec:initial1}
A good starting point is to extract some useful information from the two contractions of Chiral Theory \cite{Ponomarev:2017nrr,Krasnov:2021nsq}, which can be understood as higher spin extensions of self-dual Yang-Mills (SDYM) and self-dual gravity (SDGR). We begin by reviewing some necessary facts about free fields. Impatient readers familiar with the formalism can skip to section \ref{sec:FDA1}.

\subsection{Free fields}
\begin{figure}[h!]
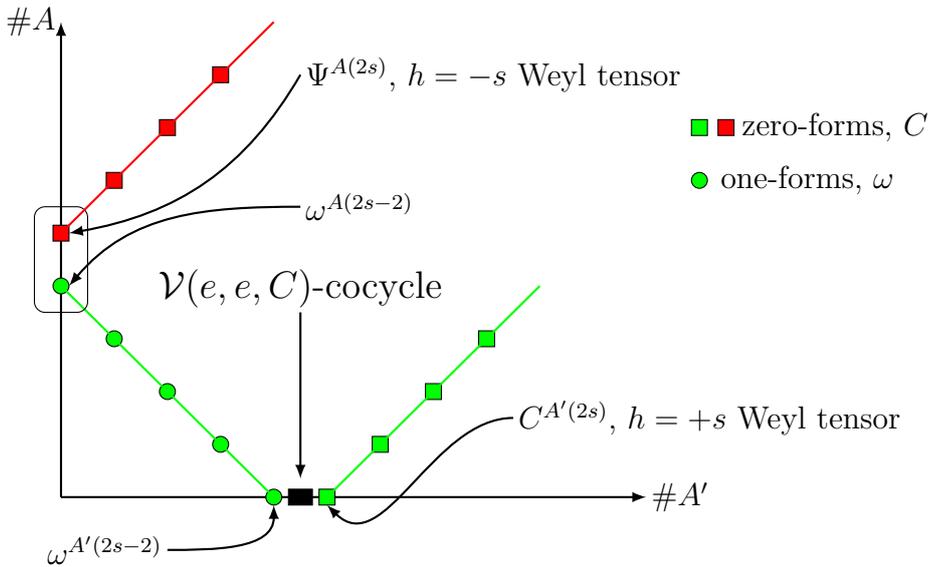

\FIELDS
  \caption{A diagram to show fields/coordinates involved into the description of higher spin fields. Along the axes we have the number of unprimed/primed indices on a spin-tensor. Degree-one coordinates are circles, degree-zero ones are rectangles. The fields that describe a helicity $\lambda=+s$ state are green. Those needed to describe helicity $\lambda=-s$ state are red. The black square shows a cocycle that links the one-form sector to zero-forms (at the free level it relates two fields for each spin's subsystem). The two fields in the rounded rectangle enter the free action. The rest of the fields encode derivatives thereof in a coordinate invariant and background independent way. The solid lines link pairwise the fields that `talk' to each other in the free equations. }\label{fig:figure1}
\end{figure}

Free massless fields of any spin can be described by equations proposed by Penrose \cite{Penrose:1965am}\footnote{We also introduce a compact notation for symmetric indices: all indices in which some tensor is symmetric or to be symmetrized are denoted by the same letter. In addition a group of $k$ symmetric indices $A_1...A_k$ can be abbreviated as $A(k)$. }
\begin{align}\label{hsA1}
    \nabla^\fdu{B}{A'} \Psi^{BA(2s-1)}&=0\,, &\nabla\fud{A}{B'} \Psi^{B'A'(2s-1)}&=0\,.
\end{align}
The equations help to separate helicity eigenstates: one of them describes, say positive, and another the negative helicity states. Twistor theory is very handy in constructing self-dual theories. It requires positive and negative helicity states be described asymmetrically \cite{Hughston:1979tq,Eastwood:1981jy,Woodhouse:1985id}
\begin{align}\label{hsB1}
    \nabla\fdu{A}{A'}\Phi^{A,A'(2s-1)}&=0\,, && \delta \Phi^{A,A'(2s-1)}=\nabla^{A A'}\xi^{A'(2s-2)}\,,
\end{align}
where $\Phi^{A_1...A_{2s-1},A'}$ is a gauge potential. For $s=1$ it coincides with the usual one $A_\mu \sim \Phi^{A,A'}$. For $s=2$ it can be identified with a component of the spin-connection. A bit more geometrically one can \cite{Hitchin:1980hp} introduce a one-form connection 
\begin{align}
\omega^{A'(2s-2)}=\Phi^{B,A'(2s-2)'B'} dx_{BB'}\,.
\end{align}
It can be decomposed into two irreducible spin-tensors
\begin{align}
    \omega^{A'(2s-2)}\equiv e_{BB'}\Phi^{B,A'(2s-2)B'}+e\fdu{B}{A'}\Theta^{B,A'(2s-3)}\,,
\end{align}
where $e^{AA'}\equiv e^{AA'}_\mm \, dx^\mm$ is the vierbein one-form. With the help of gauge transformations
\begin{align}
    \delta \omega^{A'(2s-2)}&= \nabla \xi^{A'(2s-2)} +e\fdu{C}{A'} \eta^{C,A'(2s-3)}\,,
\end{align}
we get \eqref{hsB1} for $\Phi$ and can eliminate $\Theta$. Eqs. \eqref{hsA1} and \eqref{hsB1} follow from a simple action \cite{Hitchin:1980hp,Krasnov:2021nsq}\footnote{This action also can be derived as the presymplectic AKSZ action \cite{Sharapov:2021drr}.} 
\begin{align}
    S= \int \Psi^{A'(2s)}\wedge H_{A'A'}\wedge \nabla \omega_{A'(2s-2)}\,.
\end{align}
Here $H^{A'B'}\equiv e\fdu{C}{A'}\wedge e^{CB'}$. For $s=1$ we have the action of the free SDYM theory. By replacing $\nabla \omega$ with $F=\nabla \omega-\tfrac12[\omega, \omega]$ and promoting $\omega$ and $\Psi$ to a Lie-algebra-valued one-form we get the complete SDYM action \cite{Krasnov:2021nsq}.
\paragraph{Free equations of motion as Free Differential Algebra.} Let us start\footnote{The content of this paragraph has a large overlap with original paper \cite{Vasiliev:1986td}. Apart from the self-dual subtleties the material is standard and can be found, e.g., in \cite{Didenko:2014dwa}.} with the variational equations of motion, which do not have an FDA-form yet:
\begin{align}\label{first}
    \nabla \Psi^{A'(2s)}\wedge H_{A'A'}&=0\,, && H^{A'A'}\wedge \nabla \omega^{A'(2s-2)}=0\,.
\end{align}
Indeed, we need $\nabla \Psi=...$ and $\nabla \omega=...$. The equations are equivalent to
\begin{align}
    \nabla \Psi^{A'(2s)}&= e_{BB'}\Psi^{B,A'(2s)B'}\,,
&
    \nabla \omega^{A'(2s-2)} &= e\fdu{B}{A'} \omega^{B,A'(2s-3)}\,,
\end{align}
where we introduced a zero-form $\Psi^{A,A'(2s+1)}$ and one-form $\omega^{A,A'(2s-3)}$. These fields are known to be relevant for free higher spin fields since \cite{Vasiliev:1986td}.\footnote{Indeed, since \cite{Vasiliev:1986td} introduces fields to parameterize all on-shell nontrivial derivatives of massless fields, any other covariant formulation has to employ at least some of them. Note, however, that the fields of \eqref{first} appeared first thanks to the twistor approach \cite{Penrose:1965am,Hughston:1979tq,Eastwood:1981jy,Woodhouse:1985id}.} 
Of course, we need to know what $\nabla$ of these new fields is, which encourages one to introduce other fields and so on. It is clear that the free equations are easy to write as (note that $\nabla^2=0$)
\besubeqs
\begin{align}
    \nabla\omega^{A(i),A'(n-i)}&= e\fdu{B}{A'} \omega^{A(i)B,A'(n-i-1)}\,, && i=0,...,n-1\,,\\
    \nabla\omega^{A(n)}&= H_{BB} C^{A(n)BB}\,,\\
    \nabla C^{A(n+k+2),A'(k)}&= e_{BB'} C^{BA(n+k+2),B'A'(k)}\,, && k=0,1,2,...\,,\\
    \nabla\Psi^{A(k),A'(n+k+2)}&= e_{BB'} \Psi^{BA(k),A'(n+k+2)B'}\,, && k=0,1,2,...\,,
\end{align}
\esubeqs
where $C$ and $\Psi$ are zero forms and $\omega$ are one-forms. Figure \ref{fig:figure1} illustrates the field content for a fixed spin $s$. 
The only dynamical fields are the ones that enter the action. The rest of the fields are expressed as derivatives of the dynamical ones. It may not seem very useful to introduce infinitely many fields to encode higher derivatives of the dynamical ones, especially when the fields are free, but it can be of great help later: interactions of Chiral Theory may have any number of derivatives (with helicities $\lambda_i$ on external legs fixed, the number of derivatives is $\lambda_1+\lambda_2+\lambda_3>0$, hence, finite).

It is convenient to introduce generating functions:
\begin{align}
    \omega(y,\bry)&= \sum_{n,m}\tfrac{1}{n!m!} \omega_{A(n),A'(m)}\, y^A...y^A\, \bry^{A'}...\bry^{A'}\,,  
\end{align}
\textit{idem}. for $C$, where we pack both $C^{A(k),A'(n+k+2)}$ and $\Psi^{A(n+k+2),A'(k)}$ into a single generating function $C(y,\bry)$. On top of that $C(y,\bry)$ contains $C^{A(k),A'(k)}$, which describe a free massless scalar field. Note that the scalar field is necessarily present in Chiral Theory. We can summarize the free equations as (recall that $\nabla^2=0$)
\begin{align}\label{linearizeddata3}
    \nabla\omega &= e^{BB'}y_{B'} \pl_{B} \omega +H^{BB} \pl_{B}\pl_{B}C(y,\bry=0)\,,& 
    \nabla C&= e^{BB'}\pl_B \pl_{B'} C\,.
\end{align}
These equations form a boundary condition for the non-linear theory.  

\subsection{Initial data for interactions}

It can be useful to have a look at the two contractions of Chiral Theory \cite{Ponomarev:2017nrr,Krasnov:2021nsq} in order to understand how interactions can be introduced. Both HS-SDYM and HS-SDGR \cite{Krasnov:2021nsq} operate with holomorphic fields $\omega^{A'(2s-2)}$ and $\Psi^{A'(2s)}$. It is still useful to package them into generating functions $\omega(\bry)$ and $\Psi(\bry)$.

\paragraph{HS-SDYM.} In order to construct Yang-Mills type interactions of higher spin fields, we promote $\omega$ and $\Psi$ to Lie-algebra-valued fields, e.g. $\omega^{A'(k)}\equiv \omega^{A'(k);a}\, T_a$. It is convenient to realize $T_a$ as matrices $\mathrm{Mat}_N$ for some $N$, e.g. $\omega(\bry)\equiv \omega(\bry)\fud{i}{j}$. We will omit the matrix indices and the only trace they leave is that we cannot swap various $\omega$ and $\Psi$ factors, the order is important, e.g. $\Psi\wedge \omega\neq \omega\wedge\Psi$. The action of HS-SDYM can be written as
\begin{align}\label{HSSDDYM}
    S&=\sum_{s=1}\tfrac{1}{(2s)!}\tr\int \Psi^{A'(2s)} \wedge H_{A'A'}\wedge F_{A'(2s-2)}\,,
\end{align}
where the curvature is $F(\bry)=\nabla\omega-\omega\wedge \omega$. Note that indices contracted with $\bry^{A'}$ are symmetrized automatically:
\begin{align}\label{commutatorYM}
    \omega \wedge \omega &= \sum_{n,m=0} \frac{1}{2\,n!m!}\,[\omega_{A'(n)}, \omega_{A'(m)}]\, \bry^{A'_1}...\,\bry^{A'_{n+m}}\,.
\end{align}
The action is invariant under the Yang-Mills transformations:
\begin{align}
    \delta \omega&= \nabla\xi-[\omega,\xi]\,,& 
    \delta \Psi&= [\Psi,\xi]\,.
\end{align}
It is also invariant under the algebraic symmetries (thanks to $e^{BA'}\wedge H^{A'A'}\equiv0$):
\begin{align}\label{symmetry-shifts}
    \delta \omega^{A'(k)}&= e\fdu{C}{A'} \eta^{C,A'(k-1)}\,,
\end{align}
which is vital for $\omega$ to have the right number of degrees of freedom. See \cite{Krasnov:2021nsq} for detail and \cite{Tran:2021ukl} for the twistor reformulation.

In principle, we can write down the variational equations of motion and try to represent them as an FDA. Two important hints will play a role in what follows: (a) interactions must contain \eqref{commutatorYM}, i.e. $d\omega(\bry)=\omega(\bry)\wedge \omega(\bry)+...$; (b) $\Psi(\bry)$ takes values in the module that is dual to that of $\omega$, which follows from the structure of the action.  

\paragraph{HS-SDGR.} Higher spin extension of SDGR \cite{Krasnov:2016emc} is more peculiar \cite{Krasnov:2021nsq}. Let us start with its version on constant (non-zero) curvature spacetimes. The flat-space version \cite{Krasnov:2021cva} is a simple limit. To proceed we introduce a Poisson structure on the space $\mathbb{C}[\bry]$ of functions in $\bry^{A'}$:\footnote{This algebra is also know as $w_{1+\infty}$, see e.g. \cite{Adamo:2021lrv} for the latest applications.}
\begin{align}
    \{f,g\}&= \pl^{C'} f \pl_{C'}g= \sum_{n,m} \tfrac{1}{(n-1)!(m-1)!} f\fdu{A'(n-1)}{C'}g_{A'(m-1)C'}\, \bry^{A'}...\,\bry^{A'}\,.
\end{align}
Since Poisson implies Lie, we can define a curvature as usual
\begin{align}
    F&= d\omega-\tfrac12\{\omega,\omega \}\,, & \delta \omega&=d\xi -\{\omega,\xi\}\equiv D\xi\,.
\end{align}
In particular, the Poisson bracket reproduces the standard $F^{AB}= d\omega^{AB} +\omega\fud{A}{C} \wedge\omega^{CB}$ in the spin-two sector. The action reads:
\begin{align}\label{HSSDGRAads}
   S&=\tfrac12\action{\Psi}{F\wedge F} = \sum_{n,m=0}\frac{1}{2(n+m)!}\int \Psi^{A'(n+m)}\wedge F_{A'(n)}\wedge F_{A'(m)}\,.
\end{align}
It is again important that there is a generalization of the shift symmetry that leaves the full action invariant \cite{Krasnov:2021nsq}. To this effect, one first needs to induce the module structure on $\Psi$, which is a module dual to the Poisson algebra as a Lie algebra:
\begin{align}
     \scalar{ f}{ \{\xi,g\}}&:=  \scalar{ f\circ\xi}{g}\,.
\end{align}
That it is a module structure is manifested by 
\begin{align}
    \mathcal{R}_f (\Psi)&:=-\Psi\circ f\,, && [\mathcal{R}_f,\mathcal{R}_g] (\Psi)= \mathcal{R}_{\{f,g\}} (\Psi) \,.
\end{align}
The structure of the action and of the gauge symmetries gives a strong support to the idea that $\Psi$ has to be in the dual (coadjoint) representation of the higher spin symmetry. The flat-space limit is easy to take: one just needs to drop the $\{\omega,\omega \}$-term in the curvature, which is equivalent to taking the commutative limit for $\bry$.  While we could discuss the FDA formulation of this theory, an example of SDGR gives enough information about the gauge algebra to attack the main problem.

\paragraph{SDGR in flat space.} It may be useful to recall the first few terms of the FDA for self-dual gravity \cite{Siegel:1992wd,AbouZeid:2005dg} in flat space \cite{SDFDA}. The action reads \cite{Krasnov:2021cva}
\begin{align}
    \int \Psi^{A'B'C'D'}\wedge d\omega_{A'B'} \wedge d\omega_{C'D'}\,.
\end{align} 
The equations of motion are ($F^{A'B'}\equiv d\omega^{A'B'}$)
\begin{align}
    F_{(A'B'} \wedge F_{C'D')}&=0\,, & d\Psi^{A'B'C'D'}\wedge F_{A'B'}&=0\,.
\end{align}
The first equation implies that there is no $5$-dimensional representation of $sl_2$ in the symmetric tensor product of two $F^{A'B'}$. Therefore, $F^{A'B'}$ can be represented as $e\fdu{B}{A'}\wedge e^{BA'}$ for some field $e^{AA'}$. Indeed, it is easy to see that $F^{A'A'}\wedge F^{A'A'}=0$. Now, it is not surprising that the first few equations in the FDA read
\begin{align*}
d \omega^{A'A'}&= e\fdu{B}{A'}\wedge e^{BA'}\,,
    & d e^{AA'}&=\omega\fud{A}{B}\wedge e^{BA'} \,,
   & d \omega^{AA}&=\omega\fud{A}{C}\wedge\omega^{CA}+H_{BB}C^{AABB}\,.
\end{align*}
We note that the non-abelian terms with $\omega^{A'A'}$ are missing here-above as compared to the standard curvature of $so(3,2)\sim sp(4)$. However, we do not recognize the curvature of the Poincare algebra either. As for $\Psi$, the equation can be rewritten as 
\begin{align}
    d\Psi^{A'B'C'D'}\wedge H_{A'B'}&=0\,,
\end{align}
which is equivalent to
\begin{align}
    \nabla\Psi^{A'A'A'A'}&= e_{BB'}\Psi^{B,A'A'A'A'B'}\,.
\end{align}
One can see that we employ exactly the same fields as for the full gravity, but certain structures 'abelianize'. Half of the Lorentz symmetry becomes global rather than originating from a local gauge symmetry. 

\section{FDA for Chiral Higher Spin Gravity}
\label{sec:FDA2}
After the preliminary steps above we proceed to constructing the Free Differential Algebra of Chiral Theory. Firstly, we summarize the known initial data and boundary conditions for the $L_\infty$ structure maps. 

\subsection{Initial data}

\paragraph{Coordinates/fields, on-shell jet.} The coordinates on the $Q$-manifold or, alternatively, the fields of the minimal model are exactly the same as for the free fields discussed in Section \ref{sec:initial1}. 
\besubeqs\label{spec}
\begin{align}
    h&=+s: && \omega^{A(k),A'(2s-2-k)}\,, C^{A(2s+i),A'(i)} \,, \qquad k=0,...,2s-2\,,\quad i=0,1,2,... \,,\\
    h&=-s: && C^{A(i),A'(2s+i)}\,,\qquad i=0,1,2,... \,,\\
    h&=0: && C^{A(i),A'(i)}\,,\qquad i=0,1,2,...\,.
\end{align}
\esubeqs
As before, it is convenient to keep all components of $\omega$ and $C$ confined in generating functions $\omega(y,\bry)$, $C(y,\bry)$. Chiral Theory is known to admit Yang-Mills gaugings \cite{Skvortsov:2020wtf} that, however, come in a very restricted Chan-Paton-like fashion. To be precise, one can have $U(N)$, $O(N)$ and $USp(N)$ gaugings. Therefore, we assume that $\omega$ and $C$ take values in $\mathrm{Mat}_N$.\footnote{It was shown in \cite{Sharapov:2018kjz,Sharapov:2019vyd,Sharapov:2020quq} that this assumption allows one to reduce a complicated Chevalley-Eilenberg cohomology problem to a much simpler Hochschild one. In other words, it is important to remember that usually higher spin algebras originate from associative ones. }

\paragraph{General form.} Given all the data above, we are looking for Chiral Theory in the form
\besubeqs\label{eq:chiraltheory}
\begin{align} 
    d\omega&= \mathcal{V}(\omega, \omega) +\mathcal{V}(\omega,\omega,C)+...\,,\\
    dC&= \mathcal{U}(\omega,C)+ \mathcal{U}(\omega,C,C)+... \,.
\end{align}
\esubeqs
Here, $\mathcal{V}$ and $\mathcal{U}$ are some $L_\infty$ structure maps to be determined. It would be sufficient if the expansion stops at the quartic terms. This can be justified on the basis of the light-cone action of Chiral Theory: interactions stop at the cubic level. One might argue that they have to stop then at quadratic terms for equations. However, this does not have to be the case since the light-cone gauge theory requires a background, i.e. some specific $\omega_0$. Therefore, $\mathcal{V}(\omega,\omega,C)$ is legit, as well as $\mathcal{V}(\omega,\omega,C,C)$, while higher order terms may not be necessary. One can also see that $\mathcal{V}(\omega, \omega) $ cannot account for all of the interactions, e.g. $\omega$ does not contain the scalar field at all. 

An important subtlety is that covariantization of a given theory (going from the light-cone gauge to a covariant formulation) may require more terms in the perturbation theory that are there only for the sake of covariance. Such contact terms will not give any contribution to physical amplitudes. Another subtlety is due to field redefinitions: it is easy to perform a nonlinear field redefinition in the cubic theory and generate spurious interactions. Alternatively, when looking for $\mathcal{V}$'s and $\mathcal{U}$'s one can find oneself in an unfortunate field frame with such spurious interactions all around. We check in Appendix \ref{app:} that certain cubic amplitudes are reproduced correctly. Therefore, \eqref{eq:chiraltheory} contains all the physically relevant information. Comparing the Chiral Theory FDA to those of SDYM and SDGR \cite{SDFDA} we find that the former contains only the terms essential for consistency, which fixes field redefinitions. 

\paragraph{Boundary conditions.} There are some boundary conditions for $\mathcal{V}$'s and $\mathcal{U}$'s that we learned from the free equations \eqref{linearizeddata3}:
\besubeqs\label{eq:boundaryconditions1}
\begin{align}
    \mathcal{V}(e,\omega)+\mathcal{V}(\omega,e)&=e^{CC'} \pl_C \bry_{C'} \omega\,,\\
    \mathcal{U}(e,C)+\mathcal{U}(C,e)&=e^{CC'} \pl_C \pl_{C'} C\,,\\
    \mathcal{V}(e,e,C)&= e\fud{C}{B'}e^{CB'} \pl_C \pl_C C(y,\bry=0)\,.
\end{align}
\esubeqs
To summarize we are looking for a theory with the spectrum of fields given in \eqref{spec}, in the form of FDA \eqref{eq:chiraltheory} such that it reproduces the boundary conditions \eqref{eq:boundaryconditions1}, i.e. the free equations. 

\subsection{FDA}

In what follows we will have to write down ans{\"a}tze for $L_\infty$-maps. Given that we have packaged the coordinates into generating functions $\omega(y,\bry)$ and $C(y,\bry)$, the $L_\infty$-structure maps can be represented by poly-differential operators:
\begin{align}
    \mathcal{V}(f_1,...,f_n)&= \mathcal{V}(y, \pl_1,...,\pl_2)\, f(y_1)...f(y_n) \Big|_{y_i=0}\,,
\end{align}
where $f_i$'s are $\omega$'s or $C$'s and we have explicitly indicated dependence on $y$, omitting $\bry$ which can be treated similarly. With further details on the operator calculus collected in Appendix \ref{app:}, we only note that (i) we abbreviate $\bry^{A'}\equiv p_0^{A'}$, $\pl^{\bry_i}_{A'}\equiv p_{A'}^i$, $y^A\equiv q_0^A$, $\pl^{y_i}_A\equiv q_A^i$; (ii) contractions $p_{ij}\equiv p_i \cdot p_j\equiv -\epsilon_{AB}p^A_{i}p_{j}^B=p^A_{i}p_{jA}$ are done in such a way that $\exp[p_0\cdot p_i]f(y_i)=f(y_i+y)$; (iii) all operators are Lorentz invariant in the most naive sense of having all indices contracted either with $\epsilon_{AB}$ or $\epsilon_{A'B'}$; (iv) we usually omit explicit arguments $y_i$ in $f$'s, drop $|_{y_i=0}$ and sometimes write down only the operator itself whenever it is clear what the arguments are. Of course, all poly-differential operators are assumed to be local, i.e. they map polynomials to polynomials, which, after Taylor expansion means, that the operators contract a number of Lorentz indices on the arguments.\footnote{Note that this locality is just a requirement for $\mathcal{V}$ to imply some contraction of Lorentz indices (hidden by $y$) on the arguments, which is a type of locality used in \cite{Vasiliev:1988sa}. The locality in the field theory sense is more subtle --- one has to control the number of derivatives in interactions. The interactions in the present paper are local as in Chiral Theory, i.e. vertices contain a finite number of derivatives provided the helicities of the fields at a given vertex are fixed. } To give a couple of useful examples, the usual commutative product on $\bry$ and the Moyal-Weyl star-product on $y$ correspond to the following symbols
\besubeqs
\begin{align}
    \exp{(\bar{y}(\bar{\partial}_1+\bar{\partial}_2))}&= \exp[p_0 \cdot p_1+p_0 \cdot p_2]\equiv \exp[p_{01}+p_{02}]\,, \\
    \exp{(y(\partial_1+\partial_2)+\partial_1\partial_2})&= \exp[q_0 \cdot q_1+q_0 \cdot q_2+q_1 \cdot q_2]\equiv \exp[q_{01}+q_{02}+q_{12}]\,.
\end{align}
\esubeqs
We also would like to rewrite  the boundary conditions \eqref{eq:boundaryconditions1} in the operator language:
\besubeqs
\begin{align}
    \mathcal{V}(e,\omega)+\mathcal{V}(\omega,e)&\sim p_{01}q_{12}\, e^{p_{02}+q_{02}}\, ( e^{CC'}y^1_C \bry^1_{C'})\, \omega(y_2,\bry_2)\Big|_{y_{1,2}=\bry_{1,2}=0}\label{eq:boundaryconditionsBA1}\,,\\
    \mathcal{U}(e,C)+\mathcal{U}(C,e)&\sim q_{12} p_{12}\, e^{p_{02}+q_{02}}\, ( e^{CC'}y^1_C \bry^1_{C'})\, C(y_2,\bry_2)\Big|_{y_{1,2}=\bry_{1,2}=0}\label{eq:boundaryconditionsBB1}\,,\\
    \mathcal{V}(e,e,C)&\sim  q_{13} q_{23} p_{12}\, e^{q_{03}}\,( e^{BB'}y^1_B \bry^1_{B'})( e^{CC'}y^2_C \bry^2_{C'})\, C(y_3,\bry_3)\Big|_{y_{1,2,3}=\bry_{1,2,3}=0}\,,
\end{align}
\esubeqs
where the $\sim$ sign means that in the actual FDA we only care about reproducing these structures up to an overall coefficient. The last boundary condition, if satisfied, ensures the nontriviality of the full vertex. We will also give a rigorous proof of this fact. 

\paragraph{Higher spin algebra.} The $L_\infty$-relations or the formal consistency of \eqref{eq:chiraltheory} at order $\omega^3$ imply the Jacobi identity for $\mathcal{V}(\bullet,\bullet)$
\begin{equation}
    \mathcal{V}(\mathcal{V}(\omega,\omega),\omega)-\mathcal{V}(\omega,\mathcal{V}(\omega,\omega))=0\,.
\end{equation}
The presence of the matrix factors reduces the Jacobi identity to a much simpler and more restrictive associativity condition, i.e $\mathcal{V}(a,b)$ must define an associative product, where $a,b\in \mathbb{C}[y,\bry]$. Given the nonlinear pieces of various (sub)theories there are not so many associative algebras one can think of. In fact, the only option \cite{Krasnov:2021cva} is to define\footnote{A very similar algebra in the same context, but in the light-cone gauge, appeared even before \cite{Ponomarev:2017nrr}. }
\begin{align}
    \mathcal{V}(f,g)&=c\exp{[q_{01}+q_{02}+q_{12}]} \exp{[p_{01}+p_{02}]}f(y_1,\bar{y}_1)\wedge g(y_2,\bar{y}_2)\Big|_{y_i=\bar{y}_i=0}\equiv f\star g\,,
\end{align}
with $c$ an undetermined prefactor. In words $\mathcal{V}(f,g)\equiv f\star g$ is the commutative product on $\bry$ and the star-product on $y$. Therefore, as the higher spin algebra $\hs$ we take the tensor product of the Weyl algebra in $y$ and of the commutative algebra of function in $\bry$, $\hs = A_1\otimes \mathbb{C}[\bry]$. In addition we assume the matrix factor $\mathrm{Mat}_N$. This choice for $\mathcal{V}(\omega,\omega)$ is also consistent with the boundary conditions in equation \ref{eq:boundaryconditionsBA1}:
\begin{equation}
    \mathcal{V}(e,\omega)+\mathcal{V}(\omega,e)=2c\, e^{BB'}\bar{y}_B'\partial_B\omega\,,
\end{equation}
which encourages us to set $c=\frac{1}{2}$, so that
\begin{empheq}[box=\fbox]{align} \label{V(w,w)}
     \mathcal{V}(f,g)&=\tfrac12 \exp{[q_{01}+q_{02}+q_{12}]} \exp{[p_{01}+p_{02}]}
\end{empheq}

\paragraph{Coadjoint module.} Similarly, the formal consistency implies that $U(\bullet,\bullet)$ defines a representation of the higher spin algebra:
\begin{equation}
    \mathcal{U}(\mathcal{V}(\omega,\omega),C)-\mathcal{U}(\omega,\mathcal{U}(\omega,C))=0\,.
\end{equation}
The actions of HS-SDYM and HS-SDGR strongly suggest that $C(0,\bry)$ lives in the space dual to $\omega(0,\bry)$. The action on the dual space (dual to the commutative algebra of functions in $\bry$) can be defined via $\bry_A \rightarrow \alpha \pl_{A'}$, where $\alpha$ is any number. In other words, the commutative algebra of functions in $\bry$ acts on the dual space via differential operators.\footnote{Since understanding that $C$ lives in the dual module has been important for the present paper and this idea is slightly different from the usual approach in the literature, we elaborate on it more in Appendix \ref{app:coadjoint}. } In terms of symbols of operators we can write
\begin{align}
    \omega(f)&= \exp{[p_{02}+\alpha p_{12}]}\, \omega(\bar{y}_1) f(\bar{y}_2)\Big|_{\bar{y}_i=0}\,.
\end{align}
With indices explicit we find
\begin{align}
    \omega(f)&=\sum_{i,n} \tfrac{\alpha^i}{n!}\, \omega^{B'(i) }f_{A'(n) B'(i)}\, \bry^{A'}...\bry^{A'}\,.
\end{align}
It is plausible to extend the idea with the dual space to the complete space $\mathbb{C}(y,\bry)$, as it is unclear how to induce the module structure, otherwise. Now it is time to remember about the matrix factors. We consider functionals based on their ordering of $\omega$ and $C$:
\begin{align}
    \mathcal{U}(\omega,C)&=\mathcal{U}_1(\omega,C)+\mathcal{U}_2(C,\omega)\,.
\end{align}
The consistency condition splits into the following equations.
\begin{equation}
    \begin{split}
        &\mathcal{U}_1(\mathcal{V}(\omega,\omega),C)-\mathcal{U}_1(\omega,\mathcal{U}_1(\omega,C))=0\,,\\
        &\mathcal{U}_2(\mathcal{U}_1(\omega,C),\omega)-\mathcal{U}_1(\omega,\mathcal{U}_2(C,\omega))=0\,,\\
        &\mathcal{U}_2(\mathcal{U}_2(C,\omega),\omega)+\mathcal{U}_2(C,\mathcal{V}(\omega,\omega))=0\,.
    \end{split}
\end{equation}
In words, we have a right and a left actions of the higher spin algebra on $\mathbb{C}(y,\bry)$. The actions must be compatible with each other, which is the middle equation. Given that $C$ should be in the dual module, the structure maps $\mathcal{U}_{1,2}$ are easy to fix to be:
\begin{equation}\boxed{
    \begin{aligned}
        &\mathcal{U}_1(\omega,C)=+\tfrac{1}{2}\exp{[q_{01}+q_{02}+q_{12}]}\exp{[p_{02}+p_{12}]}\, \omega(y_1,\bar{y}_1) C(y_2,\bar{y}_2)\Big|_{y_i=\bar{y}_i=0}\\
        &\mathcal{U}_2(C,\omega)=-\tfrac{1}{2}\exp{[q_{01}+q_{02}+q_{12}]}\exp{[p_{01}-p_{12}]}\, C(y_1,\bar{y}_1) \omega(y_2,\bar{y}_2)\Big|_{y_i=\bar{y}_i=0}
    \end{aligned}}
\end{equation}
It is easy to check that boundary condition \eqref{eq:boundaryconditionsBB1} is satisfied with coefficient $1$.

\paragraph{Cubic Vertex $\boldsymbol{\mathcal{V}(\omega,\omega,C)}$.} As a next step we turn to the cocycle $\mathcal{V}(\omega,\omega,C)$. It has a right to be called a cocycle. Indeed, the bilinear structure maps of any FDA (more generally, of any $L_\infty$-algebra) define a graded Lie algebra. Let us pack them into $Q_0$, $(Q_0)^2=0$. Next we look for the first order deformation $Q_1$ of $Q_0$. It is clear that $Q_1$ must be in the cohomology of $Q_0$. The action of $Q_0$ on $Q_1$ is that of the Chevalley-Eilenberg differential, according to which ${\mathcal{V}(\omega,\omega,C)}$ is a two-cocycle with values in $\hs\otimes \hs$: it takes values in $\hs$ and $C$ is in $\hs^*$. To find the equation for ${\mathcal{V}(\omega,\omega,C)}$ we evaluate the $\omega^3C$ terms after applying $d$ to \ref{eq:chiraltheory}, which leads to
\begin{align}\notag
    \mathcal{V}(\mathcal{V}(\omega,\omega,C),\omega)-\mathcal{V}(\omega,\mathcal{V}(\omega,\omega,C))&+\mathcal{V}(\mathcal{V}(\omega,\omega),\omega,C)\\
    &-\mathcal{V}(\omega,\mathcal{V}(\omega,\omega),C)+\mathcal{V}(\omega,\omega,\mathcal{U}(\omega,C))=0\,.
\end{align}
Like we did for $\mathcal{U}(\omega,C)$, we can split $\mathcal{V}(\omega,\omega,C)$ into three vertices, with different ordering of $\omega$ and $C$:
\begin{align}
    \mathcal{V}(\omega,\omega,C)=\mathcal{V}_1(\omega,\omega,C)+\mathcal{V}_2(\omega,C,\omega)+\mathcal{V}_3(C,\omega,\omega)    \,.
\end{align}
The consistency condition should now be evaluated for each ordering of $\omega$ and $C$ separately, which leads to
\begin{equation}
    \begin{split}
        &\mathcal{V}_1(\mathcal{V}(\omega,\omega),\omega,C)-\mathcal{V}(\omega,\mathcal{V}_1(\omega,\omega,C))+\mathcal{V}_1(\omega,\omega,\mathcal{U}_1(\omega,C))-\mathcal{V}_1(\omega,\mathcal{V}(\omega,\omega),C)=0\,,\\
        &\mathcal{V}(\mathcal{V}_1(\omega,\omega,C),\omega)+\mathcal{V}_1(\omega,\omega,\mathcal{U}_2(C,\omega))+\mathcal{V}_2(\mathcal{V}(\omega,\omega),C,\omega)-\mathcal{V}(\omega,\mathcal{V}_2(\omega,C,\omega))\\
        &-\mathcal{V}_2(\omega,\mathcal{U}_1(\omega,C),\omega)=0\,,\\
        &\mathcal{V}(\mathcal{V}_2(\omega,C,\omega),\omega)-\mathcal{V}_2(\omega,C,\mathcal{V}(\omega,\omega))-\mathcal{V}_2(\omega,\mathcal{U}_2(C,\omega),\omega)-\mathcal{V}(\omega,\mathcal{V}_3(C,\omega,\omega))\\
        &+\mathcal{V}_3(\mathcal{U}_1(\omega,C),\omega,\omega)=0\,,\\
        &V(\mathcal{V}_3(C,\omega,\omega),\omega)-\mathcal{V}_3(C,\omega,\mathcal{V}(\omega,\omega))+\mathcal{V}_3(C,\mathcal{V}(\omega,\omega),\omega)+\mathcal{V}_3(\mathcal{U}_2(C,\omega),\omega,\omega)=0\,.
    \end{split}
\end{equation}
In Appendix \ref{app:complex} we rewrite the equations here-above in terms of symbols of operators. In Appendix \ref{app:vertices} we find a nontrivial solution. The idea is to look for regular vertices in the form of singular field-redefinitions. In other words, if the cocycle is formally trivial but the coboundary does not belong to the required functional class, the cocycle is nontrivial. The final result can be written as
\besubeqs
\begin{empheq}[box=\fbox]{align}
    \mathcal{V}_1(\omega,\omega,C)&: && +p_{12}\,S \int_{\Delta_2}\exp[\left(1-t_1\right) p_{01}+\left(1-t_2\right) p_{02}+t_1 p_{13}+t_2 p_{23}]\,, \\
    \mathcal{V}_2(\omega,C,\omega)&: &&\begin{aligned}
       -p_{13}\,S& \int_{\Delta_2}\exp[\left(1-t_2\right) p_{01}+\left(1-t_1\right) p_{03}+t_2 p_{12}-t_1 p_{23}]+\\
       -&p_{13}\,S \int_{\Delta_2}\exp[\left(1-t_1\right) p_{01}+\left(1-t_2\right) p_{03}+t_1 p_{12}-t_2 p_{23}]\,,
    \end{aligned}
    \\
    \mathcal{V}_3(C,\omega,\omega)&: && +p_{23}\,S \int_{\Delta_2}\exp[\left(1-t_2\right) p_{02}+\left(1-t_1\right) p_{03}-t_2 p_{12}-t_1 p_{13}]\,.
\end{empheq}
\esubeqs
Here, $\Delta_n$ is an $n$-dimensional simplex $t_0=0\leq t_1\leq ...\leq t_n\leq 1$. The nontriviality of the solution is proved in Appendix \ref{app:vertices}. There is an overall factor $S$
\begin{align}\label{juststar}
    S&= \exp[q_{01}+q_{02}+q_{03}+q_{12}+q_{13}+q_{23}]
\end{align}
that computes the star-product over all $y$ variables. In other words, the vertex factorizes 
\begin{align}
    \mathcal{V}_1(a(y)\otimes \bar{a}(\bry),b(y)\otimes \bar{b}(\bry),c(y)\otimes \bar{c}(\bry))&= a\star b\star c \otimes v_1(\bar{a},\bar{b},\bar{c})\,,
\end{align}
and similarly for the other vertices. 

\textit{Remark.} As it was pointed out in \cite{Gerasimenko:2021sxj,Sharapov:2018hnl,Sharapov:2022eiy}, constructing FDA's of higher spin gravities calls for an extension of the deformation quantization of Poisson manifolds to Poisson orbifolds, which is an open problem. Nevertheless, the traces of Kontsevich and of Shoikhet-Tsygan-Kontsevich formality are sometimes visible \cite{Sharapov:2017yde}. The key point in the proof of the formality theorems is to find the right configuration space and the right closed form on it, so that the proof amounts to a simple application of the Stokes theorem. As we show in Appendix \ref{app:vertices}, one can find a closed two-form $\Omega$, $d\Omega=0$, on $\Delta_3$ such that its integral over the four boundaries of the simplex reduces to the four terms in the equation for $\mathcal{V}_1$ and similarly for other vertices. In this regard let us note that the integral form is not unique. It arises as an integral over the configuration space of ordered points on a circle. With the help of translation invariance one can (gauge) fix the times of different points and also use the reflection symmetry of the circle. Altogether there are six different forms. 

\textit{Remark.} The cubic vertex has an interesting property: if we remove for a moment the matrix factors $\mathrm{Mat}_N$, make $y$ commutative (by taking the $\hbar=0$ limit after introducing $\hbar$ into the Moyal-Weyl star-product) and bring $\omega$'s and $C$ to the same $\omega\omega C$-ordering, we get zero:
\begin{align}
    \mathcal{V}_1(\omega,\omega,C)+\mathcal{V}_2(\omega,C,\omega)+\mathcal{V}_3(C,\omega,\omega)\Big|_{\hbar=0\,, N=1} &\equiv0\,.
\end{align}
This does not have to be the case. However, erasing matrix factors together with the commutative limit in $y$ must give a trivial vertex. Indeed, there is no such truncation of Chiral Theory. Therefore, the vertices we found enjoy some kind of minimality. This is to be expected, since we are now taking the graded anti-symmetrization of a commutative structure.

\textit{Remark.} It can be shown that $\mathcal{V}_{1,2,3}\neq0$. In all the cases considered before in the literature $C$ takes values in the (twisted)-adjoint representation of a higher spin algebra. This allows one to set $\mathcal{V}_{2,3}=0$ and choose $\mathcal{V}_{1}(a,b,c)=\phi(a,b)\star c$, where $\phi(a,b)$ is a certain Hochschild two-cocycle that deforms the higher spin algebra. Indeed, assuming $\mathcal{V}_{2,3}=0$ we find
\begin{align}\label{equivar}
    \mathcal{V}(\mathcal{V}_1(a,b,c),d)+\mathcal{V}_1(a,b,\mathcal{U}_2(c,d))&=0\,.
\end{align}
Here, $\mathcal{U}_2(c,d)=-c\star d$ and $\mathcal{V}(a,b)=a\star b$ in the previously studied cases. Therefore, setting $c=1$ leads to $\mathcal{V}_1(a,b,d)=\mathcal{V}_1(a,b,1)\star d$. Moreover, $\phi(a,b)=\mathcal{V}_1(a,b,1)$ turns out to be a Hochschild two-cocycle. For Chiral Theory this cannot be true. Indeed, it is easy to see that while in the second term of \eqref{equivar} $d$ has all of its indices contracted with $c$, the same indices are free in the first term, i.e. \eqref{equivar} cannot be satisfied. 
Therefore, we have to look for a solution with all $\mathcal{V}_{1,2,3}\neq0$, as we did.

\paragraph{Cubic Vertex $\boldsymbol{\mathcal{U}(\omega,C,C)}$. } The previously found vertex $\mathcal{V}(\omega,\omega,C)$ serves as a source for $\mathcal{U}(\omega,C,C)$. As before, we split it according to different orderings:
\begin{align}
    \mathcal{U}(\omega,C,C)=\mathcal{U}_1(\omega,C,C)+\mathcal{U}_2(C,\omega,C)+\mathcal{U}_3(C,C,\omega)   \,. 
\end{align}
There are six equations that can be obtained as various $\omega^2C^2$-terms after applying $d$ to \eqref{eq:chiraltheory}. We rewrite them in terms of symbols of operators in Appendix \ref{app:complex} and solve in Appendix \ref{app:vertices}. The final form of the solution reads:\footnote{A resemblance to some of the formulas in the literature \cite{Vasiliev:1988sa} is striking, of course. However, as different from \cite{Vasiliev:1988sa}, all vertices in the present paper are local and do not contain infinite (divergent) sums over different representations of the same interactions \cite{Boulanger:2015ova,Skvortsov:2015lja}. Therefore, we are constructing an actual theory rather than the most general ansatz for interactions compatible with symmetries. }
\besubeqs
\begin{empheq}[box=\fbox]{align}
    \mathcal{U}_1(\omega,C,C)&: && +p_{01}\,S \int_{\Delta_2}\exp[\left(1-t_2\right) p_{02}+t_2 p_{03}+\left(1-t_1\right) p_{12}+t_1 p_{13}]\,, \\
    \mathcal{U}_2(C,\omega,C)&: &&\begin{aligned}
       -p_{02}\,S& \int_{\Delta_2}\exp[t_2 p_{01}+\left(1-t_2\right) p_{03}-t_1 p_{12}+\left(1-t_1\right) p_{23}]+\\
       -&p_{02}\,S \int_{\Delta_2}\exp[t_1 p_{01}+\left(1-t_1\right) p_{03}-t_2 p_{12}+\left(1-t_2\right) p_{23}]\,,
    \end{aligned}
    \\
    \mathcal{U}_3(C,C,\omega)&: && +p_{03}\,S \int_{\Delta_2}\exp[\left(1-t_1\right) p_{01}+t_1 p_{02}+\left(t_2-1\right) p_{13}-t_2 p_{23}]\,,
\end{empheq}
\esubeqs
where $S$ is the star-product over $y$'s, \eqref{juststar}.

\subsection{Summary and Discussion}

The main result of this paper are the boxed formulas above that define vertices $\mathcal{V}(\omega,\omega)$, $\mathcal{U}(\omega,C)$, $\mathcal{V}(\omega,\omega,C)$, $\mathcal{U}(\omega,C,C)$. Altogether they satisfy the $L_\infty$-relations up to order $\mathcal{O}(C^2)$. These vertices determine both the free equations and the essential interactions of Chiral Theory. By essential we mean those that contribute to the cubic amplitude and which fully determine Chiral Theory. Let us recall that one can switch on very few higher spin interactions and it is the consistency of the theory that will enforce the unique completion  \cite{Metsaev:1991mt,Metsaev:1991nb,Ponomarev:2016lrm}.  However, the covariantization may require more contact vertices, which is an interesting problem for the future. 

In Chiral Theory there is one dimensionful coupling constant, $l_P$, which is needed to compensate for higher powers of momenta in the vertices. The power of momenta equals the sum of the helicities, $\lambda_1+\lambda_2+\lambda_3$, of the fields that meet at the vertex. Given that the action of SDGR (with cosmological constant) contains  $d\omega^{A'B'}+\omega\fud{A'}{C'}\wedge \omega^{C'B'}$, it makes sense to assign mass dimension $1$ to all $\omega^{A'(2s-2)}$ and, hence, mass dimension zero to all $\Psi^{A'(2s)}$. Similarly, $e^{AA'}$ has dimension one. All $\omega^{A'(2s-2-k),A(k)}$ are expressed as derivatives of $\omega^{A'(2s-2)}$. It is then tempting to extend this to the whole $\omega$ and $C$. To recover $l_P$ we need to introduce it into $e^{AA'}$, e.g. $e^{AA'}_\mu \sim l_P^{-1} \sigma^{AA'}_\mu$ in Cartesian coordinates.

The dimensionless coupling $\kappa$ simply counts the orders of $\omega$ and $C$ in the perturbative expansion. In the light-cone gauge the expansion stops at the cubic terms. This does not have to be the case after covariantization. Let us compare the general structure of interactions in the light-cone gauge and in the FDA expanded over Minkowski vacuum $\omega_0=e$. We will be sketchy here. It is convenient to pack all positive helicity fields into $\Phi$ and all negative helicity fields plus scalar into $\Psi$. The action reads (very schematically)
\begin{align}\label{sketch}
    \mathcal{L}&= \Psi \square \Phi + c_{+++}\Phi\Phi\Phi+c_{++-}\Phi\Phi\Psi+c_{+--}\Phi\Psi\Psi\,,
\end{align}
where we drop the helicity labels and omit the detailed structure of interactions. The equations of motion would be
\begin{align}
    \square \Phi&= c_{++-}\Phi\Phi +c_{+--}\Phi\Psi\,, & \square\Psi&=c_{+++}\Phi\Phi+c_{++-}\Phi\Psi +c_{+--}\Psi\Psi\,.
\end{align}
This should be compared with ($D\equiv d-\omega_0$ is the background covariant derivative in the appropriate representations of the higher spin algebra)
\besubeqs
\begin{align} 
    D\omega&= \mathcal{V}(\omega, \omega) +\mathcal{V}(\omega_0,\omega,C)+\mathcal{V}(\omega_0,\omega_0,C,C)\,,\\
    DC&= \mathcal{U}(\omega,C)+ \mathcal{U}(\omega_0,C,C) \,,
\end{align}
\esubeqs
where we indicated all terms that can potentially contribute to the cubic amplitude. We recall that $\omega$ carries positive helicity and, hence, is a cousin of $\Phi$, while $C$ contains both $\Psi$ and descendants of $\omega$. We show in Appendix \ref{app:amplitude} that $\mathcal{V}(\omega, \omega)$ and $\mathcal{U}(\omega, C)$ give the correct amplitudes. There is a unique theory that has such amplitudes, which is a consistency check.

Another valuable consistency check is to restrict interactions to the spin-two and to the spin-one sectors to reproduce the recently obtained FDA's of SDYM and SDGR \cite{SDFDA}. To be precise, the restriction has to give FDA's that are quasi-isomorphic to those of SDYM and SDGR. Luckily, this exercise directly leads to the interactions of \cite{SDFDA}. The latter were found in the most minimal form, i.e. we have not introduced any nonlinear terms into the FDA beyond what is necessary, which fixes all field redefinitions. It is encouraging that the FDA of Chiral Theory is also minimal in this sense. 

By the same token the higher spin extensions of SDYM and SDGR \cite{Krasnov:2021nsq}, which were previously discovered as  contractions of Chiral Theory in \cite{Ponomarev:2017nrr}, must be consistent contractions of the present FDA as well. We note that in the latter two cases the FDA of this paper should provide a complete solution of the problem. Indeed, the actions of these two theories are schematically
\begin{align}
    \mathcal{L}&= \Psi \square \Phi +\Phi\Phi\Psi\,,
\end{align}
which is much simpler than the structure of interactions of Chiral Theory. Therefore, it is tempting to argue that we have determined all interaction vertices in these theories since this is the case for SDYM and SDGR. 

A very interesting observation made in \cite{Ponomarev:2017nrr} is that the coupling constants of Chiral Theory determine a certain (kinematic) algebra in the light-cone gauge and the product in this algebra is a remnant of the star-product. This statement covers all vertices. For the FDA at hand, it is the $\Phi\Phi\Psi$-vertex where the star-product structure is manifest. The other vertices correspond to the Chevalley-Eilenberg cocycles of the higher spin algebra. Nevertheless, according to \cite{Ponomarev:2017nrr}, what survives of these vertices in the light-cone gauge is the same star-product. It would be interesting to clarify this statement.

\section{Conclusions}

The main result of this paper is the covariant form that incorporates some essential interactions of Chiral Theory which was previously known in the light-cone gauge only. By essential we mean those interactions that, if present, unambiguously fix the theory. Technically, the result is the minimal model of Chiral Theory --- a Free Differential Algebra consistent to order $\mathcal{O}(C^2)$. 

The FDA of the present paper contains FDA's of SDYM, SDGR \cite{SDFDA} and of the higher spin extensions thereof \cite{Krasnov:2021nsq}. For these four cases the FDA should be complete. For Chiral Theory certain higher order vertices may still be required for formal consistency and covariantization. One can also look for supersymmetric extensions that would combine SDYM and SDGR and higher spin extensions thereof \cite{Devchand:1996gv} as well as for the full supersymmetric Chiral Theory \cite{Metsaev:2019dqt,Metsaev:2019aig}. 

Even though we found a covariant form for the essential interactions of Chiral Theory, there might still be an obstruction to getting the complete theory in a manifestly Lorentz invariant form if some of the interactions cannot be written with the help of the new field variables ($\omega^{A'(2s-2)}$ and $\Psi^{A'(2s)}$ as compared to the old $\Phi_{\mu_1...\mu_s}$) . In Appendix \ref{app:amplitude} we also show that the most problematic $V^{+--}$ amplitudes can be reproduced. Independently of that, a simple extension of the cohomological arguments along the lines of \cite{Sharapov:2020quq} indicates that there are no obstructions to the FDA of this paper. Therefore, the complete Chiral Theory can be written in a manifestly Lorentz invariant form as an FDA.  

As is well-understood \cite{Barnich:2010sw,Grigoriev:2012xg,Grigoriev:2019ojp}, the minimal model of a (gauge) field theory contains all the essential information about the theory (local BRST cohomology), e.g. actions/counterterms, anomalies, conserved charges, deformations, etc. It is a very encouraging statement given that the differential $Q$ can be extracted from classical field equations rewritten as a Free Differential Algebra. Therefore, the results of this paper should help to address the problems where having a covariant form of the theory is an advantage, i.e. all of them. Chiral Theory was shown to be one-loop finite in the light-cone gauge \cite{Skvortsov:2018jea,Skvortsov:2020wtf,Skvortsov:2020gpn}, but extending these results to higher loop orders should be simpler within a covariant approach. It would also be interesting to look for exact solutions where generalizations of Ward/Penrose/ADHM \cite{Ward:1977ta,Penrose:1976js,Atiyah:1978ri} constructions to Chiral Theory together with its twistor formulation should be of great help.

\chapter{All order vertices}
\label{chap:all}

In this chapter, we obtain all higher order vertices for chiral HiSGRA in any constant curvature background. The content is entirely based on \cite{Sharapov:2022faa}, co-authored with Alexey Sharapov, Evgeny Skvortsov, and Arseny Sukhanov and published in the \textit{Journal of High Energy Physics}; \cite{sharapov2023more}, co-authored by Alexey Sharapov and Evgeny Skvortsov and published in \textit{SciPost Physics}; \cite{sharapov2023more}, co-authored by Alexey Sharapov, Evgeny Skvortsov and Arseny Sukhanov and published in \textit{Nuclear Physics B}.

\section{Introduction}


In this chapter, the covariant form of Chiral Theory wil be constructed via the standard homological perturbation theory: there is a differential graded Lie algebra that encodes the free theory, whereas its simple deformation leads to a nontrivial $L_\infty$-algebra $\mathbb{L}$ that encodes the interaction vertices. This $L_\infty$-algebra $\mathbb{L}$ is obtained by symmetrization of a certain $A_\infty$-algebra $\hat{\mathbb{A}}$. It is the latter algebra which  structure maps (or products) we compute.

\begin{wrapfigure}{r}{0.3\textwidth}
\begin{tikzpicture}
\draw[-]  (0,4) -- (4,4)--(4,0)--cycle;
\draw[-]  (0,4) -- (4,0);
\draw[thick]  (0,0) -- (0,4)--(0.5,1.5)--(1,0.7)--(2.5,0.15)--(4,0)--cycle;
 \fill[black!5!white] (0,0) -- (0,4)--(0.5,1.5)--(1,0.7)--(2.5,0.15)--(4,0)--cycle;
\filldraw (0,4) circle (1.0 pt);
\filldraw (0.5,1.5) circle (1.5 pt);
\filldraw (1,0.7) circle (1.5 pt);
\filldraw (2.5,0.15) circle (1.5 pt);
\filldraw (0,0) circle (1 pt);
\filldraw (4,0) circle (1 pt);
\node[] at (0.4,0.7) {A};
\node[] at (1.7,1.5) {B};
\end{tikzpicture}
\caption{A convex polygon B and a swallowtail A.}\label{F0}
\end{wrapfigure}
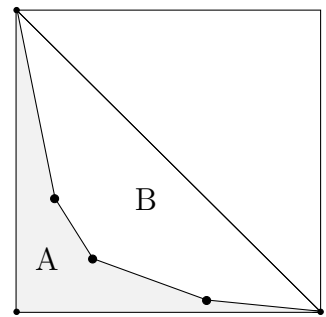

Another observation is that all nontrivial algebraic structures defining the interaction vertices are effectively low-dimensional. To put it more formally, the $A_\infty$-algebra $\hat{\mathbb{A}}$ is given by a tensor product of a smaller $A_\infty$-algebra $\mathbb{A}$ with some associative algebra $B$. While $B$ enters trivially and can be replaced with any other associative algebra, e.g. $\mathrm{Mat}_N$, the theory based on $\mathbb{A}$ is effectively low-dimensional. The effective dimension can be seen from the functional dimension of the vector space underlying $\mathbb{A}$, which is $2$. By definition, both $\hat{\mathbb{A}}$ and $\mathbb{A}$ have natural pairings that make them into cyclic $A_\infty$-algebras, or more specifically, pre-Calabi--Yau algebras of degree two \cite{kontsevich2021pre}. This cyclic structure appears to be very useful in linking different interaction vertices to each other.

The vertices that come out of homological perturbation theory can be considerably simplified by performing a certain change of variables, which, among other things, makes the  cyclic structures of $\hat{\mathbb{A}}$ and $\mathbb{A}$ explicit. Remarkably, the final result is that, pretty much like in Kontsevich \cite{Kontsevich:1997vb} and Shoikhet--Tsygan--Kontsevich Formality \cite{Shoikhet:2000gw}, the structure maps can be written as integrals over a certain configuration space. The configuration space, which will be defined in detail in Section \ref{sec:config}, can be described as the space of concave polygons (region A in Fig. \ref{F0}), which we call swallowtails. Alternatively, it is the space of convex polygons with one edge corresponding to the diagonal of the square, i.e., polygons B inscribed into a protractor triangle ($45^\circ-90^\circ-45^\circ$). The area of region A also plays a role and appears in front of the cosmological constant term in the $A_\infty$-structure maps, as we construct the formulation of chiral HiSGRA for any constant curvature background. The example in Fig. \ref{F0} corresponds to quintic structure maps that are given by an integral over the six-dimensional configuration space, the positions of the three points in between A and B. The compactness of the configuration space implies that the vertices are formally well-defined. Importantly, the vertices also obey an additional property that translates into locality from the field theory vantage point. The vertices we found turn out to be maximally local, which corresponds to a certain distinguished coordinate system from the $A_\infty/L_\infty$-perspective. 

The chapter is structured as follows. In the \ref{sec:allVertices}, we introduce homological perturbation theory and in \ref{sec:qua} we use it to obtain the quartic vertex of chiral HiSGRA. Then, in \ref{high}, we derive all higher order vertices for arbitary cosmological constant $\lambda$.

\section{Homological perturbation recipe}
\label{sec:allVertices}
As we have seen before, vertices $\mathcal{V}$ and $\mathcal{U}$ encode certain contractions of indices of their arguments. All vertices have the factorized form
\begin{align}
    \mathcal{V}(f_1,\ldots ,f_n)&= v(f_1'(y),\ldots , f_n'(y)) \otimes f_1''\ast\cdots \ast f_n''  \,,
\end{align}
where $f_i=f_i'(y) \otimes f_i''$, $f''_i\in B$, and $\ast$ denotes the product in $B$. In case $B=A_1\otimes \mathrm{Mat}_N$, all $\bry$-dependent factors are multiplied via the star-product:
\begin{align}
    f_1''(\bry)\star \cdots \star f_n''(\bry)&= \exp{\left[\sum_{0=i<j=n} q_i \cdot q_j\right]} f_1''(\bry_1)\cdots f_n''(\bry_n)\Big|_{\bry_i=0}\,.
\end{align}
Here $q$ for $\bry$ is the same as $p$ for $y$. Due to additional matrix factors, all $f''_i=\{f''_i(\bry){}\fud{\aAt}{\aBt}\}$ are also multiplied as matrices in the same order as with $\star$. In the future, we will choose to write only the factor of the structure maps that contains the differential operators on $y$, i.e.
\begin{align}
    \mathcal{V}(f_1,\ldots,f_n)&= \mathcal{V}(y, \pl_1,\ldots,\pl_n)\, f_1(y_1)\cdots f_n(y_n) \Big|_{y_i=0}\,.
\end{align}

Since Chiral Theory is local, there must be certain restrictions imposed on the structure maps containing more than one $C$. Indeed, $C$ encodes arbitrarily high derivatives of the dynamical fields. Therefore, any harmlessly looking $\mathcal{V}(...,C,...,C)$ can hide an infinite sum in derivatives. If we choose $\mathcal{V}(\omega,\omega, C,...,C)$ and $\mathcal{U}(\omega, C,...,C)$ for concreteness, the locality forbids infinite tails in $p_{ij}$ with $2<i,j$ for the first and $p_{ij}$ with $1<i,j$ for the second.\footnote{We should recall that there is a silent star-product over $\bry$'s. Therefore, all $r_{ij}$'s are present in the $\exp{[...]}$. Each pair of contracted derivatives on two fields $...\pl_{AA'}\bullet \pl^{AA'}\bullet$ comes from $r_{ij} p_{ij}$. Given that $r_{ij}$ is already present we have to forbid infinite tails in some $p_{ij}$'s. } For example, $p_{23}$ is not found in $\mathcal{U}_1(\omega,C,C)$ at all (having it in the prefactor would still be admissible). In the following, we shall see that the vertices of chiral HiSGRA are indeed constructed via a procedure that ensures locality.

\paragraph{Ingredients of the Homological Perturbation Theory soup.} The main idea is to use the technique of multiplicative resolutions and homological perturbation theory,\footnote{This is a refinement of the original idea of \cite{Vasiliev:1990cm,Vasiliev:1990vu} to introduce additional variables $z$ as to enlarge the field content and write down certain simple equations constraining the $z$-dependence in such a way that perturbative solution to the $z$-equations would reproduce the sought-for vertices. Since all (hypothetical) $4d$ HiSGRA can be cast into the FDA form and, hence, have the same $\omega$ and $C$ field content, we will try to use a notation designed to reveal similarity to \cite{Vasiliev:1990cm,Vasiliev:1990vu}, stressing the important differences along the way. } see \cite{Sharapov:2017lxr,Sharapov:2018hnl,Sharapov:2018ioy,Li:2018rnc} for the discussion and applications to higher spin gravity motivated problems. In a few words, one can try to look for an embedding of the Hochschild complex into a bigger bicomplex, where one can apply different spectral sequences to get explicit formulas for the cocycles. Higher order corrections can also be obtained via homological perturbation theory. 

A suitable  resolution is obtained by extending the commutative algebra $\mathbb{C}[y]$ to another commutative algebra $\mathbb{C}[y,z]$ of polynomials in  $y_{A}$ and $z_{A}$ equipped with a peculiar product. Assuming momentarily we are dealing with deformation quantization type problems, a quite general class of star-products on functions $\mathbb{C}[Y]$ of $Y\equiv Y^a$, $a=1,...,2n$ is determined by matrices $\Omega^{ab}$ via 
\begin{align}\label{stpr}
    (f\star g)(Y)&= \exp{[Y^a \pl^1_a +Y^a\pl^2_a +\Omega^{ab} \pl^1_a \pl^2_b]} \, f(Y_1)g(Y_2)\big|_{Y_{1,2}=0}\,.
\end{align}
The symplectic structure $C^{ab}$ is given by the anti-symmetric part of $\Omega^{ab}$ and the symmetric part is responsible for the choice of ordering (e.g. normal, anti-normal, totally-symmetric or Weyl). One can also write down an integral representation of the same star-product, which is sometimes more useful,\footnote{Here and in what follows we assume that the integrals are defined in such a way that $\int \exp[u^a\xi_a] du=\delta (\xi)$. This is the only formula that we will need to use. }
\begin{align}
    (f\star g)(Y)&=\int dU\,dV\, d\xi\,d\eta\,f(Y+U)g(Y+V) \exp{[U^m\xi_m+V^m\eta_m -\Omega^{mn}\xi_m \eta_n]}\,.
\end{align}
We choose $\Omega^{ab}$ to be symmetric matrix of the form  
\begin{align}
    \Omega=-\begin{pmatrix}
        \lambda\epsilon & \epsilon \\
        -\epsilon & 0
    \end{pmatrix}\,,\qquad \epsilon\equiv \epsilon^{AB}\,,
\end{align}
where $\lambda=\sqrt{\Lambda}$ with $\Lambda$ the cosmological constant. However, we will refer to $\lambda$ as the cosmological constant from now on.

Via the usual formulas this defines a commutative product, whose differential form has symbol
\begin{align}
    \exp{[ p_{01}+p_{02}+q_{01}+q_{02} +p_1\cdot q_2-q_1\cdot p_2 +\lambda p_{12}]}\,,
\end{align}
where we defined $z^{A}\equiv q_0^{A}$, $\pl^{z_i}_{A}\equiv q_{A}^i$. The integral representation reads 
\begin{align}
    (f\star g)(y,z)&= \int du\,dv\,dp\,dq\,f(y+u,z+v) g(y+q,z+p) \exp{[v\cdot q-u\cdot p+\lambda q\cdot u]}\,.
\end{align}
We will sometimes use notation $\mu(f,g)\equiv f\star g$. The generators $y$ and $z$ act as follows:
\begin{align*}
    \begin{aligned}
        y_{A}\star f &=  (y_{A} -\lambda\pl^y_A -\pl^z_{A})f\,,   &  z_{A}\star f&=f\star z_{A}= (z_{A}+\pl^y_{A}) f\,,\\
        f\star y_{A} &=  (y_{A} +\lambda\pl^y_A -\pl^z_{A})f\,.
    \end{aligned}
\end{align*}
The star-product has some other interesting properties, the most important being the existence of the element $\varkappa = \exp[z^{C}y_{C}]$ satisfying the relations 
\begin{align}\label{klein}
    y_{A} \star \varkappa = \varkappa \star y_{A} = z_{A}\star\varkappa =\varkappa\star z_{A}=0\,.
\end{align}
We will consider an extension of algebra $A$ by the algebra of differential forms in $d z$ with exterior differential $d_z$ and a familiar two-form $\lambda= \tfrac12 \varkappa\, d z^2$.
We will also repeatedly use the Poincare lemma in the form of
\begin{align}\label{homofor}
      f^{(1)}=h[f^{(2)}]&= d z^{A}\, z_{A} \int_0^1 t\, dt\, f^{(2)}(tz)\,, &
      f^{(0)}=h[f^{(1)}]&= z^{A} \int_0^1 dt\, f_{A}^{(1)}(tz)\,,
\end{align}
see e.g. \cite{Didenko:2014dwa}. 
The first part gives a particular solution to $d_z f^{(1)}= f^{(2)}$ for a one-form $f^{(1)}\equiv dz^{A} f^{(1)}_{A}(z)$ and a given two-form $f^{(2)}\equiv \tfrac12 f^{(2)}(z) \epsilon_{AB}dz^{A} dz^{B}$. The second part gives a particular solution to $d_z f^{(0)}= f^{(1)}$ for a closed one-form $f^{(1)}$ and a zero-form $f^{(0)}\equiv f^{(0)}(z)$. We also complete this definition with $h[f^{(0)}]=0$ for any zero-form $f^{(0)}$. 

With the definitions above we are ready to present the whole set of $\mathcal{U}$ and $\mathcal{V}$ vertices defining Chiral HiSGRA (\ref{eq:chiraltheory}). Both type of vertices are constructed as compositions of only three operations: the contracting homotopy $h$, the $\star$-product (\ref{stpr}), called $\mu$, and the product $\diamond$ that is required for the $C$ fields to enter the game; the latter will be introduced shortly. Suitable compositions are conveniently depicted by directed tree graphs, which consist of trivalent vertices, internal edges,  and external edges.  Both ends of an internal edge are on two vertices. Each vertex has two incoming and one outgoing edges. An external edge has one end on a vertex and another end is free. The graphs are supposed to be connected. All the vertices correspond to the star product $\mu$, while the internal edges depict the action of the contracting homotopy $h$:

$$ \begin{tikzcd}[column sep=small,row sep=small]
   & {}& \\
    & \mu\arrow[u]  & \\
    \arrow[ur]  & & \arrow[ul]   & 
\end{tikzcd} \;\;\;\;\;\;\;\;\;\;\;\;
\begin{tikzcd}[column sep=small,row sep=small]
    \mu & \\
    &\mu\arrow[ul, "h" ']   
\end{tikzcd}
$$
The external incoming edges (or leaves) correspond to the arguments $\omega$ and $C$ of the interaction vertices $\mathcal{V}$ and $\mathcal{U}$. Therefore, each graph may be decorated with either one or two $\omega$'s, depending on the type of a vertex. As to the  arguments $C$, all of them enter the interaction vertices through a special combination $\Lambda[C]= h[C\diamond\lambda]$, where $C=C(y)$ and $\diamond$ is defined as  $$C\diamond g(z,y)\equiv g(z, y+p_i) \,C(y_i)\,.$$ 
The expression $\Lambda[C]$ decorates the remaining leaves.
Finally, the only outgoing edge (or root) of a connected tree corresponds to the value of an interaction  vertex. 
The $\star$-product being commutative, many trees lead to analytical expressions  that differ only by a permutation of arguments. One should keep in mind that $h^2=0$ and forms of degree higher than two vanish identically. There are also certain classes of trees that vanish due to specific properties of the resolution.\footnote{At this point we can list the crucial similarities/differences as compared to \cite{Vasiliev:1990cm,Vasiliev:1990vu}. The spectrum of the fields (coordinates on $\mathcal{N}$) is exactly the same. The higher spin algebras are different: star-product in $y,\bry$ as compared to star-product in $\bry$ and commutative algebra in $y$. This entails the second difference: zero-forms are no longer in the twisted-adjoint representation \cite{Vasiliev:1999ba}, but in the coadjoint, and, like the twisted-adjoint, form a genuine module of the higher spin algebra. The vertices are all, of course, different. The ones in the present paper are local, those of \cite{Vasiliev:1990cm,Vasiliev:1990vu} form a gauge-invariant ansatz where infinitely many copies of the same interaction are present in different forms (with higher and higher derivatives), field redefinitions are not fixed and, as a result, there are infinitely many free parameters hidden. For generic choice of these parameters, e.g. the one made in \cite{Vasiliev:1990cm,Vasiliev:1990vu}, one gets nonsensical results for correlation functions \cite{Boulanger:2015ova}, which is to be expected and has little to do with the higher spin problem. The whole class of theories sought for in \cite{Vasiliev:1990cm,Vasiliev:1990vu} cannot be constructed with the help of the standard field theory tools due to the non-locality \cite{Dempster:2012vw,Bekaert:2015tva,Maldacena:2015iua,Sleight:2017pcz,Ponomarev:2017qab}, which makes it an interesting challenge to find more general principles to deal with field theories with such non-localities. The resolution, i.e. the $z$-extension we use is also different from \cite{Vasiliev:1990cm,Vasiliev:1990vu}, even though the number of $z$-variables is the same (ignoring $\brz$ that we do not need). At the most basic level the integral form of $\mu$-product contains two integrals more because matrix $\Omega^{ab}$ has rank four, while in \cite{Vasiliev:1990cm,Vasiliev:1990vu} it has rank two. As a result, there is a smooth deformation of our $\mu$-product that leads to \cite{Vasiliev:1990cm,Vasiliev:1990vu}, but not the other way around. }  All admissible trees that contribute to the interaction vertices are generated via Homological Perturbation Theory, which is detailed in Appendices \ref{app:hpt} and \ref{app:trees}. It might be well to point out that the resulting analytical expressions for the vertices $\mathcal{V}$ and $\mathcal{U}$ do not depend on $z$'s as it must if one treats them 
as elements of the higher spin algebra $\hs$ and its coadjoint module, respectively.  Below we present some explicit expressions, but we start by introducing the bilinear an trilinear structure maps with cosmological constant.

\paragraph{$\boldsymbol{\mathcal{V}(\omega,\omega)}$ and higher spin algebra.} In what follows we assume that the higher spin algebra is of the form $\hs=A_\hhbar \otimes B$. Here, $A_\hhbar$ is the Weyl algebra, that is, the algebra of a polynomial functions $f(\hat y)$ in the  operators $\hat y_A$ subject to the canonical commutation relations $[\hat y_A,\hat y_B]=-2\hhbar \epsilon_{AB}$. Note that all $A_\hhbar$ are isomorphic to each other whenever $\hhbar\neq 0$ and the commutative limit $A_{\lambda=0}$ coincides with $\mathbb{C}[y^A]$. One can also understand $A_\hhbar$ as the result of deformation quantization of the polynomial algebra $A_0$, the quantum product being the Moyal--Weyl star-product. The symbol of the star-product is defined by
\begin{align}\label{hsalgebra}
    \mathcal{V}(f,g)&= \exp{[p_{01}+p_{02}+\hhbar\, p_{12}]}f({y}_1)\, g({y}_2)\Big|_{{y}_i=0} \equiv (f\star g)(y)\,.
\end{align}
The parameter $\hhbar$ has the meaning of the cosmological constant. All vertices of Chiral Theory depend smoothly on $\hhbar$.  

\paragraph{$\boldsymbol{\mathcal{U}(\omega,C)}$ and the dual module.} Thanks to the $A_\infty$-structure this bilinear vertex splits into the sum of two vertices\footnote{By a slight abuse of notation $\mathcal{V}(\omega,
\omega, C, \ldots ,C)$ and $\mathcal{U}(\omega, C, \ldots ,C)$ denote the whole collections of vertices/$A_\infty$-products at a given order that differ by the order of the arguments. When a detailed structure is discussed we enumerate various orderings by subscripts. }
\begin{align}
    \mathcal{U}(\omega,C)&=\mathcal{U}_1(\omega,C)+\mathcal{U}_2(C,\omega)
\end{align}
The $A_\infty$-relations imply that $\mathcal{U}_1(\omega,C)$ and $\mathcal{U}_2(C,\omega)$ define an $\hs$-bimodule structure on $C$. Action \eqref{HSSDDYM} suggests that zero-forms $C$ take values in the space dual to the space of one-forms $\omega$, see \cite{Krasnov:2021nsq,Skvortsov:2022syz,Sharapov:2022faa,Sharapov:2022awp}. Therefore, we define the nondegenerate pairing
\begin{align}
    \langle \omega|C \rangle&=-\langle C|\omega \rangle= \exp[p_{12}]\,\omega(y_1)\,C(y_2) \big|_{y_i=0}
\end{align}
between the $\hs$-bimodule of fields $C$ and the higher spin algebra $\hs$ of fields $\omega$. With the help of this pairing we can define the bimodule structure by the following symbols:
\begin{equation}
    \begin{split}
        &\mathcal{U}_1(\omega,C)=+\exp{[\hhbar\, p_{01}+ p_{02}+p_{12}]}\, \omega({y}_1)\, C({y}_2)\Big|_{{y}_i=0}\,,\\
        &\mathcal{U}_2(C,\omega)=-\exp{[p_{01}-\hhbar\, p_{02}-p_{12}]}\, C({y}_1)\, \omega({y}_2)\Big|_{{y}_i=0}\,.
    \end{split}
\end{equation}
Consider, for example, the left action. At $\lambda=0$ the symbol corresponds to $\mathcal{U}_1(\omega,C)(y)=\omega(\pl_y) C(y)$, i.e., the commutative algebra $A_0=\mathbb{C}[y^A]$ acts on the dual space by differential operators.\footnote{For $\lambda=1$ one can recognize the twisted-adjoint action \cite{Vasiliev:1999ba}. The twisted-adjoint representation, however, does not admit the flat limit. It is also not very useful for Chiral Theory with cosmological constant: the zero-form should be treated differently, whereas the twisted-adjoint interpretation suggests to deal with $C$ as an element of $\hs$ and this immediately entails some problems with locality. } 

It is worth noting that the bilinear structure maps defined so far satisfy the boundary conditions imposed by the free limit \eqref{linearizeddata3}. The next vertex will generate the trilinear term in \eqref{linearizeddata3}, thereby, we do reproduce the $L_\infty$-algebra determined by the free action \eqref{actionSDYMchap}.

\paragraph{$\boldsymbol{\mathcal{V}(\omega,\omega,C)}$.} Since the $A_\infty$-algebra is concentrated in only two  degrees, $\mathbb{A}_0\oplus \mathbb{A}_{-1}$, there are three structure maps hidden in $\mathcal{V}(\omega,\omega,C)$:
\begin{align}
    \mathcal{V}(\omega,\omega,C)=\mathcal{V}_1(\omega,\omega,C)+\mathcal{V}_2(\omega,C,\omega)+\mathcal{V}_3(C,\omega,\omega)    \,.
\end{align}
For example, one of the $L_\infty$-relations reads
\begin{align}\label{Linfrel}
        &\mathcal{V}_1(\mathcal{V}(\omega,\omega),\omega,C)-\mathcal{V}(\omega,\mathcal{V}_1(\omega,\omega,C))+\mathcal{V}_1(\omega,\omega,\mathcal{U}_1(\omega,C))-\mathcal{V}_1(\omega,\mathcal{V}(\omega,\omega),C)=0\,.
\end{align}
It originates from the $A_\infty$-relation 
\begin{align}
        &\mathcal{V}_1(\mathcal{V}(a,b),c,u)-\mathcal{V}(a,\mathcal{V}_1(b,c,u))+\mathcal{V}_1(a,b,\mathcal{U}_1(c,u))-\mathcal{V}_1(a,\mathcal{V}(b,c),u)=0\,,
\end{align}
where $a,b,c\in \mathbb{A}_{-1}$ and $u\in \mathbb{A}_0$. It is, of course, much more constraining than the one of $L_\infty$. Indeed, in \eqref{Linfrel} the $\omega$'s, being one-forms, anti-symmetrize over the first three arguments. To solve the $A_\infty$-relation, it is useful to rewrite it in terms of symbols: 
\begin{align*}
0&=-\mathcal{V}_1(p_0+\hhbar\, p_1,p_2,p_3,p_4)e^{p_{01}}+\mathcal{V}_1(p_0,p_1+p_2,p_3,p_4)e^{\hhbar\, p_{12}}\\&\qquad -\mathcal{V}_1(p_0,p_1,p_2+p_3,p_4)e^{\hhbar\, p_{23}}+\mathcal{V}_1(p_0,p_1,p_2,\hhbar\, p_3+p_4)e^{p_{34}}
\end{align*}
and similarly for the rest of the $A_\infty$-relations, some of which mix $\mathcal{V}$ with different orderings of the arguments. The resulting equations are not difficult to solve directly \cite{Skvortsov:2022syz}:
\begin{align*}
     \mathcal{V}_1(\omega,\omega,C)&=+p_{12}\, \int_{\Delta_2}\exp[\left(1-u\right) p_{01}+\left(1-v\right) p_{02}+u p_{13}+v p_{23} +\hhbar (1+u-v) p_{12} ]\,, \\
     \mathcal{V}_2(\omega,C,\omega)&=-p_{13}\, \int_{\Delta_2}\exp[\left(1-v\right) p_{01}+\left(1-u\right) p_{03}+v p_{12}-u p_{23}+\hhbar (1-u-v) p_{13}]\\
       &\phantom{=}\,-p_{13}\, \int_{\Delta_2}\exp[\left(1-u\right) p_{01}+\left(1-v\right) p_{03}+u p_{12}-v p_{23}+\hhbar (1-u-v) p_{13}]\,,
    \\
    \mathcal{V}_3(C,\omega,\omega)&=+p_{23}\, \int_{\Delta_2}\exp[\left(1-v\right) p_{02}+\left(1-u\right) p_{03}-v p_{12}-u p_{13}+\hhbar (1+u-v) p_{23} ]\,.
\end{align*}
Here $\Delta_2$ denotes the $2$-simplex $0\leq u\leq v \leq 1$. From the homological perturbation theory point of view, these vertices correspond to\footnote{We refer to Appendix \ref{app:homo} for  basic definitions and to \cite{Sharapov:2022faa,Sharapov:2022awp} for more details on how homological perturbation theory works. }
$$
   \mathcal{V}_1(\omega,\omega,C)=\omega(y) \star h[ \omega(y) \star \Lambda[C] ]|_{z=0}= \begin{tikzcd}[column sep=small,row sep=small]
   & {}& \\
    & \mu\arrow[u]  & \\
    \omega\arrow[ur]  & & \mu\arrow[ul, "h" ']   & \\
    & \omega \arrow[ur]& &\Lambda[C]\arrow[ul]
\end{tikzcd}
$$
its mirror image
$$
   \mathcal{V}_3(C,\omega,\omega)= h[\Lambda[C] \star  \omega(y)  ] \star \omega(y)|_{z=0} = \begin{tikzcd}[column sep=small,row sep=small]
   && {} \\
    && \mu\arrow[u]   \\
    &\mu\arrow[ur, "h"]  & & \omega \arrow[ul]    \\
    \Lambda[C] \arrow[ur]& &\omega\arrow[ul] &&
\end{tikzcd}
$$
and the middle vertex receives contributions from two graphs
\begin{align*}
    \mathcal{V}_2(\omega,C,\omega)&=\omega(y) \star h[  \Lambda[C] \star\omega(y) ]|_{z=0}+ h[  \omega(y)\star  \Lambda[C]] \star \omega(y)|_{z=0}=
\end{align*}
$$
=\begin{tikzcd}[column sep=small,row sep=small]
   & {}& \\
    & \mu\arrow[u]  & \\
    \omega\arrow[ur]  & & \mu\arrow[ul, "h" ']   & \\
    & \Lambda[C] \arrow[ur]& &\omega \arrow[ul]
\end{tikzcd} \oplus\begin{tikzcd}[column sep=small,row sep=small]
   && {} \\
    && \mu\arrow[u]   \\
    &\mu\arrow[ur, "h"]  & & \omega \arrow[ul]    \\
    \omega \arrow[ur]& &\Lambda[C]\arrow[ul] &&
\end{tikzcd}   
$$
Let us illustrate the process of evaluation of a tree on the example of $\mathcal{V}(a,b,c)$ with $a,b\in \mathbb{A}_{-1}$ and $c\in \mathbb{A}_0$, see also \cite{Sharapov:2022faa}. One begins with (here $\varkappa=\exp{(z^Ay_A)}$)
\begin{align}\label{LambdaGuy}
    \Lambda[c]&= dz^{A}z_{A} \int_0^1 t\,dt\, \varkappa(t z, y+p_3)\, c(y_3)\,.
\end{align}
Next, we evaluate the star-product:
\begin{align*}
    b(y)\star \Lambda[c]&= dz^{A}(z_{A}+p^2_{A})\, e^{y p_2} \int_0^1 t\,dt\, \varkappa(t z+t p_2, y+p_3+\hhbar\, p_2)\, b(y_2) c(y_3)\,.
\end{align*}
This is a one-form and we apply $h$ to it:
\begin{align*}
    h[b(y)\star \Lambda[c]]&= (z\cdot p_2)\, e^{y p_2}\int_0^1 dt'\, t\,dt\, \varkappa(tt' z+t p_2, y+p_3+\hhbar\, p_2)\, b(y_2) c(y_3)\,.
\end{align*}
In the last step we evaluate one more product and set $z=0$ to find
\begin{align*}
   a\star h[b\star \Lambda[c]]|_{z=0}&= p_{12}\, e^{y p_1+y p_2}\int_0^1 dt'\, t\,dt\, \varkappa(tt' p_1+t p_2, y+p_3+\hhbar\, p_1+\hhbar\, p_2)\, a(y_1) b(y_2) c(y_3).
\end{align*}
After renaming $y\rightarrow p_0$ and changing the integration domain to the $2d$ simplex $\Delta_2$, $u=tt'$, $v=t$, we arrive at
\begin{align*}
   \mathcal{V}_1(a,b,c)&= p_{12}\, e^{p_{01}+p_{02}} \int_{\Delta_2} \varkappa(u p_1+v p_2, p_0+p_3+\hhbar\, p_1+\hhbar\, p_2)\, a(y_1) b(y_2) c(y_3)\big|_{y_i=0}\,.
\end{align*}
This coincides with $\mathcal{V}_1$ on the previous page. We will derive the result for an arbitrary tree later on.

\paragraph{$\boldsymbol{\mathcal{U}(\omega,C,C)}$.} The next group of structure maps is
\begin{align}
    \mathcal{U}(\omega,C,C)=\mathcal{U}_1(\omega,C,C)+\mathcal{U}_2(C,\omega,C)+\mathcal{U}_3(C,C,\omega)   \,.
\end{align}
The $A_\infty$-relations can also be written down and solved directly \cite{Skvortsov:2022syz,Sharapov:2022awp}. It is remarkable that one does not have to do that. There is a canonical way to generate all $\mathcal{U}$-vertices from $\mathcal{V}$-vertices. We refer to this recipe as a duality map since it relies on the fact that $\mathbb{A}_0 =(\mathbb{A}_{-1})^*$, as $\hs$-bimodules. This is a particular manifestation of a (hidden) cyclicity of the underlying $A_\infty$-algebra $\hat{\mathbb{A}}$; we discuss it in Appendix  \ref{CY}.  Given a $\mathcal{V}$-vertex at some order, one can canonically pair it with $C$ to build a scalar. By cyclicity/duality,
\begin{equation}\label{DR}
    \langle \mathcal{V}(\omega,\omega,C,\ldots,C)|C\rangle=\langle \omega|\mathcal{U}(\omega, C,\ldots,C)\rangle\,.
\end{equation}
A consistent $\mathcal{U}$-vertex can be obtained by peeling off one $\omega$, which is again a canonical operation. 

The duality map gives automatically local $\mathcal{U}$-vertices, provided that the $\mathcal{V}$-vertices are local.\footnote{There is another canonical recipe \cite{Vasiliev:1988sa} in case $\mathbb{A}_0\simeq \mathbb{A}_{-1}$. However, this one gives nonlocal $\mathcal{U}$'s out of local $\mathcal{V}$'s. This recipe is built-in into \cite{Vasiliev:1990cm} and leads to one of the open problems pointed out in  \cite{Boulanger:2015ova}.} In the simplest case we find
\besubeqs
\begin{align}
    \mathcal{U}_1(p_0,p_1,p_2,p_3)&=+ \mathcal{V}_1(-p_3,p_0,p_1,p_2)\,,\\
    \mathcal{U}_2(p_0,p_1,p_2,p_3)&=- \mathcal{V}_2(-p_1,p_2,p_3,p_0)\,,\\
    \mathcal{U}_3(p_0,p_1,p_2,p_3)&= -\mathcal{V}_3(-p_1,p_2,p_3,p_0)\,.
\end{align}
\esubeqs
For example, the first one reads
\begin{align}\label{uone}
    \mathcal{U}_1&=p_{01} \int_{\Delta_2}\exp \left[\hhbar\left(1+u-v\right)   p_{01}+u p_{02}+\left(1-u\right) p_{03}+v p_{12}+\left(1-v\right) p_{13}\right]\,.
\end{align}
It is a local vertex because no $p_{23}$ enters the exponent. For completeness, the other two read
\begin{align*}
    \mathcal{U}_2&=-p_{02} \int_{\Delta_2}\exp \left[\left(1-v\right) p_{01}+v p_{03}-\left(1-u\right) p_{12}+u p_{23}-\hhbar \left(1-u-v\right)  p_{02}\right]\\
    &\phantom{=}-p_{02} \int_{\Delta_2}\exp \left[\left(1-u\right) p_{01}+u p_{03}-\left(1-v\right) p_{12}+v p_{23}-\hhbar\left(1-u-v\right)  p_{02}\right]\,,\\
    \mathcal{U}_3&= +p_{03} \int_{\Delta_2}\exp \left[\left(1-u\right) p_{01}+u p_{02}-\left(1-v\right) p_{13}-v p_{23}-\hhbar\left(1+u-v\right)   p_{03}\right]\,.
\end{align*}

\section{Quartic vertices}
\label{sec:qua}

\paragraph{$\boldsymbol{\mathcal{V}(\omega,\omega,C,C)}$.} The brute-force approach above is less efficient starting from this vertex. First of all, there are $6$ different orderings for $\omega^2C^2$. Secondly, the defining equations ($A_\infty$-algebra relations) are inhomogeneous w.r.t. the sought-for quartic vertices. A complete all-order solution follows immediately from homological perturbation theory \cite{Sharapov:2022faa,Sharapov:2022awp}.\footnote{The light-cone approach operates only with the physical degrees of freedom and, for this reason, may allow one to see certain structures that are not self-evident in a given covariant approach, see e.g. \cite{Metsaev:1991mt,Metsaev:1991nb,Ponomarev:2016lrm,Skvortsov:2018uru}. It was shown in \cite{Metsaev:1991mt,Metsaev:1991nb,Metsaev:2018xip} that the cubic vertices can be split into chiral and anti-chiral ones. The cubic vertices from the Lagrangian point of view have overlap with a great deal of the vertices, $\mathcal{V}(\omega,\omega)$, $\mathcal{U}(\omega,C)$, $\mathcal{V}(\omega,\omega,C)$, $\mathcal{U}(\omega,C,C)$ and even $\mathcal{V}(\omega,\omega,C,C)$. Indeed, we should be looking at all vertices that have any number of background insertions $\omega_0$ and are bilinear in the fluctuations. For example, $\mathcal{V}(\omega_0,\omega_0,C,C)$ is a kind of stress-tensor's contributions. For all these vertices, it should be possible to find a split into chiral and anti-chiral ones plus, possibly, other contributions that come from higher orders in the Lagrangian. Therefore, we expect that various truncations/subsectors like chiral/self-dual/holomorphic are closely related to each other, if not identical at these orders. In this regard it is worth mentioning some partial low order results in the literature \cite{Didenko:2018fgx,Didenko:2019xzz,Didenko:2020bxd,Gelfond:2021two}. } The vertices at this order read
{\allowdisplaybreaks
\begin{align*}
    \mathcal{V}_1(\omega,\omega,C,C)&=(p_{12})^2\int_{\mathcal{D}_1}\exp((1-u_1-u_2)p_{01}+(1-v_1-v_2)p_{02}+u_1 p_{13}+u_2 p_{14}+v_1 p_{23}+v_2 p_{24}+\\
    &+\lambda p_{12}(1+u_1+u_2-v_1-v_2+u_1 v_2-u_2 v_1))\,,\\
    \mathcal{V}_2(\omega,C,\omega,C)&=-(p_{13})^2\int_{\mathcal{D}_1}\exp(p_{01}(1-u_1-u_2)+(1-v_1-v_2)p_{03}+u_2 p_{12}+u_1 p_{14}-v_2 p_{23}+v_1 p_{34}+\\
    &+\lambda p_{13}(1+u_1-u_2-v_1-v_2-u_1 v_2+u_2 v_1))\\
    &-(p_{13})^2\int_{\mathcal{D}_1}\exp(p_{01}(1-u_1-u_2)+(1-v_1-v_2)p_{03}+u_1 p_{12}+u_2 p_{14}-v_1 p_{23}+v_2 p_{34}+\\
    &+\lambda p_{13}(1-u_1+u_2-v_1-v_2+u_1 v_2-u_2 v_1)) + \\
    &-(p_{13})^2\int_{\mathcal{D}_2}\exp((1-u^R-v^L)p_{01}+(1-u^L-v^R)p_{03}+v^L p_{12}+u^R p_{14}-u^L p_{23}+v^R p_{34}\\
    &+\lambda p_{13}(1-u^L+u^R-v^L-v^R-u^L u^R+v^L v^R)) \,,\\
    \mathcal{V}_3(\omega,C,C,\omega)&=(p_{14})^2\int_{\mathcal{D}_1}\exp((1-u_1-u_2)p_{01}+(1-v_1-v_2)p_{04}+u_2 p_{12}+u_1 p_{13}-v_2 p_{24}-v_1 p_{34}+\\
    &+\lambda p_{14}(1-u_1-u_2-v_1-v_2-u_1 v_2+u_2 v_1))-\\
    &+(p_{14})^2\int_{\mathcal{D}_1}\exp((1-v_1-v_2)p_{01}+(1-u_1-u_2)p_{04}+v_1 p_{12}+v_2 p_{13}-u_1 p_{24}-u_2 p_{34}+\\
    &+\lambda p_{14}(1-u_1-u_2-v_1-v_2-u_1 v_2+u_2 v_1))-\\
    &+(p_{14})^2\int_{\mathcal{D}_2}\exp((1-u^R-v^L)p_{01}+(1-u^L-v^R)p_{04}+v^L p_{12}+u^R p_{13}-u^L p_{24}-v^R p_{34}\\
    &+\lambda p_{14}(1-u^L-u^R-v^L-v^R-u^L u^R+v^L v^R)) \,,\\
    \mathcal{V}_4(C,\omega,\omega,C)&=(p_{23})^2\int_{\mathcal{D}_2}\exp((1-u^R-v^L)p_{02}+(1-u^L-v^R)p_{03}-v^L p_{12}-u^L p_{13}+u^R p_{24}+v^R p_{34}\\
    &+\lambda p_{23}(1+u^L+u^R-v^L-v^R-u^L u^R+v^L v^R))\,,\\
    \mathcal{V}_5(C,\omega,C,\omega)&=-(p_{24})^2\int_{\mathcal{D}_1}\exp((1-v_1-v_2)p_{02}+(1-u_1-u_2)p_{04}-v_2 p_{12}-u_2 p_{14}+v_1 p_{23}-u_1 p_{34}+\\
    &+\lambda p_{24}(1-u_1+u_2-v_1-v_2+u_1 v_2-u_2 v_1))+\\
    &-(p_{24})^2\int_{\mathcal{D}_1}\exp((1-v_1-v_2)p_{02}+(1-u_1-u_2)p_{04}-v_1 p_{12}-u_1 p_{14}+v_2 p_{23}-u_2 p_{34}+\\
    &+\lambda p_{24}(1+u_1-u_2-v_1-v_2-u_1 v_2+u_2 v_1))+\\
    &-(p_{24})^2\int_{\mathcal{D}_2}\exp((1-u^R-v^L)p_{02}+(1-u^L-v^R)p_{04}-v^L p_{12}-u^L p_{14}+u^R p_{23}-v^R p_{34}\\
    &+\lambda p_{24}(1+u^L-u^R-v^L-v^R-u^L u^R+v^L v^R))\,,\\
    \mathcal{V}_6(C,C,\omega,\omega)&=(p_{34})^2\int_{\mathcal{D}_1}\exp((1-v_1-v_2)p_{03}+(1-u_1-u_2)p_{04}-v_2 p_{13}-u_2 p_{14}-v_1 p_{23}-u_1 p_{24}+\\
    &+\lambda p_{34}(1+u_1+u_2-v_1-v_2+u_1 v_2-u_2 v_1))\,,\\
\end{align*}}\noindent
where we have introduced the integration variables
\begin{align*}
    u_1&\equiv\frac{t_1t_2(1-t_3)t_4}{1-t_1t_2t_3}\,, & v_1&\equiv\frac{t_1(1-t_2t_3)}{1-t_1t_2t_3} \,,\\
    u_2&\equiv\frac{(1-t_1t_2)t_3t_4}{1-t_1t_2t_3}\,, & v_2&\equiv\frac{(1-t_1)t_3}{1-t_1t_2t_3} \,,
\end{align*}
which correspond to the domain of integration $\mathcal{D}_1$ and
\begin{align*}
    u^L&\equiv\frac{t_1t_2(1-t_3)}{1-t_1t_2t_3t_4}\,, & v^L&\equiv\frac{t_1(1-t_2t_3t_4)}{1-t_1t_2t_3t_4}\,,\\
    u^R&\equiv\frac{(1-t_1)t_3t_4}{1-t_1t_2t_3t_4}\,, & v^R&\equiv\frac{t_3(1-t_1t_2t_4)}{1-t_1t_2t_3t_4} 
\end{align*}
for the domain $\mathcal{D}_2$. All times $t_i$ are integrated over $[0,1]$. In terms of $u$'s and $v$'s the domains of integration can be found by inverting the above relations. We start with $\mathcal{D}_1$:
\begin{align*}
    t_1&=\frac{u_2v_1(1-v_1-v_2)+u_1v_2(v_1+v_2)}{u_1v_2+u_2(1-v_1-v_2)} \,, & t_3&=\frac{v_2}{1-v_1}  \,, \\
    t_2&=\frac{u_1v_2}{u_2v_1(1-v_1-v_2)+u_1v_2(v_1+v_2)} \,, & t_4&=u_1+u_2\frac{1-v_1}{v_2} \,.
\end{align*}
The fact that the $t_i$'s take values in the interval $[0,1]$ translates into restrictions on the $u$ and $v$ variables. In Appendix \ref{app:domain}, we prove that these variables belong to a subinterval of $[0,1]$. Some of these restrictions merely confirm this. The other restrictions 
\begin{align*}
    0\leq v_2 \leq 1\,,\qquad 0\leq u_1\leq v_1\leq 1-v_2\,,\qquad \frac{u_1}{v_1}\leq \frac{u_2}{v_2}\leq \frac{1-u_1}{1-v_1}
\end{align*}
define the integration domain as
\begin{align*}
    \int_{\mathcal{D}_1}&\equiv\int_0^1 dv_2 \int_0^{1-v_2} dv_1 \int_0^{v_1} du_1 \int_{\frac{u_1 v_2}{v_1}}^{v_2\frac{1-u_1}{1-v_1}} du_2 \,.
\end{align*}
For $\mathcal{D}_2$ we obtain
\begin{align*}
    t_1&=\frac{u^Lu^R-v^Lv^R+v^L}{1-v^R} \,, & t_3&=\frac{u^Lu^R-v^Lv^R+v^R}{1-v^L} \,, \\
    t_2&=\frac{u^L}{u^Lu^R-v^Lv^R+v^L} \,, & t_4&=\frac{u^R}{u^Lu^R-v^Lv^R+v^R} \,.
\end{align*}
This gives  the restrictions
\begin{align*}
    &0 \leq u^L \leq 1 \,, & &0 \leq u^L \leq v^L \leq 1-u^R \,,\\
    &\frac{u^L}{v^L} \leq \frac{1-v^R}{1-u^R} \,, & &\frac{u^R}{v^R} \leq \frac{1-v^L}{1-u^L}\,,
\end{align*}
which determine the domain of integration to be
\begin{align*}
    \int _{\mathcal{D}_2}&\equiv \int_0^1 du^L \int_0^{1-u^L}du^R \int_{u^L}^{1-u^R}dv^L \int_{u^R\frac{1-u^L}{1-v^L}}^{1-\frac{u^L(1-u^R)}{v^L}} dv^R \,.
\end{align*}
Hence, both $\mathcal{D}_1$ and $\mathcal{D}_2$ are compact and the corresponding integrals converge. In the language of trees emerging from homological perturbation theory, there are only two nontrivial topologies given by
$$
   G_1=\omega \star h[ h[ \omega \star \Lambda[C] ] \star \Lambda[C]]|_{z=0}= \begin{tikzcd}[column sep=small,row sep=small]
   &{}&\\
    & \arrow[u] \mu  & \\
    \omega\arrow[ur]&&  \arrow[ul,"h"']\mu &  \\
    & \arrow[ur,"h"]\mu &&\arrow[ul]\Lambda[C] \\
    \arrow[ur]\omega & & \arrow[ul]\Lambda[C] &
\end{tikzcd}
$$
and
$$
   G_2= h[ \omega \star \Lambda[C] ]\star  h[ \omega \star \Lambda[C]]|_{z=0}= \begin{tikzcd}[column sep=small,row sep=small]
   &&&{}&&&\\
    &&& \arrow[u]\mu  &&& \\
    & \mu\arrow[urr,"h"]& && & \arrow[ull,"h"']\mu &  \\
    \omega\arrow[ur]&& \arrow[ul]\Lambda[C]  & &  \omega\arrow[ur]&&\arrow[ul]\Lambda[C]
\end{tikzcd}
$$
All other graphs can be derived from these by swapping incoming edges at any vertex. Evaluation of all diagrams leads to the quartic vertices above. For Chiral HiSGRA with vanishing cosmological constant all quartic vertices have been written down in \cite{Sharapov:2022faa}.

\paragraph{$\boldsymbol{\mathcal{U}(\omega,C,C,C)}$.} This group of structure maps can effortlessly be obtained via the duality map. For example, 
\begin{align*}
    \mathcal{U}_1(\omega,C,&C,C)(p_0,p_1,p_2,p_3,p_4)=G_1(-p_4,p_0,p_1,p_2,p_3)=\\
    &=(p_{01})^2\int_{\mathcal{D}_1}\exp\big[u_1 p_{02}+u_2 p_{03}+(1-u_1-u_2)p_{04}+v_1 p_{12}+v_2 p_{13}+(1-v_1-v_2)p_{14}\\
    &+\lambda (1+u_1+u_2-v_1-v_2+u_1 v_2-u_2 v_1)p_{01}\big]\,.
\end{align*}
All other $\mathcal{U}$-vertices can be derived in a similar manner. This completes the low order analysis, which can be useful for a number of reasons: to get an idea of how interaction vertices look like; to compute low order holographic correlation functions; to be compared with the all order analysis that follows. The rest of this section is occupied with the evaluation of all trees coming out of the homological perturbation theory. 

\section{Higher order vertices}
\label{high}

\subsection{All vertices with vanishing cosmological constant}
\label{sec:flat}
Thanks to the duality map, it suffices to work out vertices of type ${\mathcal{V}(\omega,C,\ldots,C,\omega,C,\ldots,C)}$, but we will provide a complete description of all non-zero vertices. Given the specific nuts and bolts of  homological perturbation theory it can be shown \cite{Sharapov:2022faa} that only a very limited class of trees makes nonvanishing contributions. They can be described as `trees with two branches'. Either branch has one leaf decorated by an element of $\mathbb{A}_{-1}$ and the other leaves by elements of $\mathbb{A}_0$. Such trees can be depicted as
$$
\begin{tikzcd}[column sep=small,row sep=small]
&            & &                   &                           &       & {} &       &               &\\
&            & &                   &                           &       & \arrow[u]\mu &       &               &\\
&            & &                   &\arrow[urr,"h"]\mu    &       &                           &       &\arrow[ull,"h"']\mu&\\
&            & &\dots\arrow[ur,"h"]    &                           &\arrow[ul]\Lambda[c_{m+1}]      &                           &\dots\arrow[ur,"h"]  &              &\arrow[ul]\Lambda[c_m]\\
&           &\mu\arrow[ur,"h"]      &   & & &\mu\arrow[ur,"h"] & &   &\\
&\mu\arrow[ur,"h"] & & \arrow[ul]\Lambda[c_{m+n-1}]                  &                           &       \mu\arrow[ur,"h"]&  &\arrow[ul]\Lambda[c_{2}]                 &               &\\
a\arrow[ur]&            &\arrow[ul]\Lambda[c_{m+n}]                & &                           b\arrow[ur]&       &       \arrow[ul]\Lambda[c_{1}]& & &                           
\end{tikzcd}
$$
with $c_i \in \mathbb{A}_0$ and $a,b \in \mathbb{A}_{-1}$. As a first step we need to understand what a single branch of arbitrary length looks like, after which we can join two such branches together to obtain a tree. In general, leaves with $c_i$ can be attached at the left or at the right, which results in a variety of trees for a certain choice of the length of the branches. Our approach is to construct trees with all these leaves attached on the right and then find a recipe to derive all permutations from this.  In this section, we are only concerned with integrands and do not care about the domains of integration in terms of the new variable.  We will return to the question of domain in Section \ref{sec:config}. Otherwise, the initial integration variables that emerge from homological perturbation theory, $t_i$, are integrated over $[0,1]$.

A single branch of length $n$ has the form
$$
B_n=h[\dots h[h[a\star \Lambda[c_1]]\star  \Lambda[c_2]]\star\dots\star \Lambda[c_n]]=\hspace{-2cm}\begin{tikzcd}[column sep=small,row sep=small]
& & & & {} & &\\
& & & & \arrow[u]\mu & &\\
& & & \arrow[ur,"h"]\mu && \arrow[ul]\Lambda[c_n] &\\
& & \arrow[ur,"h"] \dots && \arrow[ul]\Lambda[c_{n-1}] & &\\
& \arrow[ur,"h"]\mu && \arrow[ul]\Lambda[c_2] & & &\\
\arrow[ur]a && \arrow[ul]\Lambda[c_1] & & & &
\end{tikzcd}
$$
The low-order considerations  suggest that such a branch is evaluated as
\begin{align} \label{ansatz}
    B_n=&(zp_1)^n\int \exp\Big[(1-V_n)(yp_1)+U_n(zy)+\sum_{i=1}^{n}u_{n,i}(zp_{i+1})+\sum_{i=1}^{n}v_{n,i}p_{1,i+1}\Big]\,,
\end{align}
where 
\begin{align*}
    U_n&=\sum_{i=1}^{n}u_{n,i} \,, &    V_n&=\sum_{i=1}^{n}v_{n,i}\,,
\end{align*}
and $u_{n,i}$, $v_{n,i}$ are the integration variables with $i=1,\dots,n$. To verify this ansatz, we attach another leaf decorated by $c_{n+1}$ to the right of the branch, creating a branch of length $n+1$, which then reads
\begin{align} \label{B_{n+1}}
    \begin{aligned}
    B_{n+1}&=\frac{t_{2n+1} t_{2n+2}^n(1-t_{2n+1})^n(1-V_n)}{(1-t_{2n+1}U_n)^{n+3}}(zp_1)^{n+1}\times\\
    &\times\int\exp\Big[\frac{(1-t_{2n+1})(1-V_n)}{1-t_{2n+1}U_n}(yp_1)+\frac{(1-t_{2n+1})U_n+t_{2n+1}(1-U_n)}{1-t_{2n+1}U_n}t_{2n+2}(zy)+\\
    &+\frac{(1-t_{2n+2})t_{2n+2}}{1-t_{2n+1}U_n}\sum_{i=1}^nu_{n,i} (zp_{i+1})+\frac{1-U_n}{1-t_{2n+1}U_n}t_{2n+1}t_{2n+2}(zp_{n+2})+\\
    &+\sum_{i=1}^n (v_{n,i}-u_{n,i}\frac{t_{2n+1}(1-V_n)}{1-t_{2n+1}U_n})p_{1,i+1}+\frac{1-V_n}{1-t_{2n+1}U_n}t_{2n+1}p_{1,n+2}\Big]\,.
    \end{aligned}
\end{align}
One can bring this into much simpler form
\begin{align} \label{B_{n+1}newcoordinates}
    B_{n+1}&=(zp_1)^{n+1}\int\exp\Big[(1-V_{n+1})(yp_1)+U_{n+1}(zy)+\sum_{i=1}^{n+1}u_{n+1,i}(zp_{i+1})+\sum_{i=1}^{n+1}v_{n+1,i}p_{1,i+1}\Big]\,,
\end{align}
where the new integration variables are given by the recurrence relations
\begin{align} \label{newcoordinates}
    \begin{aligned}
    u_{n+1,i}&\equiv\frac{(1-t_{2n+1})t_{2n+2}}{1-t_{2n+1}U_n}u_{n,i} \,, & i=0,1,\dots,n \,, \\
    u_{n+1,n+1}&\equiv\frac{1-U_n}{1-t_{2n+1}U_n}t_{2n+1}t_{2n+2} \,, \\
    v_{n+1,i}&\equiv v_{n,i}-u_{n,i}\frac{t_{2n+1}(1-V_n)}{1-t_{2n+1}U_n} \,, & i=0,1,\dots,n \,, \\
    v_{n+1,n+1}&\equiv\frac{t_{2n+1}(1-V_n)}{1-t_{2n+1}U_n}  \,,
    \end{aligned}
\end{align}
where we have to set $U_0=V_0=0$ to match our initial values
\begin{align*}
    u_{1,1}&=t_1t_2 & v_{1,1}&=t_1 \,.
\end{align*}
All the $t_i$'s run from $0$ to $1$. In Appendix \ref{app:allorders}, we prove that the Jacobian associated with the change of variables from $\{u_{n,1},v_{n,1},\dots,u_{n,n},v_{n,n},t_{2n+1},t_{2n+2}\}$ to $\{u_{n+1,1},v_{n+1,1},\dots, u_{n+1,n+1},v_{n+1,n+1}\}$ is exactly the prefactor in \eqref{B_{n+1}}. Since \eqref{B_{n+1}newcoordinates} fits the ansatz \eqref{ansatz}, we conclude that the ansatz is correct for all branches. We also make the observation that the variables satisfy the remarkable chain of inequalities,
\begin{align*}
    \frac{u_{n,1}}{v_{n,1}} \leq \frac{u_{n,2}}{v_{n,2}} \leq \dots \leq \frac{u_{n,n}}{v_{n,n}} \leq \frac{1-U_{n}}{1-V_{n}} \,,
\end{align*}
which is proven in Appendix \ref{app:fulldomain}. This pattern allows one to easily retrieve the domain of integration associated to this choice of variables.

We can now compute a tree by evaluating the star-product of two branches with length $n$ and $m$. In order to obtain the most symmetric form, assume that the left branch contains only zero-forms attached to the left and we denote this branch by $\bar{B}_n$. This does not limit the generality: for $\lambda=0$ attaching zero-forms to the left or right gives the same result since the product is commutative. An important remark is that notation eventually becomes very cumbersome if we want the labels on $p_{ij}$ to consistently refer to the position of the elements $a,b \in \mathbb{A}_0$ and $c_1,\dots,c_n \in \mathbb{A}_{-1}$, read from left to right. Therefore, it is convenient to always assign $p_1,p_2$ and $a(y_1), b(y_2)$ to the first and second one-form, respectively, and assign $p_3,\dots,p_{n+2}$ and $c_1(y_3),\dots,c_{n}(y_{n+2})$ to elements of $\mathbb{A}_0$ based on the position on the branches that they originated from, starting from the bottom of the right branch, to the top and then from the top of the left branch to the bottom. We then leave the reshuffling of the labels according to the respective positions as seen in the tree as the last step in the recipe of finding vertices. Vertices should be $z$-independent, so we set $z=0$ at the end. This gives
\begin{align} \label{tree}
    \begin{aligned}
    \bar{B}_n\star B_m|_{z=0}&=\frac{(-1)^n(1-V^R_m)^n(1-V_n^L)^m}{(1-U_n^LU^R_m)^{n+m+2}}p_{12}^{n+m}\times\\
    &\times\int\exp\Big[\frac{(1-U^R_m)(1-V_n^L)}{1-U_n^LU^R_m}p_{01}+\frac{(1-V^R_m)(1-U_n^L)}{1-U_n^LU^R_m}p_{02}\\
    &+\frac{1-V_n^L}{1-U_n^LU^R_m}\sum_{i=1}^{m}u_{m,i}^R p_{1,2+i}+\sum_{i=1}^n\big( v_{n,i}^L-u_{n,i}^L\frac{U^R_m(1-V_n^L)}{1-U_n^LU^R_m}\big)p_{1,m+n+3-i}\\
    &+\sum_{i=1}^m\big( v_{m,i}^R-u_{m,i}^R\frac{U_n^L(1-V^R_m)}{1-U_n^LU^R_m}\big) p_{2,2+i} +\frac{1-V^R_m}{1-U_n^LU^R_m}\sum_{i=1}^n u_{n,i}^L p_{2,m+n+3-i}\Big]\,.
    \end{aligned}
\end{align}
Here we distinguish between variables coming from the left and the right branch by the superscripts $L$ and $R$, as both branches have their own set of recurrence relations \eqref{newcoordinates}, in which the $t_i$'s in the left branch run from $t_{m+1}$ to $t_{n+m}$, going from top to bottom. To simplify \eqref{tree} we introduce new variables
\begin{align} \label{newcoordinatestwobranches}
    \begin{aligned}
        r^L_{n,i}&\equiv \frac{1-V^R_m}{1-U_n^LU^R_m}u_{n,i}^L \,, & r^R_{m,i}&\equiv \frac{1-V_n^L}{1-U_n^LU^R_m}u_{m,i}^R \,, \\
        s^L_{n,i}&\equiv v_{n,i}^L-u_{n,i}^L\frac{U^R_m(1-V_n^L)}{1-U_n^LU^R_m} \,, & s^R_{m,i}&\equiv v_{m,i}^R-u_{m,i}^R\frac{U_n^L(1-V^R_m)}{1-U_n^LU^R_m}\,,
    \end{aligned}
\end{align}
which allows us to rewrite \eqref{tree} as
\begin{align} \label{twobranches}
    \begin{aligned}
    \bar{B}_n\star B_m|_{z=0}&=(-1)^n p_{12}^{n+m}\int \exp\Big[\big(1-\sum_{i=1}^n s^L_{n,i}-\sum_{i=1}^m r^R_{m,i}\big)p_{01}+\big(1-\sum_{i=1}^m s^R_{m,i}-\sum_{i=1}^n r^L_{n,i}\big)p_{02}+\\
    &+\sum_{i=1}^m r^R_{m,i}p_{1,2+i}+\sum_{i=1}^n s^L_{n,i}p_{1,m+n+3-i}+\sum_{i=1}^m s^R_{m,i}p_{2,2+i}+\sum_{i=1}^n r^L_{n,i}p_{2,m+n+3-i}\Big]\,.
    \end{aligned}
\end{align}
In Appendix \ref{app:allorders}, we show that the Jacobian resulting from the change of variables from the coordinates $\{u^L_{n,1},\dots,v^L_{n,n},u^R_{m,1},\dots,v^R_{m,m}\}$ to $\{r^L_{n,1},\dots,s^L_{n,n},r^R_{m,1},\dots,s^R_{m,m}\}$ is exactly the prefactor in \eqref{tree}.

In order to specify a domain of integration in (\ref{twobranches}), we rename the variables as
\begin{align}\label{rename}
\begin{aligned}
    \{u_1,\dots,u_m,u_{m+1},u_{m+2},\dots,u_{m+n},u_{m+n+1}\} &\\= \{r^R_{m,1},\dots,r^R_{m,m},&1-\sum_{i=1}^n s^L_{n,i}-\sum_{i=1}^{m} r^R_{m,i},s^L_{n,n},\dots,s^L_{n,1}\} \,, \\
    \{v_1,\dots,v_{m},v_{m+1},v_{m+2},\dots,v_{m+n},v_{m+n+1}\}&\\ = \{s^R_{m,1},\dots,s^R_{m,m}, &1-\sum_{i=1}^n r^L_{n,i}-\sum_{i=1}^m s^R_{m,i},r^L_{n,n},\dots,r^L_{n,1}\} \,,
    \end{aligned}
\end{align}
where $u_{m+n+1}=1-\sum_{i=1}^{m+n}u_i$ and $v_{m+n+1}=1-\sum_{i=1}^{m+n}v_i$. In Appendix \ref{app:fulldomain}, we prove that these variables satisfy the inequalities 
\begin{align}\label{inequalities}
    \frac{u_1}{v_1} \leq \frac{u_2}{v_2} \leq \dots \leq \frac{u_{m+n}}{v_{m+n}} \leq \frac{u_{m+n+1}}{v_{m+n+1}} \,.
\end{align}
Now \eqref{twobranches} takes the form
\begin{align} \label{twobranches2}
    \begin{aligned}
    \bar{B}_n\star B_m|_{z=0}&=(-1)^n (p_{12})^{n+m} \int \exp\Big[ u_{m+1} p_{01} + v_{m+1} p_{02} + (1-U_{m+n}) p_{1,m+n} + \\
    &+ (1-V_{m+n}) p_{2,m+n} +\sum_{i=1}^m u_i p_{1,2+i} + \sum_{i=1}^m v_i p_{2,2+i} + \sum_{i=1}^{n-1} u_{m+1+i} p_{1,m+2+i} +\\
    &+ \sum_{i=1}^{n-1} v_{m+1+i} p_{2,m+2+i} \Big]\,.
    \end{aligned}
\end{align}

\paragraph{Constructing vertices.}

There are still a few differences between the trees that we have constructed above and the vertices that solve for the $A_\infty$-relations. Above we associated $p_1$ and $p_2$ with the two leaves decorated by elements of $\mathbb{A}_{-1}$ and the other $p_i$ acted on the $c_i$ that were labeled from bottom right to bottom left on the branches. However, in the expressions for vertices the $p_i$'s are assigned from left to right. Moreover, we have only considered trees with elements of $\mathbb{A}_0$ attached to the left(right) on the left(right) branch. Obviously, general vertices are not restricted to this choice. As will become clear in the next section, the only change as compared to \eqref{twobranches2} is by relabeling of the $p_{ij}$'s when elements of $\mathbb{A}_0$ are attached differently in the absence of a cosmological term. 

\begin{figure}[!ht]
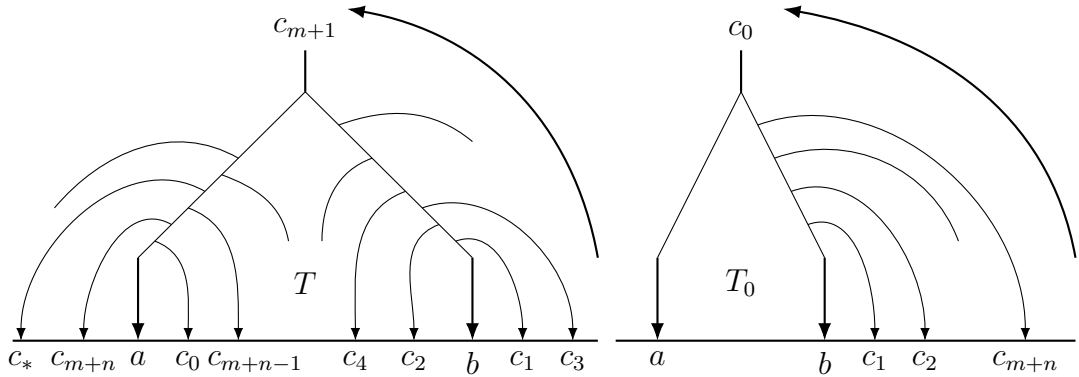

\centering

\caption{\label{fig:trees} A generic tree $T$ in the left panel with elements of $\mathbb{A}_0$ attached left and right arbitrarily and the `base' tree $T_0$ in the right panel with only elements of $\mathbb{A}_0$ attached to the right on the right branch. $T$ can be obtained form $T_0$ through flipping $c_i$'s to the left of the right branch and/or shifting them to the left branch.}
\end{figure}

To simplify the procedure of obtaining expressions for trees, let us consider the trees in Fig.\ref{fig:trees}. We assign vectors $\vec a_i=(u_i,v_i)$, $\vec r_i  =  (p_{1,i},p_{2,i})$ to $c_i$ and $\vec r_0 = (p_{01},p_{02})$, $\vec a_0 = (1-\sum_{i=1}^{m+n}u_i, 1-\sum_{i=1}^{m+n} v_i)$ to $c_0$. We also introduce the matrices 
\begin{align}
    P&=(\vec 0,\vec 0,\vec r_1,\dots,\vec r_{m+n}, \vec r_0)\,, &
    Q&= (-\vec e_1,-\vec e_2,\vec a_1,\dots \vec a_{m+n},\vec a_0)\,,
\end{align}
where $\vec e_1 = \begin{pmatrix}
    1\\
    0
\end{pmatrix}$, $\vec e_2 = \begin{pmatrix}
    0\\
    1
\end{pmatrix}$. The expression for the symbol associated with  the basic tree $T_0$ can now be written as
\begin{align*}
    B_0 \star B_{m+n}|_{z=0} = (p_{12})^{m+n}\int_{\mathbb{V}_{m+n}}\exp(\text{tr}[PQ^t]);
\end{align*}
it is understood to yield the vertex $V(\omega,\omega,C,\dots,C)$ when acting on
\begin{align} \label{abc1}
    a(y_1)b(y_2)c_1(y_3)\dots c_{m+n}(y_{m+n+2})|_{y_i}=0 \,.
\end{align}
The configuration space $\mathbb{V}_{m+n}$ is given by the chain of inequalities in \eqref{inequalities}.
A generic tree $T$ can be obtained from $T_0$ through two types of operations: (i) flipping $c_i$ to the left of the right branch and (ii) a counterclockwise shift of all $c_i$'s along the cord connecting $a$ and $b$. Importantly, in the latter case $c_0$ also moves along the cord, while another $c_i$ takes its place. To express the symbol corresponding to $T$ we define $P_T = (\vec 0,\vec 0, \vec r_1,\dots\vec{r}_m,-\vec r_{m+1},\vec{r}_{m+2},\dots,\vec r_n, -\vec r_0)$. For vertices, the labels on $p_i$ and the corresponding arguments $y_i$ of $a$, $b$ and the $c_j$'s are read off from the tree from left to right. Since we have labeled them from bottom right to top left, we require a permutation $\sigma_T$ that relabels the $p_i$'s and $y_i$'s accordingly. Moreover, $\sigma_T$ also shuffles the elements in \eqref{abc1} corresponding to their respective position in the tree $T$. In the absence of a cosmological constant, a generic tree $T$ contributes  to the vertex $\mathcal{V}(C,\dots,C,\omega,C,\dots,C,\omega,C,\dots,C)$ by
\begin{align*}
    s_T\sigma_T (p_{12})^{m+n}\int_{\mathbb{V}_{m+n}} \exp(\text{tr}[P_T Q^t]) a(y_1)b(y_2)c_1(y_3)\dots c_{y_{m+n}}(y_{m+n+2})|_{y_i=0} \,.
\end{align*}
Here, $s_T=(-1)^k$ and $k$ is the number of zero-forms $C$ in between the two $\omega$'s. The sign $\sigma_T$ is the combination of the sign factor we get by evaluating the product of two branches with a sign coming from homological perturbation theory.

\subsection{All vertices with cosmological constant}
\label{sec:ads}
All the main properties discussed in the previous section remain true if we turn on the cosmological constant, which is a smooth deformation of Chiral Theory in flat space. Most importantly, the deformation maintains locality. In particular, we have to evaluate exactly the same graphs as before. It will turn out, as the low order examples illustrate, that switching on the cosmological constant adds one term to the exponent, e.g. $\hhbar p_{12} (\ldots)$ for $\mathcal{V}(\omega,\omega,C,\ldots,C)$ vertices. More specifically, a single branch takes the form
\begin{align}\label{onecosmologicalbranch}
    \begin{aligned}
    B_n&=(zp_1)^n\int\exp\Big[(1-V_n)yp_1+U_n(zy)+\lambda (zp_1)(U_n+\sum_{i,j=1}^n\text{sign}(j-i)u_{n,i}v_{n,j})+\\
    &+\sum_{i=1}^n u_{n,i}(zp_{i+1})+\sum_{i=1}^n v_{n,i}p_{1,i+1}\Big]\,.
    \end{aligned}
\end{align}
In the presence of the cosmological constant the construction of general vertices from the trees is more complicated than on the flat background. For example, attaching a leaf decorated by $c_{n+1}$ to the left of a branch of length $n$ yields
\begin{align*}
    \begin{aligned}
    h[\Lambda[c_{n+1}],B_n]&=(zp_1)^{n+1}\int\exp\Big[(1-V_{n+1})yp_1+U_{n+1} (zy)+\sum_{i=1}^{n+1} u_{n+1,i}(zp_{i+1})+\sum_{i=1}^{n+1} v_{n+1,i}p_{1,i+1}+\\
    &+\lambda (zp_1)(\sum_{i=1}^n u_{n+1,i}-u_{n+1,n+1}+\sum_{i,j=1}^n \text{sign}(j-i)u_{n+1,i}v_{n+1,j}-\sum_{i=1}^n u_{n+1,i}v_{n+1,n+1}+\\
    &+\sum_{j=1}^n u_{n+1,n+1}v_{n+1,j})\Big] \,,
    \end{aligned}
\end{align*}
i.e., the variables $u_{n+1,n+1}$ and $v_{n+1,n+1}$ enter the cosmological term with a minus sign as opposed to when the leaf is attached to the right. Otherwise, the expression remains the same. This coincides with the statement that the ordering of the leaves is irrelevant in the absence of the cosmological constant. Since the presence of the cosmological constant does not modify the piece of the expression we found in \eqref{twobranches2}, we will only consider the cosmological term for the following discussion. For a branch with leaves attached to the left and right arbitrarily, the cosmological term reads
\begin{align}\label{mostgeneralbranch}
\begin{aligned}
    \lambda (\tilde{U}_n+\sum_{i<j}^nu_{n,i}\tilde{v}_{n,j}-\sum_{j<i}^n\tilde{u}_{n,i}v_{n,j})(zp_1)\,,
\end{aligned}
\end{align}
where  $\tilde{x}\equiv \sigma_i x$,  $i$ corresponds to the label of the element of $\mathbb{A}_0$, and
\begin{align*}
    \sigma_{i}&\equiv
    \begin{cases}
    -1 &\text{if $\Lambda[C_{i}]$ is attached to the left (right) in the right (left) branch} \,,\\
    +1 &\text{if $\Lambda[C_i]$ is attached to the right (left) in the left (right) branch} \,.
    \end{cases}\
\end{align*}
The cosmological term `remembers' how the leaves were attached. In terms of the coordinates \eqref{newcoordinatestwobranches}, the cosmological term of a generic tree reads
\begin{align} \label{mostgeneraltree}
    \begin{aligned}
    &\lambda\Big(1+\sum_{i=1}^n \tilde{r}^L_{n,i}+\sum_{i=1}^m \tilde{r}^R_{m,i}-\sum_{i=1}^n s^L_{n,i}-\sum_{i=1}^m s^R_{m,i}-\sum_{i=1}^n r^L_{n,i}\sum_{j=1}^m r^R_{m,j}+\\
    &+\sum_{i=1}^n s^L_{n,i}\sum_{j=1}^m s^R_{m,j}+\sum_{i<j}^m r^R_{m,i}\tilde{s}^R_{m,j}-\sum_{j<i}^m \tilde{r}^R_{m,i}s^R_{m,j}+\sum_{i<j}^n r^L_{n,i}\tilde{s}^L_{n,j}-\sum_{j<i}^n \tilde{r}^L_{n,i}s^L_{n,j}\Big) p_{12}\,.
    \end{aligned}
\end{align}
In order to apply the change of coordinates \eqref{rename}, we need to differentiate between two cases: $n=0$ and $n>0$. In the former case we find the cosmological term to be
\begin{align*}
    \lambda  \Big(1+\tilde{U}_{m}-V_m+\sum_{i=1}^m \text{sign}(j-i)\sigma_{\max\{i,j\}}u_iv_j\Big) p_{12}\,,
\end{align*}
where we introduced
\begin{align*}
    \sigma_{\text{max}\{i,j\}} &= \left\{ \begin{array}{ll}
         &  \sigma_i \,, \quad \text{if } i>j\,, \\
         & \sigma_j \,, \quad \text{if } i<j \,.
    \end{array}\right.
\end{align*}
In case $n > 0$, we obtain
\begin{align*}
    \lambda \Big(&\sigma_{m+n}+\sum_{i=1}^m \sigma_i u_i+u_{m+1}+\sum_{i=1}^{n-1} \sigma_{m+i}u_{m+1+i} - \sigma_{m+n}\sum_{i=1}^{m+n}v_i + \sum_{i,j=1}^m \text{sign}(j-i)\sigma_{max\{i,j\}}u_iv_j+\\
    &+ \sum_{i=1}^m u_i v_{m+1}-\sum_{i=1}^m u_{m+1}v_i +\sum_{i=1}^{n-1} \sigma_{m+i} u_{m+1}v_{m+1+i}-\sum_{i=1}^{n-1}\sigma_{m+i}u_{m+1+i}v_{m+1}+\\
    &+\sum_{i=1}^m\sum_{j=1}^{n-1}\sigma_{m+j}u_iv_{m+1+j}-\sum_{i=1}^{n-1}\sum_{j=1}^m \sigma_{m+i}u_{m+1+i}v_j + \sum_{i,j=1}^{n-1}\text{sign}(j-i)\sigma_{max\{i,j\}}u_{m+1+i}v_{m+1+j}\Big)p_{12} \,.
\end{align*}
At the end of Section \ref{sec:flat}, we expressed the contribution to a vertex in terms of the matrices $P_T$ and $Q$. It turns out that, despite its complicated form, the cosmological term can be expressed in a similar fashion that is consistent with both aforementioned cases. We define a matrix $Q_T$ by filling up its columns, starting with $\vec e_1$, corresponding to $a$ in Fig.\ref{fig:trees} and from there on with $\vec a_i$  following through the tree counterclockwise. As an example, for the tree in the left panel of Fig.\ref{fig:trees} this looks like 
\begin{align}\label{qtmatrix}
    Q_T&=(-\vec e_1,\vec{a}_{m+n+1},\vec a_{m+n-1},\dots,\vec a_4,\vec a_2,-\vec e_2,\vec a_1,\vec a_3,\dots,\vec a_{m+1},\dots,\vec a_{m+n})\,.
\end{align}
The cosmological term for a generic tree is then given by $\lambda p_{12} |Q_T|$, where $|Q_T|$ is the sum of minors of $Q_T$. A generic tree with cosmological constant contributes to a vertex by
\begin{align}\label{compactform}
    s_T\sigma_T (p_{12})^{m+n}\int_{\mathbb{V}_{m+n}} \exp(\text{tr}[P_T Q^t]+\lambda p_{12}|Q_T|) a(y_1)b(y_2)c_1(y_3)\dots c_{y_{m+n}}(y_{m+n+2})|_{y_i=0} \,.
\end{align}
The simplest example is given by a single tree contributing to $\mathcal{V}(\omega,\omega,C,\ldots, C)$, see Eq. \eqref{besttree} below.

\subsection{Duality map and homological perturbation theory}
\label{sec:uvertices}
A very helpful idea put forward in \cite{Sharapov:2022faa,Sharapov:2022awp} is that of a duality map.  This map allows one to automatically generate all $\mathcal{U}$-vertices from $\mathcal{V}$-vertices. Moreover, the duality map manifestly preserves locality. Nevertheless, it is important to check that  homological perturbation theory leads to exactly the same $\mathcal{U}$-vertices as the duality map. Additionally, the duality map also allows one to relate various $\mathcal{V}$-vertices to each other.
\paragraph{$\mathcal{U}$-vertices.}
First of all, as it is shown in Appendix \ref{app:homo}, all the trees that contribute to $\mathcal{U}$-vertices are made up of a single branch (in contrast to the $\mathcal{V}$-vertices that consist of two-branch trees).  According to \eqref{circaction} the differentials $dz^A$ annihilate the module where the zero-forms $C$ take their values, so that  $dz^A\circ C\equiv0$ and the module action $\circ$ can only appear at the very last step. For example, for the vertex  $\mathcal{U}(\omega,C,C)$ with the symbol \eqref{uone} we should have
\begin{align}
    \mathcal{U}(\omega,C,C)= h[\omega \star \Lambda[C]] \circ C\,.
\end{align}
The reason is that any expression that acts on the bare $C$ has to be $dz$-independent to be different from zero and this can only occur at the end of the branch. The symbol of a branch $B_n$ of length $n$ is given by \eqref{onecosmologicalbranch}. There is a subtlety in computing the module action \eqref{moduleaction} for 
\begin{align}\label{BC}
     B_n\circ C&\equiv (B_n\star C^\tau)^\tau \equiv e^{zy}\Big[B_n(y,z) \star e^{zy} C(z)\Big]\Big|_{y\leftrightarrow z}\,,
\end{align}
$\tau$ being the involution defined by \eqref{tauinvol}. 
The point is that the expressions (\ref{BC}) involve star-products of nonpolynomial functions like  $e^{tz\cdot y}$, which, as is well-known, are ill-defined in general. For example, the product
$$
e^{tz\cdot y}\star e^{sz\cdot y}=\frac{e^{\frac{z\cdot y(t+s-2ts)}{1-ts}}}{(1-ts)^2}
$$
features a singularity as $t,s\rightarrow 1$. As a result, the integrals corresponding to the expressions (\ref{BC}) are not absolutely convergent. This, however, does not cause much trouble since $\circ$ appears only in the very last step and can easily be  resolved with any simple regularization. Specifically, we choose the following definition:
\begin{align}\label{regaction}
     B_n\circ C&\equiv (B_n\star C^{\tau_\varepsilon})^{\tau_\varepsilon}\equiv \lim_{\varepsilon\rightarrow +0}e^{(1-\varepsilon)zy}\Big(B_n(y,z) \star e^{(1-\varepsilon)zy} C(z)\Big)\Big|_{y\leftrightarrow z}\,,
\end{align} 
which just modifies the $\tau$-involution \eqref{tauinvol}. Plugging \eqref{onecosmologicalbranch} into \eqref{regaction}, we get after a straightforward calculation
\begin{align} \label{bigbranch}
    \begin{aligned}
    &\left(\frac{1-\varepsilon}{1-U_n(1-\varepsilon)}\right)^2 \left(\frac{\varepsilon}{1-U_n(1-\varepsilon)}\right)^n (p_{01})^n\times\\ &\int\exp\Big[\hhbar \frac{(1-V_n)(1-\varepsilon)+\varepsilon(U_n+\sum_{i,j=1}^n \text{sign}(j-i) u_{n,i}v_{n,j}
    )}{1-U_n(1-\varepsilon)}p_{01} +\\
    &\qquad+\varepsilon\sum_{i=1}^n \frac{u_{n,i}}{1-U_n(1-\varepsilon)}p_{0,i+1}+\frac{1-U_n}{1-U_n(1-\varepsilon)}p_{0,n+2} +\\
    &\qquad\qquad\sum_{i=1}^n(v_{n,i}-u_{n,i}\frac{(1-V_n)(1-\varepsilon)}{1-U_n(1-\varepsilon)})p_{1,i+1} +\frac{1-V_n}{1-U_n(1-\varepsilon)}p_{1,n+2}\Big] \,.
    \end{aligned}
\end{align}
Then, by analogy with the $\mathcal{V}$-vertices, we introduce the new variables
\begin{align}\label{changezeroform}
\begin{aligned}
    T_{n,i} &\equiv u_{n,i}\frac{\varepsilon }{1-U_n(1-\varepsilon)}\,, & S_{n,i}&\equiv v_{n,i}-u_{n,i}\frac{(1-V_n)(1-\varepsilon)}{1-U_n(1-\varepsilon)}\,, \\
    T_{n,n+1} &\equiv \frac{1-U_n}{1-U_n(1-\varepsilon)} \,, & S_{n,n+1} &\equiv \frac{1-V_n}{1-U_n(1-\varepsilon)}\,.
\end{aligned}
\end{align}
Again, the determinant of the Jacobian corresponding to this change of variables (note that we do not integrate $T_{n,n+1}$, $S_{n,n+1}$) cancels the exponential prefactor, see Appendix \ref{app:Jacobians}. In terms of the new coordinates, the symbol \eqref{bigbranch} takes the form
\begin{align} \label{bigbranchnewvar}
    \begin{aligned}
   (p_{01})^n  &\int \exp\Big[\hhbar\big(1+\sum_{i=1}^{n}T_{n,i}-\sum_{i=1}^{n}S_{n,i}+\sum_{i,j=1}^{n}\text{sign}(j-i)T_{n,i}S_{n,j}+\\
   &-\varepsilon S_{n,n+1}\big) p_{01}+\sum_{i=1}^{n+1} T_{n,i} p_{0,1+i}+\sum_{i=1}^{n+1} S_{n,i} p_{1,i+1} \Big] \,.
   \end{aligned}
\end{align}
At the same time, the duality map implies 
\begin{align} \label{V1U1duality}
    \langle \mathcal{V}_1(a,b,c_1,\dots,c_n)|c_{n+1}\rangle=\langle a|\mathcal{U}_1(b,c_1,\dots,c_n,c_{n+1})\rangle\,,
\end{align}
whence $\mathcal{U}_1(p_0,p_1,\dots,p_{n+1})=\mathcal{V}_1(-p_{n+1},p_0,p_1,\dots,p_n)$ with
\begin{align*}
    &\mathcal{V}_1(p_0,p_1,\dots,p_{n+2})=(p_{12})^n \exp\Big[(1-\sum_{i=1}^n u_{n,i})p_{01} +(1-\sum_{i=1}^n v_{n,i})p_{02}+\\
    &\sum_{i=1}^n u_{n,i}p_{1,i+2}+\sum_{i=1}^n v_{n,i}p_{2,i+2} +\hhbar \big (1+\sum_{i=1}^n u_{n,i}-\sum_{i=1}^n v_{n,i}+\sum_{i,j=1}^{n}\text{sign}(j-i)u_{n,i}v_{n,j}\big)p_{12}\Big]\,.
\end{align*}
We thus conclude that \eqref{bigbranchnewvar} approaches $\mathcal{U}_1(p_0,p_1,\dots,p_{n+1})$ as $\varepsilon\rightarrow +0$. Of course, the domain of integration for the $\mathcal{U}$-vertices is correct  and coincides with that for the $\mathcal{V}$-vertices. In the same way we can  evaluate $C \circ B_n$, which gives almost the same expression as before, up to a small change in the cosmological term. The final result reads
\begin{align*}
    \begin{aligned}
    (p_{01})^n  &\int \exp\Big[-\hhbar\big(1-\sum_{i=1}^{n}T_{n,i}-\sum_{i=1}^{n}S_{n,i}-\sum_{i,j=1}^{n}\text{sign}(j-i)T_{n,i}S_{n,j}+\\
   &-\varepsilon(1-\sum_{i=1}^n S_{n,i})\big) p_{01}+\sum_{i=1}^{n+1} T_{n,i} p_{0,1+i}+\sum_{i=1}^{n+1} S_{n,i} p_{1,i+1} \Big] \,.
    \end{aligned}
\end{align*}
In the same way as before, we get rid of the $\varepsilon$-dependent term by setting $\varepsilon=0$. This coincides with the result obtained from the duality map, i.e.
\begin{align*}
    \langle \mathcal{V}(a,c_{n+1},b,c_1,\dots,c_{n-1})|c_{n}\rangle = \langle a|\mathcal{U}(c_{n+1},b,c_1,\dots,c_n)\rangle \,.
\end{align*}
To obtain a generic branch, the other elements of $\mathbb{A}_0$ could also be attached on the left. Here, as before, we have adapted the convention of labeling the $p_{i}$'s from the bottom to the top of the branch and we need to perform a permutation $\sigma_T$ to rearrange them accordingly and shuffle the elements $a(y_1)c_1(y_2),\dots c_n(y_{n+1})|_{y_i=0}$. In the limit when $\varepsilon \rightarrow +0$, a generic contribution to $\mathcal{U}$-vertex with zero-forms attached left and right arbitrarily approaches
\begin{align*}
    \begin{aligned}
    \mathcal{U}(c_{i+1},\dots,c_{n+1},a,c_1,\dots,c_i) &= \sigma_T(p_{01})^n  \int \exp\Big[\hhbar \big(\sigma_{n+1}+\sum_{i=1}^{n}\tilde{T}_{n,i}-\sigma_{n+1}\sum_{i=1}^{n}S_{n,i}+\\
    &+\sum_{i<j}^{n}T_{n,i}\tilde{S}_{n,j}-\sum_{j<i}^n \tilde{T}_{n,i}S_{n,j}\big)p_{01}+\sum_{i=1}^{n+1} T_{n,i} p_{0,1+i}+\\
    &+\sum_{i=1}^{n+1} S_{n,i} p_{1,i+1} \Big]\times a(y_1)c_1(y_2),\dots c_n(y_{n+1})|_{{y_i}=0}\,.
    \end{aligned}
\end{align*}
Naturally, a generic $\mathcal{U}$-vertex is related to a class of $\mathcal{V}$-vertices by
\begin{align}\label{genericUVduality}
    \langle \mathcal{V}(a,c_1,\dots,c_i,b,c_{i+1},\dots,c_n)|c_{n+1}\rangle = \langle a|\mathcal{U}(c_1,\dots,c_i,b,c_{i+1},\dots,c_n)\rangle \,.
\end{align}
\paragraph{$\mathcal{V}$-$\mathcal{V}$-duality.}
The duality map also operates as a map between various $\mathcal{V}$-vertices via
\begin{align*}
    &\langle \mathcal{V}(c_{j+1},\dots,c_n,a,c_1,\dots,c_i,b,c_{i+1}\dots,c_{j-1}|c_j\rangle = \\
    &\qquad\qquad\langle \mathcal{V}(c_{j-k+1},\dots,c_n,a,c_1,\dots,c_i,b,c_{i+1},\dots,c_{j-k-1}|c_{j-k}\rangle  \,,
\end{align*}
where we rotated the arguments by $k$ units and $j \geq i+1$. Through this duality all $\mathcal{V}$-vertices with the same number of elements of $\mathbb{A}_0$ between $a$ and $b$ and the same total number of $\mathbb{A}_0$ elements are related to each other, which vastly reduces the number of vertices to be computed. In particular, it suffices to only determine $\mathcal{V}$-vertices of the type $\mathcal{V}(a,c_1,\dots,c_i,b,c_{i+1},\dots,c_n)$ and one can relate all $\mathcal{V}$-vertices in the class of vertices characterized by $(n,i)$. In hindsight, some hints of this duality were hidden in the expression for $\mathcal{V}$-vertices that we presented in \eqref{compactform}, namely (i) the overall sign is determined by the number of elements of $\mathbb{A}_0$ between $a$ and $b$, which is an invariant within a class, (ii) the matrix $Q_T$ in the cosmological term is constructed similarly for all vertices belonging to the same class and (iii) the configuration space of trees that share the same number of total elements of $\mathbb{A}_0$ is identical.
\paragraph{$\mathcal{U}$-$\mathcal{U}$-dualities.}
A natural generalization of the idea discussed above is to introduce dualities between $\mathcal{U}$-vertices themselves. However, this can only be done if a $\mathcal{U}$-vertex is contracted with an element of $\mathbb{A}_{-1}$ and subsequently the other element of $\mathbb{A}_{-1}$ is stripped off. This leaves only one duality relation for the $\mathcal{U}$-vertices, namely,
\begin{align}\label{uuduality}
    \langle a|\mathcal{U}_{n+2}(c_1,\dots,c_{n+1},b)\rangle &= \langle \mathcal{U}_1(a,c_1,\dots,c_{n+1})|b\rangle = - \langle b|\mathcal{U}_1(a,c_1,\dots,c_{n+1})\rangle \,.
\end{align}
Consistency of this duality can be checked either using the explicit expressions for the relevant vertices or through various dualities. The latter method is particularly easy to implement, as its consistency implies that the following diagram commutes: 
\begin{center}
\begin{tikzpicture}
  \matrix (m)
    [
      matrix of math nodes,
      row sep    = 3em,
      column sep = 4em
    ]
    {
      \mathcal{V}(a,b,c_1,\dots,c_n)              & \mathcal{U}(b,c_1,\dots,c_{n+1}) \\
      \mathcal{V}(c_1,\dots,c_n,a,b) & \mathcal{U}(c_1,\dots,c_{n+1},b)            \\
    };
  \path
    (m-1-1) edge [<->] node [left] {$\mathcal{V}$-$\mathcal{V}$} (m-2-1)
    (m-1-1.east |- m-1-2)
      edge [<->] node [above] {$\mathcal{V}$-$\mathcal{U}$} (m-1-2)
    (m-2-1.east) edge [<->] node [below] {$\mathcal{V}$-$\mathcal{U}$} (m-2-2)
    (m-1-2) edge [<->] node [right] {$\mathcal{U}$-$\mathcal{U}$} (m-2-2);
\end{tikzpicture}
\end{center}
(the type of duality is specified on the arrows).
\paragraph{$\mathbb{Z}_2$-transformation.}
When discussing the duality between $\mathcal{V}$- and $\mathcal{U}$-vertices we only considered taking out $a$, while for the duality among $\mathcal{V}$-vertices themselves we always took out a $c_i$ that appeared at the right of $b$. There is a natural pairing $\langle a(y)|c\rangle=-\langle c|a(-y)\rangle$ between $a \in \mathbb{A}_{-1}$ and $c \in \mathbb{A}_0$, which allows us to take out $b$ or $c_i$ to the left of $a$ in the aforementioned cases. As a consequence, some of the dualities can take place through two different routes, e.g.
\begin{align*}
    \langle \mathcal{V}_2(a,c_1,b)|c_2\rangle &= \langle a|\mathcal{U}_2(c_1,b,c_2) \rangle\,,\\
    \langle \mathcal{V}_2(a,c_1,b)|c_2\rangle &= -\langle \mathcal{U}_2(c_2,a,c_1)|b(-y)\rangle \,.
\end{align*}
Both cases evaluate to different expressions, but they are related to each other by a $\mathbb{Z}_2$-transformation that preserves the domain of integration, i.e.,
\begin{align*}
    \frac{u_1}{v_1} \leq \frac{u_2}{v_2} \leq \dots \leq \frac{u_n}{v_n} \leq \frac{u_{n+1}}{v_{n+1}} \,.
\end{align*}
This maps $\{v_{n+1},\dots v_1\} \rightarrow \{u_1,\dots,u_{n+1}\}$ and $\{u_{n+1},\dots u_1\} \rightarrow \{v_1,\dots,v_{n+1}\}$. 

To summarize, we have directly checked that our $A_\infty$-algebra $\mathbb{A}$ has the remarkable property we called the duality map in \cite{Sharapov:2022faa,Sharapov:2022awp}. This implies that the $A_\infty$-algebra $\mathbb{A}$ underlying Chiral Theory is a pre-Calabi--Yau algebra \cite{IYUDU202163, kontsevich2021pre}, see Appendix \ref{CY} for more detail. In practical terms, this implies that there are few independent multi-linear products with a given number of arguments.

\section{Configuration space}
\label{sec:config}
By construction, each contracting homotopy $h$ entering an interaction vertex brings one integration variable $t_i\in[0,1]$, so that the whole integration domain appears to be the hypercube $[0,1]^{2n}$. However, in terms of `times' $t_i$ the `propagators' in front of $p_{ij}$ as well as the pre-exponential factors look ugly (see \cite{Sharapov:2022faa,Sharapov:2022awp} and the change of variables in the previous section). In addition, it is not immediately obvious that the integrals converge. In terms of the new variables $u$'s and $v$'s all integrands are obviously smooth functions and the  
question of convergence reduces to the compactness of the new integration domain. In Appendix \ref{app:domain}, we prove that the domain is compact indeed.

With the help of the new integration variables the vertices simplify a lot. In particular, the propagators are linear except for the only $\hhbar$-term in the exponent where it is no more than bilinear and the pre-exponential factor  is completely eliminated by the Jacobians of the coordinate transformations. These drastic simplifications should convince one that the variables we have chosen above are the preferred ones. It is time to describe the integration domain in more detail. Let us concentrate on vertices of type $\mathcal{V}(\omega,\omega,C,\ldots, C)$, of which the symbol is given by \cite{Sharapov:2022awp}
\begin{align}\label{besttree}
\begin{aligned}
       G=&(p_{12})^n \exp\Big[ (1-\sum_i u_i) p_{01} +(1-\sum_i v_i) p_{02} +\sum_i u_i p_{1,i+2}+\sum_i v_i p_{2,i+2}+ \\
     &\qquad \qquad \qquad \qquad +\hhbar\, \Big(1+\sum_i (u_i-v_i) +\sum_{i,j} u_iv_j \sign(j-i) \Big ) p_{12} \Big]\,. 
\end{aligned}
\end{align}
We will first consider this family of vertices at lower orders in Section \eqref{sec:configstart}, then provide a straightforward generalization to all orders with details left to Appendix \ref{app:fulldomain}. A more formal description of the configuration space together with its relation to Grassmannians is presented in Section \ref{sec:swallowtail}.

\subsection{Order by order analysis}
\label{sec:configstart}
\begin{wrapfigure}{r}{0.32\textwidth}
\begin{tikzpicture}
\draw[thick]  (0,0) -- (0,4)--(1.5,1)--(4,0)--cycle;
\draw[-]  (0,4) -- (4,4);
\draw[-]  (0,4) -- (4,0);
\fill[black!5!white] (0,0) -- (0,4)--(1.5,1)--(4,0)--cycle;
\filldraw (0,4) circle (0.7 pt);
\filldraw (1.5,1) circle (0.7 pt);
\filldraw (0,0) circle (1 pt);
\filldraw (4,4) circle (1 pt);
\filldraw (4,0) circle (0.7 pt);
\coordinate [label=below:$1$] (B) at (0,0);
\coordinate [label=right:$1$] (B) at (4,4);
\coordinate [label=below:$0$] (B) at (4,0);
\coordinate [label=below:$v$]  (B) at (1.5,0);
\coordinate [label=right:$u$] (B) at (4,1);
\draw[dashed] (4,1) -- (1.5,1); 
\draw[dashed] (1.5,0) -- (1.5,1); 
\draw[->](4,0)--(4,4.5);
\draw[->](4,0)--(-0.5,0);
\node[] at (0.7,0.9) {A};
\node[] at (1.9,1.5) {B};
\end{tikzpicture}
\caption{Cubic order.}\label{F1}
\end{wrapfigure}
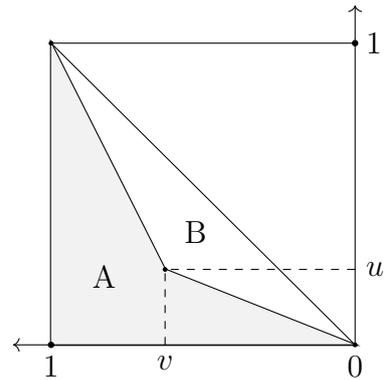
Let us start from the cubic vertex $\mathcal{V}(\omega,\omega,C)$, for which the integration domain has been identified as the simplex $0<u<v<1$ in the Cartesian plane, see Fig. \ref{F1}. The configuration space is constituted by points lying below the diagonal of the unit square. A simple plane geometry exercise identifies the  multiplier $1+u-v$ of the cosmological constant as twice the area of the shaded region $A$. The volume of the configuration space is $1/2$ since any point below the diagonal is admissible. 

At the quartic order,  $\mathcal{V}(\omega,\omega,C,C)$, the integration domain is defined  by more complicated  inequalities:
$$
0\leq v_2\leq 1\,,\qquad 0\leq u_1\leq v_1\leq 1-v_2\,,\qquad \frac{u_1}{v_1}\leq \frac{u_2}{v_2}\leq \frac{1-u_1}{1-v_1}\,.
$$
In order to clarify their geometric  meaning it is convenient to introduce the pair of new variables $v_3$ and $u_3$ subject to the relations 
\begin{equation}\label{uv}
u_1+u_2+u_3=1\,,\qquad v_1+v_2+v_3=1\,.
\end{equation}
With these variables  we can rewrite the inequalities above in a more symmetric form:
$$
0 \leq v_2\leq 1\,,\qquad 0\leq u_1\leq v_1\leq 1-v_2\,,\qquad \frac{u_1}{v_1}\leq \frac{u_2}{v_2}\leq \frac{u_3}{v_3}\,.
$$

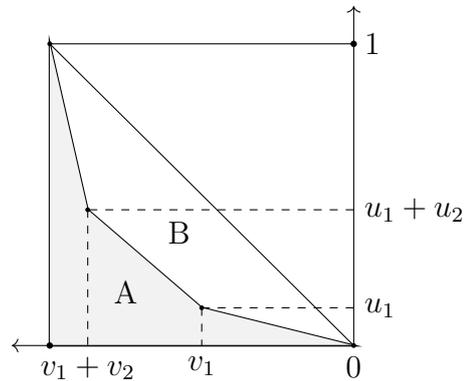
\begin{wrapfigure}{r}{0.4\textwidth}
\begin{tikzpicture}
\draw[thick]  (0,0) -- (0,4) -- (0.5,1.8) --(2,0.5)--(4,0)--cycle;
\draw[-]  (0,4) -- (4,4);
\draw[-]  (0,4) -- (4,0);
\fill[black!5!white] (0,0) -- (0,4) -- (0.5,1.8) --(2,0.5)--(4,0)--cycle;
\filldraw (0,4) circle (0.7 pt);
\filldraw (0.5,1.8) circle (0.7 pt);
\filldraw (2,0.5) circle (0.7 pt);
\filldraw (0,0) circle (1 pt);
\filldraw (4,4) circle (1 pt);
\filldraw (4,0) circle (0.7 pt);
\coordinate [label=right:$1$] (B) at (4,4);
\coordinate [label=below:$v_1+v_2$]  (B) at (0.5,0);
\coordinate [label=below:$v_1$]  (B) at (2.0,0);
\coordinate [label=right:$u_1$] (B) at (4,0.5);
\coordinate [label=right:$u_1+u_2$] (B) at (4,1.8);
\coordinate [label=below:$0$] (B) at (4,0);
\draw[dashed] (4,0.5) -- (2,0.5);
\draw[dashed] (4,1.8) -- (0.5,1.8);
\draw[dashed] (2,0.5) -- (2,0); 
\draw[dashed] (0.5,1.8) -- (0.5,0); 
\draw[->](4,0)--(4,4.5);
\draw[->](4,0)--(-0.5,0);
\node[] at (1.0,0.7) {A};
\node[] at (1.7,1.5) {B};
\end{tikzpicture}
\caption{Quartic order.}\label{F2}
\end{wrapfigure}

The last group of inequalities implies that the corresponding segments, see Fig. \ref{F2}, form a concave shape (the upper boundary of region $A$). In other words, the quadrilateral $B$ is convex and one of its edges coincides with the diagonal of the unit  square.
Again, the multiplier of the cosmological constant, $1+u_1+u_2-v_1-v_2+u_1 v_2-u_2 v_1$, can be recognized as twice the area of the shaded region $A$.  For an obvious reason we will call such concave polygons $A$ {\it swallowtails}. It is easy to see that the volume of this four-dimensional configuration space is equal to  $1/24$. 

Now the generalization to all orders is straightforward, see Appendix \ref{app:fulldomain} for the proof: vertex $\mathcal{V}(\omega,\omega,C,\ldots,C)$ with $n$ zero-forms $C$ is given by $2n$-tuple integral over the configuration space of swallowtails with $n+3$ vertices, three of which are fixed to be the corners of the unit square. 
Since the integration domain is obviously compact, the interaction vertices are well-defined at least as $A_\infty$ structure maps. The positions of the $n$ points inside the wedge, which are the actual degrees of freedom of a swallowtail, correspond to coefficients in front of $p_{1,i+2}$ and $p_{2,i+2}$ that connect the two one-forms $\omega$ to $n$ zero-forms $C$. In case $\hhbar\neq 0$, the coefficient of the cosmological term $\hhbar p_{12}$ is given by twice the area of the swallowtail. As discussed at length in \cite{Sharapov:2022faa,Sharapov:2022awp}, the fact that no other differential operators $p_{ij}$ appear that would connect pairs of zero-forms implies spacetime locality. 

Regarding trees with other topologies, first of all the configuration space is exactly the same as above, see Appendix \ref{app:fulldomain}. This, among other things, implies that the homological perturbation theory, even though yielding a solution, does not reveal all hidden symmetries of the vertices. 

\begin{figure}
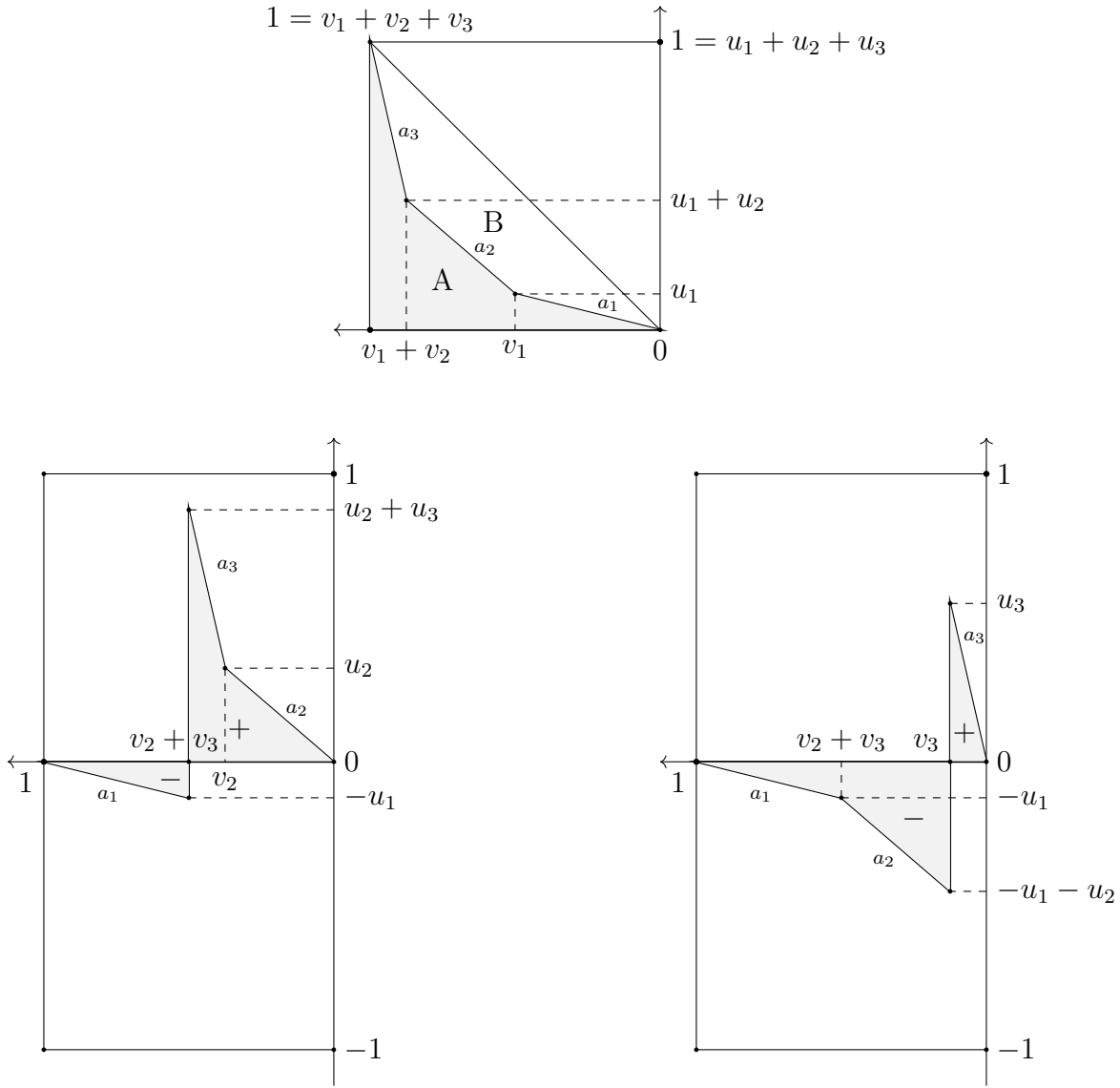

\centering\label{fig:flip}


\caption{Quartic order with various orderings of the zero-forms. On the top panel we have the swallowtail that determines $\mathcal{V}(\omega,\omega,C,C)$. The coefficient of the cosmological term is twice the area of region $A$, which is made of two segments of unit length followed by $a_1$, $a_2$, $a_3$. On the bottom right panel we flipped the position of two zero forms, which makes a contribution to $\mathcal{V}(\omega,C,C,\omega)$. Accordingly, the order of the segments is changed: $a_1$ and $a_2$ are inserted in between the first one and the second one that are of unit length. The coefficient of the cosmological term is twice the oriented area: the area below the mid-line contributes with minus sign. Similarly, on the bottom left panel one zero-form is flipped giving  a contribution to $\mathcal{V}(\omega,C,\omega,C)$. The edge $a_1$ is placed in between the segments of unit length and the coefficient of the cosmological constant again follows from twice the oriented area. }\label{quarticflipped}
\end{figure}

Trees with different ordering of zero-forms on either branch within the same topology have the same configuration space. {The term in the exponential proportional to the cosmological constant changes however by flipping one or more signs. For instance, changing the order of both zero-forms in $\mathcal{V}(\omega,\omega,C,C)$ gives a tree for  the vertex $\mathcal{V}(\omega,C,C,\omega)$. The multiplier of the cosmological constant reads $1-u_1-u_2-v_1-v_2-u_1v_2+u_2v_1$ and is equal to twice the area of region `$+$' minus region `$-$' in Fig. \ref{quarticflipped}}.
{The edges whose coordinates correspond to the flipped zero-forms create a new structure, which turns out to be a swallowtail itself. Meanwhile, these vectors are removed from the original swallowtail, which preserves the defining features of a swallowtail. Thus, for a tree with mixed ordering of its zero-forms, the term proportional to the cosmological constant is related to the difference between the area of two swallowtails, which is an oriented area. Also notice that for a mixed ordering this term can become negative.} There is a simple algebraic interpretation of these manipulations as the sum $|Q_T|$ of minors of matrix $Q_T$, \eqref{qtmatrix}, see also below.
Now we proceed to a more formal discussion of the configuration space and its relation to Grassmannians. 

\subsection{Measuring swallowtails }
\label{sec:swallowtail}
Consider a Euclidean plane $\mathbb{E}^2$ with its natural metric topology. It will be convenient on occasion to forget about Euclidean structure and treat $\mathbb{E}^2$ as an affine space with the automorphism group $\mathrm{Aff}(2,\mathbb{R})=GL(2,\mathbb{R})\ltimes \mathbb{R}^2$. By the Jordan curve theorem each simple polygon chain separates $\mathbb{E}^2$ into two disconnected regions, called exterior and interior. 
Consequently,  to each vertex of a simple polygon  one can assign exterior and interior angles.
We say that a vertex is convex (concave) if its interior angle is $\leq \pi$ ($>\pi$). 
A polygon is called convex if all its vertices are convex. By definition, a concave polygon has at least one concave vertex. 

\begin{figure}[!ht]
\center{
\begin{tikzpicture}
\draw[thick]  (0,0) -- (1,4)--(1.3,3)--(1.6,2.8)--(2.2,2.8)--(3.5,3.5)--cycle;
\fill[black!5!white] (0,0) -- (1,4)--(1.3,3)--(1.6,2.8)--(2.2,2.8)--(3.5,3.5)--cycle;
\filldraw (0,0) circle (1.0 pt);
\filldraw (1,4) circle (1.0 pt);
\filldraw (1.3,3) circle (1.0 pt);
\filldraw (1.6,2.8) circle (1.0 pt);
\filldraw (2.2,2.8) circle (1.0 pt);
\filldraw (3.5,3.5) circle (1.0 pt);
\coordinate [label=below:$A$] (B) at (0,0);
\coordinate [label=left:$B$] (B) at (1,4);
\coordinate [label=right:$C$] (B) at (3.5,3.5);
\end{tikzpicture}
}
\caption{A simple concave $6$-gon. The vertices $A$, $B$, and $C$ are convex, the other three are concave.}\label{st}
\end{figure}
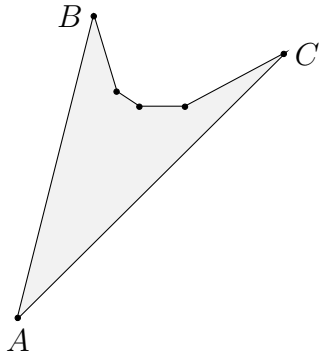

It is clear that for a simple concave polygon the minimal number of convex vertices is equal to $3$ (hence, every triangle is convex).
We are interested in simple concave polygons with exactly three convex vertices that go one after another. As in the previous section, these will be referred to as {\it swallowtails}, see Fig.\ref{st}. 
 It is known that convexity is an affine property, meaning that the affine transformations of $\mathrm{Aff}(2,\mathbb{R})$ map swallowtails to swallowtails. We say that two swallowtails are equivalent to each other if they are related by an affine transformation. 
\begin{figure}[!ht]
\center{
\begin{tikzpicture}
\draw[thick]  (0,0) -- (0,4)--(0.5,1.5)--(1,0.7)--(2.5,0.15)--(4,0)--cycle;
 \fill[black!5!white] (0,0) -- (0,4)--(0.5,1.5)--(1,0.7)--(2.5,0.15)--(4,0)--cycle;
\filldraw (0,4) circle (1.0 pt);
\filldraw (0.5,1.5) circle (1.0 pt);
\filldraw (1,0.7) circle (1.0 pt);
\filldraw (2.5,0.15) circle (1.0 pt);
\filldraw (0,0) circle (1 pt);
\filldraw (4,0) circle (1.0 pt);
\coordinate [label=below:$-1$] (B) at (4,0);
\coordinate [label=below:$0$] (B) at (0,0);
\coordinate [label=left:$-e_1$] (B) at (0,2);
\coordinate [label=below:$-e_2$] (B) at (2,0);
\draw[-Latex](0,4)--(0,0);
\draw[-Latex](0,0)--(4,0);
\draw[->](0,0)--(0,4.5);
\draw[->](4,0)--(-0.5,0);
\end{tikzpicture}
}
\caption{A canonical swallowtail with six vertices.}\label{cst}
\end{figure}
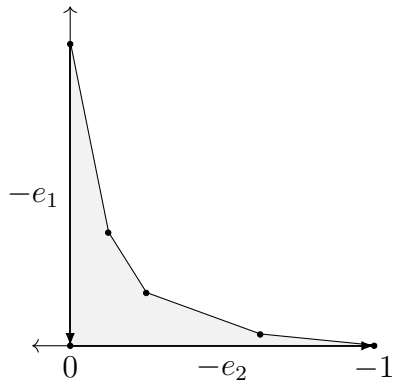

\begin{figure}[!ht]
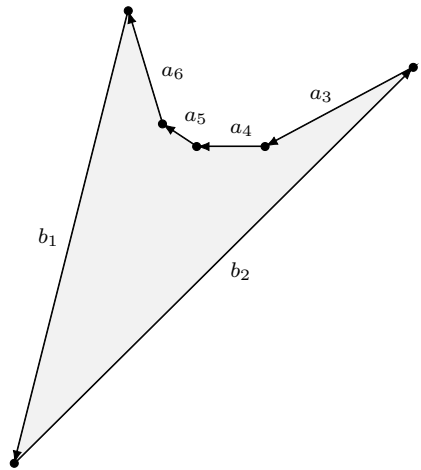

\center{

}
\caption{An oriented swallowtail.}\label{ost}
\end{figure}

In order to describe the  equivalence classes of swallowtails modulo affine transformations we fix an origin $0$ and an orthonormal basis $(e_1,e_2)$ in $\mathbb{E}^2$. Then, we translate the middle of the three convex vertices to the origin $0\in \mathbb{E}^2$. Finally, applying a linear transformation of $GL(2,\mathbb{R})$, we can match the edges forming the convex vertex with the (reversed for convenience) unit basis vectors $-e_1$ and $-e_2$. In such a way each swallowtail appears to be equivalent to one of the forms depicted in Fig. \ref{cst}. Although the last step does not specify the linear transformation uniquely, the only ambiguity  concerns the permutation of the basis vectors $e_1$ and $e_2$.  To fix this ambiguity one needs to choose an orientation in $\mathbb{E}^2$. 
We will indicate each of two  possible orientations by putting arrows on the edges of polygons as in Fig. \ref{ost}. The affine transformations that preserve either orientation form a subgroup $\mathrm{Aff}^+(2,\mathbb{R})=GL^+(2,\mathbb{R})\ltimes \mathbb{R}^2$ of the full affine group $\mathrm{Aff}(2,\mathbb{R})$. We will denote the space of all nonequivalent oriented swallowtails with $n$ vertices by $\mathbb{V}_n$.

Considering now  the oriented edges of an $n$-gon as affine vectors in $\mathbb{E}^2$, we can arrange their  coordinates into a $2\times n$ array $P\in \mathrm{Mat}_{\mathbb{R}}(2,n)$; in so doing, the order of vectors corresponds to the order of edges.\footnote{By altering the order of vectors/edges one can easily get a polygon with crossed edges as in Fig. \ref{quarticflipped}} For instance, the array corresponding to the swallowtail in Fig.\ref{ost} looks as 
$$
P=(b_1, b_2, a_3,a_4,a_5,a_6)=\left(\begin{array}{cccccc}b_{1}^1 &b_2^1&a_3^1&a_4^1&a_5^1&a_6^1\\
b_{1}^2 &b_2^2&a_3^2&a_4^2&a_5^2&a_6^2
\end{array}
\right)\,.
$$
Clearly, each array $P$ determines the corresponding polygon up to translations in $\mathbb{E}^2$ and cyclic permutations of its columns does not affect the polygon.  For oriented swallowtails we can fix the order completely by writing the coordinates of the right edge of the middle convex vertex in the  first column. Applying the transformation  $$
G=-\left(\begin{array}{cc}
    b_1^1 &b_2^1  \\
    b_2^1 & b_2^2 
\end{array}\right)^{-1}\in GL^+(2,\mathbb{R})
$$
brings the matrix  $P$ into the canonical form 
$$
GP=(-e_1,-e_2,a_3,a_4,a_5, a_6)=\left(\begin{array}{cccccc}-1 &0&a_3^1&a_4^1&a_5^1&a_6^1\\
0&-1&a_3^2&a_4^2&a_5^2&a_6^2
\end{array}
\right)
$$
that corresponds to the swallowtail in Fig. \ref{cst}. We will refer to such swallowtails as canonical representatives. Notice that the remaining entries $a$'s are not arbitrary. First of all, the closeness of the polygon chain implies that the sum of column vectors is equal to zero, i.e.,
\begin{equation}\label{ba}
b_1+b_2+a_3+\cdots+a_n=0\,.
\end{equation}
This allows us to express one of the vectors $a_i$ as the sum of the others. The concavity condition imposes further restrictions on $a$'s. Let $[a,b]$ denote the determinant of a $2\times 2$-matrix $(a,b)$. Then an array 
$$P=(b_1,b_2,a_3,\ldots,a_n)\in \mathrm{Mat}_{\mathbb{R}}(2,n)$$ 
defines a oriented swallowtail iff its entries satisfy Eq.(\ref{ba}) together with the following inequalities: 
\begin{equation}\label{aabb}
\begin{array}{c}
 \pluk_{12}=[b_1,b_2]>0\,,\qquad \pluk_{1i}= [b_1,a_i]<0\,,\qquad \pluk_{2i}=[b_2,a_i]>0\,,\\[5mm]\pluk_{ij}=[a_i,a_j]<0\,,\qquad 3\leq i<j\leq n \,.
 \end{array}
\end{equation}
The introduced variables $\pluk_{ij}$ are convenient to express the area of a swallowtail:
$$
\mathrm{Area}(P)=\frac12\sum_{i<j}\pluk_{ij}\,.
$$

Eqs. (\ref{ba}) and (\ref{aabb}) define $\mathbb{V}_n$ -- the space of all nonequivalent oriented swallowtails with $n>3$ vertices -- as a bounded domain  in $\mathbb{R}^{2(n-3)}$. The space $\mathbb{V}_n$ enjoys a natural measure given by the volume form
\begin{equation}\label{m1}
\omega_n=\prod_{k=3}^{n-1} da^1_k\wedge da^2_k\,,
\end{equation}
where the coordinates $(a^1_k, a^2_k)$ correspond to a canonical representative $P$ with $b_1=-e_1$ and $b_2=e_2$ as in Fig. \ref{cst}. With this measure one can easily find that $\mathrm{Vol}(\mathbb{V}_4)=1/2$ and $\mathrm{Vol}(\mathbb{V}_5)=1/24$.

Geometrically, there are two natural ways to look at a $2\times n$-array: either as a set of $n$ vector in $\mathbb{R}^2$ or as a pair of vectors in $\mathbb{R}^n$. So far we have followed the former interpretation; now let us try the latter. By definition, taking the quotient of full rank matrices of  $\mathrm{Mat}_{\mathbb{R}}(2,n)$ 
by the left action of $GL^+(2;\mathbb{R})$ gives the oriented Grassmannian $\widetilde G_{\mathbb{R}}(2,n)$. It  can also be visualized as the space of all oriented $2$-planes in $\mathbb{R}^n$.\footnote{More generally, one defines $\widetilde G_{\mathbb{R}}(k,n)$ to be the space of all oriented $k$-planes in $\mathbb{R}^n$. Topologically, $\widetilde G_{\mathbb{R}}(k,n)$ is just the universal double cover of the Grassmann manifold $G_{\mathbb{R}}(k,n)$.} This allows us to think of $\mathbb{V}_n$ as a subset of the oriented Grassmannian $\widetilde G_{\mathbb{R}}(2,n)$. The subset is defined by the linear equation (\ref{ba}) and inequalities (\ref{aabb}). From this perspective the variables $\{\pluk_{ij}\}$, where $i,j=1,\ldots, n$ and $i<j$, are nothing but the Pl\"ucker coordinates defining the embedding of $\widetilde G_{\mathbb{R}}(2,n)$ into the oriented projective space $\widetilde{\mathbb{P}}^{N}=\widetilde G_{\mathbb{R}}(1,N)$ of dimension $N=\frac12n(n-1)-1$. (As a smooth manifold $\widetilde{\mathbb{P}}^N$ is diffeomorphic to the standard $N$-sphere, which is the universal covering space  of $\mathbb{P}^N$.) It is known that the image of the Pl\"ucker embedding $i: \widetilde G_{\mathbb{R}}(2,n)\rightarrow \widetilde{\mathbb{P}}^N$ is given by the intersection of the projective quadrics 
\begin{equation}\label{PR}
Q_{ijkl}: \quad \pluk_{ij}\pluk_{kl}-\pluk_{ik}\pluk_{jl}+\pluk_{jk}\pluk_{il}=0\,,\qquad \forall i<j<k<l\,.
\end{equation}
These are known as the Pl\"ucker relations. Among other things the relations say that the 
inequalities (\ref{aabb}), which single out an open domain in the intersection $\bigcap Q_{ijk}$,  are highly redundant. For instance, the relation $$\pluk_{13}\pluk_{24}=\pluk_{12}\pluk_{34}+\pluk_{23}\pluk_{14}$$ implies that $\pluk_{13}<0$ whenever 
$$
\pluk_{24}>0\,,\qquad \pluk_{12}<0\,,\qquad \pluk_{23}<0\,,\qquad \pluk_{34}>0\,,\qquad  \pluk_{14}>0\,.
$$
The above geometric interpretation in terms of swallowtails suggests that it would be enough to specify the signs of 
only consecutive minors $\pluk_{i, i+1}$ and $\pluk_{1n}$ provided Eq. (\ref{ba}) holds. 
As to the remaining relation  (\ref{ba}), it is clearly equivalent to the pair of linear equations 
\begin{equation}\label{ppp}
\pluk_{12}=-\sum_{i=3}^n\pluk_{1i}=\sum_{i=3}^n\pluk_{2i}\,,
\end{equation}
which define a plane $\Pi$ of codimension two in $\widetilde{\mathbb{P}}^N$. Summarizing all of the above, we can identify the space of swallowtails $\mathbb{V}_n$ with an open region in the intersection of the projective codimension-two plane (\ref{ppp}) with the projective quadrics (\ref{PR}); the region is specified by prescribing signs (\ref{aabb}) to the Pl\"ucker coordinates.  In terms of the projective coordinates $\pluk_{ij}$ the volume form (\ref{m1}) on $\mathbb{V}_n\subset \widetilde{\mathbb{P}}^N$ is obtained as the  restriction of the form 
$$
\Omega_n=\prod_{i=3}^{n-1}\frac{d\pluk_{1i}\wedge d\pluk_{2i}}{(\pluk_{12})^2}
$$
of degree $2(n-3)$ on $\widetilde{\mathbb{P}}^N$. The closure $\overline{\mathbb{V}}_n\subset \widetilde{\mathbb{P}}^N$ defines the integration domain for the
interaction vertices of order $n$. Topologically, $\overline{\mathbb{V}}_n$ is a smooth manifold with corners. Hence, it admits a smooth stratification. For example, the stratum of codimension one corresponds to degenerate canonical swallowtails where exactly one concave (or convex) internal angle becomes $\pi$ (or $0$).

In the last decade, much attention has been paid to the what is called {\it positive Grassmannians} because of their remarkable  applications in statistical physics, integrable models, and  scattering amplitudes. For a recent account of the subject we refer the reader to \cite{PGr, williams2021positive}. By definition, a positive Grassmannians is just an open  region of  a real Grassmann manifold where all Pl\"ucker coordinates are  strictly positive. Our considerations show that other distributions of signs among the Pl\"ucker coordinates may also be of interest, at least  for some field-theoretical problems.

\section{Discussion and Conclusions}
\label{sec:conclusions1}
In this  paper, we have obtained all vertices of Chiral Theory with and without cosmological constant. As it was already pointed out in \cite{Sharapov:2022faa,Sharapov:2022awp} the final form of the vertices is remarkably simple: the exponents become linear in the new variables (or  quadratic for nonzero cosmological constant) and complicated exponential prefactors are eliminated by the corresponding Jacobians. 
Another result is an explicit description of the configuration space. It is given by what we call swallowtails --  concave polygons that have two edges coinciding with two adjacent edges of the unit square. Another way to describe the same geometric shape is to consider the space of convex polygons that can be inscribed into a unit square with one edge coinciding with the diagonal. The area of the swallowtail also has a meaning and determines the coefficient of the cosmological term.

There is an intriguing relation \cite{Sharapov:2017yde,Sharapov:2022eiy} to the formality theorems, in particular to Shoikhet--Tsygan--Kontsevich formality \cite{Kontsevich:1997vb, Shoikhet:2000gw}. This indicates that with the help of the simple configuration space we have now the $A_\infty/L_\infty$-relations can be proved via Stokes theorem, which we will address elsewhere. A more intriguing question is whether the configuration space we identified can be generalized and extended into the `bulk'. Indeed, the Poisson structure $\pi$ we begin with is just $\epsilon^{AB}$, i.e., symplectic and constant. For this reason, all Kontsevich--Shoikhet's graphs where $\pi$ is hit by derivatives  disappear. What remains of undifferentiated $\pi$ is the Moyal--Weyl star-product and the Feigin--Felder--Shoikhet cocycle \cite{FFS} that justifies the existence of cubic vertices as well as higher order vertices. These structures are also closely related to the deformation quantization of Poisson Orbifolds \cite{Sharapov:2022eiy,Sharapov:2022phg}. There should also exist an extension of our construction to Feigin's $gl_\lambda$ \cite{Feigin}. Another direction is due to a surprising appearance of pre-Calabi--Yau algebras \cite{IYUDU202163, kontsevich2021pre}, see Appendix \ref{CY}. Eventually, all of this should admit a description in terms of a certain two-dimensional topological field theory. 

Another interesting direction is to uncover what is special about the multi-linear products we found as compared to other representatives of the same $A_\infty/L_\infty$-algebra. From the viewpoint of a sigma-model $d\Phi=Q(\Phi)$, different choices of coordinates for the underlying homological vector field $Q$ translate into redefinitions of fields $\Phi$, most of which are too nonlocal to give meaningful interactions. In other words, most of coordinates for $Q$ violate the equivalence theorem. It is tempting to say that there should always exist a coordinate system that leads to maximally local interactions. For every field theory, one can think of $Q$ as a deformation of a `free' homological vector field $Q_0$ that defines a certain graded Lie algebra (via the bilinear maps of the associated $L_\infty$-algebra) and the first order deformation corresponds to a certain Chevalley--Eilenberg cocycle. Therefore, the maximal locality requirement selects one specific representative of the Chevalley--Eilenberg cohomology. It is easy to see that the vertices we found are maximally local (any field redefinition can only increase the number of derivatives). It would be interesting to find out exactly which property of the Chevalley--Eilenberg cohomology is equivalent to maximal locality in the field theory language.

The immediate  applications of the obtained results are obvious: (a) it would be interesting to look for exact solutions building upon the general tools \cite{Sezgin:2005pv,Didenko:2009td,Aros:2017ror}  worked out in the context of formal HiSGRA;\footnote{By a {\it formal HiSGRA} we mean the sigma-models above, $d\Phi=Q(\Phi)$, without taking locality into account.   Interestingly, the equations may have nicely looking solutions even for physically nonsensical vertices hidden in $Q$.  } (b) it is important to compute holographic correlation functions as to compare with (Chern--Simons) vector models (Chiral Theory should be dual to a closed subsector of Chern--Simons vector models \cite{Sharapov:2022awp}); (c) presymplectic AKSZ actions along the lines of \cite{Sharapov:2021drr} can be constructed as well as possible counterterms and anomalies can be classified \cite{Sharapov:2020quq}; (d) the study of integrability of Chiral Theory \cite{Ponomarev:2017nrr} and its relation to twistors \cite{Tran:2021ukl} should also be a fruitful direction.  

In the regard to item (d) let us point out that the $A_\infty$-algebra of Chiral Theory $\hat{\mathbb{A}}$ naturally defines a two-dimensional theory, which should be closely related to an important observation made in \cite{Ponomarev:2017nrr} that the equations of Chiral Theory in flat space can be cast into the form of the principal chiral model. Indeed, the higher spin algebra $\hs$ is given by the tensor product $A_\hhbar \otimes A_1 \otimes \mathrm{Mat}_N$, where $A_\lambda$ is the Weyl algebra, with $\hhbar$ being the parameter of noncommutativity (effective cosmological constant). Clearly, all the higher products  of $\hat{\mathbb{A}}$ owe their existence to the first factor $A_\hhbar$ and its bimodule $A^\ast_\lambda$, while the rest part, $B=A_1 \otimes \mathrm{Mat}_N$, enters via the usual associative product. What makes the system four-dimensional is the functional dimension of $\hs$. If we simply drop $A_1$ and take $B=\mathrm{Mat}_N$ (or any other associative algebra with zero functional dimension), we can write the same sigma-model $d\Phi=Q(\Phi)$,  but on a two-dimensional space.\footnote{The functional dimension of $A_\hhbar$ implies that the theory is off-shell in $2d$ or, perhaps similarly to \cite{Alkalaev:2019xuv}, can be understood as an on-shell one for infinitely-many fields. In $3d$ the theory would be on-shell to begin with. } The factor $A_\hhbar$ implies that $AdS_2$ is a natural vacuum for such a system. According to \cite{Ponomarev:2017nrr} this system (as well as the whole Chiral Theory) should be integrable. Its exact solutions can perhaps be obtained by adapting the techniques from \cite{Sharapov:2019vyd} and it would be interesting to compare it with the standard techniques from integrable models. With $B=\mathrm{Mat}_2$ one can get a $3d$ interpretation. The functional dimension of {$A_\lambda$}, which is $2$, corresponds to off-shell equations in $2d$ and to on-shell in $3d$, which seem to be the most natural dimensions for the theory underlying the Chiral one. It would be interesting to uncover the properties of this parent theory.

Lastly, let us present the Chiral HiSGRA equations of motion in a concise form.\footnote{In this regard one can mention the very recent Didenko equations \cite{Didenko:2022qga} that are claimed to give a local theory in $AdS_4$. Provided the vertices are explicitly extracted from \cite{Didenko:2022qga} it would be interesting to compare them with Chiral Theory in $AdS_4$. A closely related interesting open question is whether there are more than one local higher spin gravity in $AdS_4$. Without taking locality into account there are infinitely many formal deformations at higher orders \cite{Vasiliev:1999ba,Sharapov:2020quq}. Also, similar ambiguities are present for low spin theories. Therefore, the question of (non-)uniqueness of local theories remains open, which is also relevant for the study of quantum consistency of Chiral Theory. } As it has been discussed, the $\mathcal{V}$-vertices come from trees with two branches and $\mathcal{U}$-vertices originate from trees with just one branch. The expression for the most general branch $B_n[C,\dots, \omega,\ldots C]$ is given in \eqref{compactform}. Let us introduce the sum 
$$
B[\omega, C]=\sum_{n=0}^\infty B_n[C,\dots, \omega,\ldots C]
$$
over all possible branches and orderings of zero-forms $C$ therein. 
With this we can write the equations of motion as
\begin{align*}
    d\omega=B[\omega, C]\star B[\omega,C]\Big|_{z=0}\,,\qquad
    dC=B[\omega, C]\circ C-C\circ B[\omega, C]\,.
\end{align*}
As is seen, upon switching  on interaction, the one-form field $\omega$ on the right is just replaced  with $B=\omega + O(C)$. One  can regard the full branch $B$ as an effective  field $\omega$ `dressed' by $C$. The equations can also be understood as a Poisson sigma-model, see Appendix \ref{CY}.

\chapter{$A_\infty$-relations from Stokes' theorem}\label{chap:Stokes}

In this chapter, we prove the $A_\infty$-relations underlying chiral HiSGRA using Stokes' theorem, which indicates that an extension of the (Shoikhet-Tsygan-)Kontsevich formality theorem exists. The content is entirely based on \cite{sharapov2024strong}, co-authored with Alexey Sharapov and Evgeny Skvortsov, and published in the \textit{Journal of High Energy Physics}.

\section{Introduction and summary}
The main idea behind `higher spins' is to switch on a consistent interaction with or between the fields of spin $s>2$. For massive higher spin fields effective field theories with a single field of spin-$s$ are known to be  possible and find their applications, for instance, in the gravitational wave physics, see e.g. \cite{Cangemi:2023ysz} and references therein. Theories with massless higher spin fields, called higher spin gravities (HiSGRA) \cite{Bekaert:2022poo}, aim at constructing viable models of quantum gravity.  Each HiSGRA is strongly constrained by a certain higher spin symmetry and incorporates the graviton as a part of the (usually infinite) symmetry multiplet. Quantum gravity models still are not easy to construct along these lines: the masslessness, which simulates some features of the UV-regime already classically and higher spins in the spectrum usually come in tension with the conventional field theory paradigm, e.g. with the requirement of locality \cite{Bekaert:2015tva, Maldacena:2015iua, Sleight:2017pcz, Ponomarev:2017qab}. 

As a result, there are very few  perturbatively local theories with massless higher spin fields: topological $3d$ models \cite{Blencowe:1988gj, Bergshoeff:1989ns, Campoleoni:2010zq, Henneaux:2010xg, Pope:1989vj, Fradkin:1989xt, Grigoriev:2019xmp, Grigoriev:2020lzu}, conformal higher spin gravity \cite{Segal:2002gd, Tseytlin:2002gz, Bekaert:2010ky, Basile:2022nou}, Chiral Theory \cite{Metsaev:1991mt, Metsaev:1991nb, Ponomarev:2016lrm, Skvortsov:2018jea, Skvortsov:2020wtf} and its contractions \cite{Ponomarev:2017nrr,  Krasnov:2021nsq} (see also \cite{Tran:2021ukl, Tran:2022tft, Adamo:2022lah}). Very close to them is the higher spin IKKT-model, see e.g. \cite{Sperling:2017dts, Tran:2021ukl, Steinacker:2022jjv, Steinacker:2023cuf}, which is also a noncommutative field theory.\footnote{Another direction in the context of holographic theories is to derive the bulk dual by massaging the CFT partition function, see e.g. \cite{deMelloKoch:2018ivk,Aharony:2020omh}, which, however, leads to higher derivative free equations, has a slightly different spectrum and nonlocal interactions, the strong point being in that it does reproduce the CFT correlation functions by construction.} There are also incomplete (formal) theories \cite{Vasiliev:1990cm, Vasiliev:1999ba,Bekaert:2013zya,Bonezzi:2016ttk,Bekaert:2017bpy,Grigoriev:2018wrx} that can only be constructed at the formal level of $L_\infty$-algebras with the associated field equations suffering from nonlocality  \cite{Boulanger:2015ova}.\footnote{It should be noted that while every classical field theory (PDE) is defined by some $L_\infty$-algebra, e.g. \cite{Grigoriev:2019ojp}, not every $L_\infty$-algebra leads to a well-defined theory. Firstly, one problem is in that any $L_\infty$-algebra is defined up to canonical automorphisms and the latter correspond to very non-local field redefinitions, in general, which are not admissible. Secondly, it is not clear how to treat genuinely nonlocal field theories in this language, for a generic HiSGRA a way out would be to define a set of observables in a more algebraic terms at the level of the given $L_\infty$-algebra, e.g. \cite{Sharapov:2020quq}, or to resort to even more general ideas, e.g. \cite{Sezgin:2011hq, Bonezzi:2016ttk,de2019fronsdal,DeFilippi:2021xon}, that operate with a differential graded Lie algebra of which a given $L_\infty$ is the minimal model.  } The general problem of constructing formal theories, i.e., $L_\infty$-algebras from any higher spin algebra, was solved in \cite{Sharapov:2019vyd}. 

The Chiral Theory was first found in the light-cone gauge in flat space \cite{Metsaev:1991mt, Metsaev:1991nb, Ponomarev:2016lrm} and later covariantized at the level of equations of motion and extended to (anti-)de Sitter space in \cite{Skvortsov:2022syz, Sharapov:2022faa, Sharapov:2022awp, Sharapov:2022wpz, Sharapov:2022nps}. The perturbatively local equations of motion have the form of a nonlinear sigma-model
\begin{align}\label{pdefdaA}
    d\Phi&=Q(\Phi)\,, && d=dx^\mu \, \pl_\mu\,,
\end{align}
where the fields $\Phi$ are maps from the spacetime $Q$-manifold $(\Omega(\mathcal{X}),d)$ (the differential graded algebra of forms on $\mathcal{X})$ to another $Q$-manifold $(\mathcal{M},Q)$, $Q^2=0$. Perturbatively around a stationary point, $Q$ determines a flat $L_\infty$-algebra. Chiral HiSGRA is a happy occasion where the structure maps of the  $L_\infty$-algebra can be fine-tuned to maintain locality of \eqref{pdefdaA}. In this case, the formal approach yields a real field theory.

It was found that the $L_\infty$-algebra underlying Chiral HiSGRA originates from an $A_\infty$-algebra through symmetrization. The $A_\infty$-algebra is a very special one \cite{Sharapov:2022wpz, Sharapov:2022nps}: it is given by the tensor product of a pre-Calabi--Yau algebra of degree $2$ and an associative algebra.  The former can be viewed as a noncommutative counterpart of a Poisson structure. 

It was also known \cite{Sharapov:2017yde, Sharapov:2020quq} that the first two `floors' of the $A_\infty$-algebra are related to the (Shoikhet--Tsygan--)Kontsevich formality theorem \cite{Kontsevich:1997vb, Tsygan, Shoikhet:2000gw}. All vertices of the theory can be represented \cite{Sharapov:2022wpz,Sharapov:2022nps} as sums over certain graphs $\Gamma$, 
\begin{align}\label{mmm}
    m_n(f_1,\ldots, f_n)&=\sum_\Gamma w_{\Gamma}\, \mathcal{W}_\Gamma (f_1\otimes   \cdots \otimes f_n)\,,
\end{align}
where $w_\Gamma$ are certain weights and $\mathcal{W}_\Gamma $ are poly-differential operators acting on the fields $f_i$. The  maps $m_n$ define the vertices in \eqref{pdefdaA} via the symmetrization. The graphs are not much different from the (Shoikhet--)Kontsevich graphs, but the weights are given by the integrals over the configuration space $C_\Gamma$ of compact concave polygons. The graphs can be summed up into simple $\exp$-like generating functions, the Moyal--Weyl product being the trivial example.

In this paper, we prove the $A_\infty$-relations (aka Stasheff's identities) the way it is usually done for the formality theorems \cite{Kontsevich:1997vb,Shoikhet:2000gw}. This gives a further evidence that there might be a bigger formality behind Chiral Theories. The $A_\infty$-relations have the form 
\begin{align}\label{SI}
    \sum_{i,j} \pm m_i (\bullet, \ldots,  m_j(\bullet, \ldots,\bullet),  \ldots, \bullet)=0
\end{align}
and are thus given by the `products' of the vertices. To verify the $A_\infty$-relations, for each number of arguments in (\ref{SI}), we construct a configuration space $C$ and a closed form $\Omega$ on $C$ such that the boundary 
\begin{align*}
    \pl C&= \sum_{\Gamma, \Gamma'} C_{\Gamma} \times C_{\Gamma'}
\end{align*}
reproduces the various products of the configuration spaces of (\ref{mmm})  and $\Omega$ restricted to the boundary gives all summands in (\ref{SI}): 
\begin{align}
   0= \int_C d\Omega&= \int_{\pl C} \Omega  && \Longleftrightarrow && \text{$A_\infty$-relations}\,.
\end{align}

The earlier papers
\cite{Sharapov:2022faa, Sharapov:2022awp,Sharapov:2022wpz,Sharapov:2022nps} rely on homological perturbation theory based on a certain multiplicative resolution of the higher spin algebra. While giving the interaction vertices in some form for this specific case, the resolution does not have any invariant meaning in itself. It is also not clear how to construct such resolutions in general. After certain nontrivial transformations the vertices were found to have a form reminiscent of formality theorems, to which the first two structure maps are directly related. This raises several questions. (i) What are the constructs in the deformation quantization and noncommutative geometry that can explain the vertices of Chiral Theory without invoking ad hoc resolutions? (ii) Is there a more general structure (formality) of which (Shoikhet--Tsygan--)Kontsevich ones are particular cases and that gives the vertices of Chiral Theory? As a step towards the answers we show in this paper that the $A_\infty$-relations can be proved via Stokes theorem. 

The outline of the paper is as follows. Sec. \ref{sec:precalau}  contains some algebraic background on $A_\infty$-, $L_\infty$-, and pre-Calabi--Yau algebras. In  Sec. \ref{sec:CD}, we reformulate the vertices of \cite{Sharapov:2022wpz,Sharapov:2022nps} in a more pre-Calabi--Yau friendly way. The proof via Stokes' theorem is confined to Sec. \ref{sec:proof}. Conclusions and discussion can be found in Sec. \ref{sec:conclusions2}.

\section{Pre-Calabi--Yau algebras}
\label{sec:precalau}
One can think of a pre-Calabi--Yau (pre-CY) algebra of degree $2$ as a noncommutative analogue of a (formal) Poisson structure \cite{kontsevich2021pre}, \cite{IYUDU202163}, \cite{Kontsevuch:2006jb}.  With the hope that the algebraic structures described below may also be relevant to other HiSGRA-type models and beyond we keep our discussion as general as possible. For their string-theoretic motivations and interpretations, see \cite{Kajiura:2003ax}.

We fix $\mathbb{C}$ as the ground field so that all unadorned  Hom's and $\otimes$ would  be over $\mathbb{C}$.  
Let $V$ be a complex vector space regarded as a $\mathbb{Z}$-graded space concentrated in degree zero. Define the direct sum $W=V[-1]\oplus V^\ast$ where $V^\ast$ is the vector space dual to $V$ and $[-1]$ stands for the degree shift. In other words, $W=W_{1}\oplus W_0$ with $W_{1}=V[-1]$ and $W_0=V^\ast $. We will denote the degree of a homogeneous element $a\in W$ by $|a|$. 
The natural pairing between $V$ and $V^\ast$ gives rise to a symplectic structure of degree-one on $W$:
\begin{equation}\label{ip}
    \langle a, a'\rangle =0=\langle b,b' \rangle \,,\qquad \langle a,b\rangle= a(b)=-\langle b,a\rangle 
\end{equation}
for all $a, a'\in V^\ast$ and $b,b'\in V[-1]$.  Let $T W=\bigoplus_{n> 0} T^nW$ be the (restricted)  tensor algebra of the space $W$ with $T^nW=W^{\otimes n}$.  Besides the tensor degree, the algebra $TW$ inherits the $\mathbb{Z}$-grading from $W$, so that $TW=\bigoplus_{n\geq 0}(TW)_{n}$.  To simplify exposition, we will assume the vector space $V$ to be finite-dimensional. It should be noted, however, that all the constructions below extend to the infinite-dimensional case with appropriate modifications, see \cite{kontsevich2021pre}.

We define the space of $p$-cochains $C^p$ as the dual to the space $T^{p+1}W$, that is,
\begin{equation}
    C^p=(T^{p+1}W)^\ast=\mathrm{Hom}(T^{p+1}W, \mathbb{C})\,.
\end{equation}
The direct product $C^\bullet=\prod_{p\geq 0}C^p$ contains the  subspace of {\it cyclic cochains} $C^\bullet_{\mathrm{cyc}}$  satisfying  the cyclicity condition 
\begin{equation}\label{cyc}
    f(a_0,a_1,\ldots, a_p)=(-1)^{|a_0|(|a_1|+\cdots +|a_{p}|)}f(a_1, \ldots, a_{p}, a_0)
\end{equation}
for all $a_i \in W$ and $p\geq 0$.  
Using the symplectic structure (\ref{ip}), we can write 
\begin{equation}\label{vf}
    f (a_0,\ldots, a_p)= \langle a_0, \hat f(a_{1},\ldots, a_{p})\rangle 
\end{equation}
for some homomorphism $\hat f\in \mathrm{Hom}(T^{p}W, W)$. Since the symplectic structure is nondegenerate, the last relation determines  $\hat f$ unambiguously.

Now we can endow the  space $C^\bullet_{\mathrm{cyc}}[1]$ with the structure of a graded Lie algebra. The Lie bracket of two homogeneous cochains $f\in C^p_{\mathrm{cyc}}[1]$ and $g\in C^{n-p}_{\mathrm{cyc}}[1]$ is given by the {\it necklace bracket}
\begin{equation}\label{nb}
    [f,g] (a_0,\ldots, a_{n})=\sum_{i=0}^n(-1)^{\kappa_i}g(\hat f(a_{i+1},\ldots,a_n),a_0,\ldots,  a_{i})
\end{equation}
$$
 =\sum_{i=0}^n(-1)^{\kappa_i}  \langle\hat f(a_{i+1},\ldots,a_n),\hat g(a_0,\ldots,  a_{i})\rangle\,.
$$
Here summation is over all cyclic permutations of $a$'s and the Koszul sign 
is determined by 
$$
\kappa_i={(|a_0|+\cdots +|a_{i}|)(|a_{i+1}|+\cdots+|a_n|)}\,.
$$
By construction, the $n$-cochain $[f,g]$ satisfies the cyclicity condition (\ref{cyc}) and graded skew-symmetry,
\begin{equation}
    [f,g]=(-1)^{\bar f\bar g}[g,f]\,.
\end{equation}
Here $\bar f=|f|-1$ is the degree of $f$ as an element of the Lie algebra $C^\bullet_{\mathrm{cyc}}[1]$.

A pre-CY structure on $W$ is given now by any Maurer--Cartan  element $S$ of the graded Lie algebra $ C^\bullet_{\mathrm{cyc}}[1]$. By definition, 
\begin{equation}\label{MC}
    [S,S]=0\,.
\end{equation}
As an element of $C^\bullet_{\mathrm{cyc}}$, $S$ has degree $-2$.  Therefore, one usually refers to the pair $(W, S)$ as a pre-CY algebra of degree $2$. It also follows from the definition that $\hat S$ defines the structure of a cyclic $A_\infty$-algebra (see Appendix \ref{CY}) on $W$ with the cyclic structure given by the symplectic form (\ref{ip}). 

By degree reasons and cyclicity, each pre-CY structure of degree $2$ defines and is defined by a sequence of multilinear maps
\begin{equation}\label{smn}
    S_{n,m} : V[-1]\otimes T^{n}V^\ast \otimes V[-1]\otimes T^{m}V^\ast\rightarrow \mathbb{C}\,,
\end{equation}
with $n\geq m$. Then $S=\sum_{n\geq m}S_{n,m}$. We say that the pre-CY algebra is minimal if $S_{0,0}=0$. It is clear that minimal pre-CY algebras correspond to minimal $A_\infty$-algebras, hence the name. Geometrically, one can think of the $A_\infty$-structure $\hat S$ as a homological vector field on a noncommutative manifold $\mathcal{N}$ associated with $W$, see \cite{Kontsevuch:2006jb}.

Each multilinear map $\hat S_{n,m}$ has two components $\hat S_{n,m}^{(0)}$ and $\hat S^{(1)}_{n,m}$ taking values in $W_0$ and $W_{1}$, respectively.  It follows from the MC equation (\ref{MC}) that the map $\hat S_{1,0}^{(1)}$ of the minimal pre-CY algebra gives rise to an associative product on $V$, namely, 
\begin{equation}
    b_1\cdot b_2=\hat S^{(1)}_{1,0}(b_1,b_2)\,,\qquad \forall b_1, b_2\in V[-1]\,.
\end{equation}
Moreover, the map $\hat S^{(0)}_{1,0}$ makes $V^\ast$ into a bimodule over the associative algebra $V$:
\begin{equation}
    b\cdot a =\hat S^{(0)}_{1,0}(b,a)\,,\qquad  a\cdot b =-\hat S^{(0)}_{1,0}(a,b)\,,\qquad \forall b\in V[-1], \quad\forall a\in V^\ast\,.
\end{equation}
Regarding the elements of $W$ as `coordinates' on a noncommutative manifold $\mathcal{N}$ and $C^\bullet_{\mathrm{cyc}}$ as a ring of `functions' on $\mathcal{N}$, one can define a  graded-commutative 
submanifold $\mathcal{C}\subset \mathcal{N}$ by imposing graded-commutativity conditions. Technically, this implies factorization of the tensor algebra $TW$ by the two-sided ideal $I$  generated by the commutators: 
\begin{equation}
    a\otimes b-(-1)^{|a||b|}b\otimes a\,,\qquad \forall a,b\in W\,.
\end{equation}
This results in the symmetric tensor algebra $SW=TW/I$ of the graded vector space $W$. The formula \begin{equation}\label{ff}
     f_{sym}(a_0,a_1, \ldots, a_p)=\frac{1}{p!}\sum_{\sigma\in S_{p}} (-1)^{|\sigma|} f(a_0, a_{\sigma(1)},\ldots, a_{\sigma(p)})\,,
\end{equation}
where $(-1)^{|\sigma|}$ is the Koszul sign associated with the permutations of $a$'s, defines then  a map from 
$C^p_{\mathrm{cyc}}$ to $\mathrm{Hom}(S^pW,\mathbb{C})$. Identifying $\prod_{p>0}\mathrm{Hom}(S^pW,\mathbb{C})$ with the ring of functions on $\mathcal{C}$, one can think of ${f}_{sym}$ as the restriction to $\mathcal{C}$ of the function $f$ on $\mathcal{N}$. Writing (\ref{ff}) as
\begin{equation}
       f_{sym}(a_0,a_1, \ldots, a_p)= \langle a_0, \hat f_{sym}(a_{1},\ldots, a_{p})\rangle\,,
\end{equation}
we define the map $\hat f_{sym}: S^pW\rightarrow W$ by
\begin{equation}\label{ffs}
    \hat f_{sym}(a_1,\ldots, a_p)=\frac{1}{p!}\sum_{\sigma\in S_{p}} (-1)^{|\sigma|} \hat f(a_{\sigma(1)},\ldots, a_{\sigma(p)})\,.
\end{equation}
The multilinear maps (\ref{ff}) and (\ref{ffs}), being totally symmetric, are completely determined through polarization by their restriction on the diagonal, that is, by  the nonlinear maps from $W$ to itself defined as
\begin{equation}\label{fhf}
\begin{array}{l}
    \mathbf{f}(a)=f_{sym}(a,\ldots, a) = f(a,\ldots, a)\,,\\[3mm]
    \hat{\mathbf{f}}(a)=\hat f_{sym}(a,\ldots,a)=\hat f(a,\ldots, a)\,.
    \end{array}
\end{equation}
Geometrically, we can treat $\mathbf{f}$ and $\hat{\mathbf{f}}$, respectively, as a function and a vector field on the (formal) graded-commutative manifold $\mathcal{C}$ associated with $W$. 

It is known that the symmetrization of an $A_\infty$-structure gives an $L_\infty$-algebra structure on the same vector space, see e.g. \cite{Lada_commutators}. In particular, $\hat S_{sym}$ makes $W$ into a cyclic $L_\infty$-algebra, with the cyclic structure being given by the natural pairing (\ref{ip}).  The direct sum decomposition $W=V[-1]\oplus V^\ast$ allows for more refined geometric interpretation, namely, one can think of  $\mathcal{C}$ as the total space of the shifted cotangent bundle $T^\ast[-1]V^\ast$ of the formal  manifold $\mathcal{M}$ associated with the space $V^\ast$. Then each function $\mathbf{f}$, defined by (\ref{ffs}), gives rise to a polyvector field on $\mathcal{M}$. Upon the degree shift, the space of polyvector fields is known to carry the structure of graded Lie algebra w.r.t. the Schouten--Nijenhuis bracket. Let us denote this Lie algebra by $\mathcal{P}$. The assignment $f\mapsto \mathbf{f}$ defines then a homomorphism from $C_{\mathrm{cyc}}^\bullet[1]$ to $\mathcal{P}$, i.e.,
\begin{equation}
    [\mathbf{f},\mathbf{g}]_{SN} =[f,g] (a,\ldots,a)
\end{equation}
Here, the l.h.s.  is given by the SN bracket of the poly-vector fields $\mathbf{f}$ and $\mathbf{g}$, while the r.h.s. is obtained by the restriction on $\mathcal{C}$ of the necklace bracket of their preimages. 

The function $\mathbf{S}$ corresponding to the minimal pre-CY structure $S$ has the form
\begin{equation}\label{minchoiceS}
    \mathbf{S}(a,b)=\sum_{n=1}^\infty \mathbf{S}_n (b,b, a,\ldots, a)=\sum_{n\geq m} S_{n,m}(b, \overbrace{a,\ldots,a}^{n},b,\overbrace{a,\ldots,a}^m)
\end{equation}
for all $a\in V^\ast$ and $b\in V[-1]$. Being quadratic in $b$'s, the function  $\mathbf{S}$ can be considered as a bivector field on $\mathcal{M}$. Moreover, the Maurer--Cartan equation (\ref{MC}) implies that the bivector  $\mathbf{S}$ is Poisson.

We thus see that every pre-CY structure on $W=V[-1]\otimes V^\ast$ gives rise to a Poisson bivector on the formal manifold associated with $V^\ast$. From this perspective,  pre-CY structures of degree $2$ extend the notion of a Poisson structure to noncommutative setting. 

Finally, let us mention that given any $A_\infty$-algebra $\mathbb{A}$ on a graded vector space $W$ and any associative algebra $B$, one can construct a new $A_\infty$-algebra given by the tensor product $\mathbb{A}\otimes B$. Its vector space is the tensor product $W\otimes B$ and the structure maps read
\begin{align}\label{tensorproductmaps}
m_n(a_1\otimes b_1, \ldots, a_n\otimes b_n)&= m_n(a_1, \ldots, a_n) \otimes (b_1\cdots b_n)\,,     
\end{align}
where $a_i\in W$, $b_i\in B$, and $(b_1 \cdots b_n)$ is the product of $b_i$ in $B$. This construction will be used below for Chiral Theory where $\mathbb{A}$ is a pre-CY algebra.

\section{Vertices} \label{sec:CD}
As explained in Sec. \ref{sec:precalau}, every pre-CY structure of degree $2$ is defined by a sequence (\ref{smn}) of maps\footnote{It is convenient to define a slightly over-complete set of maps that are related to each other via cyclic symmetry. For example, with the help of the cyclic symmetry we can reach $k=0$ and further impose $m\geq n$, cf. discussion around \eqref{smn}.}
\begin{equation} \label{genericS}
    S_{N}(\alpha_1,\dots,\alpha_{k},a,\alpha_{k+1},\ldots, \alpha_{k+m}, b, \alpha_{k+m+1}, \dots, \alpha_{k+m+n+1})\,,
\end{equation}
where the $\alpha$'s are elements of $V^\ast$, while the arguments $a$ and $b$ live in $V[-1]$ and $k+m+n+1=N$. We recall that the graded space of a pre-CY algebra associated with Chiral HiSGRA is given by ${W}=V[-1]\oplus V^\ast$, where $V=\mathbb{C}[y^A]$ is the space of complex polynomials in $y^A$. Correspondingly, the dual space $V^\ast=\mathbb{C}[[y^A]]$ is given by formal power series in the same formal variables. Note that in this section we ignore other factors such as the dependence on $\bry^{A'}$ and matrix factors since they enter simply via the tensor product. 

\begin{figure}[!ht]
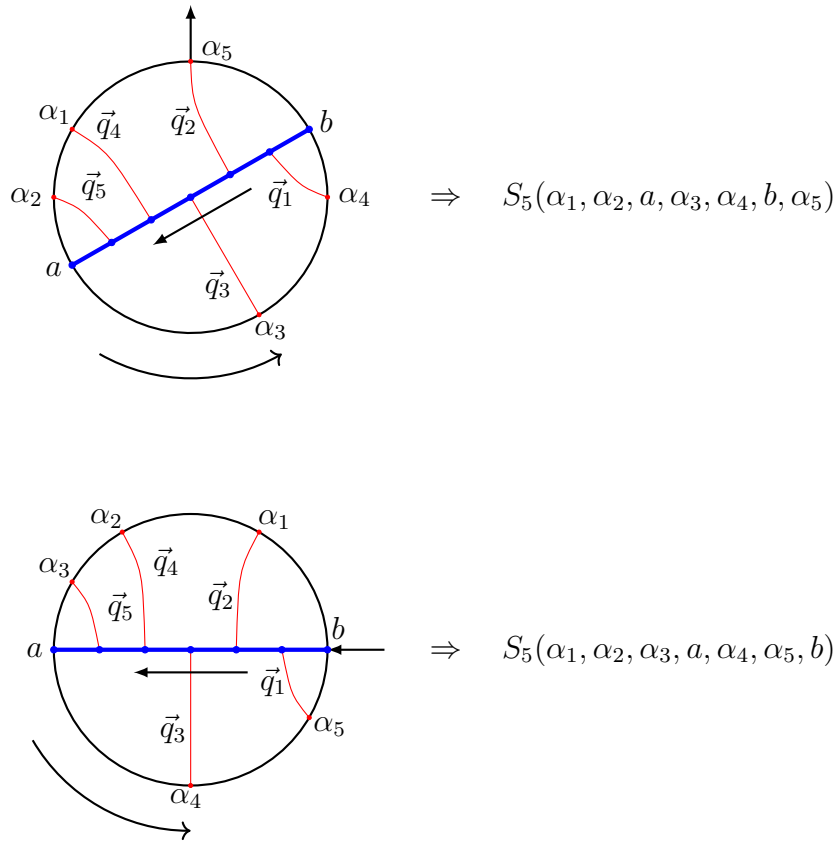

\centering

\caption{Two  disk diagrams that contribute to $S_{5}$.}\label{SD1}
\end{figure}

One can encode each map $S_N$  by disk diagrams as depicted  in Fig. \ref{SD1}. Our disk diagrams  are specific planar graphs with trivalent vertices. By definition, each disk diagram consists of a circle with a diameter, a set of (nonintersecting) lines connecting vertices on the diameter to vertices on the circle and an arrow pointing outwards from one of the $\alpha$'s in the northern semicircle, or towards the vertices $a$ or $b$, i.e. one of the points on the boundary is marked by an arrow and the direction of the arrow depends on whether the argument is in $V^\ast$ or in $V[-1]$. The direction of an arrow is to indicate the symplectic structure \eqref{ip}. No other links or vertices are allowed. As a result, each diagram is characterized by the number of vertices on the northern semicircle of the circle to the left and right of the arrow, $k$ and $n$, respectively, and the number of vertices on the southern semicircle, $m$. To avoid ambiguity we denote the set of all such  diagrams by $\mathcal{O}_{k,m,n}$. The boundary vertices are decorated either by $a$ and $b$ or by the arguments $\alpha$'s. The arguments of $S_N$ are written in the order the vertices appear on the boundary of the disk diagram in the counterclockwise direction. The starting point is taken to be to the left of the arrow. This is called the \textit{boundary ordering}. Moreover, we will always consider the semicircle that one finds when traversing from $a$ to $b$ clockwise to be the northern semicircle and its complement the southern semicircle.

To write down an analytical expression for $S_N$ we also need to decorate the red lines with $2$-vectors $\vec{q}_i = (u_i,v_i)$, one for each line. The label $i$ increases from $b$ to $a$ along the diameter. This is referred to as the \textit{bulk ordering}. The structure map $S_N$ is given then by an integral over a bounded domain $\mathbb{V}_{N-1}\subset \mathbb{R}^{2(N-1)}$ parameterized by the $u$'s and $v$'s. The definition of the integration domain involves the bulk ordering, while the definition of the integrand uses the boundary ordering. Fig. \ref{SD1} shows how labels $\alpha_i$ and vectors $\vec{q}_i$ are assigned to a disk diagram. The straight arrow illustrates the bulk ordering, while the curved arrow displays the boundary ordering.
 
\begin{figure}
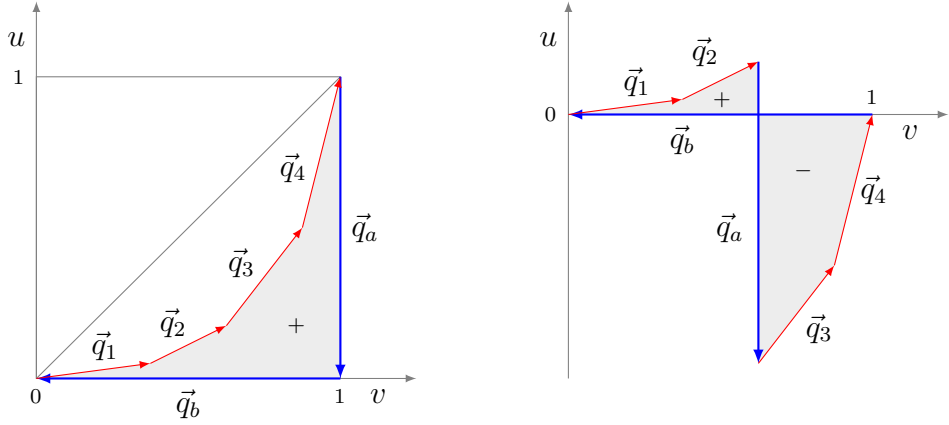

    \centering


\caption{Left panel: a swallowtail associated with the disk diagram in Fig. \ref{SD1}. The number $\frac{1}{2}|\vec{q}_a,\vec{q}_b,\vec{q}_1,\vec{q}_2,\vec{q}_3,\vec{q}_4|$ is  the area enclosed by the swallowtail. Right panel: a self-intersecting polygon  $(\vec{q}_1,\vec{q}_2,\vec{q}_a,\vec{q}_3,\vec{q}_4,\vec{q}_b)$ and its oriented area $\frac{1}{2}|\vec{q}_1,\vec{q}_2,\vec{q}_a,\vec{q}_3,\vec{q}_4,\vec{q}_b|$.}
\label{ST}
\end{figure}

\paragraph{Integration domains.} To associate an integration domain to a decorated disk diagram of $\mathcal{O}_{k,m,n}$, we treat the pairs $(u_i,v_i)$ as the coordinates of  vectors $\vec{q}_i$ on  an affine plane. We also add the pair of vectors $\vec{q}_a=(-1,0)$ and $\vec{q}_{b}=(0,-1)$ associated with the boundary vertices $a$ and $b$. The vectors are assumed to form a closed polygon chain, that is,
\begin{align}
    \vec{q}_1+\cdots +\vec{q}_N+ \vec{q}_a+\vec{q}_b=0\,,
\end{align}
or equivalently, 
\begin{align} \label{Dom1}
    u_1 + \dots + u_N = 1 =v_1 + \dots + v_N  \,.
\end{align}
We also require that 
\begin{equation}\label{Dom2}
    u_i\geq 0 \quad \mbox{and}\quad v_i\geq 0\,,\qquad i=1,\ldots,N\,.
\end{equation}
Suppose  that, moving along the diameter from $b$ to $a$ (bulk ordering) we pass through the $i$-th vertex at time $t_i=u_i/v_i$. Then the chronological ordering implies the following chain of inequalities: 
\begin{align} \label{Dom3}
   0\leq  \frac{u_1}{v_1} \leq \frac{u_2}{v_2}\leq \dots \leq \frac{u_N}{v_N} \leq \infty\,.
\end{align}
This, in turn, implies\footnote{To illustrate this, assume $u_1 > v_1$. Then, \eqref{Dom3} implies that $u_i > v_i$ for $i=1,\dots,N$, while the closure constraint \eqref{Dom1} requires $\sum_{i=1}^{N} u_i = \sum_{i=1}^{N} v_i = 1$, which leads to a contradiction. The same logic can be applied to the last time in the chain.} 
\begin{align} \label{Dom4}
    u_1 \leq v_1 \,, \quad v_N \leq u_N\,.
\end{align}
Together, Eqs. (\ref{Dom1}), (\ref{Dom2}) (\ref{Dom3}), and (\ref{Dom4}) define a bounded domain $\mathbb{V}_{N-1} \subset\mathbb{R}^{2(N-1)}$ for $N\geq 1$. In the special case that $N=0$, the domain is empty and there is no integration taking place. The integration domain admits a nice  visualization on the plane. The vectors $\vec{q}_i$ form  a maximally concave polygon inscribed into a unit square, see the left panel of Fig. \ref{ST}. The two acute angles and the right angle correspond to the fixed unit vectors $\vec{q}_a$ and $\vec{q}_{b}$; the other interior angles  of the polygon are concave.  In \cite{Sharapov:2022wpz, Sharapov:2022nps}, such polygons were called `swallowtails'. We note that $\mathbb{V}_{N-1}$ admits a $\mathbb{Z}_2$-symmetry by swapping the variables $u_i \leftrightarrow v_i$ and reversing their labels. Visually, this can be interpreted as a mirror reflection in the diagonal from $(1,0)$ to $(0,1)$ in Figure \ref{ST}.

\paragraph{Integrands.} The integrand associated with a disk diagram of $\mathcal{O}_{k,m,n}$ is given by a family of polydifferential operators with $N+2$ arguments since ${W}=V[-1]\oplus V^\ast$, where $V=\mathbb{C}[y^A]$ and $V^\ast=\mathbb{C}[[y^A]]$. The family is parametrized by the points of the integration domain $\mathbb{V}_{N-1}$. The poly-differential operators in question are  defined through compositions of elementary endomorphisms of $TW$: 
\begin{equation}\label{pij}
\begin{array}{c}
    p_{ij}(w_1\otimes \cdots \otimes w_i\otimes \cdots\otimes  w_j\otimes \cdots \otimes w_n)\\[3mm]
  \displaystyle  = \epsilon^{AB}\, w_1\otimes \cdots \otimes \frac{\partial w_i}{\partial y^A}\otimes \cdots\otimes  \frac{\partial w_j}{\partial y^B}\otimes \cdots \otimes w_n   
    \end{array}
\end{equation}
for all $w_i\in W$ and $n>1$. 
By definition, $p_{ij}=-p_{ji}$. The operators $p_{ij}$ act by partial derivatives on the  tensor factors labelled by $i$ and $j$.  Most often we will use the individual notation for the elements of $V[-1]$ and write e.g.
\begin{equation}
    p_{a, 2}(a\otimes \alpha_1\otimes b\otimes \alpha_2)=\epsilon^{AB} \frac{\partial a}{\partial y^A}\otimes \alpha_1\otimes b\otimes \frac{\partial\alpha_2}{\partial y^B}\,,
\end{equation}
where $a,b \in V[-1]$ and $\alpha_1,\alpha_2 \in V^\ast$. Finally, we let $\mu: TW\rightarrow \mathbb{C}$ denote the multilinear map that takes each homogeneous element $w_1\otimes w_2\otimes \cdots\otimes w_n$ to the complex number $w_1(0)w_2(0)\cdots w_n(0)$, where $w(0)$ is the constant term of the power series $w(y)$. The poly-differential operators associated with disk diagrams of $\mathcal{O}_{k,m,n}$ have the following general structure:
\begin{align*}
    \begin{aligned}
        F_{uv}(&\alpha_1,\dots,\alpha_k,a,\alpha_{k+1},\ldots, \alpha_{k+m},b,\alpha_{k+m+1},\dots,\alpha_{k+m+n+1})
        =\\
        &\mu\circ I(p_{a,b}, p_{a,i}, p_{b,j})(\alpha_1 \otimes \dots \otimes \alpha_n \otimes a\otimes \alpha_{n+1}\otimes\cdots\otimes \alpha_{n+m}\otimes b \otimes \alpha_{k+m+1} \otimes \alpha_{k+m+n+1})\,.
    \end{aligned}
\end{align*}
Here $I$ is a power series in the $p$'s, $u$'s, and $v$'s. Since the operators (\ref{pij}) pairwise commute, there is no ordering ambiguity.  
Notice that the operator $I$ does not involve elementary operators $p_{ij}$ that hit the pairs of  elements $\alpha_i,\alpha_j\in V^\ast$. 
Otherwise, the operator $I$ would be ill-defined on the elements of $W$.\footnote{It may be ill-defined if $I$ depends on $p_{i,j}$ because $I$ is a formal power series and elements from $V^\ast$ are series as well. That $I$ does not depend on $p_{i,j}$ implies the perturbative locality of Chiral HiSGRA.} It remains to describe the construction of the operator $I$ by a given disk diagram. The construction goes as follows. 

It is convenient to organize all the vectors in a single  $2\times (N+2)$ array, $Q_D$, for a disk diagram $D\in\mathcal{O}_{k,m,n}$. The order of column vectors in the array corresponds to the boundary ordering. This reads
\begin{align} \label{QD}
    Q_D=(\vec{q}_{i_1}, \ldots, \vec{q}_{i_{N+2}}) \,,
\end{align}
with the labels on the vectors assigned according to the boundary ordering. We will often provide the matrix $Q$ that corresponds to the {\it canonical ordering} associated with $S_{N}(a,b,\alpha_1,\dots,\alpha_N)$, given by
\begin{align}\label{Q}
    Q = \begin{pmatrix}
         -1& 0& u_1 & \dots & u_N \\
         0 & -1& v_1 & \dots & v_N
    \end{pmatrix}=(\vec{q}_a, \vec{q}_b, \vec{q}_1,\ldots, \vec{q}_{N})\,.
\end{align}
From this, one can construct the matrix $Q_D$ for a particular diagram.

Together with $Q_D$, we assign the $2\times (N+2)$ array
\begin{equation}\label{QP}
    P_D=( \vec{r}_{1},\ldots , \vec{r}_{k},\vec{r}_a,\vec{r}_{k+1},\dots,\vec{r}_{k+m},\vec{r}_b,\vec{r}_{k+m+1},\dots,\vec{r}_{k+m+n+1})
\end{equation}
to the diagram $D \in \mathcal{O}_{k,m,n}$. Here the $r$-vectors are assigned according to the boundary ordering. Whenever one encounters the vertices $a$ and $b$, one fills up $P_D$ with $\vec{r}_a=(0,0)$ and $\vec{r}_b=(0,0)$, respectively. For the $\alpha$'s one inserts the vectors $\vec{r}_i=(p_{a,i},p_{b,i})$, with $i$ increasing counterclockwise. Finally, to each matrix $Q_D$ we assign a quadratic polynomial in the $u$'s and $v$'s defined by the formula
\begin{equation}
    |Q_D|=\sum_{k<l}\det (\vec{q}_{i_k}, \vec{q}_{i_l})\,.
\end{equation}
In other words, $|Q_D|$ is the sum of all $2\times 2$ minors of the matrix $Q_D$. Like the integration domain $\mathbb{V}_{N-1}$, the number $|Q_D|$ admits a simple visualisation. The column vectors of $Q_D$, ordered from left to right, define a closed polygon chain on the affine plane. In general, the polygon is self-intersecting and splits into two regions with opposite orientations, 
see Fig. \ref{ST}. Then $|Q_D|$ is nothing but two times the oriented area of the polygon.

With the data above, we define the operator $I_D$ as 
\begin{align}\label{Iddd}
    I_D=s_{D} (p_{a,b})^{N-1} \exp\Big(\text{Tr}(P_DQ_D^T) + \lambda |Q_D| p_{a,b}\Big)\,,
\end{align}
where $\lambda$ a free parameter, which is related to the cosmological constant; $s_{D}=(-1)^m$, with $m$ the number of elements $\alpha_i$ found in the southern semicircle of the corresponding disk diagram. A structure map $S(\ldots)$ of the pre-CY algebra with a given ordering of the arguments is defined by summing operators $I_D$ over all disk diagrams with this order of the arguments:
\begin{align}
    S_{N}(\alpha_1,\dots,\alpha_{k},a,\alpha_{k+1},\ldots, \alpha_{k+m}, b, \alpha_{k+m+1}, \dots, \alpha_{k+m+n+1})&= \sum_{D\in \mathcal{O}_{k,m,n}} I_D\,,
\end{align}
where the arguments $a$, $b$ and $\alpha_i$ are implicit on the right. The argument to the left of the one marked with an arrow becomes the first argument of $S$. 

\paragraph{Vertices.}
In order to write the equations of motion  we  need the expression for the components of the structure maps $\hat{S}=(\mathcal{V}, \mathcal{U})$ defined by Eq. (\ref{vf}), using the natural pairing
\begin{align*}
    \langle\bullet,\bullet\rangle: V[-1] \otimes V^\ast \rightarrow \mathbb{C} \,,
\end{align*}
which is given by
\begin{align} \label{pairing}
    \langle f(y_1) , g(y_2) \rangle = - \langle g(y_2) , f(y_1) \rangle = \exp[p_{1,2}]f(y_1)g(y_2)|_{y_1=y_2=0}\,,
\end{align}
for $f(y_1) \in V[-1]$ and $g(y_2) \in V^\ast$. The argument on the boundary marked with an arrow is the one to be removed via the nondegenerate symplectic structure. We then find
\begin{align} \label{V}
    \begin{aligned}
        &-S_{N-1}(\alpha_1,\dots,\alpha_m,b,\alpha_{m+1},\ldots, \alpha_{m+n},a)=\\
        &=- \langle  \mathcal{U}(\alpha_{1},\ldots,\alpha_{m},b,\alpha_{m+1},\dots,\alpha_{m+n}),a\rangle = \langle a,\mathcal{U}(\alpha_{1},\ldots,\alpha_{m},b,\alpha_{m+1},\dots,\alpha_{m+n})\rangle\,,\\
        &S_{N-1}(\alpha_1,\dots,\alpha_k,a,\alpha_{k+1},\ldots, \alpha_{k+m},b)=\\
        &=\langle  \mathcal{U}(\alpha_{1},\ldots,\alpha_{k},a,\alpha_{k+1},\dots,\alpha_{k+m}),b\rangle=-\langle  b,\mathcal{U}(\alpha_{1},\ldots,\alpha_{k},a,\alpha_{k+1},\dots,\alpha_{k+m})\rangle\,,\\
        &S_{N}(\alpha_1,\dots,\alpha_k,a,\alpha_{n+1},\ldots, \alpha_{k+m},b,\alpha_{k+m+1},\dots,\alpha_{k+m+n+1})=\\
        &\langle\mathcal{V}(\alpha_1\ldots,\alpha_{k}, a,\alpha_{k+1},\ldots,\alpha_{k+m},b, \alpha_{k+m+1},\ldots,\alpha_{k+m+n}), \alpha_{k+m+n+1}\rangle\,,
    \end{aligned}
\end{align}
where we suppressed the $y$-dependence of the vertices. Notice that the minus sign was added in the first line, as to compensate for the swapping of $a$ and $b$. Using these relations, the $\mathcal{U}$-vertex and $\mathcal{V}$-vertex can be extracted by replacing
\begin{align*}
    \begin{aligned}
        p_a &\rightarrow p_0  \text{\quad or\quad}  p_b \rightarrow p_0 &\text{and} &&p_{k+m+n+1} &\rightarrow -p_0 \,, 
    \end{aligned}
\end{align*}
respectively, where $p_0^A\equiv y^A$ is the output argument. This yields
\begin{align} \label{vertices}
    \begin{aligned}
        &\mathcal{U}(\alpha_{1}\ldots,\alpha_{m}, b,\alpha_{m+1},\ldots,\alpha_{m+n}) \ni s_{D} \int_{\mathbb{V}_{m+n-1}} p_{0,b}^{m+n-1} \exp\Big(\text{Tr}(P_DQ_D^T) + \lambda |Q_D| p_{0,b}\Big) \,,\\
        &\mathcal{U}(\alpha_{1}\ldots,\alpha_{m}, a,\alpha_{m+1},\ldots,\alpha_{m+n}) \ni (-1)^{m+n}s_{D} \int_{\mathbb{V}_{m+n-1}} p_{0,a}^{m+n-1} \exp\Big(\text{Tr}(P_DQ_D^T) + \lambda |Q_D| p_{0,a}\Big) \,,\\
        &\mathcal{V}(\alpha_{1}\ldots,\alpha_{k}, a,\alpha_{k+1},\ldots,\alpha_{k+m},b, \alpha_{k+m+1},\ldots,\alpha_{k+m+n}) \ni\\
        &\qquad\qquad\qquad\qquad\qquad\qquad\qquad\ni s_{D} \int_{\mathbb{V}_{k+m+n}}p_{a,b}^{k+m+n} \exp\Big(\text{Tr}(P_DQ_D^T) + \lambda |Q_D| p_{a,b}\Big)\,,
    \end{aligned}
\end{align}
where we relabeled the $\alpha$'s in the $\mathcal{U}$-vertices for convenience. Note that we used $\ni$ instead of $=$ to indicate that what is on the r.h.s. is a specific contribution to a given vertex, while the vertex is a sum over all contributions with the same order of the arguments. The matrices $P_D$ for the $\mathcal{U}$- and $\mathcal{V}$-vertices are constructed according to the boundary ordering of the corresponding diagram from the $P$ matrices
\begin{align*}
     \begin{pmatrix}
            0 & 0 & p_{0,1} & \dots & p_{0,n+m} \\
            0 & 0 & p_{b,1} & \dots & p_{b,n+m}
        \end{pmatrix} \quad\text{and}\quad \begin{pmatrix}
            0 & 0 & p_{a,1} & \dots & p_{a,k+n+m} & p_{0,a} \\
            0 & 0 & p_{b,1} & \dots & p_{b,k+n+m} & p_{0,b}
        \end{pmatrix} \,,
\end{align*}
respectively, and $Q_D$ constructed from
\begin{align*}
    Q= (\vec{q}_a,\vec{q}_b,\vec{q}_1,\dots,\vec{q}_N)\,,
\end{align*} 
for $N=n+m$ and $N=k+n+m+1$, respectively. Notice that we provided two ways to extract a $\mathcal{U}$-vertex. Obviously, they should be identical, while at first sight they seem different. For instance, the matrix $Q_D$ is constructed differently in both cases, as the arrow is placed in a different position and therefore they have a different boundary ordering. However, if we find two diagrams that produce the same $\mathcal{U}$-vertex, while using the different methods, it is easy to see that they are just the reversed versions of each other. After renaming $b\rightarrow a$, we see that the matrix $P_D$ only differs by swapping the rows. The $\mathbb{Z}_2$-transformation on the domain allows one to exactly identify both realizations, as this is equivalent to swapping the rows in $P_D$ and reversing the order of the entries in $Q_D$. However, when $N=1$, there is no integration domain and this identification fails. Therefore, the vertices $\mathcal{U}(a,\alpha)$ and $\mathcal{U}(\alpha,a)$ are in fact different, see the examples below. This procedure of extracting vertices can also be represented diagrammatically as in Figs. \ref{SD2} and \ref{SD3}, where an element $\alpha$ or $a$, corresponding to the vertex the arrow is attached to, is removed. We refer to this arrow as the \textit{output arrow}, as it is now related to the output variables $p_0=y$. Although not necessary for most considerations, we will implicitly assume that the closure constraint is solved for the variables assigned to the output arrow. Note that while $S_N$ is a scalar, the $\mathcal{U}$- and $\mathcal{V}$-vertices are valued in $V^\ast$ and $V[-1]$, respectively.

\textit{Remark.} The vertices have an interesting property. If we symmetrize the vertices, i.e. we ignore additional tensor factors responsible for $\bry^{A'}$-dependence and for matrix extensions, and bring them to the same ordering, they satisfy 
\begin{align*}
    \begin{aligned}
        \sum_{k+m+n=N} \mathcal{V}(\alpha_{1}\ldots,\alpha_{k}, b,\alpha_{k+1},\ldots,\alpha_{k+m},a, \alpha_{k+m+1},\ldots,\alpha_{k+m+n}) &= 0 \,, \quad \text{for} \quad N \geq 1\,,\\
        \sum_{m+n=N} \mathcal{U}(\alpha_{1}\ldots,\alpha_{m}, b,\alpha_{m+1},\ldots,\alpha_{m+n}) &= 0\,, \quad \text{for} \quad N \geq 2\,,
    \end{aligned}
\end{align*}
for $\lambda = 0$. Here, the sum is over all vertices that take the same total number of arguments. In other words, if the pre-CY algebra of this section is $\mathbb{A}$ and $B$ is any associative commutative algebra then for $\lambda=0$ the structure maps $L_\infty$-algebra obtained from $\mathbb{A}\otimes B$ all vanish, except for $\mathcal{V}(\omega,\omega)$ and $\mathcal{U}(\omega,C)$, which describe the free theory.

It is easy to see why this is the case for $\mathcal{V}$-vertices: after symmetrizing, the expression for zero cosmological constant changes only by a sign when flipping a red line from the northern to the southern hemisphere and vice versa. As each red line connected to an element of $V^\ast$ can be flipped, there are as many positive as negative equal contributions, which proves the above relation. Since the $\mathcal{U}$-vertices can be derived from the $\mathcal{V}$-vertices through \eqref{V} using the cyclic property of $S$, the relation is easily seen to hold for them too.

\begin{figure}[!ht]
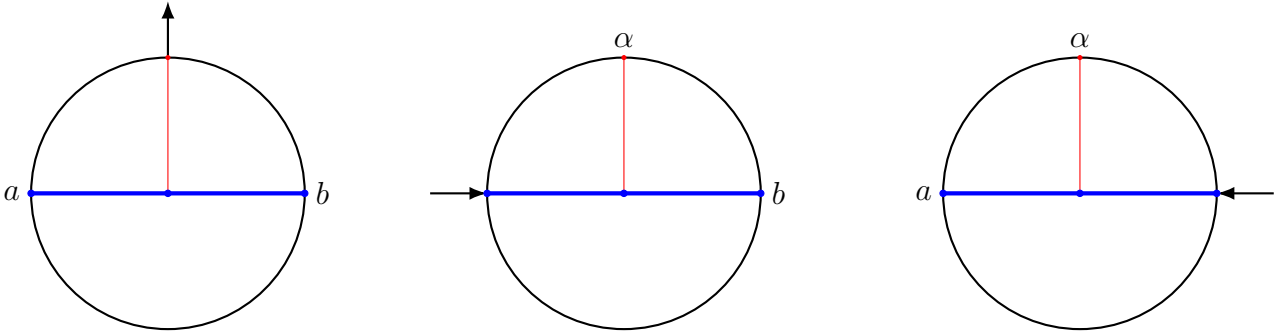

\centering
%
\caption{Disk diagrams contributing to $\mathcal{V}(a,b)$, $\mathcal{U}(b,\alpha)$ and $\mathcal{U}(\alpha,a)$, from left to right.}\label{ex1}
\end{figure}

\paragraph{Examples.} Let us illustrate the general recipe above with a few examples. We start with some lowest order vertices that will be used in Sec. \ref{sec:example} to prove the $A_\infty$-relations with zero or one $\alpha$. Fig. \ref{ex1} shows the disk diagrams contributing to the expressions
\begin{align} \label{quadV}
    \begin{aligned}
    \mathcal{V}(a,b) &= \exp[p_{0,a} + p_{0,b} + \lambda p_{a,b}] \,, \\
    \mathcal{U}(b,\alpha) &= \exp[p_{0,1} + p_{b,1} + \lambda p_{0,b} ] \,, \\
    \mathcal{U}(\alpha,a) &= - \exp[p_{a,1} + p_{0,1} - \lambda p_{0,a}] \,.
    \end{aligned}
\end{align}
In order to write these vertices as in \eqref{vertices}, one associates to them the matrices
\begin{align*}
    \begin{aligned}
        Q_D &= 
 \,,
    \end{aligned}
\end{align*}
respectively. The sign in the formula for $\mathcal{U}(\alpha,a)$ consists of two factors: $(-)^{m+n}=-1$, where $m=1$, $n=0$, see the second line of \eqref{vertices}; $s_D=1$, which is the number of red lines in the southern semicircle (none). 

The disk diagram relevant for the $\mathcal{V}$-vertices at the next order are given in Fig. \ref{ex2} and correspond to the expressions
\begin{align}\label{cubicV}
    \begin{aligned}
        \mathcal{V}(a,b,\alpha) &= \int_{\mathbb{V}_{1}} p_{a,b}  \exp[u_2 p_{0,a} + v_2 p_{0,b} + u_1 p_{a,1} +v_1 p_{b,1} +\lambda A_1 p_{a,b}] \,, \\
        \mathcal{V}(a,\alpha,b) &= -  \int_{\mathbb{V}_{1}} p_{a,b} \exp[u_2 p_{0,a} + v_2 p_{0,b} + u_1 p_{a,1} +v_1 p_{b,1} +\lambda A_2 p_{a,b}] + \\ 
        &-\int_{\mathbb{V}_{1}} p_{a,b} \exp[u_1 p_{0,a} + v_1 p_{0,b} + u_2 p_{a,1} +v_2 p_{b,1} +\lambda A_3 p_{a,b}]\,,\\
        \mathcal{V}(\alpha,a,b) &=\int_{\mathbb{V}_{1}} p_{a,b} \exp[u_1 p_{0,a} + v_1 p_{0,b} + u_2 p_{a,1} +v_2 p_{b,1} +\lambda A_4 p_{a,b}] \,,
    \end{aligned}
\end{align}
with
\begin{align*}
    \begin{aligned}
        A_1 =& 1+u_1+u_2-v_1-v_2+u_1v_2-u_2v_1 \,,& A_2 =& 1-u_1+u_2-v_1-v_2+u_1v_2-u_2v_1 \,,\\
        A_3 =& 1+u_1-u_2-v_1-v_2-u_1v_2+u_2v_1 \,, & A_4 =&1+u_1-u_2-v_1+v_2-u_1v_2+u_2v_1 \,.
    \end{aligned}
\end{align*}
To these vertices one associates the matrices
\begin{align*}
    \begin{aligned}
        Q_D &= \begin{pmatrix}
            -1 & 0 & u_1 & u_2 \\
            0 & -1 & v_1 & v_2
        \end{pmatrix} \,, & P_D & = \begin{pmatrix}
            0 & 0 & p_{a,1} & p_{0,a} \\
            0 & 0 & p_{b,1} & p_{0,b}
        \end{pmatrix} \,,\\
        Q_D &= \begin{pmatrix}
            -1 & u_1 & 0 & u_2 \\
            0 & v_1 & -1 & v_2
        \end{pmatrix} \,, & P_D & = \begin{pmatrix}
            0 & p_{a,1} & 0 & p_{0,a} \\
            0 & p_{b,1} & 0 & p_{0,b}
        \end{pmatrix} \,, \\
        Q_D &= \begin{pmatrix}
            -1 & u_2 & 0 & u_1 \\
            0 & v_2 & -1 & v_1
        \end{pmatrix} \,, & P_D & = \begin{pmatrix}
            0 & p_{a,1} & 0 & p_{0,a} \\
            0 & p_{b,1} & 0 & p_{0,b}
        \end{pmatrix} \,,\\
        Q_D &= \begin{pmatrix}
            u_2 & -1 & 0 & u_1 \\
            v_2 & 0 & -1 & v_1
        \end{pmatrix} \,, & P_D & = \begin{pmatrix}
            p_{a,1} & 0 & 0 & p_{0,a} \\
            p_{b,1} & 0 & 0 & p_{0,b}
        \end{pmatrix} \,,
    \end{aligned}
\end{align*}
respectively. Here, the second and third line contribute to $\mathcal{V}(a,\alpha,b)$. The integration domain, $\mathbb{V}_{1}$, is the $2$-simplex, which is described by
\begin{align*} 
    \begin{aligned}
        0 &\leq u_1,u_2,v_1,v_2\leq 1 \,, & 0 &\leq \frac{u_1}{v_1} \leq \frac{u_2}{v_2} \leq \infty \,, & u_1+u_2=1&=v_1+v_2\,.
    \end{aligned}
\end{align*}
The `hidden constraints' are
\begin{align} \label{2-simplex2}
    0&\leq u_1\leq v_1\leq 1 \,, & 0 &\leq v_2 \leq u_2 \leq 1
\end{align}
and can be equivalently used to describe the domain $\mathbb{V}_1$.
\begin{figure}[!ht]
\centering

\caption{Disk diagrams contributing to $\mathcal{V}(a,b,\alpha)$, $\mathcal{V}(a,\alpha,b)$ and $\mathcal{V}(\alpha,a,b)$, from left to right.}\label{ex2}
\end{figure}

Figs. \ref{SD2} and \ref{SD3} show some more involved examples, yielding the expressions 
\begin{align*}
    \mathcal{U}(\alpha_1,\alpha_2,b,\alpha_3, \alpha_4,\alpha_5) \ni \int_{\mathbb{V}_4} p_{0,b}^4 \exp[\text{Tr}[P_DQ_D^T]+\lambda p_{0,b}|Q_D|] \,,
\end{align*}
with
\begin{align*}
    \begin{aligned}
        P_D &= %
\,.
    \end{aligned}
\end{align*}

The integration domain $\mathbb{V}_4$ is described by
\begin{align*}
    \begin{aligned}
        0&\leq u_i, v_i \leq 1 \,, & 0&\leq\frac{u_1}{v_1}\leq\frac{u_2}{v_2}\leq\frac{u_3}{v_3}\leq\frac{u_4}{v_4}\leq 1\,, & \sum_{i=1}^4 u_i &= \sum_{i=1}^4 v_i = 1\,.
    \end{aligned}
\end{align*}

\begin{figure}[!ht]
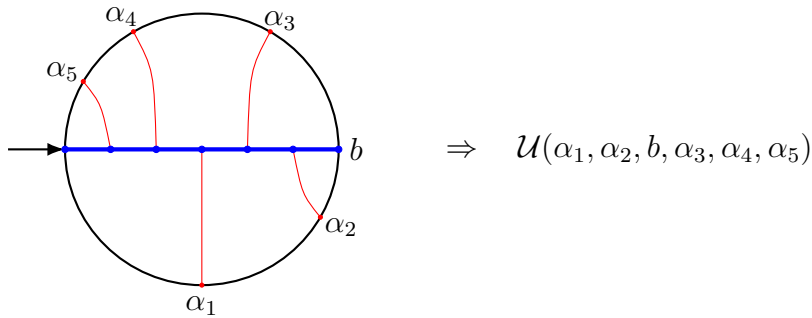

\centering
%
\caption{A disk diagram that contributes to  $\mathcal{U}$. }\label{SD2}
\end{figure}
\begin{figure}[h!]
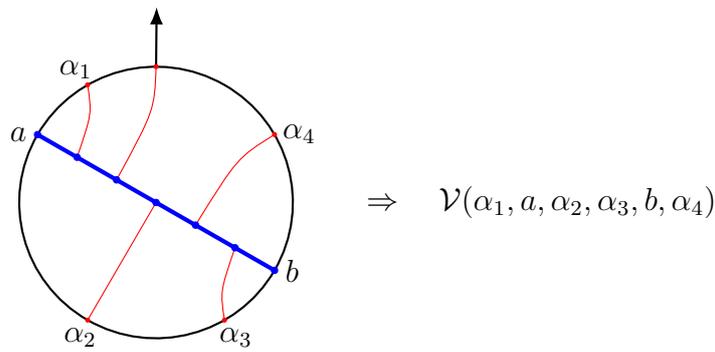

\centering
%
\caption{A disk diagram for $\mathcal{V}$. }\label{SD3}
\end{figure}

\begin{figure}[h!]
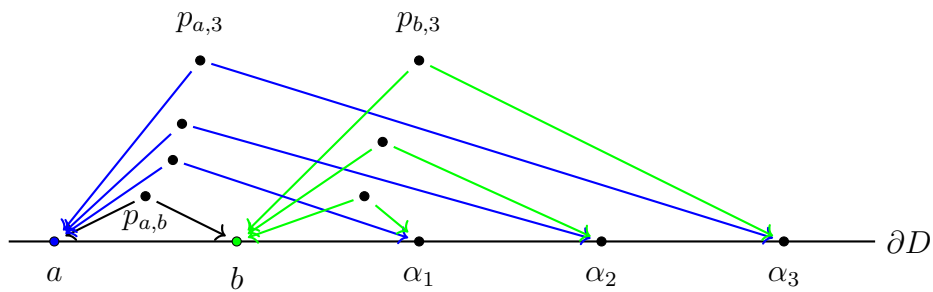

\centering

\caption{A typical Kontsevich-like graph contributing to $S(a,b,\alpha_1,\alpha_2,\alpha_3)$. $D$ is the upper half-plane and $\pl D$ is its boundary.}\label{formality}
\end{figure}

\paragraph{A more general formality?} The above construction of interaction vertices is a piece of clear evidence that (Shoikhet--Tsygan--)Kontsevich's formality can be extended further. We could only see a small piece of this hypothetical extension because our Poisson structure $\epsilon^{AB}$ is constant, nondegenerate and two-dimensional. Therefore, genuine bulk vertices of Kontsevich-like graphs are absent and all vertices have legs on the boundary. In addition, the graphs can be resumed into simple $\exp$-like generating functions $\mathcal{V}$ and $\mathcal{U}$ defined above. A generic $A_\infty$-map $m_n$, i.e. $\mathcal{V}$ or $\mathcal{U}$, can be Taylor-expanded to reveal 
\begin{align}
    m_n(f_1,\ldots,f_n)&=\sum_\Gamma w_{\Gamma}\, \mathcal{W}_\Gamma (f_1\otimes   \cdots \otimes f_n)\,, \qquad f_i\in W\,,
\end{align}
where the sum is over certain graphs $\Gamma$, $w_\Gamma$ are weights associated to $\Gamma$ and $\mathcal{W}_\Gamma$ are certain poly-differential operators (Taylor coefficients of various $I_D$, \eqref{Iddd}). Similar to the Moyal--Weyl case, the graphs $\Gamma$ are built from simple 'wedges' that represent $p_{\bullet,\bullet}$, see Fig. \ref{formality}.
What is different from the Moyal--Weyl case are the weights that are given by the integrals over the configuration space of concave polygons. The integrands are polynomials in $u_i$ and $v_i$. Also, there are no contractions between $\alpha$'s.

\section{A proof via Stokes' theorem}
\label{sec:proof}

\begin{figure}[!ht]
\centering
\begin{tikzpicture}[scale=0.3]
\draw[thick](0,0) circle (4);
\draw[thick](10,0) circle (4);
\draw [thick, -Latex] (6,0) -- (4,0);
\coordinate [label=left: { $[S,S]=0\quad \Leftrightarrow\qquad\quad $}] (B) at (-6.5,0);
 \draw (-6.5,0) node[gray, scale=5]{$\Sigma$\;} ;
 \coordinate [label=below: { $\mathcal{U}_{n}$}] (B) at (0,1);
  \coordinate [label=below: { $\mathcal{V}_{m}$}] (B) at (10,1);
   \coordinate [label=below: { ${}$}] (B) at (-7.5,-2.5);
   \draw[thick, ->] (10,5) arc (90:10:5);  
   \draw[thick, ->] (0,5) arc (90:10:5);  
    \filldraw [red] (90:4)+(10,0) circle (3pt);     
  \filldraw [red] (30:4)+(10,0) circle (3pt);  
   \filldraw [blue] (60:4)+(10,0) circle (5pt);      
   \filldraw [blue] (-120:4)+(10,0) circle (5pt);     
  \filldraw [red] (0:4)+(10,0) circle (3pt);    
   \filldraw [red] (-30:4)+(10,0) circle (3pt);    
    \filldraw [red] (-75:4)+(10,0) circle (3pt);     
\filldraw [red] (-150:4)+(10,0) circle (3pt);       
    \filldraw [red] (135:4)+(10,0) circle (3pt); 
    \filldraw [red] (180:4)+(10,0) circle (3pt); 
      \filldraw [red] (90:4) circle (3pt);     
  \filldraw [red] (30:4) circle (3pt);  
   \filldraw [red] (60:4) circle (3pt);      
   \filldraw [red] (-120:4) circle (3pt);     
  \filldraw [blue] (0:4)circle (5pt);    
   \filldraw [red] (-30:4) circle (3pt);    
    \filldraw [red] (-75:4) circle (3pt);     
\filldraw [red] (-150:4) circle (3pt);       
    \filldraw [red] (135:4) circle (3pt); 
    \filldraw [blue] (180:4) circle (5pt);   
     \draw (17,0) node[black, scale=1]{$=0$} ;
    \end{tikzpicture}
    \caption{Graphical representation for the master equation. }\label{SD4}    
    \end{figure}
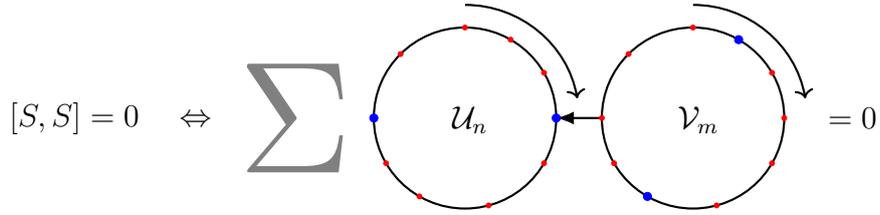

With the tools introduced in the previous section, the master equation (\ref{MC}) can be depicted as in Fig. \ref{SD4}. Here, the blue (red) vertices represent the blue (red) lines from the previous diagrams. The summation symbol accounts for collecting contributions from cyclic permutations of both disk diagrams, where we only allow a red and blue vertex to be connected to each other by the arrow between the disks. It also sums over the number of elements of red vertices in the disk diagrams.
The master equation $[S,S]=0$ looks as Fig. \ref{SD4} and
\begin{equation} \label{master}
    [S, S](a,\ldots,b,\ldots,c, \ldots)=\sum \pm S(\ldots, \mathcal{U}, \ldots)\pm S(\ldots, \mathcal{V},\ldots)\,.
\end{equation}
Using the natural pairing \eqref{pairing}, one can extract the $A_\infty$-relations that describe Chiral HiSGRA as
\begin{equation}\label{SSJ}
    [S,S](w, \ldots)= \langle w, J(\ldots)\rangle=0\quad \Rightarrow \quad J(\ldots)=0\,.
\end{equation}
Choosing $w$ to be an element of $V$, we get an infinite set of $A_\infty$-relations of the form: 
\begin{align} \label{A}
    \begin{aligned}
         &J_{N+3}(\bullet,\ldots,\bullet, a,\bullet,\ldots,\bullet, b,\bullet,\ldots,\bullet, c,\bullet,\ldots,\bullet)=\\
        =&\sum\mathcal{V}(\bullet,\dots,\bullet,\mathcal{V}(\bullet,\dots,\bullet,a,\bullet,\dots,\bullet,b,\bullet,\dots,\bullet),\bullet,\dots,\bullet,c,\bullet,\dots,\bullet)+\\
        -&\sum\mathcal{V}(\bullet,\dots,\bullet,a,\bullet,\dots,\bullet,\mathcal{V}(\bullet,\dots,\bullet,b,\bullet,\dots,\bullet,c,\bullet,\dots,\bullet),\bullet,\dots,\bullet)+\\
        +&\sum\mathcal{V}(\bullet,\dots,\bullet,\mathcal{U}(\bullet,\dots,\bullet,a,\bullet,\dots,\bullet),\bullet,\dots,\bullet,b,\bullet,\dots,\bullet,c,\bullet,\dots,\bullet)+\\
        -&\sum\mathcal{V}(\bullet,\dots,\bullet,a,\bullet,\dots, \bullet,\mathcal{U}(\bullet,\dots,\bullet,b,\bullet,\dots,\bullet),\bullet,\dots,\bullet,c,\bullet,\dots,\bullet)+\\
        +&\sum\mathcal{V}(\bullet,\dots,\bullet,a,\bullet,\dots,\bullet,b,\bullet,\dots,\bullet,\mathcal{U}(\bullet,\dots,\bullet,c,\bullet,\dots,\bullet),\bullet,\dots,\bullet)=0\,.
    \end{aligned}
\end{align}
Here $a,b,c \in V[-1]$ and bullets stand for $N$ elements $\alpha_i \in V^\ast$. 
The summations are over all possible combinations of vertices in each term, i.e., all ordered distributions of the $\alpha$'s in the arguments and contributions from all corresponding disk diagrams. A single $A_\infty$-relation consists of all terms with the same number of elements $\alpha$ before, between and behind $a,b,c$, i.e., with the same ordering and same total number of the $\alpha$'s, $N$.  
For $w\in V[-1]$, the same Eq. (\ref{SSJ}) yields one more set of $A_\infty$-relations:
\begin{align}\label{Urelations}
    \begin{aligned}
             &J_{N+2}(\bullet,\ldots,\bullet, a,\bullet,\ldots,\bullet, b,\bullet,\ldots,\bullet)=\\       
             =&\sum \mathcal{U}(\bullet,\dots,\bullet,\mathcal{U}(\bullet,\dots,\bullet,a,\bullet,\dots,\bullet),\bullet,\dots,\bullet,b,\bullet,\dots,\bullet) +\\
        + &\sum \mathcal{U}(\bullet,\dots,\bullet,\mathcal{V}(\bullet,\dots,\bullet,a,\bullet,\dots,\bullet,b,\bullet,\dots,\bullet),\bullet,\dots,\bullet) +\\
        + &\sum \mathcal{U}(\bullet,\dots,\bullet,a,\bullet,\dots,\bullet,\mathcal{U}(\bullet,\dots,\bullet,b,\bullet,\dots,\bullet),\bullet,\dots,\bullet)=0 \,.
    \end{aligned}
\end{align}   
The $A_\infty$-relations for $\mathcal{V}$-vertices \eqref{A} will be proven via Stokes' theorem in the next section, which also implies the $A_\infty$-relations for the $\mathcal{U}$-vertices \eqref{Urelations} through the master equation \eqref{master}. Therefore, we will only focus on the former from now on. The proof will follow the scheme
\begin{align} \label{StokesToA_infty}
    \begin{aligned}
        0 = \sum \int_{\mathbb{W}_{k,l,m,n}} d\Omega_{k,l,m,n}^a(y) + d&\Omega_{k,l,m,n}^c(y) = \sum \int_{\partial\mathbb{W}_{k,l,m,n}} \Omega_{k,l,m,n}^a(y) + \Omega_{k,l,m,n}^c(y)\\
        &\Updownarrow\\
        A_{\infty}&\text{-relations}
    \end{aligned}
\end{align}
where $k+m+l+n=N$. $\Omega_{k,l,m,n}^a(y)$ and $\Omega_{k,l,m,n}^c(y)$ are closed differential forms, called \textit{potentials}, with values in multilinear maps
\begin{align*}
    \Omega_{k,l,m,n}^{a,c}(y) : T^kV^\ast \otimes V[-1] \otimes T^lV^\ast \otimes V[-1] \otimes T^mV^\ast \otimes V[-1] \otimes T^nV^\ast \rightarrow V[-1]\,.
\end{align*}
In what follows,  we will suppress the $y$-dependence of the potentials. Similar to the $A_\infty$-relation, the summation in \eqref{StokesToA_infty} is over the total number of elements of $V^\ast$ and ways of distributing them, as we will see shortly. Apart from the potentials $\Omega_{k,l,m,n}^a$ and $\Omega_{k,l,m,n}^c$, we must consider bounded domains $\mathbb{W}_{k,l,m,n}\subset \mathbb{R}^{2N+1}$ with $\text{dim}(\mathbb{W}_{k,l,m,n})=2N+1$. In Sec. $\ref{sec:recipe}$, we will explain how the potentials and domains can be constructed from disk diagrams and how to evaluate the potentials on the boundary of the domains. We will also present expressions and disk diagrams for the $A_\infty$-relations. A couple of lowest order examples of \eqref{StokesToA_infty} follow in Sec. \ref{sec:example}, while in Sec. \ref{sec:leftOrdered} we provide a proof for all orders in a particular ordering, the what is called left-ordered case. Left-ordered means that the arguments, e.g. in a disk diagram or in an $A_\infty$-relation are ordered as $a,b,c,\alpha_1,\ldots,\alpha_N$, $a,b,c\in V[-1]$, $\alpha_i\in V^\ast$.  Sec. \ref{sec:allOrderings} contains the proof for a generic ordering.

The idea of the proof follows the formality theorems: since the potentials $\Omega$ are closed forms, the l.h.s. of \eqref{StokesToA_infty} obviously vanishes. At the same time, the r.h.s. of \eqref{StokesToA_infty} is given by the sum of potentials evaluated at the boundaries of $\mathbb{W}_{k,l,m,n}$. Domain $\mathbb{W}_{k,l,m,n}$ and potentials $\Omega$ are carefully designed in such a way that this sum gives the $A_\infty$-terms plus certain other terms that vanish by themselves. The genuine $A_\infty$-terms contain vertices nested into each other. Therefore, some boundaries of $\mathbb{W}_{k,l,m,n}$ reduce to the products of two configuration spaces that define vertices and $\Omega$ evaluated on such boundaries give exactly the nested vertices.

\subsection{Recipe} \label{sec:recipe}

The data $\Omega_{k,l,m,n}^a$, $\Omega_{k,l,m,n}^c$, and $\mathbb{W}_{k,l,m,n}$, entering (\ref{StokesToA_infty}), can all be encoded by disk diagrams decorated by the variables $u_i$, $v_i$, and $w_i$ that coordinatize $\mathbb{W}_{k,l,m,n}$. To understand these disk diagrams and their properties, let us first consider disk diagrams for the scalars
\begin{align} \label{potPairing}
    \begin{aligned}
        &\langle \Omega^a_{k,l,m,n},\alpha_{k+1}\rangle & \text{and} && \langle \Omega^c_{k,l,m,n},\gamma_{m+1}\rangle \,,
    \end{aligned}
\end{align}
with $\alpha_{k+1},\gamma_{m+1} \in V^\ast$ and the natural pairing defined in \eqref{pairing}. From here the disk diagrams for the potentials can be extracted in a similar fashion as before. These disk diagrams are constructed as follows:
\begin{itemize}
    \item Consider a circle. The interior will be referred to as the \textit{bulk} and the circle as the \textit{boundary}.
    \item Choose three distinct points on the boundary and label them $a,b,c$ counterclockwise. Consider the point at the center of the bulk, now called \textit{junction}, and connect this to each of the points $a,b,c$. These points correspond to elements of $V[-1]$. The lines are called $a$\textit{-leg}, $b$\textit{-leg} and $c$\textit{-leg}, correspondingly.
    \item Draw any number of lines connecting these legs to the boundary at either side of the legs. Lines are not allowed to intersect. Their endpoints at the boundary correspond to elements of $V^\ast$.
    \item Connect an arrow to one of the vertices on the boundary of the disk between $a$ and $c$, pointing away from the disk, i.e. there has to be one marked point on the boundary. If the arrow is connected to a vertex connected to the $a$-leg by a red line, the potential that can be extracted using \eqref{potPairing} is $\Omega_{k,l,m,n}^a$, while $\Omega_{k,l,m,n}^c$ can be found when it is connected to the $c$-leg.
    \item Label the points at the boundary that are connected to red lines $\alpha_i, \beta_i, \gamma_i, \delta_i$ if the lines emanate from the $a$-leg, $b$-leg, $c$-leg or are in between the red line connected to the arrow and the junction, respectively, and $i$ increases from the boundary to the junction. This way if the arrow is attached to an argument belonging to the $a$-leg, the arguments after the arrow are labelled $\delta_i$ and those in between $a$ and including the arrow are named $\alpha_i$. The subscripts $k,l,m,n$ on the potentials count the number of points with labels $\alpha_i, \beta_i, \gamma_i, \delta_i$, respectively, disregarding the label associated with the arrow.
    \item The diagram must contain at least one element of $V^\ast$, connected to the piece of the boundary between $a$ and $c$, which can be attached to either the $a$-leg or the $c$-leg. If the diagram contains more than one element of $V^\ast$, they have to be attached to at least two different legs.
\end{itemize}
\begin{figure}[!ht]
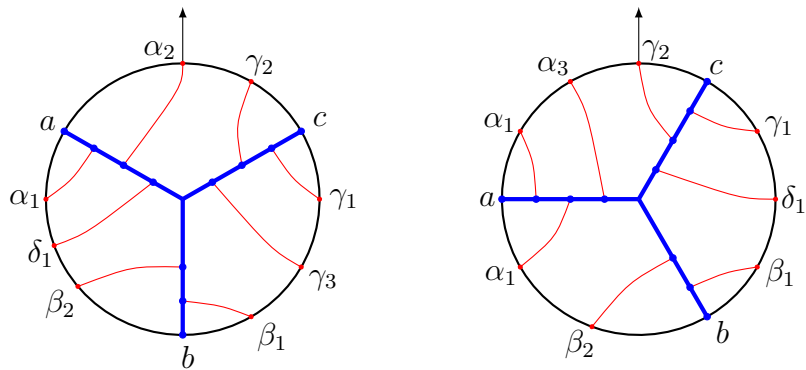

\centering
 \, 
\caption{The disk diagrams for $\langle\Omega_{1,2,3,1}^a(a,\alpha_1,\delta_1,\beta_2,b,\beta_1,\gamma_3,\gamma_1,c,\gamma_2),\alpha_2 \rangle$ and $\langle \Omega_{3,2,1,1}^c(\alpha_3,\alpha_1,a,\alpha_2,\beta_2,b,\beta_1,\delta_1,\gamma_1,c),\gamma_2\rangle$ on the left and right, respectively.}\label{SD5}
\end{figure}
Fig. \ref{SD5} shows two examples of such disk diagrams. The disk diagrams corresponding to the potentials $\Omega_{k,l,m,n}^a$ and $\Omega_{k,l,m,n}^c$ are now obtained by removing the element $\alpha_{k+1}$ or $\gamma_{m+1}$ in \eqref{potPairing}. This is visualized in the disk diagrams by removing the corresponding label, see  Fig. \ref{SD6}. 

The arrow also induces an ordering on the elements of $W$, which is counterclockwise around the circle starting from to the left of the arrow. Again, we will refer to this as the boundary ordering. The potentials $\Omega_{k,l,m,n}^{a,c}$ are poly-differential operators acting on elements of $W$. These should be read off according to the boundary ordering. Moreover, the elements of $W$ are generated by the $y_i$'s. For $a,b,c$ we write $y_a,y_b,y_c$, respectively and for elements of $V^\ast$ they are just $y_i$ with $i=1,\ldots, N$, assigned counterclockwise. Lastly, while it is cumbersome to specify the ordering of the elements of $W$ for generic potentials $\Omega_{k,l,m,n}^{a,c}$, we do specify the ordering with respect to the boundary ordering when we consider a particular potential. For general potentials we prefer to use {\it canonical ordering}, see below. The examples in Fig. \ref{SD6} correspond to poly-differential operators $\Omega_{1,2,3,1}^a(a,\alpha_1,\delta_1,\beta_2,b,\beta_1,\gamma_3,\gamma_1,c,\gamma_2)$ and $\Omega_{3,2,1,1}^c(\alpha_3,\alpha_1,a,\alpha_2,\beta_2,b,\beta_1,\delta_1,\gamma_1,c)$ acting on
\begin{align*}
    \begin{aligned}
        a(y_a) \alpha_1(y_1) \delta_1(y_2) \beta_2(y_3) b(y_b) \beta_1(y_4) \gamma_3(y_5) \gamma_1(y_6) c(y_c) \gamma_2(y_7)|_{y_{\bullet}=0} \,,\\
        \alpha_3(y_1) \alpha_1(y_2) a(y_a) \alpha_2(y_3) \beta_2(y_4) b(y_b) \beta_1(y_5) \delta_1(y_6) \gamma_1(y_7) c(y_c)|_{y_{\bullet}=0}
    \end{aligned}
\end{align*}
for the left and right disk diagrams, respectively.
The $A_\infty$-relations also consist of poly-differential operators acting on elements of $W$. The names and labels we assigned to elements of $V^\ast$ here are not the same as for the $A_\infty$-terms. This was done simply because it will be useful to keep track of what leg an element is attached to in the potentials. To relate the potentials to $A_\infty$-terms, one should rename and relabel the elements of $V^\ast$ accordingly. We now claim that an $A_\infty$-relation of a particular ordering can be rewritten as \eqref{StokesToA_infty} by considering contributions from all potentials of the same ordering, i.e. the same number of elements of $V^\ast$ before, between and after $a,b,c$. Therefore, we will no longer write the elements of $W$ as part of the expressions; they will be implied.

The procedure laid out in this section leads to two different types of disk diagrams: the arrow can be attached to the $a$-leg or $c$-leg, given rise to the disk diagrams for $\Omega_{k,l,m,n}^a$ or $\Omega_{k,l,m,n}^c$, respectively, and we refer to the diagrams as $a$-diagrams and $c$-diagrams. It is easy to see that following an $a$-diagram in the opposite boundary ordering, i.e. clockwise, and relabeling $a \leftrightarrow c$ yields a $c$-diagram. We may write
\begin{align} \label{flip}
    \Omega_{k,l,m,n}^a \leftrightarrow \Omega_{m,l,k,n}^c \,, \quad \text{with } a \leftrightarrow c 
\end{align}
and consequently, the potential $\Omega_{m,l,k,n}^c$ is accompanied by the same integration domain $\mathbb{W}_{k,l,m,n}$ as $\Omega_{k,l,m,n}^a$. A proper relabeling of the elements of $V^\ast$ according to which leg they are attached is also implied. Due to this relation between the diagrams, it will be possible to extract $c$-diagrams from $a$-diagrams and therefore we will focus mainly on the latter in the remainder of the text. We will make the above transformation more concrete, when we have all the necessary tools.

We will sometimes refer a special class of potentials as \textit{left/right-ordered}. These are the potentials with all elements of $V[-1]$ appearing before/after the elements of $V^\ast$ and they are given by $\Omega_{0,0,m,n}^a$ and $\Omega_{k,0,0,n}^c$ with the appropriate ordering, respectively. \eqref{StokesToA_infty} relates these potentials to left/right ordered $A_\infty$-terms, which are similarly defined as $A_\infty$-terms with all elements of $V[-1]$ appearing before/after the elements of $V^\ast$.

\begin{figure}
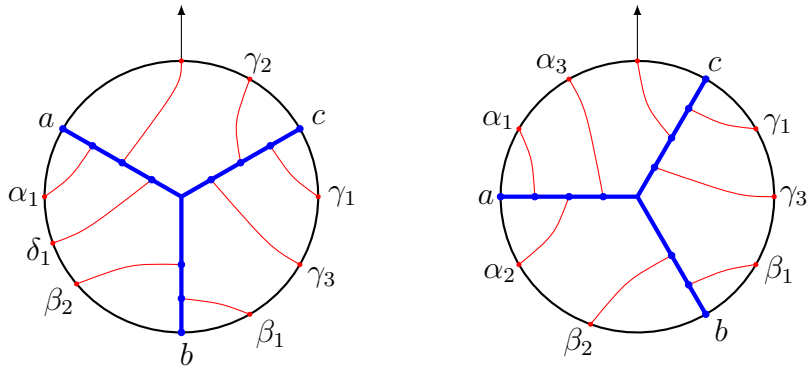

\centering

\caption{On the left a disk diagrams corresponding to $\Omega_{1,2,3,1}^a(a,\alpha_1,\delta_1,\beta_2,b,\beta_1,\gamma_3,\gamma_1,c,\gamma_2)$ and on the right a disk diagram corresponding to $\Omega_{3,2,1,1}^c(\alpha_3,\alpha_1,a,\alpha_2,\beta_2,b,\beta_1,\delta_1,\gamma_1,c)$.} \label{SD6}
\end{figure}
\paragraph{Domain.} When $k+l+m+n=N > 0$, the domain $\mathbb{W}_{k,l,m,n} \subset \mathbb{R}^{2N+1}$ can be read off from the disk diagrams, e.g. Fig. \ref{SD6}. For this purpose, one assigns a vector of variables $\vec{q}_{a,i}=(u^a_i,v^a_i,w^a_i)$, $\vec{q}_{b,i}=(u^b_i,v^b_i,w^b_i)$ and $\vec{q}_{c,i}=(u^c_i,v^c_i,w^c_i)$ to the red lines connected to the $a$-, $b$- and $c$-leg, respectively, and to the arrow, with $i$ increasing from boundary to junction. Additionally, we assign vectors $\vec{q}_a=(-1,0,0)$, $\vec{q}_b=(0,-1,0)$, $\vec{q}_c=(0,0,-1)$ to $a, b, c$ accordingly. Then one introduces the times $t^{uv}_i=u_i^\bullet/v_i^\bullet$, $t^{uw}_i=u_i^\bullet/w_i^\bullet$ and $t^{vw}_i=v_i^\bullet/w_i^\bullet$ and imposes chronological orderings along three different paths in the bulk to formulate the domain.
\begin{itemize}
    \item Path 1: one starts at $b$ and moves to $c$ and then from $c$ to $a$. This imposes a chronological ordering on the times $t^{uv}_i$.
    \item Path 2: one starts at $c$ and moves to $b$ and then from $b$ to $a$. This imposes a chronological ordering on the times $t^{uw}_i$.
    \item Path 3: one starts at $c$ and moves to $a$ and then from $a$ to $b$. This imposes a chronological ordering on the times $t^{vw}_i$.
\end{itemize}

\begin{figure}
    \centering

    \caption{Paths 1 - 3 from left to right, leading to the time ordering of times $t^{uv}_i$, $t^{uw}_i$ and $t^{vw}_i$, respectively.}
    \label{paths}
\end{figure}

As an additional condition, all $u$-, $v$- and $w$-variables take values between $0$ and $1$. The domain $\mathbb{W}_{k,l,m,n}$ is then described by
\begin{align} \label{DD3}
    \begin{aligned}
        0 \leq& u_i^\bullet,v_i^\bullet,w_i^\bullet \leq 1 \,, \quad \sum (u_i^\bullet,v_i^\bullet,w_i^\bullet)=(1,1,1) \,, \quad  \frac{v_i^a}{w_i^a}\frac{w_i^b}{u_i^b}\frac{u_i^c}{v_i^c}=1\,,\\
        0 \leq& \frac{u_1^b}{v_1^b} \leq \dots \leq \frac{u_{l}^b}{v_{l}^b} \leq \frac{u_m^c}{v_m^c} = \dots = \frac{u_1^c}{v_1^c} \leq \frac{u_{k+n+1}^a}{v_{k+n+1}^a} \leq \dots \leq \frac{u_1^a}{v_1^a}\leq \infty\,,\\
       0 \leq& \frac{u_1^c}{w_1^c} \leq \dots \leq \frac{u_m^c}{w_m^c} \leq \frac{u_{l}^b}{w_{l}^b} = \dots = \frac{u_1^b}{w_1^b} \leq \frac{u_{k+n+1}^a}{w_{k+n+1}^a} \leq \dots \leq \frac{u_1^a}{w_1^a}\leq \infty\,,\\
        0 \leq& \frac{v_1^c}{w_1^c} \leq \dots \leq \frac{v_m^c}{w_m^c} \leq \frac{v_{k+n+1}^a}{w_{k+n+1}^a} = \dots = \frac{v_1^a}{w_1^a}\leq \frac{v_{l}^b}{w_{l}^b} \leq \dots \leq \frac{v_1^b}{w_1^b} \leq \infty \,.
    \end{aligned}
\end{align}
Note that the paths described above run over some legs twice. This leads to the equalities in \eqref{DD3}, since for example
\begin{align*}
    \frac{u_m^c}{v_m^c} \leq \dots \leq \frac{u_1^c}{v_1^c} \leq \frac{u_1^c}{v_1^c} \leq \dots \leq \frac{u_m^c}{v_m^c} \quad \Rightarrow \quad  \frac{u_m^c}{v_m^c} = \dots = \frac{u_1^c}{v_1^c} \,.
\end{align*}
The second equation in the first line of \eqref{DD3} is called the \textit{closure constraint} and explicitly it reads
\begin{align*}                  \sum_{i=1}^{k+n+1}\vec{q}_{a,i}+\sum_{i=1}^l\vec{q}_{b,i}+\sum_{i=1}^m\vec{q}_{c,i} +\vec{q}_a+\vec{q}_b+\vec{q}_c= (0,0,0) \,.
\end{align*}
This condition allows one to solve for one vector $\vec{q}_{\bullet,i}$ in terms of the other vectors. The third equation in the first line deserves some further explanation. To the equalities in the chains of (in)equalities one can ascribe
\begin{align*}
    \begin{aligned}
        \alpha&=\frac{u_i^c}{v_i^c} \,, & \frac{1}{\beta} &= \frac{v_i^a}{w_i^a} \,, & \gamma &= \frac{u_i^b}{w_i^b} \,.
    \end{aligned}
\end{align*}
The vectors $\vec{q}_{a,i},\vec{q}_{b,i},\vec{q}_{c,i}\in\mathbb{R}^3$ are then restricted to planes characterized by $\alpha,\frac{1}{\beta},\gamma$, as for example one may write $\vec{q}_{c,i}=(\alpha v^c_i,v^c_i,w^c_i)$ for $i=1,\dots,m$. The constraint in \eqref{DD3} tells us that the planes are related by
\begin{align} \label{abc2}
    \gamma=\frac{\alpha}{\beta} \,.
\end{align}
However, this relation is only valid when $m,l,k+n \neq 0$. Otherwise, $\alpha,\beta,\gamma$ are independent. In practice, we will have
\begin{align*}
    \alpha &= \frac{u_1^c}{v_1^c} \,, & \beta &= \frac{1-\sum_{i=1}^{l} w_i^b-\sum_{i=1}^{m} w_i^c}{1-\sum_{i=1}^{l} v_i^b-\sum_{i=1}^{m} v_i^c} \,, & \gamma &= \frac{u_1^b}{w_1^b}\,.
\end{align*}
The second relation is obtained from the closure constraint for the $v$ and $w$ coordinates:
\begin{align*}
    \begin{aligned}
        \sum v_i^\bullet = 1 \quad &\Rightarrow \quad \sum_{i=1}^{k+n+1} v_i^a = 1 - \sum_{i=1}^{l} v_i^b - \sum_{i=1}^{m} v_i^c \,,\\
        \sum_{i=1}^{k+n+1} \beta v_i^a + \sum_{i=1}^{l}w_i^b + \sum_{i=1}^{m}w_i^c = 1 \quad &\Rightarrow \quad \beta = \frac{1-\sum_{i=1}^{l} w_i^b-\sum_{i=1}^{m} w_i^c}{\sum_{i=1}^{k+n+1}  v_i^a}=\frac{1-\sum_{i=1}^{l} w_i^b-\sum_{i=1}^{m} w_i^c}{1-\sum_{i=1}^{l} v_i^b-\sum_{i=1}^{m} v_i^c} \,.
    \end{aligned}
\end{align*}

The domain \eqref{DD3} contains a couple of `hidden' constraints on the variables. To illustrate this, suppose ${u_1^b}/{v_1^b} > 1$ and consider the first chain of (in)equalities, the $uv$-chain. This implies that $u_i^\bullet > v_i^\bullet$. Then, the closure constraint requires $\sum v_i^\bullet=1$ and we find $\sum u_i^\bullet > 1$, so the closure constraint cannot be satisfied for the $u$-variables. The same logic applied to the start and to the end of all three chains of (in)equalities leads to the additional constraints
\begin{align} \label{startEnd}
    \begin{aligned}
        0 \leq& u_1^b \leq v_1^b \leq 1 \,, \quad 0 \leq u_1^c \leq w_1^c \leq 1 \,, \quad 0 \leq v_1^c \leq w_1^c \leq 1 \,,\\
        0 \leq& v_1^a \leq u_1^a \leq 1 \,, \quad 0 \leq w_1^a \leq u_1^a \leq 1 \,, \quad 0 \leq w_1^b \leq v_1^b \leq 1 \,.
    \end{aligned}
\end{align}
However, one must be careful, as the domain changes significantly whenever $l=0$, in which case the first and last inequality are replaced by
\begin{align*}
    0 \leq& u_1^c \leq v_1^c \leq 1 \,, \quad 0 \leq w_1^a \leq v_1^a \leq 1\,,
\end{align*}
where in the latter we had some freedom to choose in which variables we express the inequalities. It turns out that the domain takes this form for left/right ordered $A_\infty$-relations, although not exclusively for this class. Since the start and end of the chains of (in)equalities yields constraints, the domain also takes a different form when $m=0$, in which case the second and third inequality in \eqref{startEnd} are replaced by
\begin{align} \label{startEnd2}
    0 \leq& u_l^b \leq w_l^b \leq 1 \,, \quad 0 \leq v_{k+n+1}^a \leq w_{k+n+1}^a \leq 1 \,.
\end{align}

In the special case $N=0$, the disk diagrams corresponding to $\Omega^{a,c}_{0,0,0,0}(a,b,c)$ contain no elements of $V^\ast$ and the above prescription for the domain breaks down. We manually define the domain $\mathbb{W}_{0,0,0,0} \subset \mathbb{R}$ to be described by
\begin{align*}
    0 \leq t \leq 1 \,.
\end{align*}

Furthermore, we note that the integration domain has a $\mathbb{Z}_2$-symmetry under swapping the $v$ and $w$ variables together with the labels $b$ and $c$ on all variables. Graphically, this means that swapping the $b$- and $c$-leg, together with all the lines attached to them, and renaming the elements, again yields a disk diagram to which a domain can be ascribed. This relates integration domains $\mathbb{W}_{k,l,m,n}$ and $\mathbb{W}_{k,m,l,n}$ by a coordinate transformation. 

A quick check shows that $\text{dim}(\mathbb{W}_{k,l,m,n})=2N+1$. The domain is described in \eqref{DD3} by a total of $3N+3$ coordinates. Not all coordinates are independent, as the description consists of some equalities; the chains of (in)equalities contain in total $N-2$ equalities, the closure constraint subtracts $3$ variables and, lastly, \eqref{abc2} adds $1$ more equality, which justifies the dimension. In case $m=0$ or $l=0$, the chains of (in)equalities yield $N-1$ equalities, the closure constraint adds $3$ and \eqref{abc2} is absent and again the right dimension is found. As the last remark, we note that the integration domain has a nice visual representation in the left-ordered case. This will be explained in Sec. \ref{sec:leftOrdered}.

\begin{figure}
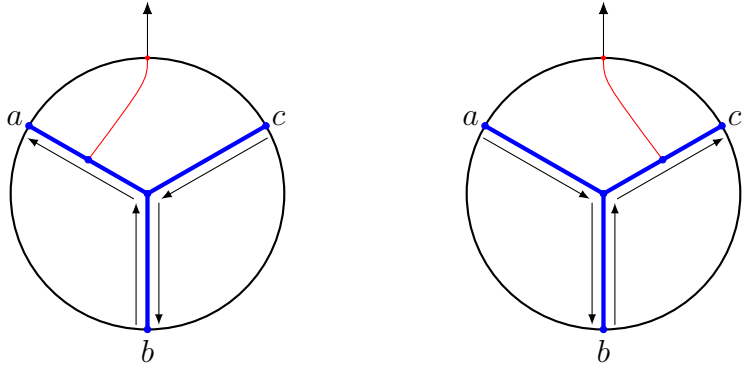

    \centering

    \caption{Canonical ordering for a potentials $\Omega_{k,l,m,n}^a$ and $\Omega_{k,l,m,n}^c$ on the left and right, respectively.}
    \label{canPot}
\end{figure}

\paragraph{Potential.} The potentials $\Omega_{k,l,m,n}^{a,c}$ are most conveniently defined through a slight detour. It is natural to define them first on a higher dimensional space and then get the actual potentials upon restricting them to $\mathbb{W}_{k,l,m,n}$. Let us first focus on the potentials $\Omega_{k,l,m,n}^a$. Consider a subspace $\mathbb{U}_{N} \subset \mathbb{R}^{3N}$ that is defined by
\begin{align} \label{masterDomain}
        0 &\leq u_i,v_i,w_i \leq 1 \,, & &\sum_{i=1}^{N+1}(u_i,v_i,w_i)=(1,1,1)\,.
\end{align}
In principle, any subspace that contains all $\mathbb{W}_{k,l,m,n}$ with dimension $2N+1$ fixed would work. We define the potential $\Omega_N^{a}$ on $\mathbb{U}_{N}$ by
\begin{align*}
    \Omega_N^{a} =\mu I_D \,,
\end{align*}
where $\mu$ is the measure and $I_D$ is the integrand. The measure reads
\begin{align} \label{measure}
    \begin{aligned}
        \mu =& \mu_1\wedge\dots\wedge\mu_N \,,\\
        \mu_{i} =&  p_{a,b} du_i \wedge dv_i + p_{a,c} du_i \wedge dw_i + p_{b,c} dv_i \wedge dw_i \,,\\
    \end{aligned}
\end{align}
and the integrand is given by
\begin{align} \label{pot}
    I_D = s_D\exp[\text{Tr}[P_D Q_D^T] + \lambda (p_{a,b}|Q_D^{12}|+p_{a,c}|Q_D^{13}|+p_{b,c}|Q_D^{23}|)] \,.
\end{align}
Here, $s_D$ is a sign that will be discussed at the end of this section. The matrix $Q_D$ is an array filled with the $q$-vectors according to the boundary ordering of the disk diagram. For example, the matrix $Q_D$ corresponding to the left disk diagram in Fig. \ref{SD6} reads
\begin{align*}
    Q_D =& (\vec{q}_a,\vec{q}_8,\vec{q}_6,\vec{q}_5,\vec{q}_b,\vec{q}_4,\vec{q}_3,\vec{q}_1,\vec{q}_c,\vec{q}_2,\vec{q}_7)\,.
\end{align*}
$|Q^{ij}_D|$ denotes the sum of all $2 \times 2$ minors composed of the $i$-th and $j$-th row of the matrix $Q_D$. A more explicit expression yields
\begin{align} \label{cosm}
    \begin{aligned}
        |Q_{D}^{12}| &= 1 + \sum_{i=1}^{N+1}(\sigma_{b,i} u_i - \sigma_{a,i} v_i + \sum_{j=1}^{N+1} \sigma_{i,j}u_i v_j) \,,\\
        |Q_{D}^{13}| &= 1 + \sum_{i=1}^{N+1}(\sigma_{c,i} u_i - \sigma_{a,i} w_i + \sum_{j=1}^{N+1}\sigma_{i,j}u_i w_j) \,,\\
        |Q_{D}^{23}| &= 1 + \sum_{i=1}^{N+1}(\sigma_{c,i} v_i - \sigma_{b,i} w_i + \sum_{j=1}^{N+1} \sigma_{i,j}v_i w_j) \,.
    \end{aligned}
\end{align}
Here, $\sigma_{I,J} = +1$ when the $J$-th element of $W$ appears before the $I$-th element in the boundary ordering and $\sigma_{I,J}=-1$ otherwise, for $I,J \in \{a,b,c,1,\dots,N+1\}$. The matrix $P_D$ is an array composed of the $r$-vectors in the following manner: going around the circle counterclockwise, starting to the left of the arrow, one fills $P_D$ with $\vec{r}_i$ for every element of $V^\ast$ or the arrow, with $i$ increasing counterclockwise, or with $\vec{r}_a=(-1,0,0)$, $\vec{r}_b=(0,-1,0)$, $\vec{r}_c=(0,0,-1)$ for the corresponding element of $V[-1]$. For example, the matrix $P_D$ corresponding to the left diagram in Fig. \ref{SD6} reads
\begin{align*}
    P_D=& (\vec{r}_a,\vec{r}_1,\vec{r}_2,\vec{r}_3,\vec{r}_b,\vec{r}_4,\vec{r}_5,\vec{r}_6,\vec{r}_c,\vec{r}_7,\vec{r}_8)\,.
\end{align*}
We will often present a matrix $Q$ when considering potentials with an unspecified ordering. We say that the entries of the matrix $Q$ are in the {\it canonical ordering}. This means that the first three entries are the vectors $\vec{q}_{a},\vec{q}_b,\vec{q}_c$, and then we insert the vectors $\vec{q}_{i}$ in the order visualized in Fig. \ref{canPot}. The matrix reads
\begin{align*}
    Q =& (\vec{q}_a,\vec{q}_b,\vec{q}_c,\vec{q}_1,\dots,\vec{q}_N) \,.
\end{align*}
This canonical ordering has the advantage to reduce to the matrix $Q_D$ for $D$ being a left ordered disk diagram with $k=l=0$. From $Q$ any matrix $Q_D$ may be constructed when the ordering is specified.

The potential $\Omega_{k,l,m,n}^{a}$ is now found by restricting $\Omega_N^{a}$ to $\mathbb{W}_{k,l,m,n} \subseteq \mathbb{U}_{N}$, i.e.,
\begin{align*}
    \Omega_{k,l,m,n}^{a} = \Omega_N^{a}\Big|_{\mathbb{W}_{k,l,m,n}} \,.
\end{align*}
Explicitly, this is achieved by renaming
\begin{align*}
    \begin{aligned}
        u_1,\dots,u_m &\rightarrow u^c_1,\dots,u_m^c \,, & u_{m+1},\dots,u_{m+l} &\rightarrow u^b_l,\dots,u^b_1 \,,\\
        u_{m+l+1},\dots,u_{k+l+m+n+1} &\rightarrow u^a_{k+n+1},\dots,u^a_{1} \,,
    \end{aligned}
\end{align*}
and similarly for the $v$- and $w$-coordinates, and setting
\begin{align} \label{R}
    \begin{aligned}
        \frac{u_1^c}{v_1^c} &= \dots = \frac{u_m^c}{v_m^c}=\alpha \,, & \frac{u_1^b}{w_1^b} &= \dots = \frac{u_l^b}{w_l^b} = \gamma\,,\\
        \frac{v_1^a}{w_1^a} &= \dots = \frac{v_{k+n+1}^a}{w_{k+n+1}^a}=\frac{1}{\beta} \,.
    \end{aligned}
\end{align}
Thus, for a potential $\Omega_{k,l,m,n}^a$ the matrix $Q$ reads
\begin{align*}
    Q =& (\vec{q}_a,\vec{q}_b,\vec{q}_c,\vec{q}_{c,1},\dots,\vec{q}_{c,m},\vec{q}_{b,l},\dots,\vec{q}_{b,1},\vec{q}_{a,k+n+1},\dots,\vec{q}_{a,1}) \,,     
\end{align*}
with the variables satisfying \eqref{R}. As an example, the left diagram in Fig. \ref{SD6} yields a potential described by the matrix
\begin{align*}
    Q_D =& (\vec{q}_a,\vec{q}_{a,1},\vec{q}_{a,3},\vec{q}_{b,2},\vec{q}_{b},\vec{q}_{b,1},\vec{q}_{c,3},\vec{q}_{c,1},\vec{q}_{c},\vec{q}_{c,2},\vec{q}_{a,2})\,,
\end{align*}
again with variables satisfying \eqref{R}. Using \eqref{cosm} and the anti-symmetric property
\begin{align*}
    \sigma_{i,j} &= - \sigma_{j,i}\,,
\end{align*}
together with the closure constraint and the Fierz identity, the potential $\Omega_N^{a}$ can easily be shown to be closed on $\mathbb{U}_{N}$, thus $\Omega_{k,l,m,n}^{a}$ is automatically closed too:
\begin{align}
    d \Omega_N^{a}&=0 && \Longrightarrow && d \Omega_{k,l,m,n}^{a}=0\,,\qquad \quad d \Omega_{k,l,m,n}^{c}=0\,.
\end{align}

As was briefly mentioned in \eqref{flip}, one can extract $\Omega_{m,l,k,n}^c$ from $\Omega_{k,l,m,n}^a$ by reversing the boundary ordering and swapping $a \leftrightarrow c$ on a disk diagram that is associated to $\Omega_{k,l,m,n}^a$. The latter implies that one has to swap $p_a \leftrightarrow p_c$ and $\vec{q}_a \leftrightarrow \vec{q}_c$ too. Reversing the boundary ordering negates the effect of swapping $\vec{q}_a$ and $\vec{q}_c$, while swapping $p_a$ and $p_c$ can be replaced by swapping the $u$ and $w$ coordinates. This tells us all we need to know to construct potentials $\Omega_{m,l,k,n}^c$ from their disk diagrams: the only difference is that the vectors 
\begin{align*}
    \begin{aligned}
        \vec{q}_{a,i}&\rightarrow\vec{q}_{c,i}^{\,\,'}=(w_i^c,v_i^c,u_i^c) \,, & \vec{q}_{b,i}&\rightarrow\vec{q}_{b,i}^{\,\,'}=(w_i^b,v_i^b,u_i^b) \,, & \vec{q}_{c,i}\rightarrow\vec{q}^{\,\,'}_{a,i}&=(w_i^a,v_i^a,u_i^a)
    \end{aligned}
\end{align*}
and that the canonical ordering for the matrix $Q$ is reversed, as can be seen in Fig. \ref{canPot}, but we keep $\vec{q}_a,\vec{q}_b,\vec{q}_c$ as its first entries. Thus, for $\Omega_{m,l,k,n}^c$ the matrix reads
\begin{align} \label{Q'}
    Q' =& (\vec{q}_a,\vec{q}_b,\vec{q}_c,\vec{q}_{c,1}^{\,\,'},\dots,\vec{q}_{c,m}^{\,\,'},\vec{q}_{b,l}^{\,\,'},\dots,\vec{q}_{b,1}^{\,\,'},\vec{q}_{a,k+n+1}^{\,\,'},\dots,\vec{q}_{a,1}^{\,\,'},)
\end{align}
and the matrix $Q'_D$ for the right disk diagram in Fig. \ref{SD6} is
\begin{align*}
    Q_D' =(\vec{q}_{c,3}^{\,\,'},\vec{q}_{c,1}^{\,\,'},\vec{q}_{a},\vec{q}_{c,2}^{\,\,'},\vec{q}_{b,2}^{\,\,'},\vec{q}_{b},\vec{q}_{b,1}^{\,\,'},\vec{q}_{a,3}^{\,\,'},\vec{q}_{a,1}^{\,\,'},\vec{q}_{c},\vec{q}_{a,2}^{\,\,'}) \,.
\end{align*}
Then, the expression for the potentials $\Omega^c_N$ becomes
\begin{align*}
    \Omega_{N}^c = \mu'I_D' \,,
\end{align*}
with 
\begin{align*}
    \begin{aligned}
        \mu' =& \mu'_1\wedge\dots\wedge\mu'_N \,,\\
        \mu'_{i} =&  p_{b,c} du_i \wedge dv_i + p_{a,c} du_i \wedge dw_i + p_{a,b} dv_i \wedge dw_i \,,\\
    \end{aligned}
\end{align*}
and 
\begin{align} \label{potc}
    I'_D = s'_D\exp[\text{Tr}[P_D Q_{D}^{\prime \, T}] + \lambda (p_{a,b}|Q_D^{\prime \, 12}|+p_{a,c}|Q_D^{\prime \, 13}|+p_{b,c}|Q_D^{\prime \, 23}|)] \,.
\end{align}
One obtains a potential $\Omega_{m,l,k,n}^c$ through
\begin{align*}
    \Omega_{m,l,k,n}^c=\Omega_{N}^c|_{\mathbb{W}_{k,l,m,n}} \,.
\end{align*}

\begin{figure}
    \centering
    \begin{tikzpicture}[scale=0.3]
        \draw[thick](0,0) circle (6);
        \draw[ultra thick, blue](0,0) -- (150:6);
        \draw[ultra thick, blue](0,0) -- (30:6);
        \draw[ultra thick, blue](0,0) -- (270:6);
        \draw[red,thin] (150:3) .. controls (90:5) .. (90:6);
        \draw[dashed, red](0,0) -- (0,6);
        \draw[dashed, red](0,0) -- (-30:6);
        \draw[dashed, red](0,0) -- (210:6);
        \draw[-Latex] (0,6) -- (0,8.5);
        \filldraw [red!45!white, fill opacity=0.7] (0:0cm) -- (30:6cm) arc(30:-30:6cm) -- cycle;
        \filldraw [red!45!white, fill opacity=0.7] (0:0cm) -- (-90:6cm) arc(-90:-150:6cm) -- cycle;
        \filldraw [red!45!white, fill opacity=0.7] (0:0cm) -- (150:6cm) arc(150:210:6cm) -- cycle;
        \coordinate[label=above left : $a$] (B) at (150:6);
        \coordinate[label=above right : $c$] (B) at (30:6);
        \coordinate[label=below : $b$] (B) at (270:6);

        \begin{scope}[xshift=20cm]
        \draw[thick](0,0) circle (6);
        \draw[ultra thick, blue](0,0) -- (150:6);
        \draw[ultra thick, blue](0,0) -- (30:6);
        \draw[ultra thick, blue](0,0) -- (270:6);
        \draw[red,thin] (30:3) .. controls (90:5) .. (90:6);
        \draw[dashed, red](0,0) -- (0,6);
        \draw[dashed, red](0,0) -- (-30:6);
        \draw[dashed, red](0,0) -- (210:6);
        \draw[-Latex] (0,6) -- (0,8.5);
        \filldraw [red!45!white, fill opacity=0.7] (0:0cm) -- (-30:6cm) arc(-30:30:6cm) -- cycle;
        \filldraw [red!45!white, fill opacity=0.7] (0:0cm) -- (-30:6cm) arc(-30:-90:6cm) -- cycle;
        \filldraw [red!45!white, fill opacity=0.7] (0:0cm) -- (150:6cm) arc(150:210:6cm) -- cycle;
        \coordinate[label=above left : $a$] (B) at (150:6);
        \coordinate[label=above right : $c$] (B) at (30:6);
        \coordinate[label=below : $b$] (B) at (270:6);
        \end{scope}
    \end{tikzpicture}
    \caption{The sign $s_D$ of $\Omega_{k,l,m,n}^a$ and $\Omega_{k,l,m,n}^c$ is determined by the number of red lines in the shaded regions in the left and right diagram, respectively.}
    \label{sign}
\end{figure}
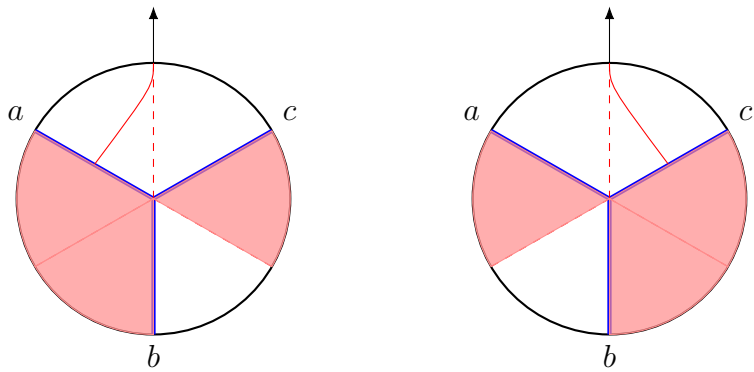

\paragraph{Signs.} The signs $s_D$ and $s'_D$ in the expression for the potentials \eqref{pot} and \eqref{potc} are determined by the orientations of the red lines, like for the vertices, where we counted the number of red lines in the southern semicircle. Here the counting rule is a bit more involved: for potentials $\Omega_{k,l,m,n}^a$, one sums the number of red lines attached to $b$- and $c$-leg in the clockwise direction and to the $a$-leg in the anticlockwise direction, see Fig. \ref{sign}. For potentials $\Omega_{k,l,m,n}^c$ this is mirrored. We call the sum $M$ and the sign is
\begin{align*}
    \begin{aligned}
        s_D &= (-1)^M  & \text{and} & & s'_D &= (-1)^{M+1}
    \end{aligned}
\end{align*}
for the potentials $\Omega_{k,l,m,n}^a$ and $\Omega_{k,l,m,n}^c$, respectively. Note that throughout the text the red lines of diagrams are not always drawn precisely in the shaded regions as in Fig. \ref{sign} for the sake of convenience, but it should be clear from the context which region they belong to. To avoid cluttering the text with minus signs, we will discuss the proof of the $A_\infty$-relations up to a sign. However, with the conventions discussed here, \eqref{StokesToA_infty} yields the correct signs.

\paragraph{Boundaries.} Stokes' theorem requires one to evaluate the potentials at the codimension-one boundary (just the boundary) of the domain $\mathbb{W}_{k,l,m,n}$. This boundary, $\partial\mathbb{W}_{k,l,m,n}$ is the union of many boundary components $P_i$, i.e., $\partial\mathbb{W}_{k,l,m,n}=\cup_i P_i$. Each $P_i$ is obtained by saturating an inequality in \eqref{DD3}-\eqref{startEnd2}. However, sometimes saturating an inequality results in a higher codimension component. By abuse of nomenclature we also refer to these as boundaries and include them into the set of $P_i$. Each boundary may have interesting characteristics, together with the potential evaluated on the boundary. In the subsequent section, we will categorize the various types of boundaries. They belong to the following classes:
\begin{itemize}
    \item At some boundaries one finds, after a change of variables, $A_\infty$-terms contributing to the proof of $\eqref{StokesToA_infty}$. The change of variables is necessary to recognize the vertices as described in Section \ref{sec:CD}.
    \item The potentials may also yield nonzero results at certain boundaries, without contributing to $A_\infty$-relations. Fortunately, this does not spoil the proof, as these terms always come in pairs and consequently cancel each other. The pairs are always formed by terms arising from different potentials and the terms are therefore called $\textit{gluing terms}$ as they `glue' together different potentials.
    \item Since potentials are differential forms, they consist of a measure and an integrand. When the measure evaluates to zero at a boundary, we refer to this as a \textit{zero measure term}.
    \item As mentioned above, we describe boundaries of $\mathbb{W}_{k,l,m,n}$ simply by saturating inequalities in \eqref{DD3}. However, this will sometimes give rise to a \textit{higher codimension boundary}, as saturating one inequality requires other inequalities to be saturated at the same time. This yields a boundary that is parameterized by less than $2N$ variables. As Stokes' theorem only requires codimension-one boundaries, higher codimension boundaries do not contribute to the proof.
\end{itemize}

So far we have established a visual representation for the potentials and their corresponding domains through disk diagrams. With some amount of hindsight, we will provide a way to visualize the evaluation of the potentials on the boundaries throughout the later sections. It is known from earlier work \cite{Sharapov:2022nps} that a parameter is associated with all line segments in the disk diagrams, except the line connected to the arrow and the segments of the legs that are directly connected to the points $a,b,c$. These parameters are coordinates of a hypercube and are related to the $u,v,w$ coordinatizing $\mathbb{W}_{k,l,m,n}$ by a smooth coordinate transformation. In fact, $\mathbb{W}_{k,l,m,n}$ is a subspace of the hypercube. We will not use the exact coordinate transformations. Still, we borrow this knowledge to realize that evaluating potentials at a boundary of $\mathbb{W}_{k,l,m,n}$ coincides with evaluating them at the upper and lower bound of these parameters. We will visualize this by drawing a green/red region on the line under evaluation for the upper/lower bound. 

\begin{figure}
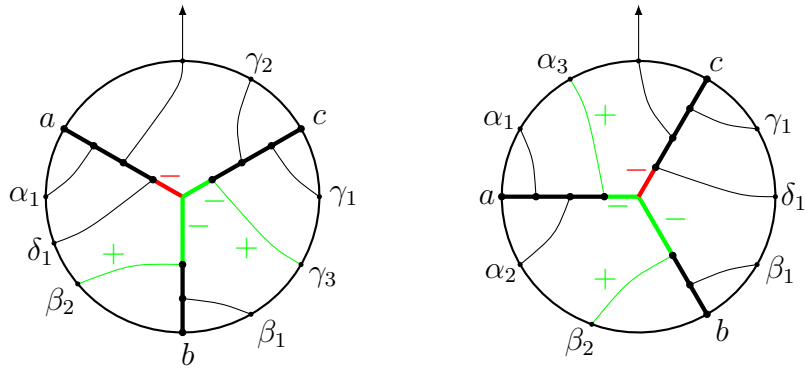

\centering

\caption{Disk diagrams that show which line segments correspond to boundaries that yield non-zero expressions for potentials of the type $\Omega_{k,l,m,n}^a$ and $\Omega_{k,l,m,n}^c$ on the left and right, respectively. The diagrams also show the signs that the boundaries are accompanied with.} \label{signs}
\end{figure}

The final result is easy to formulate. Fig. \ref{signs} shows on which lines in disk diagrams a boundary leads to a non-vanishing expression. These are the line segments connected to the junction and the bulk-to-boundary lines closest to the junction on the legs that are not connected to output arrow. The color of the line refers to type of boundary that yields this result and each line is accompanied by a sign that should be taken into account when considering the boundary. Throughout the following sections, we will show how these disk diagrams can be understood as $A_\infty$-terms or other terms.

Let us note that the pictures below are to display which part of the analytical expression for potential $\Omega$ is being affected by evaluating it at a certain boundary, i.e. the pictures are only to help visualize certain analytical manipulations. Given a disk diagram, there are, roughly speaking, two boundaries per each line. 

\paragraph{$A_\infty$-terms.} So far we have sketched the picture of how to employ Stokes' theorem to prove the $A_\infty$-relations. It only remains to provide a recipe for constructing the $A_\infty$-terms. As can be seen from \eqref{A}, the $A_\infty$-terms are nested vertices, so they can be visualized using the disk diagrams for vertices introduced in section \ref{sec:CD}. Examples of the different types of $A_\infty$-terms found in \eqref{A} are given in Fig. \ref{A_inftyDiagrams}. Here we visualize the nesting of vertices by inserting one vertex into the other through a \textit{nesting arrow}.

\begin{figure}
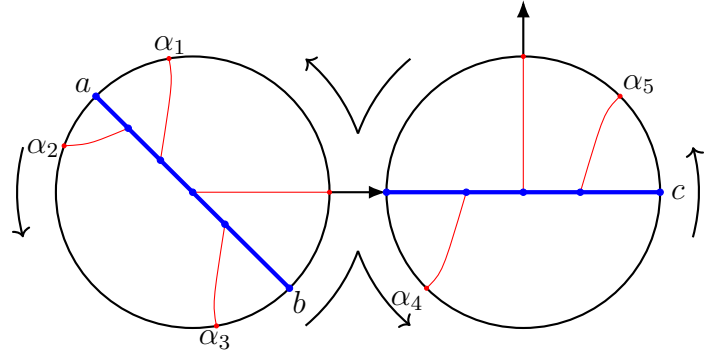

    \centering

    \caption{The nested boundary ordering for the diagram corresponding to the $A_\infty$-term $\mathcal{V}(\mathcal{V}(\alpha_1,a,\alpha_2,\alpha_3,b),\alpha_4,c,\alpha_5)$. One follows the arrow, starting from the ouput arrow. When the nesting arrow is reached, one follows the arrows around the nested vertex and when the nesting arrow is reached again, the path around the outer disk diagram is continued.}
    \label{nestedBO}
\end{figure}

To relate an expression with a nested disk diagram, one assigns vectors $\vec{q}_a=(-1,0,0),\vec{q}_b=(0,-1,0),\vec{q}_c=(0,0,-1)$ to $a,b,c\in V[-1]$, $\vec{q}_{i}^{\,\,1}=(u^1_i,v^1_i,w^1_i)$ to $\alpha_i \in V^\ast$ in the inner vertex and $\vec{q}_i^{\,\,2}=(u^2_i,v^2_i,w^2_i)$ to $\alpha_j \in V^\ast$ in the outer vertex and to the output arrow according to the bulk ordering for the two vertices separately, as was explained in section \ref{sec:CD}. For the moment, we leave the values of these vectors undefined. One also assigns vectors $\vec{r}_{a}=\vec{r}_{b}=\vec{r}_{c}=(0,0,0)$ to the elements $a,b,c$ and $\vec{r}_i=(p_{a,i},p_{b,i},p_{c,i})$ to $\alpha_i\in V^\ast$ and to the output arrow. This is done according to the \textit{nested boundary ordering}, just like the labeling of the $\alpha$'s. This ordering starts to the left of the output arrow, follows around the circle counterclockwise until it hits the nesting arrow. It then follows the nested circle counterclockwise, after which it completes the counterclockwise path around the outer circle, see Fig. \ref{nestedBO} for an example. We then construct the $3 \times (N+4)$ matrices $Q_D$ and $P_D$ by filling it up with the vectors $q$- and $r$-vectors, respectively, according to the nested boundary ordering. Like before, for the matrix $P_D$ this means that one enters the vectors $\vec{r}_i$ in increasing order, while one inserts $\vec{r}_a,\vec{r}_b,\vec{r}_c$ for the elements $a,b,c$, respectively. 

To determine the vectors $\vec{q}_i^{\,\,1}$ and $\vec{q}_i^{\,\,2}$, let us demonstrate how to insert vertices into each other at the level of expressions. Remember that $\mathcal{V}(\bullet,\dots,\bullet,a,\bullet,\dots,\bullet,b,\bullet,\dots,\bullet) \in V[-1]$ and $\mathcal{U}(\bullet,\dots,\bullet,a,\bullet,\dots,\bullet) \in V^\ast$. This means that an element of $V[-1]$ or $V^\ast$ can be replaced by a $\mathcal{V}$- or $\mathcal{U}$-vertex. For example, one can use the vertices in \eqref{quadV} and \eqref{cubicV} to compute
\begin{align*}
    \begin{aligned}
        \mathcal{V}(\mathcal{V}(a,b),c)&=\exp[p_{0,d}+p_{0,c}+\lambda p_{d,c}]\exp[y_dp_a+y_dp_b+\lambda p_{a,b}] =\\
        &= \exp[p_{0,a}+p_{0,b}+p_{0,c}+\lambda(p_{a,b}+p_{a,c}+p_{b,c})]
    \end{aligned}
\end{align*}
and (here and below, $y_d$ is an auxiliary variable to deal with the insertion of one vertex into another one, which is set to zero in the end)
\begin{align*}
    \begin{aligned}
        \mathcal{V}(a,\mathcal{V}(b,c,\alpha_1),\alpha_2) &= \int_{\mathbb{V}_1}p_{a,d}\exp[u_2^2 p_{0,a} + v_2^2 p_{0,d} + u_1^2p_{a,2} + v_1^2 p_{d,2} + \lambda p_{a,d} A_2]\times\\
        &\times\int_{\mathbb{V}_1}p_{b,c}\exp[u_2^1 y_d p_b + v_2^1 y_d p_c + u_1^1p_{b,1} + v_1^1 p_{c,1} + \lambda p_{b,c}A_1] =\\
        &=\int_{\mathbb{V}_1}\int_{\mathbb{V}_1}(u_2^1p_{a,b}+v_2^1p_{a,c})p_{b,c}\exp[u_2^2 p_{0,a} + v_2^2 u_2^1 p_{0,b} + v_2^2 v_2^1 p_{0,c} + u_1^2 p_{a,2} +\\
        &+ u_1^1 p_{b,1} + u_2^1v_1^2 p_{b,2} + v_1^1 p_{c,1} + v_2^1 v_1^2 p_{c,2} + \lambda(u_2^1 A_2 p_{a,b} + v_2^1 A_2 p_{a,c} + A_1 p_{b,c})] \,,
    \end{aligned}
\end{align*}
with $A_i = 1+u_1^i+u_2^i-v_1^i-v_2^i+u_1^iv_2^i-u_2^iv_1^i$. 
The five different types of $A_\infty$-terms are given by
\begin{align} \label{generalA_infty}
    \begin{aligned}
        &\mathcal{V}(\bullet,\dots,\bullet,\mathcal{V}(\bullet,\dots,\bullet,a,\bullet,\dots,\bullet,b,\bullet,\dots,\bullet),\bullet,\dots,\bullet,c,\bullet,\dots,\bullet) =\\
        &= s_{D_1}s_{D_2}\int_{\mathbb{V}_{s}}\int_{\mathbb{V}_{r}}p_{a,b}^{r}(u^1_t p_{a,c}+v^1_t p_{b,c})^{s}  I_1 \,,\\
        &\mathcal{V}(\bullet,\dots,\bullet,a,\bullet,\dots,\bullet,\mathcal{V}(\bullet,\dots,\bullet,b,\bullet,\dots,\bullet,c,\bullet,\dots,\bullet),\bullet,\dots,\bullet) =\\
        &=s_{D_1}s_{D_2}\int_{\mathbb{V}_{s}}\int_{\mathbb{V}_{r}}(u^1_t p_{a,b}+v^1_t p_{a,c})^{s} p_{b,c}^{r} I_2 \,,\\
        &\mathcal{V}(\bullet,\dots,\bullet,\mathcal{U}(\bullet,\dots,\bullet,a,\bullet,\dots,\bullet),\bullet,\dots,\bullet,b,\bullet,\dots,\bullet,c,\bullet,\dots,\bullet) =\\ 
        &=(-1)^{r-1}s_{D_1}s_{D_2}\int_{\mathbb{V}_{s+1}}\int_{\mathbb{V}_{r-1}}(u_{t}^2 p_{a,b}+ v_{t}^2 p_{a,c})^{r-1} p_{b,c}^{s+1} I_3 \,,\\
        &\mathcal{V}(\bullet,\dots,\bullet,a,\bullet,\dots,\bullet,\mathcal{U}(\bullet,\dots,\bullet,b,\bullet,\dots,\bullet),\bullet,\dots,\bullet,c,\bullet,\dots,\bullet) =\\ &=s_{D_1}s_{D_2}\int_{\mathbb{V}_{s+1}}\int_{\mathbb{V}_{r-1}} p_{a,c}^{s+1}(u_{t}^2 p_{a,b}-v_{t}^2 p_{b,c})^{r-1}I_4 \,,\\
        &\mathcal{V}(\bullet,\dots,\bullet,a,\bullet,\dots,\bullet,b,\bullet,\dots,\bullet,\mathcal{U}(\bullet,\dots,\bullet,c,\bullet,\dots,\bullet),\bullet,\dots,\bullet)=\\
        &=s_{D_1}s_{D_2}\int_{\mathbb{V}_{s+1}}\int_{\mathbb{V}_{r-1}} p_{a,b}^{s+1}(u_{t}^2 p_{a,c}+v_{t}^2 p_{b,c})^{r-1} I_5 \,,
    \end{aligned}
\end{align}
where $r$ and $s$ are the total number of elements of $V^\ast$ in the inner and outer vertex, respectively, and $t$ is the label of the $q$-vector corresponding to the line connected to the nesting arrow. $s_{D_1}$ and $s_{D_2}$ are the signs associated to the inner and outer vertex, respectively. The functions $I_i$ read
\begin{align*}
    I_i =& \exp[\text{Tr}[P_D(Q_{D}^i)^T]+\lambda(p_{a,b}|(Q_{D}^i)^{12}|+p_{a,c}|(Q_D^i)^{13}|+p_{b,c}|(Q_D^i)^{23}|)] \,.
\end{align*}
Like for the potentials, we provide matrices
\begin{align*}
    \begin{aligned}
        Q^i =& (
        \vec{q}_a , \vec{q}_b , \vec{q}_c , \vec{q}^{\,\,1}_1 , \dots , \vec{q}^{\,\,1}_r , \vec{q}^{\,\,2}_1 , \dots , \vec{q}^{\,\,2}_s
        )
    \end{aligned}
\end{align*}
that are said to be in the canonical ordering. For diagrams that admit a left-ordering, the matrices $Q^i$ reduce to the matrices $Q_D^i$ for left-ordered diagrams $D$. These matrices $Q$ are given by
\begin{align} \label{Qs}
    \begin{aligned}
        Q^1 =& 

    \caption{Examples of nested disk diagrams for each type of $A_\infty$-terms. }
    \label{A_inftyDiagrams}
\end{figure}

\subsection{First examples}  \label{sec:example}
To get acquainted with the methods used in the proof that is going to follow, let us start with the lowest order examples, i.e., $N=0$ and $N=1$. 

\paragraph{$\boldsymbol{N=0}$.} This first example is perhaps a bit too simple, but it allows us to get a feel for some of the methods used for higher orders. The $A_\infty$-relation reads
\begin{align*}
    \mathcal{V}(\mathcal{V}(a,b),c) - \mathcal{V}(a,\mathcal{V}(b,c)) = 0
\end{align*}
and the only relevant vertex is the star-product $\mathcal{V}(a,b)\equiv a\star b$, given by
\begin{align*}
    \mathcal{V}(a,b) = \exp[p_{0,a}+p_{0,b}+\lambda p_{a,b}] \,.
\end{align*}

\begin{figure}[!ht]
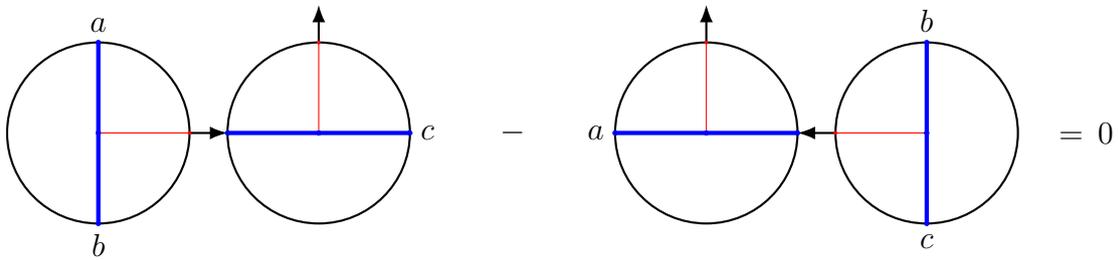

\centering


\caption{The disk diagram representation of the $A_\infty$-relation for $N=0$.}\label{nested1}
\end{figure}

\begin{figure}
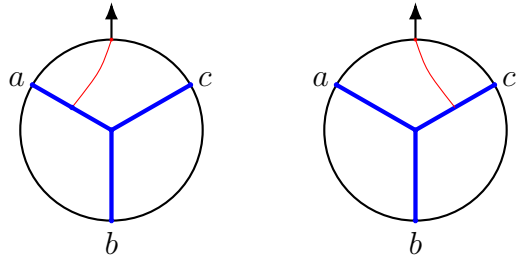

    \centering
    %
    \caption{On the left the disk diagram corresponding to the potential $\Omega_{0,0,0,0}^{a}(a,b,c)$ and on the right the one that corresponds to $\Omega_{0,0,0,0}^c(a,b,c).$}
    \label{N0}
\end{figure}

The $A_\infty$-relation is diagrammatically represented in Fig. \ref{nested1}. The relevant potentials are
\begin{align}\label{t}
    \Omega_{0,0,0,0}^a(a,b,c) = -\Omega_{0,0,0,0}^c(a,b,c) = \exp[p_{0,a} + p_{0,b} + p_{0,c} + \lambda(p_{a,b} + p_{a,c} + p_{b,c})]
\end{align}
and their disk diagrams are shown in Fig. \ref{N0}. The forms are closed, as they are constants. By definition, $\mathbb{W}_{0,0,0,0}=[0,1]$ and the boundary $\partial \mathbb{W}_{0,0,0,0}=\{0,1\}$ consists of the pair of points. This can also be written as
\begin{align*} 
    \partial\mathbb{W}_{0,0,0,0} = (\mathbb{V}_0 \times \mathbb{V}_0)\cup (\mathbb{V}_0 \times \mathbb{V}_0) \,,
\end{align*}
$\mathbb{V}_0$ being a one-point set. We stressed that the boundary components are the products of the configuration spaces of  vertices involved. The $A_\infty$-relation can now be recast in terms of Stokes' theorem \eqref{StokesToA_infty} through
\begin{align}\label{firstStokes}
    \begin{aligned}
        0 =& \int_0^1 (d\Omega_{0,0,0,0}^a(a,b,c)+d\Omega_{0,0,0,0}^c(a,b,c)) = [\Omega_{0,0,0,0}^a(a,b,c)+\Omega_{0,0,0,0}^c(a,b,c)]_0^1 \,.
    \end{aligned}
\end{align}
This 'too simple' example may look confusing since identical contributions are assigned different meaning and the integrand vanishes identically. Nevertheless, it showcases various features of the general proof. For instance, the last expression in \eqref{firstStokes} consists of four terms, while the $A_\infty$-relation contains only two terms. It turns out that whenever the proof requires contributions from multiple potentials, in this case $\Omega_{0,0,0,0}^a$ and $\Omega_{0,0,0,0}^c$, we find more terms than one would expect from the $A_\infty$-relation, but the extra terms from different potentials cancel each other. These are the gluing terms that were mentioned above.

\begin{figure}
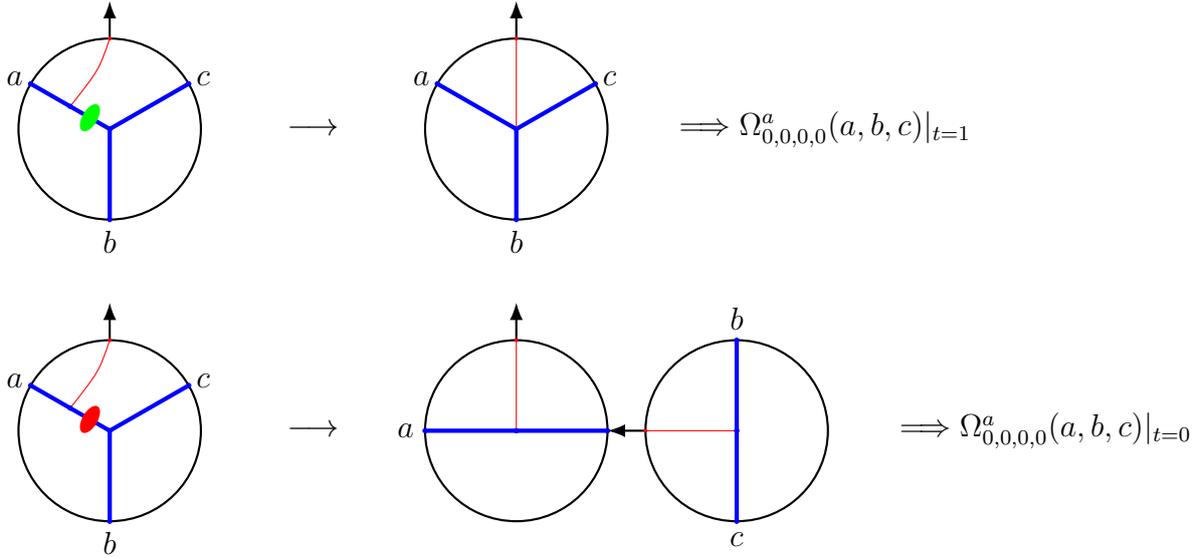

    \centering

    \caption{$\Omega_{0,0,0,0}^a(a,b,c)$ evaluated at its boundaries. The green region highlights the boundary $t=1$, while the red region exhibits the boundary $t=0$.}
    \label{boundaries1a}
\end{figure}

\begin{figure}
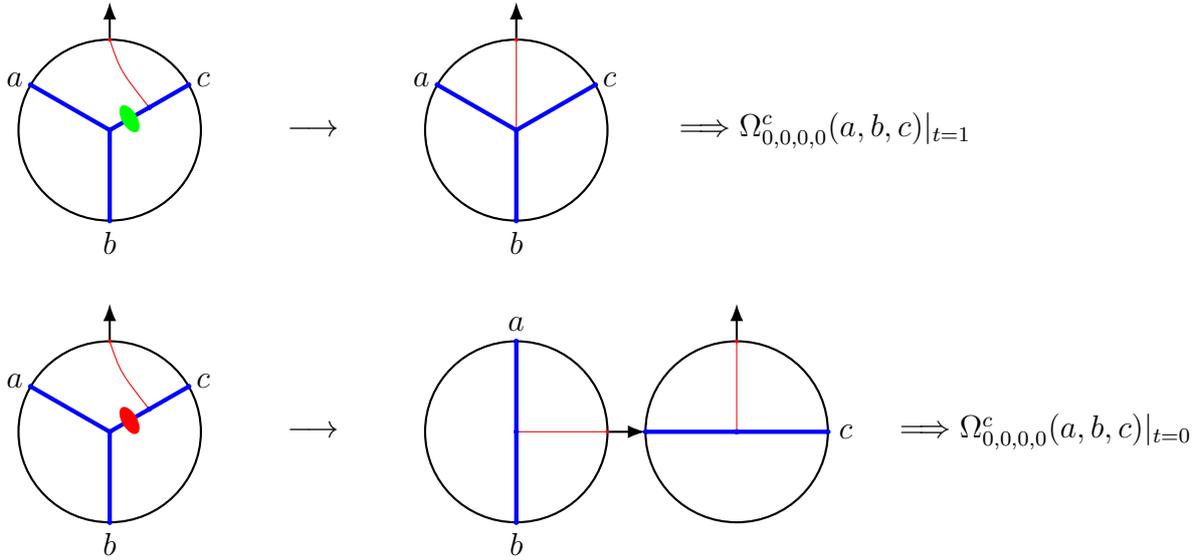

    \centering
    %
    \caption{$\Omega_{0,0,0,0}^c(a,b,c)$ evaluated at its boundaries. The green region highlights the boundary $t=1$, while the red region exhibits the boundary $t=0$.}
    \label{boundaries1c}
\end{figure}

Although the analytic expression \eqref{firstStokes} still looks manageable, the proof becomes increasingly more complicated at higher orders. It will be invaluable to have a visual representation of what happens to the disk diagram when the potentials are evaluated at a boundary. This will provide a quick way of identifying which $A_\infty$-term or gluing term is generated. Figs. \ref{boundaries1a} and \ref{boundaries1c} display how the disk diagrams can be interpreted as $A_\infty$-terms or gluing terms. The green boundary on the blue line shrinks the line to a point, giving rise to a disk diagram with a four-point vertex. This is a gluing term. The red boundary on the blue line separates the disk diagram in two disk diagrams with the blue line replaced by the nesting arrow. This is recognized as an $A_\infty$-term. Figs. \ref{boundaries1a} and \ref{boundaries1c} allow one to immediately observe that the gluing terms are identical and they will cancel each other. The remaining terms are easily read off to be the $A_\infty$-terms $\mathcal{V}(a,\mathcal{V}(b,c))$ and $\mathcal{V}(\mathcal{V}(a,b),c)$.

\paragraph{$\boldsymbol{N=1}$.} At this order there are four $A_\infty$-relations: one for each ordering of the elements of $V[-1]$ and $V^\ast$. However, some of these are related by the natural pairing and only two orderings are independent, i.e., the left-ordered case $a,b,c,\alpha$ and the almost-left-ordered case $a,b,\alpha,c$. The vertices relevant to the $A_\infty$-relations for $N=1$ are given in \eqref{quadV} and \eqref{cubicV}, while the domain is described in \eqref{2-simplex2}. The recipe of Sec. \ref{sec:recipe} tells us to use the vectors $\vec{q}_{a,1}$, $\vec{q}_{b,1}$ and/or $\vec{q}_{c,1}$ to construct expressions for the potentials. However, for notational simplicity, we will replace
\begin{align*}
    \begin{aligned}
        \vec{q}_{b,1},\vec{q}_{c,1} &\rightarrow \vec{q}_1=(u_1,v_1,w_1) \,, & \vec{q}_{a,1} &\rightarrow \vec{q}_2=(u_2,v_2,w_2)  \,.
    \end{aligned}
\end{align*}
The former makes sense, since every potential contains $\vec{q}_{b,1}$ or $\vec{q}_{c,1}$ and not both. Also note that this example will not show the full behaviour of \eqref{Qs}, as either the inner or the outer vertex has no integration domain for $N=1$, such that $u_1^1=v_1^1=1$ or $u_1^2=v_1^2=1$. As a result, the fact that some entries of the $Q$-matrices are composed of products of variables, is not visible in this example.

\paragraph{Left-ordering.} The $A_\infty$-relation for this ordering reads
\begin{align}\label{firstAinfty}
    \mathcal{V}(\mathcal{V}(a,b),c,\alpha) - \mathcal{V}(a,\mathcal{V}(b,c),\alpha) - \mathcal{V}(a,\mathcal{V}(b,c,\alpha)) + \mathcal{V}(a,b,\mathcal{U}(c,\alpha)) = 0 \,.
\end{align}

\begin{figure}
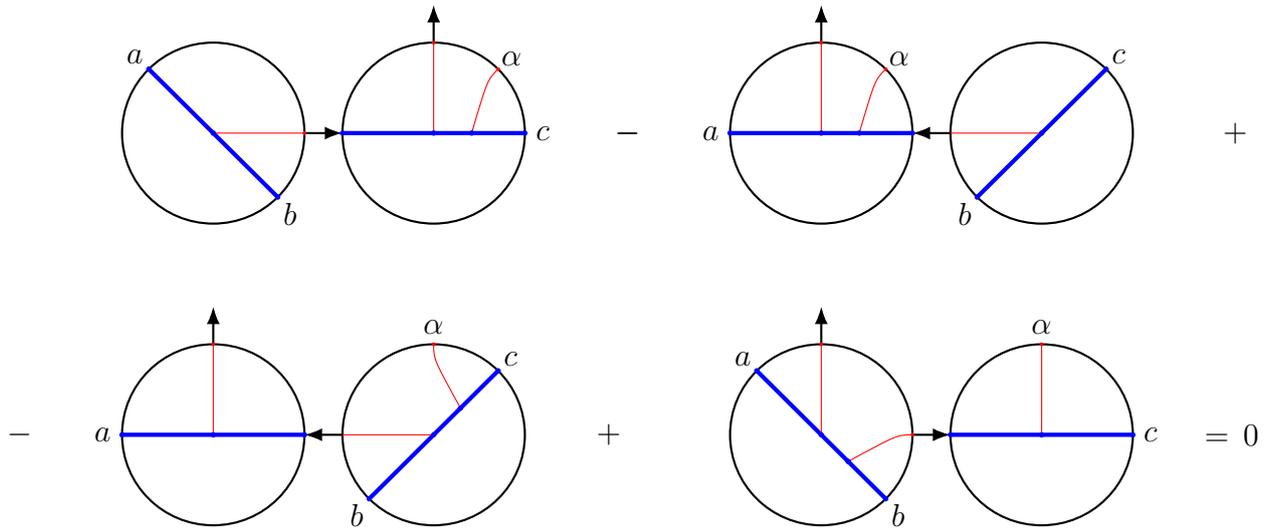

    \centering

    \caption{Graphical representation of the left-ordered $A_\infty$-relations for $N=1$.}
    \label{leftN1}
\end{figure}

This is an example of a left-ordered $A_\infty$-relation, i.e., all elements of $V[-1]$ appear before the element of $V^\ast$. A visualization in terms of disk diagrams is given in Fig. \ref{leftN1}. The order in which the elements of $W$ appear in the $A_\infty$-terms coincides with the nested boundary ordering in the disk diagram.
The left-ordered disk diagram for a potential with $N=1$ is shown in Fig. \ref{wwwCDiagram}. The potential $\Omega_{0,0,1,0}^a(a,b,c,\gamma)$ and domain $\mathbb{W}_{0,0,1,0}$ can be constructed from this diagram. We introduce the $3 \times 5$ matrices
\begin{align*}
    Q =\begin{pmatrix}
        -1 & 0 & 0 & u_1 & u_2 \\
        0 & -1 & 0 & v_1 & v_2 \\
        0 & 0 & -1 & w_1 & w_2
    \end{pmatrix} \,, \quad P &= \begin{pmatrix}
            0 & 0 & 0 & p_{a,1} & p_{0,a} \\
            0 & 0 & 0 & p_{b,1} & p_{0,b} \\
            0 & 0 & 0 & p_{c,1} & p_{0,c}
            \end{pmatrix}
\end{align*}
according to the recipe in section \ref{sec:recipe}. We also define the integrand
\begin{align*}
    I =& \exp[\text{Tr}[P Q^T]+\lambda( p_{a,b} |Q^{1,2}| +  p_{a,c} |Q^{1,3}| +  p_{b,c} |Q^{2,3}|)]
\end{align*}
and the measure
\begin{align*}
    \mu &= p_{a,b}du_1\wedge dv_1 + p_{a,c} du_1 \wedge dw_1 + p_{b,c} dv_1 \wedge dw_1\,.
\end{align*}
The potential $\Omega_{0,0,1,0}^a(a,b,c,\gamma)$ is then given by
\begin{align*}
\begin{aligned}
    \Omega_{0,0,1,0}^a(a,b,c,\gamma) =&[p_{a,b} du_1 \wedge dv_1 + p_{a,c} du_1 \wedge dw_1 + p_{b,c} dv_1 \wedge dw_1] \times \\
    \times&\exp[\text{Tr}[P Q^T]+\lambda (p_{a,b} |Q^{1,2}| + p_{a,c} |Q^{1,3}| + p_{b,c} |Q^{2,3}|)] \,.
\end{aligned}
\end{align*}
After solving the closure constraint for $\vec{q}_2$, i.e., $\vec{q}_2=1-\vec{q}_1$, we see that the potential is a closed form, since
\begin{align*}
    \begin{aligned}
    d\Omega_{0,0,1,0}^a(a,b,c,\gamma) &= (p_{a,b}p_{c,1} - p_{a,c}p_{b,1} + p_{b,c}p_{a,1} - p_{0,c}p_{a,b} + p_{0,b}p_{a,c} - p_{0,a}p_{b,c}) \times\\
    & \times du_1 \wedge dv_1 \wedge dw_1 I = 0\,,
    \end{aligned}
\end{align*}
where in the last step the Fierz identity is used twice:
\begin{align*}
    \begin{aligned}
        p_{a,b}p_{c,1} - p_{a,c}p_{b,1} + p_{b,c}p_{a,1} &= 0 \,, &
        p_{0,c}p_{a,b} - p_{0,b}p_{a,c} + p_{0,a}p_{b,c} &= 0 \,.
    \end{aligned}
\end{align*}
The terms in the $A_\infty$-relation \eqref{firstAinfty} are now conveniently written as
\begin{align} \label{A_infty terms}
    \begin{aligned}
    \mathcal{V}(\mathcal{V}(a,b),c,\alpha) &= \int_{\mathbb{V}_{1}} \, (p_{a,c} + p_{b,c}) I|_{u_\bullet=v_\bullet} \,, &
    \mathcal{V}(a,\mathcal{V}(b,c),\alpha) &= \int_{\mathbb{V}_{1}} \, (p_{a,b} + p_{a,c}) I|_{v_\bullet=w_\bullet} \,, \\
    \mathcal{V}(a,\mathcal{V}(b,c,\alpha)) &= \int_{\mathbb{V}_{1}} \, p_{b,c} I|_{u_1=0,u_2=1} \,, &
    \mathcal{V}(a,b,\mathcal{U}(c,\alpha)) &= \int_{\mathbb{V}_{1}} \, p_{a,b} I|_{w_1=1,w_2=0} \,.
    \end{aligned}
\end{align}
Here, $u_\bullet=v_\bullet$ means that the equality holds for all $u$'s and $v$'s, i.e., $u_1=v_1$ and $u_2=v_2$. 

The domain $\mathbb{W}_{0,0,1,0}$ is parameterized by the variables $u_i,v_i,w_i$, for $i=1,2$, that satisfy
\begin{align*}
    \begin{aligned}
        &0 \leq \frac{u_1}{v_1} \leq \frac{u_2}{u_2} \leq \infty \,, & 0 &\leq \frac{u_1}{w_1} \leq \frac{u_2}{w_2} \leq \infty \,, & 0 &\leq \frac{v_1}{w_1} \leq \frac{v_2}{w_2} \leq \infty \,,\\
        0 &\leq u_1,u_2,v_1,v_2 \leq 1 \,, &  u_1+u_2&=v_1+v_2=w_1+w_2=1\,.
    \end{aligned}
\end{align*}
This is a $3$-simplex and can equivalently be described by its `hidden constraints'
\begin{align} \label{simplex}
    0 \leq& u_1 \leq v_1 \leq w_1 \leq 1 \,, & 0 \leq& w_2 \leq v_2 \leq u_2 \leq 1 \,,
\end{align}
together with the closure constraint. The boundary of a $3$-simplex is composed of four $2$-simplices:
\begin{align}
    \partial \mathbb{W}_{0,0,1,0} \sim \bigcup_{i=1}^4 \mathbb{V}_1 \times \mathbb{V}_0 \,.
\end{align}%
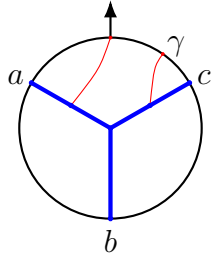
\begin{figure}
    \centering
    \begin{tikzpicture}[scale=0.2]
        \draw[thick](0,0) circle (6);
        \draw[ultra thick, blue] (0:0) -- (270:6);
        \draw[ultra thick, blue] (0:0) -- (150:6);
        \draw[ultra thick, blue] (0:0) -- (30:6);
        \draw[thick, Latex-] (0,8.5) -- (0,6);
        \draw[-][red, thin](90:6) .. controls (100: 4)..  (150:3); 
        \draw[-][red, thin](55:6) .. controls (55: 5)..  (30:3); 

        \filldraw[blue] (0,0) circle (4pt);
        \filldraw[blue] (270:6) circle (4pt);
        \filldraw[blue] (150:6) circle (4pt);
        \filldraw[blue] (30:6) circle (4pt);
        \filldraw[blue] (30:3) circle (4pt);
        \filldraw[blue] (150:3) circle (4pt);
        \filldraw[red] (90:6) circle (2.5 pt);
        \filldraw[red] (55:6) circle (2.5 pt);

        \coordinate[label=left : {$a$}] (B) at (146:6);
        \coordinate[label=below : {$b$}] (B) at (270:6);
        \coordinate[label=right : {$c$}] (B) at (34:6);
        \coordinate[label=right : {$\gamma$}] (B) at (61:6.2);
    \end{tikzpicture}
    \caption{Disk diagram corresponding to $\Omega_{0,0,1,0}^a(a,b,c,\gamma)$.}
    \label{wwwCDiagram}
\end{figure}%
The boundaries of $\mathbb{W}_{0,0,1,0}$ are reached by saturating the inequalities in \eqref{simplex}. Due to the closure conditions, saturating an inequality in the first chain of inequalities forces an inequality in the second chain to be saturated at the same time. Explicit evaluation shows
\allowdisplaybreaks{
\begin{align*}
    \begin{aligned}
        \int_{\partial\mathbb{W}_{0,0,1,0}}\Omega_{0,0,1,0}^a(a,b,c,\gamma)|_{u_1=0,u_2=1} &= \int_{\mathbb{V}_{1}}  p_{b,c} I|_{u_1=0,u_2=1} &\sim \mathcal{V}(a,\mathcal{V}(b,c,\alpha)) \,, \\[3mm]
         \int_{\partial\mathbb{W}_{0,0,1,0}}\Omega_{0,0,1,0}^a(a,b,c,\gamma)|_{u_\bullet=v_\bullet} &= \int_{\mathbb{V}_{1}}  (p_{a,c} + p_{b,c}) I|_{u_\bullet=v_\bullet} &\sim \mathcal{V}(\mathcal{V}(a,b),c,\alpha) \,, \\[3mm]
         \int_{\partial\mathbb{W}_{0,0,1,0}}\Omega_{0,0,1,0}^a(a,b,c,\gamma)|_{v_\bullet=w_\bullet} &= \int_{\mathbb{V}_{1}}  (p_{a,b} + p_{a,c}) I|_{v_\bullet=w_\bullet} &\sim \mathcal{V}(a,\mathcal{V}(b,c),\alpha) \,, \\[3mm]
        \int_{\partial\mathbb{W}_{0,0,1,0}}\Omega_{0,0,1,0}^a(a,b,c,\gamma)|_{w_1=1,w_2=0} &= \int_{\mathbb{V}_{1}}  p_{a,b} I|_{w_1=1,w_2=0} &\sim \mathcal{V}(a,b,\mathcal{U}(c,\alpha)) \,.
    \end{aligned}
\end{align*}}\noindent
For each term it is indicated which $A_\infty$-term it corresponds to, up to possible a change of integration variables and a change of the elements of $V^\ast$, as the labeling for potentials differs from the labeling for $A_\infty$-terms. The latter means in this case that $\gamma$ is replaced by $\alpha$. This proves that all $A_\infty$-terms \eqref{A_infty terms} are correctly recovered using Stokes' theorem \eqref{StokesToA_infty}. 

The above evaluation has been visualized in Fig. \ref{N0Boundaries}. Please note that the labeling of elements of $V^\ast$ is different for potentials and $A_\infty$-terms. Therefore, we change the labeling when considering a boundary, which in this case means that we replace the $\gamma$ by an $\alpha$. In the first row we again observe that a red boundary on a blue line separates the disk diagram into two disks, with the nesting arrow replacing this blue line. In the second row the green boundary shrinks the blue line to a point, like before. However, we then observe that if two legs with no red lines connected to them meet, they can be split off in a separate disk diagram. This interpretation arises from the fact that the expression for the potential produces the depicted $A_\infty$-term at this boundary. The same happens in the third row: the green boundary shrinks the blue line to a point and the $b$- and $c$-leg split off in a separate disk diagram. The red line connected to $\gamma$ is connected to the junction in the intermediate diagram and then migrates to the $a$-leg. As we will see more often, two legs with no red lines attached to them will split off as its own disk diagram. In the last row we encounter a combination we have not seen before: a green boundary on a red line. This splits off the entire leg to which this red line is attached and creates two disk diagrams, with the nesting arrow at the end of the leg in question. As turns out later, this may only happen to the last red line on a leg.

\begin{figure}[h!]
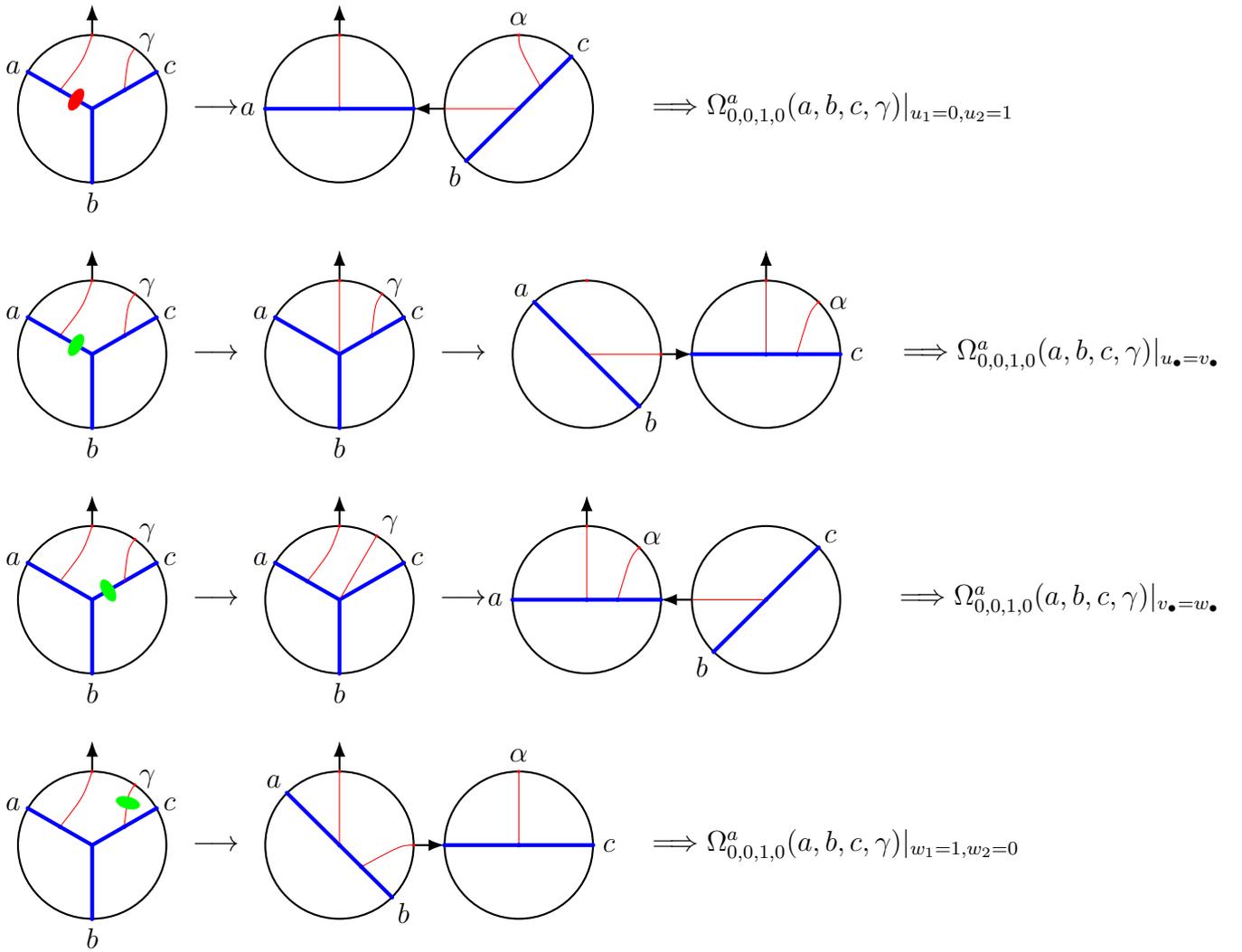

    \centering

    \caption{Visual representation of $\Omega_{0,0,1,0}^a(a,b,c,\gamma)$ evaluated at the boundaries of $\mathbb{W}_{0,0,1,0}$.} 
    \label{N0Boundaries}
\end{figure}

\paragraph{Almost-left-ordered.} 
\begin{figure}[h!]
    \centering
    %
    \caption{Graphical representation of the almost-left-ordered $A_\infty$-relations for $N=1$.}
    \label{secondOrderingN1}
\end{figure}
The $A_\infty$-relation for this ordering reads
\begin{align} \label{wwCwAinfty}
    \begin{aligned}
         \mathcal{V}(\mathcal{V}(a,b),\alpha,c)+\mathcal{V}(\mathcal{V}(a,b,\alpha),c) - \mathcal{V}(a,\mathcal{V}(b,\alpha,c)) - \mathcal{V}(a,\mathcal{U}(b,\alpha),c) + \mathcal{V}(a,b,\mathcal{U}(\alpha,c)) = 0
    \end{aligned}
\end{align}
and are visualized in Fig. \ref{secondOrderingN1}. We introduce the $3 \times 5$ matrices
\begin{align*}
    \begin{aligned}
        Q_1 =& \begin{pmatrix}
            -1 & 0 & u_1 & 0 & u_2 \\
            0 & -1 & v_1 & 0 & v_2 \\
            0 & 0 & w_1 & -1 & w_2 
        \end{pmatrix} \,, &
        Q_2 =& \begin{pmatrix}
            -1 & 0 & w_1 & 0 & w_2 \\
            0 & -1 & v_1 & 0 & v_2 \\
            0 & 0 & u_1 & -1 & u_2 
        \end{pmatrix} \,,\\
        P =& \begin{pmatrix}
            0 & 0 & p_{a,1} & 0 & p_{0,a} \\
            0 & 0 & p_{b,1} & 0 & p_{0,b} \\
            0 & 0 & p_{c,1} & 0 & p_{0,c}
        \end{pmatrix}
    \end{aligned}
\end{align*}
and define
\begin{align*}
    \begin{aligned}
        I_i =& \exp[\text{Tr}[PQ_{i}^T]+\lambda (p_{a,b} |Q_{i}^{1,2}|+ p_{a,c} |Q_i^{1,3}| + p_{b,c} |Q_{i}^{2,3}|)]\,.
    \end{aligned}
\end{align*}
The $A_\infty$-terms take the form
\allowdisplaybreaks{
\begin{align} \label{wwCwNested}
    \begin{aligned}
        \mathcal{V}(\mathcal{V}(a,b),\alpha,c) =& - (p_{a,c}+p_{b,c})(\int_0^1 dw_1\int_0^{w_1} du_1  I_1|_{u_\bullet=v_\bullet}+\int_0^1 dv_1\int_0^{v_1} du_1I_2|_{v_\bullet=w_\bullet}) \,, \\
        \mathcal{V}(\mathcal{V}(a,b,\alpha),c) =&   p_{a,b} \int_0^1 dv_1\int_0^{v_1} dw_1 I_2|_{u_1=0,u_2=1} \,,\\
        \mathcal{V}(a,\mathcal{V}(b,\alpha,c)) =& - p_{b,c} (\int_0^1 dw_1\int_0^{w_1} dv_1I_1|_{u_1=0,u_2=1}+\int_0^1 dv_1\int_0^{v_1} dw_1I_1|_{u_1=0,u_1=2}) \,,\\
        \mathcal{V}(a,\mathcal{U}(b,\alpha),c) =& -p_{a,c} (\int_0^1 dw_1\int_0^{w_1} du_1I_1|_{v_1=1,v_2=0} + \int_0^1 dw_1\int_0^{w_1} du_1I_2|_{v_1=1,v_2=0}) \,,\\
        \mathcal{V}(a,b,\mathcal{U}(\alpha,c)) =& -  p_{a,b} \int_0^1 dv_1\int_0^{v_1} du_1I_1|_{w_1=1,w_2=0} \,.    
    \end{aligned}
\end{align}}\noindent
Here, the $\mathbb{Z}_2$-symmetry of the domain $\mathbb{V}_n$ was used in several terms. The disk diagrams corresponding to potentials with the almost-left-ordering are shown in Fig. \ref{wwCwDiagram}. Please note that only one disk diagram is included with the arrow connected to the $c$-leg. Indeed, diagrams with all lines connected to a single leg are prohibited. The expressions corresponding to the disk diagrams in Fig. \ref{wwCwDiagram} are
\begin{align} \label{cubicForms}
    \begin{aligned}
    \int_{\partial\mathbb{W}_{0,0,1,0}} \Omega_{0,0,1,0}^a(a,b,\gamma,c) =& \int_{\partial\mathbb{W}_{0,0,1,0}}(p_{a,b} du_1 \wedge dv_1 + p_{a,c} du_1 \wedge dw_1 + p_{b,c} dv_1 \wedge dw_1)I_1 \,,\\
    \int_{\partial\mathbb{W}_{0,1,0,0}}\Omega_{0,1,0,0}^a(a,b,\beta,c) =& \int_{\partial\mathbb{W}_{0,1,0,0}}(p_{a,b} du_1 \wedge dv_1 + p_{a,c} du_1 \wedge dw_1 + p_{b,c} dv_1 \wedge dw_1)I_1 \,,\\
    \int_{\partial\mathbb{W}_{0,1,0,0}}\Omega_{0,1,0,0}^c(a,b,\beta,c) =& \int_{\partial\mathbb{W}_{0,1,0,0}}(-p_{b,c} du_1 \wedge dv_1 - p_{a,c} du_1 \wedge dw_1 - p_{a,b} dv_1 \wedge dw_1)I_2 \,,
    \end{aligned}
\end{align}
from left to right. The domain $\mathbb{W}_{0,0,1,0}$ is given in \eqref{simplex} and the closure constraint, while $\mathbb{W}_{0,1,0,0}$ is described by
\begin{align}\label{otherSimplex}
    \begin{aligned}
        &0 \leq \frac{u_1}{v_1} \leq \frac{u_2}{v_2} \leq \infty \,, \quad 0 \leq \frac{u_1}{w_1} \leq \frac{u_2}{w_2} \leq \infty \,, & 0 &\leq \frac{v_2}{w_2} \leq \frac{v_1}{w_1} \leq \infty \,,\\
        0 &\leq u_1,u_2,v_1,v_2 \leq 1 \,, & u_1+u_2&=v_1+v_2=w_1+w_2=1 \,,
    \end{aligned}
\end{align}
which simplifies to
\begin{align*} 
    0 \leq& u_1 \leq w_1 \leq v_1 \leq 1 \,, & 0 \leq& v_2 \leq w_2 \leq u_2 \leq 1\,.
\end{align*}
At this point, one has to be careful when describing the boundaries of $\mathbb{W}_{0,1,0,0}$. If one considers the boundary $u_1=0$, \eqref{otherSimplex} reduces to
\begin{align*}
    0 \leq& \frac{v_2}{w_2} \leq \frac{v_1}{w_1} \leq \infty \,, & u_1+u_2=v_1+v_2=w_1+w_2=1 \,,
\end{align*}
which is not the $2$-simplex in the way we usually describe it: one has to swap $v_1 \leftrightarrow v_2$ and $w_1 \leftrightarrow w_2$ to restore the correct description of the domain, which is with the labels in increasing order from left to right in the chain of inequalities and with the numerator and denominator in alphabetical order. This is related to the fact that the $q$-vectors were assigned differently for potentials than for nested vertices. Moreover, the potentials can be checked to be closed in a similar way as in the left-ordering.
\begin{figure}
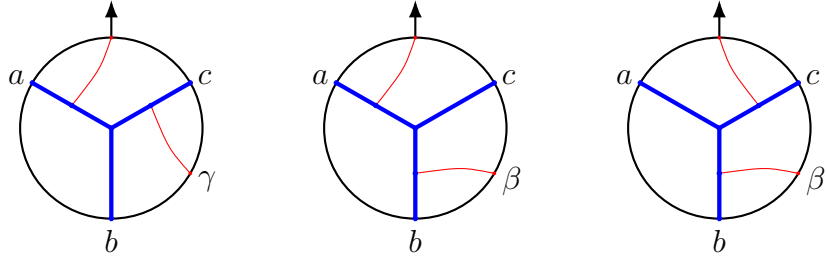

    \centering

    \caption{The disk diagrams corresponding to $\Omega_{0,0,1,0}^a(a,b,\gamma,c)$, $\Omega_{0,1,0,0}^a(a,b,\beta,c)$ and $\Omega_{0,1,0,0}^c(a,b,\beta,c)$, respectively.}
    \label{wwCwDiagram}
\end{figure}

Finally, evaluating Stokes' theorem yields the $A_\infty$-terms through
\begin{align*}
    \begin{aligned}
        \int_{\partial\mathbb{W}_{0,0,1,0}}\Omega_{0,0,1,0}^a(a,b,\gamma,c)|_{u_1=0,u_2=1} &= p_{b,c}\int_0^1dw_1\int_0^{w_1}dv_1 I_1|_{u_1=0,u_2=1} &\sim\mathcal{V}(a,\mathcal{V}(b,\alpha,c))\,, \\
        \int_{\partial\mathbb{W}_{0,0,1,0}}\Omega_{0,0,1,0}^a(a,b,\gamma,c)|_{u_\bullet=v_\bullet} &= (p_{a,c}+p_{b,c})\int_0^1dw_1\int_0^{w_1}du_1  I_1|_{u_\bullet=v_\bullet} &\sim \mathcal{V}(\mathcal{V}(a,b),\alpha,c) \,, \\ 
        \int_{\partial\mathbb{W}_{0,0,1,0}}\Omega_{0,0,1,0}^a(a,b,\gamma,c)|_{w_1=1,w_2=0} &= p_{a,b}\int_0^1dv_1\int_0^{v_1}du_1  I_1|_{w_1=1,w_2=0} &\sim \mathcal{V}(a,b,\mathcal{U}(\alpha,c)) \,, \\ 
        \int_{\partial\mathbb{W}_{0,1,0,0}}\Omega_{0,1,0,0}^a(a,b,\beta,c)|_{u_1=0,u_2=1} &= p_{b,c}\int_0^1dv_1\int_0^{v_1}dw_1 I_1|_{u_1=0,u_2=1} &\sim \mathcal{V}(a,\mathcal{V}(b,\alpha,c)) \,, \\
        \int_{\partial\mathbb{W}_{0,1,0,0}}\Omega_{0,1,0,0}^a(a,b,\beta,c)|_{v_1=1,v_2=0} &=p_{a,c}  \int_0^1dw_1\int_0^{w_1}du_1  I_1|_{v_1=1,v_2=0} &\sim \mathcal{V}(a,\mathcal{U}(b,\alpha),c) \,, \\ 
        \int_{\partial\mathbb{W}_{0,1,0,0}}\Omega_{0,1,0,0}^c(a,b,\beta,c)|_{u_1=0,u_2=1} &= - p_{a,b}\int_0^1dv_1\int_0^{v_1}dw_1  I_2|_{u_1=0,u_2=1} &\sim \mathcal{V}(\mathcal{V}(a,b,\alpha),c) \,, \\ 
        \int_{\partial\mathbb{W}_{0,1,0,0}}\Omega_{0,1,0,0}^c(a,b,\beta,c)|_{v_\bullet=w_\bullet} &= -(p_{a,c}+p_{b,c}) \int_0^1dv_1\int_0^{v_1}du_1 I_2|_{v_\bullet=w_\bullet} &\sim \mathcal{V}(\mathcal{V}(a,b),\alpha,c) \,, \\ 
        \int_{\partial\mathbb{W}_{0,1,0,0}}\Omega_{0,1,0,0}^c(a,b,\beta,c)|_{v_1=1,v_2=0} &= -p_{a,c}\int_0^1dw_1\int_0^{w_1}du_1 I_2|_{v_1=1,v_2=0} &\sim \mathcal{V}(a,\mathcal{U}(b,\alpha),c) \,.
    \end{aligned}
\end{align*}
\begin{figure}
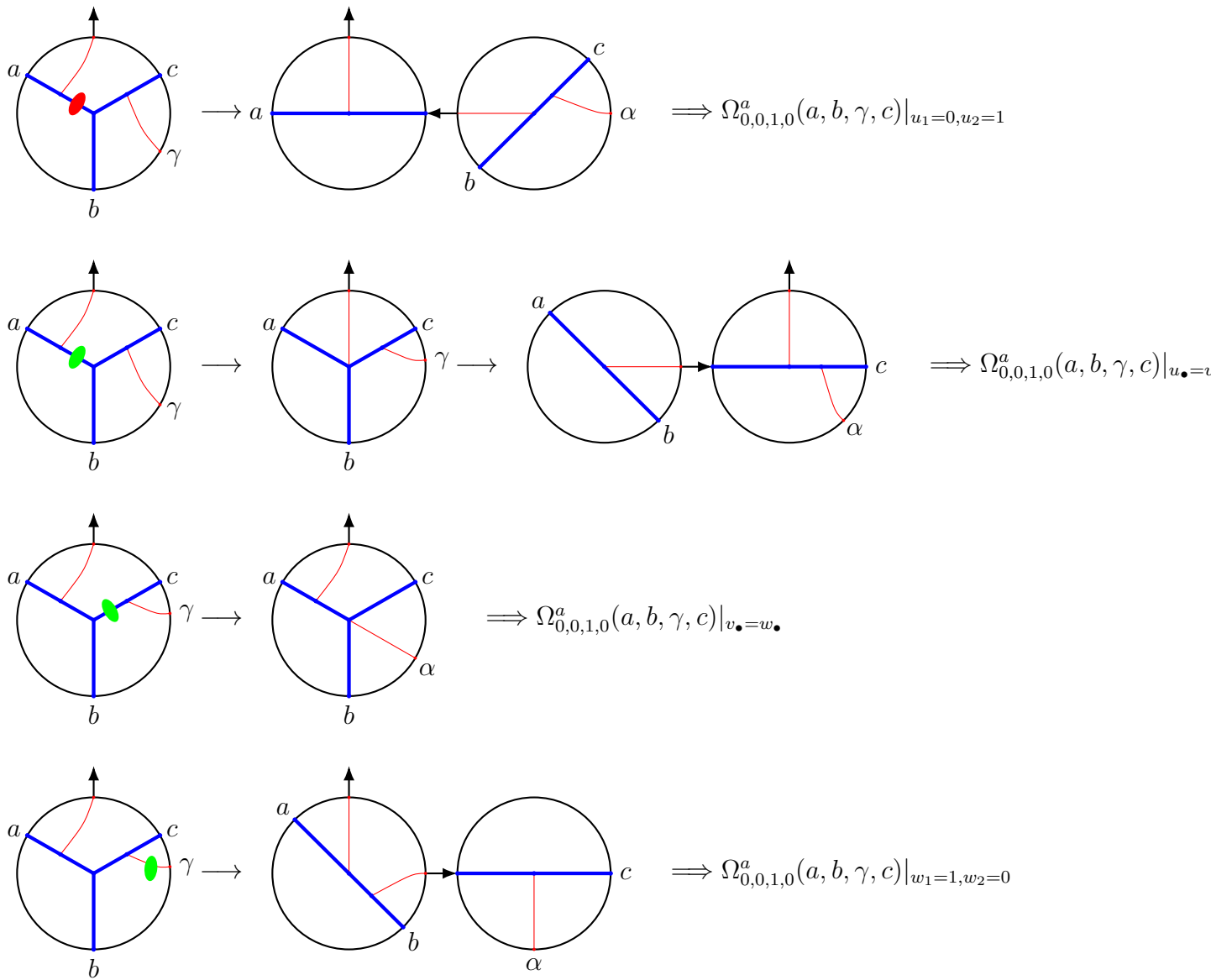

    \centering

    \caption{The disk diagrams of $\Omega_{0,0,1,0}^a(a,b,\gamma,c)$ evaluated at the boundaries of $\mathbb{W}_{0,0,1,0}$.}
    \label{1stboundaries}
\end{figure}

\begin{figure}[h!]
    \centering
    %
    \caption{The disk diagrams of $\Omega_{0,1,0,0}^a(a,b,\beta,c)$ evaluated at the boundaries of $\mathbb{W}_{0,1,0,0}$.}
    \label{2ndboundaries}
\end{figure}

\begin{figure}
    \centering
    %
    \caption{The disk diagrams of $\Omega_{0,1,0,0}^c(a,b,\beta,c)$ evaluated at the boundaries of $\mathbb{W}_{0,1,0,0}$.}
    \label{3rdboundaries}
\end{figure}

On the remaining boundaries, one finds the gluing terms
\begin{align*}
    \begin{aligned}
        \int_{\partial\mathbb{W}_{0,0,1,0}}\Omega_{0,0,1,0}^a(a,b,\gamma,c)|_{v_\bullet=w_\bullet} = (p_{a,b}+p_{a,c})\int_0^1 dw_1 \int_0^{w_1}du_1  I_1|_{v_\bullet=w_\bullet} \sim\\
        \sim\int_{\partial\mathbb{W}_{0,1,0,0}}\Omega_{0,1,0,0}^a(a,b,\beta,c)|_{v_\bullet=w_\bullet}\,,\quad\quad\hspace{9pt} \\ 
        \int_{\partial\mathbb{W}_{0,1,0,0}}\Omega_{0,1,0,0}^a(a,b,\beta,c)|_{u_\bullet=w_\bullet} = (p_{a,b}-p_{b,c})\int_0^1dv_1\int_0^{v_1}du_1 I_1|_{u_\bullet=w_\bullet} \sim\quad\quad\quad\quad\quad\hspace{2pt}\\
        \sim(p_{a,b}-p_{b,c})\int_0^1dv_1\int_0^{v_1}du_1I_2|_{u_\bullet=w_\bullet}= \int_{\partial\mathbb{W}_{0,1,0,0}}\Omega_{0,1,0,0}^c(a,b,\beta,c)|_{u_\bullet=w_\bullet} \,.\quad\quad\hspace{6pt}
    \end{aligned}
\end{align*}
All boundaries are visualized in Figs. \ref{1stboundaries} - \ref{3rdboundaries}. It is easy to see that all $A_\infty$-terms in \eqref{wwCwNested} are produced, together with gluing terms that cancel each other.

\subsection{All order generalization: left-ordered} \label{sec:leftOrdered}

It was already mentioned that the left and right-ordered cases are special: the domain is different and, in particular, for $N=1$ there are no gluing terms. Although gluing terms will appear for higher orders, there will be fewer for the left and right-ordered cases, making them easier to evaluate. In this section, we prove the $A_\infty$-relations through Stokes' theorem at all orders in the left-ordered case, from which the right-ordered case can easily be inferred.

\paragraph{$A_\infty$-terms.}
The left-ordered $A_\infty$-relations for $r+s=N \geq 1$ read
\begin{align} \label{A_infty}
    \begin{aligned}
        &\mathcal{V}(\mathcal{V}(a,b),c,\alpha_1,\dots,\alpha_{N})-\sum_{r+s=N}\mathcal{V}(a,\mathcal{V}(b,c,\alpha_1,\dots,\alpha_r),\alpha_{r+1},\dots,\alpha_{r+s})+\\
        &+\sum_{r+s=N}\mathcal{V}(a,b,\mathcal{U}(c,\alpha_1,\dots,\alpha_r),\alpha_{r+1},\dots,\alpha_{r+s})=0 \,.
    \end{aligned}
\end{align}
The individual $A_\infty$-terms then read
\begin{align} \label{nestedVertices}
    \begin{aligned}
    \mathcal{V}(\mathcal{V}(a,b),c,\alpha_1,\dots,\alpha_{N}) =& \int_{\mathbb{V}_{N}} (p_{a,c} + p_{b,c})^{N} I_1 \,,\\
    \mathcal{V}(a,\mathcal{V}(b,c,\alpha_1,\dots,\alpha_r),\alpha_{r+1},\dots,\alpha_{r+s}) =& \int_{\mathbb{V}_{s}}\int_{\mathbb{V}_{r}} (u^1_t p_{a,b}+v^1_t p_{a,c})^s p_{b,c}^r I_2 \,,\\
    \mathcal{V}(a,b,\mathcal{U}(c,\alpha_1,\dots,\alpha_r),\alpha_{r+1},\dots,\alpha_{r+s}) =& \int_{\mathbb{V}_{s+1}}\int_{\mathbb{V}_{r-1}} p_{a,b}^{s+1}(u^2_t p_{a,c} + v^2_t p_{b,c})^{r-1} I_3 \,,
    \end{aligned}
\end{align}
where
\begin{align*}
    I_i =& \exp[\text{Tr}[PQ_i^T]+\lambda (|Q_i^{12}|p_{a,b} + |Q_i^{13}| p_{a,c} + |Q_i^{23}| p_{b,c})]
\end{align*}
and
\begin{align*}
    \begin{aligned}
            Q_1 =& 

    \caption{Disk diagram corresponding to $\Omega_{0,0,3,2}^a(a,b,c,\gamma_1,\gamma_2,\gamma_3,\delta_2,\delta_1)$.}
    \label{leftOrdered}
\end{figure}

\paragraph{Domain.}
An example of a disk diagram for a potential with a left-ordering is shown in Fig. \ref{leftOrdered}. The relevant potentials and domain for $m+n=N$ are $\Omega_{0,0,m,n}^a(a,b,c,\gamma_1,\dots,\gamma_m,\delta_n,\dots,\delta_1)$ and $\mathbb{W}_{0,0,m,n}$, respectively. The domain $\mathbb{W}_{0,0,m,n}$ is described by
\begin{align} \label{leftOrderedDomain}
    \begin{aligned}
        0 &\leq u_i^\bullet,v_i^\bullet,w_i^\bullet \leq 1 \,, \quad  \sum_i(u_i^\bullet,v_i^\bullet,w_i^\bullet)=(1,1,1)\,,\\
        0 &\leq u_1^c \leq v_1^c \leq w_1^c \leq 1 \,, \quad 0 \leq w_1^a \leq v_1^a \leq u_1^a \leq 1\,,\\
        0 &\leq \frac{u_m^c}{v_m^c} = \dots = \frac{u_1^c}{v_1^c} \leq \frac{u_{n+1}^a}{v_{n+1}^a} \leq \dots \leq \frac{u_1^a}{v_1^a} \leq \infty\,,\\
        0 &\leq \frac{u_1^c}{w_1^c} \leq \dots \leq \frac{u_m^c}{w_m^c} \leq \frac{u_{n+1}^a}{w_{n+1}^a} \leq \dots \leq \frac{u_1^a}{w_1^a} \leq \infty\,,\\
        0 &\leq \frac{v_1^c}{w_1^c} \leq \dots \leq \frac{v_m^c}{w_m^c} \leq \frac{v_{n+1}^a}{w_{n+1}^a} = \dots = \frac{v_1^a}{w_1^a} \leq \infty \,.
    \end{aligned}
\end{align}
and has a special visualization in $\mathbb{R}^3$ in terms of the vectors $\vec{q}_a, \vec{q}_b, \vec{q}_c$ and $\vec{q}_{\bullet,i}$. In Fig. \ref{R3} we see that they form a closed polygon in $\mathbb{R}^3$. The domain is described by three chains of (in)equalities that obey the same chronological ordering, as the $uv$-chain starts with equalities. Therefore, the projection of the closed polygon on the $uv$-, $uw$- and $vw$-plane are swallowtails, each described by one of these chains, as shown in Fig. \ref{R3}. We refer to these polygons $(\vec{q}_{a},\vec{q}_b,\vec{q}_c,\vec{q}_{c,1},\dots,\vec{q}_{c,m},\vec{q}_{a,n+1},\dots,\vec{q}_{a,1})$ in $\mathbb{R}^3$ as maximally concave polygons. The equalities in the $uv$-plane ensure that the vectors $\vec{q}_{c,i}$ and $\vec{q}_{c,j}$ are coplanar, whereas the equalities in the $vw$-chain imply that vectors $\vec{q}_{a,i}$ and $\vec{q}_{a,j}$ are coplanar. This is depicted in Fig. \ref{coplan} for the domain $\mathbb{W}_{0,0,2,2}$. The blue arrows $\vec{q}_{c,1}$ and $\vec{q}_{c,2}$ lie in the same plane, highlighted by the blue shaded region, while the vectors $\vec{q}_{a,1}$, $\vec{q}_{a,2}$ and $\vec{q}_{a,3}$ are coplanar in the red shaded plane.

\begin{figure}
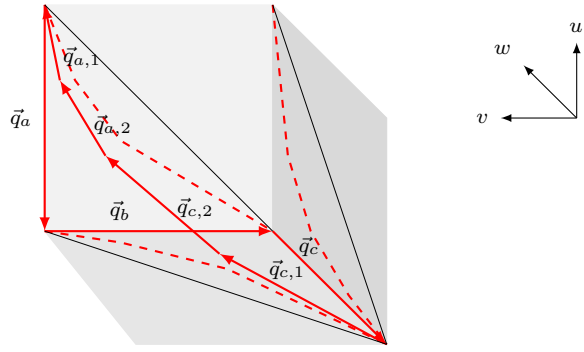

    \centering

    \caption{A visual representation of the left-ordered integration domain $\mathbb{W}_{0,0,2,1}$. The vectors $(\vec{q}_a,\vec{q}_b,\vec{q}_c,\vec{q}_{c,1},\vec{q}_{c,2},\vec{q}_{a,2},\vec{q}_{a,1})$ form a maximally concave polygon.}
    \label{R3}
\end{figure}

\begin{figure}[h!]

\begin{center}


%

\end{center}

\caption{On the left (right) the blue (red) shaded region depicts the plane in which the vectors $\vec{q}_{c,i}$ ($\vec{q}_{a,i}$) lie in $\mathbb{W}_{0,0,2,2}$. The scale in the $v$-direction was doubled to accentuate the details in the pictures.}\label{coplan}
\end{figure}

\paragraph{Potential.}
Following the recipe in section \ref{sec:recipe}, we construct a $2N$-form 
\begin{align*}
\Omega_{m+n}^a(a,b,c,\gamma_1,\dots,\gamma_m,\delta_n,\dots,\delta_1)
\end{align*}
on the space $\mathbb{U}_{m+n} \supseteq \mathbb{W}_{0,0,m,n}$. 
We construct a potential that reads
\begin{align*}                          \Omega_{m+n}^a(a,b,c,\gamma_1,\dots,\gamma_m,\delta_n,\dots,\delta_1) =& \mu I_{m+n} \,,
\end{align*}
with
\begin{align}
    I_{m+n} = & \exp[\text{Tr}[PQ^T] + \lambda(|Q^{12}|p_{a,b} + |Q^{13}|p_{a,c} +|Q^{23}|p_{b,c})]
\end{align}
and
\begin{align} \label{leftQ}
    Q =& (\vec{q}_{a},\vec{q}_{b},\vec{q}_{c},\vec{q}_{c,1},\dots,\vec{q}_{c,m},\vec{q}_{a,n},\dots,\vec{q}_{a,1})\,.
\end{align}
The measure $\mu$ is given in \eqref{measure}. The restriction of the potential to $\mathbb{W}_{0,0,m,n}$ is effectuated by requiring
\begin{align*}
    \begin{aligned}
        u_1,\dots,u_m &\rightarrow u_1^c,\dots,u_m^c \,, & u_{m+1},\dots,u_{m+n+1} &\rightarrow u_{n+1}^a,\dots,u_1^a
    \end{aligned}
\end{align*}
and
\begin{align} \label{restrictionLeftOrdered}
    \alpha &= \frac{u_1^c}{v_1^c} = \dots = \frac{u_m^c}{v_m^c} \,, & \frac{1}{\beta} &=\frac{v_{n+1}^a}{w_{n+1}^a} = \dots = \frac{v_1^a}{w_1^a} \,.
\end{align}
Requiring that no obvious singularities arise,\footnote{As an example, $\mathbb{W}_{0,0,m,n}$ contains the subspace attained by setting $u_1^c=0$. We can solve \eqref{restrictionLeftOrdered} by $v_i^c=\frac{v_1^c}{u_1^c}u_i^c$, but this looks singular at $u_1^c = 0$, while $u_i^c=\frac{u_1^c}{v_1^c}v_i^c$ behaves nicely as $0 \leq u_1^c \leq v_1^c \leq 1$.} we choose
\begin{align} \label{identification1}
    \begin{aligned}
        u_i^c &= \alpha v_i^c \,,& w_i^a &= \beta v_i^a \,.
    \end{aligned}
\end{align}
Explicitly, we write
\begin{align*}
    \alpha =& \frac{u_1^c}{v_1^c} \,, & \beta =&  \frac{1 - \sum_{i=1}^{m}w^c_i}{1 - \sum_{i=1}^m v^c_i} \,.
\end{align*}
Finally, the potential we are interested in reads
\begin{align*}
\Omega_{0,0,m,n}^a(a,b,c,\gamma_1,\dots,\gamma_m,\delta_n,\dots,\delta_1) &= \Omega_{m+n}^a(a,b,c,\gamma_1,\dots,\gamma_m,\delta_n,\dots,\delta_1)\Big|_{\mathbb{W}_{0,0,m,n}} = \mu_{m,n}I_{m,n} \,,
\end{align*}
with
\begin{align*}
    \begin{aligned}
        \mu_{m,n} &= \mu\Big|_{\mathbb{W}_{0,0,m,n}}\,, & I_{m,n} &= I_{m+n}\Big|_{\mathbb{W}_{0,0,m,n}}\,.
    \end{aligned}
\end{align*}
Solving the closure constraint for $\vec{q}_{a,1}$, i.e.
\begin{align*}
    \vec{q}_{a,1}=1-\sum_{i=1}^m \vec{q}_{c,i} - \sum_{i=1}^n \vec{q}_{a,i} \,,
\end{align*}
yields the measure
\begin{align*}
    \begin{aligned}
        \mu_{m,n} &= (\alpha p_{a,c} + p_{b,c})^{m-1}(p_{a,b} + \beta p_{a,c})^n \bigg[ \sum_{j=2}^{n+1} \frac{1-v_{1}^a}{v_{1}^a v_1^c} v_j^a  \\
         &\times(p_{a,c} du_1^c \wedge dv_1^c \wedge dw_1^c \wedge \dots \wedge \widehat{dv_j^c} \wedge \dots dv_m^c \wedge dw_m^c \wedge du_{2}^a \wedge dv_{2}^a \wedge \dots \wedge du_{n+1}^a \wedge dv_{n+1}^a+\\
         &+ p_{b,c} du_1^c \wedge dv_1^c \wedge dw_1^c \wedge \dots \wedge \widehat{du_j^c} \wedge \dots dv_m^c \wedge dw_m^c \wedge du_{2}^a \wedge dv_{2}^a \wedge \dots \wedge du_{n+1}^a \wedge dv_{n+1}^a) + \\
         &+p_{b,c} dv_1^c \wedge dw_1^c \wedge \dots dv_m^c \wedge dw_m^c \wedge du_{2}^a \wedge dv_{2}^a \wedge \dots \wedge du_{n+1}^a \wedge dv_{n+1}^a \bigg] \,,
    \end{aligned}
\end{align*}
where the symbol $\,\,\widehat{.}\,\,$ denotes omission and

\begin{align*}
    I_{m,n} =& \exp[\text{Tr}[P Q_{m,n}^T] + \lambda(|Q_{m,n}^{12}|p_{a,b} + |Q_{m,n}^{13}|p_{a,c} + |Q_{m,n}^{23}|p_{b,c})] \,,
\end{align*}
for
\begin{align} \label{QLeftRestriction}
    Q_{m,n} =& \begin{pmatrix}
        -1 & 0 & 0 & \alpha v_1^c  & \dots & \alpha v_m^c& u_{n+1}^a & \dots & u_1^a \\
            0 & -1 & 0 & v_1^c  & \dots &v_m^c & v_{n+1}^a & \dots & v_1^a  \\
            0 & 0 & -1 & w_1^c  & \dots & w_m^c& \beta v_{n+1}^a & \dots & \beta v_1^a
    \end{pmatrix}
\end{align}
and $P$ as defined in \eqref{PQ}. From now on we will omit the arguments of $\Omega_{0,0,m,n}^a$, as they should be clear from the subscript.

The domain $\mathbb{W}_{0,0,m,n}$ \eqref{leftOrderedDomain} is vastly more complicated than the domains discussed in the lower order examples. In particular, the chains of (in)equalities introduce new types of boundaries. In the following, we will categorize the boundaries according to whether the differential form $\Omega_{0,0,m,n}^a$ evaluates to an $A_\infty$-term, gluing term, zero, or the boundary turns out to be a higher codimension boundary. Of course, the latter does not play a role in Stokes' theorem and therefore does not contribute to the proof. In principle, saturating any inequality in \eqref{leftOrderedDomain} leads to a boundary, but some inequalities might seem to be missing in this categorization. This is simply because they are already accounted for in some other boundary. For example, the $uw$-chain can almost entirely be derived from the $uv$- and $vw$-chain.

\paragraph{$A_\infty$-terms.} The same boundaries that were present in the left-ordered $N=1$ example yield $A_\infty$-terms, with the boundary $w_1^c=1$ taking a more general form, see boundary 5.
\begin{itemize}
    \item Boundary 1: At this boundary
    \begin{align*}
        u_i^\bullet=0 \,.
    \end{align*}
    \begin{figure}[h!]
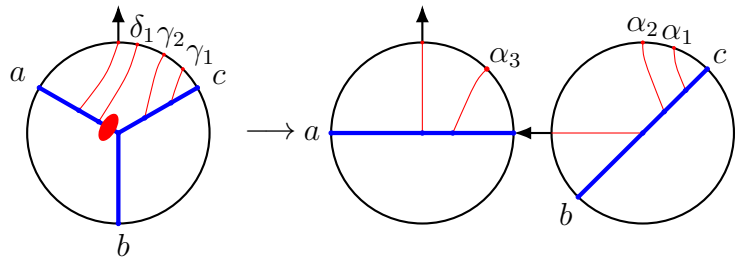

        \centering

        \caption{An example of boundary 1 contributing to $\mathcal{V}(a,\mathcal{V}(b,c,\alpha_1,\alpha_2),\alpha_3)$.}
        \label{B1}
    \end{figure}
    If $u_i^a=0$, the $uv$-chain becomes
    \begin{align*}
        0 \leq \frac{u_m^c}{v_m^c} \leq \dots \leq \frac{u_1^c}{v_1^c} \leq \frac{u_{n+1}^a}{v_{n+1}^a} \leq \dots \leq \frac{0}{v_{i}^a} \leq \dots \leq \frac{u^a_1}{v_1^a} \leq \infty \,,
    \end{align*}
    which forces $u_j^c=0$ for $j=1,\dots,m$ and $u_j^a=0$ for $j \geq i$. This leads to a higher codimension boundary. However, if $u_i^c=0$ for $i=1,\dots,m$, leading to $\alpha=0$, and the other $u$-variables nonzero, we find an $A_\infty$-term. After the change of coordinates
    \begin{align*}
    \begin{aligned}
        v_i^c &\rightarrow u^1_i\,, & w_i^c &\rightarrow v^1_{i} \,, & & \text{for } i=1,\dots,m\,,\\
        u_i^a &\rightarrow u^2_{n+2-i} \,, & v_{i}^a &\rightarrow u^1_{m+1} v^2_{n+2-i} \,, & w_i^a \rightarrow v_{m+1}^1 v_{n+2-i}^2\,, \, & \text{for } i=1,\dots,n+1\,,
    \end{aligned}
    \end{align*}
    this boundary is identified as $\mathbb{V}_{n} \times \mathbb{V}_{m}$ and 
\begin{align*}
        \boxed{\int_{\partial\mathbb{W}_{0,0,m,n}}\Omega_{0,0,m,n}^a|_{u_i^c=0} \sim \mathcal{V}(a,\mathcal{V}(b,c,\alpha_1,\dots,\alpha_m),\alpha_{m+1},\dots,\alpha_{m+n})}
\end{align*}
    on this boundary, with the exception of $\mathcal{V}(a,\mathcal{V}(b,c),\alpha_1,\dots,\alpha_{n})$, since the recipe required at least one element of $V^\ast$ to be attached to the $c$-leg. An example of a disk diagram at this boundary is shown in Fig. \ref{B1}.

    \item Boundary 2: At this boundary
    \begin{align*}
        u_1^c=v_1^c \,.
    \end{align*}
    The $uv$-chain of (in)equalities then becomes
    \begin{align*}
        1 \leq& \frac{u_{n+1}^a}{v_{n+1}^a} \leq \dots \leq \frac{u_1^a}{v_1^a} \leq \infty
    \end{align*}
    and the closure constraint requires $u_i^c=v_i^c$ for $i=1,\dots,m$ and $u_i^a=v_i^a$ for $i=1,\dots, n+1$. However, this describes a higher codimension boundary, except for $n=0$. In this case, we find, after the change of coordinates
    \begin{align*}
        \begin{aligned}
            v_i^c &\rightarrow u^2_i \,, & w_i^c &\rightarrow v^2_i \,, & & \text{for } i=1,\dots, m \,,\\
            u_1^a &\rightarrow u_{m+1}^2 \,, & v_1^a &\rightarrow u_{m+1}^2 \,, & w_1^a &\rightarrow v_{m+1}^2 \,,
        \end{aligned}
    \end{align*}
    that the boundary is identified as $\mathbb{V}_{m}\times\mathbb{V}_0$ and 
    \begin{align*}
        \boxed{\int_{\partial\mathbb{W}_{0,0,m,0}}\Omega_{0,0,m,0}^a|_{u_1^c=v_1^c} \sim \mathcal{V}(\mathcal{V}(a,b),c,\alpha_1,\dots,\alpha_{m})}
    \end{align*}
    on this boundary. An example of a disk diagram at this boundary is shown in Fig. \ref{B2}.
        \begin{figure}[h!]
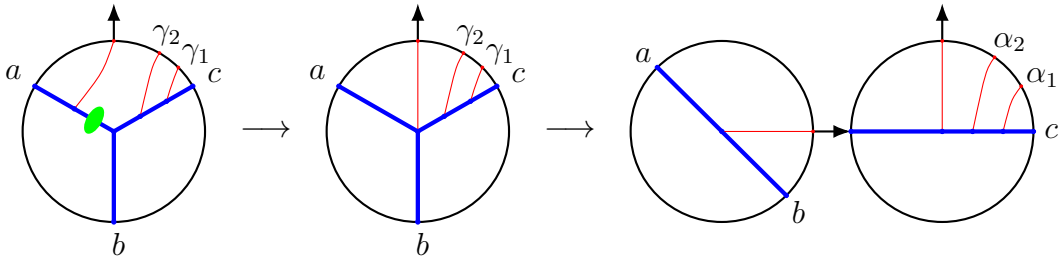

        \centering

        \caption{An example of boundary 2 contributing to $\mathcal{V}(\mathcal{V}(a,b),c,\alpha_1,\alpha_2)$.}
        \label{B2}
    \end{figure}

    \item Boundary 3: At this boundary
    \begin{align*}
        v_1^c&=w_1^c \,.
    \end{align*}
    The $vw$-chain of (in)equalities then becomes
    \begin{align*}
        1 &\leq \frac{v_2^c}{w_2^c} \leq \dots \leq \frac{v_m^c}{w_m^c} \leq \frac{v_{n+1}^a}{w_{n+1}^a} = \dots = \frac{v_1^a}{w_1^a} \leq \infty
    \end{align*}
    and the closure constraint requires $v_i^c=w_i^c$ for $i=1,\dots, m$ and $v_i^a=w_i^a$ for $i=1,\dots,n+1$. However, this describes a higher codimension boundary, except when $m=1$. In that case, we find after the change of coordinates
    \begin{align*}
        \begin{aligned}
            u_1^c &\rightarrow u_1^2 \,, & v_1^c &\rightarrow v_1^2 \,, & w_1^c &\rightarrow v_1^2 \,,\\
            u_i^a &\rightarrow u^2_{n+3-i} \,, & v_i^a &\rightarrow v^2_{n+3-i} \,, & & &\text{for } i=1,\dots,n+1 \,,\\
        \end{aligned}
    \end{align*}
    that the boundary is identified as $\mathbb{V}_{n+1} \times \mathbb{V}_{0}$ and  
    \begin{align*}
        \boxed{\int_{\partial\mathbb{W}_{0,0,1,n}}\Omega_{0,0,1,n}^a|_{v_1^c=w_1^c} \sim \mathcal{V}(a,\mathcal{V}(b,c),\alpha_1,\dots,\alpha_{n+1})}
    \end{align*}
    on this boundary. An example of a disk diagram at this boundary is shown in Fig. \ref{B3}.
\begin{figure}[h!]
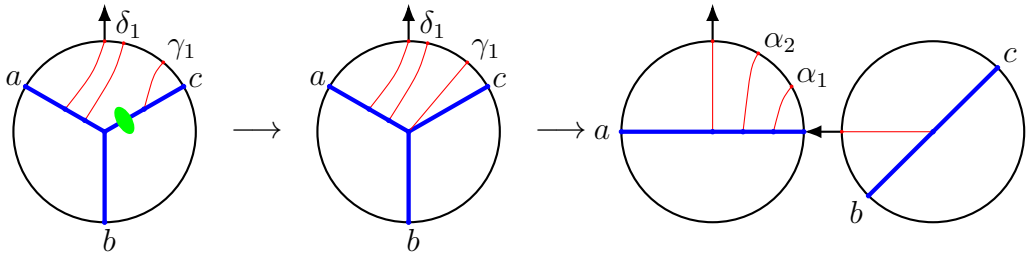

    \centering

    \caption{An example of boundary 3 contributing to $\mathcal{V}(a,\mathcal{V}(b,c),\alpha_1,\alpha_2)$.}
    \label{B3}
\end{figure}

    \item Boundary 4: At this boundary
    \begin{align*}
        w_i^\bullet=1 \,.
    \end{align*}
    If $w_i^a=1$, this requires both $\beta=1$ and $v_i^a=1$ and yields a higher codimension boundary. Next, we consider $w_i^c=1$. Then, the closure condition requires all other $ w$ variables to be zero, which leads to $\beta=0$. This yields a higher codimension boundary, except for $m=1$. After the change of coordinates
    \begin{align*}
        \begin{aligned}
            u_1^c &\rightarrow u^2_{1} \,, & v_1^c &\rightarrow v^2_{1} \,,\\
            u_i^a &\rightarrow u^2_{n+3-i} \,, & v_i^a &\rightarrow v^2_{n+3-i} \,, & \text{for } i=1,\dots,n+1 \,,            
        \end{aligned}
    \end{align*}
    the boundary is identified as $\mathbb{V}_{n+1}\times\mathbb{V}_{0}$ and 
    \begin{align*}
        \boxed{\int_{\partial\mathbb{W}_{0,0,1,n}}\Omega_{0,0,1,n}^a|_{w_1^c=1} \sim \mathcal{V}(a,b,\mathcal{U}(c,\alpha_1),\alpha_2,\dots,\alpha_{n+1})}
    \end{align*}
    on this boundary. An example of a disk diagram at this boundary is shown in Fig. \ref{B4}.
    \begin{figure}[h!]
        \centering

        \caption{An example of boundary 4 contributing to $\mathcal{V}(a,b,\mathcal{U}(c,\alpha_1),\alpha_2)$.}
        \label{B4}
    \end{figure}

    \item Boundary 5: At this boundary
    \begin{align*}
        w_i^\bullet &= 0\,.
    \end{align*}
        \begin{figure}[h!]
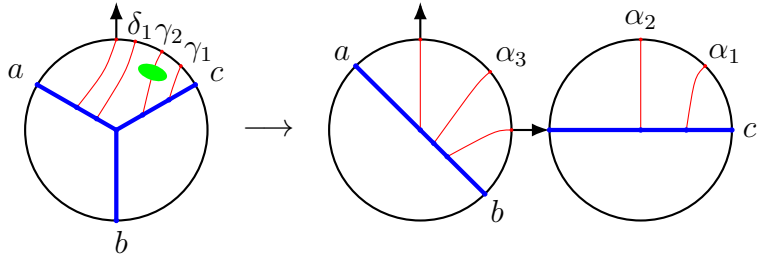

        \centering
        %
        \caption{An example of boundary 5 contributing to $\mathcal{V}(a,b,\mathcal{U}(c,\alpha_1,\alpha_2),\alpha_3)$.}
        \label{B5}
    \end{figure}
    If $w_i^c=0$, the $uw$-chain is equivalent to
    \begin{align*}
        \infty \geq \frac{w_1^c}{u_1^c} \geq \dots \geq \frac{0^{+}}{u_i^c} \geq \dots \geq \frac{w_m^c}{u_m^c} \geq \frac{w_{n+1}^a}{u_{n+1}^a} \geq \dots \geq \frac{w_1^a}{u_1^a} \geq 0\,,
    \end{align*}
    which forces $w_j^c=0$ for $j \leq i$ and $w_j^a=0$ for $j=1,\dots,n+1$. This leads to a higher codimension boundary. Next we consider $w_i^a=0$, which leads to $\beta=0$. For $m=1$ this is equivalent to boundary 4. After the change of coordinates
    \begin{align*}
    \begin{aligned}
        u_1^c &\rightarrow u^2_{1} u^1_1 \,,\\
        v_i^c &\rightarrow v^2_{1} u^1_i\,, & w_i^c &\rightarrow v^1_i \,, & \text{for } i=1,\dots,m \,,\\
        u_{i}^a &\rightarrow u^2_{n+3-i} \,, & v_{i}^a &\rightarrow v^2_{n+3-i} \,, & \text{for } i=1,\dots,n+1
    \end{aligned}
    \end{align*}
    the boundary is identified as $\mathbb{V}_{n+1}\times\mathbb{V}_{m-1}$ and 
    \begin{align*}
        \boxed{\int_{\partial\mathbb{W}_{0,0,m,n}}\Omega_{0,0,m,n}^a|_{w_{i}^a=0} \sim \mathcal{V}(a,b,\mathcal{U}(c,\alpha_,\dots,\alpha_{m}),\alpha_{m+1}\dots,\alpha_{m+n})}
    \end{align*}
    on this boundary, with the exception of $\mathcal{V}(a,b,\mathcal{U}(c,\alpha_1),\alpha_{2},\dots,\alpha_{n+1})$. An example of a disk diagram at this boundary is shown in Fig. \ref{B5}.
\end{itemize}

\paragraph{Gluing terms.}
\begin{itemize}
\item Boundary 6: At this boundary
\begin{align*}
    \frac{v_m^c}{w_m^c} &= \frac{v_{n+1}^a}{w_{n+1}^a} \,.
\end{align*}
For $m=1$ the $vw$-chain becomes
\begin{align*}
    0 \leq \frac{v_1^c}{w_1^c} = \frac{v_{n+1}^a}{w_{n+1}^a} = \dots = \frac{v_1^a}{w_1^a}  \leq \infty \,.
\end{align*}
\begin{figure}[h!]
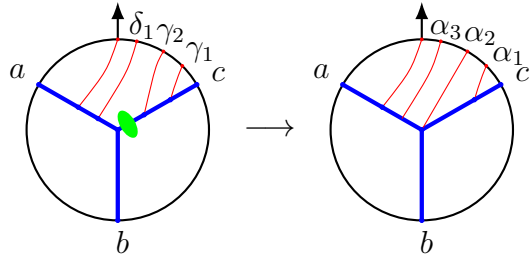

    \centering

    \caption{An example of boundary 6 contributing to a gluing term.}
    \label{B6}
\end{figure}
The closure constraint then requires $v_i^\bullet=w_i^\bullet$, so this boundary is equivalent to boundary 3 when $m=1$. For $m \neq 1$, $\int_{\partial\mathbb{W}_{0,0,m,n}}\Omega_{0,0,m,n}^a$ does not yield a familiar $A_\infty$-term and it does not vanish either. An example of a
disk diagram at this boundary is shown in Fig. \ref{B6}.

\item Boundary 7: At this boundary
\begin{align*}
    \frac{u_1^c}{v_1^c} &= \frac{u_{n+1}^a}{v_{n+1}^a}\,.
\end{align*}
For $n=0$ the $uv$-chain becomes
\begin{align*}
    0 \leq \frac{u_1^c}{v_1^c} = \dots = \frac{u_{m}^c}{v_{m}^c} = \frac{u_1^a}{v_1^a} \leq \infty \,.
\end{align*}
The closure constraint then requires $u_i^\bullet=v_i^\bullet$, so this boundary is equivalent to boundary 2 when $n=0$. For $n \neq 0$, $\int_{\partial\mathbb{W}_{0,0,m,n}}\Omega_{0,0,m,n}^a$ does not yield a familiar $A_\infty$-term and it does not vanish either. An example of a
disk diagram at this boundary is shown in Fig. \ref{B7}.
\begin{figure}[h!]
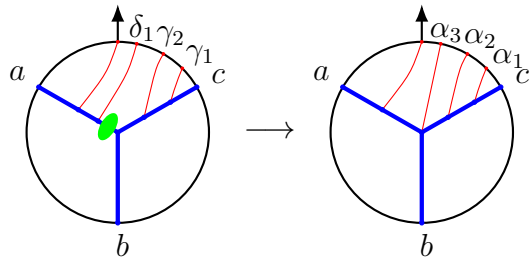

    \centering

    \caption{An example of boundary 7 contributing to a gluing term.}
    \label{B7}
\end{figure}
\end{itemize}

Fortunately, it turns out not to be necessary to explicitly evaluate $\Omega_{0,0,m,n}^a$ on boundary 6 and 7, as the contributions can be seen to cancel each other by construction. $\Omega_{0,0,m,n}^a$ on boundary 6 is found by restricting $\Omega_{m+n}^a$ to a submanifold by requiring
\begin{align*}
    \frac{u_1^c}{v_1^c}=&\dots=\frac{u_{m}^c}{v_m^c} \,, & \frac{v_m^c}{w_m^c}=\frac{v_{n+1}^a}{w_{n+1}^a}=\dots=\frac{v_1^a}{w_1^a} \,,
\end{align*}
while on boundary 7 one invokes
\begin{align*}
    \frac{u_1^c}{v_1^c}=&\dots=\frac{u_m^c}{v_m^c}=\frac{u_{n+1}^a}{v_{n+1}^a} \,, & \frac{v_{n+1}^a}{w_{n+1}^a}=\dots=\frac{v_1^a}{w_1^a} \,.
\end{align*}
Clearly, boundary 6 is equivalent to boundary 7 after the shift $m \rightarrow m+1$ and $n \rightarrow n-1$ and relabeling of the variables. Since \eqref{StokesToA_infty} sums over all $m,n$, such that $m+n=N$, all gluing terms cancel pairwise.

\paragraph{Zero measure terms.} Remember the definition $\vec{q}_{\bullet,i}=(u_i^\bullet,v_i^\bullet,w_i^\bullet)$ for the vectors that fill up the matrix $Q_D$. From \eqref{measure} it is clear that if two $q$-vectors are colinear , e.g. $\vec{q}_{c,i} = \xi \vec{q}_{c,j}$, with $\xi\in\mathbb{R}$, then $\mu_i^c \wedge \mu_j^c = 0$.
\begin{itemize}
\item Boundary 8: At this boundary
\begin{align*}
    \frac{v_i^c}{w_i^c} &= \frac{v_{i+1}^c}{w_{i+1}^c} \,, & \text{for } i=1,\dots ,m-1 \,,
\end{align*}
which makes the vectors $\vec{q}_{c,i}$ and $\vec{q}_{c,i+1}$ colinear and the measure in $\Omega_{0,0,m,n}^a$ vanishes on this boundary.

\item Boundary 9: At this boundary
\begin{align*}
    \frac{u_{i}^a}{v_{i}^a} &= \frac{u_{i+1}^a}{v_{i+1}^a} \quad \text{for } i=1,\dots, n \,,
\end{align*}
which makes the vectors $\vec{q}_{a,i}$ and $\vec{q}_{a,i+1}$ colinear. As a result the measure in $\Omega_{0,0,m,n}^a$ vanishes on this boundary.
\end{itemize}

\paragraph{Higher codimension boundaries.} Some types of boundaries are necessarily higher codimension boundaries: saturating one inequality leads to saturation of more inequalities and hence the resulting submanifold is parameterized by less than $2(m+n)$ independent coordinates. We have seen examples of this for the boundaries that also produce $A_\infty$-terms for specific values of $m$ and $n$. Here we present the higher codimension boundaries that are not discussed yet.

\begin{itemize}

\item Boundary 10: At this boundary
\begin{align*}
    v_i^\bullet=0 \,.
\end{align*}
The $vw$-chain becomes
\begin{footnotesize}
\begin{align*}
    \begin{aligned}
        0 \leq& \frac{v_1^c}{w_1^c} \leq \dots \leq \frac{0}{w_i^c} \leq \dots \leq \frac{v_m^c}{w_m^c} \leq \frac{v_{n+1}^a}{w_{n+1}^a} = \dots = \frac{v_1^a}{w_1^a} \leq \infty \,, \quad \text{or} \\
        0 \leq& \frac{v_1^c}{w_1^c} \leq \dots \leq \frac{v_m^c}{w_m^c} \leq \frac{v_{n+1}^a}{w_{n+1}^a} = \dots = \frac{0}{w_{i}^a} = \dots = \frac{v_1^a}{w_1^a}  \leq \infty \,, 
    \end{aligned}
\end{align*}
\end{footnotesize}\noindent
which implies $v_j^c=0$ for $j < i$ or all $v$-variables zero, respectively. In both cases one finds a higher codimension boundary. Only for $v_1^c=0$, the $vw$-chain is not responsible for a higher codimension boundary, but the condition
\begin{align*}
    0 \leq u_1^c \leq v_1^c \leq w_1^c \leq 1
\end{align*}
is, as it imposes $u_1^c=v_1^c=0$.

\item Boundary 11: At this boundary
\begin{align*}
    u_i^\bullet=1 \,.
\end{align*}
The closure constraint implies that all other $u$-variables vanish. If $m=1$ and $n=0$ and the boundary is given by $u_1^c=1$, the condition
\begin{align*}
    0 \leq u_1^c \leq v_1^c \leq w_1^c \leq 1
\end{align*}
 implies $u_1^c=v_1^c=w_1^c=1$.

\item Boundary 12: At this boundary
\begin{align*}
    v_i^\bullet=1 \,.
\end{align*}
The closure constraint implies that all other $v$-variables vanish. If $m=1$ and $n=0$ and the boundary is given by $v_1^c=1$, the condition
\begin{align*}
    0 \leq u_1^c \leq v_1^c \leq w_1^c \leq 1
\end{align*}
implies $v_1^c=w_1^c=1$.

\item Boundary 13: At this boundary
\begin{align*}
    \frac{u_m^c}{w_m^c}=\frac{u_{n+1}^a}{w_{n+1}^a}\,.
\end{align*}
Since
\begin{align*}
    \frac{u_m^c}{v_m^c}\frac{v_m^c}{w_m^c} =& \frac{u_m^c}{w_m^c} \leq \frac{u_{n+1}^a}{w_{n+1}^a} =\frac{u_{n+1}^a}{v_{n+1}^a}\frac{v_{n+1}^a}{w_{n+1}^a}
\end{align*}
and
\begin{align*}
    \frac{u_m^c}{v_m^c} \leq& \frac{u_{n+1}^a}{v_{n+1}^a} \,, \quad \frac{v_m^c}{w_m^c} \leq \frac{v_{n+1}^a}{w_{n+1}^a} \,,
\end{align*}
this boundary implies $\frac{u_m^c}{v_m^c} = \frac{u_{n+1}^a}{v_{n+1}^a}$ and $\frac{v_m^c}{w_m^c} = \frac{v_{n+1}^a}{w_{n+1}^a}$ and thus yields a higher codimension boundary.
\end{itemize}

With the above categorization of boundaries, we have established the equivalence of Stokes' theorem for $\Omega_{0,0,m,n}^a$ on $\mathbb{W}_{0,0,m,n}$ and the left-ordered $A_\infty$-relation, \eqref{StokesToA_infty},
that is,
\begin{align} \label{refinedStokes}
    0=\sum_{m+n=N}\int_{\mathbb{W}_{0,0,m,n}} d\Omega_{0,0,m,n}^a = \sum_{m+n=N}\int_{\partial\mathbb{W}_{0,0,m,n}} \Omega_{0,0,m,n}^a  \quad \Longleftrightarrow \quad A_\infty\text{-relations} \,. 
\end{align}
The same proof for the right-ordered $A_\infty$-relations can easily be inferred from \eqref{flip}, which relates potentials of the type $\Omega_{k,l,m,n}^a$ with $\Omega_{k,l,m,n}^c$. The disk diagram of the potential $\Omega_{m,0,0,n}^c$ is just the mirror image of the disk diagram for $\Omega_{0,0,m,n}$. It is easy to see from \eqref{generalA_infty} that the right-ordered $A_\infty$-terms can be obtained from the left-ordered ones by mirror symmetry as well.
\subsection{All order generalization: arbitrary ordering} \label{sec:allOrderings}
In the previous section, we considered the $A_\infty$-relations with a specific ordering. Now we turn to the $A_\infty$-relations with arbitrary ordering. These $A_\infty$-relations are given in \eqref{A} and the $A_\infty$-terms are presented in \eqref{generalA_infty}. The corresponding $Q$ matrices are given in \eqref{Qs}. As we explained in Sec. \ref{sec:recipe}, there are two types of diagrams and potentials: the $a$- and $c$- diagrams that correspond to potentials $\Omega_{k,l,m,n}^a$ and $\Omega_{k,l,m,n}^c$, respectively. Two examples of these types of diagrams are shown in Fig. \ref{SD6}. We also illustrated how the potentials related to the latter type of diagram can be obtained from the former in \eqref{flip}. Hence, we will thoroughly discuss the $a$-diagrams, after which we only state the results for the $c$-diagrams that contribute to the proof.

\begin{figure}
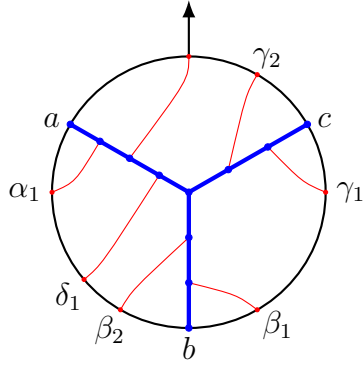

    \centering

    \caption{A disk diagram that corresponds to $\Omega_{1,2,2,1}^a(a,\alpha_1,\delta_1,\beta_2,b,\beta_1,\gamma_1,c,\gamma_2)$.}
    \label{generic}
\end{figure}

We consider the potentials $\Omega_{k,l,m,n}^a$ that we constructed in Sec. \ref{sec:recipe}. We note that the restriction \eqref{R} to $\mathbb{W}_{k,l,m,n} \subset \mathbb{U}_{N}$ can be solved in a multitude of ways. It will be a matter of convenience. In the left-ordered case, we required the coordinates to be nonsingular on all of $\mathbb{W}_{0,0,m,n}$, but it turns out that there is not a single chart that covers all of $\mathbb{W}_{k,l,m,n}$. This is easy to see from the equalities in the $uw$-chain that allows one to solve, for instance, $u_{l}^b=\frac{\alpha}{\beta}w_{l}^b$ or $w_{l}^b=\frac{\beta}{\alpha}u_{l}^b$, which both can become singular. Therefore, we will have to use various charts to describe the boundaries. The charts are constructed by choosing
\begin{align*}
    \begin{aligned}
        u_i^c &= \alpha v_i^c \quad\text{or}\quad v_i^c = \frac{u_i^c}{\alpha} \,, & \text{for } i=1,\dots,m \,,\\
        u_{i}^b &= \frac{\alpha}{\beta} w_{i}^b \quad\text{or}\quad w_{i}^b = \frac{\beta}{\alpha} u_{i}^b \,, & \text{for } i=1,\dots l \,,\\
        w_{i}^a &= \beta v_{i}^a \quad\text{or}\quad v_{i}^a = \frac{w_{i}^a}{\beta} \,, & \text{for } i=1,\dots,k+n+1 \,.
    \end{aligned}
\end{align*}
It will be clear from the context which chart was chosen. 

Like in the left-ordered case, here follows a categorization of the potentials evaluated on all boundaries of $\mathbb{W}_{k,l,m,n}$. Remember that for the left-ordered case $k=l=0$. These cases will be included in the following discussion, but not only as the left-ordered case: potentials with no lines attached to the $a$- and $b$-leg can still contribute to different orderings. One difference with the left-ordered case is that on the boundaries where we find $A_\infty$-terms, we find gluing terms as well, depending on the orientation of the red lines in the diagrams. Moreover, in the following categorization of $a$-diagrams we only consider potentials $\Omega_{k,l,m,n}^a$ and it is therefore not sufficient to prove \eqref{StokesToA_infty}. This categorization, however, is followed by a recipe for extracting the same information for the potentials $\Omega_{k,l,m,n}^c$ and a brief categorization of the $c$-diagrams that contribute to Stokes' theorem. This completes the proof.

\paragraph{$a$-diagrams.} The boundaries that yield $A_\infty$-terms are a bit more subtle than before. Namely, some boundaries that produce $A_\infty$-terms equally produce gluing terms, depending on the orientation of a particular line attached to one of the legs. This happens for boundaries where the green region is drawn on either one of the legs near the junction and for those diagrams we will show both options.
\begin{itemize}
    \item Boundary 1: At this boundary
    \begin{align*}
        u_{i}^\bullet=0 \,.
    \end{align*}
    If $u_i^a=0$, the $uv$-chain becomes
    \begin{align*}
        0 \leq& \frac{u_1^b}{v_1^b} \leq \dots \leq \frac{u_{l}^b}{v_{l}^b} \leq \frac{u_m^c}{v_m^c} = \dots = \frac{u_1^c}{v_1^c} \leq \frac{u_{k+n+1}^a}{v_{k+n+1}^a} \leq \dots \leq \frac{0}{v_i^a} \leq \dots \leq \frac{u_1^a}{v_1^a}\leq \infty\,,
    \end{align*}
    which leads to $u_j^c=0$ for $j=1,\dots,m$, $u_j^b=0$ for $j=1,\dots,l$ and $u_j^a=0$ for $j\geq i$. This yields a higher codimension boundary. Next we consider $u_i^c=0$ for $i=1,\dots,m$ and $u_i^b=0$ for $i=1,\dots,l$, which leads to $\alpha=0$. After the change of coordinates
    \begin{align*}
        \begin{aligned}
            v_i^c &\rightarrow u^1_i \,, & w_i^c &\rightarrow v^1_i \,, & & \text{for } i=1,\dots,m \,,\\
            v_i^b &\rightarrow u^1_{m+l+2-i} \,, & w_i^b &\rightarrow v^1_{m+l+2-i} \,, & & \text{for } i=1,\dots, l \,,\\
            u_{i}^a &\rightarrow u^2_{n+k+2-i} \,, & v_{i}^a &\rightarrow u^1_{m+1}v^2_{n+k+2-i} \,, & w_i^a &\rightarrow v_{m+1}^1v_{n+k+2-i}^2\,, & \text{for } i=1,\dots,k+n+1 \,,
        \end{aligned}
    \end{align*}
    the boundary is identified as $\mathbb{V}_{k+n}\times\mathbb{V}_{m+l}$ and
    \begin{align*}
        \boxed{
        \begin{aligned}
        &\int_{\partial\mathbb{W}_{k,l,m,n}}\Omega_{k,l,m,n}^{a}|_{u_i^c=u_i^b=0} \sim\\
        &\sim\mathcal{V}(\bullet,\dots,\bullet,a,\bullet,\dots,\bullet,\mathcal{V}(\bullet,\dots,\bullet,b,\bullet,\dots,\bullet,c,\bullet,\dots,\bullet),\bullet,\dots,\bullet)
        \end{aligned}}
    \end{align*}
    on this boundary, with the exception of $\mathcal{V}(\bullet,\dots,\bullet,a,\bullet,\dots,\bullet,\mathcal{V}(b,c),\bullet,\dots,\bullet)$. An example of a disk diagram at this boundary is shown in Fig. \ref{BB1}.
    \begin{figure}[h!]
        \centering

        \caption{An example of boundary 1 contributing to $\mathcal{V}(a,\alpha_1,\alpha_2,\mathcal{V}(\alpha_3,b,\alpha_4,\alpha_5,c,\alpha_6))$.}
        \label{BB1}
    \end{figure}

    \item Boundary 2: At this boundary
    \begin{align*}
        u_1^c=v_1^c \,.
    \end{align*}
    \begin{figure}[h!]
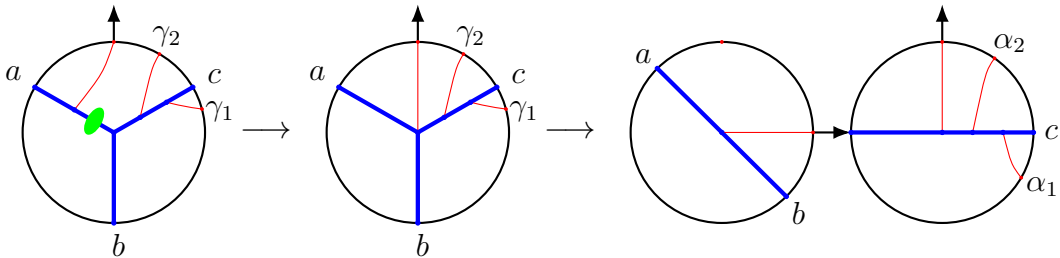

        \centering
        %
        \caption{An example of boundary 2 contributing to $\mathcal{V}(\mathcal{V}(a,b),\alpha_1,c,\alpha_2)$.}
        \label{BB2}
    \end{figure}
    This is only a boundary if $l=0$, in which case it is a higher codimension boundary unless $k=n=0$, like in the left-ordered case. The $uv$-chain then reads
    \begin{align*}
        1= \frac{u_m^c}{v_m^c} \leq \dots \leq \frac{u_1^c}{v_1^c} \leq \frac{u_1^a}{v_1^a}\,.
    \end{align*}
    The closure constraint now forces $u_i^\bullet=v_i^\bullet$. Then, after a change of coordinates
    \begin{align*}
        \begin{aligned}
            v_i^c &\rightarrow u^2_i \,, & w_i^c &\rightarrow v^2_i \,, & &\text{for } i=1,\dots,m \,,\\
            u_1^a &\rightarrow u^2_{m+1} \,, & w_1^a &\rightarrow v_{m+1}^2 \,,
        \end{aligned}
    \end{align*}
    the boundary is identified as $\mathbb{V}_m\times\mathbb{V}_0$ and
    \begin{align*}
        \boxed{\int_{\partial\mathbb{W}_{0,0,m,0}}\Omega_{0,0,m,0}^{a}|_{u_1^c=v_1^c} \sim \mathcal{V}(\mathcal{V}(a,b),\bullet,\dots,\bullet,c,\bullet,\dots,\bullet)}
    \end{align*}
    on this boundary. Since the line connected to the output arrow can only have one orientation, there is only one disk diagram at this boundary shown in Fig. \ref{BB2}.

    \item Boundary 3: At this boundary
    \begin{align*}
        v_1^c=w_1^c\,.
    \end{align*}
    \begin{figure}[h!]
    \centering

    \caption{An example of boundary 3 contributing to $\mathcal{V}(a,\alpha_1,\alpha_2,\mathcal{V}(b,c),\alpha_3)$.}
    \label{BB3}
\end{figure}
    The $vw$-chain becomes
    \begin{align*}
        1 =& \frac{v_1^c}{w_1^c} \leq \dots \leq \frac{v_m^c}{w_m^c} \leq \frac{v_{k+n+1}^a}{w_{k+n+1}^a} = \dots = \frac{v_1^a}{w_1^a}\leq \frac{v_{l}^b}{w_{l}^b} \leq \dots \leq \frac{v_1^b}{w_1^b} \leq \infty \,.
    \end{align*}
    and implies $v_i^\bullet \geq w_i^\bullet$. The closure constraint then leads to $v_i^\bullet=w_i^\bullet$, which gives a higher codimension boundary, except for $l=0, m=1$. Then after the change of coordinates
    \begin{align*}
        \begin{aligned}
            u_1^c &\rightarrow u^2_1 \,, & v_1^c &\rightarrow v^2_1 \,,\\
            u_i^a &\rightarrow u^2_{k+n+3-i} \,, & v_i^a &\rightarrow v^2_{k+n+3-i} \,, & \text{for } i=1,\dots,k+n+1\,,\\
        \end{aligned}
    \end{align*}
    the boundary is identified as $\mathbb{V}_{k+n+1}\times\mathbb{V}_0$ and  
    \begin{align*}
        \boxed{\int_{\partial\mathbb{W}_{k,0,1,n}}\Omega_{k,0,1,n}^{a}|_{v_1^c=w_1^c} \sim \mathcal{V}(\bullet,\dots,\bullet,a,\bullet,\dots,\bullet,\mathcal{V}(b,c),\bullet,\dots,\bullet)}
    \end{align*}    
    on this boundary. An example of a disk diagram at this boundary contributing to the $A_\infty$-term is shown in Fig. \ref{BB3}, while Fig. \ref{BB3G} shows a disk diagram contributing to a gluing term.

\begin{figure}[h!]
    \centering

    \caption{An example of boundary 3 contributing to a gluing term.}
    \label{BB3G}
\end{figure}

    \item Boundary 4: At this boundary
    \begin{align*}
        v_1^b &= w_1^b \,.
    \end{align*}
    \begin{figure}[h!]
        \centering
        %
        \caption{An example of boundary 4 contributing to $\mathcal{V}(a,\alpha_1,\alpha_2,\alpha_3,\mathcal{V}(b,c))$.}
        \label{BB4}
    \end{figure}
    The $vw$-chain becomes
    \begin{align*}
         0 \leq& \frac{v_1^c}{w_1^c} \leq \dots \leq \frac{v_m^c}{w_m^c} \leq \frac{v_{k+n+1}^a}{w_{k+n+1}^a} =\dots= \frac{v_1^a}{w_1^a}\leq \frac{v_{l}^b}{w_{l}^b} \leq \dots \leq \frac{v_1^b}{w_1^b} =1
    \end{align*}
    and implies $v_i^\bullet \leq w_i^\bullet$ for $i=1,\dots,N+1$. The closure constraint then leads to $v_i^\bullet=w_i^\bullet$, which leads to a higher codimension boundary, except for $l=1$, $m=0$. Then, after the change of coordinates
    \begin{align*}
        \begin{aligned}
            u_1^b &\rightarrow u^2_1 \,, & v_1^b &\rightarrow v^2_1 \,,\\
            u_i^a &\rightarrow u^2_{k+n+3-i} \,, & v_i^a &\rightarrow v^2_{k+n+3-i} \,, & \text{for } i=1,\dots,k+n+1 \,,\\
        \end{aligned}
    \end{align*}
    the boundary is identified as $\mathbb{V}_{k+n+1}\times\mathbb{V}_0$ and  
    \begin{align*}
        \boxed{\int_{\partial\mathbb{W}_{k,1,0,n}}\Omega_{k,1,0,n}^{a}|_{v_1^b=w_1^b} \sim \mathcal{V}(\bullet,\dots,\bullet,a,\bullet,\dots,\bullet,\mathcal{V}(b,c),\bullet,\dots,\bullet)}
    \end{align*}    
    on this boundary. An example of a disk diagram at this boundary contributing to the $A_\infty$-term is shown in Fig. \ref{BB4}, while Fig. \ref{BB4G} shows a disk diagram contributing to a gluing term.
    \begin{figure}[h!]
        \centering

        \caption{An example of boundary 4 contributing to a gluing term.}
        \label{BB4G}
    \end{figure}

    \item Boundary 5: At this boundary
    \begin{align*}
        w_i^\bullet=1\,.
    \end{align*}
    The closure constraint forces all other $w$-variables to be zero. This yields a higher codimenion boundary, except for $m=1$ and $w_1^c=1$, which leads to $\beta=0$. After the change of coordinates
    \begin{align*}
        \begin{aligned}
            u_1^c &\rightarrow u^2_{l+1} \,, & v_1^c &\rightarrow v^2_{l+1}\,,\\
            u_i^b & \rightarrow u^2_{i} \,, & v_i^b &\rightarrow v^2_{i} \,, & \text{for } i=1,\dots,l\,,\\
            u_{i}^a &\rightarrow u^2_{k+n+l+3-i} \,, & v_i^a &\rightarrow v^2_{k+n+l+3-i} \,, & \text{for } i=1,\dots,k+n+1 \,.
        \end{aligned}
    \end{align*}
    the boundary is identified as $\mathbb{V}_{k+l+n+1}\times\mathbb{V}_0$. The canonical ordering of the $q$-vectors in $Q$ in the resulting nested disk diagram is different than the canonical ordering of the corresponding $A_\infty$-term. We therefore rewrite the matrix $Q$ as
    \begin{align*}
        Q &= (
            \vec{q}_a , \vec{q}_b , \vec{q}_c , \vec{q}_{c,1} \vec{q}_{b,_1} , \dots , \vec{q}_{b,l} , \vec{q}_{a,k+n+1}, \dots , \vec{q}_{a,1}
        ) \,.
    \end{align*}
    It can now be seen that 
    \begin{align*}
        \boxed{\int_{\partial\mathbb{W}_{k,l,1,n}}\Omega_{k,l,1,n}^{a}|_{w_1^c=1} \sim \mathcal{V}(\bullet,\dots,\bullet,a,\bullet,\dots,\bullet,b,\bullet,\dots,\bullet,\mathcal{U}(c,\bullet),\bullet,\dots,\bullet)}
    \end{align*}
    and
    \begin{align*}
        \boxed{\int_{\partial\mathbb{W}_{k,l,1,n}}\Omega_{k,l,1,n}^{a}|_{w_1^c=1} \sim \mathcal{V}(\bullet,\dots,\bullet,a,\bullet,\dots,\bullet,b,\bullet,\dots,\bullet,\mathcal{U}(\bullet,c),\bullet,\dots,\bullet)}
    \end{align*}
    on this boundary. An example of a disk diagram at this boundary is shown in Fig. \ref{BB5}.

    \begin{figure}[h!]
        \centering

        \caption{An example of boundary 5 contributing to $\mathcal{V}(a,\alpha_1,\alpha_2,\alpha_3,b,\alpha_4,\mathcal{U}(\alpha_5,c))$.}
        \label{BB5}
    \end{figure}
    
    \item Boundary 6: At this boundary
    \begin{align*}
        w_i^\bullet=0 \,.
    \end{align*}
    For $m=1$ and $w_1^c=1$, this is the same as boundary 5. Otherwise, if $w_i^c=0$, the $uw$-chain is equivalent to
    \begin{align*}
        0 \geq& \frac{w_1^a}{u_1^a} \geq \dots \geq \frac{w_{k+n+1}^a}{u_{k+n+1}^a} \geq \frac{w_{1}^b}{u_{1}^b} = \dots = \frac{w_l^b}{u_l^b} \geq \frac{w_{m}^a}{u_{m}^c} \geq \dots \geq \frac{0^+}{u_i^c} \geq \dots \geq \frac{w_1^c}{u_1^c}\geq \infty \,,
    \end{align*}
    which forces $w_j^c=0$ for $j>i$ and $w_k^a=w_k^b=0$ for any $k$. This leads to a higher codimension boundary. Only when we consider $w_i^a=0$ for $i=1,\dots,k+n+1$ and $w_i^b=0$ for $i=1,\dots,l$, we find an $A_\infty$-term. Then, after the change of coordinates
    \begin{align*}
        \begin{aligned}
            u_1^c &\rightarrow u^2_{l+1} u^1_1 \,,\\
            v_i^c &\rightarrow v^2_{l+1} u^1_i \,, & w_i^c & \rightarrow v^1_i \,, & \text{for } i=1,\dots,m\\
            u_i^b &\rightarrow u^2_{i} \,, & v_i &\rightarrow v^2_{i} \,, & \text{for } i=1,\dots, l \,,\\
            u_{i}^a & \rightarrow u^2_{k+l+n+3-i} \,, & v_{i}^a & \rightarrow v^2_{k+l+n+3-i} \,, & \text{for } i=1,\dots,k+n+1
        \end{aligned}
    \end{align*}
    the boundary is identified as $\mathbb{V}_{k+l+n+1}\times\mathbb{V}_{m-1}$. We also change the matrix $Q$, such that it corresponds to the canonical ordering for nested vertices. It then reads
    \begin{align*}
        Q &= (
            \vec{q}_a , \vec{q}_b , \vec{q}_c , \vec{q}_{c,1} , \dots , \vec{q}_{c,m} , \vec{q}_{b,1} , \dots , \vec{q}_{b,l} , \vec{q}_{a,k+n+1} , \dots , \vec{q}_{a,1}
        ) \,.
    \end{align*}
    It can now be seen that 
    \begin{align*}
        \boxed{\int_{\partial\mathbb{W}_{k,l,m,n}}\Omega_{k,l,m,n}^{a}|_{w_{i}^a=w_i^b=0} \sim \mathcal{V}(\bullet,\dots,\bullet,a,\bullet,\dots,\bullet,b,\bullet,\dots,\bullet,\mathcal{U}(\bullet,\dots,\bullet,c,\bullet,\dots,\bullet),\bullet,\dots,\bullet)}
    \end{align*}
    on this boundary, with the exception of $\mathcal{V}(\bullet,\dots,\bullet,a,\bullet,\dots,\bullet,b,\bullet,\dots,\bullet,\mathcal{U}(c,\bullet),\bullet,\dots,\bullet)$ and $\mathcal{V}(\bullet,\dots,\bullet,a,\bullet,\dots,\bullet,b,\bullet,\dots,\bullet,\mathcal{U}(\bullet,c),\bullet,\dots,\bullet)$. An example of a disk diagram at this boundary is shown in Fig. \ref{BB6}.

    \begin{figure}[h!]
        \centering

        \caption{An example of boundary 6 contributing to $\mathcal{V}(a,\alpha_1,\alpha_2,\alpha_3,b,\alpha_4,\mathcal{U}(\alpha_5,c,\alpha_6))$.}
        \label{BB6}
    \end{figure}

    \item Boundary 7: At this boundary
    \begin{align*}
        v_i^\bullet=1\,.
    \end{align*}
    \begin{figure}[h!]
        \centering
        %
        \caption{An example of boundary 7 contributing to $\mathcal{V}(a,\alpha_1,\alpha_2,\mathcal{U}(b,\alpha_3),\alpha_4,c,\alpha_5)$.}
        \label{BB7}
    \end{figure}

    The closure constraint forces all other $v$-variables to be zero. This yields a higher codimension boundary, except for $l=1$ and $v_1^b=1$. Then, after the change of variables
    \begin{align*}
        \begin{aligned}
            u_i^c &\rightarrow u^2_i \,, & w_i^c &\rightarrow v^2_i \,, & \text{for } i=1,\dots,m \,,\\
            u_1^b &\rightarrow u^2_{m+1} \,, & w_1^b &\rightarrow v^2_{m+1} \,,\\
            u_i^a &\rightarrow u^2_{k+m+n+3-i} \,, & w_i^a &\rightarrow v^2_{k+m+n+3-i} \,, & \text{for } i=1,\dots,k+n+1
        \end{aligned}
    \end{align*}
    the boundary is identified as $\mathbb{V}_{k+m+n+1}\times\mathbb{V}_0$. We also change the matrix $Q$, such that it corresponds to the canonical ordering for nested vertices. It then reads
    \begin{align*}
        Q &= (
            \vec{q}_a , \vec{q}_b , \vec{q}_c , \vec{q}_{b,1}, \vec{q}_{c,1} , \dots , \vec{q}_{c,m} , \vec{q}_{a,k+n+1} , \dots , \vec{q}_{a,1}
        ) \,.
    \end{align*}
    It can now be seen that 
    \begin{align*}
        \boxed{\int_{\partial\mathbb{W}_{k,1,m,n}}\Omega_{k,1,m,n}^{a}|_{v_{1}^b=1} \sim \mathcal{V}(\bullet,\dots,\bullet,a,\bullet,\dots,\bullet,\mathcal{U}(b,\bullet),\bullet,\dots,\bullet,c,\bullet,\dots,\bullet)}
    \end{align*}
    and
    \begin{align*}
        \boxed{\int_{\partial\mathbb{W}_{k,1,m,n}}\Omega_{k,1,m,n}^{a}|_{v_{1}^b=1} \sim \mathcal{V}(\bullet,\dots,\bullet,a,\bullet,\dots,\bullet,\mathcal{U}(\bullet,b),\bullet,\dots,\bullet,c,\bullet,\dots,\bullet)}
    \end{align*}
    on this boundary. An example of a disk diagram at this boundary is shown in Fig. \ref{BB7}.

    \item Boundary 8: At this boundary 
    \begin{align*}
        v_i^\bullet=0 \,.
    \end{align*}
    \begin{figure}[h!]
        \centering

        \caption{An example of boundary 8 contributing to $\mathcal{V}(a,\alpha_1,\alpha_2,\mathcal{U}(\alpha_3,b,\alpha_4),\alpha_5,c,\alpha_6)$.}
        \label{BB8}
    \end{figure}
    The $vw$-chain becomes
    \begin{align*}
        \begin{aligned}
            0 &\leq \frac{v_1^c}{w_1^c} \leq \dots \leq \frac{v_m^c}{w_m^c} \leq \frac{v_{k+n+1}^a}{w_{k+n+1}^a} = \dots =\frac{0^+}{w_i^a} = \dots = \frac{v_1^a}{w_1^a}\leq \frac{v_{l}^b}{w_{l}^b} \leq \dots \leq \frac{v_1^b}{w_1^b} \leq \infty \,,\\
            0 &\leq \frac{v_1^c}{w_1^c} \leq \dots \leq \frac{v_m^c}{w_m^c} \leq \frac{v_{k+n+1}^a}{w_{k+n+1}^a} = \dots = \frac{v_1^a}{w_1^a}\leq \frac{v_{l}^b}{w_{l}^b} \leq \dots \leq \frac{0^+}{w_i^b} \dots \leq \frac{v_1^b}{w_1^b} \leq \infty
        \end{aligned}
    \end{align*}
    if $v_i^a=0$ and $v_i^b=0$, respectively. This forces all $v_i^a=0$ and $v_i^c=0$ in both cases, while in the latter we also have $v_j^b=0$ for $j>i$.  This leads to a higher codimension boundary. It is only when $v_i^c=0$ for $i=1,\dots,m$ and $v_i^a=0$ for $i=1,\dots,k+n+1$, that one finds an $A_\infty$-term. After the change of variables
     \begin{align*}
        \begin{aligned}
            u_{i}^c &\rightarrow u^2_{i} \,, & w_{i}^c &\rightarrow v^2_{i}\,, &&& \text{for } i=1,\dots,m \,,\\
            u_i^b &\rightarrow u^2_{m+1}u^1_i \,,&
            v_{i}^b &\rightarrow v^1_i \,, & w_{i}^b &\rightarrow v^2_{m+1}u^1_1\,, & \text{for } i=1,\dots,l \,,\\
            u_{i}^a &\rightarrow u^2_{k+m+n+3-i} \,, & w_{i}^a &\rightarrow v^2_{k+m+n+3-i}\,, &&& \text{for } i=1,\dots,k+n+1 \,,\\
        \end{aligned}
    \end{align*}
    the boundary is identified as $\mathbb{V}_{k+m+n+1}\times\mathbb{V}_{l-1}$. We also change the matrix $Q$, such that it corresponds to the canonical ordering for nested vertices. It then reads
    \begin{align*}
        Q &= (
            \vec{q}_a , \vec{q}_b , \vec{q}_c , \vec{q}_{b,1} , \dots , \vec{q}_{b,l} , \vec{q}_{c,1} , \dots , \vec{q}_{c,m} , \vec{q}_{a,k+n+1} , \dots , \vec{q}_{a,1}
        ) \,.
    \end{align*}
    It can now be seen that 
    \begin{align*}
        \boxed{\int_{\partial\mathbb{W}_{k,l,m,n}}\Omega_{k,l,m,n}^{a} \sim \mathcal{V}(\bullet,\dots,\bullet,a,\bullet,\dots,\bullet,\mathcal{U}(\bullet,\dots,\bullet,b,\bullet,\dots,\bullet),\bullet,\dots,\bullet,c,\bullet,\dots,\bullet)}
    \end{align*}
    on this boundary, with the exception of $\mathcal{V}(\bullet,\dots,\bullet,a,\bullet,\dots,\bullet,\mathcal{U}(b,\bullet),\bullet,\dots,\bullet,c,\bullet,\dots,\bullet)$ and $\mathcal{V}(\bullet,\dots,\bullet,a,\bullet,\dots,\bullet,\mathcal{U}(\bullet,b),\bullet,\dots,\bullet,c,\bullet,\dots,\bullet)$. An example of a disk diagram at this boundary is shown in Fig. \ref{BB8}.

\end{itemize}

\paragraph{Gluing terms.}
\begin{itemize}

    \item Boundary 9: 
    
    At this boundary
    \begin{align*}
        \frac{u_m^c}{v_m^c}=\frac{u_{l}^b}{v_{l}^b}\,.
    \end{align*}
    \begin{figure}[h!]
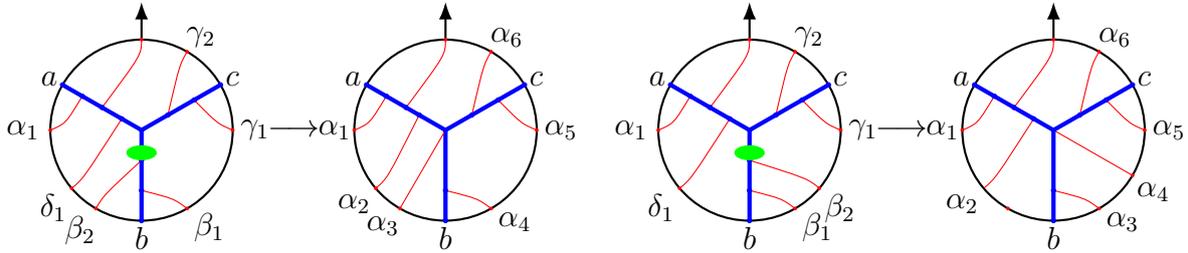

       \centering

       \caption{Two examples of boundary 9 contributing to a gluing term, with both orientations of $\beta_2$.}
       \label{BB9}
   \end{figure}
   $\int_{\partial\mathbb{W}_{k,l,m,n}}\Omega_{k,l,m,n}^{a}$ does not yield a familiar $A_\infty$-term on this boundary and it does not vanish either. An example of a disk diagram at this boundary is shown in Fig. \ref{BB9}.

   \item Boundary 10: At this boundary
    \begin{align*}
        \frac{u_1^c}{v_1^c}=\frac{u_{k+n+1}^a}{v_{k+n+1}^a} \,.
    \end{align*}
    For $k=l=n=0$ the $uv$-chain becomes
    \begin{align*}
        0 \leq \frac{u_m^c}{v_m^c} = \dots = \frac{u_1^c}{v_1^c} = \frac{u_{1}^a}{v_{1}^a} \leq \infty\,.
    \end{align*}
    The closure constraint then gives $u_i^\bullet=v_i^\bullet$, so this boundary is equivalent to boundary 2 when $k=l=n=0$ and does not yield  a gluing term, but an $A_\infty$-term instead. Otherwise, $\int_{\partial\mathbb{W}_{k,l,m,n}}\Omega_{k,l,m,n}^{a}$ does not yield a familiar $A_\infty$-term on this boundary and it does not vanish either. An example of a disk diagram at this boundary is shown in Fig. \ref{BB10}. Gluing terms corresponding to the same type as the left diagram cancel with gluing terms belonging to the type of diagrams on the left of Fig. \ref{BB9}. A special type of gluing term with no elements between the junction and the output arrow, i.e. $n=0$, is shown in Fig. \ref{BB10S}. This is the only type of gluing term that does not cancel with any other gluing term from the potential $\Omega_{k,l,m,n}^a$, but rather from the type $\Omega_{k,l,m,n}^c$, `gluing' them together.

       \begin{figure}[h!]
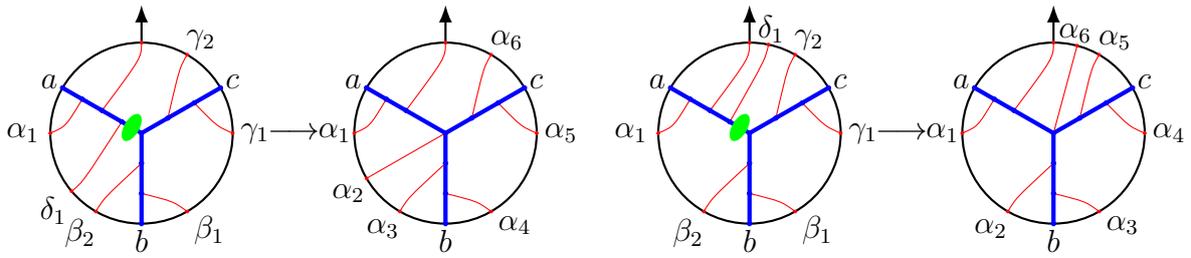

       \centering

       \caption{Two examples of boundary 10 contributing to a gluing term, with both orientations of $\delta_1$.}
       \label{BB10}
   \end{figure}

   \begin{figure}[h!]
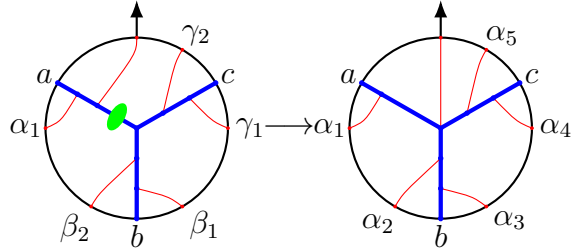

       \centering
       %
       \caption{Special case of a gluing term coming from boundary 10. This gluing term is cancelled by a gluing term coming from a potential of the type $\Omega_{k,l,m,n}^c$.}
       \label{BB10S}
   \end{figure}

    \item Boundary 11: At this boundary
    \begin{align*}
        \frac{v_m^c}{w_m^c} =& \frac{v_{k+n+1}^a}{w_{k+n+1}^a}  \,.
    \end{align*}
    For $l=0$, $m=1$ the $vw$-chain reads
    \begin{align*}
         0 \leq& \frac{v_1^c}{w_1^c} = \frac{v_{k+n+1}^a}{w_{k+n+1}^a} = \dots = \frac{v_{1}^a}{w_{1}^a} \leq \infty\,.
    \end{align*}
    The closure constraint then gives $v_i^\bullet=w_i^\bullet$, so this boundary is equivalent to boundary 3 when $l=0$ and $m=1$, producing either a gluing term or an $A_\infty$-term. Otherwise, $\int_{\partial\mathbb{W}_{k,l,m,n}}\Omega_{k,l,m,n}^{a}$ does not yield a familiar $A_\infty$-term on this boundary and it does not vanish either. An example of a disk diagram at this boundary is shown in Fig. \ref{BB11}. Gluing terms corresponding to the type of diagrams on the left cancel with gluing terms belonging to the type of diagrams on the right of Fig. \ref{BB10}, while the type of diagrams on the right cancel with the diagrams of the type on the right of Fig. \ref{BB9}.

    \begin{figure}[h!]
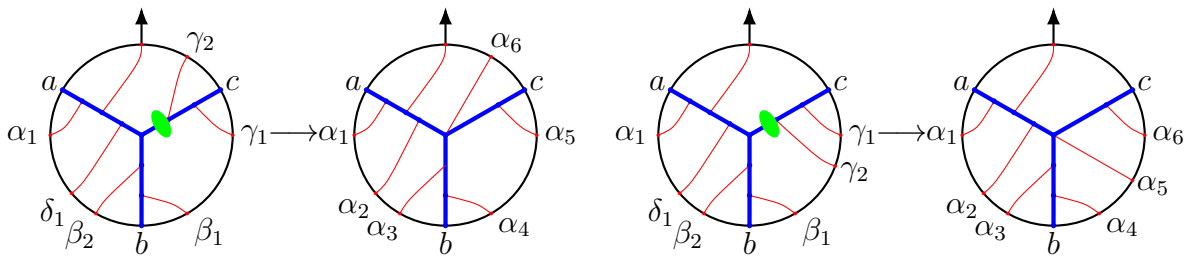

       \centering

       \caption{Two examples of boundary 11 contributing to a gluing term, with both orientations of $\gamma_2$.}
       \label{BB11}
   \end{figure}

    \item Boundary 12: At this boundary
    \begin{align*}
        \frac{v_1^a}{w_1^a} &= \frac{v_{l}^b}{w_{l}^b} \,.
    \end{align*}
    This implies $w_{l}^b=\beta v_{l}^b$ and from 
    \begin{align*}
        \frac{u_{l}^b}{w_{l}^b} =& \frac{1}{\beta}\frac{u_{l}^b}{v_{l}^b} = \frac{\alpha}{\beta}
    \end{align*}
    we get $\frac{u_{l}^b}{v_{l}^b}=\alpha=\frac{u_m^c}{v_m^c}$, so we find that this boundary is equivalent to boundary 9. It is only when $m=0$ that boundary 9 does not exist and we have to consider this one. $\int_{\partial\mathbb{W}_{k,l,m,n}}\Omega_{k,l,m,n}^{a}$ does not yield a familiar $A_\infty$-term on this boundary and it does not vanish either. The disk diagrams at this boundary resemble the ones in Fig. \ref{BB9} when there are no lines attached to the $c$-leg. An example of a disk diagram at this
    boundary is shown in Fig. \ref{BB12}. Gluing terms corresponding to the type of diagrams on the right cancel with gluing terms belonging to the type of diagrams on the right of Fig. \ref{BB11}.

    \begin{figure}[h!]
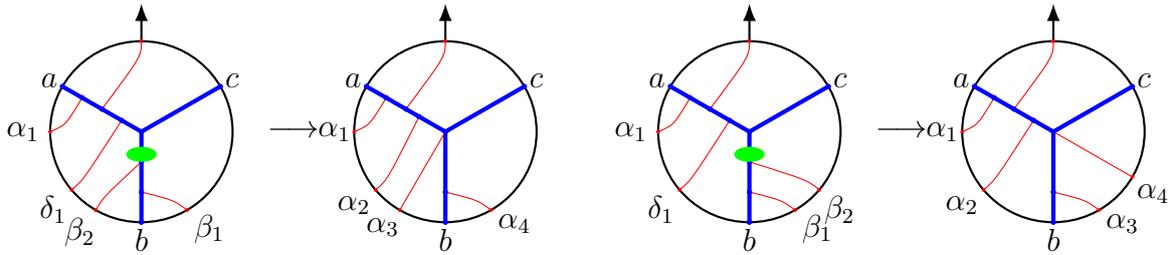

       \centering

       \caption{Two examples of boundary 12 contributing to a gluing term, with both orientations of $\beta_2$.}
       \label{BB12}
   \end{figure}

    \item Boundary 13: At this boundary
    \begin{align*}
        \frac{u_1^b}{w_1^b} =& \frac{u_{k+n+1}^a}{w_{k+n+1}^a}\,.
    \end{align*}
    This implies $w_{k+n+1}^a=\frac{\beta}{\alpha}u_{k+n+1}^a$ and from
    \begin{align*}
        \frac{v_{k+n+1}^a}{w_{k+n+1}^a}=\frac{\alpha}{\beta}\frac{v_{k+n+1}^a}{u_{k+n+1}^a}=\frac{1}{\beta}
    \end{align*}
    we get $\frac{u_{k+n+1}^a}{v_{k+n+1}^a}=\alpha=\frac{u_1^c}{v_1^c}$, so we find that this boundary is equivalent to boundary 10. It is only when $m=0$ that boundary 10 does not exist and we have to consider this one. $\int_{\partial\mathbb{W}_{k,l,m,n}}\Omega_{k,l,m,n}^{a}$ does not yield a familiar $A_\infty$-term on this boundary and it does not vanish either. The disk diagrams at this boundary resemble the ones in Fig. \ref{BB10} when there are no lines attached to the $c$-leg. An example of a disk diagram at this
    boundary is shown in Fig. \ref{BB13}. Gluing terms corresponding to the type of diagrams on the left cancel with gluing terms belonging to the type of diagrams on the left of Fig. \ref{BB12}, while the type of diagrams on the right cancel with gluing terms corresponding to the left of Fig. \ref{BB11}.
       \begin{figure}[h!]
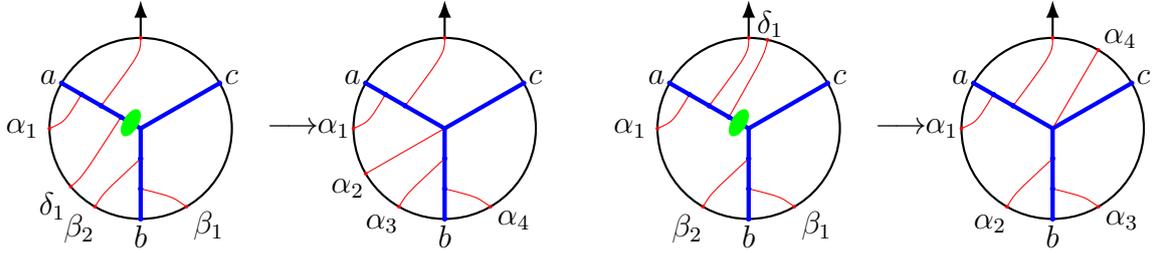

       \centering

       \caption{Two examples of boundary 13 contributing to a gluing term, with both orientations of $\delta_1$.}
       \label{BB13}
   \end{figure}

    \item Boundary 14: At this boundary
    \begin{align*}
        \frac{u_{m}^c}{w_{m}^c} =& \frac{u_{l}^b}{w_{l}^b}\,.
    \end{align*}
    This implies $u_{m}^c=\frac{\alpha}{\beta}w_{m}^c$ and from
    \begin{align*}
        \frac{u_m^c}{v_m^c}=\frac{\alpha}{\beta}\frac{w_m^c}{v_m^c}=\alpha
    \end{align*}
    we get $\frac{v_m^c}{w_m^c}=\frac{1}{\beta}=\frac{v_{k+n+1}^a}{w_{k+n+1}^a}$, so we find that this boundary is equivalent to boundary 11. Boundary 11 does not exist when $m=0$, but neither does this one. This means that this boundary is always equivalent to boundary 11.

\end{itemize}

\paragraph{Zero measure terms.}

\begin{itemize}
    \item Boundary 15: At this boundary
    \begin{align*}
        \frac{v_{i}^c}{w_{i}^c}=\frac{v_{i+1}^c}{w_{i+1}^c}\,, \quad \text{for } i=1,\dots,m-1 \,,
    \end{align*}
    which makes the vectors $\vec{q}_{c,i}$ and $\vec{q}_{c,i+1}$ colinear and the measure in $\Omega_{k,l,m,n}^{a}$ vanishes on this boundary.
    
    \item Boundary 16: At this boundary
    \begin{align*}
        \frac{v_{i}^b}{w_i^b}=\frac{v_{i+1}^b}{w_{i+1}^b}\,, \quad \text{for } i=1,\dots,l-1 \,,
    \end{align*}
    which makes the vectors $\vec{q}_{b,i}$ and $\vec{q}_{b,i+1}$ colinear and the measure in $\Omega_{k,l,m,n}^{a}$ vanishes on this boundary.
    
    \item Boundary 17: At this boundary
    \begin{align*}
        \frac{v_i^a}{w_i^a}=\frac{v_{i+1}^a}{w_{i+1}^a}\,, \quad \text{for } i=1,\dots,k+n\,,
    \end{align*}
    which makes the vectors $\vec{q}_{a,i}$ and $\vec{q}_{a,i+1}$ colinear and the measure in $\Omega_{k,l,m,n}^{a}$ vanishes on this boundary.
\end{itemize}

\paragraph{Higher codimension boundaries.}

\begin{itemize}
\item Boundary 18: At this boundary
\begin{align*}
    u_i^\bullet=1 \,.
\end{align*}
The closure constraint implies that all other $u$-variables are zero. If $m=1$ and $k=l=n=0$, the boundary is given by $u_1^c=1$, which implies $v_1^c=w_1^c=1$ through
\begin{align*}
    0 \leq u_1^c \leq v_1^c \leq w_1^c \leq 1\,.
\end{align*}

\item Boundary 19: At this boundary
\begin{align*}
    u_1^c =& w_1^c \,.
\end{align*}
The $uw$-chain becomes
\begin{align*}
    1=\frac{u_1^c}{w_1^c} \leq \dots \leq \frac{u_m^c}{w_m^c} \leq \frac{u_{l}^b}{w_{l}^b} = \dots = \frac{u_1^b}{w_1^b} \leq \frac{u_{k+n+1}^a}{w_{k+n+1}^a} \leq \dots \leq \frac{u_1^a}{w_1^a}  \,.
\end{align*}
The closure constraint implies that this is a higher codimension boundary, except when $m=1$ and $k+n=0$, in which case $u_i^\bullet=w_i^\bullet$. The $uv$-chain and $vw$-chain then read
\begin{align*}
    0 \leq \frac{u_1^b}{v_1^b} \leq \dots \leq \frac{u_l^b}{v_l^b} \leq \frac{u_1^c}{v_1^c} \leq \frac{u^a_{1}}{v_{1}^a} \leq \infty \,,\\
    0 \leq \frac{v_1^c}{w_1^c} \leq \frac{v_{1}^a}{w_{1}^a} \leq \frac{v_l^b}{w_l^b} \leq \dots \leq \frac{v_{1}^b}{w_{1}^b} \leq \infty
\end{align*}
and contradict each other when setting $u_i^\bullet=w_i^\bullet$, thus we find a higher codimension boundary.

\item Boundary 20: At this boundary
\begin{align*}
    u_{l}^b =& w_{l}^b \,.
\end{align*}
This is only a boundary when $m=0$. The $uw$-chain becomes
\begin{align*}
    1=& \frac{u_{l}^b}{w_{l}^b} = \dots = \frac{u_1^b}{w_1^b} \leq \frac{u_{k+n+1}^a}{w_{k+n+1}^a} \leq  \dots \leq \frac{u_1^a}{w_1^a}\leq \infty\,,
\end{align*}
This yields a higher codimension boundary, unless $k=n=0$, in which case this boundary is equivalent to boundary boundary 13.

\item Boundary 21: At this boundary
    \begin{align*}
        u_1^b &= v_1^b \,.
    \end{align*}
    The $uv$-chain becomes
    \begin{align*}
        1 = \frac{u_1^b}{v_1^b} = \dots = \frac{u_l^b}{v_l^b} \leq \frac{u_m^c}{v_m^c} = \dots = \frac{u_1^c}{v_1^c} \leq \frac{u_{k+n+1}^a}{v_{k+n+1}^a} \leq \dots \leq \frac{u_1^a}{v_1^a}\leq \infty\,.
    \end{align*}
    which implies $u_i^\bullet \geq v_i^\bullet$. The closure constraint leads to a higher codimension boundary, except when $l=1$, $k=n=0$. However, in this case the $uw$-chain and $vw$-chain become
    \begin{align*}
        0 &\leq \frac{u_1^c}{w_1^c} \leq \dots \leq \frac{u_{m}^c}{w_{m}^c} \leq \frac{u_1^b}{w_1^b} \leq \frac{u_{1}^a}{w_1^a} \leq \infty \,,\\
        0 &\leq \frac{v_1^c}{w_1^c} \leq \dots \leq \frac{v_m^c}{w_m^c} \leq \frac{v_{1}^a}{w_{1}^a} \leq \frac{v_{1}^b}{w_{1}^b} \leq \infty
    \end{align*}
   and contradict each other when setting $u_i^\bullet=v_i^\bullet$. Thus, we find a higher codimension boundary.

\item Boundary 22: At this boundary
\begin{align*}
    \frac{u_m^c}{w_m^c}=\frac{u_{k+n+1}^a}{w_{k+n+1}^a}\,.
\end{align*}
This is a boundary only if $l=0$. Since
\begin{align*}
    \frac{u_m^c}{v_m^c}\frac{v_m^c}{w_m^c} =& \frac{u_{m}^c}{w_{m}^c} \leq \frac{u_{k+n+1}^a}{w_{k+n+1}^a} =\frac{u_{k+n+1}^a}{v_{k+n+1}^a}\frac{v_{k+n+1}^a}{w_{k+n+1}^a}
\end{align*}
and
\begin{align*}
    \frac{u_m^c}{v_m^c} \leq& \frac{u_{k+n+1}^a}{v_{k+n+1}^a} \,, \quad \frac{v_m^c}{w_m^c} \leq \frac{v_{k+n+1}^a}{w_{k+n+1}^a} \,,
\end{align*}
this implies $\frac{u_m^c}{v_m^c} = \frac{u_{k+n+1}^a}{v_{k+n+1}^a}$ and $\frac{v_m^c}{w_m^c} = \frac{v_{k+n+1}^a}{w_{k+n+1}^a}$ and thus yields a higher codimension boundary.

\item Boundary 23: At this boundary
\begin{align*}
    \frac{u_{l}^b}{v_{l}^b}=\frac{u_{k+n+1}^a}{v_{k+n+1}^a}\,.
\end{align*}
This is only a boundary for $m=0$. Since
\begin{align*}
    \frac{u_l^b}{w_l^b}/\frac{v_l^b}{w_l^b}=\frac{u_l^b}{v_l^b}\leq \frac{u_{k+n+1}^a}{v_{k+n+1}^a}=\frac{u_{k+n+1}^a}{w_{k+n+1}^a}/\frac{v_{k+n+1}^a}{w_{k+n+1}^a}
\end{align*}
and
\begin{align*}
    \begin{aligned}
        \frac{u_l^b}{w_l^b} &\leq \frac{u_{k+n+1}^a}{w_{k+n+1}^a} \,, & \frac{v_{k+n+1}^a}{w_{k+n+1}^a} &\leq\frac{v_l^b}{w_l^b} \,,
    \end{aligned}
\end{align*}
this implies $\frac{u_l^b}{w_l^b} = \frac{u_{k+n+1}^a}{w_{k+n+1}^a} $ and  $\frac{v_{k+n+1}^a}{w_{k+n+1}^a} = \frac{v_l^b}{w_l^b}$ and thus yields a higher codimension boundary.
\end{itemize}

As can be seen from the disk diagrams for $A_\infty$-relations in Fig. \ref{A_inftyDiagrams} or in the corresponding expression \eqref{generalA_infty}, \eqref{Qs}, 
\begin{align*}
    \mathcal{V}(\bullet,\dots,\bullet,\mathcal{V}(\bullet,\dots,\bullet,a,\bullet,\dots,\bullet,b,\bullet,\dots,\bullet),\bullet,\dots,\bullet,c,\bullet,\dots,\bullet)
\end{align*}
and
\begin{align*}
    \mathcal{V}(\bullet,\dots,\bullet,a,\bullet,\dots,\bullet,\mathcal{V}(\bullet,\dots,\bullet,b,\bullet,\dots,\bullet,c,\bullet,\dots,\bullet),\bullet,\dots,\bullet)
\end{align*}
are related to each other by reversing the nested boundary ordering and swapping $a \leftrightarrow c$ and so are 
\begin{align*}
    \mathcal{V}(\bullet,\dots,\bullet,a,\bullet,\dots,\bullet,b,\bullet,\dots,\bullet,\mathcal{U}(\bullet,\dots,\bullet,c,\bullet,\dots,\bullet),\bullet,\dots,\bullet)
\end{align*}
and
\begin{align*}
    \mathcal{V}(\bullet,\dots,\bullet,\mathcal{U}(\bullet,\dots,\bullet,a,\bullet,\dots,\bullet),\bullet,\dots,\bullet,b,\bullet,\dots,\bullet,c,\bullet,\dots,\bullet) \,.
\end{align*}
One can take the vectors $\vec{q}^{\,\,1}_i=(q^1_{u,i},q^1_{v,i},q^1_{w,i})$ and $\vec{q}^{\,\,2}_i=(q^2_{u,i},q^2_{v,i},q^2_{w,i})$ from the first $A_\infty$-term of both of the pairs mentioned and replace them by $\vec{q}^{\,\,'1}_i=(q^1_{w,1},q^1_{v,i},q^1_{u,i})$ and $\vec{q}^{\,\,'2}=(q^2_{w,i},q^2_{v,i},q^2_{u,i})$. The expressions for the entries of these vectors can be found in \eqref{Qs}. Since, after reversing the nested boundary ordering, the labeling of the vectors $\vec{q}^{\,\,'1}$ and $\vec{q}^{\,\,'2}$ does not match the ordering of the corresponding $A_\infty$-terms in the matrix $Q$, the ordering needs to be adjusted. The matrix becomes
\begin{align*}
    Q &= (
         \vec{q}_a , \vec{q}_b , \vec{q}_c , \vec{q}^{\,\,'1}_{r} , \dots , \vec{q}^{\,\,'1}_1 , \vec{q}^{\,\,'2}_{s} , \dots , \vec{q}^{\,\,'2}_1
    )\,.
\end{align*}
Lastly, one applies the $\mathbb{Z}_2$-transformation on both $u^1$ and $v^1$ and $u^2$ and $v^2$ variables in the first pair, while in the second pair one only applies the transformation on the $u^2$ and $v^2$ variables. One could say that the $A_\infty$-terms in these pairs are mirror images of each other. Similarly,
\begin{align*}
    \mathcal{V}(\bullet,\dots,\bullet,a,\bullet,\dots,\bullet,\mathcal{U}(\bullet,\dots,\bullet,b,\bullet,\dots,\bullet),\bullet,\dots,\bullet,c,\bullet,\dots,\bullet)
\end{align*}
simply returns to an $A_\infty$-term of the same type after applying these operations, so these $A_\infty$-terms are mirror images of themselves.

The potentials $\Omega_{k,l,m,n}^a$ and $\Omega_{m,l,k,n}^c$ are related through almost the same operations: one reverses the boundary ordering and swaps $a \leftrightarrow c$ and $\vec{q}_{a,i} \leftrightarrow \vec{q}_{c,i}$. This is not yet the same as for the $A_\infty$-terms, but after evaluating the potentials on the boundary, one can apply a $\mathbb{Z}_2$-transformation on the variables $u^1,v^1$ and/or $u^2,v^2$. It is now easy to see that the boundaries in the categorization above that gave $A_\infty$-terms, will give the $A_\infty$-terms related by the above relation on the same boundaries, but for the potential $\Omega_{m,l,k,n}^c$. Moreover, the gluing terms coming from $\Omega_{m,l,k,n}^c$ can be related in a similar way and can be easily seen to vanish among themselves or with the gluing terms from $\Omega_{k,l,m,n}^a$. The boundaries that yielded zero measure terms and higher codimension boundary terms will do so again and they will therefore not contribute. We will now list the boundaries that produce $A_\infty$-terms and gluing terms for $\Omega_{m,l,k,n}^c$. For this, remember that the matrix $Q$ is given by \eqref{Q'} for these potentials.

\paragraph{$c$-diagrams.}
\begin{itemize}
    \item Boundary 1: At this boundary
    \begin{align*}
        u_{i}^\bullet=0 \,.
    \end{align*}
    This leads to a higher codimension boundary, except when $u_i^c=0$ for $i=1,\dots,m$ and $u_i^b=0$ for $i=1,\dots,l$, in which case it leads to $\alpha=0$. After the change of coordinates
    \begin{align*}
        \begin{aligned}
            v_i^c &\rightarrow u^1_i \,, & w_i^c &\rightarrow v^1_i \,, & & \text{for } i=1,\dots,m \,,\\
            v_i^b &\rightarrow u^1_{m+l+2-i} \,, & w_i^b &\rightarrow v^1_{m+l+2-i} \,, & & \text{for } i=1,\dots, l \,,\\
            u_{i}^a &\rightarrow  u^2_{n+k+2-i} \,, & v_{i}^a &\rightarrow u^1_{m+1}v^2_{n+k+2-i} \,, & w_i^a &\rightarrow v_{m+1}^1v_{n+k+2-i}^2\,, \, & \text{for } i=1,\dots,k+n+1 \,,
        \end{aligned}
    \end{align*}
    and a $\mathbb{Z}_2$-transformation on both integration domains, the boundary is identified as $\mathbb{V}_{k+n}\times\mathbb{V}_{m+l}$. We also change the matrix $Q$,such that it corresponds to the canonical ordering for nested vertices. It then reads
    \begin{align*}
        Q =& (\vec{q}_a,\vec{q}_b,\vec{q}_c,\vec{q}_{b,1}^{\,\,'},\dots,\vec{q}_{b,l}^{\,\,'},\vec{q}_{c,m}^{\,\,'},\dots,\vec{q}_{c,1}^{\,\,'},\vec{q}_{a,1}^{\,\,'},\dots,\vec{q}_{a,k+n+1}^{\,\,'}) \,.
    \end{align*}
    It can now be seen that
    \begin{align*}
        \boxed{
        \begin{aligned}
        &\int_{\partial\mathbb{W}_{m,l,k,n}}\Omega_{m,l,k,n}^c|_{u_i^c=u_i^b=0} \sim\\
        &\sim\mathcal{V}(\bullet,\dots,\bullet,\mathcal{V}(\bullet,\dots,\bullet,a,\bullet,\dots,\bullet,b,\bullet,\dots,\bullet),\bullet,\dots,\bullet,c,\bullet,\dots,\bullet)
        \end{aligned}}
    \end{align*}
    on this boundary, with the exception of $\mathcal{V}(\bullet,\dots,\bullet,\mathcal{V}(a,b),\bullet,\dots,\bullet,c,\bullet,\dots,\bullet)$. An example of a disk diagram at this boundary is shown in Fig. \ref{BB1C}.
    \begin{figure}[h!]
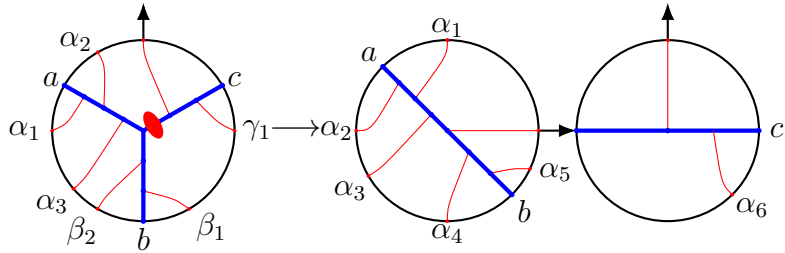

        \centering

        \caption{An example of boundary 1 contributing to $\mathcal{V}(\mathcal{V}(\alpha_1,a,\alpha_2,\alpha_3,\alpha_4,b,\alpha_5),\alpha_6,c)$.}
        \label{BB1C}
    \end{figure}

    \item Boundary 2: At this boundary
    \begin{align*}
        u_1^c=v_1^c \,.
    \end{align*}
    \begin{figure}[h!]
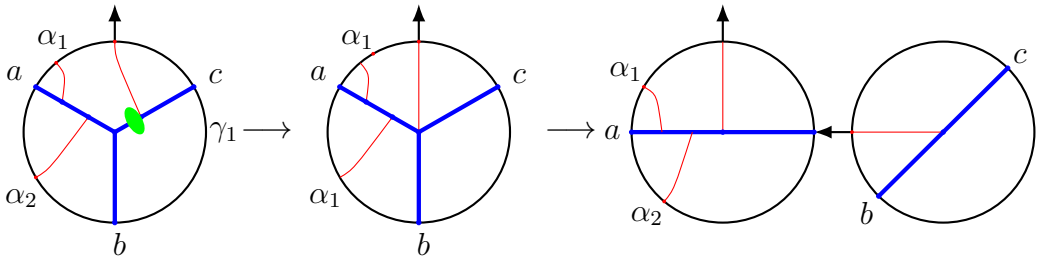

        \centering
        %
        \caption{An example of boundary 2 contributing to $\mathcal{V}(\alpha_1,a,\alpha_2,\mathcal{V}(b,c))$.}
        \label{BB2C}
    \end{figure}
    This is only a boundary if $l=0$, in which case it is a higher codimension boundary unless $k=n=0$, like in the left-ordered case. The $uv$-chain then reads
    \begin{align*}
        1= \frac{u_m^c}{v_m^c} \leq \dots \leq \frac{u_1^c}{v_1^c} \leq \frac{u_1^a}{v_1^a}\,.
    \end{align*}
    The closure constraint now forces $u_i^\bullet=v_i^\bullet$. Then, after a change of coordinates
    \begin{align*}
        \begin{aligned}
            v_i^c &\rightarrow u^2_i \,, & w_i^c &\rightarrow v^2_i \,, & \text{for } i=1,\dots,m \,,\\
            u_1^a &\rightarrow u^2_{m+1} \,, & w_1^a &\rightarrow v^2_{m+1} \,,
        \end{aligned}
    \end{align*}
    and a $\mathbb{Z}_2$-transformation on the remaining variables, the boundary is identified as $\mathbb{V}_m\times\mathbb{V}_0$ . We also change the matrix $Q$, such that it corresponds to the canonical ordering for nested vertices. It then reads
    \begin{align*}
        Q =& (\vec{q}_a,\vec{q}_b,\vec{q}_c,\vec{q}_{a,1}^{\,\,'},\vec{q}_{c,m}^{\,\,'},\dots,\vec{q}_{c,1}^{\,\,'}) \,.
    \end{align*}
    It can now be seen that
    \begin{align*}
        \boxed{\int_{\partial\mathbb{W}_{0,0,m,0}}\Omega_{m,0,0,0}^c|_{u_1^c=v_1^c} \sim \mathcal{V}(a,\bullet,\dots,\bullet,\mathcal{V}(b,c),\bullet,\dots,\bullet)}
    \end{align*}
    on this boundary. Since the line connected to the output arrow can only have one orientation, there is only one disk diagram at this boundary shown in Fig. \ref{BB2C}.

    \item Boundary 3: At this boundary
    \begin{align*}
        v_1^c=w_1^c\,.
    \end{align*}
    \begin{figure}[h!]
    \centering

    \caption{An example of boundary 3 contributing to $\mathcal{V}(\alpha_1,\mathcal{V}(a,b),\alpha_2,c)$.}
    \label{BB3C}
\end{figure}
    The $vw$-chain becomes
    \begin{align*}
        1 =& \frac{v_1^c}{w_1^c} \leq \dots \leq \frac{v_m^c}{w_m^c} \leq \frac{v_{k+n+1}^a}{w_{k+n+1}^a} = \dots = \frac{v_1^a}{w_1^a}\leq \frac{v_{l}^b}{w_{l}^b} \leq \dots \leq \frac{v_1^b}{w_1^b} \leq \infty \,.
    \end{align*}
    and implies $v_i^\bullet \geq w_i^\bullet$. The closure constraint then leads to $v_i^\bullet=w_i^\bullet$, which gives a higher codimension boundary, except for $l=0, m=1$. Then after the change of coordinates
    \begin{align*}
        \begin{aligned}
            u_1^c &\rightarrow u^2_1 \,, & v_1^c &\rightarrow v^2_1 \,,\\
            u_i^a &\rightarrow u^2_{k+n+3-i} \,, & v_i^a &\rightarrow v^2_{k+n+3-i} \,, & \text{for } i=1,\dots,k+n+1\,,\\
        \end{aligned}
    \end{align*}
    and a $\mathbb{Z}_2$-transformation on the remaining variables, the boundary is identified as $\mathbb{V}_{k+n+1}\times\mathbb{V}_0$. We also change the matrix $Q$, such that it corresponds to the canonical ordering for nested vertices. It then reads
    \begin{align*}
        Q =& (\vec{q}_a,\vec{q}_b,\vec{q}_c,\vec{q}_{a,1}^{\,\,'},\dots,\vec{q}_{a,k+n+1}^{\,\,'},\vec{q}_{c,1}^{\,\,'}) \,.
    \end{align*}
    It can now be seen that
    \begin{align*}
        \boxed{\int_{\partial\mathbb{W}_{1,0,k,n}}\Omega_{1,0,k,n}^c|_{v_1^c=w_1^c} \sim \mathcal{V}(\bullet,\dots,\bullet,\mathcal{V}(a,b),\bullet,\dots,\bullet,c,\bullet,\dots,\bullet)}
    \end{align*}    
    on this boundary. An example of a disk diagram at this boundary contributing to the $A_\infty$-term is shown in Fig. \ref{BB3C}, while Fig. \ref{BB3GC} shows a disk diagram contributing to a gluing term.

\begin{figure}[h!]
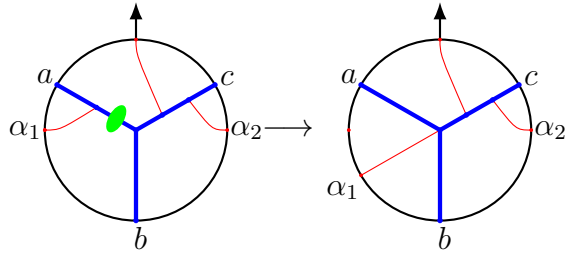

    \centering

    \caption{An example of boundary 3 contributing to a gluing term.}
    \label{BB3GC}
\end{figure}

    \item Boundary 4: At this boundary
    \begin{align*}
        v_1^b &= w_1^b \,.
    \end{align*}
    \begin{figure}[h!]
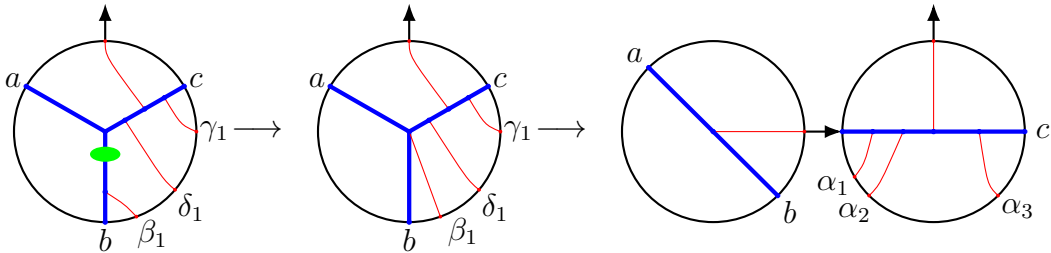

        \centering
        %
        \caption{An example of boundary 4 contributing to $\mathcal{V}(\mathcal{V}(a,b),\alpha_1,\alpha_2,\alpha_3)$.}
        \label{BB4C}
    \end{figure}
    The $vw$-chain becomes
    \begin{align*}
         0 \leq& \frac{v_1^c}{w_1^c} \leq \dots \leq \frac{v_m^c}{w_m^c} \leq \frac{v_{k+n+1}^a}{w_{k+n+1}^a} =\dots= \frac{v_1^a}{w_1^a}\leq \frac{v_{l}^b}{w_{l}^b} \leq \dots \leq \frac{v_1^b}{w_1^b} =1
    \end{align*}
    and implies $v_i^\bullet \leq w_i^\bullet$. The closure constraint then leads to $v_i^\bullet=w_i^\bullet$, which leads to a higher codimension boundary, except for $l=1$, $m=0$. Then, after the change of coordinates
    \begin{align*}
        \begin{aligned}
            u_1^b &\rightarrow u^2_1 \,, & v_1^b &\rightarrow v^2_1 \,,\\
            u_i^a &\rightarrow u^2_{k+n+3-i} \,, & v_i^a &\rightarrow v^2_{k+n+3-i} \,, & \text{for } i=1,\dots,k+n+1 \,,\\
        \end{aligned}
    \end{align*}
    and a $\mathbb{Z}_2$-transformation on the remainging coordinates, the boundary is identified as $\mathbb{V}_{k+n+1}\times\mathbb{V}_0$. We also change the matrix $Q$, such that it corresponds to the canonical ordering for nested vertices. It then reads
    \begin{align*}
        Q =& (\vec{q}_a,\vec{q}_b,\vec{q}_c,\vec{q}_{a,1}^{\,\,'},\dots,\vec{q}_{a,k+n+1}^{\,\,'},\vec{q}_{b,1}^{\,\,'}) \,.
    \end{align*}
    It can now be seen that
    \begin{align*}
        \boxed{\int_{\partial\mathbb{W}_{0,1,k,n}}\Omega_{0,1,k,n}^c|_{v_1^b=w_1^b} \sim \mathcal{V}(\bullet,\dots,\bullet,\mathcal{V}(a,b),\bullet,\dots,\bullet,c,\bullet,\dots,\bullet)}
    \end{align*}    
    on this boundary. An example of a disk diagram at this boundary contributing to the $A_\infty$-term is shown in Fig. \ref{BB4C}, while Fig. \ref{BB4GC} shows a disk diagram contributing to a gluing term.
    \begin{figure}[h!]
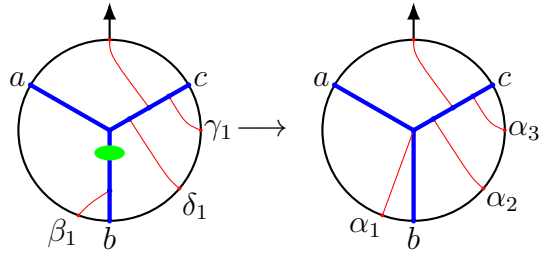

        \centering

        \caption{An example of boundary 4 contributing to a gluing term.}
        \label{BB4GC}
    \end{figure}

    \item Boundary 5: At this boundary
    \begin{align*}
        w_i^\bullet=1\,.
    \end{align*}
    The closure constraint forces all other $w$-variables to be zero. This yields a higher codimenion boundary, except for $m=1$ and $w_1^c=1$, which leads to $\beta=0$. After the change of coordinates
    \begin{align*}
        \begin{aligned}
            u_1^c &\rightarrow u^2_{l+1} \,, & v_1^c &\rightarrow v^2_{l+1}\,,\\
            u_i^b & \rightarrow u^2_{i} \,, & v_i^b &\rightarrow v^2_{i} \,, & \text{for } i=1,\dots,l\,,\\
            u_{i}^a &\rightarrow u^2_{k+n+l+3-i} \,, & v_i^a &\rightarrow v^2_{k+n+l+3-i} \,, & \text{for } i=1,\dots,k+n+1 \,.
        \end{aligned}
    \end{align*}
    and a $\mathbb{Z}_2$-transformation on both integration domains, the boundary is identified as $\mathbb{V}_{k+l+n+1}\times\mathbb{V}_0$. We also change the matrix $Q$, such that it corresponds to the canonical ordering for nested vertices. It then reads
    \begin{align*}
        Q =& (\vec{q}_a,\vec{q}_b,\vec{q}_c,\vec{q}_{c,1}^{\,\,'},\vec{q}_{a,1}^{\,\,'},\dots,\vec{q}_{a,k+n+1}^{\,\,'},\vec{q}_{b,l},\dots,\vec{q}_{b,1}^{\,\,'}) \,.
    \end{align*}
    It can now be seen that
    \begin{align*}
        \boxed{\int_{\partial\mathbb{W}_{1,l,k,n}}\Omega_{1,l,k,n}^c|_{w_1^c=1} \sim \mathcal{V}(\bullet,\dots,\bullet,\mathcal{U}(a,\bullet),\bullet,\dots,\bullet,b,\bullet,\dots,\bullet,c,\bullet,\dots,\bullet)}
    \end{align*}
    and
    \begin{align*}
        \boxed{\int_{\partial\mathbb{W}_{1,l,k,n}}\Omega_{1,l,k,n}^c|_{w_1^c=1} \sim \mathcal{V}(\bullet,\dots,\bullet,\mathcal{U}(\bullet,a),\bullet,\dots,\bullet,b,\bullet,\dots,\bullet,c,\bullet,\dots,\bullet)}
    \end{align*}
    on this boundary. An example of a disk diagram at this boundary is shown in Fig. \ref{BB5C}.

    \begin{figure}[h!]
        \centering

        \caption{An example of boundary 5 contributing to $\mathcal{V}(\mathcal{U}(a,\alpha_1),\alpha_2,b,\alpha_3,\alpha_4,c)$.}
        \label{BB5C}
    \end{figure}
    
    \item Boundary 6: At this boundary
    \begin{align*}
        w_i^\bullet=0 \,.
    \end{align*}
    For $m=1$ and $w_1^c=1$, this is the same as boundary 5. Otherwise, if $w_i^c=0$, the $uw$-chain is equivalent to
    \begin{align*}
        0 \geq& \frac{w_1^a}{u_1^a} \geq \dots \geq \frac{w_{k+n+1}^a}{u_{k+n+1}^a} \geq \frac{w_{1}^b}{u_{1}^b} = \dots = \frac{w_l^b}{u_l^b} \geq \frac{w_{m}^a}{u_{m}^c} \geq \dots \geq \frac{0^+}{u_i^c} \geq \dots \geq \frac{w_1^c}{u_1^c}\geq \infty \,,
    \end{align*}
    which forces $w_j^c=0$ for $j>i$ and $w_k^a=w_k^b=0$ for any $k$. This leads to a higher codimension boundary. Only when we consider $w_i^a=0$ for $i=1,\dots,k+n+1$ and $w_i^b=0$ for $i=1,\dots,l$, we find an $A_\infty$-term. Then, after the change of coordinates
    \begin{align*}
        \begin{aligned}
            u_1^c &\rightarrow u^2_{l+1} u^1_1 \,,\\
            v_i^c &\rightarrow v^2_{l+1} u^1_i \,, & w_i^c & \rightarrow v^1_i \,, & \text{for } i=1,\dots,m\\
            u_i^b &\rightarrow u^2_{i} \,, & v_i &\rightarrow v^2_{i} \,, & \text{for } i=1,\dots, l \,,\\
            u_{i}^a & \rightarrow u^2_{k+l+n+3-i} \,, & v_{i}^a & \rightarrow v^2_{k+l+n+3-i} \,, & \text{for } i=1,\dots,k+n+1
        \end{aligned}
    \end{align*}
    and a $\mathbb{Z}_2$-transformation on the remaining variables, the boundary is identified as $\mathbb{V}_{k+l+n+1}\times\mathbb{V}_{m-1}$. We also change the matrix $Q$, such that it corresponds to the canonical ordering for nested vertices. It then reads
    \begin{align*}
        Q =& (\vec{q}_a,\vec{q}_b,\vec{q}_c,\vec{q}_{c,m}^{\,\,'},\dots,\vec{q}_{c,1}^{\,\,'},\vec{q}_{a,1}^{\,\,'},\dots,\vec{q}_{a,k+n+1}^{\,\,'},\vec{q}_{b,l}^{\,\,'},\dots,\vec{q}_{b,1}^{\,\,'}) \,.
    \end{align*}
    It can now be seen that
    \begin{align*}
        \boxed{\int_{\partial\mathbb{W}_{m,l,k,n}}\Omega_{m,l,k,n}^c|_{w_{i}^a=w_i^b=0} \sim \mathcal{V}(\bullet,\dots,\bullet,\mathcal{U}(\bullet,\dots,\bullet,a,\bullet,\dots,\bullet),\bullet,\dots,\bullet,b,\bullet,\dots,\bullet,c,\bullet,\dots,\bullet)}
    \end{align*}
    on this boundary, with the exception of $\mathcal{V}(\bullet,\dots,\bullet,\mathcal{U}(a,\bullet),\bullet,\dots,\bullet,b,\bullet,\dots,\bullet,c,\bullet,\dots,\bullet)$ and $\mathcal{V}(\bullet,\dots,\bullet,\mathcal{U}(\bullet,a),\bullet,\dots,\bullet,b,\bullet,\dots,\bullet,c,\bullet,\dots,\bullet)$. An example of a disk diagram at this boundary is shown in Fig. \ref{BB6C}.

    \begin{figure}[h!]
        \centering

        \caption{An example of boundary 6 contributing to $\mathcal{V}(\mathcal{V}(a,\alpha_1,\alpha_2),\alpha_3,b,\alpha_4,\alpha_5,c)$.}
        \label{BB6C}
    \end{figure}

    \item Boundary 7: At this boundary
    \begin{align*}
        v_i^\bullet=1\,.
    \end{align*}
    \begin{figure}[h!]
        \centering
        %
        \caption{An example of boundary 7 contributing to $\mathcal{V}(a,\alpha_1,\alpha_2,\mathcal{U}(b,\alpha_3),\alpha_4,c)$.}
        \label{BB7C}
    \end{figure}

    The closure constraint forces all other $v$-variables to be zero. This yields a higher codimension boundary, except when $l=1$ and $v_1^b=1$. Then, after the change of variables
    \begin{align*}
        \begin{aligned}
            u_i^c &\rightarrow u^2_i \,, & w_i^c &\rightarrow v^2_i \,, & \text{for } i=1,\dots,m \,,\\
            u_1^b &\rightarrow u^2_{m+1} \,, & w_1^b &\rightarrow v^2_{m+1} \,,\\
            u_i^a &\rightarrow u^2_{k+m+n+3-i} \,, & w_i^a &\rightarrow v^2_{k+m+n+3-i} \,, & \text{for } i=1,\dots,k+n+1
        \end{aligned}
    \end{align*}
    and after a $\mathbb{Z}_2$-transformation on the remaining variables, the boundary is identified as $\mathbb{V}_{k+m+n+1}\times\mathbb{V}_0$. We also change the matrix $Q$, such that it corresponds to the canonical ordering for nested vertices. It then reads
    \begin{align*}
        Q =& (\vec{q}_a,\vec{q}_b,\vec{q}_c,\vec{q}_{b,1}^{\,\,'},\vec{q}_{a,1}^{\,\,'},\dots,\vec{q}_{a,k+n+1}^{\,\,'},\vec{q}_{c,m}^{\,\,'},\dots,\vec{q}_{c,1}^{\,\,'}) \,.
    \end{align*}
    It can now be seen that
    \begin{align*}
        \boxed{\int_{\partial\mathbb{W}_{m,1,k,n}}\Omega_{m,1,k,n}^c|_{v_{1}^b=1} \sim \mathcal{V}(\bullet,\dots,\bullet,a,\bullet,\dots,\bullet,\mathcal{U}(b,\bullet),\bullet,\dots,\bullet,c,\bullet,\dots,\bullet)}
    \end{align*}
    and
    \begin{align*}
        \boxed{\int_{\partial\mathbb{W}_{m,1,k,n}}\Omega_{m,1,k,n}^c|_{v_{1}^b=1} \sim \mathcal{V}(\bullet,\dots,\bullet,a,\bullet,\dots,\bullet,\mathcal{U}(\bullet,b),\bullet,\dots,\bullet,c,\bullet,\dots,\bullet)}
    \end{align*}
    on this boundary. An example of a disk diagram at this boundary is shown in Fig. \ref{BB7C}.

    \item Boundary 8: At this boundary 
    \begin{align*}
        v_i^\bullet=0 \,.
    \end{align*}
    \begin{figure}[h!]
        \centering

        \caption{An example of boundary 8 contributing to $\mathcal{V}(a,\alpha_1,\alpha_2,\mathcal{V}(\alpha_3,b,\alpha_4),\alpha_5,c)$.}
        \label{BB8C}
    \end{figure}
    The $vw$-chain becomes
    \begin{align*}
        \begin{aligned}
            0 &\leq \frac{v_1^c}{w_1^c} \leq \dots \leq \frac{v_m^c}{w_m^c} \leq \frac{v_{k+n+1}^a}{w_{k+n+1}^a} = \dots =\frac{0^+}{w_i^a} = \dots = \frac{v_1^a}{w_1^a}\leq \frac{v_{l}^b}{w_{l}^b} \leq \dots \leq \frac{v_1^b}{w_1^b} \leq \infty \,,\\
            0 &\leq \frac{v_1^c}{w_1^c} \leq \dots \leq \frac{v_m^c}{w_m^c} \leq \frac{v_{k+n+1}^a}{w_{k+n+1}^a} = \dots = \frac{v_1^a}{w_1^a}\leq \frac{v_{l}^b}{w_{l}^b} \leq \dots \leq \frac{0^+}{w_i^b} \dots \leq \frac{v_1^b}{w_1^b} \leq \infty
        \end{aligned}
    \end{align*}
    if $v_i^a=0$ and $v_i^b=0$, respectively. This forces all $v_i^a=0$ and $v_i^c=0$ in both cases, while in the latter we also have $v_j^b=0$ for $j>i$.  This leads to a higher codimension boundary. It is only when $v_i^c=0$ for $i=1,\dots,m$ and $v_i^a=0$ for $i=1,\dots,k+n+1$, that one finds an $A_\infty$-term. After the change of variables
     \begin{align*}
        \begin{aligned}
            u_{i}^c &\rightarrow u^2_{i} \,, & w_{i}^c &\rightarrow v^2_{i}\,, &&& \text{for } i=1,\dots,m \,,\\
            u_i^b &\rightarrow u^2_{m+1}u^1_i \,,&
            v_{i}^b &\rightarrow v^1_i \,, & w_{i}^b &\rightarrow v^2_{m+1}u^1_1\,, & \text{for } i=1,\dots,l \,,\\
            u_{i}^a &\rightarrow u^2_{k+m+n+3-i} \,, & w_{i}^a &\rightarrow v^2_{k+m+n+3-i}\,, &&& \text{for } i=1,\dots,k+n+1 \,,\\
        \end{aligned}
    \end{align*}
    and a $\mathbb{Z}_2$-transformation on the $u^2$- and $v^2$-variables, the boundary is identified as $\mathbb{V}_{k+m+n+1}\times\mathbb{V}_{l-1}$. We also change the matrix $Q$, such that it corresponds to the canonical ordering for nested vertices. It then reads
    \begin{align*}
        Q =& (\vec{q}_a,\vec{q}_b,\vec{q}_c,\vec{q}_{b,1}^{\,\,'},\dots,\vec{q}_{b,l}^{\,\,'},\vec{q}_{a,1}^{\,\,'},\dots,\vec{q}_{a,k+n+1}^{\,\,'},\vec{q}_{c,m}^{\,\,'},\dots,\vec{q}_{c,1}^{\,\,'}) \,.
    \end{align*}
    It can now be seen that
    \begin{align*}
        \boxed{\int_{\partial\mathbb{W}_{m,l,k,n}}\Omega_{m,l,k,n}^{c} \sim \mathcal{V}(\bullet,\dots,\bullet,a,\bullet,\dots,\bullet,\mathcal{U}(\bullet,\dots,\bullet,b,\bullet,\dots,\bullet),\bullet,\dots,\bullet,c,\bullet,\dots,\bullet)}
    \end{align*}
    on this boundary, with the exception of $\mathcal{V}(\bullet,\dots,\bullet,a,\bullet,\dots,\bullet,\mathcal{U}(b,\bullet),\bullet,\dots,\bullet,c,\bullet,\dots,\bullet)$ and $\mathcal{V}(\bullet,\dots,\bullet,a,\bullet,\dots,\bullet,\mathcal{U}(\bullet,b),\bullet,\dots,\bullet,c,\bullet,\dots,\bullet)$. An example of a disk diagram at this boundary is shown in Fig. \ref{BB8C}.

\end{itemize}

\paragraph{Gluing terms.}
\begin{itemize}

    \item Boundary 9: 
    
    At this boundary
    \begin{align*}
        \frac{u_m^c}{v_m^c}=\frac{u_{l}^b}{v_{l}^b}\,.
    \end{align*}
    \begin{figure}[h!]
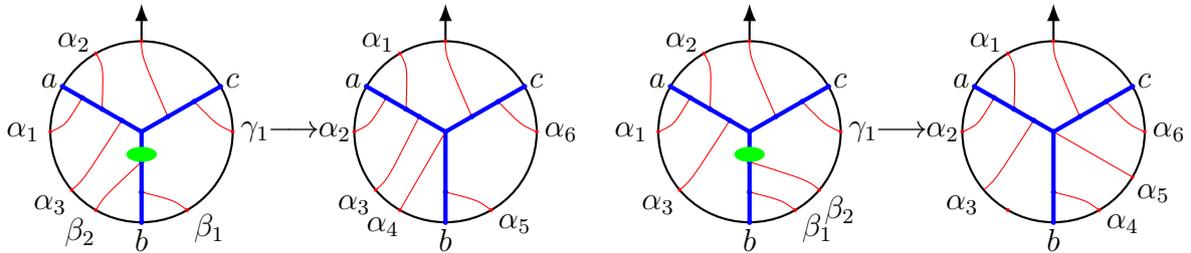

       \centering

       \caption{Two examples of boundary 9 contributing to a gluing term, with both orientations of $\beta_2$.}
       \label{BB9C}
   \end{figure}
   $\int_{\partial\mathbb{W}_{k,l,m,n}}\Omega_{k,l,m,n}^c$ does not yield a familiar $A_\infty$-term on this boundary and it does not vanish either. An example of a disk diagram at this boundary is shown in Fig. \ref{BB9C}.

   \item Boundary 10: At this boundary
    \begin{align*}
        \frac{u_1^c}{v_1^c}=\frac{u_{k+n+1}^a}{v_{k+n+1}^a} \,.
    \end{align*}
    For $k=l=n=0$ the $uv$-chain becomes
    \begin{align*}
        0 \leq \frac{u_m^c}{v_m^c} = \dots = \frac{u_1^c}{v_1^c} = \frac{u_{1}^a}{v_{1}^a} \leq \infty\,.
    \end{align*}
    The closure constraint then gives $u_1^\bullet=v_1^\bullet$, so this boundary is equivalent to boundary 2 when $k=l=n=0$ and does not yield  a gluing term, but an $A_\infty$-term instead. Otherwise, $\int_{\partial\mathbb{W}_{m,l,k,n}}\Omega_{m,l,k,n}^c$ does not yield a familiar $A_\infty$-term on this boundary and it does not vanish either. An example of a disk diagram at this boundary is shown in Fig. \ref{BB10C}. Gluing terms corresponding to the type of diagrams on the right cancel with gluing terms
    belonging to the type of diagrams on the right of Fig. \ref{BB9C}. A special type of gluing term with no elements between the junction and the output arrow, i.e. $n=0$, is shown in Fig. \ref{BB10SC}. This is the only type of gluing term that does not cancel with any other gluing term from the potential $\Omega_{m,l,k,n}^c$, but rather from the type $\Omega_{k,l,m,n}^a$, `gluing' them together. In particular, it cancels with the type of diagrams depicted in Fig. \ref{BB10S}.

       \begin{figure}[h!]
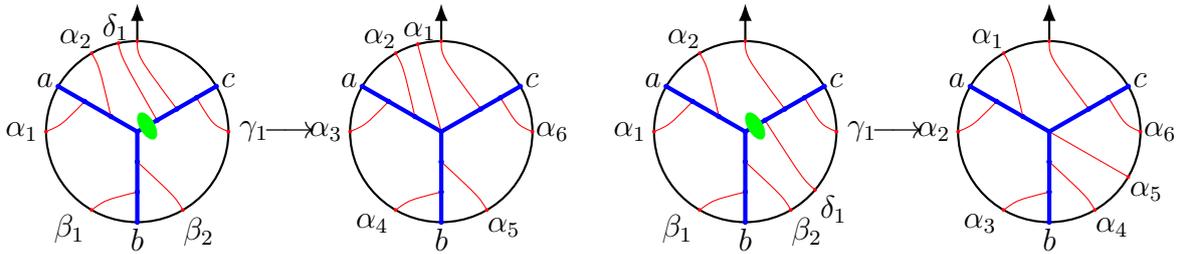

       \centering

       \caption{Two examples of boundary 10 contributing to a gluing term, with both orientations of $\delta_1$.}
       \label{BB10C}
   \end{figure}

   \begin{figure}[h!]
       \centering
       %
       \caption{Special case of a gluing term coming from boundary 10. This gluing term is cancelled by a gluing term coming from a potential of the type $\Omega_{k,l,m,n}^a$.}
       \label{BB10SC}
   \end{figure}

    \item Boundary 11: At this boundary
    \begin{align*}
        \frac{v_m^c}{w_m^c} =& \frac{v_{k+n+1}^a}{w_{k+n+1}^a}  \,.
    \end{align*}
    For $l=0$, $m=1$ the $vw$-chain reads
    \begin{align*}
         0 \leq& \frac{v_1^c}{w_1^c} = \frac{v_{k+n+1}^a}{w_{k+n+1}^a} = \dots = \frac{v_{1}^a}{w_{1}^a} \leq \infty\,.
    \end{align*}
    The closure constraint then gives $v_i^\bullet=w_i^\bullet$, so this boundary is equivalent to boundary 3 when $l=0$ and $m=1$, producing either a gluing term or an $A_\infty$-term. Otherwise, $\int_{\partial\mathbb{W}_{m,l,k,n}}\Omega_{m,l,k,n}^c$ does not yield a familiar $A_\infty$-term on this boundary and it does not vanish either. An example of a disk diagram at this boundary is shown in Fig. \ref{BB9C}. Gluing terms corresponding to the type of diagrams on the left cancel with gluing terms belonging to the type of diagrams on the left of Fig. \ref{BB11C}, while gluing terms corresponding to the type of diagrams on the right cancel with gluing terms belonging to the type of diagrams on the left of Fig. \ref{BB10C}.

    \begin{figure}[h!]
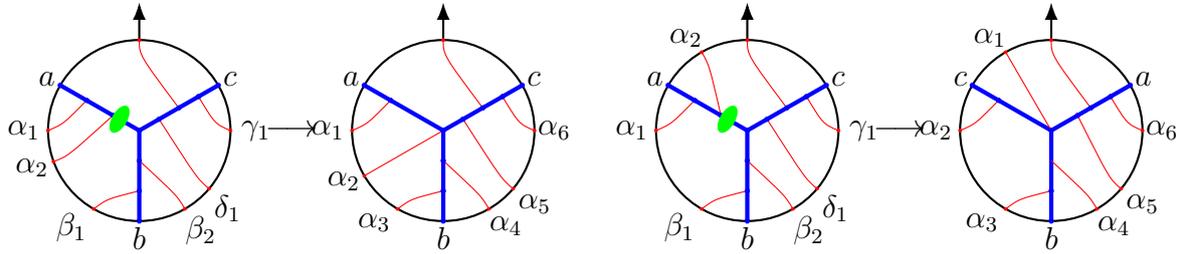

       \centering

       \caption{Two examples of boundary 11 contributing to a gluing term, with both orientations of $\gamma_2$.}
       \label{BB11C}
   \end{figure}

    \item Boundary 12: At this boundary
    \begin{align*}
        \frac{v_1^a}{w_1^a} &= \frac{v_{l}^b}{w_{l}^b} \,.
    \end{align*}
    This implies $w_{l}^b=\beta v_{l}^b$ and from 
    \begin{align*}
        \frac{u_{l}^b}{w_{l}^b} =& \frac{1}{\beta}\frac{u_{l}^b}{v_{l}^b} = \frac{\alpha}{\beta}
    \end{align*}
    we get $\frac{u_{l}^b}{v_{l}^b}=\alpha=\frac{u_m^c}{v_m^c}$, so we find that this boundary is equivalent to boundary 9. It is only when $m=0$ that boundary 9 does not exist and we have to consider this one. $\int_{\partial\mathbb{W}_{m,l,k,n}}\Omega_{m,l,k,n}^c$ does not yield a familiar $A_\infty$-term on this boundary and it does not vanish either. The disk diagrams at this boundary resemble the ones in Fig. \ref{BB9C} when there are no lines attached to the $c$-leg. An example of a disk diagram at this
    boundary is shown in Fig. \ref{BB12C}. Gluing terms corresponding to the type of diagrams on the left cancel with gluing terms belonging to the type of diagrams on the left of Fig. \ref{BB11C}.

    \begin{figure}[h!]
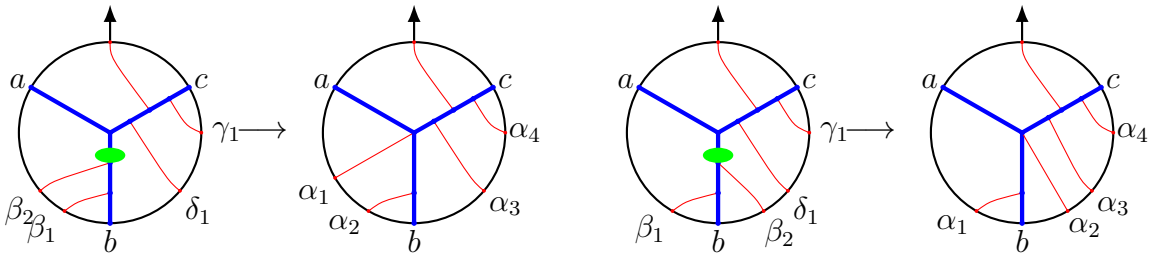

       \centering

       \caption{Two examples of boundary 12 contributing to a gluing term, with both orientations of $\beta_2$.}
       \label{BB12C}
   \end{figure}

    \item Boundary 13: At this boundary
    \begin{align*}
        \frac{u_1^b}{w_1^b} =& \frac{u_{k+n+1}^a}{w_{k+n+1}^a}\,.
    \end{align*}
    This implies $w_{k+n+1}^a=\frac{\beta}{\alpha}u_{k+n+1}^a$ and from
    \begin{align*}
        \frac{v_{k+n+1}^a}{w_{k+n+1}^a}=\frac{\alpha}{\beta}\frac{v_{k+n+1}^a}{u_{k+n+1}^a}=\frac{1}{\beta}
    \end{align*}
    we get $\frac{u_{k+n+1}^a}{v_{k+n+1}^a}=\alpha=\frac{u_1^c}{v_1^c}$, so we find that this boundary is equivalent to boundary 10. It is only when $m=0$ that boundary 10 does not exist and we have to consider this one. $\int_{\partial\mathbb{W}_{m,l,k,n}}\Omega_{m,l,k,n}^c$ does not yield a familiar $A_\infty$-term on this boundary and it does not vanish either. The disk diagrams at this boundary resemble the ones in Fig. \ref{BB10C} when there are no lines attached to the $c$-leg. An example of a disk diagram at this
    boundary is shown in Fig. \ref{BB13C}. Gluing terms corresponding to the type of diagrams on the left cancel with gluing terms belonging to the type of diagrams on the right of Fig. \ref{BB11C}, while gluing terms corresponding to the type of diagrams on the right cancel with gluing terms belonging to the type of diagrams on the right of Fig. \ref{BB12C}.
       \begin{figure}[h!]
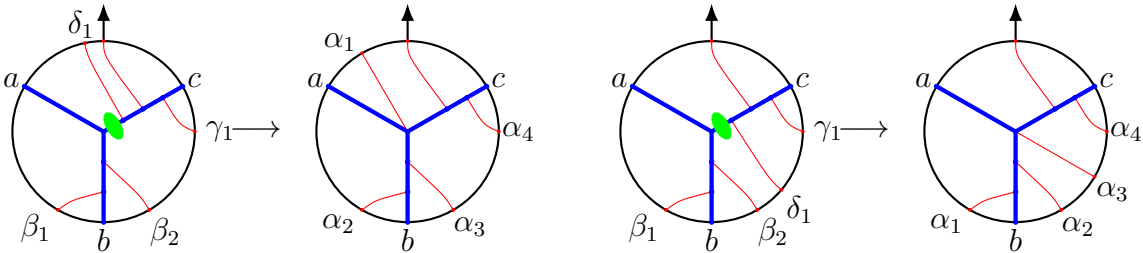

       \centering

       \caption{Two examples of boundary 13 contributing to a gluing term, with both orientations of $\delta_1$.}
       \label{BB13C}
   \end{figure}

    \item Boundary 14: At this boundary
    \begin{align*}
        \frac{u_{m}^c}{w_{m}^c} =& \frac{u_{l}^b}{w_{l}^b}\,.
    \end{align*}
    This implies $u_{m}^c=\frac{\alpha}{\beta}w_{m}^c$ and from
    \begin{align*}
        \frac{u_m^c}{v_m^c}=\frac{\alpha}{\beta}\frac{w_m^c}{v_m^c}=\alpha
    \end{align*}
    we get $\frac{v_m^c}{w_m^c}=\frac{1}{\beta}=\frac{v_{k+n+1}^a}{w_{k+n+1}^a}$, so we find that this boundary is equivalent to boundary 11. Boundary 11 does not exist when $m=0$, but neither does this one. This means that this boundary is always equivalent to boundary 11.
\end{itemize}

\section{Conclusions and discussion}
\label{sec:conclusions2}
It has already been understood that there are tight links between higher spin gravities and deformation quantization. The first link is obvious: higher spin algebras are associative algebras resulting from deformation quantization of coadjoint orbits that correspond to irreducible representations of the spacetime symmetry algebras (often-times, $so(d,2)$). Therefore, the product is given by the Kontsevich formula (in fact, Fedosov's construction suffices since the coadjoint orbits are symplectic manifolds). 

The second link to deformation quantization is more subtle. Any higher spin algebra determines the free equations of motion. The interactions are due to a certain Hochschild cohomology group being nontrivial and the next one, which contains obstructions, being trivial. In the simplest case of the Weyl algebra $A_1$, it is the group $HH^2(A_1, A^*_1)$ that leads to Chiral Theory. The group $HH^2(A_1, A^*_1)$ is one dimensional and the cocycle can be obtained from Shoikhet--Tsygan--Kontsevich's formality \cite{Kontsevich:1997vb, Tsygan,Shoikhet:2000gw}. However, a nontrivial cocycle is not yet a vertex, it only justifies its existence. There does not seem any simple way to generate any vertices directly from the formality.\footnote{The cubic vertex can formally be written in a factorized form with the help of the cocycle \cite{Sharapov:2019vyd}, but this form is forbidden by additional physical assumptions, e.g. the existence of the smooth flat limit. } 

Another link to formality is the very form of the vertices: they are represented by graphs similar to Kontsevich's ones with certain weights. Since the Poisson structure $\epsilon^{AB}$ is constant for our case, there are no genuine bulk vertices and all the graphs have legs on the boundary. These graphs can be re-summed to give the final result presented in Sec. \ref{sec:CD}, see also \cite{Sharapov:2022awp,Sharapov:2022wpz,Sharapov:2022nps}. Lastly, in this paper, we managed to prove the $A_\infty$-relations via the Stokes theorem, which is a method typical for formality theorems thanks to Kontsevich. 

The arguments here and above suggest that there is a bigger picture where (Shoikhet--Tsygan--)Kontsevich formality occupies the first two floors. While this structure is yet to be found, a more specific problem is to construct new theories (or recast the old ones, e.g. conformal higher spin gravity) along the lines of this paper, i.e., to find appropriate configuration spaces for vertices and $A_\infty$-relations.

The observation \cite{Sharapov:2022wpz,Sharapov:2022nps} that Chiral Theory is essentially a Poisson sigma-model\footnote{That the $A_\infty$-algebra of Chiral theory is a pre-Calabi-Yau one implies that its symmetrization (essentially, by inserting the fields $\omega$ and $C$ into the structure maps) is a usual commutative Poisson structure (note that pre-Calabi-Yau structure is a noncommutative analog of the Poisson structure, see e.g. \cite{IYUDU202163, kontsevich2021pre}). Therefore, the equations of motion have automatically the form of those of a Poisson sigma-model, i.e. $d\omega_k=\tfrac12\partial_k\pi^{ij}(C)\,\omega_i\,\omega_j$, $dC^i=\pi^{ij}(C)\,\omega_j$, where we introduced notation $\omega_k$ and $C^i$ for the fields, which also stresses that they live in the spaces dual to each other. } -- determined by a (noncommutative) Poisson structure -- may also lead to new insight into the problem of higher spin theories. It is plausible that all of them are Poisson sigma-models too, at least at the formal level. In this regard let us note that Poisson sigma-models of a different kind have already appeared in the higher spin literature, see e.g. \cite{Engquist:2005yt,Arias:2015wha,arias2016differential}.

Thus, the main conclusions of the paper are: (i) there has to exist a formality that extends (Shoikhet--Tsygan--)Kontsevich formality; (ii) Chiral Theory's vertices are its elementary consequences; (iii) there should exist a two-dimensional topological model that explains all of the above at the physics' level of rigour, similar to how the Poisson sigma-model is related to the Kontsevich formality theorem \cite{Cattaneo:1999fm}. It would be interesting to give these observations more solid support in the future.

\chapter{Conclusion}
\label{chap:conclusion}

The first part of this thesis presented the construction of chiral Higher Spin Gravity (HiSGRA): the first Lorentz covariant and coordinate-independent formulation of a HiSGRA with massless propagating degrees of freedom in both flat space and (A)dS backgrounds. This was achieved by deriving the equations of motion from the underlying $L_\infty$-algebra structure. The same formulation was also constructed independently for self-dual Yang–Mills (SDYM) and self-dual gravity (SDGRA), which are embedded as closed subsectors of chiral HiSGRA, and was shown to be consistent with the full $L_\infty$-algebra. Furthermore, the covariant formulation reproduces known flat-space amplitudes obtained in light-cone gauge. Through the AdS/CFT correspondence, the existence of this formulation establishes that the $O(N)$ vector models -- including notable examples such as the Ising model -- contain a well-defined chiral and anti-chiral subsector.

A remarkable feature of the formulation is that the resulting $n$-point vertices are manifestly local for all $n \geq 2$. These vertices are expressed as integrals of simple exponential functions over a class of configuration spaces referred to as swallowtails, which correspond to convex polygons in $\mathbb{R}^2$. This points to a surprising connection between chiral HiSGRA, convex geometry, and positive Grassmannians. The underlying $L_\infty$-algebra is derived from an $A_\infty$-algebra of the form $\hat{\mathbb{A}}=\mathbb{A}\otimes B$, where $\mathbb{A}=A_\lambda[-1]\oplus A_\lambda^\star$ and $B=A_1\otimes \text{Mat}_N$ is associative. This $A_\infty$-algebra is of pre-Calabi–Yau type of degree 2, suggesting that chiral HiSGRA may admit a description as a two-dimensional topological field theory \cite{kontsevich2021pre}. 

The second line of research explored the connection between chiral HiSGRA and the celebrated (Shoikhet–Tsygan-)Kontsevich formality theorems. We showed that the $A_\infty$-relations underlying chiral HiSGRA can be derived via an application of Stokes’ theorem — a hallmark feature of known formality proofs. To this end, we introduced a class of carefully constructed configuration spaces whose boundaries decompose into products of swallowtails. These spaces generalize convex polygons to three dimensions and can be viewed as higher-dimensional analogues in $\mathbb{R}^3$. The resulting Stokes-based proof hints at an extension of formality in the framework of non-commutative deformation quantization.

A notable observation is that among all quasi-isomorphic $A_\infty$-algebras, the minimal model produces vertices that are manifestly local. A quasi-isomorphism will typically induce a non-local field redefinition of the vertices. In a sense, this suggests that $A_\infty$/$L_\infty$-algebra that represent field theories have preferred bases where the vertices are maximally local. It would be interesting to find how locality is exactly tied to a choice of representative in an equivalence classes of quasi-isomorphic $A_\infty$-algebras.

Another surprising feature is that the most economical representation of the vertices is through integrals over convex polygons and that the Stokes-based proof requires a generalization of these objects, wherein the boundaries of the configuration spaces are themselves products of convex polygons. This leads to the question whether other HiSGRAs -- or field theories in general -- may be constructed from similar vertices with associated configuration spaces. The requirement of a Stokes-based proof might offer a set of conditions, e.g. the existence of spaces with boundaries that share the same structural properties, that allows one to obtain HiSGRAs in a completely novel way.

Ongoing work is focused on computing the three- and four-point correlation functions for SDYM and its higher spin extension in $AdS_4$. This will provide a novel realization of AdS/CFT in terms of first-order theories and it aims to carefully study the results in various gauges and compare them to known results in Yang-Mills. This work will be extended to a series of projects, in which the same will be computed for SDGR and its higher spin extension and finally for chiral HiSGRA. This long-term goal is to obtain all higher order correlation functions by identifying all higher spin invariants.

The results obtained in this thesis give rise to many more interesting research questions for future work. On the more mathematical side, for instance, one might study (i) how the (Shoiket-Tsygan-)Kontsevich can be extended in light of the chiral HiSGRA construction, (ii) what topological model describes chiral HiSGRA, (iii) what is the exact relation between chiral HiSGRA and positive Grassmanians (also including the larger configuration spaces in the Stokes-based proof), (iv) can the admissible configuration spaces be classified to generate other field theories?

Leaning more towards the physics side of the spectrum, it would be interesting study chiral HiSGRA on twistor space by (i) using HPT to transfer the $A_\infty$-algebra of chiral HiSGRA to twistor space and construct its equations of motion on twistor space, (ii) constructing a parity-invariant extension of chiral HiSGRA, perhaps following the recently found HS Yang-Mills extension of HS-SDYM on twistor space \cite{Adamo:2022lah}, (iii) constructing an action for chiral HiSGRA, (iv) proving that chiral HiSGRA is integrable.

Quantum effects of the theory can also be studied. This includes (i) computing quantum corrections to two-point and three-point functions and studying the anomalous dimensions of dual operators and their OPE coefficients, (ii) proving that chiral HiSGRA is UV-finite at loop-level by extending the results obtained in \cite{Skvortsov:2020wtf,Skvortsov:2018jea} to all $n$-point functions.

In addition, it would be interesting to study chiral HiSGRA in relation to broader formalisms such as the double copy and soft theorems in celestial holography.

\appendix

\chapter{Notation and conventions}
\label{app:notation}
The most important conventions and definitions that were used in the main text are introduced here. A short discussion on the spinor formalism is given, a more elaborate treatment can be found in \cite{penroserindler}.

It is useful for $4d$ theories to express all spacetime indices in terms of spinor indices, using $so(3,1)\sim sl(2,\mathbb{C})$. This isomorphism allows to map $4$-dimensional spacetime vectors to $2\times2$ Hermitian matrices, which can be extended to tensors. As basis for the $2\times2$ Hermitian matrices in flat spacetime we choose the Pauli matrices and the unit matrix $\sigma^\mu_{AB'}=(1\!\!1,\sigma^i)$. The Greek letters run over spacetime indices, whereas the lower case Latin letters are space indices and capitals are the matrix indices. The Pauli matrices satisfy $\mathrm{Tr}\,\sigma^i=0$ and $\{\sigma^\mu,\sigma^\nu\}_{AA'}=2\eta^{\mu\nu}1\!\!1_{AA'}$, with $\eta^{\mu\nu}=\text{diag}(-1,1,1,1)$ the Minkowski metric. We define
\begin{align*}
    x_{AA'}=x^\mu(\sigma_{\mu})_{AA'}=\begin{pmatrix}
    x^0+x^3 & x^1-ix^2\\
    x^1+ix^2 & x^0-x^3
    \end{pmatrix} \,,
\end{align*}
which is Hermitian. We also introduce a dual set $(\bar{\sigma}^{\mu})^{   AA'}=(\mathbf{1},-\sigma^i)$, such that
\begin{align*} 
    v^\mu=-\tfrac{1}{2}x_{AA'}\bar{\sigma}^{\mu A'A} \,.
\end{align*}
The two sets are related by $\sigma_\mu^{AA'}=\bar{\sigma}_\mu^{A'A}$. We also introduce raising and lowering rules for the primed (and similarly for unprimed indices):
\begin{align*}
    y_A&=y^B\epsilon_{BA},& y^A&=\epsilon^{AB}y_B \,.
\end{align*}
The inner product in spinor indices is defined as $(xy)=x^Ay^B\epsilon_{AB}=x_Ay^A=-x^Ay_A$. We define the $\epsilon$'s as
\begin{align*}
    \epsilon^{AB}=\epsilon^{A'B'}=i(\sigma^2)^{AB}=\begin{pmatrix}
    0 & 1\\
    -1 & 0
    \end{pmatrix}
\end{align*}
and their inverse $-\epsilon_{AB}=(-\sigma^2)^{-1}=-i\sigma^2$. $\epsilon_{AB}$ is anti-symmetric and $\epsilon_{AC}\epsilon^{BC}=\delta_{A}^{\hspace{5pt}B}$. Inner products in spinor indices look slightly different from inner products in spacetime indices:
\begin{align*}
    x_{AA'}y^{AA'}&=-2x_\mu y^\mu\,, & z_{A}z^{A}&=z^{A}z^{B}\epsilon_{AB}=0\,.
\end{align*}
Any bi-spinor $T_{AB}$ can be decomposed into symmetric and anti-symmetric parts:
\begin{align} \label{eq:spinordecomp}
T_{AB}=\tfrac{1}{2}(T_{AB}+T_{BA}+T_{AB}-T_{BA})=T_{(AB)}+\tfrac{1}{2}\epsilon_{AB}T_C^{\hspace{5pt}C} \,,
\end{align}
where $T_{(AB)}=\tfrac{1}{2!}(T_{AB}+T_{BA})$ denotes the symmetric part of $T_{AB}$. From now on, we will use the convention that if a tensor carries identical indices, it is implied that the tensor is symmetric in them, e.g. $T^{AA}=\tfrac{1}{2!}(T^{A_1A_2}+T^{A_2A_1})$ and in a more condensed notation, $T^{A(n)}$ is symmetrized over the $n$ indices in a similar fashion. Tensors can carry two types of unrelated indices, primed and unprimed. In the most general case we write $T^{A(m),A'(n)}=\tfrac{1}{m!n!}\sum_{permutations}T^{A_1\dots A_m,A'_1\dots A'_n}$.
\\
An object that we will often use is the vierbein $e^{AA'}\equiv e_\mu^{AA'}dx^\mu$, which is a one-form. A direct consequence of the decomposition of \eqref{eq:spinordecomp} leads to the important identity
\begin{align} \label{e-wedge-e}
    e_{AA'}\wedge e_{BB'}=\tfrac{1}{2}(\epsilon_{A'B'}H_{AB}+\epsilon_ {AB}H_{A'B'}) \,,
\end{align}
where $H_{AB}=e_{AC'}\wedge e\fdu B{C'}$ and $H_{A'B'}=e_{CA'}\wedge e\fud C{B'}$. This identity allows one, for example, to rewrite the Yang-Mills field strength in terms of its (anti)-self-dual parts
\begin{align*}
    F=F_{AA'|BB'}e^{AA'}\wedge e^{BB'}=H^{BB}F_{BB}+H^{B'B'}F_{B'B'}\,,
\end{align*}
with $F_{AB}=\tfrac{1}{2}F\fdu{AC'|B}{C'}$ and $F_{A'B'}=\tfrac{1}{2}F\fdud{CA'|}{C}{B'}$.

Another useful feature of the spinor formalism is the Fierz identity. Given three spinors, $\phi_A$, $\chi_B$, $\psi_C$, the anti-symmetrization over their indices equals zero, as their indices only run over two values $A,B,C=0,1$.  The Fierz identity is obtained by contracting this anti-symmetrized product with $\epsilon^{BC}$, which leads to
\begin{align} \label{Fierz}
    3\epsilon^{BC}\phi_{[A}\chi_{B}\psi_{C]}=\phi_{A}(\chi\psi)+\chi_{A}(\psi\phi)+\psi_{A}(\phi\chi)\equiv0 \,.
\end{align}

\chapter{Self-dual theories}
\label{app:selfdual}

\section{Technicalities: SDYM}\label{app:sdym}
The calculations in the main text have been highly compacted for the sake of brevity. In this appendix we aim to present some proofs and additional details to the reader.
\subsection{\texorpdfstring{$\Psi$}{psi}-sector} \label{app:SDYMPsi}
The calculations of the $\Psi$-sector have been moved to this appendix as they are very much similar to the $F$-sector. The approach is as follows: we apply a covariant derivative to the ansatz \eqref{ansatzpsi} and we also contract $e^{BB'}$ with $\Psi_{A(k+1),A'(k+3)}$ as to obtain two expressions for $e^{BB'}D\Psi_{A(k)B,A'(k+2)B'}$, which we then compare. The former yields
\begin{align} \label{DsquaredPsi}
    \begin{aligned}
        D^2\Psi_{A(k),A'(k+2)}=&-H^{BB}[F_{BB},\Psi_{A(k),A'(k+2)}]=-e^{CC'}\wedge D\Psi_{A(k)C,A'(k+2)C'} \\
        &-e\fud{C}{A'}\wedge\sum_{n=0}^{k-1}\beta_{nk}[DF_{A(n+1)C,A'(n)},\Psi_{A(k-n-1),A'(k-n+1)}] \\
        &-e\fud{C}{A'}\wedge\sum_{n=0}^{k-1}\beta_{nk}[F_{A(n+1)C,A'(n)},D\Psi_{A(k-n-1),A'(k-n+1)}] \\
        &-e\fud{C}{A'}\wedge\sum_{n=0}^{k-1}\gamma_{nk}[DF_{A(n+2),A'(n)},\Psi_{A(k-n-2)C,A'(k-n+1)}] \\
        &-e\fud{C}{A'}\sum_{n=0}^{k-1}\gamma_{nk}[F_{A(n+2),A'(n)},D\Psi_{A(k-n-2),A'(k-n+1)}] \,.
    \end{aligned}
\end{align}
Considering only quadratic terms in the fields gives
{\allowdisplaybreaks
\begin{align} \label{nablasquaredpsi}
    \begin{aligned}
        e^{BB'}\wedge D\Psi_{A(k)C,A'(k+2)C'}&=H^{BB}[F_{BB},\Psi_{A(k),A'(k+2)}]\\
        &-\sum_{n=1}^{k}\tfrac{\beta_{(n-1)k}}{2}H^{BB}[F_{A(n)BB,A'(n)},\Psi_{A(k-n),A'(k-n+2)}]\\
        &-\sum_{n=0}^{k}\tfrac{\beta_{nk}+\gamma_{(n-1)k}}{2}H^{BB}[F_{A(n+1)B,A'(n)},\Psi_{A(k-n-1)B,A'(k-n+2)}]\\
        &-\sum_{n=0}\tfrac{\gamma_{nk}}{2}H^{BB}[F_{A(n+2),A'(n)},\Psi_{A(k-n-2)BB,A'(k-n+2)}]\\
        &+\sum_{n=0}^{k}\tfrac{\gamma_{(n-1)k}}{2}H\fdu{A'}{B'}[F\fdud{A(n+1)}{B}{,A'(n-1)B'},\Psi_{A(k-n-1)B,A'(k-n+2)}]\\
        &-\sum_{n=0}^{k}\tfrac{\beta_{nk}}{2}H\fdu{A'}{B'}[F\fdud{A(n+1)}{B}{,A'(n)},\Psi_{A(k-n-1)B,A'(k-n+1)B'}]\,.
    \end{aligned}
\end{align}}\noindent
We have renamed the dummy indices in some terms in order to match the summation limits with the expression for $e^{BB'}\Psi_{A(k)B,A'(k+2)B'}$ that we will derive next. This makes some coefficients show up that were not present in the minimal ansatz, so we have to set them to zero by hand: $\beta_{kk}=0$, $\gamma_{(-1)k}=0$. Contracting $e^{BB'}$ with $D\Psi_{A(k),A'(k+4)}$ gives
\begin{align*}
    \begin{aligned}
        &e^{BB'}\wedge D\Psi_{A(k)B,A'(k+2)B'}=\\
        &-\sum_{n=0}^{k}\tfrac{(n+1)(k+4)}{(k+1)(k+3)}\tfrac{\beta_{n(k+1)}}{2}H^{BB}[F_{A(n)BB,A'(n)},\Psi_{A(k-n),A'(k-n+2)}]\\
        &-\sum_{n=0}^{k}(\tfrac{(k-n)(k+4)}{(k+1)(k+3)}\tfrac{\beta_{n(k+1)}}{2}+\tfrac{(n+2)(k+4)}{(k+1)(k+3)}\tfrac{\gamma_{n(k+1)}}{2})H^{BB}[F_{A(n+1)B,A'(n)},\Psi_{A(k-n-1)B,A'(k-n+2)}]\\
        &-\sum_{n=0}^{k}(\tfrac{n(k-n)}{(k+1)(k+3)}\tfrac{\beta_{n(k+1)}}{2}-\tfrac{n(n+2)}{(k+1)(k+3)}\tfrac{\gamma_{n(k+1)}}{2})H\fdu{A'}{B'}[F\fdud{A(n+1)}{B}{,A'(n-1)B'},\Psi_{A(k-n-1)B,A'(k-n+2)}]\\
        &-\sum_{n=0}^{k}(\tfrac{(k-n)(k-n+2)}{(k+1)(k+3)}\tfrac{\beta_{n(k+1)}}{2}-\tfrac{(n+2)(k-n+2)}{(k+1)(k+3)}\tfrac{\gamma_{n(k+1)}}{2})H\fdu{A'}{B'}[F\fdud{A(n+1)}{B}{,A'(n)},\Psi_{A(k-n-1)B,A'(k-n+1)B'}]\\
        &-\sum_{n=0}^{k}\tfrac{(k-n-1)(k+4)}{(k+1)(k+3)}\tfrac{\gamma_{n(k+1)}}{2}H^{BB}[F_{A(n+2),A'(n)},\Psi_{A(k-n-2)BB,A'(k-n+2)}]\,.
    \end{aligned}
\end{align*}
Comparing this expression to \eqref{nablasquaredpsi}, one obtains the recurrence relations
\besubeqs
    \begin{align*}
        0&=\beta_{0k}+\tfrac{2k(k+2)}{k+3} \,, \\
        0&=\tfrac{(n+2)(k+4)}{(k+1)(k+3)}\tfrac{\beta_{(n+1)(k+1)}}{2}-\tfrac{\beta_{nk}}{2} \,, \\
        0&=\tfrac{(k-n)(k+4)}{(k+1)(k+3)}\tfrac{\beta_{n(k+1)}}{2}+\tfrac{(n+2)(k+4)}{(k+1)(k+3)}\tfrac{\gamma_{n(k+1)}}{2}-\tfrac{\beta_{nk}+\gamma_{(n-1)k}}{2} \,, \\
        0&=\tfrac{(k-n)(k-n+2)}{(k+1)(k+3)}\tfrac{\beta_{n(k+1)}}{2}-\tfrac{(n+2)(k-n+2)}{(k+1)(k+3)}\tfrac{\gamma_{n(k+1)}}{2}-\tfrac{\beta_{nk}}{2} \,, \\
        0&=\tfrac{(k-n-1)(n+1)}{(k+1)(k+3)}\tfrac{\beta_{(n+1)(k+1)}}{2}-\tfrac{(n+3)(n+1)}{(k+1)(k+3)}\tfrac{\gamma_{(n+1)(k+1)}}{2}+\tfrac{\gamma_{nk}}{2} \,, \\
        0&=\tfrac{(k-n-1)(k+4)}{(k+1)(k+3)}\tfrac{\gamma_{n(k+1)}}{2}-\tfrac{\gamma_{nk}}{2} \,.
    \end{align*}
\esubeqs
The system is solved by
\begin{align*}
    \beta_{nk}&=-\tfrac{2}{(n+1)!}\tfrac{k-n+2}{k+3}\tfrac{k!}{(k-n-1)!} \,, & \gamma_{nk}&=\tfrac{2}{(n+2)!}\tfrac{n+1}{k+3}\tfrac{k!}{(k-n-2)!} \,.
\end{align*}

\subsection{Absence of higher order corrections} \label{app:sdymTruncation}
In section \ref{sec:flatSDYM} it was mentioned that the obtained solutions for $DF_{A(k+2),A'(k)}$ and $D \Psi_{A(k),A'(k+2)}$ ensured that no higher order corrections were needed. This result is equivalent to the consistency of the $L_\infty$-relation in \eqref{SDYMstasheffF3} and \eqref{SDYMstafhessPsi3}. Here we shall present the proof.

\paragraph{$\boldsymbol{F}$-sector.} 
As a starting point we take the solution from \eqref{spin1sol} and plug it into \eqref{DsquaredF}, from which we only consider only the cubic terms. This gives us the l.h.s. of the $L_\infty$-relation \eqref{SDYMstasheffF3}:
\begin{align*}
\begin{aligned}
    &l_3(e,l_3(e,F,F),F)+l_3(e,F,l_3(e,F,F))=\\
    &-\tfrac{1}{2}H_{A'A'}\sum_{n=1}^{k-1}\sum_{m=0}^{n-1}\tfrac{n-m+1}{n+2}\alpha_{nk}\alpha_{mn}[F_{A(k-n+1),A'(k-n-1)},[F_{A(m+1)B,A'(m)},F\fdud{A(n-m)}{B}{,A'(n-m-1)}]]\\
    &+\tfrac{1}{2}H_{A'A'}\sum_{n=0}^{k-2}\sum_{m=0}^{k-n-2}\alpha_{nk}\alpha_{m(k-n-1)}[F\fdud{A(n+1)}{B}{,A'(n)},[F_{A(m+1)B,A'(m)},F_{A(k-n-m),A'(k-n-m-2)}]]\\
    &=-\tfrac{1}{2}H_{A'A'}\sum_{n=1}^{k-1}\sum_{m=0}^{n-1}\tfrac{n-m+1}{n+2}\alpha_{nk}\alpha_{mn}[F\fdud{A(m+1)}{B}{,A'(m)},[F_{A(n-m)B,A'(n-m-1)},F_{A(k-n+1),A'(k-n-1)}]]\\
    &-\tfrac{1}{2}H_{A'A'}\sum_{n=1}^{k-1}\sum_{m=0}^{n-1}\tfrac{n-m+1}{n+2}\alpha_{nk}\alpha_{mn}[F\fdud{A(n-m)}{B}{,A'(n-m-1)},[F_{A(m+1)B,A'(m)},F_{A(k-n+1),A'(k-n-1)}]]\\
    &+\tfrac{1}{2}H_{A'A'}\sum_{n=0}^{k-2}\sum_{m=0}^{k-n-2}\alpha_{nk}\alpha_{m(k-n-1)}[F\fdud{A(n+1)}{B}{,A'(n)},[F_{A(m+1)B,A'(m)},F_{A(k-n-m),A'(k-n-m-2)}]]\,,
\end{aligned}
\end{align*}
where we applied the Jacobi identity on the very first term. In order to compare the three terms on the r.h.s., the nested commutators must be cast into the same form, which can be achieved by renaming the dummy indices. The final result allows all terms to be collected into one and evaluates to
\begin{align*}
\begin{aligned}
    &\tfrac{1}{2}H_{A'A'}\sum_{n=0}^{k-2}(\sum_{m=0}^{n}\alpha_{mk}\alpha_{(n-m)(k-m-1)}-\tfrac{m+2}{n+3}\alpha_{(n+1)k}\alpha_{(n-m)(n+1)}-\tfrac{n-m+2}{n+3}\alpha_{(n+1)k}\alpha_{m(n+1)})\\
    &\times[F\fdud{A(m+1)}{B}{,A'(m)},[F_{A(n-m+1)B,A'(n-m)},F_{A(k-n),A'(k-n-2)}]]=0\,,
\end{aligned}
\end{align*}
for which the solution for $\alpha_{nk}$ was used. This proves the $L_\infty$-relation \eqref{SDYMstasheffF3}.

\paragraph{$\boldsymbol{\Psi}$-sector.}
We isolate the terms cubic in the fields in \eqref{DsquaredPsi} and we plug in \eqref{spin1sol} and \eqref{solpsi}, which yields the l.h.s. of $L_\infty$-relation \eqref{SDYMstafhessPsi3} and reads{\allowdisplaybreaks
\begin{align*}
    \begin{aligned}
        &l_3(e,l_3(e,F,F),\Psi)+l_3(e,F,l_3(e,F,\Psi))=\\
        &H_{A'A'}\sum_{n=1}^{k-1}\sum_{m=0}^{n-1}\tfrac{\alpha_{mn}\beta_{nk}}{2}\tfrac{n-m+1}{n+2}[[F_{A(m+1)B,A'(m)},F\fdud{A(n-m)}{B}{,A'(n-m-1)}],\Psi_{A(k-n-1),A'(k-n+1)}]\\
        &+H_{A'A'}\sum_{n=0}^{k-2}\sum_{m=0}^{k-n-2}\tfrac{\beta_{m(k-n-1)}\beta_{nk}}{2}[F\fdud{(n+1)}{B}{,A'(n)},[F_{A(m+1)B,A'(m)},\Psi_{A(k-n-m-2),A'(k-n-m))}]]\\
        &+H_{A'A'}\sum_{n=0}^{k-3}\sum_{m=0}^{k-n-3}\tfrac{\beta_{nk}\gamma_{m(k-n-1)}}{2}[F\fdud{A(n+1)}{B}{,A'(n)},[F_{A(m+2),A'(m)},\Psi_{A(k-n-m-3)B,A'(k-n-m)}]]\\
        &+H_{A'A'}\sum_{n=1}^{k-2}\sum_{m=0}^{n-1}\tfrac{\alpha_{mn}\gamma_{nk}}{2}[[F_{A(m+1)B,A'(m)},F_{A(n-m+1),A'(n-m-1)}],\Psi\fdud{A(k-n-2)}{B}{,A'(k-n+1)}]\\
        &-H_{A'A'}\sum_{n=0}^{k-2}\sum_{m=0}^{k-n-3}(\tfrac{\beta_{m(k-n-1)\gamma_{nk}}}{2}\tfrac{k-n-m-2}{k-n-1}-\tfrac{\gamma_{m(k-n-1)\gamma_{nk}}}{2}\tfrac{m+2}{k-n-1})\\
        &     \times[F_{A(n+2),A'(n)},[F\fdud{A(m+1)}{B}{,A'(m)},\Psi_{A(k-n-m-3)B,A'(k-n-m)}]].
    \end{aligned}
\end{align*}}\noindent
Our approach is similar to the one for the $F$-sector: we aim to reduce the equations as much as possible by casting the nested commutators into a similar form. A particular technicality in this case is that a contraction can be either between two $F$'s or between $F$ and $\Psi$. The Fierz identity is used to convert all contractions into the latter type. However, one must be careful, as the Fierz identity requires some free indices on the available spinors, which might not be present in all terms of the summation. Hence we isolate these cases and check that their contribution vanishes.
\begin{align*}
    \begin{aligned}
        &H_{A'A'}\sum_{n=0}^{k-2}\tfrac{\beta_{(k-1)k}\alpha_{n(k-1)}}{2}\tfrac{k-n}{k+1}[[F_{A(n+1)B,A'(n)},F\fdud{A(k-n-1)}{B}{,A'(k-n-2)}],\Psi_{A'A'}]\\
        &+H_{A'A'}\sum_{n=0}^{k-2}\tfrac{\beta_{nk}\beta_{(k-n-2)(k-n-1)}}{2}[F\fdud{A(n+1)}{B}{,A'(n)},[F_{A(k-n-1)B,A'(k-n-2)},\Psi_{A'A'}]]\\
        &=H_{A'A'}\sum_{n=0}^{k-2}(-\tfrac{\beta_{(k-1)k}\alpha_{n(k-1)}}{2}\tfrac{k-n}{k+1}+\tfrac{\beta_{nk}\beta_{(k-n-2)(k-n-1)}}{2}-\tfrac{\beta_{(k-1)k}\alpha_{(k-n-2)(k-1)}}{2}\tfrac{n+2}{k+1})\\
        &   \times[F\fdud{A(n+1)}{B}{,A'(n)},[F_{A(k-n-1)B,A'(k-n-2)},\Psi_{A'A'}]]=0\,,
    \end{aligned}
\end{align*}
where we used the solution for $\beta_{nk}$ and $\alpha_{nk}$. Finally, applying the Fierz identity, Jacobi identity and renaming of dummy indices allows one to cast the remaining terms into a more practical form that reads
\begin{align*}
    \begin{aligned}
        &H_{A'A'}\sum_{ n=0}^{k-3}\sum_{m=0}^{n}(\tfrac{\beta_{(n+1)k}\alpha_{m(n+1)}}{2}\tfrac{n-m+2}{n+3}+\tfrac{\beta_{(n+1)k}\alpha_{(n-m)(n+1)}}{2}\tfrac{m+2}{n+3}+\tfrac{\beta_{mk}\beta_{(n-m)(k-m-1)}}{2}\\
        &+\tfrac{\beta_{mk}\gamma_{(n-m)(k-m-1)}}{2}-\tfrac{\gamma_{(n+1)k}\alpha_{m(n+1)}}{2})[F\fdud{A(m+1)}{B}{,A'(m)},[F_{A(n-m+2),A'(n-m)},\Psi_{A(k-n-3)B,A'(k-n)}]]\\
        &+H_{A'A'}\sum_{n=0}^{k-3}\sum_{m=0}^{n}(\tfrac{\beta_{(n+1)k}\alpha_{(n-m)(n+1)}}{2}\tfrac{m+2}{n+3}+\tfrac{\beta_{(n+1)k}\alpha_{m(n+1)}}{2}\tfrac{n-m+2}{n+3}-\tfrac{\beta_{mk}\beta_{(n-m)(k-m-1)}}{2}\\
        &+\tfrac{\gamma_{(n+1)k}\alpha_{(n-m)(n+1)}}{2}-\tfrac{\gamma_{mk}\beta_{(n-m)(k-m-1)}}{2}\tfrac{k-n-2}{k-m-1}+\tfrac{\gamma_{mk}\gamma_{(n-m)(k-m-1)}}{2}\tfrac{n-m+2}{k-m-1})\\
        &\times[F_{A(m+2),A'(m)},[F\fdud{A(n-m+1)}{B}{,A'(n-m)},\Psi_{A(k-n-3)B,A'(k-n)}]]=0 \,,
    \end{aligned}
\end{align*}
which is obtained by plugging in the solutions for $\alpha_{nk}$, $\beta_{nk}$ and $\gamma_{nk}$ were used. This implies the consistency of $L_\infty$-relation \eqref{SDYMstafhessPsi3}.

\subsection{Higher gravitational corrections} \label{app:sdymLambda}
In section \ref{sec:SDYMcurved} we mentioned that the correction due to the constant gravitational background to the linear term in $D F_{A(k+2),A'(k)}$ and $D\Psi_{A(k),A'(k+2)}$ does not propagate to the quadratic term or higher. This appendix is dedicated to prove this.

\paragraph{$\boldsymbol{F}$-sector.}
The $L_\infty$-relations are modified on a constant curvature background according to \eqref{SDYMstasheffLambda}. It was mentioned in \eqref{SDYMstasheffLambda} that the gravitational contribution decouples and vanishes independently. We shall present a proof here.

We are interested in checking consistency of $L_\infty$-relation \eqref{stasheffLambdaF}. We do so by taking the covariant derivative of \eqref{curvedFsol}, which gives
\begin{align} \label{DsquaredFLambda}
    \begin{aligned}
        D^2F_{A(k+2),A'(k)}&=-H^{BB}[F_{BB},F_{A(k+2),A'(k)}]+(k+2)H\fdu{A}{B}F_{A(k+1)B,A'(k)}\\
        &+k H\fdu{A'}{B'}F_{A(k+2),A'(k-1)B'}=-e^{BB'}\wedge DF_{A(k+2)B,A'(k)B'}\\
        &-e\fud{B}{A'}\wedge\sum_{n=0}^{k-1}\alpha_{nk}[DF_{A(n+1)B,A'(n)},F_{A(k-n+1),A'(k-n-1)}] \\
        &-e\fud{B}{A'}\wedge\sum_{n=0}^{k-1}\alpha_{nk}[F_{A(n+1)B,A'(n)},DF_{A(k-n+1),A'(k-n-1)}]\\
        &-k(k+2)e_{AA'}DF_{A(k+1),A'(k-1)}\,.
    \end{aligned}
\end{align}
Considering only the terms coming from gravitational contributions gives the l.h.s of $L_\infty$-relation \eqref{stasheffLambdaF} and reads after introducing $f_{k}=k(k+2)$:
\begin{align*}
    \begin{aligned}
        &\tilde{l}_2(e,l_3(e,F,F))+l_3(e,\tilde{l}_2(e,F),F)+l_3(e,F,\tilde{l}_2(e,F))=\\
        &H_{A'A'}\sum_{n=0}^{k-2}\tfrac{\tfrac{1}{2}n+2}{n+3}f_{n+1}\alpha_{(n+1)k}[F_{A(n+2),A'(n)},F_{A(k-n),A'(k-n-2)}]\\
        &+H_{A'A'}\sum_{n=0}^{k-2}\tfrac{\alpha_{nk}}{2}f_{k-n-1}[F_{A(n+2),A'(n)},F_{A(k-n),A'(k-n-2)}]\\
        &-H_{A'A'}\sum_{n=0}^{k-2}\tfrac{\alpha_{n(k-1)}}{2}f_k[F_{A(n+2),A'(n)},F_{A(k-n),A'(k-n-2)}]\\
        &= \sum_{n=0}^{k-2}\tfrac{1}{2}H_{A'A'}(\tfrac{\tfrac{1}{2}n+2}{n+3}f_{n+1}\alpha_{(n+1)k}-\tfrac{\tfrac{1}{2}(k-n)+1}{k-n+1}f_{k-n-1}\alpha_{(k-n-1)k}+\tfrac{\alpha_{nk}}{2}f_{k-n-1}-\tfrac{\alpha_{(k-n-2)k}}{2}f_{n+1}\\
        &-\tfrac{\alpha_{n(k-1)}}{2}f_k+\tfrac{\alpha_{(k-n-2)(k-1)}}{2}f_k)[F_{A(n+2),A'(n)},F_{A(k-n),A'(k-n-2)}]=0\,,
    \end{aligned}
\end{align*}
where the anti-symmetry of the commutator has been made explicit and the solution for $\alpha_{nk}$ was applied. Thus, the modification to the second $L_\infty$-relation of the $F$-sector vanishes, which means that the gravitational background only modifies $DF_{A(k+2),A'(k)}$ on the linear level, identically to the free equations. This is equivalent to the consistency of \eqref{stasheffLambdaF}.

\paragraph{$\boldsymbol{\Psi}$-sector.}
The second $L_\infty$-relation for $\Psi$ on a gravitational background is modified according to \eqref{stasheffLambdaPsi}. This gives
\begin{align} \label{DsquaredPsiLambda}
    \begin{aligned}
        D^2\Psi_{A(k),A'(k+2)}=&-H^{BB}[F_{BB},\Psi_{A(k),A'(k+2)}]+kH\fdu{A}{B}\Psi_{A(k-1)B,A'(k+2)}\\
        &+(k+2)H\fdu{A'}{B'}\Psi_{A(k),A'(k+1)B'}=-e^{CC'}\wedge D\Psi_{A(k)C,A'(k+2)C'} \\
        &-e\fud{C}{A'}\wedge\sum_{n=0}^{k-1}\beta_{nk}[DF_{A(n+1)C,A'(n)},\Psi_{A(k-n-1),A'(k-n+1)}] \\
        &-e\fud{C}{A'}\wedge\sum_{n=0}^{k-1}\beta_{nk}[F_{A(n+1)C,A'(n)},D\Psi_{A(k-n-1),A'(k-n+1)}] \\
        &-e\fud{C}{A'}\wedge\sum_{n=0}^{k-1}\gamma_{nk}[DF_{A(n+2),A'(n)},\Psi_{A(k-n-2)C,A'(k-n+1)}] \\
        &-e\fud{C}{A'}\sum_{n=0}^{k-1}\gamma_{nk}[F_{A(n+2),A'(n)},D\Psi_{A(k-n-2),A'(k-n+1)}] \\
        &-k(k+2)e_{AA'}D\Psi_{A(k-1),A'(k+1)}\,.
    \end{aligned}
\end{align}
Considering only the terms containing a gravitational contribution, one obtains the l.h.s. of the $L_\infty$-relation \eqref{stasheffLambdaPsi}:
\begin{align}
    \begin{aligned}
         &\tilde{l}_2(e,l_3(e,F,\Psi))+l_3(e,\tilde{l}_2(e,F),\Psi)+l_3(e,F,\tilde{l}_2(e,\Psi))\\
         &=H_{A'A'}\sum_{n=0}^{k-2}(f_{n+1}\beta_{(n+1)k}\tfrac{\tfrac{1}{2}n+2}{n+3}+\tfrac{1}{2}\beta_{nk}f_{k-n-1}+\tfrac{1}{2}f_{n+1}\gamma_{(n+1)k}\\
        &+\tfrac{\tfrac{1}{2}k-\tfrac{1}{2}n}{k-n-1}\gamma_{nk}f_{k-n-1}-f_k\tfrac{\beta_{n(k-1)}}{2}-f_k\tfrac{\gamma_{n(k-1)}}{2})[F_{A(n+2),A'(n)},\Psi_{A(k-n-2),A'(k-n)}]=0\,,
    \end{aligned}
\end{align}
which we obtain by plugging in the solutions for $\alpha_{nk}$, $\beta_{nk}$ and $\gamma_{nk}$. This proves the consistency of \eqref{stasheffLambdaPsi}.

The results in this appendix prove that the gravitational contribution to the $L_\infty$-relations in both sectors decouples and vanishes independently, which is equivalent to consistency of \eqref{stasheffLambdaF} and \eqref{stasheffLambdaPsi}. Thus, the gravitational background only modifies
$DF_{A(k+2),A'(k)}$ and $D\Psi_{A(k),A'(k+2)}$ on the linear level, identically to the free equations.

\section{Technicalities: SDGR}
Several technicalities have been left out from the main text. In this section we aim to present the calculation of the $\Psi$-sector, as well as the proofs of the truncation of $\nabla C$ and $\nabla\Psi$, as promised in section \ref{sec:SDGRflat}
\subsection{\texorpdfstring{$\Psi$}{psi}-sector} \label{app:SDGRPsi}
We have left the details of the calculation of the $\Psi$-sector of section \ref{sec:SDGRflat} to this appendix, as it bears a lot of resemblance to the $C$-sector.

The approach is similar to before: we take the covariant derivative of the ansatz \eqref{eq:SDGRPsiansatz} and we also contract $e^{BB'}$ with $\nabla\Psi_{A(k+1),A'(k+5)}$ as this will give two expressions for $e^{BB'}\wedge\nabla\Psi_{A(k)B,A'(k+4)B'}$, so we can compare them. This will unveil its structure. The former yields{\allowdisplaybreaks
\begin{align}\label{nablasquaredPsi}
    \begin{aligned}
        \nabla^2\Psi_{A(k),A'(k+4)}&=kH^{BB}C\fdu{ABB}{D}\Psi_{A(k-1)D,A'(k+4)}=-e^{CC'}\wedge\nabla\Psi_{A(k)C,A'(k+4)C'}\\
        &-e\fud{C}{A'}\wedge\sum_{n=0}^{k}b_{nk}\nabla C\fdud{A(n+2)C}{D}{,A'(n)}\Psi_{A(k-n-2)D,A'(k-n+3)}\\
        &-e\fud{C}{A'}\wedge\sum_{n=0}^{k}b_{nk} C\fdud{A(n+2)C}{D}{,A'(n)}\nabla\Psi_{A(k-n-2)D,A'(k-n+3)}\\
        &-e\fud{C}{A'}\wedge\sum_{n=0}^{k}c_{nk}\nabla C\fdud{A(n+3)}{D}{,A'(n)}\Psi_{A(k-n-3)CD,A'(k-n+3)}\\
        &-e\fud{C}{A'}\wedge\sum_{n=0}^{k}c_{nk} C\fdud{A(n+3)}{D}{,A'(n)}\nabla\Psi_{A(k-n-3)CD,A'(k-n+3)} \,.
    \end{aligned}
\end{align}}\noindent
Isolating the terms quadratic in the fields gives
\begin{align*}
    \begin{aligned}
        e^{BB'}\wedge\nabla\Psi_{A(k)C,A'(k+4)C'}&= -kH^{BB}C\fdu{ABB}{D}\Psi_{A(k-1)D,A'(k+4)}\\
        &-\frac{1}{2}H^{BB}\sum_{n=0}^{k}b_{nk}C\fdud{A(n+2)BB}{D}{,A'(n+1)}\Psi_{A(k-n-2)D,A'(k-n+3)}\\
        &-\frac{1}{2}H^{BB}\sum_{n=0}^{k}(b_{nk}+c_{(n-1)k})C\fdud{A(n+2)B}{D}{,A'(n)}\Psi_{A(k-n-2)BD,A'(k-n+4)}\\
        &+\frac{1}{2}H\fdu{A'}{B'}\sum_{n=0}^{k}b_{nk}C\fdud{A(n+2)}{BD}{,A'(n)}\Psi_{A(k-n-2)BD,A'(k-n+3)B'}\\
        &-\frac{1}{2}H\fdu{A'}{B'}\sum_{n=0}^{k}c_{nk}C\fdud{A(n+3)}{BD}{,A'(n)B'}\Psi_{A(k-n-3)BD,A'(k-n+3)}\\
        &-\frac{1}{2}H^{BB}\sum_{n=0}^{k}c_{nk}C\fdud{A(n+3)}{D}{,A'(n)}\Psi_{A(k-n-3)BBD,A'(k-n+4)} \,,
    \end{aligned}
\end{align*}
whereas the latter gives
{\allowdisplaybreaks
\begin{align*}
        &e^{BB'}\wedge\nabla\Psi_{A(k)B,A'(k+4)B'}=-H^{BB}b_{0(k+1)}\tfrac{k+6}{(k+1)(k+5)}C\fdu{ABB}{D}\Psi_{A(k-1)D,A'(k+4)}\\
        &-\frac{1}{2}H^{BB}\sum_{n=0}^kb_{(n+1)(k+1)}\tfrac{(k+6)(n+3)}{(k+1)(k+5)}C\fdud{A(n+2)BB}{D}{,A'(n+1)}\Psi_{A(k-n-2)D,A'(k-n+3)}\\
        &-\tfrac{1}{2}H^{BB}\sum_{n=0}^{k}(b_{n(k+1)}\tfrac{(k+6)(k-n-1)}{(k+1)(k+5)}+c_{n(k+1)}\tfrac{(k+6)(n+3)}{(k+1)(k+5)})\\
        &\times C\fdud{A(n+2)B}{D}{,A'(n)}\Psi_{A(k-n-2)BD,A'(k-n+4)}\\
        &+\frac{1}{2}H\fdu{A'}{B'}\sum_{n=0}^{k}(b_{(n+1)(k+1)}\tfrac{(n+1)(k-n-2)}{(k+1)(k+5)}-c_{(n+1)(k+1)}\tfrac{(n+1)(n+4)}{(k+1)(k+5)})\\
        &\times C\fdud{A(n+3)}{BD}{,A'(n)B'}\Psi_{A(k-n-3)BD,A'(k-n+3)}\\
        &+\frac{1}{2}H\fdu{A'}{B'}\sum_{n=0}^{k}(b_{n(k+1)}\tfrac{(k-n+4)(k-n-1)}{(k+1)(k+5)}-c_{n(k+1)}\tfrac{(k-n+4)(n+3)}{(k+1)(k+5)})\\
        &\times C\fdud{A(n+2)}{BD}{,A'(n)}\Psi_{A(k-n-2)BD,A'(k-n+3)B'}\\
        &-\frac{1}{2}H^{BB}\sum_{n=0}^{k}c_{n(k+1)}\tfrac{(k+6)(k-n-2)}{(k+1)(k+5)}C\fdud{A(n+3)}{D}{,A'(n)}\Psi_{A(k-n-3)BBD,A'(k-n+4)} \,.
\end{align*}}\noindent
Comparing them gives the system of recurrence relations
{\allowdisplaybreaks
\besubeqs
    \begin{align*}
        0&=\tfrac{k+6}{(k+1)(k+5)}b_{0(k+1)}-k\,,\\
        0&=b_{nk}-b_{(n+1)(k+1)}\tfrac{(k+6)(n+3)}{(k+1)(k+5)}\,,\\
        0&=b_{nk}+c_{nk}-b_{n(k+1)}\tfrac{(k+6)(k-n-1)}{(k+1)(k+5)}-c_{n(k+1)}\tfrac{(k+6)(n+3)}{(k+1)(k+5)}\,,\\
        0&=b_{nk}-b_{n(k+1)}\tfrac{(k-n+4)(k-n-1)}{(k+1)(k+5)}+c_{n(k+1)}\tfrac{(k-n+4)(n+3)}{(k+1)(k+5)}\,,\\
        0&=c_{nk}+b_{(n+1)(k+1)}\tfrac{(n+1)(k-n-2)}{(k+1)(k+5)}-c_{(n+1)(k+1)}\tfrac{(n+1)(n+4)}{(k+1)(k+5)}\,,\\
        0&=c_{nk}-c_{n(k+1)}\tfrac{(k+6)(k-n-2)}{(k+1)(k+5)}\,,
    \end{align*}
\esubeqs}\noindent
which is solved by
    \begin{align*}
        b_{nk}&=\tfrac{2}{(n+2)!}\tfrac{k!}{(k-n-2)!}\tfrac{k-n+4}{k+5}&
        c_{nk}&=-\tfrac{2}{(n+2)!}\tfrac{k!}{(k-n-3)!}\tfrac{n+1}{(k+5)(n+3)}\,.
    \end{align*}
\subsection{Absense of higher order corrections} \label{app:sdgrTruncation}
\paragraph{$\boldsymbol{C}$-sector.} We consider \eqref{nablasquaredC} and isolate the terms cubic in $C$. Plugging in the solution from \eqref{Csol} yields the l.h.s. of the $L_\infty$-relation \eqref{stasheffC3} given by
{\allowdisplaybreaks
\begin{align} \label{Ccubed}
    \begin{aligned}
        &l_3(e,l_3(e,C,C),C)+l_3(e,C,l_3(e,C,C))=\\
        &\tfrac{1}{2}H_{A'A'}\sum_{n=1}^{k-1}\sum_{m=0}^{n-1}a_{nk}a_{mn}\tfrac{n-m+2}{n+4}\tfrac{m+2}{n+3} C\fdud{A(m+1)B}{DE}{,A'(m)}C\fdud{A(n-m+1)E}{B}{,A'(n-m-1)}C_{A(k-n+2)D,A'(k-n-1)}\\
        &+\tfrac{1}{2}H_{A'A'}\sum_{n=1}^{k-1}\sum_{m=0}^{n-1}a_{nk}a_{mn}\tfrac{n-m+2}{n+4}\tfrac{n-m+1}{n+3}\\
        & \times C\fdud{A(m+2)B}{E}{,A'(m)}C\fdud{A(n-m)E}{BD}{,A'(n-m-1)}C_{A(k-n+2)D,A'(k-n-1)}\\
        &+\tfrac{1}{2}H_{A'A'}\sum_{n=0}^{k-2}\sum_{m=0}^{k-n-2}a_{nk}a_{m(k-n-1)}\tfrac{m+2}{k-n+3}\\
        &  C\fdud{A(n+2)}{BD}{,A'(n)}C\fdud{A(m+1)BD}{E}{,A'(m)}C_{A(k-n-m+1)E,A'(k-n-m-2)}\\
        &+\tfrac{1}{2}H_{A'A'}\sum_{n=0}^{k-2}\sum_{m=0}^{k-n-2}a_{nk}a_{m(k-n-1)}\tfrac{k-n-m+1}{k-n+3}\\
        &\times C\fdud{A(n+2)}{BD}{,A'(n)}C\fdud{A(m+2)B}{E}{,A'(m)}C_{A(k-n-m)DE,A'(k-n-m-2)}\,.
    \end{aligned}
\end{align}}\noindent
The first three terms can be collected into
\begin{align*}
    \begin{aligned}
        &\tfrac{1}{2}H_{A'A'}\sum_{n=1}^{k-1}\sum_{m=0}^{n-1}(a_{nk}a_{mn}\tfrac{n-m+2}{n+4}\tfrac{m+2}{n+3}+a_{nk}a_{(n-m-1)n}\tfrac{m+3}{n+4}\tfrac{m+2}{n+3}-a_{(n-m-1)k}a_{m(k-n+m)}\tfrac{m+2}{k-n+m+4})\\
        &\times C\fdud{A(m+1)B}{DE}{,A'(m)}C\fdud{A(n-m+1)}{B}{,A'(n-m-1)}C_{A(k-n+2)D,A'(k-n-1)}=0\,,
    \end{aligned}
\end{align*}
for which the solution for $a_{nk}$ is applied. The last term in \eqref{Ccubed} may be rewritten as
\allowdisplaybreaks{
\begin{align*}
    \begin{aligned}
        &\tfrac{1}{2}H_{A'A'}\sum_{n=0}^{k-2}\sum_{m=0}^{n}a_{mk}a_{(n-m)(k-m-1)}\tfrac{k-n+1}{k-m+3} C\fdud{A(m+2)}{BD}{,A'(m)}C\fdud{A(n-m+2)B}{E}{,A'(m)}C_{A(k-n)DE,A'(k-n-2)}\\
        &=\tfrac{1}{4}H_{A'A'}\sum_{n=0}^{k-2}\sum_{m=0}^{n}(a_{mk}a_{(n-m)(k-m-1)}\tfrac{k-n+1}{k-m+3}-a_{(n-m)k}a_{m(k-n+m-1)}\tfrac{k-n+1}{k-n+m+3})\\
        &\times C\fdud{A(m+2)}{BD}{,A'(m)}C\fdud{A(n-m+2)B}{E}{,A'(m)}C_{A(k-n)DE,A'(k-n-2)}=0\,,
    \end{aligned}
\end{align*}}\noindent
where again we used the solution for $a_{nk}$. This proves that the $C$-sector truncates at quadratic order. This confirms the consistency of \eqref{stasheffC3}.

\paragraph{$\boldsymbol{\Psi}$-sector.}
We consider the cubic terms in \eqref{nablasquaredPsi} and we assume the solutions \eqref{Csol} and \eqref{psisolSDGR}. This gives the l.h.s. of the $L_\infty$-relation in \eqref{stasheffSDGRPsi3} and reads
{\allowdisplaybreaks
\begin{align*}
        &l_3(e,l_3(e,C,C),\Psi)+l_3(e,C,l_3(e,C,\Psi))=\\
        &\tfrac{1}{2}H_{A'A'}\sum_{n=0}^{k-3}\sum_{m=0}^{n}(a_{m(n+1)}b_{(n+1)k}\tfrac{(n-m+3)(m+2)}{(n+5)(n+4)}+a_{(n-m)(n+1)}b_{(n+1)k}\tfrac{(m+3)(m+2)}{(n+5)(n+4)}\\
        &-b_{m(k-n+m-1)}b_{(n-m)k}\tfrac{m+2}{k-n+m-1})\\
        &\times C\fdud{A(m+1)B}{DE}{,A'(m)}C\fdud{A(n-m+2)E}{B}{,A'(n-m)}\Psi_{A(k-n-3)D,A'(k-n+2)}\\
        &+\tfrac{1}{4}H_{A'A'}\sum_{n=0}^{k-4}\sum_{m=0}^{n}(b_{(n-m)(k-m-1)}b_{mk}\tfrac{k-n-3}{k-m-1}+c_{(n-m)(k-m-1)}b_{mk}\tfrac{n-m+3}{k-m-1}\\
        &-a_{m(n+1)}c_{(n+1)k}\tfrac{n-m+3}{n+5}-b_{m(k-n+m-1)}b_{(n-m)k}\tfrac{k-n-3}{k-n+m-1}\\
        &-c_{m(k-n+m-1)}b_{(n-m)k}\tfrac{m+3}{k-n+m-1}+a_{(n-m)(n+1)}c_{(n+1)k}\tfrac{m+3}{n+5})\\
        &\times C\fdud{A(m+2)}{BD}{,A'(m)}C\fdud{A(n-m+2)B}{E}{,A'(n-m)}\Psi_{A(k-n-4)DE,A'(k-n+2)}\\
        &+\tfrac{1}{2}H_{A'A'}\sum_{n=0}^{k-4}\sum_{m=n}^{n}(b_{(n-m)(k-m-1)}c_{mk}\tfrac{(k-n-3)(n-m+2)}{(k-m-1)(k-m-2)}\\
        &-c_{(n-m)(k-m-1)}c_{mk}\tfrac{(n-m+3)(n-m+2)}{(k-m-1)(k-m-2)}-a_{(n-m)(n+1)}c_{(n+1)k}\tfrac{n-m+2}{n+5})\\
        &\times C\fdud{A(m+3)}{D}{,A'(m)}C\fdud{A(n-m+1)BD}{E}{,A'(n-m)}\Psi\fdud{A(k-n-4)E}{B}{,A'(k-n+2)}\\
        &+\tfrac{1}{2}H_{A'A'}\sum_{n=0}^{k-5}\sum_{m=0}^{n}(c_{m(k-n+m-1)}b_{(n-m)k}\tfrac{k-n-4}{k-n+m-1}\\
        &-b_{(n-m)(k-m-1)}c_{mk}\tfrac{(k-n-3)(k-n-4)}{(k-m-1)(k-m-2)}+c_{(n-m)(k-m-1)}c_{mk}\tfrac{(n-m+3)(k-n-4)}{(k-m-1)(k-m-2)})\\
        &C\fdud{A(m+3)}{C}{,A'(m)}C\fdud{A(n-m+2)}{BD}{,A'(n-m)}\Psi_{A(k-n-5)BDE,A'(k-n+2)}=0\,,
\end{align*}}\noindent
which is obtained by plugging in the results for $a_{nk}$, $b_{nk}$ and $c_{nk}$. This proves consistency of \eqref{stasheffSDGRPsi3}.

\chapter{Cubic interactions}
\label{app:cubic}

\section{Cubic Amplitude}
\label{app:amplitude}
A useful check for a given interaction is to compute the amplitude. The amplitudes of Chiral HiSGRA are known up to one-loop \cite{Skvortsov:2018jea,Skvortsov:2020wtf,Skvortsov:2020gpn}. We do not have to go that far and should just check if the cubic amplitude is nontrivial. Let us first construct the plane wave solutions. We recall that the free equations in Minkowski space read
\besubeqs\label{linearizeddataA}
\begin{align}
    d\omega &= e^{BB'}\bry_{B'} \pl_{B} \omega +H^{BB} \pl_{B}\pl_{B}C(y,\bry=0)\,, &
    d C&= e^{BB'}\pl_B \pl_{B'} C\,,
\end{align}
\esubeqs
where $\Psi(0,\bry)=C(0,\bry)$ describes negative helicity and $\omega(0,\bry)$ describes positive helicity:\footnote{Note that we use the Moyal-Weyl star-product without $i$. Therefore, the fields need to obey less natural reality conditions. This is not an obstacle to compute the amplitude. In particular, the plane wave exponents are taken without $i$ (for appropriate $x$). What matters is the helicity structure. }
\begin{align}
    \Psi^{A'(2s)}&= a_{-s}\, k^{A'}...k^{A'} \exp{[\pm x^{AA'} k_A k_{A'}]}\,,\\
    \omega^{A'(2s-2)}&= a_{+s}\, \frac{1}{(q^{C'}k_{C'})^{2s-1}} e^{BB'} k_B q_{B'} q^{A'} ...q^{A'}\exp{[\pm x^{AA'} k_A k_{A'}]}\,.
\end{align}
Here $a_\lambda$ is a normalization factor. Eq. \eqref{linearizeddataA} is solved by
\begin{align*}
    \omega(x|y,\bry)&=e^{BB'}\frac{k_B q_B'}{\bar{q}\bar{k}+\bar{y}\bar{q}}\exp(\pm x^{AA'}k_Ak_A'+yk)\,, & C(x|y,\bry)=\frac{1}{2}\exp(\pm x^{AA'}k_Ak_A'+yk+\bry\bar{k}) \,.
\end{align*}
Laplace transform allows us to rewrite the solution for $\omega(x|y,\bry)$ as
\begin{align*}
    \omega(x|,y,\bry)=e^{BB'}k_B q_B'\int_{0}^{\infty}d\omega\exp(\pm x^{AA'}k_Ak_A'+yk-(\bar{q}\bar{k}+\bar{y}\,\bar{q})\omega) \,.
\end{align*}
In order to compute  cubic amplitudes we can isolate the equation for $\omega^{A'(2s-2)}$ and $\Psi^{A'(2s)}$:
\begin{align} 
    D\omega&= \mathcal{V}(\omega, \omega)\Big|_{y=0}\,, &
    DC&= \mathcal{U}(\omega,C) \Big|_{y=0}\,.
\end{align}
Let us have a look at the first term $V(\omega, \omega)$ contracted with $\Psi^{A'(2s_1-2)}H_{A'A'}$ to get an on-shell cubic vertex:
\begin{align}
    \frac{1}{l!}\int \Psi_{A'(2s_1)}H^{A'A'}\, \omega^{B(l),A'(n)}\wedge \omega\fdu{B(l)}{,A'(m)}\,.
\end{align}
Here we assume $l+n=2s_2-2$, $m+l=2s_3-2$ and, of course, $m+n=2s_1-2$. The coefficient in front of the action originates from the star-product. Plugging in the on-shell plane-wave values for $\Psi$ and $\omega$ we find
\begin{align} \label{cubicVertex}
    V_{-s_1,+s_2,+s_3}&\sim \frac{1}{\Gamma[-s_1+s_2+s_3]} [12]^{-s_1+s_2-s_3}[23]^{s_2+s_3+s_1}[13]^{-s_1+s_3-s_2}\,,
\end{align}
which, up to normalization of each of the plane-waves, is the right structure for Chiral Theory. It corresponds to $\Psi\Phi\Phi$-vertex of sketch \eqref{sketch}. The presence of the simplest self-interaction $V_{-s,+s,+s}$ leads unambiguously to the Chiral Theory class since it requires all other spins (at least even) together with all other possible interactions that enter with weight $1/\Gamma[\lambda_1+\lambda_2+\lambda_3]$.

Similarly, we can extract the amplitudes corresponding to $\Phi\Phi\Phi$ and $\Phi\Phi\Psi$ vertices from $\mathcal{U}(\omega,C)$. Note that since $C$ contains both positive and negative (as well as zero) helicities, we get an access to two types of vertices. The final amplitude is
\begin{align*}
    V_{+s_1,\lambda_2,+s_3}&\sim 
       \frac{1}{\Gamma[s_1+\lambda_2+s_3]}[12]^{s_1+\lambda_2-s_3}[23]^{-s_1+\lambda_2+s_3}[13]^{s_1-\lambda_2+s_3}\,.
\end{align*}
Let us also comment on the possibility to reproduce $V_{+s_1,-s_2,-s_3}$ amplitudes, $s_1-s_2-s_3>0$. From the standard covariant approach vantage point, where the dynamical variables are $\Phi_{\mu_1...\mu_s}$, these vertices are the most problematic ones \cite{Conde:2016izb}. They cannot be written at all as local expressions. Fortunately, it is easy to write down the candidate on-shell cubic vertices in terms of the new variables, where the dynamical fields are $\omega^{A'(2s-2)}$ and $\Psi^{A'(2s)}$. For example, any of the following two expressions 
\begin{align*}
   &  \omega^{A'(s_1-2)}\Psi_{A(k)B,A'(m),B'}\Psi\fud{A(k)}{A'(n)}\hat{h}^{BB'}\,,  && \omega^{A'(s_1-2)}\Psi_{A(k)B,A'(m)}\Psi\fud{A(k)}{A'(n)B'}\hat{h}^{BB'},
\end{align*}
leads to the correct amplitude
\begin{align*}
    \begin{aligned}
       [12]^{s_1-s_2+s_3}[13]^{s_1+s_2-s_3}[23]^{-s_1-s_2-s_3}\,.
    \end{aligned}
\end{align*}
Therefore, all possible types of cubic vertices/amplitudes present in Chiral Theory can be written in a manifestly Lorentz invariant way. This eliminates the very last obstruction and we can claim that Chiral Theory admits a manifestly Lorentz invariant formulation.

\section{Coadjoint vs. twisted-adjoint}
\label{app:coadjoint}
Let us make a historical remark on representations of higher spin symmetries. It was known since \cite{Vasiliev:1986td} that the FDA of free massless fields in (anti)-de Sitter space contains the following subsystem
\begin{align}
    \nabla C&= e^{AA'}(y_A \bry_{A'} -\pl_A \pl_{A'}) C(y,\bry)\,.
\end{align}
It splits according to spin into an infinite set of (still infinite) subsystems. For a given $s>0$ the subsystem splits further into one for helicity $+s$ and another one for helicity $-s$. The very first equations in these subsystems are equivalent to \cite{Penrose:1965am}
\begin{align}
    \nabla^\fdu{B}{A'} C^{BA(2s-1)}&=0\,, &\nabla\fud{A}{B'} C^{B'A'(2s-1)}&=0\,.
\end{align}
Operator $P_{AA'}=(y_A \bry_{A'} -\pl_A \pl_{A'})$ realizes the action of $(A)dS_4$ translations, which commute to a Lorentz transformation. Since the equations are assumed to be derived by linearizing a nonlinear theory, where the higher spin symmetry is manifest, it is important to understand where such $P_{AA'}$ can come form. It originates from the twisted-adjoint action \cite{Vasiliev:1999ba}:
\begin{align}\label{twad}
    a(f)&= a\star f-f\star \tilde{a}\,,
\end{align}
where $\tilde{a}$ is an automorphism of the Weyl algebra that flips the sign of $\bry$, $\tilde{a}(\bry)=a(-\bry)$. In fact, the action arises as a typical coadjoint action. Indeed, there is a nondegenerate pairing between $A_1$ and $A_1^\star$: $\langle a|f\rangle=\text{str}[a\star f]=\text{str} [f\star \tilde{a}]$, where $\text{str}[a]=a(\bry=0)$ is a supertrace, see also \cite{Iazeolla:2008ix}. The canonical bimodule structure of the higher spin algebra on itself (left/right actions) induces the twisted-adjoint representation \eqref{twad} on the dual module. What the results of the present paper show is that the coadjoint interpretation seems to be correct even for such a strange case as Chiral Theory, while the twisted-adjoint interpretation is no longer valid.\footnote{In AdS, the coadjoint and twisted-adjoint representation yield the same equations. It is only in flat space where the latter fails to describe the correct FDA.}

\section{Operator calculus}
\label{app:}
As was already sketched at the beginning of Section \ref{sec:FDA2}, we work with poly-differential operators that are represented as symbols. Let us illustrate all operations with $\bry$ and $\pl_{A'}^{\bry_i}\equiv p^i_{A'}$. The translation operator is $\exp{[\bry\cdot p_i]}f(\bry_i)= f(\bry_i+\bry)$. Operators acting on $n$ functions $a_i(\bry)$ are understood as functions of $p_0=\bry$, $p_1=\bar{\pl}_1$, ..., $p_n=\bar{\pl}_n$:
\begin{align}
V(a_1,...,a_n)=v(\bry,\bar{\pl}_1,...,\bar{\pl}_2)a_1(\bry_1)...a_n(\bry_n)\Big|_{\bry_i=0}
\end{align}
Therefore, the commutative product $f(\bry)g(\bry)$ and the Moyal-Weyl star-product $f(y)\star g(y)$ are represented by the following symbols:\besubeqs
\begin{align}
    \exp[p_0 \cdot p_1+p_0 \cdot p_2]\equiv \exp[p_{01}+p_{02}]\,, \\
    \exp[q_0 \cdot q_1+q_0 \cdot q_2+q_1 \cdot q_2]\equiv \exp[q_{01}+q_{02}+q_{12}]\,.
\end{align}
\esubeqs
Then we need the following identifications for symbols of the operators:
\begin{align*}
a_1\star V(a_2,...,a_{n+1})&\rightarrow  v(q_0+q_1,q_2,...,q_{n+1})e^{+ q_{0}\cdot q_1}\,,\\
V(a_1,...,a_n)\star a_{n+1}&\rightarrow v(q_0-q_{n+1},q_1,...,q_n)e^{+ q_{0}\cdot q_{n+1}}\,,\\
V(a_1,...,a_k\star a_{k+1},...,a_{n+1})&\rightarrow v(q_0,...,q_{k-1},q_k+q_{k+1},q_{k+2},...,q_{n+1})e^{+ q_{k}\cdot q_{k+1}}\,,
\end{align*}
\begin{align*}
a_1 V(a_2,...,a_{n+1})&\rightarrow  v(p_0,p_2,...,p_{n+1})e^{+ p_{0}\cdot p_1}\,,\\
V(a_1,...,a_n)a_{n+1}&\rightarrow v(p_0,p_1,...,p_n)e^{+ p_{0}\cdot p_{n+1}}\,,\\
V(a_1,...,a_k a_{k+1},...,a_{n+1})&\rightarrow v(p_0,...,p_{k-1},p_k+p_{k+1},p_{k+2},...,p_{n+1})\,,
\end{align*}
\begin{align*}
u_1(a_1, V(a_2,...,a_{n+1}))&\rightarrow  v(p_0+p_1,p_2,...,p_{n+1})\,,\\
u_1(V(a_1,...,a_n),a_{n+1})&\rightarrow v(-p_{n+1},p_1,...,p_n)e^{+ p_{0}\cdot p_{n+1}}\,,\\
V(a_1,...,u_1(a_k, a_{k+1}),...,a_{n+1})&\rightarrow v(p_0,...,p_{k-1},p_{k+1},p_{k+2},...,p_{n+1})e^{+p_{k}\cdot p_{k+1}}\,,
\end{align*}
\begin{align*}
u_2(a_1, V(a_2,...,a_{n+1}))&\rightarrow  v(-p_1,p_2,...,p_{n+1})e^{+ p_{0}\cdot p_1}\,,\\
u_2(V(a_1,...,a_n),a_{n+1})&\rightarrow v(p_0+p_{n+1},p_1,...,p_n)\,,\\
V(a_1,...,u_2(a_k, a_{k+1}),...,a_{n+1})&\rightarrow v(p_0,...,p_{k-1},p_k,p_{k+2},...,p_{n+1})e^{-p_{k}\cdot p_{k+1}}\,,
\end{align*}
where we defined
\begin{align}
        &u_1(a,b)=\exp{[p_{02}+p_{12}]}\,, &
        &u_2(a,b)=\exp{[p_{01}-p_{12}]}\,.
\end{align}

\section{Cochain complex}
\label{app:complex}
For completeness let us rewrite the $L_\infty$-relations in terms of symbols of operators. We do so for the $\bry$-part only since the dependence on $y$ is captured by the star-product and factorizes out. The l.h.s. of the equations for $\mathcal{V}_{1,2,3}$ read:
{\footnotesize
\begin{align*}
    &-e^{p_{01}} V_1\left(p_0,p_2,p_3,p_4\right)-V_1\left(p_0,p_1,p_2+p_3,p_4\right)+V_1\left(p_0,p_1+p_2,p_3,p_4\right)+e^{p_{34}} V_1\left(p_0,p_1,p_2,p_4\right) \,, \\
    &-e^{p_{01}} V_2\left(p_0,p_2,p_3,p_4\right)+e^{p_{04}} V_1\left(p_0,p_1,p_2,p_3\right)-e^{-p_{34}} V_1\left(p_0,p_1,p_2,p_3\right)+V_2\left(p_0,p_1+p_2,p_3,p_4\right)-e^{p_{23}} V_2\left(p_0,p_1,p_3,p_4\right) \,, \\
   & -e^{p_{01}} V_3\left(p_0,p_2,p_3,p_4\right)+e^{p_{04}} V_2\left(p_0,p_1,p_2,p_3\right)-V_2\left(p_0,p_1,p_2,p_3+p_4\right)+e^{-p_2\cdot p_3} V_2\left(p_0,p_1,p_2,p_4\right)+e^{p_{12}} V_3\left(p_0,p_2,p_3,p_4\right) \,, \\
   & e^{p_{04}} V_3\left(p_0,p_1,p_2,p_3\right)-e^{-p_{12}} V_3\left(p_0,p_1,p_3,p_4\right)-V_3\left(p_0,p_1,p_2,p_3+p_4\right)+V_3\left(p_0,p_1,p_2+p_3,p_4\right) \,.
\end{align*}}\noindent
Similarly, for $\mathcal{U}_{1,2,3}$ we find 
{\footnotesize
\begin{align*}
  & U_1\left(p_0,p_1+p_2,p_3,p_4\right)-U_1\left(p_0+p_1,p_2,p_3,p_4\right)-e^{p_{23}} U_1\left(p_0,p_1,p_3,p_4\right)+e^{p_{04}} V_1\left(-p_4,p_1,p_2,p_3\right) \,, \\
    &e^{-p_{23}} U_1\left(p_0,p_1,p_2,p_4\right)-U_2\left(p_0+p_1,p_2,p_3,p_4\right)-e^{p_{34}} U_1\left(p_0,p_1,p_2,p_4\right)+e^{p_{12}} U_2\left(p_0,p_2,p_3,p_4\right)+e^{p_{04}} V_2\left(-p_4,p_1,p_2,p_3\right),\\
   & e^{-p_{34}} U_1\left(p_0,p_1,p_2,p_3\right)-U_1\left(p_0+p_4,p_1,p_2,p_3\right)-U_3\left(p_0+p_1,p_2,p_3,p_4\right)+e^{p_{12}} U_3\left(p_0,p_2,p_3,p_4\right),\\
    &-e^{-p_{12}} U_3\left(p_0,p_1,p_3,p_4\right)+e^{-p_{34}} U_2\left(p_0,p_1,p_2,p_3\right)-U_2\left(p_0+p_4,p_1,p_2,p_3\right)+e^{p_{23}} U_3\left(p_0,p_1,p_3,p_4\right)-e^{p_{01}} V_2\left(-p_1,p_2,p_3,p_4\right),\\
   & -e^{-p_{12}} U_2\left(p_0,p_1,p_3,p_4\right)+U_2\left(p_0,p_1,p_2+p_3,p_4\right)-e^{p_{34}} U_2\left(p_0,p_1,p_2,p_4\right)-e^{p_{01}} V_1\left(-p_1,p_2,p_3,p_4\right)+e^{p_{04}} V_3\left(-p_4,p_1,p_2,p_3\right),\\
   & -e^{-p_{23}} U_3\left(p_0,p_1,p_2,p_4\right)+U_3\left(p_0,p_1,p_2,p_3+p_4\right)-U_3\left(p_0+p_4,p_1,p_2,p_3\right)-e^{p_{01}} V_3\left(-p_1,p_2,p_3,p_4\right) \,.
\end{align*}
}\noindent
When looking for nontrivial solutions, it is important to understand which ones are trivial. The latter are given by field redefinitions that act as follows on  $\mathcal{V}_{1,2,3}$
\begin{align*}
    \delta V_1&=e^{p_{01}} g_1\left(p_0,p_2,p_3\right)-g_1\left(p_0,p_1+p_2,p_3\right)+e^{p_{23}} g_1\left(p_0,p_1,p_3\right)\,,\\
   \delta V_2&= e^{p_{01}} g_2\left(p_0,p_2,p_3\right)+e^{p_{03}} g_1\left(p_0,p_1,p_2\right)-e^{-p_{23}} g_1\left(p_0,p_1,p_2\right)-e^{p_{12}} g_2\left(p_0,p_2,p_3\right)\,,\\
   \delta V_3&= e^{p_{03}} g_2\left(p_0,p_1,p_2\right)+e^{-p_{12}} g_2\left(p_0,p_1,p_3\right)-g_2\left(p_0,p_1,p_2+p_3\right)\,,
\end{align*}
and on $\mathcal{U}_{1,2,3}$
\begin{align*}
    \delta U_1&=h\left(p_0+p_1,p_2,p_3\right)-e^{p_{12}} h\left(p_0,p_2,p_3\right)+e^{p_{03}} g_1\left(-p_3,p_1,p_2\right)\,,\\
   \delta U_2&= e^{-p_{12}} h\left(p_0,p_1,p_3\right)-e^{p_{23}} h\left(p_0,p_1,p_3\right)-e^{p_{01}} g_1\left(-p_1,p_2,p_3\right)+e^{p_{03}} g_2\left(-p_3,p_1,p_2\right)\,,\\
   \delta U_3&= e^{-p_{23}} h\left(p_0,p_1,p_2\right)-h\left(p_0+p_3,p_1,p_2\right)-e^{p_{01}} g_2\left(-p_1,p_2,p_3\right)\,,
\end{align*}
It can easily be checked that the redefinitions lead to solutions of the equations. The expressions above define a particular realization of the Chevalley-Eilenberg complex, but we do not extend the action of the differential to cochains with more arguments. At the bottom level we find
\begin{align*}
    \delta g_1&= e^{p_{12}} \xi \left(p_0,p_2\right)-e^{p_{01}} \xi \left(p_0,p_2\right)\,,\\
    \delta g_2&= e^{p_{02}} \xi \left(p_0,p_1\right)-e^{-p_{12}} \xi \left(p_0,p_1\right)\,,\\
    \delta h&= e^{p_{02}} \xi \left(-p_2,p_1\right)-e^{p_{01}} \xi \left(-p_1,p_2\right)\,,
\end{align*}
which leads to redefinitions that yield vanishing vertices. 

\section{Vertices}
\label{app:vertices}
In order to find nontrivial cubic vertices we employ a number of ideas, see also \cite{Vasiliev:1988sa,Sharapov:2017yde} that were used for inspiration. Firstly, Lorentz symmetry has to be preserved, i.e., in practice, we cannot mix primed and unprimed indices. The higher spin algebra is the tensor product of two algebras, which via the K{\"u}nneth theorem suggests to look for the two-cocycle as a tensor product of two, one of them being trivial. The free equations, in particular the boundary condition for $\mathcal{V}(e,e,C)$, reveal that something interesting should happen on the $\bry$ side. Therefore, for homogeneous arguments $a(y,\bry)=a(y)\otimes \bar{a}(\bry)$, etc. we assume that all vertices have the star-product over the $y$-dependent factors:
\begin{align}
    \mathcal{V}_1(a(y)\otimes \bar{a}(\bry),b(y)\otimes \bar{b}(\bry),c(y)\otimes \bar{c}(\bry))&= a\star b\star c \otimes v_1(\bar{a},\bar{b},\bar{c})\,.
\end{align}
As a result, all terms in the cocycle equations have the same overall factor for the $y$-dependence and we can concentrate on $\bry$ only. The cocycle conditions for the $\bry$-part are collected in Appendix \ref{app:complex}. 

Now, we need to solve the equations in Appendix \ref{app:complex}. It is clear that the solution should contain some $\exp[p_{ij}]$-factors, otherwise they cannot cancel the $\exp{[p_{ij}]}$ already present in the cocycle condition. The boundary condition for $\mathcal{V}(\omega,\omega,C)$ restrict the exponents a little bit. For example, we cannot allow for $\exp{p_{03}}$ in $\mathcal{V}_1(\omega,\omega,C)$. The crucial step is to look for $\mathcal{V}$ and $\mathcal{U}$ as singular field redefinitions, i.e. we look for $g_{1,2}$ and $h$, see Appendix \ref{app:complex}. For any $g_{1,2}$ and $h$, the vertices solve the cocycle equations. We just need to make sure that (i) the vertices are regular, i.e. Taylor expandable in $p_{ij}$; (ii) the redefinitions themselves, i.e. $g_{1,2}$ and $h$, are irregular. Irregular field redefinitions are not allowed. Therefore, if (i) and (ii) are satisfied, we have a nontrivial cocycle. Let us note that the singularity of $g_{1,2}$ and $h$ must be essential and cannot be removed with the help of "redefinitions for redefinitions" with $\xi$. Looking for singular redefinitions is more economic than looking for nontrivial vertices since they depend on less arguments. Long story short, we arrived at the following redefinitions:
\besubeqs
\begin{align}
    g_1&= \frac{p_{01} e^{p_{12}}}{p_{02} \left(p_{01}-p_{12}\right)}-\frac{e^{p_{01}} p_{01}}{p_{02} \left(p_{01}-p_{12}\right)}\,,\\
    g_2&= \frac{e^{p_{02}} p_{02}}{p_{01} \left(p_{02}+p_{12}\right)}-\frac{p_{02} e^{-p_{12}}}{p_{01} \left(p_{02}+p_{12}\right)}\,,\\
    h&= \frac{e^{p_{01}} p_{01}}{p_{12} \left(p_{01}-p_{02}\right)}-\frac{e^{p_{02}} p_{02}}{p_{12} \left(p_{01}-p_{02}\right)}\,.
\end{align}
\esubeqs
The vertices, which we do not write here as fractions, have a similar structure and their regularity is not obvious. It is very important to take advantage of the Fierz/Schouten/Pl{\"u}cker identities 
\begin{equation}
    (a\cdot b)(c\cdot d)+(b\cdot c)(a\cdot d)-(a\cdot c)(b\cdot d)=0\,,
\end{equation}
which are a consequence of the fact that any three vectors in two dimensions are linearly dependent. Still, the regularity is not manifest. One can prove it by showing that the numerator and denominator have the same zeros. 

A more convenient way to make the regularity manifest to write the vertices as integrals over the $2d$-simplex, as in the main text. The nontriviality of the cocycles is then less obvious. A simple way to check if the cocycle is nontrivial is to extract the boundary condition $\mathcal{V}(e,e,C)$ since this part cannot be redefined away. Therefore, once the boundary condition is satisfied we can be certain that the cocycle is worthy. It would be interesting to compute the Chevalley-Eilenberg cohomology following the techniques of \cite{Sharapov:2020quq}, which would give a rigorous answer regarding the number of independent vertices within the covariant approach. 

Given the relation between the algebraic structures of higher spin gravities and deformation quantization and formality, it is also possible to recast the proof into the familiar language of Stokes theorem. For example, to check that the equation for $\mathcal{V}_1$ is satisfied we can construct a closed two-form $\Omega_1$
\begin{align}
\begin{aligned}
        \Omega_1&= ( p_{12}\, dt_1 \wedge dt_2 +p_{23}\, dt_2\wedge dt_3 + p_{13}\,dt_1 \wedge dt_3)F_1 \,,\\
    F_1&=\exp{[\left(1-t_1\right) p_{01}+\left(1-t_2\right) p_{02}+\left(1-t_3\right) p_{03}+t_1 p_{14}+t_2 p_{24}+t_3 p_{34}]}\,.
\end{aligned}
\end{align}
With the help of Stokes theorem we get
\begin{align}
    0&=\int_{\Delta_3} d\Omega_1= \int_{\pl\Delta_3} \Omega_1\,.
\end{align}
There are four boundaries that correspond to "collisions of points" on the circle: $t_1=0$, $t_1=t_2$, $t_2=t_3$ and $t_3=1$. It can be seen that $\Omega_1$ at these boundaries reduces to exactly the four terms in the equation for $\mathcal{V}_1$. Similar arguments are true for the rest of the equations. The closed two-form for the other equations are
\besubeqs
\begin{align}
    \begin{aligned}
        \Omega_2&= ( p_{12}\, dt_1 \wedge dt_2  + p_{14}\,dt_1 \wedge dt_3 +p_{24}\, dt_2\wedge dt_3)F_2 \,+\\
        &- ( p_{14}\, dt_1 \wedge dt_2 +p_{12}\, dt_1\wedge dt_3 - p_{24}\,dt_2 \wedge dt_3)F_3 \,+\\
        &-(p_{14} dt_1 \wedge dt_2 + p_{24}\, dt_1 \wedge dt_3 -p_{12}\, dt_2 \wedge dt_3) F_4 \,,\\
    F_2&=\exp{[\left(1-t_1\right) p_{01}+\left(1-t_2\right) p_{02}+\left(1-t_3\right) p_{04}+t_1 p_{13}+t_2 p_{23}-t_3 p_{34}]} \,, \\
    F_3&=\exp{[\left(1-t_1\right) p_{01}+\left(1-t_3\right) p_{02}+\left(1-t_2\right) p_{04}+t_1 p_{13}+t_3 p_{23}-t_2 p_{34}]} \,, \\
    F_4&=\exp{[\left(1-t_2\right)p_{01}+\left(1-t_3\right)p_{02}+\left(1-t_1\right)p_{04}+t_2p_{13} + t_3 p_{23} - t_1 p_{34}]} \,,
    \end{aligned}
\end{align}
for the second and for the third we need
\begin{align}
        \begin{aligned}
        \Omega_3&= - ( p_{34}\, dt_1 \wedge dt_2 +p_{14}\, dt_1\wedge dt_3 + p_{13}\,dt_1 \wedge dt_3)F_5 \, + \\
        &+ ( -p_{34}\, dt_1 \wedge dt_2 +p_{13}\, dt_1\wedge dt_3 + p_{14}\,dt_2 \wedge dt_3)F_6 \, +\\
        &+ ( -p_{13}\, dt_1 \wedge dt_2 +p_{34}\, dt_1\wedge dt_3 + p_{14}\,dt_2 \wedge dt_3)F_7 \,, \\
    F_5&=\exp{[\left(1-t_3\right) p_{01}+\left(1-t_2\right) p_{03}+\left(1-t_1\right) p_{04}+t_3 p_{12}-t_2 p_{23}-t_1 p_{24}]} \,, \\
    F_6&=\exp{[\left(1-t_3\right) p_{01}+\left(1-t_1\right) p_{03}+\left(1-t_2\right) p_{04}+t_3 p_{12}-t_1 p_{23}-t_2 p_{24}]} \,, \\
    F_7&=\exp{[\left(1-t_2\right) p_{01}+\left(1-t_1\right) p_{03}+\left(1-t_3\right) p_{04}+t_2 p_{12}-t_1 p_{23}-t_3 p_{24}]} \,.
    \end{aligned}
\end{align}
Note that the 2nd and 3rd equations have more terms since they mix vertices with different orderings and for this reason three closed forms are required. There is some mutual cancellation between them. For the last equation we have 
\begin{align}
\begin{aligned}
        \Omega_4&= - ( p_{34}\, dt_1 \wedge dt_2 + p_{24}\,dt_1 \wedge dt_3 +p_{23}\, dt_2\wedge dt_3 )F_8\,, \\
    F_8&=\exp{[\left(1-t_3\right) p_{02}+\left(1-t_2\right) p_{03}+\left(1-t_1\right) p_{04}-t_3 p_{12}-t_2 p_{13}-t_1 p_{14}]} \,.
\end{aligned}
\end{align}
\esubeqs
The two-forms show some similarities: $\Omega_1$ and $\Omega_4$ are related by swapping $1234 \rightarrow 4321$, i.e. they are each other's mirror image. The same holds for $\Omega_2$ and $\Omega_3$. Additionally, the terms in $\Omega_2$ are also obtained from $\Omega_1$ by permutations of $1234$. This can be understood from the different orderings of $\omega,\omega,\omega,C$ in the equations. The signs are chosen such that the $L_\infty$-relations come out correctly.

\chapter{All order vertices}
\label{app:all}

\section{Homological perturbation theory}
\label{app:hpt}
In this section, we show how to obtain all the interaction vertices of Chiral HiSGRA by means  of {\it homological perturbation theory} (HPT). A detailed account of the theory can be found in \cite{HK}, \cite{GLS}, \cite{Kr} (see also \cite{Li:2018rnc} for a similar discussion of HPT in the context of formal HiSGRA).

As in the main text, we start with the cochain complex  $A=\mathbb{C}[y, z, dz]$  of differential forms with polynomial coefficients. The coboundary operator $d_{z}: A_{n}\rightarrow A_{n+1}$ is given by the usual 
exterior differential on $z$'s.  Combining the exterior product of the basis differentials $dz^{A}$ with the $\star$-product 
\begin{equation}\label{pp}
a\star b=a(y,z)\exp\Big({\frac{\stackrel{\leftarrow}{\partial}}{\partial z^{A}}\frac{\stackrel{\rightarrow}{\partial}}{\partial y_A}}-\frac{\stackrel{\leftarrow}{\partial}}{\partial y^{A}}\frac{\stackrel{\rightarrow}{\partial}}{\partial z_{A}}\Big) b(y,z)
\end{equation}
of polynomials in the $y$'s and $z$'s, we get a commutative dg-algebra $(A, d_{z})$. 
Actually, the $\star$-product above is equivalent to the conventional (dot) product on polynomials:
$$
a\star b=e^{-\Delta}\left((e^{\Delta}a)\cdot(e^{\Delta}b)\right)\,,\qquad \Delta=\frac{\partial^2}{\partial y^{A}\partial z_{A}}\,.
$$
The dual space $A^\ast$ carries the canonical structure of a graded bimodule over $A$. In particular, $A^\ast_0$ is clearly isomorphic to the space of formal power series $\mathbb{C}[[y,z ]]$. The left/right action of $A$ on $A_0^\ast$ is given by
$$
a\circ m=m\circ a= (e^{\Delta}a)(\partial_{y},\partial_{z},0)m(y, z)\,,\qquad \forall a=a(y, z,dz)\in A,\quad m\in A^\ast_0\,.
$$
Indeed, 
$$
(a\star b)\circ m=e^{-\Delta}\left((e^{\Delta}a)\cdot(e^{\Delta}b)\right)\circ m
$$
$$
=\left((e^{\Delta}a)\cdot(e^{\Delta}b)\right)(\partial_{y},\partial_{z}, 0) m(y,z)=a\circ(b\circ m)\,,
$$
and the same for the right action. The bimodule $A_0^\ast$ is too big and highly reducible. In the following we will deal with its submodule $M=\mathbb{C}[[y]]\subset A^\ast_0$ constituted by formal power series in $y$'s. Extending the action of $d_{z}$ to $M$ by zero, we can think of $M$ as a differential bimodule over $A$. Furthermore, it is convenient to combine the differential bimodule structure into  a single dg-algebra $\mathcal{A}=A\oplus M$ for the following $\ast$-product and differential:
$$
(a, m)\ast (a', m')=(a\star a', a\circ m'+a'\circ m)\,,\qquad d_{z}(a,m)=(d_{z}a, 0)\,.
$$
By definition, the degree of an elements $a$ of $A$ coincides with its form-degree, while all the elements of $M$ have degree $1$. 

In addition to $d_{z}$ we can endow the algebra $\mathcal{A}$ with one more differential $\delta$ of degree $1$. This is defined as   
\begin{equation}\label{dd}
\delta (a,m)=\big(m(-z)e^{z^{A}y_{A}}dz^{B}\wedge dz_{B}, 0\big)\,,\qquad \forall\; (a,m)\in \mathcal{A}\,.
\end{equation}
(Notice the change of the argument in $m$.) It is clear that $\delta^2=0$. The differential $\delta$ will be a derivation of the $\ast$-product above if and only if the following identities hold: 
$$
\delta(a\circ m)=(-1)^{|a|}a\star \delta m\,,\qquad \delta m\circ m'=m\circ \delta m'\,.
$$
The first equality is enough to check only for the generators  $y^{A}$, $z^{A}$'s and $dz^{A}$. We find
$$
y^{A}\star \delta m=(y^{A}-\partial^{A}_{z})m(-z)e^{ {z}^{B} y_{B}}dz^{C}\wedge dz_{C}=e^{y^{B} z_{B}}(-\partial_{z}^{A} m(-z))dz^{C}\wedge dz_{C}=\delta(y^{A}\circ m)\,,
$$
$$
z^{A}\star \delta m=(z^{A}+\partial^{A}_{y})m(-z)e^{z^{B} y_{B}}dz^{C}\wedge dz_{C}=0=\delta(z^{A}\circ m)\,,
$$
$$
dz^{A}\star \delta m=m(-z)e^{ {z}^{B} y_{B}}dz^{A}\wedge dz^{C}\wedge dz_{C}=0=-\delta(dz^{A}\circ m)\,.
$$
The second identity is also satisfied because $\delta m$ is a two-form and $(\delta m)(y, z,0)=0$.

Since the differentials trivially commute to each other, $d_{z}\delta+\delta d_{z}=0$, we can combine them into the total differential $D=d_{z}+\delta$ of degree $1$. Given now  a dg-algebra $(\mathcal{A}, D)$, one can ask about its minimal model. In general, constructing a minimal model of a dg-algebra is quite  a difficult  problem. What helps us a lot are two things: $(i)$ the differential $d_{z }$, being the exterior differential on polynomial forms, admits an explicit contracting homotopy $h$  and $(ii)$ one may regard $D$ as a `small perturbation' of $d_{z}$ by $\delta$. Under these circumstances, homological perturbation theory offers the most efficient way to build the minimal model in question. As we will see, this minimal model yields  exactly the $A_\infty$-algebra defining the r.h.s. of the field equations in Chiral HiSGRA.
Below we recall some basic definitions and statements. 

\begin{definition}

A {\it strong deformation retract} (SDR) is given by a pair of complexes $(V,d_V)$ and $(W,d_W)$ together with chain maps $p:V\rightarrow W$ and $i:W\rightarrow V$
such that $pi=1_W$ and $ip$ is homotopic to $1_V$. The last property implies the  existence of a map $h: V\rightarrow V$ such that 
$$
d_Vh+hd_V=ip-1_V\,.
$$
Without loss in generality, one may also assume the following {\it annihilation properties}:  
$$
hi=0\,,\qquad ph=0\,,\qquad h^2=0\,.
$$
\end{definition}
All the data above can be summarized by a single  diagram 
\begin{equation}\label{SDR}
\xymatrix{
*{\hspace{5ex}(V,d_V)\;}\ar@(ul,dl)[]_{h} \ar@<0.5ex>[r]^-p
& (W, d_W) \ar@<0.5ex>[l]^-i}\,.
\end{equation}
Let us mention a special case of this construction where  $W=H(V,d_V)$ is the cohomology group of the complex $(V,d_V)$ and $d_W=0$. 

The main concern of HPT is  transferring various algebraic structures form one object to another through a homotopy equivalence. Whenever applicable, the theory provides effective algorithms and explicit formulas, as distinct from classical homological algebra.  The cornerstone of HPT is 
the following statement, often called the Basic Perturbation Lemma. 
\begin{lemma}[\cite{B}]\label{BPL}
For any SDR data (\ref{SDR}) and a small perturbation $\delta$ of  $d_V$ such that $(d_V+\delta)^2=0$ and $1-\delta h$ is invertible, there is a new SDR 
$$
\xymatrix{
*{\hspace{9ex}(V,d_V+\delta )\;}\ar@(ul,dl)[]_{h'} \ar@<0.5ex>[r]^-{p'}
&(W, d'_W) \ar@<0.5ex>[l]^-{i'}}\,,
$$
where the maps are given by 
$$
\begin{array}{ll}
     p'=p+p(1-\delta h)^{-1}\delta h\,,&\quad i'=i+h(1-\delta h)^{-1}\delta i\,, \\[3mm]
    h'=h+h(1-\delta h)^{-1}\delta h\,, & \quad d'_W = d_W+p(1-\delta h)^{-1}\delta i\,.
\end{array}
$$
\end{lemma}
One can think of the operator $A=(1-\delta h)^{-1}$ as being defined by a geometric series 
\begin{equation}\label{GS}
A=\sum_{n=0}^\infty (\delta h)^n\,. 
\end{equation}
In many practical cases its convergence is ensured by a suitable filtration on $V$. 

We are concerned with transferring  $A_\infty$-structures on $V$ to its cohomology space $W$. 
To put this transference problem into the framework of HPT one first applies the tensor-space functor $T$  to the vector spaces $V$ and $W$. Recall that, in addition to the associative algebra structure, the space $T(V)=\bigoplus_{n\geq 1}V^{\otimes n}$ carries the structure of a coassociative  coalgebra with respect to the coproduct 
$$
\Delta: T(V)\rightarrow T(V)\otimes T(V)\,,
$$
$$
\Delta (v_1\otimes \cdots\otimes v_n)=\sum_{i=1}^{n-1}(v_1\otimes\cdots\otimes v_i)\otimes (v_{i+1}\otimes\cdots\otimes v_n)\,.$$
Coassociativity is expressed by the relation $(1\otimes \Delta)\Delta=(\Delta\otimes 1)\Delta$. A linear map $F: T(V)\rightarrow T(V)$ is called a {\it coderivation}, if it obeys the co-Leibniz rule $$\Delta F=(F\otimes 1+1\otimes F)\Delta\,.$$ 
The space of coderivations is known to be  isomorphic to the space of linear maps $\mathrm{Hom}(T(V),V)$, so that  any homomorphism $f: T(V)\rightarrow V$ induces a coderivation $\hat{f}: T(V)\rightarrow T(V)$ and vice versa:  if $f\in \mathrm{Hom}(T^m(V),V)$, then  
\begin{equation}\label{f1}
\begin{array}{rl}
\hat{f}(v_1\otimes\cdots\otimes v_n)=\displaystyle \sum_{i=1}^{n-m+1}&(-1)^{|f|(|v_1|+\cdots+ |v_{i-1}|)}v_1\otimes\cdots\otimes v_{i-1}\\[4mm]
&\otimes f(v_i\otimes \cdots \otimes v_{i+m-1})\otimes v_{i+m}\otimes \cdots\otimes v_n
\end{array}
\end{equation}
for $n\geq m$ and zero otherwise. 

The notion of a coderivation provides an alternative definition of an $A_\infty$-algebra: An $A_\infty$-algebra structure on a graded vector space $V$ is given by an element $m\in \mathrm{Hom}(T(V),V)$ of degree one such that the corresponding  coderivation $\hat m$ squares to zero. Every such $\hat m$ is called a {\it codifferential}.  The condition $\hat m^2=0$ is equivalent to the equation $m\circ m=0$, where $\circ$ stands for the Gerstenhaber product
\begin{align}\label{gersproduct}
    f\circ g&= \sum_i (-1)^\kappa f(a_1,\ldots,a_i, g(a_{i+1},\ldots,a_{i+k_g}),a_{i+k_g+1},\ldots, a_{k_f+k_g-1})   \,.
\end{align}
Here $\kappa$ is the usual Koszul sign: $\kappa=|g|(|a_1|+\cdots+|a_i|)$.
Expanding $m$ into the sum $m=m_1+m_2+\cdots$ of homogeneous multi-linear maps $m_n\in \mathrm{Hom}(T^n(V),V)$ and substituting it back into $m\circ m=0$ gives an infinite sequence of homogeneous  relations on $m$'s, known as Stasheff's identities \cite{St}. In particular, the first structure map $m_1: V_l\rightarrow V_{l+1}$ squares to zero, $m_1^2=0$, making $V$ into a complex of vector spaces.  An  $A_\infty$-algebra is called {\it minimal} if $m_1=0$. For minimal algebras the second structure map  $m_2: V\otimes V\rightarrow V$ makes the space $V[-1]$ into a graded associative algebra with respect to the $\ast$-product\footnote{By definition, $V[-1]_n=V_{n-1}$. On shifting degree by one unit, the $\ast$-product acquires degree $0$.}
\begin{equation}\label{bbb}
a\ast b=(-1)^{|a|}m_2(a\otimes b)\,.
\end{equation}
Associativity is encoded by the Stasheff identity $ m_2\circ m_2=0$. From this perspective, a graded associative algebra is just an $A_\infty$-algebra with $m=m_2$.  More generally, an $A_\infty$-algebra with $m=m_1+m_2$ is equivalent to a  differential graded algebra $(V[-1],\ast, d)$ with the product (\ref{bbb}) and the differential $d=m_1$. Again, the Leibniz rule 
$$
d(a\ast b)=da\ast b+(-1)^{|a|-1}a\ast db
$$
is equivalent to the Stasheff identity $m_1\circ m_2+m_2\circ m_1=0$. 

The next statement, called the {\it tensor trick}, allows one to transfer SDR data from spaces to their tensor (co)algebras. 
\begin{lemma}[\cite{GL}]\label{TT}
With any SDR data (\ref{SDR}) one can associate a new SDR 
$$
\xymatrix{{(\,}\ar@(ul,dl)[]_-{\hat{h}}&
{\hspace{-6ex}T(V),\hat{d}_V )\;} \ar@<0.5ex>[r]^-{\hat{p}}
& (T(W), \hat{d}_W) \ar@<0.5ex>[l]^-{\hat{i}}}\,,
$$ 
where the new differentials $\hat{d}_V$ and $\hat{d}_W$ are defined by the rule (\ref{f1}),
$$
\hat{p}=\sum_{n=1}^\infty p^{\otimes n}\,,\qquad \hat{i}= \sum_{n=1}^\infty i^{\otimes n}\,,
$$
and the new homotopy is given by
$$
\hat{h}=\sum_{n= 1}^\infty \sum_{k=0}^{n-1} 1^{\otimes k}\otimes h\otimes (ip)^{\otimes n-k-1}\,.
$$
\end{lemma}

After reminding the basics of HPT let us return to our deformation problem.
Consider first the case of dg-algebra $\mathcal{A}$ with respect to the unperturbed differential $d_{z}$. 
By the algebraic Poincar\'e Lemma, $H(\mathcal{A}, d_{z})\simeq \mathbb{C}[y]\oplus \mathbb{C}[[y]]$. Here  the first summand corresponds to the differential forms of  $A$ that are independent of $z$'s and $dz$'s, while the second summand is given by the elements of the module $M$. To streamline our notation we will write $\mathcal{H}$ for the algebra of cohomology $H(\mathcal{A}, d_{z})$.  Clearly, the natural inclusion $i :\mathcal{H}\rightarrow \mathcal{A}$ is an algebra homomorphism. This leads us immediately to SDR (\ref{SDR}) with 
$$
V=\mathcal{A}[1] \,,\qquad W=\mathcal{H}[1] \,,\qquad d_V=d_{z}\,, \qquad d_W=0\,,
$$
$$
p (a,m)=(a(y,0,0), m)\,,\qquad h(a, m)=(h(a), 0)\,,
$$
and $h(a)$ was defined in \eqref{homofor} as the standard contracting homotopy for the de Rham complex. 
Applying the tensor trick yields then an SDR for the corresponding tensor (co)algebras
$$
\xymatrix{{(\,}\ar@(ul,dl)[]_-{\hat{h}}&
{\hspace{-6ex} T(\mathcal{A}[1]),\hat{d}_{z})\;} \ar@<0.5ex>[r]^-{\hat{p}}
& (T(\mathcal{H}[1]), 0 ) \ar@<0.5ex>[l]^-{\hat{i}}}\,.
$$
Let $\mu$  denote multiplication (the $\ast$-product) in $\mathcal{A}$. It defines the coderivation\footnote{A coderivation $Q$ on $\bar{S}^c(\mathfrak{g}[1])$ satisfying $Q^2=0$ encodes an $L_\infty$-structure, c.f. \cite{li2018homotopy}.} $\hat \mu$ such that $(\hat d_{z}+\hat \mu)^2=0$.\footnote{A coderivation that squares to zero yields an $A\infty$-algebra. In our case, the latter contains the star-product.} 
This allows us to treat $\hat\mu$ as a small perturbation of the differential $\hat{d}_{z}$. By making use of the Basic Perturbation Lemma \ref{BPL}, we  obtain the new SDR 
$$
\xymatrix{{(\,}\ar@(ul,dl)[]_-{\hat{h}'}&
{\hspace{-6ex}T(\mathcal{A}[1]),\hat{d}_{z}+\hat\mu)\;} \ar@<0.5ex>[r]^-{\hat{p}'}
& \big(T(\mathcal{H}[1]), \hat m_2 \big ) \ar@<0.5ex>[l]^-{\hat{i}'}}\,,
$$
where the codifferential on the right is given by 
\begin{equation}\label{hatmt}
  \hat m_2 = \hat{p}(1-\hat\mu \hat{h})^{-1}\hat\mu\hat{i}\,.
\end{equation}
Notice that $\hat h \hat \mu \hat i=0$, because $h$ vanishes on the subalgebra $ i(\mathcal H)\subset \mathcal A$. Hence, 
$\hat m_2=\hat p\hat \mu\hat i$ and the dg-algebra $(\mathcal{A}, d_{z})$ is formal: its minimal model $\mathcal{H}$ involves no higher multiplication operations in addition to the $\ast$-product (\ref{bbb}).

Finally, let us turn to the dg-algebra $(\mathcal{A}, D=d_{z}+\delta)$. This yields the SDR data
$$
\xymatrix{{(\,}\ar@(ul,dl)[]_-{\hat{h}'}&
{\hspace{-6ex}T(\mathcal{A}[1]),\hat{d}_{z}+\hat \delta+\hat\mu)\;}
 \ar@<0.5ex>[r]^-{\hat{p}'}
& \big(T(\mathcal{H}[1]), \hat m \big ) \ar@<0.5ex>[l]^-{\hat{i}'}}\,.
$$
Again, we can regard  the sum $\hat \mu+\hat \delta$ as a small perturbation of the basic differential $\hat d_{z}$. Lemma \ref{BPL} gives then the formal expression for the codifferential $\hat m$ on the right:
\begin{equation}\label{ps}
  \hat m = \hat{p}\big(1-(\hat\mu+\hat \delta) \hat{h}\big)^{-1}(\hat\mu+\hat \delta)\hat{i}=
  \hat m_2+\hat{p}\big(1-(\hat\mu+\hat \delta) \hat{h}\big)^{-1}\hat \delta\hat{i}\,.
\end{equation}
 One can simplify various terms of this formula  by noting that $\hat p\hat \delta=0$ and  $\delta h=0$.  Using (\ref{GS}), one can also find that the deformed codifferential starts as 
$$
\hat m=\hat m_2+\hat p \hat \mu\hat h\hat \mu\hat h\hat \delta \hat i+\cdots\,.
$$
The second term on the right defines the third structure map $m_3$ of the $A_\infty$-algebra $\mathcal{H}[1]$. 
Thus, the dg-algebra $(\mathcal{A}, D)$ is not formal. The diagrams in the main text and Appendices give just a pictorial representation for various terms of the perturbation series (\ref{ps}); in so doing, the inclusion and projection maps $i$ and $p$ correspond to incoming and outgoing edges, respectively. 

\section{Homological perturbation theory: a recipe}
\label{app:homo}
We recall that the Poincar\'e Lemma gives solutions to the equations $d_z f^{(1)}=f^{(2)}$ and $d_z f^{(0)}=f^{(1)}$ for any closed one-form $f^{(1)}=dz^A f^{(1)}_A(z)$ and a two-form $f^{(2)}=\frac{1}{2}\epsilon_{AB}f^{(2)}(z)dz^Adz^B$. They read
\begin{align*}
    f^{(1)}&=h[f^{(2)}]=dz^A z_{A}\int_0^1 t dt f^{(2)}(tz) \,, & f^{(0)}&=h[f^{(1)}]=z^A\int_{0}^{1} dt f_{A}^{(1)}(tz) \,.
\end{align*}
We also set $h[f^{(0)}]=0$ for any zero-form $f^{(0)}$. These relations define $h$ as the standard contracting homotopy for the de Rham complex of polynomial differential forms:\footnote{See for example \eqref{ch}.}
\begin{equation}\label{dh}
d_zh+hd_z=1-\pi\,,
\end{equation}
$\pi$ being the natural projection onto the subspace of $z$ independent zero-forms. The form degree and the exterior differential $d_z$ give $\mathfrak{R}$ the structure of a differential graded algebra (or dg-algebra for short). 
Rel. (\ref{dh}) implies that the cohomology of the dg-algebra $(\mathfrak{R}, d_z)$ is concentrated in degree zero and is described by $z$- and $dz$-independent polynomials. Hence, $H(\mathfrak{R}, d_z)\simeq A_\lambda$ and $(\mathfrak{R}, d_z)$ define a multiplicative resolution (aka  {\it model}) of the algebra $A_\lambda$. Starting with the differential graded algebra  $\mathfrak{R}$ one can systematically construct resolutions for many other algebras. For example, taking the tensor product of $\mathfrak{R}$ with an associative algebra  $B$ yields the dg-algebra $\mathfrak{R}\otimes B$, where $d_z$ extends to $B$ by zero. 
The algebra $\mathfrak{R}\otimes B$ defines then a model of the tensor product algebra $A_\lambda\otimes B$. Another possibility is to consider the trivial extension of $\mathfrak{R}$ by a differential $\mathfrak{R}$-bimodule ${M}$ concentrated in a single degree.  The result is given by a dg-algebra $\mathfrak{R}\oplus{M}$ with the product 
\begin{equation}\label{prod}
(b,a)(\tilde b,\tilde a)=(b\tilde b, b\tilde a+a\tilde b)\qquad \forall b,\tilde b\in \mathfrak{R}, \quad \forall a,\tilde a\in M\,.
\end{equation}
Since the differential necessarily annihilates ${M}$, the algebra $\mathfrak{R}\oplus{M}$ defines a model for the trivial extension $A_\lambda\oplus {M}$. In application to Chiral Theory we combine both the operations $\otimes$ and $\oplus$. Specifically, we take $B=A_1\otimes \mathrm{Mat}_N$ and define the bimodule structure on the space of formal power series $M=\mathbb{C}[[y^A]]$ by setting\footnote{The quickest way to check the bimodule axioms is with the $\tau$-involution introduced in \cite[App. A]{Sharapov:2022awp}.}
\begin{equation}\label{circaction}
\begin{aligned}
     y^A\circ a=(-\partial_y^A+\hhbar y^A)a\,, && z^A\circ a=a\circ z^A=0\,,\\
     a\circ y^A=(-\partial_y^A-\hhbar y^A)a\,,  &&\qquad dz^A\circ a=a\circ dz^A=0
\end{aligned}
\end{equation}
for all $a\in M$. As is seen the left and right actions of $\mathfrak{R}$ on ${M}$ are different unless $\hhbar\neq 0$. The quickest way to check the bimodule axioms is with the $\tau$-involution introduced in \cite[App. A]{Sharapov:2022awp}. For any function $a(y,z)$ we set
\begin{equation}\label{tauinvol}
a^\tau(y,z)=a(z,y)e^{z^Ay_A}\,.
\end{equation}
Clearly, $\tau^2=1$. Then one can equivalently define  the above $\circ$-product by the relation
\begin{equation}\label{moduleaction}
b\circ a\circ \tilde b= (b\star a^\tau\star \tilde b)^\tau\,,\qquad \forall b,\tilde b\in \mathfrak{R}, \quad a\in M\,,
\end{equation}
and the condition that $dz^A \circ a=0=a\circ dz^A$.  In this form, the bimodule axioms for the $\circ$-product hold due to the associativity of the star-product.  

The elements of the bimodule $M$ are assigned the degree one. Then the differential graded algebra 
$$
\mathcal{R}=(\mathfrak{R}\oplus M)\otimes A_1\otimes \mathrm{Mat}_N=\mathfrak{R}\otimes A_1\otimes\mathrm{Mat}_N\bigoplus M\otimes A_1\otimes\mathrm{Mat}_N=\mathbf{R}\bigoplus \mathfrak{M}
$$
defines a multiplicative resolution of the algebra 
\begin{equation}\label{hsA3}
\mathcal{A}=H(\mathfrak{R},d_z)=A_\lambda\otimes A_1\otimes\mathrm{Mat}_N\bigoplus M\otimes A_1\otimes\mathrm{Mat}_N=\mathfrak{A}\bigoplus\mathfrak{M}\,.
\end{equation}
The left summand $\mathfrak{A}$ is given by the matrix extension of the higher spin algebra $\hs=A_\lambda\otimes A_1$, the algebra where the one-form field $\omega$ assumes its values. The right summand $\mathfrak{M}$ defines then a bimodule over the algebra $\hs\otimes \mathrm{Mat}_N$, the target space of the zero-form field $C$. 
Recall that the differential in the algebra $\mathcal{R}$ is given by the trivial extension of the exterior differential $d_z$. As observed in \cite{Sharapov:2022awp}, the differential $d_z$ admits a nontrivial perturbation by another differential $\delta$ of degree one. The latter is defined as 
\begin{equation}\label{del}
\delta(b,a)=(\delta a, 0)\,,\qquad \delta a = a^\tau dz^1 dz^2\qquad\forall b\in \mathbf{R},\quad \forall a\in\mathfrak{M}\,.
\end{equation}
It is clear that $\delta^2=0$ and  $d_z\delta=-\delta d_z=0$. Eq. (\ref{klein}) ensures the graded Leibniz identity for the differential (\ref{del}) and the product (\ref{prod}). Therefore, the sum $D=d_z+\delta$ endows the algebra $\mathcal{R}$ with a new differential of degree one. It is not hard to see that the cohomology of the perturbed differential is given by the same algebra (\ref{hsA3}), that is, $H(\mathcal{R},D)\simeq H(\mathcal{R}, d_z)=\mathcal{A}$. 
Having the same cohomology, the dg-algebras $(\mathcal{R}, d_z)$ and $(\mathcal{R}, D)$ are not quasi-isomorphic to each other: the former algebra  is formal, i.e. it is quasi-isomorphic to it cohomology with trivial differential, whereas the latter is not -- although their cohomologies are isomorphic, the canonical map of underlying graded vector spaces does not define a morphism of dg-algebras. This fact implies that in addition to the binary product $m_2$ (induced by that in $\mathcal{R}$) the cohomology space  $H(\mathcal{R}, D)$ enjoys higher  multi-linear products $m_k$ making it into an $A_\infty$-algebra. (For the definition of an $A_\infty$-algebra see e.g. \cite{stasheff2018linfty}, \cite{IYUDU202163}.) This $A_\infty$-algebra, let us denote it by $\hat{\mathbb{A}}$, is called the {\it minimal model} of the dg-algebra $(\mathfrak{R}, D)$. By definition, the binary product $m_2$ coincides with the associative product in $\mathcal{A}$ and the triple product $m_3$ is given by a nontrivial  Hochschild cocycle representing a class of $HH^3(\mathcal{A},\mathcal{A})$.


Homological perturbation theory (which details can be found in Refs. \cite{HK,GLS, merkulov1999strongly}) provides explicit formulas for the multi-linear products $m_2, m_3, m_4, \ldots$ of the $A_\infty$-algebra $\hat{\mathbb{A}}$. All the products are constructed  as compositions of two basic operations: the contracting homotopy $h$ and the associative product in the multiplicative resolution $\mathcal{R}$. The latter gives rise to the coderivation $\mu$ defined by
$$
\mu (b,\tilde b)=(-1)^{\deg b-1}b\star \tilde b\,,\qquad \mu(b,a)=(-1)^{\deg b-1}b\circ a\,,\qquad \mu(a,b)=-a\circ b\,,
$$
for all $b,\tilde b\in \mathbf{R}$ and $a\in \mathfrak{M}$.
Suitable compositions are conveniently depicted by rooted planar trees. Each such a tree graph consists of  vertices, internal edges,  and external edges.  Both ends of an internal edge are on two vertices. All edges are oriented and orientation is indicated by an arrow. Each vertex has two incoming and one outgoing edge. An external edge has one end on a vertex and another end is free. The graphs are supposed to be connected. All the vertices correspond to the product $\mu$, whereas the internal edges depict the action of the contracting homotopy $h$:
$$ \begin{tikzcd}[column sep=small,row sep=small]
   & {}& \\
    & \mu\arrow[u]  & \\
    \arrow[ur]  & & \arrow[ul]   & 
\end{tikzcd} \;\;\;\;\;\;\;\;\;\;\;\;
\begin{tikzcd}[column sep=small,row sep=small]
    \mu & \\
    &\mu\arrow[ul, "h" ']   
\end{tikzcd}
$$
By definition, the algebra $\mathfrak{A}$ and the $\mathfrak{A}$-bimodule $\mathfrak{M}$ have degrees $-1$ and $0$, respectively, whereas all the products $m_k$ of the $A_\infty$-algebra $\hat{\mathbb{A}}$ are of degree one\footnote{In \cite{Sharapov:2022wpz} we used a different convention according to which all $m_k$'s are of degree $-1$. In that case, the elements of $\mathfrak{M}$ have still  degree $0$, whereas $\mathfrak{A}$ is placed in degree $1$.}. By degree considerations, each nonzero product $m_k$ may have either one or two  arguments in $\mathfrak{A}$ and the other in $\mathfrak{M}$. In the first case the image of $m_k$ belongs to the algebra $\mathfrak{A}$, whereas in the second to the bimodule $\mathfrak{M}$.  In field-theoretical terms, these two components of the product $m_k$ correspond to the interaction vertices of $\mathcal{V}$ and $\mathcal{U}$ types. Let us consider them separately. 

\paragraph{Two arguments in $\mathfrak{A}$.} The corresponding component of $m_k$ is described by the sum of trees with two branches, see left panel in Fig.  \ref{PT}. 
The incoming external edges (or leaves) correspond to the arguments of  $m_k$. More precisely, the arguments $b_1, b_2\in \mathfrak{A}$ may decorate only the four end leaves on {\it different} branches. The other leaves are decorated by the expressions 
$\Lambda[a_i]=h\delta a_i$ for $a_i\in \mathfrak{M}$. The only outgoing external edge (or root) corresponds to the value of the product $m_{k}(a_1,\ldots,b_1, \ldots, b_2, \ldots, a_{k-2})$ that arises after setting $z=0$.  The order of arguments is determined  by the natural order of incoming edges at each vertex of a planar tree. The contributions of different trees are added up (with unit weight) to obtain the desired $m_k$.

\paragraph{One argument in $\mathfrak{A}$.} The product $m_k(a_1,\ldots, b, \ldots,a_{k-1})$ is obtained by summing  up the one-branch trees; an example of such a tree is shown in the right panel of Fig. \ref{PT}. The only argument $b$ of  $\mathfrak{A}$ decorates one of the two end leaves, whereas the leaf incoming the root vertex is decorated by a `bare' element $a\in \mathfrak{M}$.  As  above, the order of arguments is determined  by the natural order of incoming edges at each vertex and the root edge symbolizes setting $z=0$ in the final expression for $m_k$. Unlike the previous case, the integrals defining the corresponding analytical expressions require a minor regularization as explained in the main text.
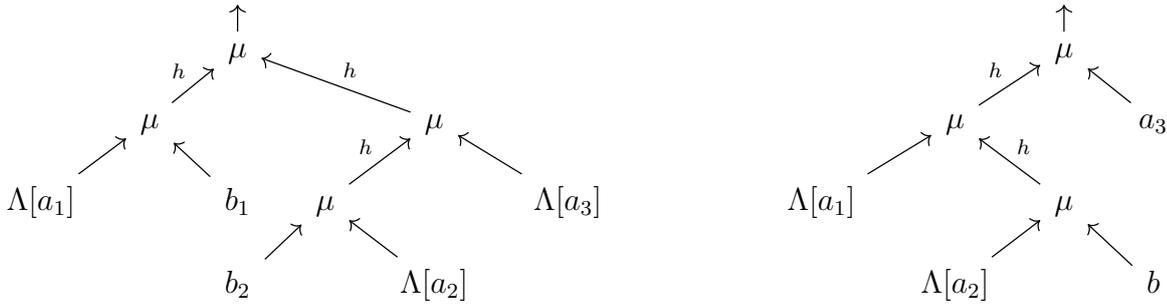
\begin{figure}
    \begin{tikzcd}[column sep=small,row sep=small]
   &&{}&&&&\\
    && \arrow[u]\mu  &&& \\
    & \mu\arrow[ur,"h"]& &&  \arrow[ull,"h"']\mu & & \\
    \Lambda[a_1]\arrow[ur]&& \arrow[ul]b_1 &   \mu\arrow[ur, "h" ]&&\arrow[ul]\Lambda[a_3]&\\
    &&b_2\arrow[ur]&&\Lambda[a_2]\arrow[ul]&&
\end{tikzcd}
   \begin{tikzcd}[column sep=small,row sep=small]
   &&&&&{}&&&&\\
   && &&&\mu \arrow[u]  &&& \\
    &&&&\mu\arrow[ru, "h" ]&&a_3\arrow[lu]&\\
    &&&\Lambda[a_1]\arrow[ur]&&\mu\arrow[lu, "h"']&&&\\
    &&&&\Lambda[a_2]\arrow[ur]&&b\arrow[ul]&&&
\end{tikzcd}
\caption{A planar rooted tree on the left panel corresponds to the analytical expression $h(h\delta a_1\star b_1)\star h(h(b_2\star h\delta a_2) \star h\delta a_3)|_{z=0}$ contributing to  $m_5(a_1,b_1,b_2,a_2,a_3)$; here $b_1,b_2\in \mathfrak{A}$ and $a_1,a_2,a_3\in \mathfrak{M}$. The right panel shows a planar  tree for the expression $h(h\delta a_1\star h(h\delta a_2\star b))\circ a_3|_{z=0}$, which contributes to $m_4(a_1,a_2,b,a_3)$; here $b\in \mathfrak{A}$ and $a_1,a_2,a_3\in \mathfrak{M}$. Notice the `bare' argument $a_3$.
}\label{PT}
\end{figure}

Finally, performing graded symmetrization  of the arguments of the products $m_k$  makes our $A_\infty$-algebra $\hat{\mathbb{A}}$ into a minimal $L_\infty$-algebra $\mathbb{L}$. At the level of interaction vertices such symmetrization is automatically achieved by substituting the form fields $\omega$ and $C$ instead of the arguments $b$'s and $a$'s. It is the $L_\infty$-algebra $\mathbb{L}$ that governs the interaction in  Chiral Theory. 

\section{Higher orders}
\label{app:trees}
Let us elaborate a bit more on the structure of HPT under consideration. Notice that the image of the differential (\ref{dd}) is not a polynomial function and the $\star$-product (\ref{pp}) of non-polynomial functions in $y$'s and $z$'s is ill-defined.  Therefore, one needs to make sure that the perturbation series (\ref{ps}) does make sense when applied to polynomial functions.

As it was already mentioned, there are many symmetries in flat space thanks to the commutativity of the $\star$-product (we denote it $\mu$), if $\lambda=0$, that the resolution is based on. The permutation symmetry over the legs attached to $\mu$-vertices is obvious. We would like to show that all nontrivial trees that contribute can be depicted as 
$$
\begin{tikzcd}[column sep=small,row sep=small]
&            & &                   &                           &       & {} &       &               &\\
&            & &                   &                           &       & \arrow[u]\mu &       &               &\\
&            & &                   &\arrow[urr,"h"]\mu    &       &                           &       &\arrow[ull,"h"']\mu&\\
&            & &\dots\arrow[ur,"h"]    &                           &\arrow[ul]\Lambda[u_i]      &                           &\dots\arrow[ur,"h"]  &              &\arrow[ul]\Lambda[u_{n-2}]\\
&           &\mu\arrow[ur,"h"]      &   & & &\mu\arrow[ur,"h"] & &   &\\
&\mu\arrow[ur,"h"] & & \arrow[ul]\Lambda[u_2]                  &                           &       \mu\arrow[ur,"h"]&  &\arrow[ul]\Lambda[u_{i+2}]                 &               &\\
a\arrow[ur]&            &\arrow[ul]\Lambda[u_1]                 & &                           b\arrow[ur]&       &       \arrow[ul]\Lambda[u_{i+1}]& & &                           
\end{tikzcd}
$$
In words, the tree consists of two branches, each having one leaf with an argument from the algebra $A_1$, $a$ or $b$ here. Apart from $a$, or $b$ each of the two branches has only simple leaves with $\Lambda[u_i]=h\delta u_i$, where $u_i$ belong to the module $A_0$. The branches may have different lengths. The graph above is a contribution to the  $A_\infty$-map $m_n$ with $n$ arguments in total:
\begin{align}
    m(a,u_1,\ldots,u_i,b,u_{i+1},\ldots,u_{n-2})\,.
\end{align}
There is a number of simple observations that reduce the variety of trees to the class we described (we introduce one-form $A$ in $z$-space as $A=\Lambda[u_i]$): (i) $h$ cannot be the last operation on a tree since we can set $z=0$ at the end and $h$ has $z$-factor; (ii) $h^2\equiv0$ is obvious; (iii) there are no three-forms, hence, $A\star A\star A \equiv0$; (iv) one can also see that $h(A\star A)\equiv0$; (v) it follows from (i) that the final result should be the product $\mu(T_1,T_2)$ of two sub-trees $T_{1,2}$, each of which being zero-form. In particular, each $T_i$ must have all $A$ balanced by $h$. Given the rules above, it is impossible to construct a zero-form tree only with $A$'s. Therefore, each $T_i$ must have one and only one of the arguments in the algebra, e.g. $a$ belongs to $T_1$ and $b$ belongs to $T_2$. Let us zoom in on one of the two sub-trees. We could see two pictures: \\
\phantom{a}\hspace{2cm}\begin{tikzcd}[column sep=small, row sep=small]
                &                           &\dots\\
                &   \mu\arrow[ur,"h"] & \\
    f(a,\dots)\arrow[ur]  &                           &\arrow[ul]A
\end{tikzcd}\hspace{2cm}
\begin{tikzcd}[column sep=small, row sep=small]
                &                           &\dots           &\\
                &   \mu\arrow[ur,"h"] &                           &\\
    f(a,\dots)\arrow[ur]  &                           &\mu\arrow[ul]    &\\
                &A\arrow[ur]                          &                           &A\arrow[ul]
\end{tikzcd}\\
In fact, the second option is inconsistent. It gives a one-form and we have to find a way to make the whole sub-tree be zero-form at the end. We cannot attach a zero-form sub-tree (and apply $h$ afterwards) since $b$ is in another sub-tree. We can only attach $A$ or any other one-form sub-tree, but this leads to a two-form, i.e. to the original problem we are trying to solve. We are in a vicious circle. Therefore, the second option cannot be realized. $\blacksquare$

\paragraph{Locality.} It is important to prove that the vertices are local in the sense of not having $p_{ij}$ in the exponent that contract some of the zero-form arguments. Given the result above, we can have a look at the general structure of one of the branches. It is easy to see that it has the following general form:
\begin{align}\label{induct}
    h&\left(\cdots h\left(a\star \Lambda[c_2]\right)\dots\star \Lambda[c_n]\right)=\nonumber\\
    &=\eta\left(z p_1\right)^{n-1}\exp\left[\gamma_0z y+\gamma_1y p_1+\gamma_2z p_2+\dots+\gamma_nz p_n+\zeta_2p_{12}+\dots+\zeta_np_{1n}\right]\times\\
    &\times a\left(y_1\right)c_2\left(y_2\right)\cdots c_n\left(y_n\right)\,,\notag
\end{align}
where $\eta$, $\gamma_i$ and $\zeta_i$ are certain functions of the integration variables $t_k$ that originate from multiple applications of $h$. The integral sign is omitted. Indeed, we begin with the lowest possible expression to start the induction
\begin{align*}
    h\left(a\star \Lambda[c]\right)=\int_0^1dt_1\int_0^{t_1}dk_1\left(z p_1\right)\exp\left[y p_1\left(1-t_1\right)+z y k_1+z p_2k_1+p_{12}t_1\right] a\left(y_1\right)c\left(y_2\right)\,.
\end{align*}
Assuming the structure is as in \eqref{induct} we attempt to proceed to the next order to find
\begin{align*}
    &\eqref{induct}\star \Lambda[c_{n+1}]=dz^{A}\,\eta'\left(z p_1\right)^{n-1}\left(\alpha_0z_{A}+\alpha_1p_{A}^1\right)\times\nonumber\\
    \times&\exp\left[\gamma'_0z y+\gamma_1y p_1+\gamma'_2z p_2+\dots+\gamma'_nz p_n+\gamma'_{n+1}z p_{n+1}+\right.\nonumber\\
    +&\left.\zeta'_2p_{12}+\dots+\zeta'_np_{1n}+\zeta'_{n+1}p_{1\ n+1}\right]\times a\left(y_1\right)c_2\left(y_2\right)\cdots c_n\left(y_n\right)c_{n+1}\left(y_{n+1}\right)\,.
\end{align*}
Applying $h$ to the expression here-above we clearly reproduce \eqref{induct}. Now, we can compute the $\mu$-product of two expressions of type \eqref{induct} to see that the final answer has the desired property of being local. $\blacksquare$

\section{Pre-Calabi--Yau algebras and duality map}\label{CY}

The above construction of the $A_\infty$-algebra $\hat{\mathbb{A}}$ by means of homological perturbation theory is  absolutely insensitive to the  choice of the tensor factor $B=A_1\otimes \mathrm{Mat}_N$. For any associative algebra $B$ we get $\hat{\mathbb{A}}=\mathbb{A}\otimes B$, where the minimal $A_\infty$-algebra $\mathbb{A}$ extends the binary product in $A_\lambda\oplus M$. Furthermore, the $A_\lambda$-bimodule $M$ is actually dual to the algebra $A_\lambda$ viewed as the natural bimodule over itself, i.e., $M\simeq A_\lambda^\ast$.  The corresponding nondegenerate pairing is given by 
\begin{align}\label{cpar}
     \langle a| u\rangle =e^{p_{12}}a(y_1)u(y_2)|_{y_i=0}\,,\qquad \forall a\in A_\lambda\,,\quad \forall u\in M\,.
\end{align}
One can easily verify that $ \langle b\star a\star c| u\rangle =  \langle a| c\circ u\circ b\rangle$. Recall that the elements of the algebra $A_\lambda$ are prescribed, by definition, the degree $-1$, whereas the elements of the bimodule $M$ live in degree $0$. With this convention all the products $m_k$ in $\mathbb{A}$ have degree one. By the above isomorphism, we can write\footnote{Dualization inverts the $\mathbb{Z}$-degree, while the symbol $[1]$ shifts the degree of the dual module by one. } $A_\lambda\oplus M\simeq A_\lambda\oplus A^\ast_\lambda[1]$. The pairing (\ref{cpar}) gives rise to a canonical symplectic form $\omega$ on the graded vector space $A_\lambda\oplus A^\ast_\lambda[1]$. This is defined as
\begin{align}\label{sf}
   \omega ( a+u ,\tilde a +\tilde u)= \langle a| \tilde u\rangle - \langle \tilde{a}|u\rangle \,.
\end{align}
Clearly, $\deg \omega =1$. Define the sequence of multi-linear forms 
\begin{align}\label{Sk}
    S_k(\alpha_0, \alpha_1,\ldots,\alpha_k)=\omega\big(\alpha_0, m_{k}(\alpha_1,\ldots,\alpha_k)\big)\,,\qquad k=2,3,\ldots\,,
\end{align}
where $\alpha=a+u\in A_\lambda\oplus A^\ast_\lambda[1]$. By definition, the $A_\infty$-algebra $\mathbb{A}$  is called {\it cyclic} (w.r.t. $\omega$) if
\begin{align}\label{Scycle}
      S_k(\alpha_0,\alpha_1,\ldots,\alpha_k)=(-1)^{\bar{\alpha}_0(\bar\alpha_1+\cdots+\bar\alpha_{k})}   S_k(\alpha_1,\ldots,\alpha_{k},\alpha_0)\,,
\end{align}
where $\bar\alpha=\deg \alpha-1$. A direct verification shows that the above identities are indeed satisfied. Hence, $\mathbb{A}$ is a cyclic $A_\infty$-algebra.  The other two properties of $\mathbb{A}$ -- shifted duality $M=A^\ast_\lambda[1]$ and the fact that $A_\lambda$ is a subalgebra of $\mathbb{A}$ -- allows us to classify   $\mathbb{A}$ as a $2$-pre-Calabi--Yau algebra \cite{IYUDU202163},
\cite{kontsevich2021pre}. The general definition is as follows. 
\begin{definition}
A $d$-pre-Calabi--Yau structure on an $A_\infty$-algebra $A$ is a cyclic $A_\infty$-structure on $A\oplus A^\ast[1-d]$, associated with the natural pairing between $A$ and $A^\ast[d-1]$,  such that $A$ is an $A_\infty$-subalgebra in $A\oplus A^\ast[1-d]$.
\end{definition}

In our case, $d=2$ and the role of an $A_\infty$-algebra $A$ is played by the associative algebra $A_\lambda$.  The latter is clearly a subalgebra in $\mathbb{A}$.  The cyclicity property (\ref{Scycle}) relates various structure maps $m_k$ among themselves. In particular, it connects the components of  the $m_k$'s with one and two arguments in $A_\lambda$:
$$
\langle a_1|m_{k+1}(u_1,\ldots, a_2,\ldots, u_k)\rangle=-\langle m_{k+1}( u_2,\ldots, a_2,\ldots, u_{k},a_1)|u_1\rangle\,.
$$
In the main text, we use these relations to express the $\mathcal{U}$-vertices via $\mathcal{V}$-vertices. 

In the case that the associative algebra $B$ enjoys a trace, one can easily extend the 2-pre-Calabi--Yau structure from $\mathbb{A}$ to the tensor product $\hat{\mathbb{A}}=\mathbb{A}\otimes B$. The symplectic structure  extends as
\begin{equation}\label{ww}
\Omega(\alpha\otimes b,\tilde \alpha\otimes\tilde b)=\omega(\alpha,\tilde\alpha)\mathrm{Tr}(b\tilde b)\qquad \forall \alpha,\tilde \alpha\in \mathbb{A}\,,\quad \forall b,\tilde b\in B\,,
\end{equation}
and the multi-linear functions (\ref{Sk}) take the form
\begin{equation}\label{SSk}
{\bf{S}}_k(\alpha_0\otimes b_0,\ldots, \alpha_k\otimes b_k)=S_k(\alpha_0,\ldots,\alpha_k)\mathrm{Tr}(b_0\cdots b_k)\,.
\end{equation}
The cyclic invariance (\ref{Scycle}) of the ${\bf S}_k$'s is obvious. 

Following the ideas of noncommutative geometry \cite{Kontsevuch:2006jb}, one can regard the cyclic forms (\ref{SSk}) as functions on a noncommutative manifold associated with $\hat{\mathbb{A}}$. The constant symplectic structure (\ref{ww}) gives then rise to a kind of Gerstenhaber bracket on the space of such functions, called {\it necklace bracket} \cite{kontsevich2021pre}.  This can be viewed as a noncommutative counterpart of the Schouten--Nijenhuis bracket on polyvector fields. It is convenient to combine the functions (\ref{SSk})  into a single non-homogeneous function $\mathbf {S}=\sum_{k=2}^\infty \mathbf{S}_k$. With the help of the necklace bracket all  $A_\infty$-structure relations for $\hat{\mathbb{A}}$ can be compactly encoded by the equation
$
    [\mathbf{S},\mathbf{S}]_{\mathrm{nec}}=0
$\,.
On passing from the $A_\infty$-algebra $\hat{\mathbb{A}}$ to the associated $L_\infty$-algebra $\mathbb{L}$, the last equation turns into the Batalin--Vilkovisky equation for the `classical master action' $\mathbf{S}(\omega, C)$ of ghost number 2 on the target space of form fields $\omega$ and $C$.
Geometrically, one can regard $\mathbf{S}(\omega, C)$ as a Poisson bivector on the space of fields $C$. Upon this interpretation the field equations  (\ref{toSolve}) define a  Poisson sigma-model in four dimensions. Schematically, 
\begin{align}
    d C^i &= \pi^{ij}(C)\, \omega_j\,, &d\omega_k&= \tfrac12\pl_k \pi^{ij}(C)\, \omega_i\, \omega_j\,,
\end{align}
where the Poisson bivector $\pi^{ij}(C)$ is read off from $\mathbf{S}=\pi^{ij}(C)\omega_i\omega_j$. 

\section{Consistency at NNLO}
\label{app:tests}
It is reassuring to check the consistency of the quartic vertices directly, which, in particular, makes sure that the signs/coefficients are correct. We add the quartic term to $d\omega$ and $dC$:
\begin{align*}
    \begin{aligned}
    d\omega&=V(\omega,\omega)+\mathcal{V}_1(\omega,\omega,C)+\mathcal{V}_2(\omega,C,\omega)+\mathcal{V}_3(C,\omega,\omega)+\mathcal{V}_1(\omega,\omega,C,C)+\mathcal{V}_2(\omega,C,\omega,C)\,,\\
    &+\mathcal{V}_3(\omega,C,C,\omega)+\mathcal{V}_4(C,\omega,C,\omega)+\mathcal{V}_5(C,\omega,\omega,C)+\mathcal{V}_6(C,C,\omega,\omega)\\
    dC&=\mathcal{U}_1(\omega,C)+\mathcal{U}_2(C,\omega)+\mathcal{U}_1(\omega,\omega,C)+\mathcal{U}_2(\omega,C,\omega)+\mathcal{U}_3(C,\omega,\omega)+\mathcal{U}_1(\omega,C,C,C)\\
    &+\mathcal{U}_2(C,\omega,C,C)+\mathcal{U}_3(C,C,\omega,C)+\mathcal{U}_4(C,C,C,\omega)\,.
    \end{aligned}
\end{align*}
The consistency condition $0\equiv d^2\omega$ can be split into $10$ equations for different ordering of $\omega\omega\omega C C$. Let us have a look at some of them. We begin with the final answer --- expressions for the vertices in terms of the two graphs $G_1$ and $G_2$ that contribute at the second order:
\begin{align*}
    \mathcal{V}_1(\omega,\omega,C,C)&=G_1(\omega,\omega,C,C)\,,\\
    \mathcal{V}_2(\omega,C,\omega,C)&=-(\sigma_{(423)} G_1)(\omega,C,\omega,C) - (\sigma_{(23)}G_1)(\omega,C,\omega,C) + G_2(\omega,C,\omega,C)\,,\\
    V_3(\omega,C,C,\omega)&=(\sigma_{(24)}G_1)(\omega,C,C,\omega)+(\sigma_{(1432)}G_1)(\omega,C,C,\omega)-(\sigma_{(34)}G_2)(\omega,C,C,\omega)\,,\\
    \mathcal{V}_4(C,\omega,C,\omega)&=-(\sigma_{(124)}G_1)(C,\omega,C,\omega)-(\sigma_{(1243)}G_1)(C,\omega,C,\omega)+(\sigma_{(12)(34)}G_2)(C,\omega,C,\omega)\\
    \mathcal{V}_5(C,\omega,\omega,C)&=-(\sigma_{(12)}G_2)(C,\omega,\omega,C)\,,\\
    \mathcal{V}_6(C,C,\omega,\omega)&=(\sigma_{(14)(23)}G_1)(C,C,\omega,\omega)\,,
\end{align*}
where $\sigma_{(...)...(...)}$ is the standard notation for the decomposition of a given permutation into disjoint cycles. The $\mathcal{U}$-vertices can be obtained via the duality. For example, 
\begin{align*}
    \mathcal{U}_1(\omega,C,C,C)=G_1(\omega,C,C,C)(-p_4,p_0,p_1,p_2,p_3)\,.
\end{align*}

Now let us check directly that some of the $A_\infty$-relations must be satisfied. One of the simplest integrability conditions reads 
\begin{align*}
    -\omega \mathcal{V}_1(\omega,\omega,C,C)+\mathcal{V}_1(\omega^2,\omega,C,C) -\mathcal{V}_1(\omega,\omega^2,C,C)+\mathcal{V}_1(\omega,\omega,\mathcal{U}(\omega, C),C)\\
    \qquad+\mathcal{V}_1(\omega,\omega,\mathcal{U}_1(\omega,C,C))-\mathcal{V}_1(\omega,\mathcal{V}_1(\omega,\omega,C),C)=0\,,
\end{align*}
which can be rewritten in terms of symbols as 
\begin{align*}
    \begin{aligned}
    &-\exp(p_{01})V_1(p_0,p_2,p_3,p_4,p_5)+V_1(p_0,p_1,p_2,p_6)\mathcal{U}_1(y_6,p_3,p_4,p_5)-V_1(p_0,p_1,p_2+p_3,p_4,p_5)\\
    &+\exp(p_{34})V_1(p_0,p_1,p_2,p_4,p_5)-V_1(p_0,p_1,p_6,p_5)V_1(y_6,p_2,p_3,p_4)+V_1(p_0,p_1+p_2,p_3,p_4,p_5)=0\,.
    \end{aligned}
\end{align*}
Nesting one vertex into another is easy to evaluate thanks to the exponential form of the vertices. We find
\begin{align*}
    \begin{aligned}
    &V_1(p_0,p_1,p_2,p_6)\mathcal{U}_1(y_6,p_3,p_4,p_5)=\int_{0}^{1}dt_2\int_{0}^{t_2}dt_1\int_{0}^{1}ds_2\int_{0}^{s_2}ds_1p_{12}(t_1p_{13}+t_2p_{23})\\
    &\times\exp((1-t_1)p_{01}+(1-t_2)p_{02}+t_1(1-s_2)p_{14}+t_1s_2p_{15}+t_2(1-s_2)p_{24}+t_2s_2p_{25}+(1-s_1)p_{34}+s_1p_{35}),\\
    &V_1(p_0,p_1,p_6,p_5)V_1(y_6,p_2,p_3,p_4)=\int_{0}^{1}dt_2\int_{0}^{t_2}dt_1\int_{0}^{1}ds_2\int_{0}^{s_2}ds_1p_{23}((1-s_1)p_{12}+(1-s_2)p_{13})\\
    &\times\exp((1-t_1)p_{01}+(1-t_2)(1-s_1)p_{02}+(1-t_2)(1-s_2)p_{03}+t_1p_{15}+t_2(1-s_1)p_{25}+t_2(1-s_2)p_{35}\\
    &+s_1p_{24}+s_2p_{34}).
    \end{aligned}
\end{align*}
It is now easy to see that the consistency condition is satisfied order by order with $\mathcal{V}_1(\omega,\omega,C,C)=G_1(\omega,\omega,C,C)$. In particular, the integrals, after Taylor expansion, can easily be done and lead to simple rational numbers. We have checked enough consistency relations to support the expressions for the vertices.

\section{All order vertices}
\label{app:allorders}

\subsection{Jacobians}
\label{app:Jacobians}
Here we compute the Jacobians introduced in sections \ref{sec:flat} and \ref{sec:uvertices}.
\paragraph{Single branch.} Eqs. \eqref{B_{n+1}} and \eqref{B_{n+1}newcoordinates} are related by a change of variables \newline $\{u_{n,1},v_{n,1},\dots,u_{n,n},v_{n,n},t_{2n+1},t_{2n+2}\}$ to $\{u_{n+1,1},v_{n+1,1},\dots u_{n+1,n+1},v_{n+1,n+1}\}$ with the Jacobian
\begin{align} \label{det}
    |J_n|=
    \begin{vmatrix}
        \tfrac{(1-t_{2n+1})t_{2n+2}}{1-t_{2n+1}U_n}\delta_{ij} & 0 & 0 & u_{n,i} \delta_{jj} \\
        -\tfrac{t_{2n+1}(1-V_n)}{1-t_{2n+1}U_n}\delta_{ij} & \delta_{ij} & 0 & 0 \\
        \tfrac{t_{2n+1}t_{2n+2}(t_{2n+1}-1)}{(1-t_{2n+1}U_n)^2} \delta_{jj} & 0 & \tfrac{t_{2n+2}(1-U_n)}{(1-t_{2n+1}U_n)^2} & \frac{t_{2n+1}(1-U_n)}{1-t_{2n+1}U_n} \\
        \tfrac{t_{2n+1}^2 (1-V_n)}{(1-t_{2n+1}U_n)^2}\delta_{jj} & -\tfrac{t_{2n+1}}{1-t_{2n+1}U_n}\delta_{jj} & \tfrac{1-V_n}{1-t_{2n+1}U_n} & 0
    \end{vmatrix} \,,
\end{align}
where $i,j=1,\dots,n$. Keeping in mind that some entries are vectors or matrices, Gaussian elemination allows one to find a diagonal form. To give an example of the steps taken during this process, one can multiply the matrix in the second row of \eqref{det} by $\frac{t_{2n+1}}{1-t_{2n+1}U_n}$ and add each of its rows to the last row in \eqref{det}. After a few manipulations, one arrives at
\begin{align*}
    |J_n|&=\left| \text{diag}\Big(\tfrac{(1-t_{2n+1})t_{2n+2}}{1-t_{2n+1}U_n}\delta_{ij} + \tfrac{(1-t_{2n+1})t_{2n+1}t_{2n+2}}{(1-t_{2n+1}U_n)^2} u_{n,i}\delta_{jj},\;\delta_{ij}+\tfrac{t_{2n+1}}{1-t_{2n+1}U_n}u_{n,i}\delta_{jj},\;\tfrac{1-V_n}{1-t_{2n+1}U_n}, \;t_{2n+1}\Big)\right| \,.
\end{align*}
Notice that the matrix is not completely diagonal as not all of its blocks are proportional to $\delta_{ij}$. We obtain
\begin{align*}
    |J_n|&=\frac{t_{2n+1}(1-V_n)}{1-U_n}\text{det}\Big(\tfrac{(1-t_{2n+1})t_{2n+2}}{1-t_{2n+1}U_n}\delta_{ij} + \tfrac{(1-t_{2n+1})t_{2n+1}t_{2n+2}}{(1-t_{2n+1}U_n)^2} u_{n,i}\delta_{jj}\Big) \text{det}\Big(\delta_{ij}+\tfrac{t_{2n+1}}{1-t_{2n+1}U_n}u_{n,i}\delta_{jj}\Big) \,.
\end{align*}
Applying Sylvester's determinant theorem, $\text{det}(I+xy^T)=1+x^Ty$, gives
\begin{align*}
    |J_n|=\frac{t_{2n+1}}{(1-t_{2n+1}U_n)^2}\left(\frac{(1-t_{2n+1})t_{2n+2}}{1-t_{2n+1}U_n}\right)^n\frac{1-V_n}{1-t_{2n+1}U_n}\,,
\end{align*}
which is exactly the prefactor in \eqref{B_{n+1}}.
\paragraph{Two branches.}
Another change of coordinates was applied to go from  \eqref{tree} to \eqref{twobranches}. Here the coordinates  $\{u^L_{n,1},\dots,v^L_{n,n},u^R_{m,1},\dots,v^R_{m,m}\}$ were replaced with $\{r^L_{n,1},\dots,s^L_{n,n},r^R_{m,1},\dots,s^R_{m,m}\}$. The corresponding Jacobian reads
\begin{align*}
    |J_n|=
    \left\lvert
    \begin{matrix}
        \tfrac{1-V_m}{1-U_mU_n}\delta_{ij}+\tfrac{(1-V_m)U_m}{(1-U_mU_n)^2}u_{n,i}\delta_{jj} & 0 & \tfrac{(1-V_m)U_n}{1-U_mU_n}u_{n,i}\delta_{jj} &                              -\tfrac{u_{n,i}\delta_{jj}}{1-U_mV_n} \\
        -\tfrac{U_m(1-V_n)}{1-U_mU_n} \delta_{ij} & \delta_{ij} & 0 & 0 \\
        \tfrac{U_m(1-V_n)}{(1-U_mU_n)^2}u_{m,i}\delta_{jj} & -\tfrac{u_{m,i}\delta_{jj}}{1-U_mU_n} & \tfrac{1-V_n}{1-U_mU_n}\delta_{ij}+\tfrac{U_n(1-V_n)}{(1-U_mU_n)^2}u_{m,i}\delta_{jj} & 0 \\
        0 & 0 & -\tfrac{U_n(1-V_m)}{1-U_mU_n}\delta_{ij} & \delta_{ij}
    \end{matrix}
    \right\rvert\,.
\end{align*}
Gaussian elimination allows one to rewrite this as
\begin{align*}
    |J_n|=
    \begin{vmatrix}
    A & 0 & 0 & 0 \\
    B & C & 0 & 0 \\
    0 & 0 & D & 0 \\
    0 & 0 & E & F
    \end{vmatrix}
    =|A||C||D||F| \,,
\end{align*}
where
\begin{align*}
    A&=\frac{1-V_m}{1-U_mU_n}\delta_{ij}+\frac{(1-V_m)U_m}{(1-U_mU_n)^2}u_{n,i}\delta_{jj}\,, &  D&=\frac{1-V_n}{1-U_mU_n}\delta_{ij}+\frac{U_n(1-V_n)}{(1-U_mU_n)^2}u_{m,i}\delta_{jj}\,,\\
    B&=-\frac{U_m(1-V_n)}{1-U_mU_n} \delta_{ij} \,, &E&=-\frac{U_n(1-V_m)}{1-U_mU_n}\delta_{ij}\,,  \\
    C&=\delta_{ij}\,, &   F&=\delta_{ij}\,.
\end{align*}
Sylvester's determinant theorem now states that
\begin{align*}
    |J_n|=|A||D|=\frac{1}{(1-U_mU_n)^2}\left(\frac{1-V_m}{1-U_mU_n}\right)^n\left(\frac{1-V_n}{1-U_mU_n}\right)^m\,,
\end{align*}
which is the prefactor in \eqref{tree} up to the alternating minus sign.

\paragraph{$\mathcal{U}$-vertices.} The determinant of the Jacobian corresponding to the change of variables \eqref{changezeroform} reads
\begin{align*}
    |J| &=
    \begin{vmatrix}
        +\frac{\epsilon}{1-U_n(1-\epsilon)}\delta_{ij}+\frac{\epsilon(1-\epsilon)}{(1-U_n(1-\epsilon))^2}u_{n,i}\delta_{jj} & 0 \\
        -\frac{(1-V_n)(1-\epsilon)}{1-U_n(1-\epsilon)}\delta_{ij} - \frac{(1-V_n)(1-\epsilon)^2}{(1-U_n(1-\epsilon))^2}u_{n,i}\delta_{jj} & \delta_{ij} + \frac{1-\epsilon}{1-U_n(1-\epsilon)}u_{n,i}\delta_{jj}
    \end{vmatrix} \,.
\end{align*}
Using Sylvester's determinant theorem yields
$|J| = (\frac{1}{1-U_n(1-\epsilon)})^2(\frac{\epsilon}{1-U_n(1-\epsilon)})^n$. This is identified with the prefactor in \eqref{bigbranch} in the limit $\varepsilon \rightarrow 0$.

\subsection{Compactness of integration domain}
\label{app:domain}
\paragraph{Single branch.} The change of variable \eqref{newcoordinates} determines the domain of integration in \eqref{B_{n+1}newcoordinates}. We are interested in knowing if this domain is compact or not. It is useful to start with deriving some properties of $U_n$ and $V_n$. The $t_i$'s run from $0$ to $1$, hence
\begin{align*} 
    1-U_{n+1}& \geq \frac{(1-t_{2n+1})(1-U_n)}{1-t_{2n+1}U_n} \geq 0
\end{align*}
whenever $U_n \leq 1$. Since $U_1=t_1t_2 \leq 1$, it follows that $U_n \leq 1$ for all $n \geq 1$. Using this result, we find
\begin{align*}
    U_{n+1}&=\frac{(1-t_{2n+1})U_n+(1-U_n)t_{2n+1}}{1-t_{2n+1}U_n} t_{2n+2} \geq 0 \,.
\end{align*}
Similarly,
\begin{align*}
    1-V_{n+1}&=\frac{(1-V_n)(1-t_{2n+1})}{1-t_{2n+1}U_n} \geq 0
\end{align*}
if $V_n \leq 1$. Since $V_1=t_1$, we conclude that $V_n \leq 1$ for all $n \geq 1$.
\begin{align*}
    V_{n+1}&=\frac{V_n(1-t_{2n+1})+t_{2n+1}(1-U_n)}{1-t_{2n+1}U_n} \geq 0
\end{align*}
for $V_n \geq 0$. As $V_1=t_1$ we conclude that $V_n \geq 0$ for all $n \geq 1$. Using the above result we see that
\begin{align*}
    V_{n+1}-U_{n+1}& = \frac{(V_n-U_n t_{2n+1})(1-t_{2n+1})+t_{2n+1}(1-U_n)(1-t_{2n+2})}{1-t_{2n+1}U_n} \geq 0
\end{align*}
provided that $V_{n} \geq U_n$. Since $U_1=t_1t_2$, $V_1=t_1$, and  $V_1 \geq U_1$, we conclude by induction that $V_n \geq U_n$ for all $n \geq 1$.

Now the restrictions on the individual variables should be more obvious. It is useful to think of the variable $u_{n+m,n}$, with $m \geq 1$, as originating from $u_{n,n}$ when the first relation in \eqref{newcoordinates} is applied  $m$ times. The same is true for $v_{n+m,n}$. It is therefore convenient to first study the properties of $u_{n,n}$ and $v_{n,n}$. It is easy to see that
\begin{align*}
    0 &\leq u_{n+1,n+1}=\frac{1-U_n}{1-t_{2n+1}U_n}t_{2n+1}t_{2n+2} \leq 1
\end{align*}
and
\begin{align*}
   0 &\leq  v_{n+1,n+1}=\frac{1-V_n}{1-t_{2n+1}U_n}t_{2n+1} \leq 1 \,.
\end{align*}
Then
\begin{align*}
    0 &\leq u_{n+m,n}=\frac{1-U_{n+m-1}}{1-t_{2(n+m)-1}U_{n+i-1}} u_{n+m-1,n} \leq 1
\end{align*}
whenever $0 \leq u_{n+m-1,n} \leq 1$. By induction we find that $0 \leq u_{n+m,n} \leq 1$, as 0 $\leq u_{n,n} \leq 1$. As a result all $u$ variables belong to the interval $[0,1]$. For the $v$ variables we find
\begin{align*}
    v_{n+m,n}&=v_{n+m-1,n}-u_{n+m-1,n}\frac{t_{2(n+m)-1}(1-V_{n+m-1})}{1-t_{2(n+m)-1}U_{n+m-1}} \leq v_{n+m-1,n} \,.
\end{align*}
Again,  proceeding by induction and using the fact that $v_{n,n} \leq 1$ we conclude that $v_{n+m,n} \leq 1$. To prove that these variables are also nonnegative requires a bit more work. We will use the relation
\begin{align} \label{centersofmass}
    \frac{1-U_n}{1-U_{n-1}}& \geq \frac{1-t_{2n-1}}{1-t_{2n-1}U_{n-1}} = \frac{1-V_n}{1-V_{n-1}} \,.
\end{align}
We have
\begin{align*}
    v_{n+m,n}&\geq v_{n+m-1,n}-u_{n+m-1}\frac{1-V_{n+m-1}}{1-U_{n+m-1}}=\\
    &=v_{n+m-2,n}-u_{n+m-2}(\frac{t_{2(n+m)-3}(1-V_{n+m-2})}{1-t_{2(n+m)-3}U_{n+m-2}}+\frac{(1-t_{2(n+m)-3})t_{2(n+m)-2}}{1-t_{2(n+m)-3}U_{n+m-2}}\frac{1-V_{n+m-1}}{1-U_{n+m-1}}) \geq \\
    & \geq v_{n+m-2,n}-u_{n+m-2,n}\frac{1-V_{n+m-2}}{1-U_{n+m-2}} \geq \dots \geq v_{n,n}-u_{n,n}\frac{1-V_n}{1-U_n}\,.
\end{align*}
The equalities arise from setting $t_i=1$ for even $i$ and going from the second to the third line we used \eqref{centersofmass}. It only remains to show that
\begin{align*}
    v_{n,n}-u_{n,n}\frac{1-V_n}{1-U_n}& \geq \frac{t_{2n-1}(1-V_{n-1})}{1-t_{2n-1}U_{n-1}}(1-t_{2n+2}) \geq 0 \,,
\end{align*}
which proves that $v_{n+m,n} \geq 0$. Ultimately, we have shown that all $u$ and $v$ variables belong to the interval $[0,1]$, although they obey even stricter restrictions, which will be discussed in the next section. Moreover, the $U_n$ and $V_n$ are restricted to the interval $[0,1]$ as well and the domain of integration for a single branch is thus a subspace of the hypercube $[0,1]^{2n}$.

\paragraph{Trees.} Another change of variables is proposed in \eqref{newcoordinatestwobranches}. As we know that all $u$ and $v$ variables and their sums $U_n$ and $V_n$ belong to the interval $[0,1]$ and that $V_n \geq U_n$, it is not hard to see that
\begin{align*}
    0 &\leq r^L_{n,i}=\frac{1-V^L_n}{1-U^L_{n}U^R_m}u^L_{n,i} \leq 1
\end{align*}
and
\begin{align*}
    s^L_{n,i} \leq v^L_{n,i} \leq 1 \,.
\end{align*}
We also find that
\begin{align*}
    s^L_{n,i}&\geq v^L_{n,i}-u^L_{n,i}\frac{1-V^L_n}{1-U^L_n} \geq 0 \,,
\end{align*}
where the latter relation coincides with $v_{n+1,i} \geq 0$ for a single branch. Obviously, the same properties hold for $r^R_{m,i}$ and $s^R_{m,i}$ and consequently the domain of integration for a tree consisting of two branches with length $n$ and $m$ belongs to the interval $[0,1]^{2(n+m)}$.

\subsection{Full domain of integration}
\label{app:fulldomain}
We will establish an analogous configuration space for branches of arbitrary length and eventually for all trees.

\paragraph{A single branch.}
Let us start by considering Rel. \eqref{newcoordinates}. Keeping in mind that all $u$ and $v$ variables and their sums $U_n$ and $V_{n}$ belong to the interval $[0,1]$, some relations may be derived. It is, however, hard to prove any relations between the variables at the same level $n$. It is therefore useful to think of $u_{n,i}$ as originating from $u_{i,i}$ and having moved up $n-i$ levels using the first relation in \eqref{newcoordinates}. The same is true for $v_{n,i}$. Thus, we first start by evaluating
\begin{align*}
    \frac{u_{n+1,n+1}}{v_{n+1,n+1}}&=\frac{t_{2n+2}(1-U_n)}{1-V_n} \leq \frac{1-U_n}{1-V_n} \leq \frac{1-U_{n+1}}{1-V_{n+1}} \,,
\end{align*}
where we have used \eqref{centersofmass} and equality is obtained for $t_{2n+2}=1$. Next, we consider
\begin{align} \label{eq0}
    \frac{v_{n+1,i}}{u_{n+1,i}} = \frac{1-t_{2n+1}U_n}{(1-t_{2n+1})t_{2n+2}}\frac{v_{n,i}}{u_{n,i}}-\frac{t_{2n+1}(1-V_n)}{1-t_{2n+1}U_n} \geq \frac{1}{t_{2n+2}}\frac{v_{n,i}}{u_{n,i}}
\end{align}
and inverting this gives
\begin{align*}
    \frac{u_{n+1,i}}{v_{n+1,i}} \leq t_{2n+2}\frac{u_{n,i}}{v_{n,i}} \leq \frac{u_{n,i}}{v_{n,i}} \,.
\end{align*}
By induction we find
\begin{align*}
    \frac{u_{n+1,i}}{v_{n+1,i}}& \leq t_{2n+2}\frac{u_{n,i}}{v_{n,i}} \leq \dots \leq t_{2n+2} \frac{u_{i,i}}{v_{i,i}} \leq \\
    &\leq t_{2n+2} \frac{1-U_{i-1}}{1-V_{i-1}} \leq t_{2n+2} \frac{1-U_n}{1-V_{n}} = \frac{u_{n+1,n+1}}{v_{n+1,n+1}}\,,
\end{align*}
where again we made use of \eqref{centersofmass}. Equality is obtained if $t_{2k+1}=0$ and $t_{2k}=1$ for all $k\in[i,n]$. Now we consider the relation between $\frac{u_{n+1,i}}{v_{n+1,i}}$ and $\frac{u_{n+1,j}}{v_{n+1,j}}$ for $i<j<n+1$. Following \eqref{eq0} we can bring the latter down to the level where it emanated from, which can be written schematically as
\begin{align} \label{eq1}
    \frac{v_{n+1,j}}{u_{n+1,j}} &= A \frac{v_{j,j}}{u_{j,j}} - B = \frac{A}{t_{2j}} \frac{1-V_{j-1}}{1-U_{j-1}} - B\,,
\end{align}
with $A \geq 1$ and $B \geq 0$. We then bring the former to the same level, which reads
\begin{align} \label{eq2}
    \frac{v_{n+1,i}}{u_{n+1,i}} = A \frac{v_{j,i}}{u_{j,i}} - B \,.
\end{align}
Since $i<j$, we have not reached the lowest level yet, so continuing this process yields
\begin{align*}
    \frac{v_{j,i}}{u_{j,i}} \geq \frac{1}{t_{2j}} \frac{v_{i,i}}{u_{i,i}} \geq \frac{1}{t_{2j}} \frac{1-V_{i-1}}{1-U_{i-1}} \geq \frac{1}{t_{2j}}\frac{1-V_{j-1}}{1-U_{j-1}}\,,
\end{align*}
and altogether
\begin{align*}
    \frac{v_{n+1,i}}{u_{n+1,i}} & \geq \frac{A}{t_{2j}}\frac{1-V_{j-1}}{1-U_{j-1}}  - B = \frac{v_{n+1,j}}{u_{n+1,j}} \,.
\end{align*}
Thus, we find
\begin{align*}
    \frac{u_{n+1,i}}{v_{n+1,i}} \leq \frac{u_{n+1,j}}{v_{n+1,j}} \,, \quad i<j \,.
\end{align*}
In particular, the equality sign occurs when $t_{2k}=1$ and $t_{2k+1}=0$ for all $k \in [i,j-1]$. Summarizing the above results, we can write
\begin{align*}
    \frac{u_{n+1,1}}{v_{n+1,1}} \leq \frac{u_{n+1,2}}{v_{n+1,2}} \leq \dots \leq \frac{u_{n+1,n}}{v_{n+1,n}} \leq \frac{u_{n+1,n+1}}{v_{n+1,n+1}} \leq \frac{1-U_{n+1}}{1-V_{n+1}} \,.
\end{align*}
Lastly, we derive a relation between the first $u$ and $v$ variable at each level. Consider
\begin{align} \label{fractions}
    v_{n+1,i}-u_{n+1,i}&=v_{n,i}-u_{n,i}(\frac{t_{2n+1}(1-V_n)}{1-t_{2n+1}U_n}+\frac{(1-t_{2n+1}t_{2n+2})}{1-t_{2n+1}U_n}) \geq\\
    & \geq v_{n,i}-u_{n,i}\frac{1-t_{2n+1}V_n}{1-t_{2n+1}U_n} \geq v_{n,i}-u_{n,i} \,.
\end{align}
Hence, if $v_{n,i} \leq u_{n,i}$, then $v_{n+1,i} \geq u_{n+1,i}$. From the initial values we know that $v_{1,1} \geq u_{1,1}$, which then extends through first terms to all orders, i.e., $u_{n+1,1} \leq v_{n+1,1}$. Together with \eqref{fractions} this determines the domain of integration for a branch of arbitrary length.
\paragraph{Trees.}
For the construction of trees we performed the coordinate transformation \eqref{newcoordinatestwobranches}. In the following discussion the statements for $r^L_{n,i}, s^L_{n,i}$ and $r^R_{m,i}, s^R_{m,i}$ are mostly the same. When both sets of variables obey a similar relation, we will mention only the former. In Appendix \ref{app:domain}, we have already shown that $r^L_{n,i} \leq 1$ and $s^L_{n,i} \leq 1$, so we can introduce new variables $r^L_n,s^L_n$ that satisfy
\begin{align*}
    \sum_{i=1}^{n} r^L_{n,i} + r^L_n &= 1 \,, &  \sum_{i=1}^{n} s^L_{n,i} + s^L_n &= 1 \,.
\end{align*}
From the analysis of a single branch we know that $\frac{v_{n,i}}{u_{n,i}} \geq \frac{v_{n,j}}{u_{n,j}}$ if $i < j$. Hence
\begin{align*}
    \frac{s^L_{n,i}}{r^L_{n,i}} &= \frac{1-U^L_n U^R_m}{1-V^R_m} \frac{v^L_{n,i}}{u^L_{n,i}} - U^R_m \frac{1-V^L_n}{1-V^R_m} \geq \frac{s^L_{n,j}}{r^L_{n,j}} \,, \quad \text{if } i<j \,, 
\end{align*}
with equality occurred for $\frac{v_{n,i}}{u_{n,i}} = \frac{v_{n,j}}{u_{n,j}}$.  Setting $v_0 \equiv r^L_n s^R_m$, $u_0 \equiv r^R_m s^L_n$, we find
\begin{align*}
    \frac{s^L_{n,n}}{r^L_{n,n}} &= \frac{1-U^L_n U^R_m}{1-V^R_m} \frac{1}{t_{2n}} \frac{1-V^L_{n-1}}{1-U^L_{n-1}} - U^R_m \frac{1-V^L_n}{1-V^R_m} \geq \frac{1-U^L_n U^R_m}{1-V^R_m} \frac{1-V^L_{n}}{1-U^L_{n}} - U^R_m \frac{1-V^L_n}{1-V^R_m} = \\
    &=\frac{(1-U^R_m)(1-V^L_n)}{(1-U^L_n)(1-V^R_m)} = \frac{1-\sum_{i=1}^m r^R_{m,i}-\sum_{i=1}^n s^L_{n,i}}{1-\sum_{i=1}^m s^R_{m,i}-\sum_{i=1}^n r^L_{n,i}} \,,
\end{align*}
with equality for $t_{2n}=1$, and
\begin{align*}
    \frac{s^R_{m,m}}{r^R_{m,m}} \geq \frac{1-\sum_{i=1}^m r^R_{m,i}-\sum_{i=1}^n s^L_{n,i}}{1-\sum_{i=1}^m s^R_{m,i}-\sum_{i=1}^n r^L_{n,i}} \,.
\end{align*}
Combining the above results yields
\begin{align}
    \frac{u_1}{v_1} \leq \frac{u_2}{v_2} \leq \dots \leq \frac{u_{m+n}}{v_{m+n}} \leq \frac{u_{m+n+1}}{v_{m+n+1}} \,.
\end{align}
where we used the variables defined in \eqref{rename}. Moreover, $u_1=r_{m,1}^R$, $v_1=s_{m,1}^R$, so we have
\begin{align*}
    \frac{s^R_{m,1}}{r^R_{m,1}} = \frac{1-U^L_n U^R_m}{1-V^L_n} \frac{v^R_{m,1}}{u^R_{m,1}}-\frac{U^L_{n}(1-V^R_m)}{1-V^L_n} \geq \frac{1-U^L_n(1+U^R_m-V^R_m)}{1-V^L_n} \geq \frac{1-U^L_n}{1-V^L_n} \geq 1 \,,
\end{align*}
where in the first inequality we used that $u^R_{m,1} \leq v^R_{m,1}$ and in the second we used $V^R_m \geq U^R_m$, which were both previously derived. This leads to the inequalities $0 \leq u_1 \leq v_1 \leq 1$. This collection of inequalities defines the configuration space for a tree. Notice that the configuration space of a general tree looks very similar to the configuration space of a `single-branch' tree. In fact, up to relabeling, the configuration space of a `two-branch' tree with the lengths of branches $n_1$ and $ n_2$ coincides with the configuration space of a single branch of length $n_1+n_2$. It follows that the domain of integration of trees can be related between different topologies by relabeling of variables.

\end{document}